\documentclass[12pt, letterpaper, final]{iopart}
\pdfoutput=1

\usepackage[pdftex,bookmarks,hidelinks]{hyperref}
\usepackage[utf8]{inputenc}
\usepackage{multicol}
\usepackage{multirow,bigdelim}
\RequirePackage{cite}
\usepackage[dvipsnames]{xcolor}
\usepackage{tikz}
\usepackage[compat=1.1.0]{tikz-feynman}
\usepackage{feynmf}
\usepackage{slashed}
\expandafter\let\csname equation*\endcsname\relax
\expandafter\let\csname endequation*\endcsname\relax
\usepackage{amsmath}
\usepackage{amssymb}
\usepackage{gensymb}
\usepackage{cleveref}
\usepackage{mathrsfs}
\usepackage{enumerate}
\usepackage{url}
\usepackage{cancel}
\usepackage[normalem]{ulem}
\usepackage{gensymb}
\usepackage{bm}
\usepackage{pdfpages}
\usepackage{lineno}
\usepackage{lscape}
\usepackage{pifont}
\usepackage{rotating}
\usepackage{siunitx}
\usepackage{xspace}
\usepackage{comment}
\usepackage{adjustbox}
\usepackage{footnote}
\usepackage{placeins}
\usepackage[margin=2.2cm]{geometry}

\RequirePackage{titlesec}
\RequirePackage{etoolbox}
\makeatletter
\patchcmd{\ttlh@hang}{\parindent\z@}{\parindent\z@\leavevmode}{}{}
\patchcmd{\ttlh@hang}{\noindent}{}{}{}
\makeatother

\titleclass{\subsubsubsection}{straight}[\subsection]

\newcounter{subsubsubsection}[subsubsection]
\renewcommand\thesubsubsubsection{\thesubsubsection.\arabic{subsubsubsection}}
\renewcommand\theparagraph{\thesubsubsubsection.\arabic{paragraph}} 

\titleformat{\subsubsubsection}
  {\normalfont\normalsize\itshape}{\thesubsubsubsection}{1em}{}
\titlespacing*{\subsubsubsection}
{0pt}{3.25ex plus 1ex minus .2ex}{1.5ex plus .2ex}

\makeatletter
\renewcommand\paragraph{\@startsection{paragraph}{5}{\z@}%
  {3.25ex \@plus1ex \@minus.2ex}%
  {-1em}%
  {\normalfont\normalsize\itshape}}
\renewcommand\subparagraph{\@startsection{subparagraph}{6}{\parindent}%
  {3.25ex \@plus1ex \@minus .2ex}%
  {-1em}%
  {\normalfont\normalsize\itshape}}
\def\toclevel@subsubsubsection{4}
\def\toclevel@paragraph{5}
\def\toclevel@paragraph{6}
\def\l@subsubsubsection{\@dottedtocline{3}{7em}{4em}}
\def\l@paragraph{\@dottedtocline{5}{10em}{5em}}
\def\l@subparagraph{\@dottedtocline{6}{14em}{6em}}
\makeatother

\graphicspath{ {/graphics/} }

\let\OLDthebibliography\thebibliography
\renewcommand\thebibliography[1]{
  \OLDthebibliography{#1}
  \setlength{\parskip}{0pt}
  \setlength{\itemsep}{0pt plus 0.3ex}
}

\newif\ifdp
\newif\ifsp

\def\expshort{WP NF02\xspace}
\def\explong{White Paper on Light Sterile Neutrinos\xspace}

\def\thedocsubtitle{Light Sterile Neutrinos}

\newcommand{\refsec}[2]{Volume~\csname volnumber#1\endcsname \xspace Section~#2}
\newcommand{\refch}[2]{Volume~\csname volnumber#1\endcsname \xspace Chapter~#2}
\newcommand{\refinch}[2]{#2 in Volume~\csname volnumber#1\endcsname \xspace}

\newcommand{\numu}{\ensuremath{\nu_\mu}\xspace}
\newcommand{\nue}{\ensuremath{\nu_e}\xspace}
\newcommand{\nutau}{\ensuremath{\nu_\tau}\xspace}

\newcommand{\anue}{\ensuremath{\bar\nu_e}\xspace}

\newcommand{\dm}[1]{\ensuremath{\Delta m^2_{#1}}\xspace} 

\newcommand{\sinst}[1]{\ensuremath{\sin^2\theta_{#1}}\xspace} 
\newcommand{\sinstt}[1]{\ensuremath{\sin^22\theta_{#1}}\xspace}  

\newcommand{\nova}{NO$\nu$A\xspace}

\newcommand{\Losc}{L_{\text{osc}}}
\newcommand{\Lcoh}{L_{\text{coh}}}
\def\argon40{$^{40}$Ar}  
\def\Ar39{$^{39}$Ar}
\def\Cl40{$^{40}$Cl}
\def\K40{$^{40}$K}
\def\B8{$^{8}$B}
\newcommand{\uFive}{$^{235}$U}
\newcommand{\uEight}{$^{238}$U}
\newcommand{\pNine}{$^{239}$Pu}
\newcommand{\pOne}{$^{241}$Pu}
\newcommand{\gsim}{{\;\raise0.3ex\hbox{$>$\kern-0.75em\raise-1.1ex\hbox{$\sim$}}\;}}
\newcommand{\beq}{\begin{equation}}
\newcommand{\eeq}{\end{equation}}
\newcommand{\bea}{\begin{eqnarray}}
\newcommand{\eea}{\end{eqnarray}}

\mathchardef\minus="002D

\begin{document}



\begin{titlepage}


# White Paper on Light Sterile Neutrino Searches and Related Phenomenology

M. A. Acero,[2, *] C. A. Argüelles,[23, *] M. Hostert,[58, 59, 70, *] D. Kalra,[13, *] G. Karagiorgi,[13, *] K. J. Kelly,[11, *] B. R. Littlejohn,[31, *] P. Machado,[19, *] W. Pettus,[33, *] M. Toups,[19, *] M. Ross-Lonergan,[13, *] A. Sousa,[12, *] P. T. Surukuchi,[106, *] Y. Y. Y. Wong,[66, *]

W. Abdallah,[108, †] A. M. Abdullahi,[19, 41, †] R. Akutsu,[90, †] L. Alvarez-Ruso,[29, †] D. S. M. Alves,[51, †] A. Aurisano,[12, †] A.B. Balantekin,[109, †] J. M. Berryman,[5, 38, †] T. Bertólez-Martínez,[4, †] J. Brunner,[110, †] M. Blennow,[48, 53, 68, †] S. Bolognesi,[79, †] M. Borusinski,[22, †] T.Y. Chen,[13, †] D. Cianci,[13, †] G. Collin,[1, †] J.M. Conrad,[62, †] B. Crow,[22, †] P. B. Denton,[8, †] M. Duvall,[22, †] E. Fernández-Martinez,[53, †] C. S. Fong,[97, †] N. Foppiani,[23, †] D. V. Forero,[56, †] M. Friend,[27, 45, †] A. García-Soto,[23, 29, †] C. Giganti,[40, †] C. Giunti,[36, †] R. Gandhi,[111, †] M. Ghosh,[26, †] J. Hardin,[62, †] K. M. Heeger,[106, †] M. Ishitsuka,[89, †] A. Izmaylov,[32, 37, †] B. J. P. Jones,[100, †] J. R. Jordan,[3, †] N. W. Kamp,[62, †] T. Katori,[46, †] S. B. Kim,[83, †] L. W. Koerner,[25, †] M. Lamoureux,[35, †] T. Lasserre,[79, †] K.G. Leach,[57, †] J. Learned,[22, †] Y. F. Li,[30, 112, †] J. M. Link,[102, †] W. C. Louis,[51, †] K. Mahn,[64, †] P. D. Meyers,[74, †] J. Maricic,[22, †] D. Markoff,[65, †] T. Maruyama,[27, †] S. Mertens,[63, 88, †] H. Minakata,[102, †] I. Mocioiu,[71, †] M. Mooney,[14, †] M.H. Moulai,[105, †] H. Nunokawa,[76, †] J. P. Ochoa-Ricoux,[42, †] Y. M. Oh,[28, †] T. Ohlsson,[68, 48, †] H. Päs,[15, †] D. Pershey,[17, †] R. G. H. Robertson,[104, †] S. Rosauro-Alcaraz,[39, †] C. Rott,[83, 99, †] S. Roy,[113, †] J. Salvado,[4, †] M. Scott,[32, †] S. H. Seo,[28, †] M. H. Shaevitz,[13, †] M. Smiley,[5, 52, †] J. Spitz,[3, †] J. Stachurska,[62, †] M. Tammaro,[114, †] T. Thakore,[12, †] C.A. Ternes,[36, †] A. Thompson,[61, †] S. Tseng,[86, †] B. Vogelaar,[102, †] T. Weiss,[106, †] R. A. Wendell,[49, 87, †] R.J. Wilson,[14, †] T. Wright,[102, †] Z. Xin,[30, 112, †] B. S. Yang,[81, †] J. Yoo,[81, †] J. Zennamo,[19, †] J. Zettlemoyer,[19, †] J. D. Zornoza,[29, †] J. Zupan,[12]

S. Ahmad,[72] E. Arrieta-Diaz,[54] V. S. Basto-Gonzalez,[69] N. S. Bowden,[50] B. C. Cañas,[69] D. Caratelli,[93] C. V. Chang,[84] C. Chen,[84] T. Classen,[50] M. Convery,[115] G. S. Davies,[60] S. R. Dennis,[9] Z. Djurcic,[116] R. Dorrill,[31] Y. Du,[117] J.J. Evans,[55] U. Fahrendholz,[88] J. A. Formaggio,[62] B. T. Foust,[106] H. Frandini Gatti,[19] D. Garcia-Gamez,[21] S. Gariazzo,[36] J. Gehrlein,[8] C. Grant,[7] R. A. Gomes,[20] A. B. Hansell,[82] F. Halzen,[105] S. Ho,[47] J. Hoefken Zink,[6, 34] R. S. Jones,[80] P. Kunkle,[7] J.-Y. Li,[18] S. C. Li,[102, 77] X. Luo,[93] Yu. Malyshkin,[37, 44] C.J. Martoff,[85] D. Massaro,[6, 95, 34] A. Mastbaum,[78] R. Mohanta,[26] H.P. Mumm,[67] M. Nebot-Guinot,[18] R. Neilson,[16] K. Ni,[94] J. Nieves,[29] G. D. Orebi Gann,[5, 52] V. Pandey,[96] S. Pascoli,[6, 34] G. Paz,[103] A. A. Petrov,[103, 98] X. Qian,[8] M. Rajaoalisoa,[12] S. H. Razafinime,[12] C. Roca,[50] G. Ron,[24] B. Roskovec,[73] E. Saul-Sala,[29] L. Saldaña,[106] D. W. Schmitz,[91] K. Scholberg,[17] B. Shakya,[118] P. L. Slocum,[106] E.L. Snider,[19] H. Th. J. Steiger,[43, 75, 88] A. F. Steklain,[101] M. R. Stock,[88] F. Sutanto,[50] V. Takhistov,[87] R. Tayloe,[33] Y.-D. Tsai,[92] Y.-T. Tsai,[115] D. Venegas-Vargas,[21] M. Wallbank,[12] E. Wang,[107] P. Weatherly,[16] S. Westerdale,[74] E. Worcester,[8] W. Wu,[19] G. Yang,[5] and B. Zamorano[21]

[1] Department of Physics, University of Adelaide, Adelaide, SA, Australia
[2] Universidad del Atlántico, Puerto Colombia, Atlántico, Colombia
[3] University of Michigan, Ann Arbor, MI, 48109, USA
[4] Departament de Física Quàntica i Astrofísica and Institut de Ciències del Cosmos, Universitat de Barcelona, Diagonal 647, E-08028 Barcelona, Spain
[5] Department of Physics, University of California, Berkeley, CA 94720, USA
[6] Dipartimento di Fisica e Astronomia, Università di Bologna, via Irnerio 46, 40126 Bologna, Italy
[7] Department of Physics, Boston University, Boston, MA 02215, USA
[8] Physics Department, Brookhaven National Laboratory, Upton, NY, USA
[9] Cavendish Laboratory, University of Cambridge, Cambridge, United Kingdom
[10] Department of Physics, Case Western Reserve University, Cleveland, OH 44106, USA
[11] CERN, Esplanade des Particules, 1211 Geneva 23, Switzerland
[12] Department of Physics, University of Cincinnati, Cincinnati, OH 45221, USA
[13] Columbia University, New York, NY, USA
[14] Colorado State University, Fort Collins, CO, USA
[15] TU Dortmund, Department of Physics, D-44221, Dortmund, Germany
[16] Department of Physics, Drexel University, Philadelphia, Pennsylvania 19104, USA
[17] Department of Physics, Duke University, Durham, NC 27708, USA
[18] University of Edinburgh, Edinburgh EH8 9YL, United Kingdom
[19] Fermi National Accelerator Laboratory, Batavia, IL, USA

[20]*Instituto de Física, Universidade Federal de Goiás, Goiânia, GO 74690-900, Brazil*
[21]*University of Granada & CAFPE, Granada, Spain*
[22]*University of Hawaii, 2505 Correa Rd, Honolulu, HI 96822, USA*
[23]*Department of Physics & Laboratory for Particle Physics and Cosmology, Harvard University, Cambridge, MA 02138, USA*
[24]*Hebrew University of Jerusalem, Israel*
[25]*Department of Physics, University of Houston, Houston, Texas 77204, USA*
[26]*School of Physics, University of Hyderabad, Hyderabad 500046, India*
[27]*High Energy Accelerator Research Organization (KEK), Tsukuba, Ibaraki 305-0801, Japan*
[28]*Center for Underground Physics, Institute for Basic Science (IBS), Daejeon 34126 Korea*
[29]*Instituto de Física Corpuscular (IFIC) and Departamento de Física Teórica, Consejo Superior de Investigaciones Científicas (CSIC) and Universidad de Valencia (UV), E-46980, Valencia, Spain*
[30]*Institute of High Energy Physics, Chinese Academy of Sciences, Beijing 100049, China*
[31]*Illinois Institute of Technology, Chicago, IL, USA*
[32]*Imperial College London, London, United Kingdom*
[33]*Indiana University, Bloomington, IN, USA*
[34]*INFN, Sezione di Bologna, viale Berti Pichat 6/2, 40127 Bologna, Italy*
[35]*INFN Sezione di Padova, 35131 Padova, Italy*
[36]*Istituto Nazionale di Fisica Nucleare (INFN), Sezione di Torino, Via P. Giuria 1, I–10125 Torino, Italy*
[37]*Institute for Nuclear Research of the Russian Academy of Sciences, Moscow, Russia*
[38]*Institute for Nuclear Theory, University of Washington, Seattle, WA 98195, USA*
[39]*Pôle Théorie, Laboratoire de Physique des 2 Infinis Irène Joliot Curie (UMR 9012) CNRS/IN2P3, 91400 Orsay, France*
[40]*Laboratoire de Physique Nucleaire et de Hautes Energies, IN2P3/CNRS, Sorbonne Universitè, Paris, France*
[41]*Institute for Particle Physics Phenomenology, Durham University, Durham DH1 3LE, UK*
[42]*Department of Physics and Astronomy, University of California at Irvine, Irvine, California 92697*
[43]*Johannes Gutenberg Universität zu Mainz, Mainz, Germany*
[44]*Joint Institute for Nuclear Research, Dubna, 141980, Russia*
[45]*J-PARC Center, Tokai, Japan*
[46]*King's College London, WC2R 2LS, London, UK*
[47]*Korea Institute for Advanced Study, Seoul 02455, Republic of Korea*
[48]*Department of Physics, School of Engineering Sciences, KTH Royal Institute of Technology, AlbaNova University Center, SE-106 91 Stockholm, Sweden*
[49]*Kyoto University, Department of Physics, Kyoto, Japan*
[50]*Lawrence Livermore National Laboratory, Livermore, CA USA*
[51]*Los Alamos National Laboratory, Los Alamos, NM, USA*
[52]*Lawrence Berkeley National Laboratory, Berkeley, CA 94720, USA*
[53]*University Autonoma Madrid, Department of Theoretical Physics, Madrid, Spain*
[54]*Universidad del Magdalena, Carrera 32 Nº 22 - 08 Santa Marta, Colombia*
[55]*University of Manchester, Manchester M13 9PL, United Kingdom*
[56]*Universidad de Medellín, Carrera 87 Nº 30 - 65 Medellín, Colombia*
[57]*Department of Physics, Colorado School of Mines, Golden, CO, USA*
[58]*School of Physics and Astronomy, University of Minnesota, Minneapolis, MN 55455, USA*
[59]*William I. Fine Theoretical Physics Institute, School of Physics and Astronomy, University of Minnesota, Minneapolis, MN 55455, USA*
[60]*University of Mississippi, University, MS 38677, USA*
[61]*Mitchell Institute for Fundamental Physics and Astronomy, Department of Physics and Astronomy, Texas A&M University, College Station, TX 77845, USA*
[62]*Massachusetts Institute of Technology, Cambridge, MA, USA*
[63]*Max-Planck-Institut für Physik, München, Germany*
[64]*Michigan State University, East Lansing, MI, USA*
[65]*North Carolina Central University, 1801 Fayetteville Street, Durham, NC 27707, USA*
[66]*School of Physics, The University of New South Wales, Sydney NSW 2052, Australia*
[67]*National Institute of Standards and Technology, Gaithersburg, MD 20899, USA*
[68]*The Oskar Klein Centre, AlbaNova University Center, SE-106 91 Stockholm, Sweden*
[69]*Universidad de Pamplona, km 1, vía salida a Bucaramanga, Campus Universitario, 543050 Pamplona, Colombia*
[70]*Perimeter Institute for Theoretical Physics, Waterloo, ON N2J 2W9, Canada*
[71]*Department of Physics, Pennsylvania State University, University Park, PA, 16802, USA*
[72]*HEPG, Pakistan Institute of Nuclear Science and Technology (PINSTECH), Islamabad, Pakistan*
[73]*Faculty of Mathematics and Physics, Charles University, Prague, V Holešovičkách 2, 18000 Prague, Czech Republic*
[74]*Princeton University, Princeton, NJ, USA*
[75]*Cluster of Excellence PRISMA+, Mainz, Germany*
[76]*Departamento de Física, Pontifícia Universidade Católica do Rio de Janeiro, C. P. 38097, 22451-900, Rio de Janeiro, Brazil*
[77]*Department of Physics and Astronomy, Purdue University, West Lafayette, IN 47907, USA*
[78]*Rutgers University, Piscataway, NJ, USA*
[79]*Université Paris-Saclay, Gif-sur-Yvette, France*

[80]Department of Physics and Astronomy, University of Sheffield, Sheffield, United Kingdom
[81]Department of Physics and Astronomy, Seoul National University, Seoul 08826, Korea
[82]Department of Physics, Susquehanna University, Selinsgrove, PA, USA
[83]Department of Physics, Sungkyunkwan University, Suwon 16419, Korea
[84]Department of Physics, National Taiwan Normal University, Taipei 116, Taiwan
[85]Temple University, Philadelphia, PA 19122, USA
[86]Department of Physics, Faculty of Science, The University of Tokyo, Bunkyo-ku, Tokyo 113-0033, Japan
[87]Kavli Institute for the Physics and Mathematics of the Universe (WPI), UTIAS
The University of Tokyo, Kashiwa, Chiba 277-8583, Japan
[88]Technische Universität München, Garching, Germany
[89]Tokyo University of Science, Department of Physics, Chiba, Japan
[90]TRIUMF, Vancouver, British Columbia, Canada
[91]Enrico Fermi Institute, University of Chicago, Chicago, IL, 60637, USA
[92]University of California, Irvine, CA, 92697, USA
[93]University of California Santa Barbara, Santa Barbara, CA, 93106, USA
[94]University of California San Diego, La Jolla, CA, 92093, USA
[95]Centre for Cosmology, Particle Physics and Phenomenology (CP3),
Université Catholique de Louvain, B-1348 Louvain-la-Neuve, Belgium
[96]Department of Physics, University of Florida, Gainesville, FL 32611, USA
[97]Centro de Ciências Naturais e Humanas, Universidade Federal do ABC, 09.210-170, Santo André, SP, Brazil
[98]Department of Physics and Astronomy, University of South Carolina, Columbia, South Carolina 29205, USA
[99]Department of physics, The University of Utah, 201 Presidents' Cir, Salt Lake City, UT 84112, USA
[100]Dept. of Physics, University of Texas at Arlington, 502 Yates
St., Science Hall Rm 108, Box 19059, Arlington, TX 76019, USA
[101]Department of Mathematics, Universidade Tecnológica Federal do Paraná, 80.230-901, Curitiba, PR, Brazil
[102]Center for Neutrino Physics, Department of Physics, Virginia Tech, Blacksburg, Virginia 24061, USA
[103]Department of Physics and Astronomy, Wayne State University, Detroit, Michigan 48201, USA
[104]University of Washington, Seattle, WA, USA
[105]Dept. of Physics and Wisconsin IceCube Particle Astrophysics
Center, University of Wisconsin-Madison, Madison, WI 53706, USA
[106]Wright Laboratory, Department of Physics, Yale University, New Haven, CT, USA
[107]School of Physics and Microelectronics, Zhengzhou University, Zhengzhou, Henan 450001, China
[108]Department of Mathematics, Faculty of Science, Cairo University, Giza 12613, Egypt
[109]Department of Physics, University of Wisconsin, Madison, WI 53706, USA
[110]Aix Marseille Univ, CNRS/IN2P3, CPPM, Marseille, France
[111]Harish-Chandra Research Institute (A CI of the Homi Bhabha
National Institute), Chhatnag Road, Jhunsi, Allahabad 211019, India
[112]School of Physical Sciences, University of Chinese Academy of Sciences, Beijing 100049, China
[113]Institute of Physics, Sachivalaya Marg, Sainik School Post, Bhubaneswar 751005, India
[114]Jozef Stefan Institute, Jamova 39, Ljubljana, Slovenia
[115]SLAC National Accelerator Laboratory
[116]Argonne National Laboratory, Lemont, IL, 60439, USA
[117]CAS Key Laboratory of Theoretical Physics, Institute of Theoretical
Physics, Chinese Academy of Sciences, Beijing 100190, P. R. China
[118]Deutsches Elektronen-Synchrotron DESY, Notkestrasse 85, 22607 Hamburg, Germany
(Dated: May 17, 2023)

---

\* Editor.
† Contributor.


\end{titlepage}

\renewcommand{\familydefault}{\sfdefault}
\renewcommand{\thepage}{\roman{page}}
\setcounter{page}{0}

\pagestyle{plain} 
\clearpage
\textsf{\tableofcontents}





\renewcommand{\thepage}{\arabic{page}}
\setcounter{page}{1}

\clearpage

\section*{Executive Summary}
\label{sec:summary}



Several decades of a rich and diverse program of experimental neutrino measurements have provided an increasingly clearer picture of the elusive neutrino sector, and uncovered physics not predicted by the Standard Model, such as the existence of nonzero neutrino masses implied by the surprising discovery of neutrino flavor mixing.
This foundational discovery represented a welcome resolution to decades-long experimental anomalies associated with solar and atmospheric neutrino measurements. 

Alongside this foundational discovery, {\it experimental neutrino anomalies} have been observed that still remain unresolved, and have served as primary drivers in the development of a vibrant short-baseline neutrino program, and in the launch of a multitude of complementary probes within a large variety of other experiments. Two of these anomalies arise from the apparent oscillatory appearance of electron (anti)neutrinos in relatively pure muon-(anti)neutrino beams originating from charged-pion decay-at-rest, specifically the {\it LSND Anomaly}, and from charged-pion decay-in-flight, the {\it MiniBooNE Low-Energy Excess}. Two other anomalies are associated with an overall normalization discrepancy of electron (anti)neutrinos expected both from conventional fission reactors, the {\it Reactor Neutrino Anomaly}, and in the radioactive decay of Gallium-71, the {\it Gallium Anomaly}. In these two latter cases, no oscillatory signature is observed, but the overall normalization deficit can be ascribed to rapid oscillations that are averaged out and appear as an overall deficit.

Historically, these anomalies were first interpreted as oscillations due to the existence of light sterile neutrinos that mix with the three Standard Model neutrinos. This interpretation requires an oscillation frequency $\Delta m^2\gtrsim 1$~eV$^2$, implying the addition of at least one neutrino to the three-flavor mixing paradigm. This new neutrino would have to be a Standard Model gauge singlet, thus it is referred to as {\it sterile}, as LEP measurements of the invisible decay width of the $Z$ boson show only three neutrinos couple to the $Z$ boson. However, this purely oscillatory interpretation is disfavored by several other direct and indirect experimental tests. Consequently, recent years have 
seen accelerating theoretical interest in more complex Beyond the Standard Model (BSM) flavor transformation and hidden-sector particle production as explanations for the anomalies. Experimental interest in testing a more diverse set of interpretations has also been growing, as well as motivation to probe deeper into potential conventional explanations. The discovery of new physics associated with these anomalies would be groundbreaking, and would have profound implications not only for particle physics but also for astrophysics and cosmology.

This white paper provides a comprehensive review of our present understanding of the experimental neutrino anomalies, charting the progress achieved over the last decade at the experimental and phenomenological level, and sets the stage for future programmatic prospects in addressing the anomalies. In a similar spirit to the ``Light Sterile Neutrinos: A White Paper'' document from a decade ago~\cite{Abazajian:2012ys}, this new white paper is purposed to serve as a guiding and motivational ``encyclopedic'' reference, with emphasis on needs and options for future exploration that may lead to the ultimate resolution of the anomalies. 

\subsection*{Developments Over The Past Decade}
Following the requirements identified in Ref.~\cite{Abazajian:2012ys} has led to a broader understanding of viable interpretations of the anomalies and strengthened experimental efforts -- and experimental capabilities -- in that direction.
Notably, the requirement to probe the anomalies with multiple and orthogonal approaches (accelerator-based short/long-baseline, reactor-based short-baseline, atmospheric neutrinos, and radioactive source) in the same spirit as employed for neutrino oscillations has been realized through recent, ongoing, or impending experimental programs: 

\begin{itemize}
    \item The development of new radioactive sources and detectors for improved tests of the Gallium Anomaly has been pursued and realized in the form of the BEST experiment.  
    \item The Reactor Antineutrino Anomaly and subsequent reactor-based activities and new results have placed a required emphasis on oscillation-testing short-baseline reactor experiments and on improved understanding of reactor neutrino fluxes.  
    \item The community has just begun a comprehensive multi-channel/multi-baseline accelerator-based short-baseline program to search for 3+$N$ oscillations while directly addressing the MiniBooNE anomaly both in regards to oscillatory and non-oscillatory solutions.  
    \item Recent searches for smoking-gun signatures of light sterile neutrinos with high-energy atmospheric neutrinos, such as the one performed by the IceCube Neutrino Observatory.
    \item A direct test of the LSND Anomaly using an improved decay-at-rest beam facility and experimental arrangement has just begun in the form of the JSNS$^2$ experiment.  
    \item Beyond direct anomaly tests, many alternate techniques/facilities, including direct neutrino mass measurements, long-baseline oscillation experiments, and atmospheric and astrophysical neutrino experiments, have been applied to the sterile neutrino explanation of the anomalies. 
\end{itemize}

\subsection*{Primary Focuses for the Next Decade}

As the question of light sterile neutrino oscillations is further explored over the next several years, the community's efforts should be directed toward disentangling the plethora of possibilities that have been identified over the past ten years as viable interpretations of the experimental anomalies in the neutrino sector.  
The goal of these collective efforts will be to validate and solidify our understanding of the neutrino sector.  
Regardless of what the ongoing and upcoming experiments observe --- be it a deviation from the three-neutrino picture or otherwise --- the community should be prepared to address how to put these anomalies to test or adequately distinguish between different interpretations.  
We see the main experimental, analysis, and theory-driven thrusts that will be essential to achieving this goal being: 

\begin{itemize}
\item{\textbf{Cover all anomaly sectors:} Given the fundamentally unresolved nature of all four canonical anomalies, it is imperative to support all pillars of a diverse experimental portfolio -- source, reactor, decay-at-rest, decay-in-flight, and other methods/sources -- to provide complementary probes of and increased precision for new physics explanations.
}
\item{\textbf{Pursue diverse signatures:} Given the diversity of possible experimental signatures associated with allowed anomaly interpretations, it is imperative that experiments make design and analysis choices that maximize sensitivity to as broad an array of these potential signals as possible.
}
\item{\textbf{Deepen theoretical engagement:} Priority in the theory community should be placed on the development of new physics models relevant to all four canonical short-baseline anomalies and the development of tools for enabling efficient tests of these models with existing and future experimental datasets.
}
\item{\textbf{Openly share data:} Fluid communication between the experimental and theory communities will be required, which implies that both experimental data releases and theoretical calculations should be publicly available.
In particular, as it is most likely that a combination of measurements will be needed to resolve the anomalies, global fits should be made public, as well as phenomenological fits and constraints to specific data sets.
}
\item{\textbf{Apply robust analysis techniques:} Appropriate statistical treatment is crucial to quantify the compatibility of data sets within the context of any given model, and in order to test the absolute viability of a given model. 
Accurate evaluation of allowed parameter space is also an important input to the design of future experiments.
}
\end{itemize}

The white paper is organized as follows.
Section~\ref{sec:introduction} provides an overall introduction and motivation for seeking resolution of the experimental neutrino anomalies, and Section~\ref{sec:expt_landscape} introduces each of the anomalies in detail, placing them within historical context. Section~\ref{sec:th_landscape} delves into the theoretical interpretation of the anomalies, detailing phenomenological consequences of various scenarios that have been or are being pursued. Section~\ref{sec:null} goes over the broader experimental landscape, discussing the impact of null results, as well as potential conventional explanations for the anomalies, while Section~\ref{sec:astro_cosmo} covers results from astrophysical and cosmological indirect probes. Section~\ref{sec:future} reviews the very diverse landscape of future experimental prospects that will be capable of addressing the anomalies. Finally, Section~\ref{sec:reqs} reiterates our vision for a path that will lead to the ultimate resolution of the anomalies, providing further discussion and elaboration of the focal points for the next decade listed in this executive summary. 
\cleardoublepage


\section{Introduction and Motivation}
\label{sec:introduction}


The Nobel prize-winning discovery of neutrino oscillation~\cite{Fukuda:1998mi,SNO:2001kpb,SNO:2002tuh} has led to a three-neutrino mixing picture that is now established as a minimal extension to the Standard Model (SM), and which is only empirically motivated. This picture prescribes an ``Extended SM'' (ESM), in which the neutrino sector includes three distinct neutrino mass states that are each an independent linear combination of the three neutrino weak eigenstates: $\nu_e$, $\nu_\mu$, and $\nu_\tau$~\cite{ParticleDataGroup:2020ssz}. This discovery stands as one of few indisputable pieces of evidence for new physics ``beyond the SM'' (BSM). 

Generating neutrino masses is qualitatively different from the mass generation for any other fermions in the SM.
The Higgs mechanism for neutrinos would require the existence of a right-handed neutrino, which would carry no SM gauge quantum number.
This in turn would allow for Majorana masses of these right-handed fields, opening up the possibility of a seesaw mechanism~\cite{Minkowski:1977sc, Gell-Mann:1979vob, Yanagida:1979as, Mohapatra:1979ia, Schechter:1980gr}.
Several other scenarios, involving different particle content, could also explain the origin of neutrino masses, such as type-II and type-III seesaw models~\cite{Magg:1980ut, Mohapatra:1980yp, Lazarides:1980nt, Ma:1998dx, Foot:1988aq, Ma:1998dn}.
In general, the mechanism of neutrino masses would require the addition of particle content to the SM that has never been observed.
The lack of experimental indication of the scale of this new physics makes the neutrino sector a promising portal to new physics.
Many neutrino mass models would predict observable deviations from the ESM and could lead to a rich phenomenology.
In particular, the existence of new states or gauge interactions associated with neutrinos could affect neutrino experiments in a variety of ways, for example as effects on oscillation phenomenology or new particles produced in neutrino beams or in neutrino detectors.

In particular, interest in this direction has been fanned by a series of anomalous experimental measurements, especially since the mid-1990's, which suggested the existence of new neutrino states. It is expected that these states should be ``sterile'', i.e.~non-weakly-interacting, in order to avoid experimental constraints from invisible $Z\rightarrow\nu\bar{\nu}$ decay measurements~\cite{ALEPH:2005ab}. 
There is now a series of indications of neutrino phenomena deviating from the three-neutrino (ESM) paradigm, many of which have the commonality of being observed primarily in association with electron neutrino observations, from either electron neutrino or muon neutrino sources, with either Cherenkov or scintillator detectors, and at relatively ``short baselines'' from the neutrino sources\footnote{We introduce and note that the term ``short baseline'' is used qualitatively, and more specifically it refers to neutrino propagation distances of $\mathcal{O}$(1~km) for measurements performed with neutrino energy of $\mathcal{O}$(1~GeV). More broadly, it refers to a ratio of neutrino propagation distance relative to neutrino energy of $\sim1$~km/GeV, corresponding to a neutrino oscillation frequency $\Delta m^2$ of $\sim1$~eV$^2$.}. These ``short-baseline experimental anomalies'', and the expansive and dedicated scientific program that has been launched over the past two decades to address them, is the focus of this paper.

One of the most widely examined theoretical frameworks considered for the interpretation of these anomalies is that of light ($\sim1$~eV) sterile neutrino oscillations ~\cite{Abazajian:2012ys}.  
This framework generally extends the three-neutrino paradigm of the ESM to accommodate (3+$N$) light neutrino masses and (3+$N$) neutrino flavor states, where $N$ refers to additional neutrino mass states with masses of order $\sim1$~eV. The latter are a linear combination of primarily $N$ sterile neutrino eigenstates but contain a small admixture of weak neutrino eigenstates so that they can participate in neutrino oscillations. This framework generally leads to observable neutrino oscillations with appearance oscillation amplitudes of order 1\% or less, and disappearance oscillation amplitudes of up to tens of percent. The relatively small oscillation amplitudes are constrained by unitarity considerations~\cite{Parke:2015goa}, and the known oscillation amplitudes from ``medium-'' and ``long-baseline'' neutrino oscillation measurements. 

While the light sterile neutrino oscillation framework can, in theory accommodate all short-baseline experimental anomalies to date, it fails to accommodate the lack of corresponding oscillations in other short-baseline, long-baseline, and atmospheric neutrino measurements. 
One particular experimental anomaly, contributed by the MiniBooNE experiment, exacerbates this issue. The need to interpret compelling experimental results, on the other hand, has given rise to an extensive experimental neutrino program, as well as a substantial body of related phenomenological work, including many viable interpretations, from modifications of three-flavor neutrino mixing to potential couplings to hidden sectors, which we review here. 

A consistent picture of short-baseline experimental anomalies has not yet formed, as will be discussed in this paper. On the other hand, new experiments launched over the past decade or about to be launched, with the goal of independently investigating either specific experimental anomalies or specific theoretical interpretations, promise to deliver new and invaluable information that will either identify the underlying source(s) of these anomalies or guide future scientific endeavors to better understand them. The paper further discusses theoretical developments over the past decade, as well as future research programs with the ability to further elucidate this picture, with attention to synergy and complementarity of both planned and proposed programs.  


Sterile neutrino states, if they exist, would reveal a new, unexpected form of a fundamental particle and possibly new types of interactions in nature. This possibility, or other BSM physics that may be the source(s) of the current experimental neutrino anomalies, has and will continue to compel particle physicists toward further experimentation in this area in the foreseeable future. The Discovery of new physics associated with these signals would be groundbreaking and would have profound implications not only for particle physics but also for astrophysics and cosmology. Additionally, in several scenarios, new physics associated with these anomalies can have a significant impact on measurements of three-neutrino oscillation parameters planned with ongoing and future long-baseline experiments, as well as on absolute measurements of neutrino mass, further necessitating their resolution. Similarly, new physics associated with these anomalies could connect to searches for neutrinoless double $\beta$ decay, and direct or indirect probes for dark matter or other dark sector particle searches. Alternatively, a clear null result, or an SM explanation for the current experimental anomalies, would bring a welcome resolution to a longstanding puzzle and greatly clarify the current picture in neutrino physics.  



\section{Experimental Anomalies}
\label{sec:expt_landscape}




There are four long-standing anomalies in the neutrino sector that have served as primary drivers in the development of a vibrant short-baseline neutrino program over the last decade.  
Two come from the apparent oscillatory appearance of electron (anti)neutrinos in relatively pure muon-(anti)neutrino beams originating from charged-pion decay-at-rest, Sec.~\ref{sec:anomaly_dat}, and charged-pion decay-in-flight, Sec.~\ref{sec:anomaly_dif}. Two more anomalies are associated with an overall normalization discrepancy of electron (anti)neutrinos expected both from conventional fission reactors, Sec.~\ref{sec:anomaly_reactor}, and in the radioactive decay of Gallium-71, Sec.~\ref{sec:anomaly_source}. In these two cases, no oscillatory signature is observed, but the overall normalization deficit can be ascribed to rapid oscillations at a high $\Delta m^2$ that are averaged out and appear as an overall deficit. This section will describe all four of these anomalies in detail, presenting both their experimental arrangements as well as the experimental (anomalous) results.  

Historically, the results have been discussed primarily in the context of a 3+1 scenario, with a single sterile neutrino. As such, the results in this section are presented in this manner; however, we emphasize that the current theoretical landscape strives to explore a much broader set of possible interpretations of the anomalies, as we describe in detail in Sec.~\ref{sec:th_landscape}. Those include more exotic flavor conversions, Sec.~\ref{sec:theory:flavorconv}, dark sector particles produced in neutrino scattering or in the neutrino source/beam itself, Sec.~\ref{sec:th_landscape:darksectors}, as well as more conventional explanations due to background mismodeling or underestimation, Sec.~\ref{sec:th_landscape:conventional}.  


\subsection{Pion Decay-at-Rest Accelerator Experiments}
\label{sec:anomaly_dat}
Pion decay-at-rest accelerator experiments provide a well-understood muon antineutrino flux of a mean energy of $\sim30$~MeV, and negligible electron antineutrino flux contamination. As such, detectors placed at relatively short baselines ($\sim30$~m) with positron identification capability offer sensitivity to $\overline{\nu}_\mu\rightarrow\overline{\nu}_e$ oscillations. Past pion decay-at-rest experiments include the Los Alamos Neutrino Detector (LSND)~\cite{Aguilar:2001ty} and the KArlsruhe Rutherford Medium Energy Neutrino (KARMEN)~\cite{KARMEN:2002zcm} experiments. Among the two, LSND observed evidence for $\overline{\nu}_\mu\rightarrow\overline{\nu}_e$ oscillations that could be parametrized by two-neutrino oscillations with a $\Delta m^2$ of $\sim1$~eV$^2$ and an oscillation amplitude $\sin^2 2\theta_{\mu e}$ of less than 1\%. Although less sensitive, KARMEN observed no evidence of such oscillations and therefore has been historically referred to as a ``null'' experiment in terms of this framework~\cite{Abazajian:2012ys}. The LSND anomalous result, which motivated a number of follow-up experimental searches for short-baseline neutrino oscillations over the past nearly three decades, is described below.

The LSND detector~\cite{LSND:1996jxj} at Los Alamos National Lab consisted of a cylindrical tank, 8.3~m long with a 5.7~m in diameter, located 29.8~m from the neutrino source. LSND was designed to search for oscillations $\overline{\nu}_\mu \rightarrow \overline{\nu}_e$. Neutrinos were produced from the decay chains of charged pions to muons decaying at rest, with the charged pions produced using 798~MeV protons on a target at the Los Alamos Neutron Science Center (LANSCE). Muon antineutrinos were produced by the sequence of $\pi^+ \rightarrow \mu^+ + \nu_\mu$ and $\mu^+ \rightarrow e^+ + \nu_e + \overline{\nu}_\mu $. The related decay of $\pi^-$ that would produce $\overline{\nu}_e$ is highly suppressed through pion capture on heavy nuclei in the vicinity of the beam target.  As a result, the intrinsic $\overline{\nu}_e$ contamination was expected to be $7.8 \times 10^{-4}$ smaller than the $\overline{\nu}_\mu$ flux. 

\begin{figure}[th]
    \centering
    \includegraphics[width=0.49\textwidth]{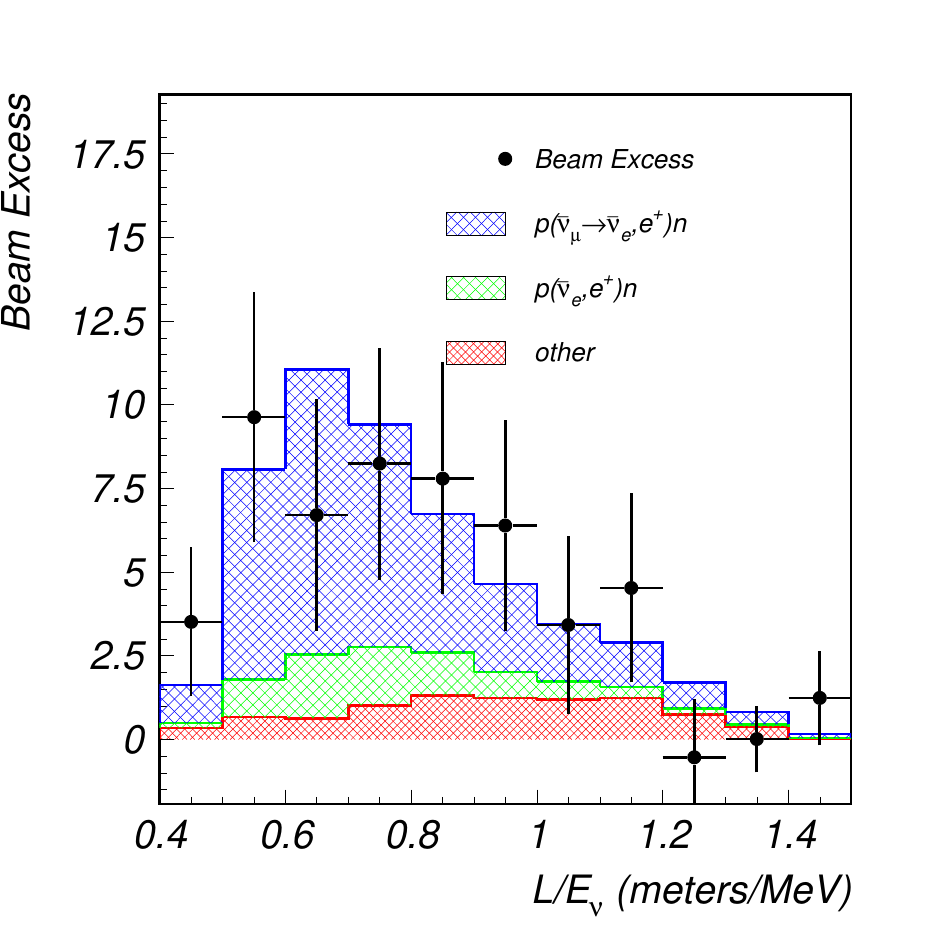}
    \includegraphics[width=0.49\textwidth]{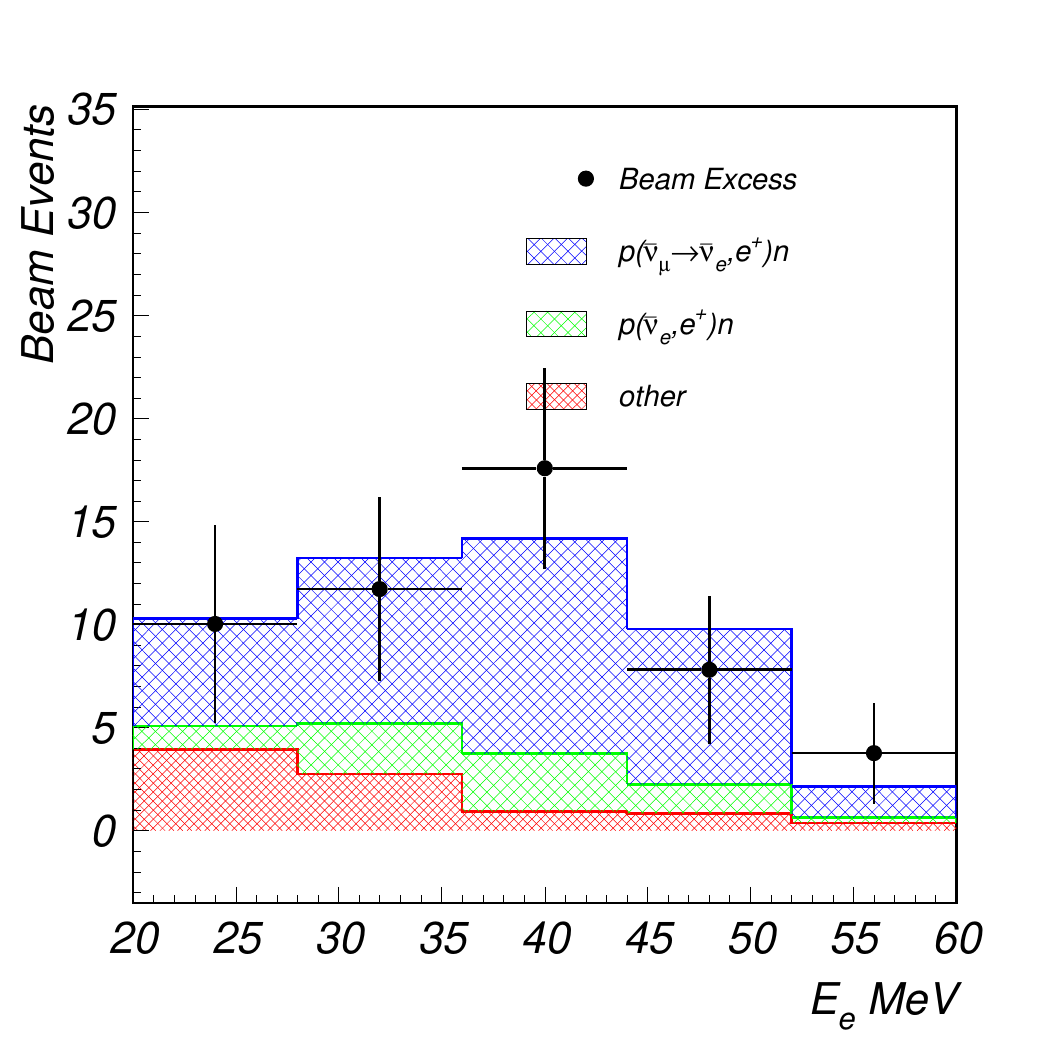}
    \caption{\label{fig:lsnd_1} The LSND anomalous events as a function of both $L/E_\nu$ (left) and observable electron energy (right), for the subset of total selected events with $R_\gamma > 10$ and $20 < E_e < 60$~MeV. Note the blue shaded region is for a best fit two-neutrino oscillation fit of $\sin^2 2\theta = 0.003$ and $\Delta m^2= 1.2\text{eV}^2$. Figure from Ref.~\cite{Aguilar:2001ty}.}
\end{figure}

The signal selection proceeded via the identification of a positron from inverse beta decay, $\overline{\nu}_e + p \rightarrow e^+ + n$, followed by detection of a 2.2~MeV photon from subsequent neutron capture that is correlated with the positron both in position and time. The interactions of $\nu_e$ inherent in the beam via $\nu_e + C_{12} \rightarrow N_{12} + e^-$ were not a major contributing background, as there was no correlated neutron capture accompanying these events, except for accidental coincidences. The target consisted of 167~metric tons of mineral oil, which was lightly doped with scintillator allowing for both the detection of Cherenkov light and isotropic scintillation light. This light was detected with 1220 8'' photomultiplier tubes (PMTs) spaced uniformly around the inner surface of the tank. As the Cherenkov ring could be detected, this allowed for the determination of both energy and angle of the outgoing positron.

Cosmic rays, although abundant, were not a major source of backgrounds for LSND, being removed by timing cuts and usage of an active and optically isolated veto shield surrounding the detector. True correlated 2.2~MeV photons were separated from coincident photons from radioactivity using a likelihood ratio $R_\gamma$ variable cut; this was defined to be the likelihood that the photon was correlated divided by the likelihood that the photon was accidental. 

\begin{figure}[th]
    \centering
    \includegraphics[width=0.6\textwidth]{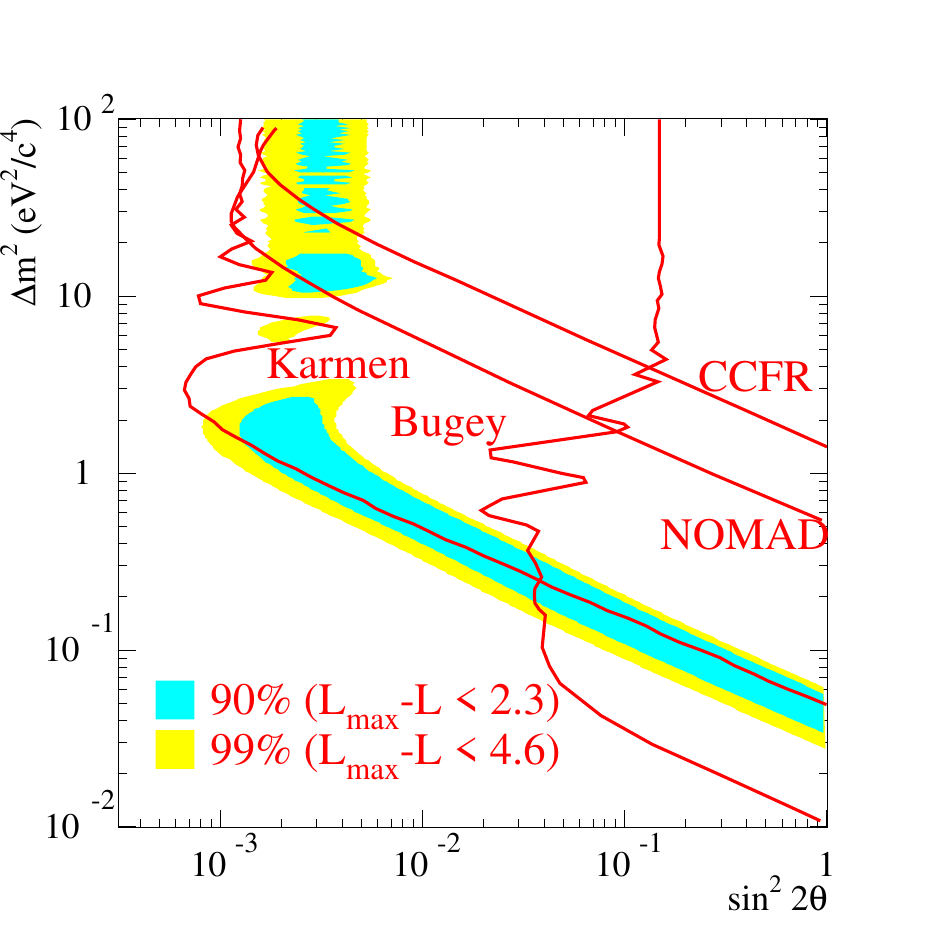}
    \caption{\label{fig:lsnd_fit} The results of a two-neutrino oscillation fit performed by the LSND collaboration for all data with reconstructed electron energy $20 < E_e < 200$~MeV, showing the resulting 90 and 99\% confidence level allowed regions for $\sin^22\theta_{\mu e}$ and $\Delta m^2$. Shown also are the 90\% CL limits from other contemporary experiments.  Figure from Ref.~\cite{Aguilar:2001ty}. }
\end{figure}

A series of LSND measurements were published, all in support of an excess of events observed over that expected from beam-off and beam-on neutrino background. The final results published in 2001~\cite{Aguilar:2001ty} concluded that an excess of events was observed, consistent with two-neutrino oscillations, corresponding to a background-subtracted excess of $87.9 \pm 22.4 \text{ (stat)} \pm 6.0 \text{ (sys)}$ events. Distributions of the observed excess are reproduced in Fig.~\ref{fig:lsnd_1} as a function of both the observed positron energy and the reconstructed $L/E_\nu$, for the subset of total selected events with $R_\gamma > 10$ (described as a clean sample of oscillation candidate events); this selection corresponds to an excess of $49.1 \pm 9.4$ events. 

The 3+1 sterile neutrino fit, which tests an effective two-flavor $\nu_\mu\rightarrow\nu_e$ appearance probability hypothesis under the short-baseline (SBL) approximation $\Delta m^2_{21}\approx\Delta m^2_{31}\equiv 0$,  was not performed simply in reconstructed neutrino energy, but as a likelihood in four dimensional $E_e$ (electron energy), $R_\gamma$ (coincidence variable), $z $ (electron distance along tank axis) and $\cos\theta$ (electron angle w.r.t neutrino beam) space. The best-fit point was found to be at an oscillation amplitude of $\sin^2 2\theta_{\mu e} = 0.003$ with a mass splitting of $\Delta m^2 = 1.2~\text{eV}^2$. The resulting allowed regions are shown in Fig.~\ref{fig:lsnd_fit} alongside then-contemporary experiments that did not see a positive signal.

\subsection{Pion Decay-in-Flight Accelerator Experiments}
\label{sec:anomaly_dif}
The LSND evidence for two-neutrino oscillation beginning in the late 1990's prompted the need for an independent follow-up experiment to test the result. Such test needed to rely on different systematics and methodology, while preserving sensitivity to the same $\Delta m^2$ and $\sin^2 2\theta_{\mu e}$. A $\pi^+$ decay-in-flight accelerator beam can produce a muon-neutrino-dominated flux with higher mean energy, providing an opportunity for an independent test at a longer baseline, and through different detection methods. This independent test was realized with the Booster Neutrino Beamline (BNB) at the U.S. Fermi National Accelerator Laboratory (FNAL, or Fermilab)~\cite{AguilarArevalo:2008yp}, providing a $\sim99.5\%$-pure muon neutrino beam with a mean neutrino energy of $\sim600$~MeV, sampled by the MiniBooNE Cherenkov detector~\cite{AguilarArevalo:2008qa}. 


Both MiniBooNE and BNB at Fermilab were designed in such a manner so as to have comparable $L/E$ to that of LSND ($L/E \approx 0.4 - 1.0$~m/MeV) but a longer baseline (540~m relative to LSND's 30~m) and higher energy (peak energy $\sim 700$~MeV relative to LSND's $\sim50$~MeV). Although initially envisaged as being a two-detector experimental setup, with a near detector at $L\approx 0$~m and a far detector at $L/E \approx 0.4 - 1.0$~m/MeV, the final MiniBooNE experiment consisted of a single detector, which was a spherical tank, 12.2~m in diameter, filled with 818 tons of mineral oil. The interior surface of the tank, including an outer veto spherical shell region, was lined with 1520  8'' PMTs, including the recycled usage of all 1220 PMTs from the LSND experiment. Cherenkov and scintillation photons emitted by particles produced in neutrino interactions were used to differentiate electrons vs.~muons produced in $\nu_e$ vs.~$\nu_\mu$ interactions, respectively. 

The primary reconstruction method in MiniBooNE uses the Cherenkov rings detected on the inside surface of the detector to differentiate between electrons, muons and charged pions, and neutral pions. Protons that fall below the Cherenkov threshold in mineral oil, $\sim350$~MeV kinetic energy, cannot be observed by their Cherenkov rings. Prompt scintillation light, however, can be used to estimate the energy of particles below the threshold. One crucial point to understand MiniBooNE's backgrounds is the fact that a single lone photon (which subsequently pair-produces a collimated $e^+e^-$ pair) is indistinguishable from a single electron in terms of their Cherenkov ring reconstruction. The separation of neutral-current (NC) $\pi^0 \rightarrow \gamma\gamma$ events thus relies entirely on reconstructing two separate Cherenkov rings. As such, the main backgrounds to searching for $\nu_e$ from $\nu_\mu \rightarrow \nu_e$ oscillations at MiniBooNE were:
\begin{itemize}
    \item Intrinsic $\nu_e$ in the BNB. Although an extremely pure $\nu_\mu$ beam, the $\mathcal{O}(0.5\%)$ $\nu_e$ and $\overline{\nu}_e$ in the beam provide an irreducible background. These are constrained by the high-statistics $\nu_\mu$ sample due to their common origin in meson decay chains. 
    \item NC $\pi^0$ events. In the scenario where one of the daughter photons from a NC $\pi^0$ decay is missed the event is indistinguishable from a single electron. This can occur due to overlapping Cherenkov cones, one photon exiting the detector before pair converting, or extremely low energy secondary photons. The NC $\pi^0$'s were constrained by a high-statistics in-situ measurement. 
    \item NC $\Delta \rightarrow N \gamma$. Radiative decay of the $\Delta$ baryon is a predicted SM process that produces a single photon, mimicking single-electron production in MiniBooNE.  
    \item ``Dirt'' events. The so-called ``Dirt'' events correspond to neutrino-induced events in which the scattering takes place in the material surrounding the detector, but some particles scatter inside the detector and are reconstructed. The majority of these are photons scattering in from $\pi^0$ decays.   
    \end{itemize}
Although the primary observable corresponds to the reconstructed outgoing electron itself, the results are often presented and interpreted in terms of quasi-elastic reconstructed neutrino energy, defined as 
\[
E^{QE}_{\nu} = \frac{1}{2} \left(\frac{2(M_n - \mathcal{B})E_e - ((M_n-B)^2 +M_e^2-M_p^2)}{(M_n-\mathcal{B})-E_e+\sqrt{E_e^2-M_e^2}\cos\theta_e}\right),
\]
where $M_n,M_p$ and $M_e$ are the masses of neutron, proton and electron respectively, $E_e$ is the reconstructed energy of the electron, $\cos\theta_e$ is the angle the reconstructed electron makes relative to the neutrino beam, and $\mathcal{B}$ is the binding energy of the target nucleus.

The first results from MiniBooNE, published in 2007, used approximately a third of the total data set collected by MiniBooNE, corresponding to a BNB proton beam delivery of $5.58 \times 10^{20}$ protons-on-target (POT). The result reported no evidence for oscillations within a two-neutrino $\nu_\mu\rightarrow\nu_e$ appearance paradigm~\cite{AguilarArevalo:2007it}, thus placing a 90\% CL limit covering the majority of the allowed LSND ($\sin^22\theta_{\mu e}$,$\Delta m^2$) parameter space. Crucially, this oscillation result was performed only for the region of reconstructed neutrino energy of $E^{QE}_\nu > 475$~MeV (assuming quasi-elastic scattering)\footnote{This restriction was decided upon as part of the data unblinding process followed by the MiniBooNE collaboration, supported by the findings that spectral information in this background-dominated region did not contribute significantly to two-neutrino oscillation sensitivity, and furthermore a data to Monte Carlo prediction discrepancy was observed with both the best-fit two-neutrino oscillation hypothesis and the SM prediction.}. While this first result contained no significant excess in the $E^{QE}_\nu > 475$~MeV region, below this energy, an excess of events was observed. This excess, further examined by the MiniBooNE collaboration in a subsequent analysis with higher statistics, is often referred to as the MiniBooNE ``Low-Energy Excess'' (or LEE), and consisted of $ 128.8\pm20.4\pm38.3$ excess events above predicted backgrounds, corresponding to $3.0\sigma$, as reported in~\cite{MiniBooNE:2008yuf}. 

Unlike LSND, MiniBooNE was designed with the ability to switch from a neutrino- to an antineutrino-dominated beam, by switching the charged-pion focusing magnetic field polarity, preferentially focusing $\pi^-$ mesons produced in proton-Be interactions toward the detector, resulting in a $\overline{\nu}_{\mu}$-dominated neutrino flux. Although the intrinsic $\nu_e$ and $\overline{\nu}_e$ contamination remains very small in antineutrino mode (0.6\%), the wrong sign contamination is not negligible (with 83.73\% $\overline{\nu}_\mu$ and 15.71\% $\nu_\mu$ components).  As such, MiniBooNE repeated its search for two-neutrino oscillations in antineutrino running mode in 2010, using data corresponding to $5.66 \times 10^{20}$ POT. The antineutrino search was motivated by findings supporting large observable CP violation in short-baseline oscillations involving two additional, mostly sterile, neutrino mass states with masses of order $\sim1$~eV~\cite{Karagiorgi:2006jf}, as well as CPT- or Lorenz-violating models suggested as alternative interpretations of LSND at the time~\cite{Hollenberg:2009tr,Kostelecky:2003cr,Katori:2006mz}. Given that LSND's result was obtained with antineutrinos, an independent antineutrino search---albeit less sensitive due to reduced statistics expected from a factor-of-two suppression in flux---was motivated by the need to provide an independent test of LSND regardless of CP or other symmetry violation assumption, and as a further probe of the LEE anomaly. The results from MiniBooNE's first $\overline{\nu}_\mu\rightarrow\overline{\nu}_e$ search~\cite{MiniBooNE:2010idf} followed in 2010, and showed an excess extending both at low energy and in the oscillation signal region of $ 475 < E^{QE}_\nu < 300 $~MeV. The results were found to be consistent with two-neutrino $\overline{\nu_\mu} \rightarrow \overline{\nu_e}$ oscillations with a $\chi^2$ probability of 8.7\% compared to 0.5\% for background only~\cite{MiniBooNE:2010idf}, with a best fit at $\Delta m^2 = 0.064 \text{eV}^2$ and $\sin^2 2\theta = 0.96$. When the fit was expanded to the whole energy range, $E^{QE}_\nu > 200 $~MeV, the best fit was found to be at $\Delta m^2 = 4.42 \text{eV}^2$ and $\sin^2 2\theta = 0.0066$, which although the best fit itself lies outside LSND's 99\% allowed contour, there was still significant overlap in the low $\Delta m^2$ allowed regions at the 90\% CL. 

\begin{figure}[!ht]
    \centering
    \includegraphics[width=0.4\textwidth]{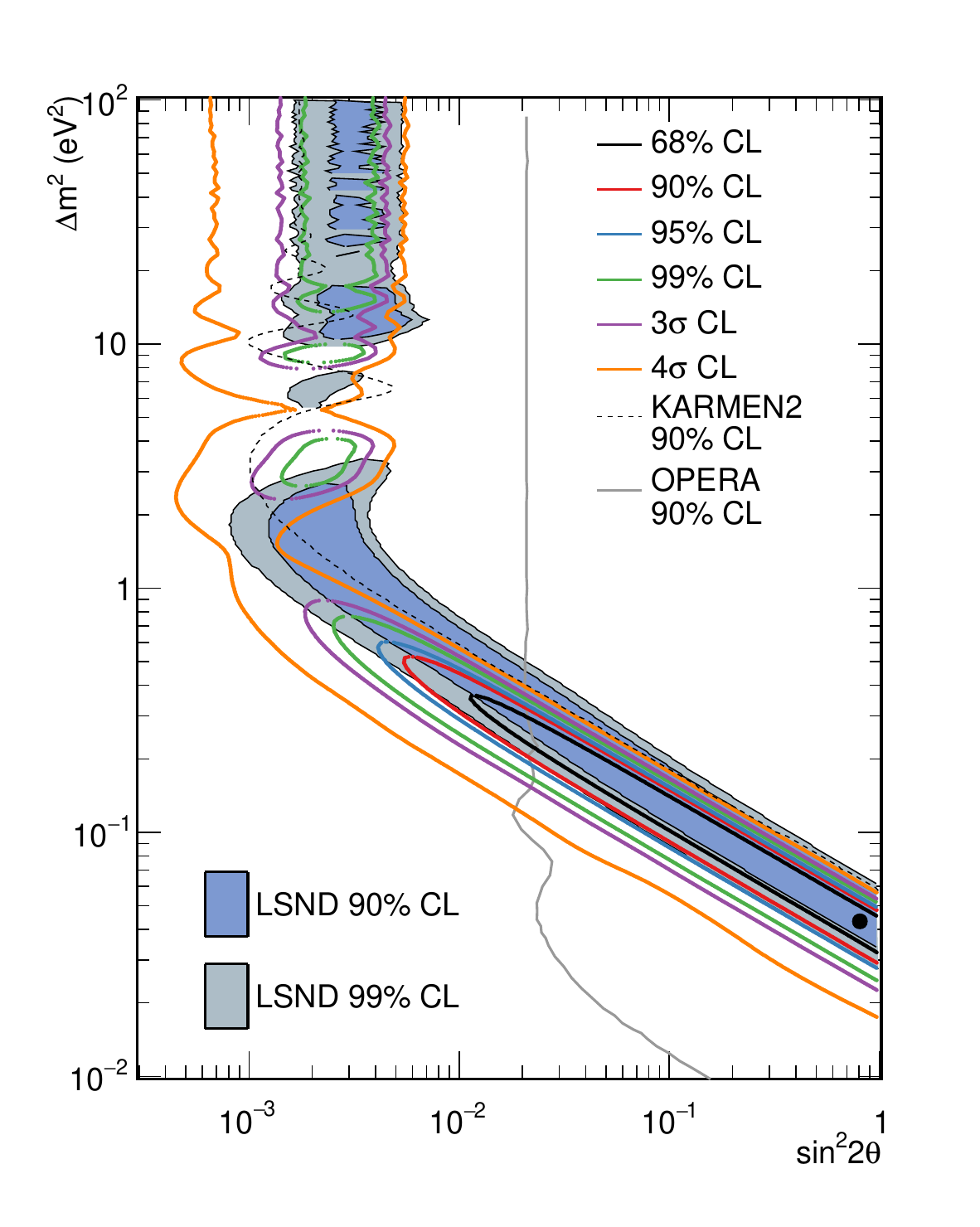}
    \caption{\label{fig:miniboone_allowed} 
    The final MiniBooNE allowed regions for the full fit of all neutrino and antineutrino running mode data. Figure from Ref.~\cite{MiniBooNE:2020pnu}. }
\end{figure}

\begin{figure}[ht]
    \centering
    \includegraphics[clip, trim=0.2cm 0cm 1cm 0cm, width=0.49\textwidth]{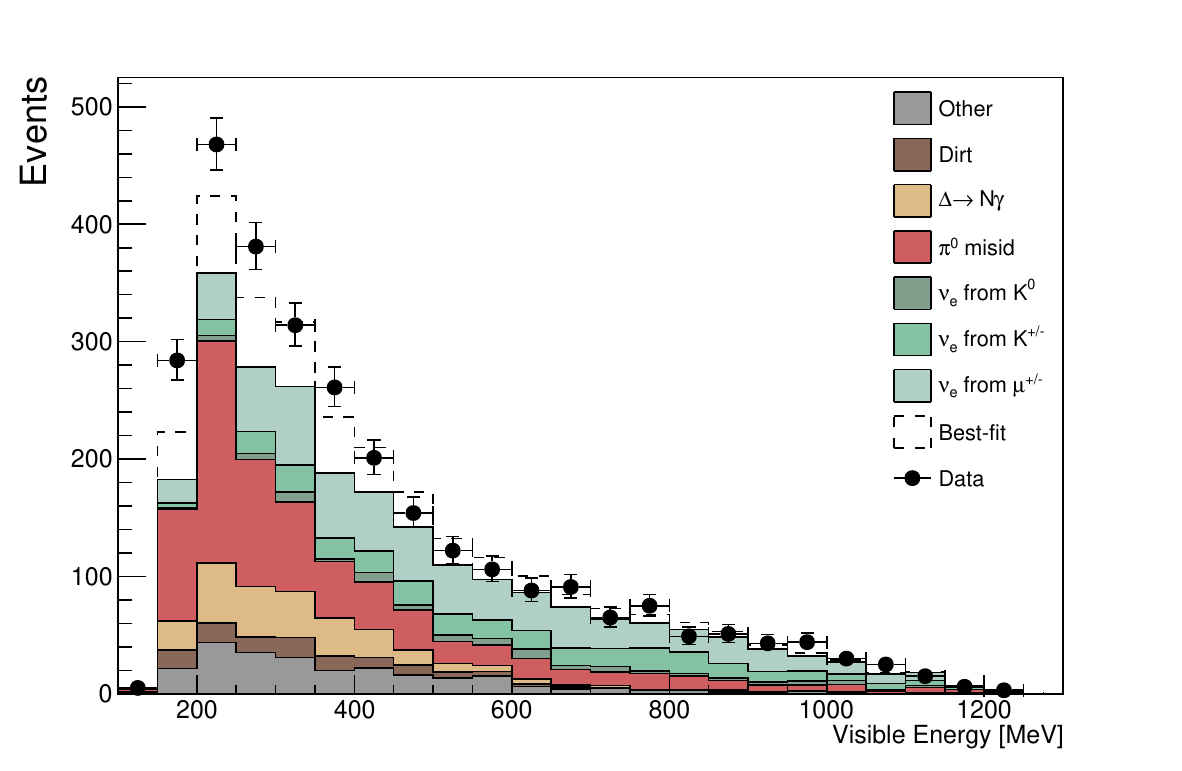}
    \includegraphics[clip, trim=0.2cm 0cm 1cm 0cm, width=0.49\textwidth]{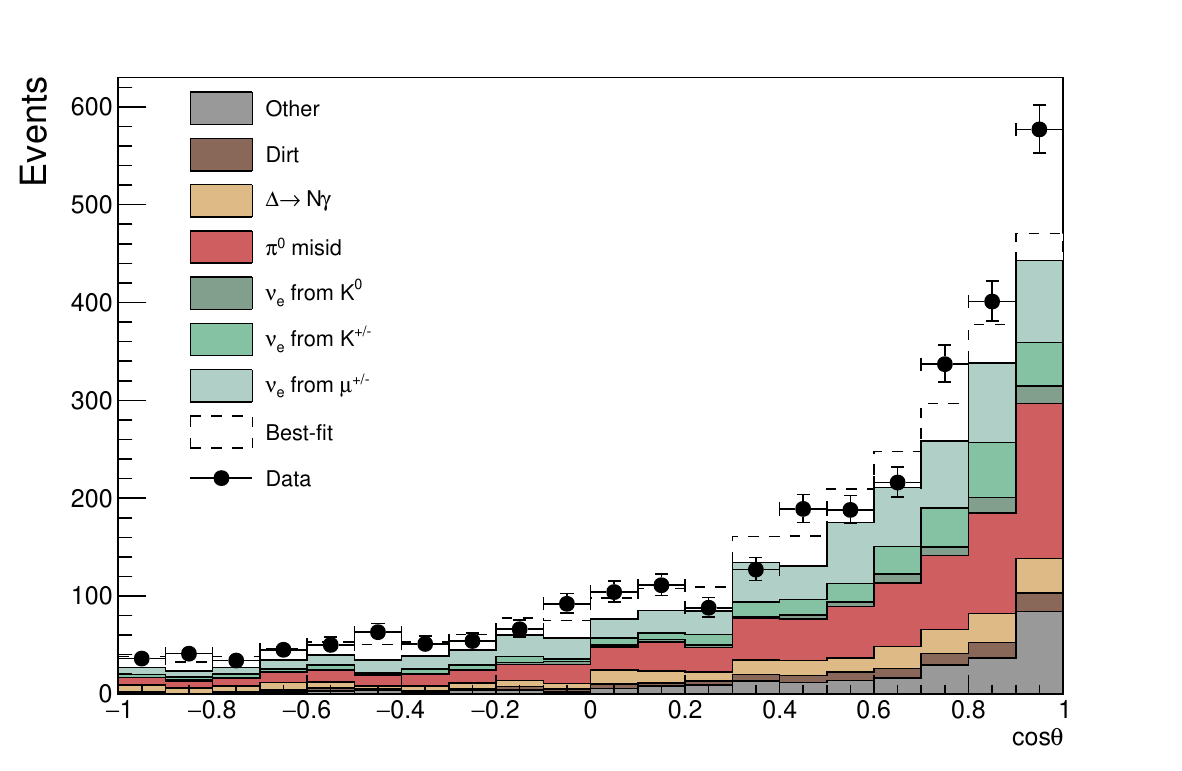}
    \includegraphics[clip, trim=0.1cm 0cm 0.1cm 9.6cm,width=0.8\textwidth]{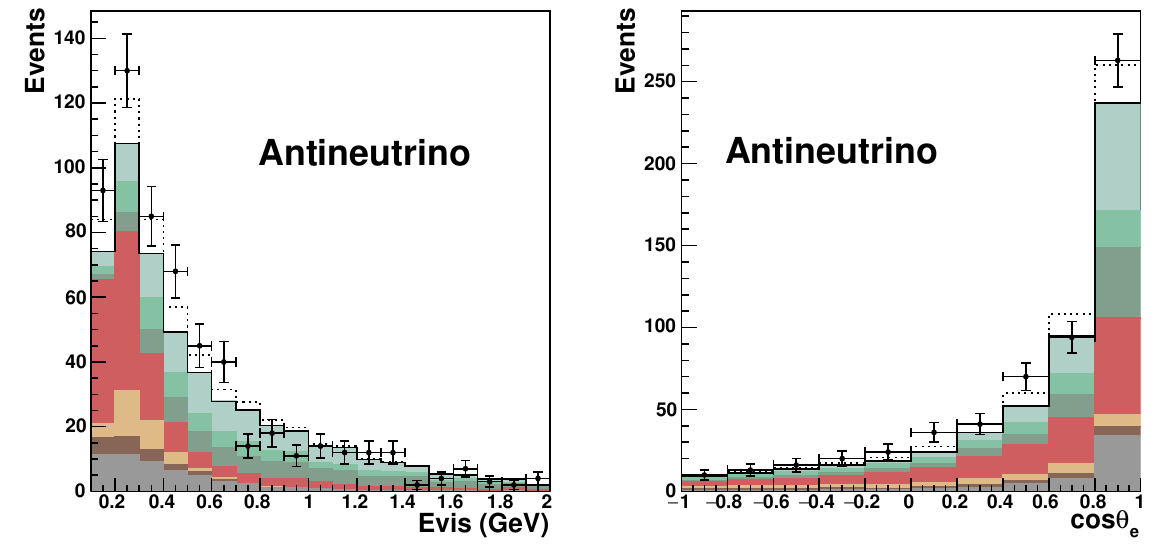}
    \caption{\label{fig:miniboone_1} 
    The final MiniBooNE results corresponding to $18.75 \times 10^{20}$ POT in neutrino mode (top figures) and $11.27 \times 10^{20}$ POT in antineutrino mode (bottom figures) for both the reconstructed visible energy (left) and the reconstructed angle that the Cherenkov cone makes with respect to the neutrino beam (right). Note that as the top two figures corresponding to neutrino mode are from Ref.~\cite{MiniBooNE:2020pnu}, and the bottom two for antineutrino running more are from Ref.~\cite{Aguilar-Arevalo:2018gpe}, the best-fit line does not correspond to the exact same point in sterile parameter space.}
\end{figure}

Since those first results, MiniBooNE ran for approximately ten more years, collecting BNB data corresponding to a total of $18.75 \times 10^{20}$ POT in neutrino running mode, and $11.27 \times 10^{20}$ POT in antineutrino running mode. Major updates were published in 2013~\cite{MiniBooNE:2013uba}, 2018~\cite{Aguilar-Arevalo:2018gpe} and 2020~\cite{MiniBooNE:2020pnu}, and for the remainder of this summary, we will concentrate on the final 2020 results, unless noted otherwise. MiniBooNE's final results are reproduced in Fig.~\ref{fig:miniboone_1}, showing an excess of data over background prediction in both neutrino and antineutrino data sets, as a function of the reconstructed electron candidate energy and reconstructed electron angle with respect to the beam. The excess is predominately evident below 600~MeV and has an overall significance of 4.8$\sigma$ (combining neutrino and antineutrino running mode data sets). This significance is almost entirely systematics-limited in nature, and corresponds to $560.6\pm119.6$ and $77.4\pm28.5$ excess events in neutrino and antineutrino running modes, respectively. The final best-fit parameters for the full neutrino and antineutrino data sets, for a fit over the entire energy range $200 < E^{QE}_\nu
< 3000$~MeV were found to be at
\[
    \Delta m^2 = 0.043\,\,\text{eV}^2, \text{ and  } \sin^2 2\theta =0.807,
\]
with the best-fit $\chi^2/\text{ndof}$ for the energy range $200 < E^{QE}_\nu
< 1250$~MeV being 21.7/15.5  (probability of 12.3\%) compared to the background-only $\chi^2/\text{ndof}$
of 50.7/17.3 (probability of 0.01\%). While this best fit is close to maximal mixing and is ruled out by a number of experiments (e.g.~OPERA~\cite{OPERA:2019kzo}), the 1$\sigma$ allowed regions stretches to much smaller mixing angles as can be seen in Fig.~\ref{fig:miniboone_allowed} overlapping substantially with the allowed LSND regions. None of the upper portion of LSND's allowed regions, the island at higher $\Delta m^2 > 10 \text{eV}^{2}$, is within the combined MiniBooNE 95\% CL. 

\begin{figure}[ht]
    \centering
    \includegraphics[clip, trim=0.0cm 0cm 0cm 0cm, width=0.47\textwidth]{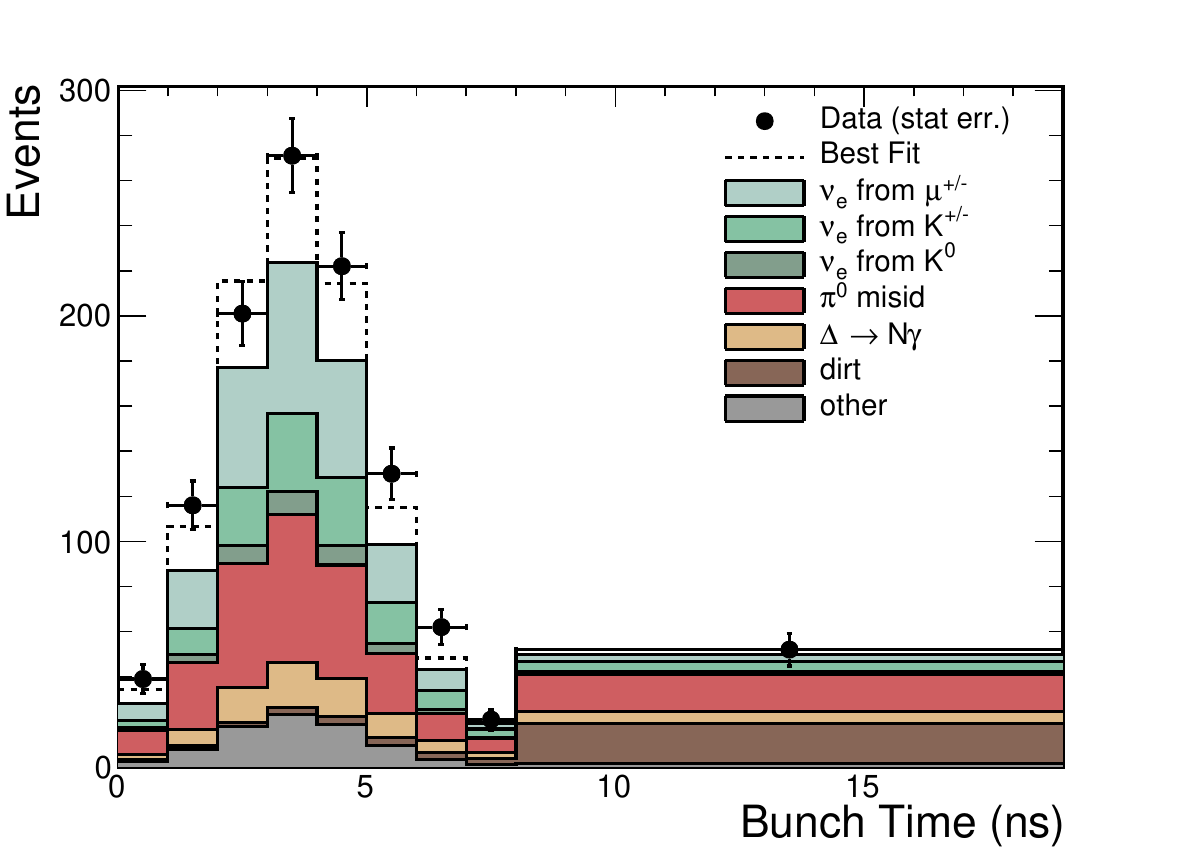}
    \includegraphics[clip, trim=0.0cm 0cm 0cm 0cm, width=0.52\textwidth]{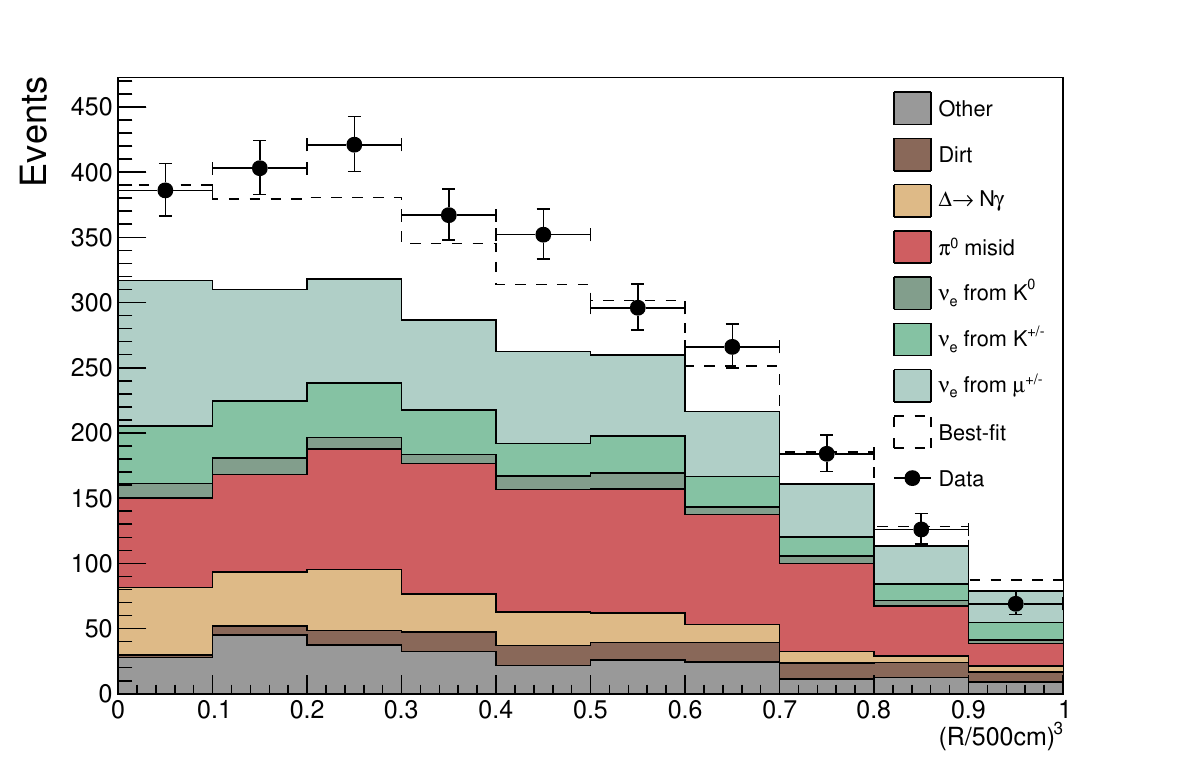}
    \caption{\label{fig:miniboone_2} 
    The final MiniBooNE results in neutrino mode in terms of both the timing of the events relative to the beam (right) and the reconstructed radial position of the spherical detector. By studying the excess in terms of additional distributions like these, a better understanding of the excess as well as the backgrounds has begun to emerge. In this example, both the beam timing and radial distributions heavily disfavor an underestimation of the ``Dirt'' component normalization being the source of the excess. Figure from Ref.~\cite{MiniBooNE:2020pnu}.}
\end{figure}

While the overall number of excess events is consistent with that expected from two-neutrino oscillations driven by LSND's best-fit parameters, a key feature of these results, highlighted in Fig.~\ref{fig:miniboone_1}, is that the predicted signal from $\nu_{\mu}\rightarrow\nu_e$ oscillations corresponding to MiniBooNE's two-neutrino oscillation best-fit parameters cannot accommodate the shape of the excess. This is particularly the case in neutrino mode, especially in the most forward region of the outgoing electron $\cos\theta$ distribution (with $\theta$ representing the angle of the electron relative to the incoming neutrino beam direction). A similar feature exists as a function of the reconstructed neutrino energy ($E^{QE}_\nu$) distribution in MiniBooNE. While historically the MiniBooNE excess was presented almost exclusively in terms of the reconstructed neutrino energy ($E^{QE}_\nu$), this was primarily due to the most common contemporary interpretation being that of a 3+1 oscillatory effect. 
It is worth noting that if the origin of the excess is not oscillatory in nature then additional information can be gained by studying the excess as a function of other reconstructed variables. 
Reconstructed visible shower energy and shower angle as shown in Fig.~\ref{fig:miniboone_1} are two such examples, as is the reconstructed radial position of the event in the detector and its timing relative to the beam, shown for neutrino mode running in Fig.~\ref{fig:miniboone_2}.

These observations have helped motivate and understand ``conventional'' interpretations involving energy misreconstruction due to mismodeled nuclear effects~\cite{Ericson:2016yjn}, mis-estimated backgrounds~\cite{Brdar:2021cgb}, as well as other beyond-SM physics~\cite{Kostelecky:2003cr,Pas:2005rb,Katori:2006mz,Harvey:2007rd,Nelson:2007yq,Gninenko:2010pr} as the source of the MiniBooNE anomaly. Many of these interpretations are discussed in Sec.~\ref{sec:th_landscape}. On the experimental front, the MicroBooNE experiment was proposed in 2008 to provide a direct test of the MiniBooNE anomaly; MicroBooNE recent results are discussed in Sec.~\ref{sec:conventional}.

\clearpage
\subsection{Reactor Experiments}
\label{sec:anomaly_reactor}
Even in the simplest light sterile neutrino oscillation framework (see Sec.~\ref{sec:3+1}), where one additional, mostly-sterile neutrino mass state is assumed, non-zero $\nu_\mu\rightarrow\nu_e$ oscillations with 1\%-level amplitudes imply that both $\nu_\mu$ and $\nu_e$ disappearance must occur at short baselines, and at a level that should be observable with current and upcoming reactor experiments, atmospheric neutrino measurements, or measurements carried out with near-only or near+far detectors of long-baseline facilities at accelerators.  
In apparent consonance with this interpretation, measurements of \anue fluxes performed at short (of order 10-100~m) reactor-detector distances were indeed found to be anomalously low.    
This energy-integrated flux deficit, observed over a broad range of short baselines, is referred to as the ``reactor antineutrino anomaly''~(RAA).  

Reactors have played an important role in neutrino physics since their discovery because they are prodigious generators of electron-flavor antineutrinos~(\anue).  
Reactor \anue are produced from beta decays of neutron-rich fission fragments generated by the heavy fissionable isotopes \uFive, \uEight, \pNine, and \pOne.  
After their production in the reactor core, these MeV-scale \anue are emitted isotropically.  
(For a broader introduction to this neutrino source, see the excellent reviews provided in Ref.~\cite{Bemporad:2001qy,Huber:2016fkt,Hayes:2016qnu}).

A typical reactor \anue spectrum is composed of \anue produced by hundreds of fission isotopes whose yields and beta decay pathways are sometimes poorly understood.  
Modeling of this complex neutrino source is thus extremely challenging.  
For this reason, 
before discussing anomalies at reactor neutrino experiments, we should briefly overview reactor antineutrino flux modeling methods.  
Modeling of reactor \anue spectrum uses two state-of-the-art approaches: the summation or \emph{ab-initio} method, and the beta conversion method.  
Both methods build~\anue predictions for individual fissioning isotopes and aggregate them for a given reactor fuel composition.  

The summation method uses $\beta$-decay information from nuclear databases to first estimate \anue contributions of individual $\beta$-decay branches~\cite{Estienne:2019ujo,Dwyer:2014eka}.
Individual beta branch contributions are then summed together, with weights based on fission yields and branching fractions, to obtain a total flux and spectrum for each fissioning isotope. 
Since nuclear databases are incomplete and sometimes inaccurate~\cite{Sonzogni:2016yac}, the inferred reactor antineutrino spectra have potentially large and ill-defined uncertainties.  
Nonetheless, significant work is being done in improving the inputs to~\cite{fijalkowska2017impact,Guadilla:2019gws,Guadilla:2019zwz,IGISOL:2015ifm,rasco2016decays,Rice:2017kfj,Valencia:2016rlr} and accuracy of~\cite{Sonzogni:2016yac} these databases.

The beta conversion method instead sums together beta particle contributions from virtual decay branches that empirically add up to a measured aggregate beta spectrum for each individual fissioning isotope. 
After defining each virtual branch's contribution in this manner, the individual beta spectrum from each branch can be converted to an \anue spectrum and summed to generate a model of the aggregate \anue spectrum specific to the fissioning isotope.  
The modern converted \anue spectra by Mueller~\textit{et al.}~\cite{Mueller:2011nm} and Huber~\cite{Huber:2011wv}~(HM model) are based on the $\beta$-decay measurements of the isotopes \uFive,~\pNine,~\pOne~performed at the Institut Laue-Langevin~(ILL) in the early 1980s~\cite{VonFeilitzsch:1982jw,Schreckenbach:1985ep,Hahn:1989zr}. An aggregate
fission beta spectrum measurement of \uEight~was not performed at ILL\footnote{Even though a \uEight~$\beta$-decay measurement was performed in 2013~\cite{Haag:2013raa}, the presented~\anue energy had a lower limit of 2.875~MeV, and consequently \emph{ab-initio} has been the model of choice for this isotope.}, since fission of \uEight~required neutrons with energies higher than those available at the thermal ILL facility.  
Ref.~\cite{Hayes:2016qnu} article provides an in-depth survey of both these methods.  

With modern reactor~\anue experiments performing flux and spectrum measurements with percent-level precision, sources of \anue~not accounted for in these two models play are also important to consider.  
The \anue arising from non-equilibrium effects~\cite{Mueller:2011nm,Kopeikin:2004cn} of the beta-decaying isotopes represent one such source that is not included in the conversion method since the $\beta$-decay measurements are done on shorter timescales where the off-equilibrium effects do not have enough time to manifest.  
Reactor and site-specific \anue are additional such sources that have to be included for each individual experiment separately. 
These may include~\anue~originating from the neutrons interacting with non-fuel reactor elements~\cite{PROSPECT:2020vcl} and spent nuclear fuel~\cite{An:2009zz,Zhou:2012zzc} placed in close proximity to the detectors.  


Reactor neutrino experiments have typically used inverse beta decay~(IBD), \anue$ + p \rightarrow e^{+} + n$, as the detection mechanism of choice due to its relatively high cross-section and the substantial background rejection capability made possible by the time-coincident signature it produces--prompt positron energy deposition followed by the delayed spatially-correlated capture of the thermalized neutron.  
Since neutrons are significantly heavier than $e^{+}$, the positron energies measured by IBD detectors are used as a high-fidelity measure of the interacting neutrino energy.
The \anue spectrum as a function of energy $E_{\anue}$ measured by a detector using the IBD mechanism is:
\begin{equation}
    \frac{dN}{dE_{\anue}}(E_{\anue})= \frac{1}{4 \pi L^2}\epsilon N_{p} \sigma_{E_{\anue}} \sum_{i} \frac{dS_i(E_{\anue})}{dE_{\anue}},
    \label{eq:IBD_spectrum}
\end{equation}
where L is the mean detector baseline from the reactor, $\epsilon$ is the (typically energy-dependent) efficiency, $N_{p}$ is the number of target protons, $\sigma_{E_{\anue}}$ is the IBD cross-section, and $dS_{i}(E_{\anue})$ is the antineutrino flux from isotope $i$. 
Whereas $dS(E_{\anue})$ decreases with energy, IBD cross section increases with energy.
Eq.~\ref{eq:IBD_spectrum} has to be modified if the detector samples neutrinos from multiple reactors, the presence of non-fissioning sources of neutrinos~\cite{conant2019antineutrino}, or in the presence of neutrino oscillations~\cite{Vogel:2015wua}.  
The uncertainties on $dS_{i}(E_{\anue})$ and $N_{p}$ are typically the dominating source of reactor-specific and detector-specific uncertainties.  

The event rate measured in reactor antineutrino experiments can be expressed in a convenient way as a ``cross section per fission'' $\sigma_{f,a}$, often called ``IBD yield'':
\begin{equation}
\sigma_{f,a}= \sum_{i} f_i^a \sigma_{i},
\label{eq.sfa}
\end{equation}
where $a$ is the experiment label, $\sigma_{i}$ is the IBD yield for the fissionable isotope $i$ (with $i=235$, $238$, $239$, and $241$ for \uFive, \uEight, \pNine, and \pOne, respectively), and $f_i^a$ is the effective fission fraction of the isotope $i$ for the experiment $a$. For each fissionable isotope $i$, the individual IBD yield is given by
\begin{equation}
\sigma_{i} = \int_{E_{\nu}^{\text{thr}}}^{E_{\nu}^{\text{max}}}
d E_{\nu} \, \Phi_{i}(E_{\nu}) \,\sigma_{\text{IBD}}(E_{\nu}),
\label{eq.si}
\end{equation}
where $E_{\nu}$ is the neutrino energy, $\Phi_{i}(E_{\nu})$ is the neutrino flux generated by the fissionable isotope $i$, and $\sigma_{\text{IBD}}(E_{\nu})$ is the detection cross-section. The neutrino energy is integrated from the threshold energy $ E_{\nu}^{\text{thr}} = 1.806 \, \text{MeV} $.
The numerical values of the $\sigma_{i}$'s predicted by a theoretical model depend on the way in which the integral in Eq.~\ref{eq.si} is performed, taking into account that the neutrino fluxes are given in tabulated bins.  

\begin{figure}
    \centering
    \includegraphics[width=1\textwidth]{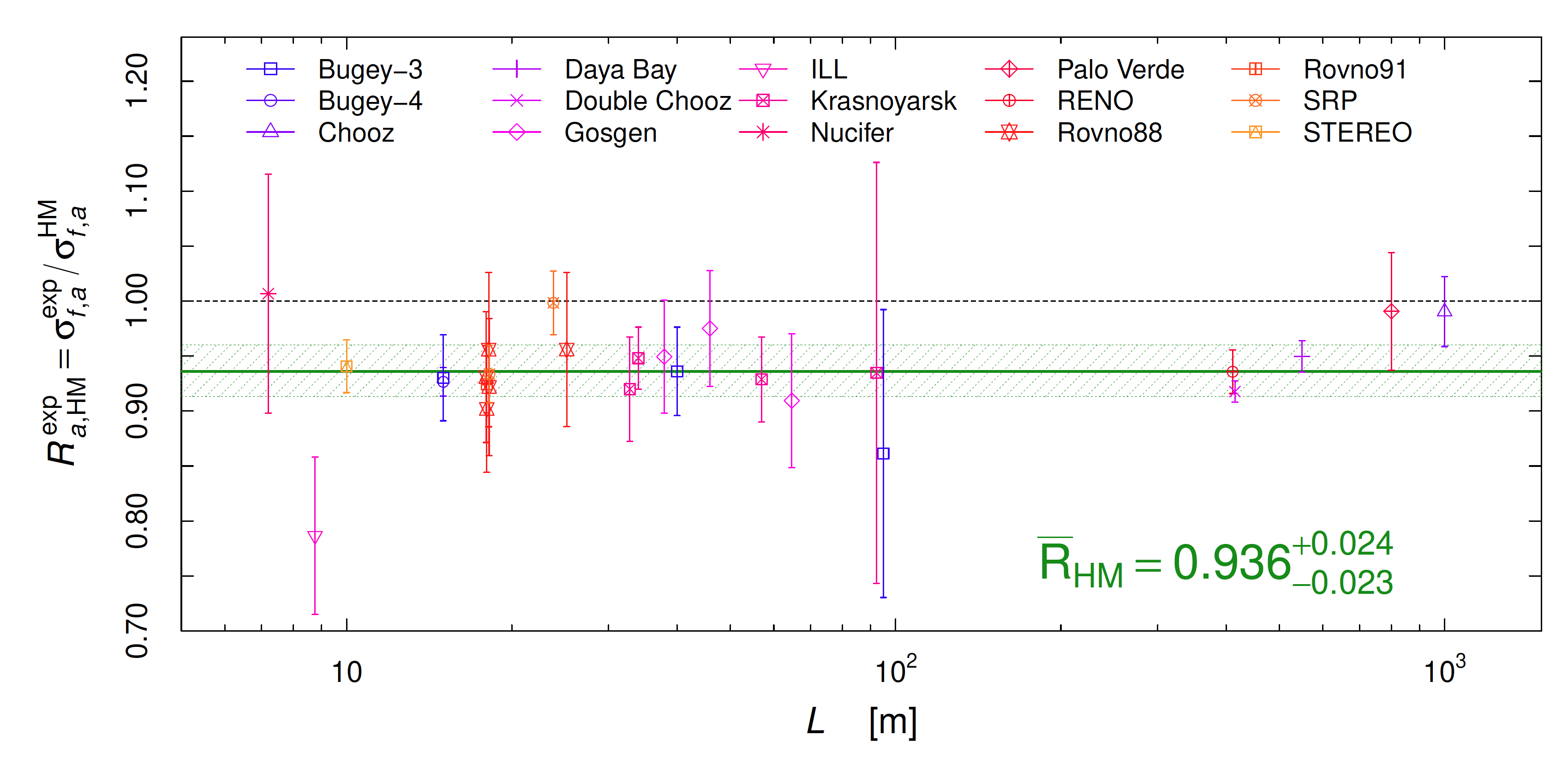}
    \caption{Ratio ($R_{a, HM}$) of the measured to the predicted IBD yields as a function of baseline. HM model is used for the predicted IBD yields. Each data point corresponds to an experiment with the error bar representing the experimental uncertainty. The green line and band show the average of $R_{a, HM}$ and average uncertainty respectively. Figure from Ref.~\cite{Giunti:2021kab}.} 
    \label{fig:RAA}
\end{figure}

\begingroup
\newcounter{ExpNum}
\begin{table*}
\centering
\resizebox{\textwidth}{!}{
\begin{tabular}{c c *{4}c c *{4}c c c c}
\hline
$a$
&
Experiment
&
$f^{a}_{235}$
&
$f^{a}_{238}$
&
$f^{a}_{239}$
&
$f^{a}_{241}$
&
$\sigma_{f,a}^{\text{exp}}$
&
$R_{a,\text{HM}}^{\text{exp}}$
&
$R_{a,\text{EF}}^{\text{exp}}$
&
$R_{a,\text{HKSS}}^{\text{exp}}$
&
$R_{a,\text{KI}}^{\text{exp}}$
&
$\delta_{a}^{\text{exp}}$ [\%]
&
$\delta_{a}^{\text{cor}}$ [\%]
&
$L_{a}$ [m]
\\
\hline
\stepcounter{ExpNum} $\theExpNum$ &
Bugey-4 &
$0.538$ &
$0.078$ &
$0.328$ &
$0.056$ &
$5.75$ &
$0.927$ &
$0.962$ &
$0.916$ &
$0.962$ &
$1.4$ &
\rdelim\}{2}{20pt}[1.4] &
$15$
\\
\stepcounter{ExpNum} $\theExpNum$ &
Rovno91 &
$0.614$ &
$0.074$ &
$0.274$ &
$0.038$ &
$5.85$ &
$0.924$ &
$0.965$ &
$0.914$ &
$0.962$ &
$2.8$ &
                        &
$18$
\\
\hline
\stepcounter{ExpNum} $\theExpNum$ &
Rovno88-1I &
$0.607$ &
$0.074$ &
$0.277$ &
$0.042$ &
$5.70$ &
$0.902$ &
$0.941$ &
$0.892$ &
$0.939$ &
$6.4$ &
\rdelim\}{2}{20pt}[3.1] \rdelim\}{5}{20pt}[2.2] &
$18$
\\
\stepcounter{ExpNum} $\theExpNum$ &
Rovno88-2I &
$0.603$ &
$0.076$ &
$0.276$ &
$0.045$ &
$5.89$ &
$0.931$ &
$0.971$ &
$0.920$ &
$0.969$ &
$6.4$ &
                                                &
$17.96$
\\
\stepcounter{ExpNum} $\theExpNum$ &
Rovno88-1S &
$0.606$ &
$0.074$ &
$0.277$ &
$0.043$ &
$6.04$ &
$0.956$ &
$0.997$ &
$0.945$ &
$0.995$ &
$7.3$ &
\rdelim\}{3}{45pt}[3.1]                         &
$18.15$
\\
\stepcounter{ExpNum} $\theExpNum$ &
Rovno88-2S &
$0.557$ &
$0.076$ &
$0.313$ &
$0.054$ &
$5.96$ &
$0.956$ &
$0.994$ &
$0.945$ &
$0.993$ &
$7.3$ &
                                                &
$25.17$
\\
\stepcounter{ExpNum} $\theExpNum$ &
Rovno88-3S &
$0.606$ &
$0.074$ &
$0.274$ &
$0.046$ &
$5.83$ &
$0.922$ &
$0.962$ &
$0.911$ &
$0.960$ &
$6.8$ &
                                                &
$18.18$
\\
\hline
\stepcounter{ExpNum} $\theExpNum$ &
Bugey-3-15 &
$0.538$ &
$0.078$ &
$0.328$ &
$0.056$ &
$5.77$ &
$0.930$ &
$0.966$ &
$0.920$ &
$0.966$ &
$4.2$ &
\rdelim\}{3}{20pt}[4.0]                         &
$15$
\\
\stepcounter{ExpNum} $\theExpNum$ &
Bugey-3-40 &
$0.538$ &
$0.078$ &
$0.328$ &
$0.056$ &
$5.81$ &
$0.936$ &
$0.972$ &
$0.926$ &
$0.972$ &
$4.3$ &
                                                &
$40$
\\
\stepcounter{ExpNum} $\theExpNum$ &
Bugey-3-95 &
$0.538$ &
$0.078$ &
$0.328$ &
$0.056$ &
$5.35$ &
$0.861$ &
$0.895$ &
$0.852$ &
$0.894$ &
$15.2$ &
                                                &
$95$
\\
\hline
\stepcounter{ExpNum} $\theExpNum$ &
Gosgen-38 &
$0.619$ &
$0.067$ &
$0.272$ &
$0.042$ &
$5.99$ &
$0.949$ &
$0.992$ &
$0.939$ &
$0.988$ &
$5.4$ &
\rdelim\}{3}{20pt}[2.0] \rdelim\}{4}{20pt}[3.8] &
$37.9$
\\
\stepcounter{ExpNum} $\theExpNum$ &
Gosgen-46 &
$0.584$ &
$0.068$ &
$0.298$ &
$0.050$ &
$6.09$ &
$0.975$ &
$1.016$ &
$0.964$ &
$1.014$ &
$5.4$ &
                                                &
$45.9$
\\
\stepcounter{ExpNum} $\theExpNum$ &
Gosgen-65 &
$0.543$ &
$0.070$ &
$0.329$ &
$0.058$ &
$5.62$ &
$0.909$ &
$0.945$ &
$0.899$ &
$0.944$ &
$6.7$ &
                                                &
$64.7$
\\
\stepcounter{ExpNum} $\theExpNum$ &
ILL &
$1.000$ &
$0.000$ &
$0.000$ &
$0.000$ &
$5.30$ &
$0.787$ &
$0.843$ &
$0.777$ &
$0.827$ &
$9.1$ &
                                                &
$8.76$
\\
\hline
\stepcounter{ExpNum} $\theExpNum$ &
Krasnoyarsk87-33 &
$1$ &
$0$ &
$0$ &
$0$ &
$6.20$ &
$0.920$ &
$0.986$ &
$0.909$ &
$0.967$ &
$5.2$ &
\rdelim\}{2}{20pt}[4.1] &
$32.8$
\\
\stepcounter{ExpNum} $\theExpNum$ &
Krasnoyarsk87-92 &
$1$ &
$0$ &
$0$ &
$0$ &
$6.30$ &
$0.935$ &
$1.002$ &
$0.924$ &
$0.983$ &
$20.5$ &
                        &
$92.3$
\\
\stepcounter{ExpNum} $\theExpNum$ &
Krasnoyarsk94-57 &
$1$ &
$0$ &
$0$ &
$0$ &
$6.26$ &
$0.929$ &
$0.995$ &
$0.918$ &
$0.977$ &
$4.2$ &
0                       &
$57$
\\
\stepcounter{ExpNum} $\theExpNum$ &
Krasnoyarsk99-34 &
$1$ &
$0$ &
$0$ &
$0$ &
$6.39$ &
$0.948$ &
$1.016$ &
$0.937$ &
$0.997$ &
$3.0$ &
0                       &
$34$
\\
\hline
\stepcounter{ExpNum} $\theExpNum$ &
SRP-18 &
$1$ &
$0$ &
$0$ &
$0$ &
$6.29$ &
$0.934$ &
$1.000$ &
$0.923$ &
$0.982$ &
$2.8$ &
0 &
$18.2$
\\
\stepcounter{ExpNum} $\theExpNum$ &
SRP-24 &
$1$ &
$0$ &
$0$ &
$0$ &
$6.73$ &
$0.998$ &
$1.070$ &
$0.987$ &
$1.050$ &
$2.9$ &
0 &
$23.8$
\\
\hline
\stepcounter{ExpNum} $\theExpNum$ &
Nucifer &
$0.926$ &
$0.008$ &
$0.061$ &
$0.005$ &
$6.67$ &
$1.007$ &
$1.074$ &
$0.995$ &
$1.056$ &
$10.8$ &
0 &
$7.2$
\\
\stepcounter{ExpNum} $\theExpNum$ &
Chooz &
$0.496$ &
$0.087$ &
$0.351$ &
$0.066$ &
$6.12$ &
$0.990$ &
$1.025$ &
$0.979$ &
$1.027$ &
$3.2$ &
0 &
$\approx1000$
\\
\stepcounter{ExpNum} $\theExpNum$ &
Palo Verde &
$0.600$ &
$0.070$ &
$0.270$ &
$0.060$ &
$6.25$ &
$0.991$ &
$1.033$ &
$0.980$ &
$1.031$ &
$5.4$ &
0 &
$\approx800$
\\
\stepcounter{ExpNum} $\theExpNum$ &
Daya Bay &
$0.564$ &
$0.076$ &
$0.304$ &
$0.056$ &
$5.94$ &
$0.950$ &
$0.988$ &
$0.939$ &
$0.987$ &
$1.5$ &
0 &
$\approx550$
\\
\stepcounter{ExpNum} $\theExpNum$ &
RENO &
$0.571$ &
$0.073$ &
$0.300$ &
$0.056$ &
$5.85$ &
$0.936$ &
$0.974$ &
$0.925$ &
$0.973$ &
$2.1$ &
0 &
$\approx411$
\\
\stepcounter{ExpNum} $\theExpNum$ &
Double Chooz &
$0.520$ &
$0.087$ &
$0.333$ &
$0.060$ &
$5.71$ &
$0.918$ &
$0.952$ &
$0.907$ &
$0.953$ &
$1.1$ &
0 &
$\approx415$
\\
\stepcounter{ExpNum} $\theExpNum$ &
STEREO &
$1$ &
$0$ &
$0$ &
$0$ &
$6.34$ &
$0.941$ &
$1.008$ &
$0.930$ &
$0.989$ &
$2.5$ &
0 &
$9-11$
\\
\hline
\end{tabular}
}
\caption{\label{tab:reactor_expts}
List of the experiments which measured the absolute reactor antineutrino flux~\cite{Giunti:2021kab}.
For each experiment numbered with the index $a$:
$f^{a}_{235}$,
$f^{a}_{238}$,
$f^{a}_{239}$, and
$f^{a}_{241}$
are the effective fission fractions of the four isotopes
$^{235}\text{U}$,
$^{238}\text{U}$,
$^{239}\text{Pu}$, and
$^{241}\text{Pu}$,
respectively;
$\sigma_{f,a}^{\text{exp}}$
is the experimental IBD yield
in units of $10^{-43} \text{cm}^{2}/\text{fission}$;
$R_{a,\text{HM}}^{\text{exp}}$,
$R_{a,\text{EF}}^{\text{exp}}$,
$R_{a,\text{HKSS}}^{\text{exp}}$, and
$R_{a,\text{KI}}^{\text{exp}}$,
are the ratios of measured and predicted rates for the IBD yields of the models in Tab.~\ref{tab:model_2020}; $\delta_{a}^{\text{exp}}$ is the total relative experimental statistical plus systematic uncertainty, $\delta_{a}^{\text{cor}}$ is the part of the relative experimental systematic uncertainty which is correlated in each group of experiments indicated by the braces; $L_{a}$ is the source-detector distance.}
\label{tab:RAA}
\end{table*}
\endgroup

The IBD yields $\sigma_{f,a}$ have been measured in a broad array of reactor antineutrino experiments spanning three continents and nearly four decades.  
A full list of experimental measurements is provided in Tab.~\ref{tab:RAA}.  
Some measurements were performed at compact, highly $^{235}$U-enriched research reactors, while others were performed at high-powered low-enriched commercial core reactors.  
Experimental reactor-detector baselines in these experiments ranged from less than 10~m to more than 1~km.  
Implemented IBD interaction detection technologies also varied widely between experiments. In some, IBD neutron detection was enabled using $^3$He counters, while in others, metal-doped liquid scintillators ($^6$Li or Gd) were used.  
Some efforts used large-volume scintillator regions to detect the prompt IBD positron signal, while others possessed no capability to detect this signal.  
Despite the broad range of employed technologies, baselines, and reactor types, experiments from the 1980s to the late 2000s were generally deemed to be consistent with state-of-the-art conversion and summation predictions available at that time.  


In 2011, new antineutrino flux calculations by Mueller \textit{et al.}~\cite{Mueller:2011nm} and Huber~\cite{Huber:2011wv} using the conversion method for \uFive, \pNine, and \pOne~and the summation method for \uEight~predicted detection rates substantially different than previous estimates.
In conjunction with the reduction in the measured neutron lifetime, as well as the inclusion of the off-equilibrium corrections, predicted IBD yields increased, leading to a 5--6\% discrepancy between this need prediction and the average of existing measurements~\cite{Mueller:2011nm}.  
This discrepancy has come to be known as the `reactor antineutrino anomaly' (RAA)~\cite{Mention:2011rk,Abazajian:2012ys}.  
Subsequent flux measurements performed using blind analyses in reactor-based $\theta_{13}$ experiments following the inception of the RAA observed a similar flux deficit~\cite{DoubleChooz:2011ymz,DayaBay:2012fng,RENO:2012mkc}.   
This development reduced the likelihood of the RAA arising from historical neutrino measurements being biased toward an agreement with contemporaneous flux predictions.  

Figure~\ref{fig:RAA} shows the ratio of the measured to the predicted IBD yields as a function of the distance between the reactors and the detectors.
A model involving sterile neutrinos that mix with active neutrinos has been invoked as a potential source for the discrepancy. 
Under the sterile neutrino hypothesis, a portion of the \anue from the nuclear reactor oscillate at frequencies of \dm{new}$\sim$1~eV$^2$--much higher than the active neutrino oscillation frequencies--into sterile states that go undetected, leading to a deficit in the measurements.  
Since the oscillation length of eV-scale-mediated oscillations is much shorter than the baselines of most of the measurements in question, the RAA would almost entirely be reflected in a common fixed IBD yield deficit in all relevant experiments~\cite{Mention:2011rk}.  
The lack of an $L/E$ character in this anomaly lends credence to a variety of alternate theories for its existence aside from sterile neutrino oscillations, including flux modeling inaccuracies; an in-depth discussion of possible flux model issues and recent modeling improvements is given in Sec.~\ref{sec:th_land_rx_models}.  
The most straightforward way to conclusively affirm a BSM origin for the RAA is to observe its $L/E$ dependence; Sec.~\ref{sec:expt_landscape_reactors} describes recent experimental efforts that probe this behavior via searches for baseline-dependent \anue energy spectrum distortions.  

\subsection{Radioactive Source Experiments}


\label{sec:anomaly_source}
A complementary probe of electron neutrino disappearance to that of reactors is provided by intense radioactive sources producing copious amounts of \nue, such as those employed by past radiochemical experiments. 

Radiochemical experiments were originally designed to detect neutrinos coming from the sun, making use of a reaction where neutrinos weakly interact with the detector chemical, converting the initial element into a radioactive isotope through the reaction
\begin{equation}\label{eq:radioreaction}
    \nu_e + N(A, Z-1) \to e^- + N(A, Z),
\end{equation}
where $Z$ and $A$ are the atomic and mass numbers, respectively. This very same neutrino detection method was implemented by Ray Davis in the Homestake Gold Mine (Lead, SD), using $^{37}$Cl~\cite{Davis:1964hf}, which allowed him and his collaborators to successfully detect the neutrinos predicted by the Standard Solar Model (SSM), an observation which led him to win the Nobel Prize in Physics in 2002.

Later on, two other solar neutrino experiments were constructed, GALLEX~\cite{Hampel:1997fc} and SAGE~\cite{SAGE:1998fvr}, this time using $^{71}$Ga as the detector medium, as suggested initially in Ref.~\cite{Kuzmin:1965zza}. In this case, the interaction of electron neutrinos with the gallium atoms leads to the emission of electrons and the creation of $^{71}$Ge atoms, which are then extracted and counted by means of chemical techniques, giving information about the neutrino flux.

\begin{center}
\begin{figure}[ht]
    \centering
    \includegraphics[width=0.35\textwidth]{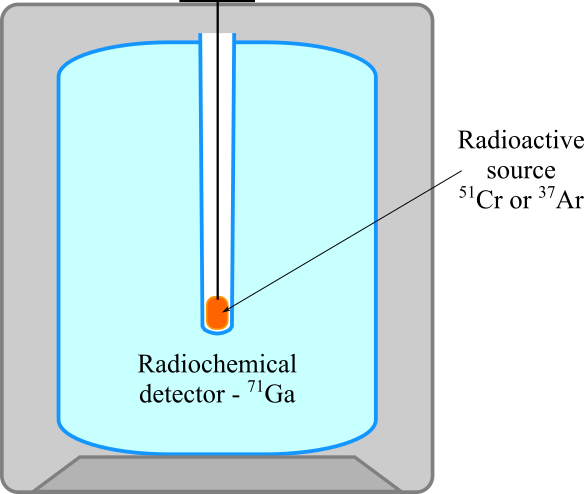}
    \caption{Generic scheme of the radioactive source experiments. The radioactive source ($^{37}\rm{Ar}$ or $^{51}\rm{Cr}$) is located inside the tank containing liquid gallium.}
    \label{fig:radioactiveExp}
\end{figure}
\end{center}

The relevance of the physics being scrutinized (the flux of neutrinos coming from the sun as a test of the SSM), made it necessary to ensure the detection technique was correctly understood. To do so, the GALLEX and SAGE Collaborations performed experiments under controlled conditions, exposing the detectors to specific neutrino calibration sources. These set of experiments are the so--called ``Gallium radioactive source experiments'', which used artificial $^{51}\rm{Cr}$ and $^{37}\rm{Ar}$ sources located inside the detectors, as schematically shown in Fig.~\ref{fig:radioactiveExp}.

\begin{figure}[ht!]
    \centering
    \includegraphics[width=0.37\textwidth]{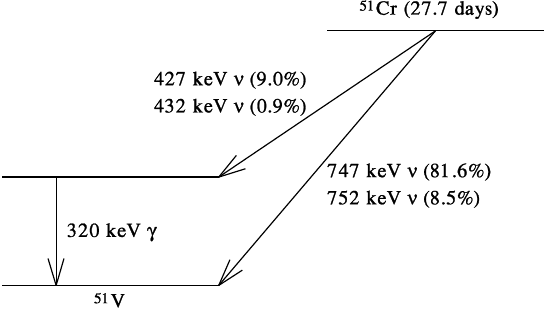}\hspace{1.0cm}
    \includegraphics[width=0.4\textwidth]{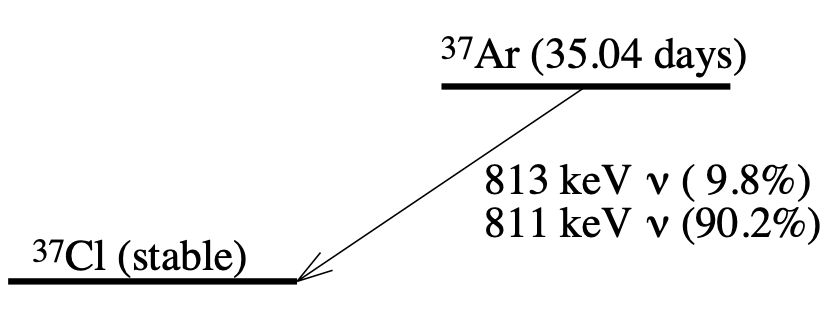}
    \caption{Nuclear levels for the $^{51}\rm{Cr}$ and $^{37}\rm{Ar}$ radioactive sources decay, according to Eq.~\ref{eq:CrAr_elCapture}. Figures from \cite{SAGE:1998fvr} and \cite{Abdurashitov:2005tb}.}
    \label{fig:SourceDecay}
\end{figure}

In these experiments, electron neutrinos are emitted during the electron capture decay of the radioactive isotopes in the sources:
\begin{align}\label{eq:CrAr_elCapture}
    e^- + \,^{51}\rm{Cr} &\to \,^{51}\rm{V}  + \nu_e,\nonumber \\
    e^- + \,^{37}\rm{Ar} &\to \,^{37}\rm{Cl} + \nu_e,
\end{align}
with neutrino energies and branching ratios as shown in Tab.~\ref{tab:nue_emission}, and the decay nuclear levels shown in Fig.~\ref{fig:SourceDecay}. The electron neutrinos interact with the main component of the detectors through the process described by Eq.~\ref{eq:radioreaction}, which for GALLEX and SAGE becomes:
\begin{equation}\label{eq:nueGa}
\nu_e + \,^{71}\rm{Ga} \to \,^{71}\rm{Ge} + e^-,
\end{equation}
with the cross sections for each emitted neutrino energy as given in Tab.~\ref{tab:nue_emission}.
\begin{table}[ht]
    \centering
    \begin{tabular}{|l|cccc|cc|}\hline
     & \multicolumn{4}{|c|}{$^{51}\rm{Cr}$} & \multicolumn{2}{|c|}{$^{37}\rm{Ar}$} \\\hline
    $E_{\nu}$ (keV) & 747.3  & 752.1. & 427.2  & 432.0  & 811   & 813 \\
    B.R.            & 0.8163 & 0.0849 & 0.0895 & 0.0093 & 0.902 & 0.098 \\
    $\sigma$ ($10^{-46}$ cm$^2$) & 60.8 & 61.5 & 26.7 & 27.1 & 70.1 & 70.3 \\\hline
    \end{tabular}
    \caption{Energies, branching ratios and cross sections for the reaction in Eq.~\ref{eq:nueGa}, for neutrinos produced by each radioactive source ($^{51}\rm{Cr}$, $^{37}\rm{Ar}$) decay. Info from Ref.~\cite{Acero:2007su} and \cite{Frekers:2011zz}. In particular, the cross section values are extracted by interpolating the calculations of J.~N.~Bahcall in Ref.~\cite{Bahcall:1997eg}. 
    }
    \label{tab:nue_emission}
\end{table}

The experimental setup for both GALLEX and SAGE radioactive source experiments were very similar, mainly consisting of a cylindrical tank containing the chemical component acting as the detector ($^{71}\rm{Ga}$), and the radioactive source located inside this tank, as schematically depicted in Fig.~\ref{fig:radioactiveExp}. 

In order to determine the number of neutrinos produced by the radioactive source and interacting with the detector, the $^{71}\rm{Ge}$ atoms produced by reaction Eq.~\ref{eq:nueGa} are extracted from the gallium by chemical mechanisms and specific cuts are applied to select the events of interest (details of these procedures can be found in~\cite{GALLEX:1994rym,Hampel:1997fc,SAGE:1998fvr,Abdurashitov:2005tb}), including the relevant information about the $\nu_e$--${}^{71}\rm{Ga}$ cross section. Uncertainties on this cross section may significantly impact the final result and its interpretation in terms of neutrino oscillations, as discussed in Sec.~\ref{sec:GaAnomaly}.

After the counting procedure, the activity of the source is computed and compared to the previously directly measured activity. The ratio between the two numbers is reported in Tab.~\ref{tab:Ge71_Rate} for the four performed experiments. It is important to note that the cross sections for reactions (Eq.~\ref{eq:nueGa}) used to compute these numbers were the ones calculated by Bahcall in Ref.~\cite{Bahcall:1997eg} and that, as pointed out in~\cite{Giunti:2010zu}, the corresponding uncertainties were not considered.  Further investigation of cross-section calculations, and their prospects for providing a `conventional' explanation for the Gallium Anomaly, is provided in Sec.~\ref{sec:null}.  

\begin{table}[ht]
    \centering
    \begin{tabular}{|c|c|}\hline
    GALLEX                  & SAGE                  \\\hline
    $0.953 \pm 0.11$        & $0.95 \pm 0.12$       \\
    $0.812^{+0.10}_{-0.11}$ & $0.791^{+0.084}_{-0.078}$\\\hline
    \end{tabular}
    \caption{Ratio of predicted and observed $^{71}\rm{Ge}$ event rates as measured by GALLEX (using $^{51}\rm{Cr}$ twice) and SAGE ($^{51}\rm{Cr}$ and $^{37}\rm{Ar}$).}
    \label{tab:Ge71_Rate}
\end{table}

As the main purpose of these experiments was to prove the experimental techniques used for the detection of solar neutrinos, the obtained results allowed the two collaborations to conclude that their setup and procedures were well understood and that the solar neutrino measurements -- a very large observed deficit on the neutrino flux when compared against the SSM -- were not due to any experimental artifact and were highly reliable.  

\begin{figure}[ht!]
    \centering
    \includegraphics[width=0.6\textwidth]{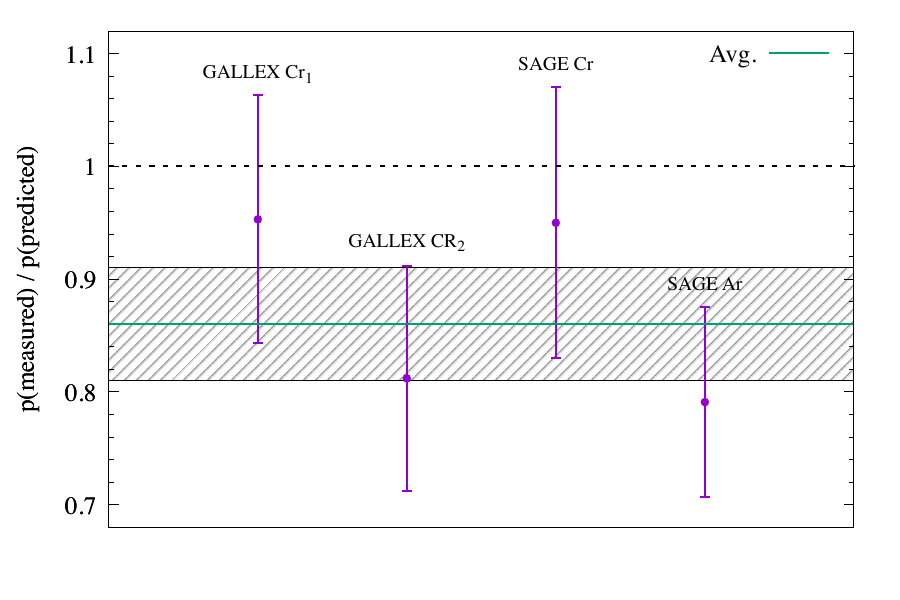}
    \caption{Ratio of the observed and the predicted event rates as measured by the different radioactive source experiments GALLEX and SAGE. The shadowed area corresponds to the $1\sigma$ region around the weighted average, $R_{\rm{avg}} = 0.86 \pm 0.05$.}
    \label{fig:ratioData}
\end{figure}

More recently, however, the difference between measured capture rates (Tab.~\ref{tab:Ge71_Rate})
and theoretical calculations was re-examined, especially in light of other indications of anomalous flavor transition from the LSND and MiniBooNE experiments in the 1990s and 2000s.  
The differences observed during this reiteration established what is known as the Gallium Anomaly. 
Figure~\ref{fig:ratioData} shows the experimental results mentioned above, together with the global average, $R_{\rm{avg}} = 0.86 \pm 0.05$, which is $\sim3\sigma$ less than unity.  

This discrepancy is usually interpreted as an anomalous disappearance of electron neutrinos trough short-baseline oscillations to sterile neutrinos in a framework of 3+1 mixing neutrinos. In the scheme in which one sterile neutrino at the eV mass scale is added to the standard three-neutrino framework, the survival probability of electron (anti)neutrinos is
\begin{equation}\label{eq:nueSurvP}
    P(\nu_e \to \nu_e) = 1 - 4\left|U_{e4}\right|^2\left(1 - \left|U_{e4}\right|^2\right) \sin^2\left(\frac{\Delta m_{41}^2 L}{4E}\right),
\end{equation}
where $L$ is the distance from the source to the detector, $E$ is the neutrino energy, $U$ is the $4\times4$ PNMS mixing matrix, and $\Delta m_{41}^2 = m^2_4 - m_1^2$, is the squared-mass difference between the heavy (mostly sterile) neutrino $\nu_4$ and the (standard) light neutrino $\nu_1$ (considering that $\Delta m_{41}^2 \approx \Delta m_{42}^2 \approx \Delta m_{43}^2$).

\begin{figure}[ht!]
    \centering
    \includegraphics[width=0.55\textwidth]{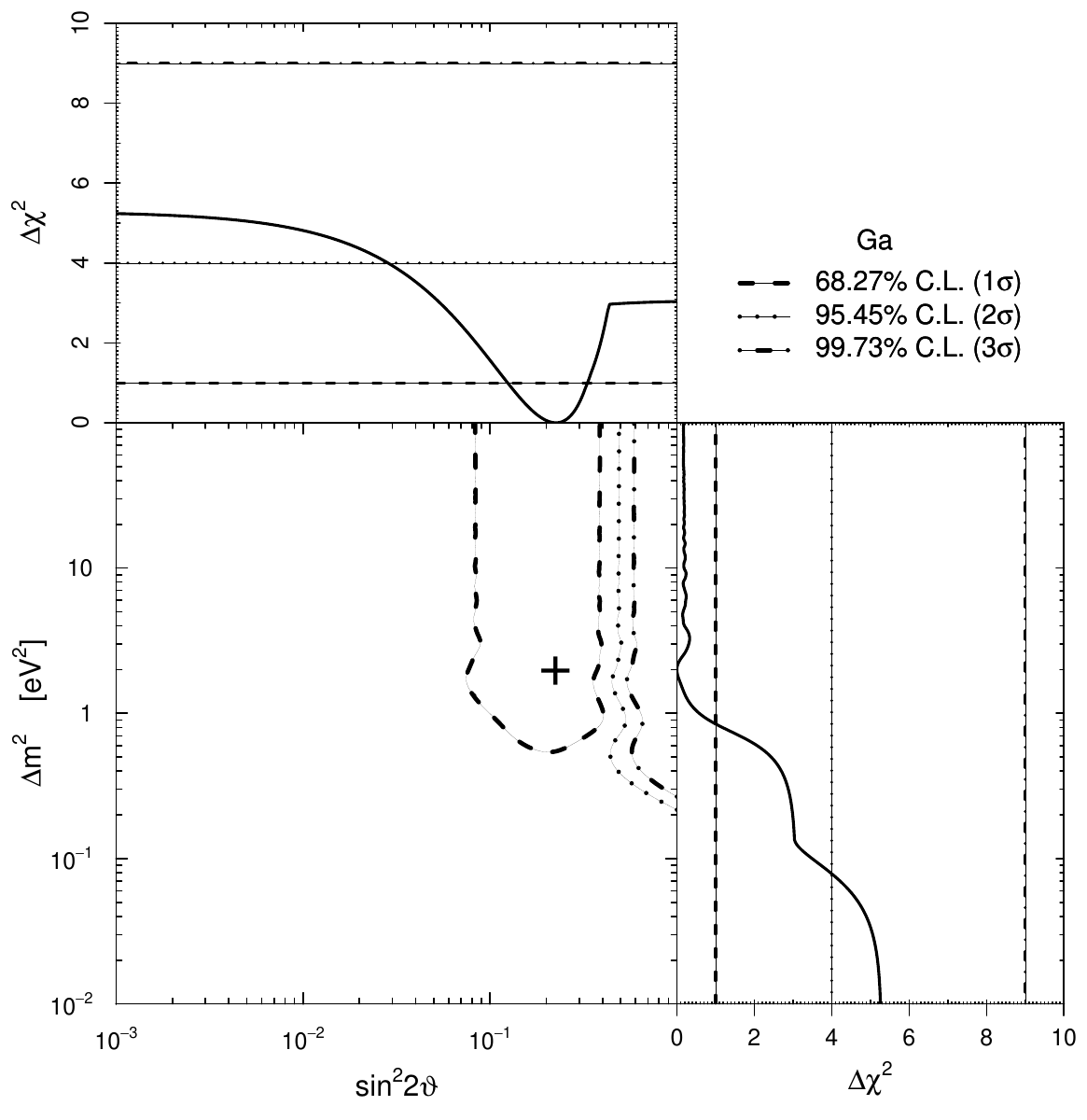}
    \caption{Allowed regions in the $\Delta m^2$--$\sin^22\theta$ parameter space obtained from the combination of the GALLEX and SAGE radioactive source experiments. Figure from Ref.~\cite{Acero:2007su}. Notice that $\sin^22\theta = 4\left|U_{e4}\right|^2\left(1 - \left|U_{e4}\right|^2\right)$.}
    \label{fig:GallexSage_contours}
\end{figure}

This model is implemented, for instance, in Ref.~\cite{Acero:2007su}, where studies of the GALLEX and SAGE results (with the cross sections listed on Tab.~\ref{tab:nue_emission}) revealed a possible indication of electron neutrino disappearance due to neutrino oscillations with $\Delta m^2_{41} \gsim 0.1$~eV$^2$, as the contour plots in Fig.~\ref{fig:GallexSage_contours} show.

The relevance of these results has led the neutrino community to search for alternative explanations, such as possible effects arising from cross section uncertainties, not considered in the analysis leading to the contours in Fig.~\ref{fig:GallexSage_contours} (described in Sec.~\ref{sec:GaAnomaly}), and to perform new experiments (e.g.~BEST, described in Sec.~\ref{sec:BExpST}) to test the deficit of electron neutrinos observed by the gallium radioactive experiments as described here.

The following section provides details on a diverse range of viable interpretations for these anomalies, from modifications of three-flavor neutrino mixing to potential couplings to hidden sectors.


\clearpage

\section{Interpretations of the Anomalies}
\label{sec:th_landscape}

In this section, we discuss the theoretical interpretations of the experimental anomalies discussed above.
While seemingly compatible when presented within the empirical picture of two-neutrino oscillations, the underlying source of the anomalies may or may not be connected.
In what follows, we describe three different categories of resolutions put forth in the literature, including those that can explain some---but not necessarily all---of the anomalies.

\subsection{Flavor Conversion}\label{sec:theory:flavorconv}

In this subsection, we discuss flavor-conversion-based explanations to the short-baseline anomalies discussed in Sec.~\ref{sec:expt_landscape}. 

We begin by discussing perhaps the simplest model that leads to flavor change in short baselines: an extension of the ESM by the inclusion of a new light sterile neutrino, referred to as the well-studied 3+1 sterile neutrino model. 
In the most simple realization of the 3+1 model, the sterile neutrino has no gauge interactions. It should be noted that, historically, experimental collaborations such as LSND and MiniBooNE have analyzed their data sets primarily under a two-neutrino oscillation hypothesis most closely represented by the 3+1 model. 
Additionally, in this model, provided that the two-neutrino oscillation approximation is valid, no observable CP violation effects are expected. Therefore, light sterile neutrino and antineutrino oscillation searches are effectively sensitive to the same oscillation parameters.
Finally, we prepare the reader in advance, in that, the 3+1 model has been shown to provide an insufficient description to the globally available experimental data that have sensitivity to its observable effects. 
Nevertheless, it has and will likely continue to be instructively used within the community as a simple ``measure'' for developing, optimizing, and comparing sensitivities of various experimental searches and comparing compatibility of different experimental results, albeit with several caveats.

After discussing this simple model in some detail, we then consider extensions and variations to this model, all of which can lead to flavor transitions at short baselines. 
The most straightforward extension beyond 3+1 is represented by the 3+$N$ model, where $N=2,3,...$ light sterile neutrinos are introduced and associated with similarly light neutrino mass states. 
Other extensions often introduce non-standard interactions and neutrino propagation effects.

\subsubsection{3+1 Light Sterile Neutrino Oscillations}
\label{sec:3+1}

\subsubsubsection{Sterile Neutrinos}

Sterile neutrinos provide one of the simplest extensions of the Standard Model that explain the non-zero mass of neutrinos. The right-handed, gauge-singlet fields $(N)_R$, sometimes denoted as $(\nu_s)_R$, provide the missing chiral partners for the left-handed, interacting SM neutrino fields $(\nu_\alpha)_L$. Neutrinos would then acquire a mass just like any other SM fermion via the Higgs mechanism, $m_D \overline{\nu_\alpha}N$, with $m_D = y v/\sqrt{2}$ and $v$ the Higgs vaccum expectation value. 
While this observation is sufficient to resolve the puzzle of neutrino masses, it raises new questions.
In particular, since $N_R$ would carry no charges under any SM gauge symmetries, it could also have a Majorana mass, $M \overline{N^c_R}N_R$.
In that case, the physical spectrum could contain the light, mostly-active neutrinos, as well as potentially heavier, mostly-sterile neutrinos that interact very weakly through mixing.
Note that since Majorana masses violate any one of the quantum numbers associated with $N_R$, it would also indicate that lepton number is violated by two units.
Conversely, if lepton number is conserved, then Majorana masses are not allowed, and neutrinos are purely Dirac particles, like any other SM fermion.

Most importantly, Majorana masses need not be related to the electroweak scale. In principle, can take any value up to the Planck scale or the scale of gauge unification. 
If it takes values larger than the electroweak scale or, more precisely, the scale of Dirac neutrino masses, it would trigger the canonical Type-I seesaw mechanism~\cite{Minkowski:1977sc,Mohapatra:1979ia,GellMann:1980vs,Yanagida:1979as,Lazarides:1980nt,Mohapatra:1980yp,Schechter:1980gr,Cheng:1980qt,Foot:1988aq}. The seesaw Lagrangian reads, 
\begin{equation}
    \mathcal{L}_{\nu {\rm -mass}} \supset - Y_{\alpha i} \overline{L_\alpha}\widetilde{H} N_j - \frac{M_{ij}}{2} \overline{N^c_i}N_j + \text{h.c.},
\end{equation}
where $\widetilde{H} = i \sigma_2 H^*$ is the conjugate of the Higgs doublet and $L$ is the lepton doublet. After electroweak symmetry breaking,
\begin{equation}
    M_\nu \simeq - \frac{v^2}{2}Y^T M^{-1} Y,
\end{equation}
where $M_\nu$ is the mass matrix for the light, mostly-active neutrinos. 
In its simplest realization, the seesaw mechanism explains the smallness of the neutrino masses using a hierarchy of scales between Dirac and Majorana masses. 
Mixing between the heavy neutrinos and the active SM neutrinos, $|U_{\alpha i}|$, is typically of order  $\mathcal{O}(M_D/M)$. 

Other variations of the Type-I seesaw mechanism exist, including low-scale models where the lightness of neutrino masses is explained instead by the approximate conservation of lepton number. Seesaw models with pairs of heavy neutral leptons, $N$ and $S$, with opposite lepton number, are often called extended seesaws. One of which, the inverse seesaw, has these particles form pseudo-Dirac pairs with a small mass splitting given by $\mu$, a lepton-number-violating parameter. Light neutrino masses, in this case, are proportional to $\mu$, which in the limit $\mu \to 0$, parametrically recovers lepton number conservation and massless neutrinos. 

In extended seesaws, it is also possible that some number of the sterile neutrinos remain relatively light~\cite{Zhang:2011vh}. This may be due to cancellations, new symmetries, or because the number of fields exceeds the number of large scales in the theory. In such cases, heavy neutrinos can ``seesaw" not only the light, mostly-active neutrinos but also some of the sterile neutrinos, rendering them light as well. These models predict the existence of light sterile neutrinos that mix with active neutrinos with large mixing angles. Finally, we note that while the seesaw mechanism provides an elegant solution to the smallness of neutrino masses, we cannot rule out the possibility that lepton number is indeed conserved and that neutrinos are Dirac. A fourth neutrino, in this case, would not necessarily be related to the origin of neutrino masses but could exist based on a new pair of left and right-handed singlet fermions. 

Sterile neutrinos can also solve other open problems in the SM. 
Heavy neutrinos above the electroweak scale can generate the observed baryon asymmetry of the Universe via their decays or oscillations. 
In this scenario, the lepton asymmetry generated by CP violation and out-of-equilibrium processes involving heavy neutrinos is converted into a baryon asymmetry via Sphalerons, non-perturbative SM processes. 
This scenario is referred to as Leptogenesis~\cite{Fukugita:1986hr}. 
In addition, light sterile neutrinos with masses of $\mathcal{O}(1 - 100)$~keV can provide a warm dark matter candidate~\cite{Dodelson:1993je,Shi:1998km,Dolgov:2000ew,Abazajian:2001nj,Boyarsky:2009ix}, produced out-of-equilibrium by oscillations in the early Universe. 
It is also possible that both of these issues are addressed by a whole new sector of sterile neutrinos, such as in the proposed $\nu$-minimal Standard Model~\cite{Asaka:2005pn,Shaposhnikov:2008pf}

In this section, we discuss short-baseline oscillations generated by sterile neutrinos with masses of order $\mathcal{O}(1-10)$~eV. 
These sterile neutrinos could very well play the role of the seesaw partners, albeit triggering the mechanism at a relatively low scale~\cite{deGouvea:2005er,Mohapatra:2005wk}.
Unfortunately, they would not provide direct evidence for Leptogenesis or sterile-neutrino dark matter. Still, their discovery would be the first laboratory observation of a particle beyond the SM and strongly motivate sterile-neutrino solutions to all open problems in particle physics.
Studying their properties and looking for potential lower and higher-scale partners would be of great importance in this case.

\subsubsubsection{3+1 Oscillation Probabilities}

In the simplest 3+1 model, the standard neutrino sector is extended by an extra neutrino flavor $\nu_s$ which is a gauge singlet and does not experience weak interactions.
The three neutrino flavors and the sterile neutrino are admixtures of four neutrino mass eigenstates, where $m_4$ is assumed to be of order $\sim1$~eV, motivated by the LSND observation. 
Parametrically, one can extend the $3\times3$ leptonic mixing matrix to a $4\times4$ matrix $U_{\alpha i}$, with $\alpha=e,\mu,\tau,s$ and $i=1,...,4$.
Notice, however, that the last row of this extended matrix is not related to experimental observables as it pertains to the amount of sterile neutrino flavor in different beams.
This matrix can be parametrized by the usual three mixing angles and CP violation phase, plus three extra mixing angles and two extra CP phases.

To avoid parametrization dependence, it is often helpful to work with the mixing matrix elements $U_{e4}$, $U_{\mu 4}$, and $U_{\tau 4}$ directly.
We assume that the fourth mass eigenstate is mostly sterile and much heavier than the other ones, so that $\Delta m^2_{41}\gg |\Delta m^2_{31}|,\Delta m^2_{21}$, allowing for the approximation that $\Delta m^2_{31}$ and $\Delta m^2_{21}$ are degenerate and at zero. 
Furthermore, this new, large mass splitting allows for short-baseline neutrino oscillations. 
In the limit where oscillations due to $\Delta m_{31}^2$ and $\Delta m_{21}^2$  --- the atmospheric and solar mass-squared splittings, respectively --- are negligible, short-baseline oscillations can be approximated by
\begin{align}
    P(\nu_e \to \nu_e) &\simeq 1 - 4\left\lvert U_{e4}\right\rvert^2\left( 1 - \left\lvert U_{e4}\right\rvert^2\right) \sin^2\left(\frac{\Delta m_{41}^2 L}{4E_\nu}\right) \equiv 1 - \sin^2\left(2\theta_{ee}\right) \sin^2\left(\frac{\Delta m_{41}^2 L}{4E_\nu}\right), \label{eq:Pee}\\
    P(\nu_\mu \to \nu_\mu) &\simeq 1 - 4\left\lvert U_{\mu4}\right\rvert^2\left( 1 - \left\lvert U_{\mu4}\right\rvert^2\right) \sin^2\left(\frac{\Delta m_{41}^2 L}{4E_\nu}\right) \equiv 1 - \sin^2\left(2\theta_{\mu\mu}\right) \sin^2\left(\frac{\Delta m_{41}^2 L}{4E_\nu}\right), \label{eq:Pmm}\\
    P(\nu_\mu \to \nu_e) &\simeq 4\left\lvert U_{\mu4}\right\rvert^2 \left\lvert U_{e4}\right\rvert^2 \sin^2\left(\frac{\Delta m_{41}^2 L}{4E_\nu}\right) \equiv \sin^2\left(2\theta_{\mu e}\right) \sin^2\left(\frac{\Delta m_{41}^2 L}{4E_\nu}\right).\label{eq:Pme}
\end{align}

Note that the above makes no explicit assumption about $U_{\tau4}$ (or, consequently, $U_{s4}$); however, by unitarity considerations, $|U_{e4}|^2+|U_{\mu4}|^2+|U_{\tau 4}|^2+|U_{s4}|^2=1$.
Note also that in the above, we have focused explicitly on electron- and muon-neutrino flavors, given that these are the transition channels that have been studied most extensively and where the short-baseline anomalies occur. 

The above three oscillation probabilities further define the effective mixing angles $\sin^2\left(2\theta_{\alpha\beta}\right)$ often used in the literature: $4\left\lvert U_{e4}\right\rvert^2\left( 1 - \left\lvert U_{e4}\right\rvert^2\right)=\sin^22\theta_{ee}$, $4\left\lvert U_{\mu4}\right\rvert^2\left( 1 - \left\lvert U_{\mu4}\right\rvert^2\right)=\sin^22\theta_{\mu\mu}$, and $4\left\lvert U_{\mu4}\right\rvert^2 \left\lvert U_{e4}\right\rvert^2=\sin^22\theta_{\mu e}$. The two new CP violating phases from the extended $4\times4$ mixing matrix do not lead to observable effects unless effects from both $\Delta m_{41}^2$ and either $\Delta m_{21}^2$ or $\Delta m_{31}^2$ are simultaneously relevant. 
In scenarios with more than one light sterile neutrino (see Sec.~\ref{subsubsec:3plusN}), CP-violating phases associated with the $(3+$N$)\times(3+$N$)$ mixing matrix may be accessible. 
Returning to the 3+1 scenario, oscillations are relevant for $L/E_\nu \approx $m/MeV or km/GeV if $\Delta m_{41}^2 \approx \SI{1}\eV^2$.
Most notably, the relationships among Eqs.~(\ref{eq:Pee}-\ref{eq:Pme}) imply that, if both $|U_{e4}|^2$ and $|U_{\mu 4}|^2$ are nonzero, then electron-neutrino disappearance, muon-neutrino disappearance, and muon-to-electron-neutrino appearance must \textit{all} occur at the same $L/E_\nu$.
More explicitly, the oscillation amplitudes for appearance and disappearance are related by $\sin^22\theta_{\mu e}\leq1/4\sin^22\theta_{\mu\mu}\sin^22\theta_{ee}$. 
This relation allows for combinations of experiments to over-constrain the 3+1 model, a feature that global fits take advantage of when performing combined analyses to experimental data sets on neutrino appearance and disappearance.

Finally, when an explicit parametrization of the $4\times4$ unitary mixing matrix $U$ is needed for oscillations, six rotation angles and three CP phases are required. In addition to the three-neutrino mixing parameters, three new angles, $\theta_{14}$, $\theta_{24}$, and $\theta_{34}$, and two new phases, $\delta_{24}$ and $\delta_{14}$, are defined, in accordance with Ref.~\cite{Harari:1986xf}. The full mixing matrix is then given by 
\begin{equation}
U_{4\times 4} = R_{34}S_{24}S_{14}R_{23}S_{13}R_{12},
\end{equation}
where $R_{ij}$ is a real matrix of rotation by an angle $\theta_{ij}$, and $S_{ij}$ is a complex matrix of rotation by $\theta_{ij}$ with a CP phase of $\delta_{ij}$. The relationship between this parametrization to the effective mixing angles as well as to  the full matrix elements is shown in Tab.~\ref{tab:relation_between_angles}.

\renewcommand{\arraystretch}{1.2}
\begin{table}[tbp] \centering
\begin{tabular}{lllll}\hline
effective angle && full 3+1 model rotation angles && mixing elements
\\\hline\hline
$\sin^2 2 \theta_{ee}$ &=& 
$\sin^2 2 \theta_{14}$ &=& 
$ 4 (1-|U_{e4}|^2)|U_{e4}|^2$\\
$\sin^2 2 \theta_{\mu\mu}$ &=& 
$4 \cos^2 \theta_{14} \sin^2 \theta_{24} (1 - \cos^2 \theta_{14} \sin^2 \theta_{24})$ &=& 
$4 (1-|U_{\mu4}|^2)|U_{\mu4}|^2$ \\
$\sin^2 2 \theta_{\tau\tau}$ &=& 
$4 \cos^2 \theta_{14} \cos^2 \theta_{24} \sin^2 \theta_{34}(1 - \cos^2 \theta_{14} \cos^2 \theta_{24} \sin^2 \theta_{34})$ &=& 
$4 (1-|U_{\tau4}|^2)|U_{\tau4}|^2$ \\
$\sin^2 2 \theta_{e\mu}$ &=& 
$\sin^2 2 \theta_{14} \sin^2 \theta_{24}$ &=& 
$4|U_{\mu 4}|^2 |U_{e 4}|^2$ \\
$\sin^2 2 \theta_{e \tau}$ &=& 
$\sin^2 2 \theta_{14} \cos^2 \theta_{24} \sin^2 \theta_{34}  $&= &
$4|U_{e 4}|^2 |U_{\tau 4}|^2$\\
$\sin^2 2 \theta_{\mu \tau}$ &=& 
$\sin^2 2 \theta_{24} \cos^4 \theta_{14} \sin^2 \theta_{34} $&= &
$4|U_{\mu 4}|^2 |U_{\tau 4}|^2$\\ \hline 
\end{tabular}
\caption{Relation between the different parametrizations of neutrino mixing parameters in a 3+1 model. Modified from Ref.~\cite{Diaz:2019fwt}.\label{tab:relation_between_angles}}
\end{table}
\renewcommand{\arraystretch}{1}

\subsubsubsection{Global Analysis of 3+1 Oscillations}

The present status of the 3+1 model is best examined through the lens of a global analysis. 
This allows each of the myriad of short-baseline experiments to contribute to a single statistical model according to the strength of their results.
Global fits have been performed independently by several groups (see, \textit{e.g.}, Ref.~\cite{Sorel:2003hf,Karagiorgi:2009nb,Kopp:2011qd,Giunti:2011cp,Giunti:2011gz,Conrad:2012qt,Kopp:2013vaa,Collin:2016aqd,Gariazzo:2017fdh,Dentler:2018sju, Diaz:2019fwt, Giunti:2019aiy, Boser:2019rta, Dasgupta:2021ies}).
While all groups find a strong preference for a 3+1 model compared to the SM, driven mainly by LSND and MiniBooNE, a significant tension among data sets is also consistently found.
The tension lies in a simple fact: large enough mixings required to explain the LSND and MiniBooNE anomalies simultaneously lead to a large disappearance amplitude, particularly of muon neutrinos, and this is in tension with $\nu_\mu\to\nu_\mu$ data, which provide strong limits on the value of $\sin^22\theta_{\mu\mu}$.

\begin{figure}[ht]
    \centering
    \includegraphics[width=0.49\textwidth]{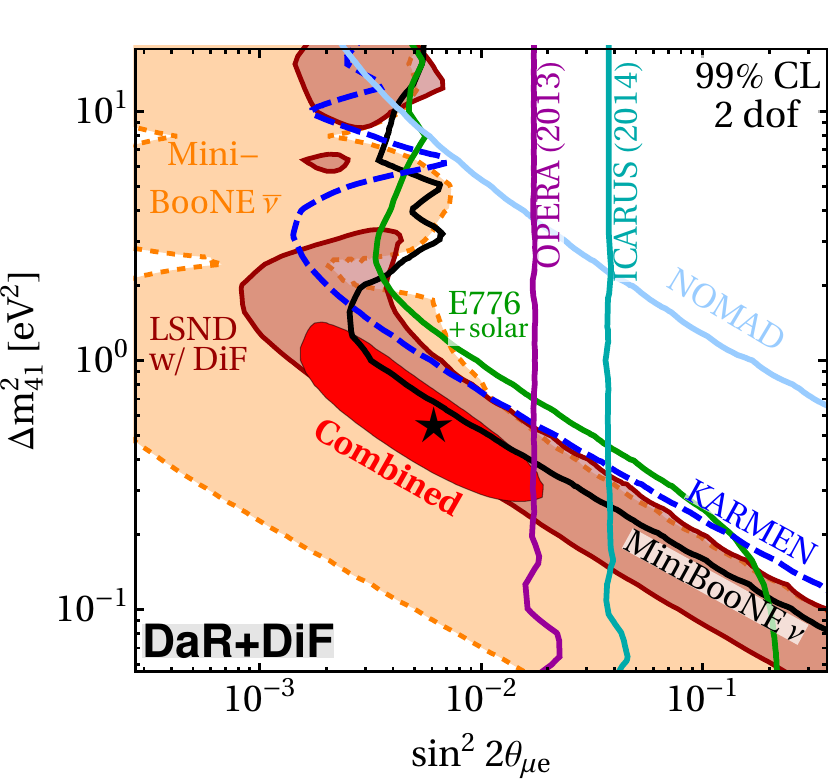}
    \includegraphics[width=0.49\textwidth]{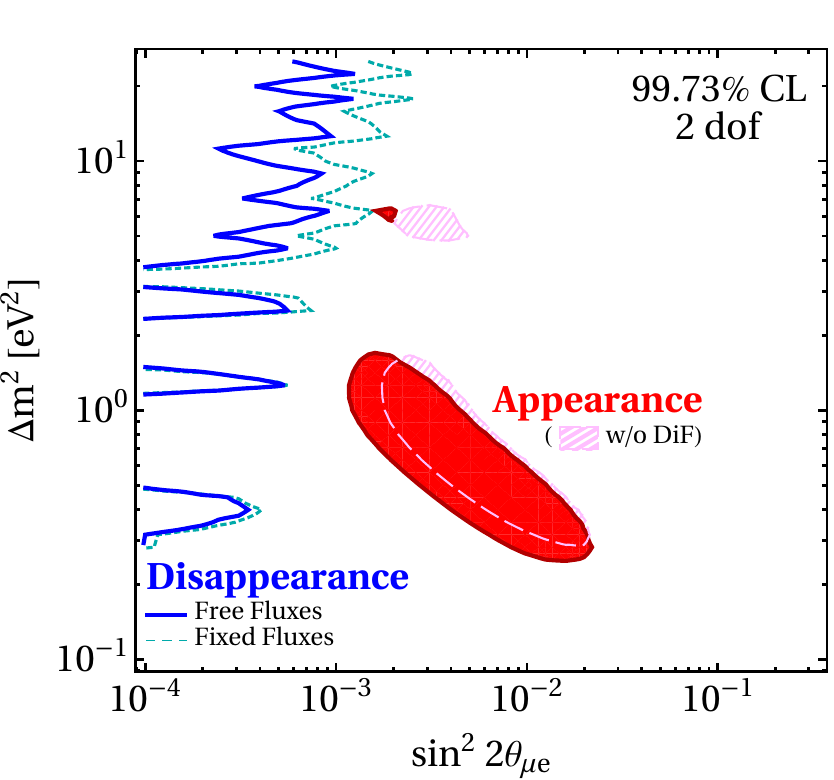}
    \caption{\label{fig:global_appearance} Left: Preferred regions by several $\nu_\mu\to\nu_e$ appearance experiments in the 3+1 scenario at 99\% CL for 2 degrees of freedom.
    Right: Preferred region of short-baseline appearance experiments (red region), compared to the region excluded by disappearance experiments (blue line) at 99.73\%  CL for 2 degrees of freedom. The figure is taken from Ref.~\cite{Dentler:2018sju}.}
\end{figure}

Figure~\ref{fig:global_appearance} shows the preferred region in the 3+1 model parameter space of several short-baseline appearance experiments, including the combination of all of them, at 99\% CL for two degrees of freedom (left panel), as well as, the regions preferred by all short-baseline appearance experiments (right panel, red region) compared to the excluded region by all disappearance experiments (blue line) at 99.73\%  CL for two degrees of freedom~\cite{Dentler:2018sju}.
Note that the region to the right of the blue line is excluded, and that encompasses the entirety of the appearance allowed region.
While the left panel clearly indicates a strong preference for the 3+1 model over the usual three neutrino framework, the right panel clearly shows the tension between appearance and disappearance data: the appearance and the disappearance $99.73\%$~CL preferred regions are disjoint.

One can be more precise and quantify the amount of tension using the parameter goodness-of-fit (PG) test~\cite{Maltoni:2003cu}, which compares the minimum chi-square values of the full data set to the sum of minimum values of the individual data sets, that is,
\begin{equation}
   \chi^2_{\rm PG}\equiv\chi^2_{\rm global}-\chi^2_{\rm app}-\chi^2_{\rm dis}.
\end{equation}
In Ref.~\cite{Dentler:2018sju}, for example, this tension is found to yield a $p$-value of $3.7\times10^{-7}$ when assuming that the $\chi^2_{\rm PG}$ follows a chi-square distribution.
Moreover, removing any individual null experiment from the fit does not lead to significant improvements in the $p$-value, evidencing that the tension is robust.
Similar conclusions have been drawn by other global fits~\cite{Diaz:2019fwt, Giunti:2019aiy, Boser:2019rta, Dasgupta:2021ies}.  

While this demonstrates the shortcomings of the 3+1 scenario as an explanation of the short-baseline anomalies, there are important caveats that should be highlighted.
First, all global fits to date have been performed assuming the validity of Wilks' theorem~\cite{Wilks:1938dza}.  
While in many cases, Wilks' theorem is valid and, therefore, the test statistic follows a chi-square distribution, this is not obvious in neutrino oscillation experiments.
Ref.~\cite{Coloma:2020ajw} has shown that these considerations are relevant for the interpretation of short-baseline reactor experiments in terms of sterile neutrinos, and the assumption of Wilks' theorem can have a significant quantitative impact on the statistical interpretation of the anomaly.
Not only would this change the CL of the allowed regions, but it would also affect the outcome of the parameter goodness-of-fit test, and thus the ``amount of tension'' between appearance and disappearance data.
Therefore, the $p$-values quoted above should be taken with a grain of salt.

\begin{figure}[ht]
    \centering
    \includegraphics[width=0.49\textwidth]{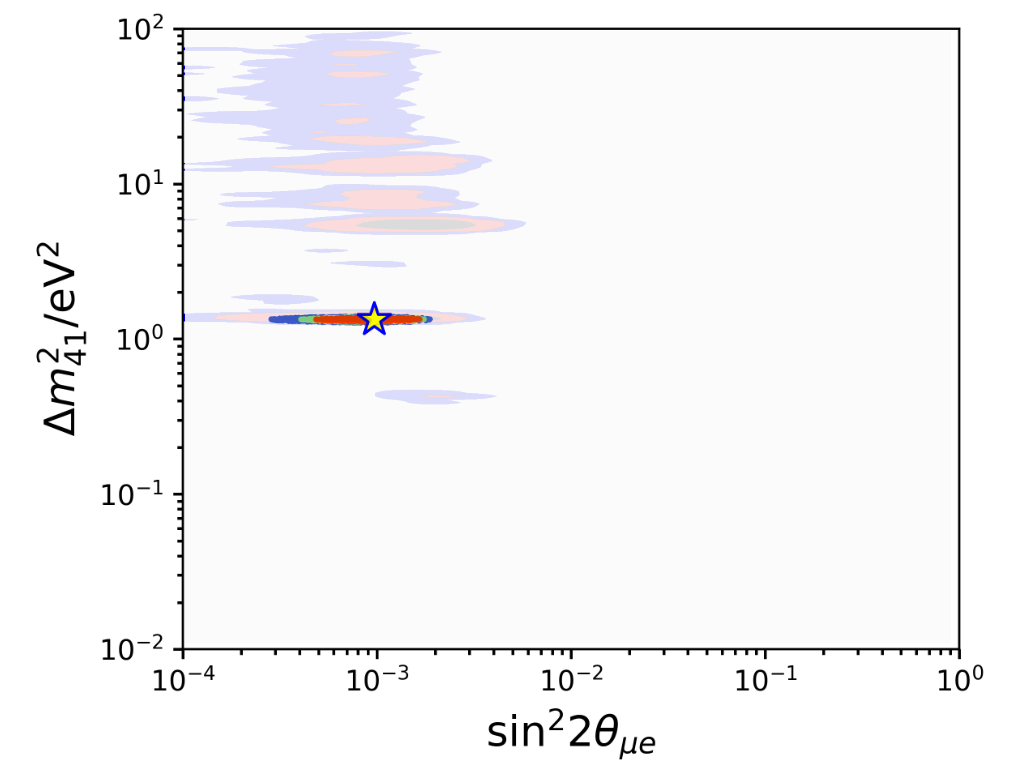}
    \includegraphics[width=0.49\textwidth]{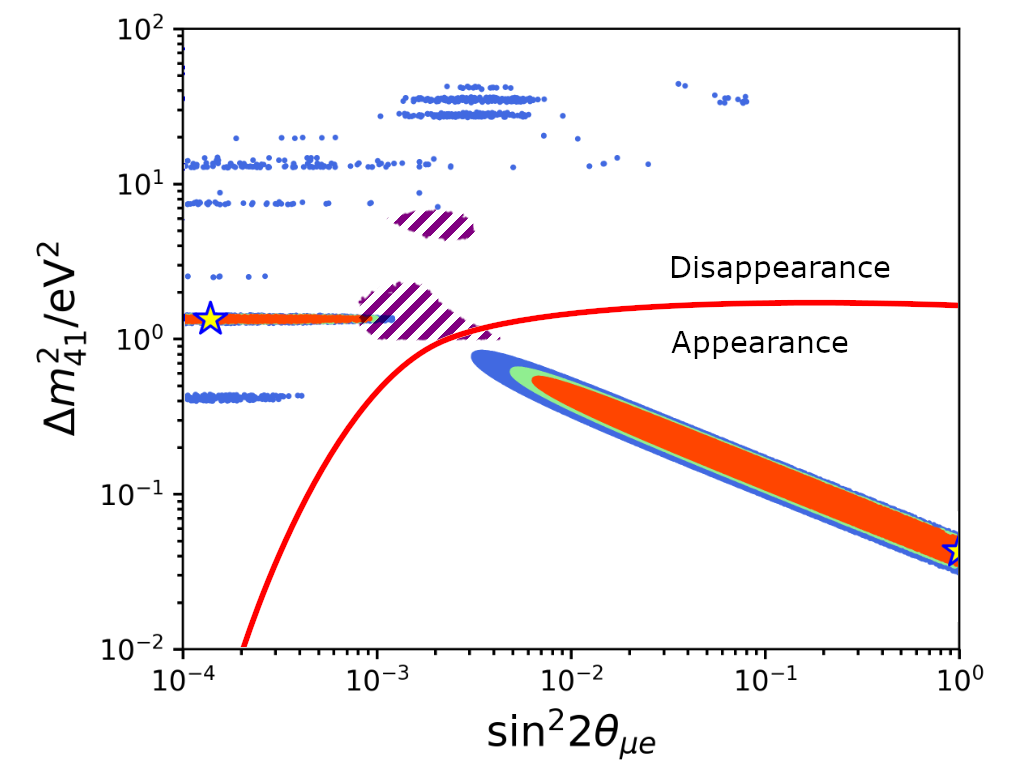}
    \caption{\label{fig:collin_fig1} Left: 3+1 global confidence regions (solid) are compared to Bayesian highest posterior density credible regions (68\%, 90\% and 99\% credible regions in black, red and blue). Right: Appearance-only (below red line) confidence regions compared to disappearance-only (above red line), with $\Delta m^2_{41} > 1$ eV${}^2$ appearance-only regions shown in hatched purple. All confidence regions are shown in red, green, and blue for 90\%, 95\%, and 99\%. The figures are taken from Ref.~\cite{Diaz:2019fwt}.}
\end{figure}

One example of the importance of the statistical treatment is shown in the left panel of Fig.~\ref{fig:collin_fig1} in which a Bayesian analysis is compared to the outcome of a frequentist one~\cite{Diaz:2019fwt}.
For the Bayesian analysis, the translucent black, red and blue regions represent the 68\%, 90\%, and 99\% highest posterior density credible regions, respectively, while the brighter colors refer to the frequentist analysis.
The allowed regions are quite different, which shows how the statistical treatment can affect the identification of promising parameter space and the statement of the tension between data sets.
Another example can be found in the right panel of Fig.~\ref{fig:collin_fig1}, where the allowed regions for the appearance (below the red line) and disappearance (above the red line) are shown~\cite{Diaz:2019fwt}. 
If a $\Delta m^2_{41}>1$ eV${}^2$ cut is applied to the appearance parameter space, the 99\% confidence region moves to the hatched purple region.
The points within this region are still preferred with respect to the null hypothesis at the $99\%$ level and, additionally, overlap with the disappearance regions, despite being disfavored when compared to the best-fit point in the 3+1 scenario.

Besides all that, since reactor experiments drive the preference for nonzero $U_{e4}$ mixing, there is an important caveat with respect to the reactor anomaly that should be discussed.
The reactor anomaly has originated in a discrepancy between data and theoretical expectations based on the calculations of reactor antineutrino fluxes. 
Nevertheless, a large, unexpected feature in the flux around 5~MeV has been identified, the so-called \emph{5~MeV bump}.
This outstanding feature lies outside the proposed theoretical uncertainties and puts in question the anomaly itself. 
Flux ratios can be used to mitigate the impact of the 5~MeV bump~\cite{Berryman:2020agd} but at the price of reducing the statistical power of the analysis.
Therefore, more precise calculations of the reactor antineutrino fluxes would help to further understand the reactor anomaly. A detailed discussion of this issue can be found in \cref{sec:expt_landscape_conventional_rates}.

\subsubsection{3+1 Light Sterile Neutrino Oscillations and Decoherence}

When considering the 3+1 model discussed above, one has assumed that neutrinos are always coherent.
However, as pointed out in Ref.~\cite{Arguelles:2022bvt}, due to the lack of detailed calculations of the neutrino production and detection mechanism, or from additional BSM effects, this is not guaranteed. 
In this section, we follow the discussion given in Ref.~\cite{Arguelles:2022bvt} and point out that when interpreting experimental data in the context of 3+1 this possibility has been overlooked. 
This fact could partially or completely resolve the existing tension between appearance and disappearance data sets.

Currently, when deriving constraints or preferred regions on the 3+1 model, the experiments assume that the neutrino state is a plane wave.
It is well-known that the plane-wave (PW) theory of neutrino oscillations~\cite{Eliezer:1975ja,Fritzsch:1975rz,Bilenky:1976yj} is a simplified framework that, upon careful inspection, contains apparent paradoxes~\cite{Akhmedov:2019iyt,Giunti:2003ax,Akhmedov:2009rb}.
These can be resolved by introducing the wave packet (WP) formalism~\cite{Nussinov:1976uw,Kayser:1981ye,Kiers:1995zj,Beuthe:2001rc,Akhmedov:2012uu,Akhmedov:2017mcc}.
The applicability of the PW approximation has been studied in detail for the standard mass-squared differences~\cite{Akhmedov:2009rb,Beuthe:2001rc,Giunti:2007ry,Bernardini:2004sw,Naumov:2010um} and has been shown to be a good approximation for current and future neutrino experiments.
However, this has not been shown to be the case for mass-squared differences relevant to the LSND observation.

In the WP formalism, the oscillation probability is given by
\begin{align} \label{eq:general_prob}
P_{\alpha\beta} =& \sum_{i=1}^n |U_{\alpha i}|^2|U_{\beta i}|^2 + 2\text{Re} \sum_{j>i}U_{\alpha i}U_{\alpha j}^*U_{\beta i}^*U_{\beta j}\exp\left\{-2\pi i\frac{L}{\Losc^{ij}}-2\pi^2\left(\frac{\sigma_x}{\Losc^{ij}}\right)^2 - \left(\frac{L}{\Lcoh^{ij}}\right)^2\right\},
\end{align} where $U_{\alpha i}$ are the neutrino mixing matrix elements and $L$ the experiment baseline.
Here the damping of the oscillations is parametrized by a length scale $\sigma_x$ that can be referred to as the wave packet size~\cite{Giunti:1997wq, Nussinov:1976uw, Kayser:1981ye, Kiers:1995zj, Beuthe:2001rc,Akhmedov:2012uu,Akhmedov:2017mcc} and depends on the neutrino production and detection mechanisms.
These lengths are defined as
\begin{equation} \label{eq:characteristic_lengths}
    \Losc^{ij} = \frac{4\pi E}{\Delta m^2_{ji}}\quad \text{and}\quad
    \Lcoh^{ij} = \frac{4\sqrt{2}E^2\sigma_x}{\Delta m^2_{ji}},
\end{equation} the oscillation and coherence lengths, respectively.
Here, $E$ is the energy of the neutrino and $\Delta m_{ji}^2$ the mass-squared difference between the $\nu_j,\,\nu_i$ mass eigenstates.
The two last terms in the exponential of Eq.~(\ref{eq:general_prob}) smear the oscillation.

Most experiments fulfill $\sigma_x \ll \Losc^{ij}$ and thus the first dampening term can be neglected.
This is not the case for the second one, which describes the decoherence arising from the separation of the mass eigenstates during their propagation at different velocities.
Notice that under a stationary situation, all the relevant information should be in the energy spectrum.
Therefore, this effect can be equivalently understood as an additional quantum uncertainty in the measurement of the true neutrino energy by the experiment~\cite{Stodolsky:1998tc}.

\begin{figure}[ht]
    \centering
    \includegraphics[width = 0.7\linewidth]{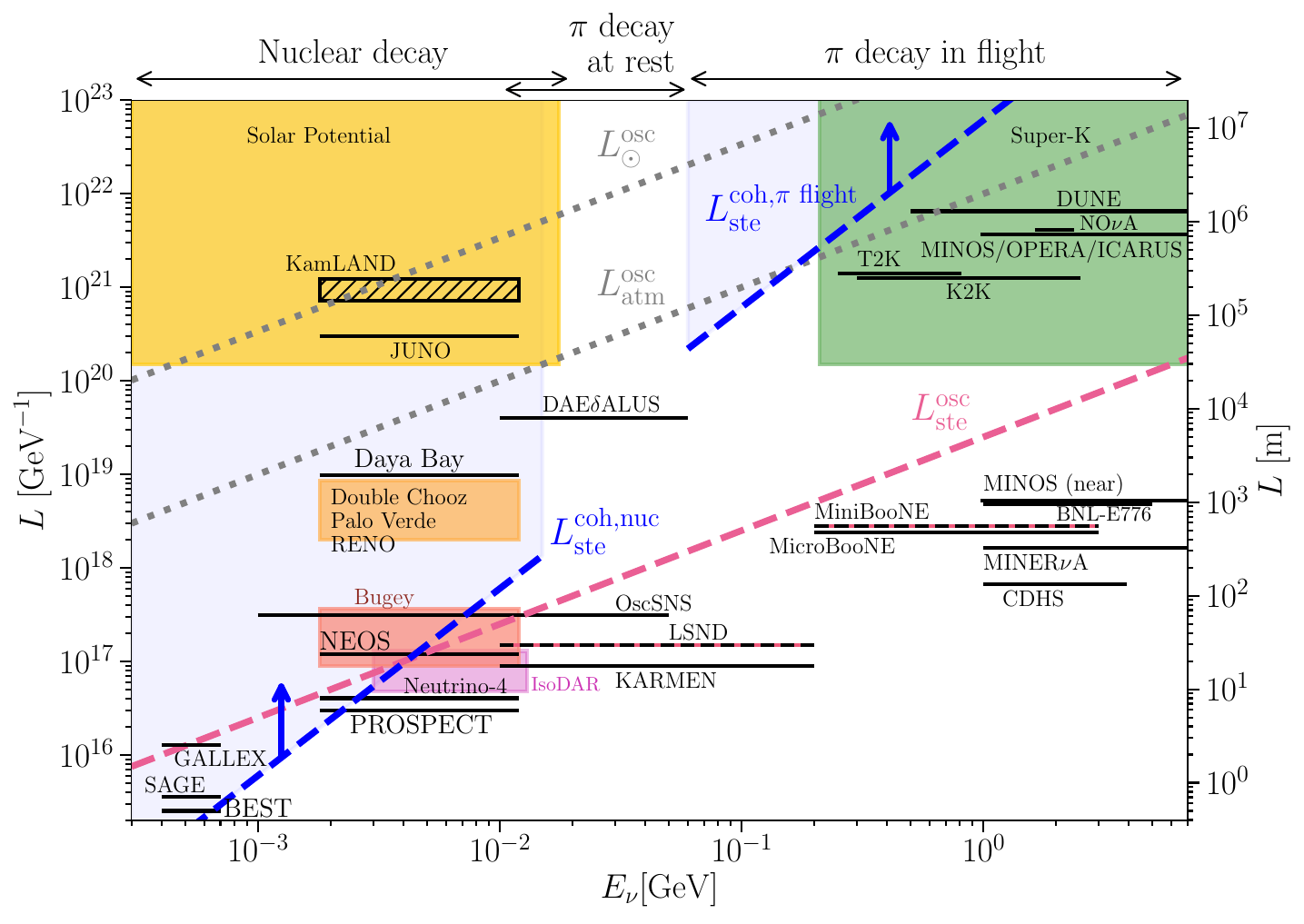}
    \caption{\textbf{\textit{Overview of the solar potential, neutrino experiments, and relevant scales.}} $L^{\text{osc}}$ (dotted gray and dashed pink) and $L^{\text{coh}}$ (dashed blue) are computed from Eq.~(\ref{eq:characteristic_lengths}) using $\Delta m_{41}^2 = \SI{1}\eV^2$ and $\sigma_x = 2.1\times 10^{-4}\text{ nm}$ for $L^{\text{coh,nuc}}_{\text{ste}}$~\cite{deGouvea:2021uvg}, and $\sigma_x = 10^{-11}$ m for $L^{\text{coh,}\pi\text{ flight}}_{\text{ste}}$~\cite{Jones:2014sfa}.
    Decoherence effects are expected at $L\gtrsim L^{\text{coh}}$.
    Figure is taken from Ref.~\cite{Arguelles:2022bvt}.}
    \label{fig:all-experiments}
\end{figure}

Figure~\ref{fig:all-experiments} shows several oscillation experiments compared to the sterile oscillation scale ($L_{\text{ste}}^{\text{osc}}$) and the decoherence scale ($L_{\text{ste}}^{\text{coh}}$).
For experiments with baselines smaller than $L_{\text{ste}}^{\text{coh}}$, decoherence can be neglected, while experiments with larger baselines will experience complete decoherence.
Take into account that the effect of not resolving fast oscillations experimentally is from an observational point of view identical to a decoherence effect.
Consequently, an experiment far above the $L_{\text{ste}}^{\text{osc}}$ line would also be effectively decoherent, and no effect due to $L_{\text{ste}}^{\text{coh}}$ would be manifest.
This narrows the region of interest for the decoherence of light sterile neutrinos to the low-energy region and in particular to the reactor and radioactive sources experiments. 

To show the impact of the wave packet separation~\cite{Arguelles:2022bvt} chooses the smallest value allowed by oscillation experiments for the wave packet size, $\sigma_x=2.1\times 10^{-4}{\rm nm}$~\cite{deGouvea:2021uvg, DayaBay:2016ouy}, and performs analyses searching for sterile neutrinos with and without the PW approximation.
Notice that this value is far from some first-order estimations of the wave packet in various contexts~\cite{Nussinov:1976uw, Kiers:1995zj,Akhmedov:2012uu,Akhmedov:2017mcc}; however, Ref.~\cite{Arguelles:2022bvt} decided to be agnostic and use an experimental result that should be robust even in more exotic scenarios.
 The global analysis performed in Ref.~\cite{Arguelles:2022bvt} considers the null results from Daya Bay~\cite{DayaBay:2016ggj, DayaBay:2016qvc}, NEOS~\cite{NEOS:2016wee}, and PROSPECT~\cite{PROSPECT:2018dtt,PROSPECT:2020sxr} and the anomalous results observed from radioactive sources by BEST~\cite{Barinov:2021asz}.

\begin{figure}[ht]
    \centering
    \includegraphics[width = 0.495\linewidth]{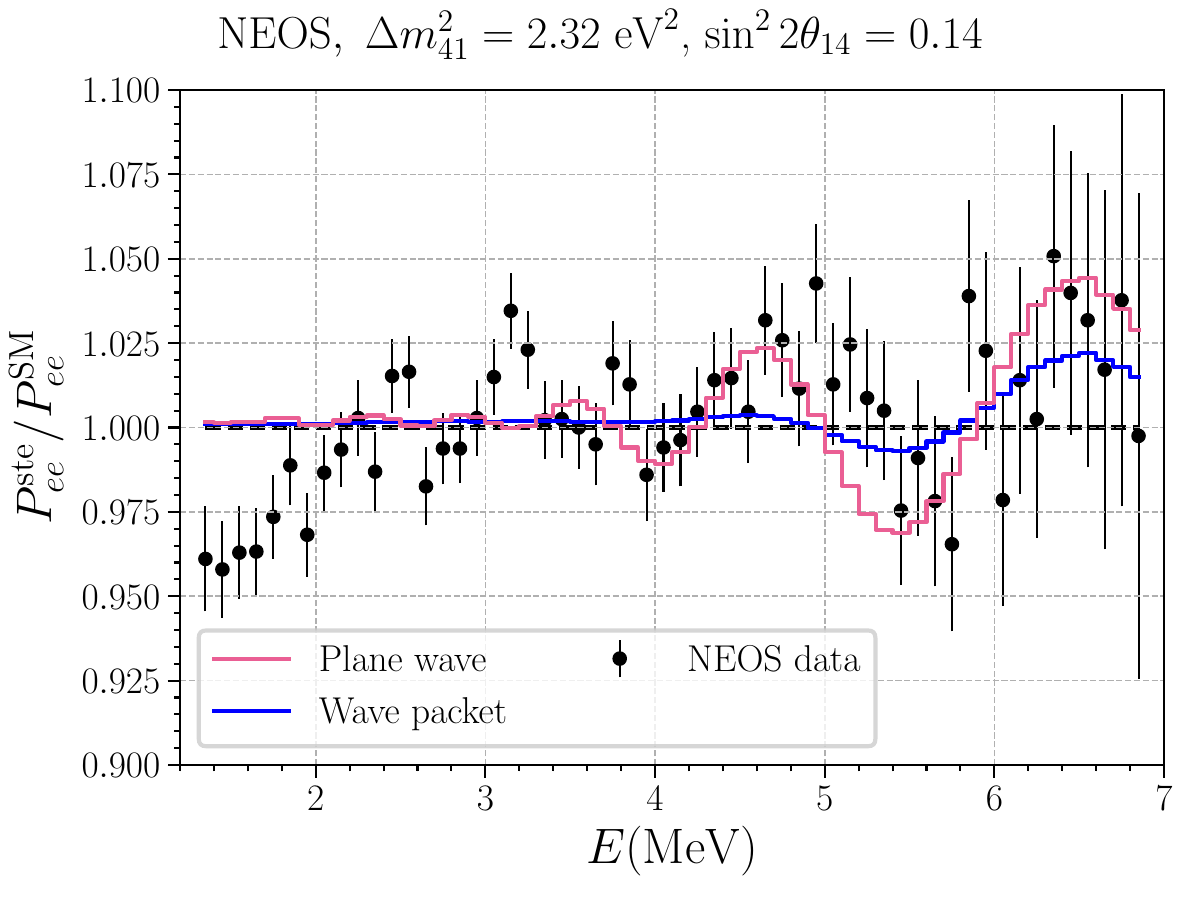}
    \includegraphics[width = 0.495\linewidth]{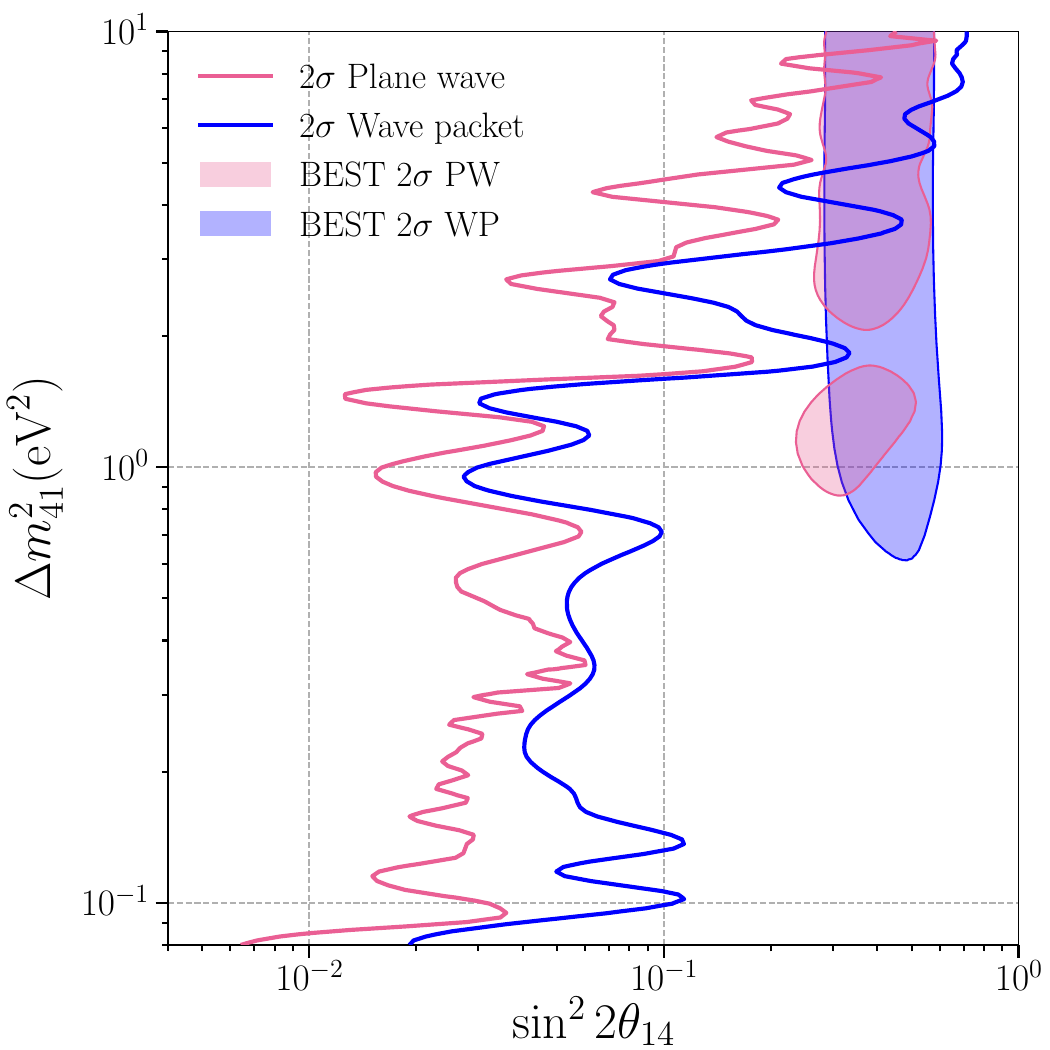}   
    \caption{\textbf{\textit{Impact of the smearing.}} On the left, the $y$-axis represents the ratio between the 3+1 and the 3 expected events in the NEOS experiment, for the reactor antineutrino anomaly (RAA) best-fit parameters~\cite{NEOS:2016wee}: $\Delta m^2_{41} = \SI{2.32}\eV^2$ and $\sin^22\theta_{14} = 0.14$. On the right, the solid pink and solid blue contours bound the exclusion region from Daya Bay, NEOS and PROSPECT; at two sigma for the PW approximation and WP formalism, respectively.
    The preferred region at two sigma for the BEST experiment is shaded for the PW approximation (pink) and the WP formalism (blue). Both figures are obtained for $\sigma_x = 2.1\times 10^{-4}\si\nm$.
    Figure is taken from Ref.~\cite{Arguelles:2022bvt}.}
    \label{fig:NEOSData}
\end{figure}

The main result of Ref.~\cite{Arguelles:2022bvt} is presented in Fig.~\ref{fig:NEOSData}, which shows the two-sigma exclusion contours for these experiments and the positive hint regions at two sigma by BEST, in both formalisms.
The WP results become compatible not only at large values of $\Delta m^2_{41}$ but also at the region around $\Delta m^2_{41} = \SI{2}\eV^2$.

Reference~\cite{Arguelles:2022bvt} finds that the damping of the oscillations due to the wave packet size may have important consequences for low-energy light sterile neutrino searches, accommodating apparently contradictory results. 
The result strongly motivates further studies to improve our understanding of the physics involved in the production and detection of the nuclear reactor and radioactive source neutrino experiments. 

\subsubsection{3+$N$ Light Sterile Neutrinos}\label{subsubsec:3plusN}

The generic model with three active and $N$ sterile neutrino states can be considered a viable explanation of the anomaly seen experimentally.
But it also provides a sterile-sector model-independent framework for non-unitarity~\cite{Fong:2016yyh,Fong:2017gke} tests.

We define the unitary mixing matrix ${\bf U}$ in the whole $(3+$N$) \times (3+$N$)$ space, and denote its $3 \times 3$ active space sub-matrix as $N$.
Then, the probability of active neutrino oscillation $P(\nu_\beta \rightarrow \nu_\alpha)$ in matter can be written in the simple form as~\cite{Fong:2017gke}
%
\begin{eqnarray}
\hskip -1cm P(\nu_\beta \rightarrow \nu_\alpha) &=& 
\mathcal{C}_{\alpha \beta} 
+ \left| \sum_{j=1}^{3} N_{\alpha j} N^{*}_{\beta j} \right|^2 
- 2 \sum_{j \neq k} 
\mbox{Re} 
\left[ (NX)_{\alpha j} (NX)_{\beta j}^* (NX)_{\alpha k}^* (NX)_{\beta k} \right] \nonumber\\
\times \sin^2 \frac{ ( h_{k} - h_{j} ) x  }{ 2 }
&-&
\sum_{j \neq k} \mbox{Im} 
\left[ (NX)_{\alpha j} (NX)_{\beta j}^* (NX)_{\alpha k}^* (NX)_{\beta k} \right] 
\sin ( h_{k} - h_{j} ) x, 
\label{eq:P-beta-alpha-final}
\end{eqnarray}
where $\alpha, \beta = e,\mu,\tau$ denote the active neutrino flavor indices, $i,j,k=1,2,3$ are the indices for the light mass eigenstates, and all oscillations involving heavier mass eigenstates with $m^2_{J} \gsim \SI{0.1}\eV^2$, which are dominantly sterile, are averaged out.

In Eq.~(\ref{eq:P-beta-alpha-final}), $P(\nu_\beta \rightarrow \nu_\alpha)$ is the leading term in a expansion by the active-sterile transition sub-matrix $W$ in ${\bf U}$~\cite{Fong:2017gke}.
The zeroth order Hamiltonian contains the $3 \times 3$ active space sub-matrix with the kinetic term plus matter potential given by $\text{diag} (a-b, -b, -b)$, where $a (b)$ denotes the Wolfenstein matter potential due to CC (NC) interactions, with a decoupled $N \times N$ sterile block.
Here, $h_{i}$ $(i=1,2,3)$ denote the energy eigenvalues of zeroth-order active states and $X$ is the unitary matrix which diagonalizes the zeroth-order active part Hamiltonian.
$\mathcal{C}_{\alpha \beta}$ in Eq.~(\ref{eq:P-beta-alpha-final}) is a probability leaking term which takes the same form in vacuum and in matter~\cite{Fong:2016yyh,Fong:2017gke} as 
\begin{eqnarray} 
\mathcal{C}_{\alpha \beta} \equiv 
\sum_{J=4}^{3+N}
\vert W_{\alpha J} \vert^2 \vert W_{\beta J} \vert^2.
\label{Cab} 
\end{eqnarray}

\begin{figure}[ht]
	\centering
	\includegraphics[width=0.7\textwidth]{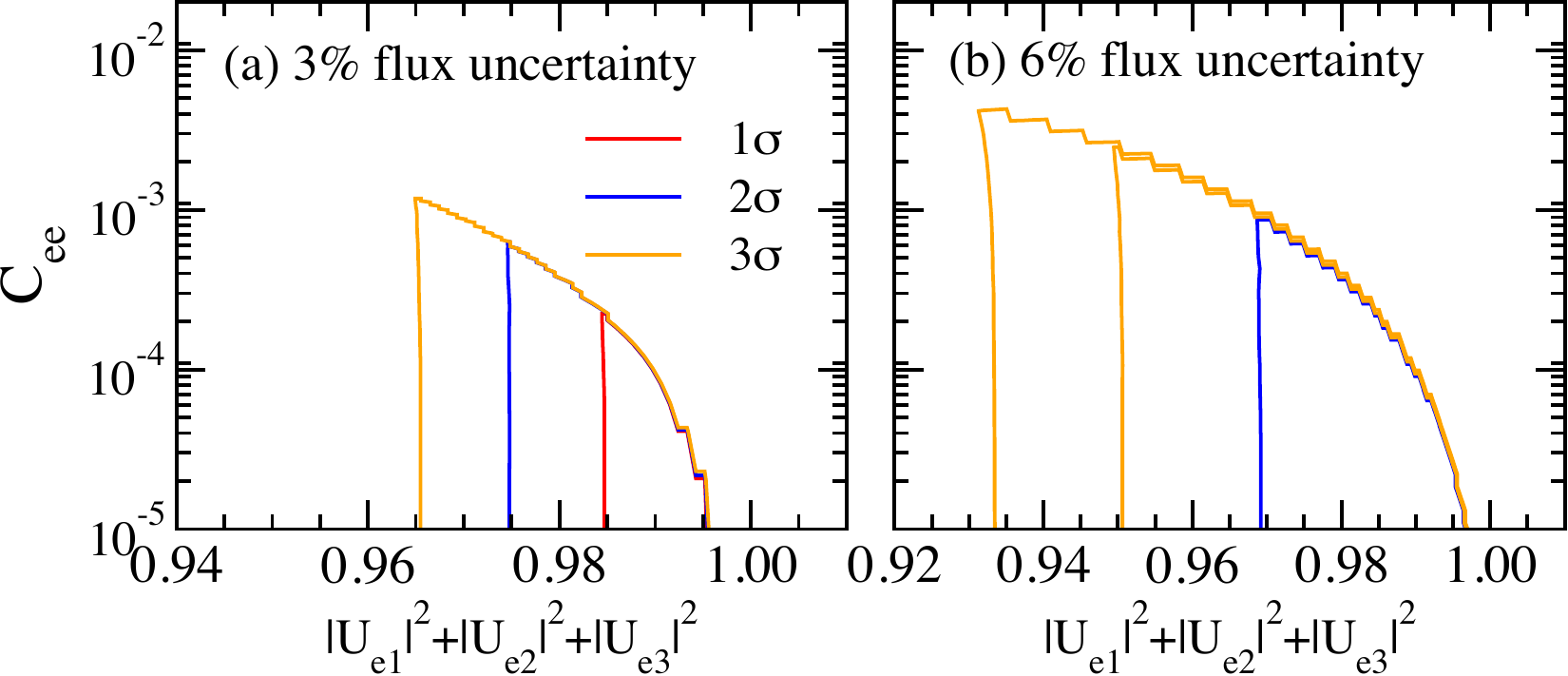}
\caption{
  Constraint in $\sum_{i=1}^3 |U_{ei}|^2- \mathcal{C}_{ee}$ space at 1$\sigma$, 2$\sigma$ and 3$\sigma$ CL expected to be
  obtained by JUNO like setting
  assuming the flux uncertainty of 3\% (left panel) and
    6\% (right panel).
  For details see Ref.~\cite{Fong:2016yyh}.
\label{fig:Cee-UV}
}
\end{figure}

\paragraph{The probability leaking and $W^2$ correction terms} 
\label{sec:correction-terms} 
In contrast to high-scale unitarity violation, observation of $\mathcal{C}_{\alpha \beta}$ in Eq.~(\ref{Cab}) would testify for low-scale unitarity violation.
Unfortunately, a detailed study of the sensitivity to high-scale unitarity, namely the constraints on $\mathcal{C}_{\alpha \beta}$, has only been performed for JUNO~\cite{Fong:2016yyh}; see Fig.~\ref{fig:Cee-UV}.

\begin{figure}[ht]
	\centering
	\includegraphics[width=0.50\textwidth]{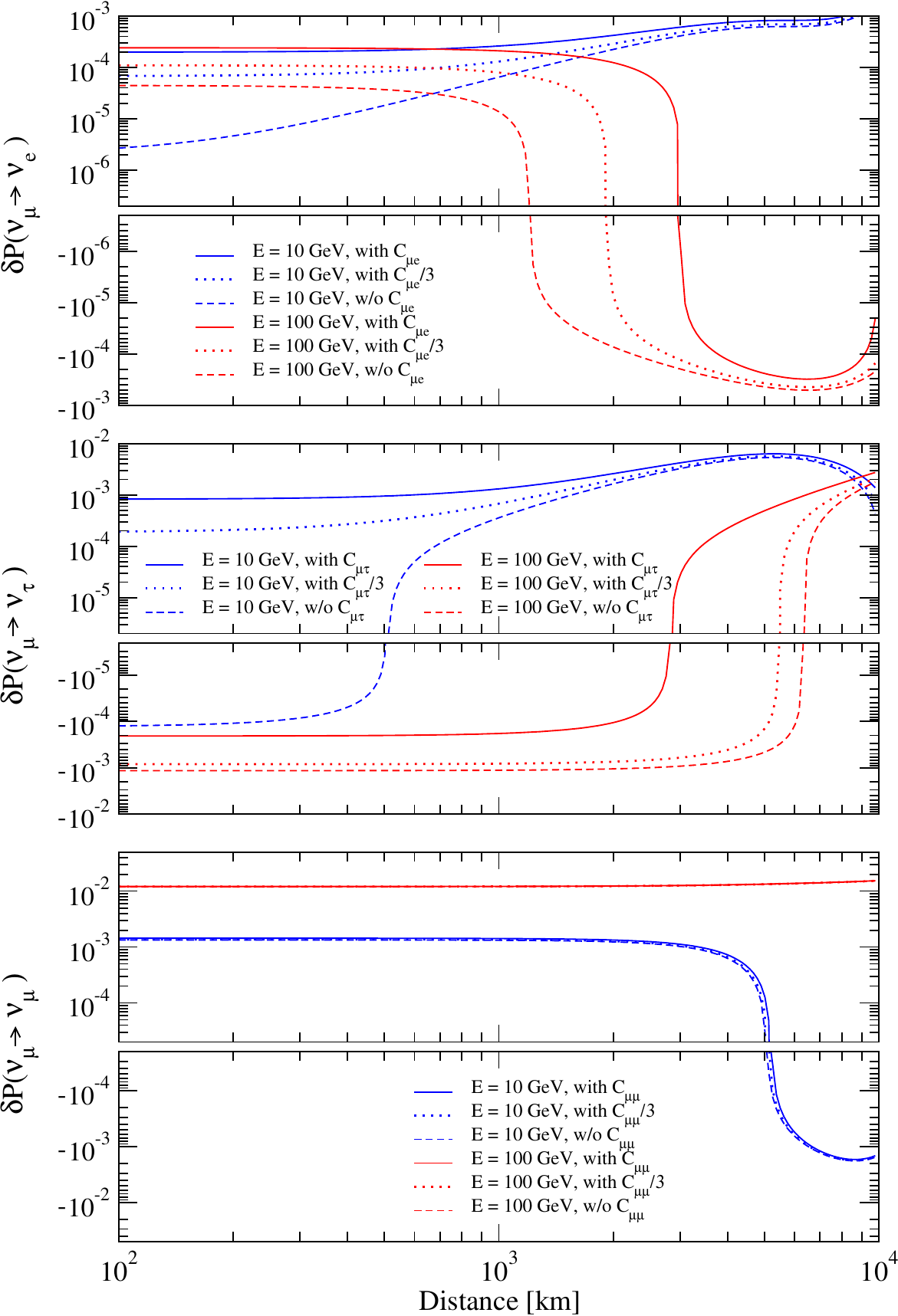}
\caption{ 
  The order $W^2$ correction terms, $\delta P(\nu_{\mu} \rightarrow \nu_{\alpha})  \equiv \mathcal{C}_{\mu \alpha} + P(\nu_{\mu} \rightarrow \nu_{\alpha})^{(2)}$, are shown as a function of the distance traveled by neutrinos in the Earth assuming a common $m_J^2 = \SI{0.1}\eV^2$. 
  The top, middle and bottom panels are for $\alpha = e, \tau$, and $\mu$, respectively.  
  In each panel the three cases are shown: $N=1$ with maximal $\mathcal{C}_{\mu \alpha}$ (solid line), the universal scaling model with $N=3$ (see Ref.~\cite{Fong:2017gke}, dotted line), and the order $W^2$ correction terms only (dashed line). 
  The blue (red) lines are for $E=10~(100)$~GeV, and the leaking term is taken as $(\mathcal{C}_{e \mu}, \mathcal{C}_{\tau \mu},\mathcal{C}_{\mu \mu}) = (20, 95, 9.6) \times 10^{-5}$ for $N=1$. For details see Ref.~\cite{Fong:2017gke}.
\label{fig:W-correction}
}
\end{figure}

Another unique prediction of the 3+$N$ model with low-mass sterile neutrinos is the presence of higher-order $W^2$ corrections.
These can be in principle measured and provide a way to differentiate this scenario from high-scale unitarity violation, where the mostly-sterile mass states are assumed to be kinematically forbidden and do not participate in neutrino oscillations.
The term is explicitly evaluated and plotted in Fig.~\ref{fig:W-correction}~\cite{Fong:2017gke}.
Notice the peculiar energy- and zenith-angle dependence of the term shown in Fig.~\ref{fig:W-correction}. 
The relevant energy region of $\rho E = 50 - 1000 \text{ (g/cm}^3) \text{GeV}$ may be explored by beam or atmospheric neutrino experiments; for example, Super-K, Hyper-K/HKK, DUNE, IceCube, or KM3NeT-ORCA, can probe this parameter space using low- to high-energy observables. 

Finally, another marked difference between low- and high-scale unitarity violation is that there is no deviation in the production and detection cross sections of neutrino.
This is because in the low-scale scenario both CC and NC vertices are not modified since all mass eigenstates, including the mostly-sterile states, are kinematically allowed in the processes.

\subsubsection{Light Sterile Neutrino Oscillations and Non-Standard Interactions}
\label{sec:steriles-and-NSI}

While many attempted solutions to the hints of anomalous $\nu_\mu\to\nu_e$ appearance at LSND and MiniBooNE have focused on those experiments, it is also conceivable to see if the strong constraints from MINOS/MINOS+~\cite{MINOS:2017cae} and IceCube~\cite{IceCube:2017ivd} could be alleviated.
As these constraints are at larger energies and over longer baselines, they would be subject to modification by non-standard neutrino interactions~\cite{Wolfenstein:1977ue} in either the active or sterile sector.
Thus a scenario with the usual sterile neutrinos explaining the short-baseline accelerator hints along with a new matter effect to modify the imprint of that sterile neutrino in the long-baseline accelerator and atmospheric constraints could be consistent with all the data.
This was investigated in~\cite{Karagiorgi:2012kw,Liao:2016reh,Liao:2018mbg,Denton:2018dqq}, which found that, while it could be possible to simultaneously explain some of the data sets in this fashion, explaining all seems to be impossible, even with both a sterile neutrino and a new interaction.
In particular, \cite{Denton:2018dqq} found that a model with a new interaction between sterile neutrinos and baryons provides an excellent fit to LSND, MiniBooNE, and IceCube data; but cannot simultaneously fit MINOS+ data due to a disagreement in the preferred values of $\theta_{34}$. Other approaches, such as \cite{Asaadi:2017bhx}, used beam neutrinos forward scattering off of a locally overdense relic neutrino background to give rise to a matter effect with an energy-specific resonance that can reproduce the MiniBooNE observed excess. 

More recently, \cite{Alves:2022vgn} reiterated that these tensions with long-baseline experiments occur because new matter effects generically distort the active (anti-)neutrino mixing and mass spectrum.  A dark sector model with both neutrino and vector portals was then proposed in \cite{Alves:2022vgn} that avoids these large active spectrum distortions and is fully compatible with long-baseline experiments, including T2K, NOvA, MINOS/MINOS+, IceCube/DeepCore, and KamLAND.  In this model, quasi-sterile neutrinos from a dark sector are charged under a light vector mediator with feeble couplings to SM fermions. This leads to new matter effects that generate resonant active-to-quasi-sterile neutrino oscillations within a narrow window of energy, $\sim$250–350~MeV, to explain the MiniBooNE low energy excess. The MiniBooNE excess at mid-to-high energies, $E_{\nu} \gtrsim 400$~MeV, as well as the LSND and Gallium anomalies, are explained by an additional ‘vanilla’ sterile neutrino which does not participate in the resonant oscillations. Besides being fully testable by the SBN program, the new matter effects in this model have interesting implications for solar neutrinos.

\subsubsection{Decaying Light Sterile Neutrinos}
\label{sec:light-sterile-decay}

As discussed above, global fits to the neutrino data show that the 3+1 sterile neutrino model suffers from internal inconsistency amongst the datasets~\cite{Maltoni:2003cu, Diaz:2019fwt}.
This ``tension'' in the global fits motivates considering more complicated physics scenarios.
More complicated physics scenarios could involve neutrino decay.
Neutrinos are not protected from decay in the SM, \textit{i.e.} radiative neutrino decay of the two heavier of the three known neutrino mass states ($\nu_1$, $\nu_2$, and $\nu_3$) can occur, albeit, extremely slowly~\cite{Pal:1981rm,Nieves:1982bq}.
Scenarios that include decay of the eV-scale neutrino mass state, $\nu_4$, are referred to as ``3+1+decay.''

The class of 3+1+decay models can be divided into two scenarios: visible decay and invisible decay.
In the visible decay scenario, one of the decay daughters is an active neutrino, which could be detected, while the other is undetectable, Beyond the Standard Model (BSM).
In the invisible decay scenarios, all decay daughters are BSM and invisible, $i.e.$ undetectable.
In either scenario, an additional degree of freedom is introduced to those from a 3+1 model: strength of the coupling that mediates the decay, which determines the $\nu_4$ lifetime.
Decay of the $\nu_4$ state causes a dampening of oscillations, resulting in different neutrino transition probabilities than in the 3+1 model.

\paragraph{Invisible decay}

The 3+1+decay model involving \textit{visible} and \textit{invisible} neutrino decay was explored in the case of the IceCube Neutrino Observatory~\cite{Moss:2017pur}.
IceCube can search for anomalous muon-neutrino disappearance due to the existence of eV-scale sterile neutrinos~\cite{IceCube:2020phf, IceCube:2020tka}.
It was shown that this model can change the interpretation of the IceCube one-year null result, which had set a strong constraint on the 3+1 model~\cite{IceCube:2016rnb}.

The 3+1+decay model with \textit{invisible} decay was fit to short-baseline data in Ref.~\cite{Diaz:2019fwt}, and subsequently, fits to the IceCube one-year dataset were combined with the short-baseline fits~\cite{Moulai:2019gpi}.
This model improves over the 3+1 model with a $\Delta \chi^2$ of 9.0, with one additional degree of freedom (DOF). 
The aforementioned tension in the global fits can be quantified with a \textit{parameter goodness-of-fit}~\cite{Maltoni:2003cu}.
The 3+1+decay model reduces the tension from a $\chi^2$/DOF of 28/2 to 19/3.
 
A search for the invisible 3+1+decay model using eight years of IceCube has found a preference for this model over either the three-neutrino or 3+1 models~\cite{Moulai:2021zey}.
Under the assumption of Wilks' theorem, the three-neutrino model is disfavored with a $p$-value 3\% and the 3+1 model is disfavored with a $p$-value of 5\%.
Incorporation of this result into global fits is expected to further reduce the tension from what was found in Ref.~\cite{Moulai:2019gpi}.
The Short-Baseline Neutrino Program at Fermi National Accelerator Laboratory offers another opportunity to search for this model of sterile neutrinos~\cite{Machado:2019oxb}.

\paragraph{Visible decay} Visible decays were discussed in Ref.~\cite{PalomaresRuiz:2005zbh} as an explanation of the LSND results. There, the authors proposed that a mostly-sterile neutrino $\nu_4$, of either Dirac of Majorana nature, could be produced in $\mu^+$ decays and decay into $\nu_e$ and $\overline{\nu}_e$ in between the source and the detector, thereby leading to an effective flavor transition. The decay was fast due to the interactions of $\nu_4$ with a massless scalar particle, $\phi$. Subsequently, Refs.~\cite{Bai:2015ztj,deGouvea:2019qre,Dentler:2019dhz} expanded on this scenario and argued that it could explain the MiniBooNE excess as well. 

In Ref.~\cite{Dentler:2019dhz}, a SU$(2)$-invariant model is proposed where a Dirac sterile neutrino $\nu_s$ interacts with a massive scalar particle $\phi$. In the mass basis, the mostly-sterile state $\nu_4$ mixes with the electron- and muon-neutrinos, and therefore can be produced in both $\pi^+$ and $\mu^+$ decays. The relevant interaction Lagrangian is given by,
\begin{equation}\label{eq:sterile_decay_vis}
    \mathcal{L} \supset - g_s U_{si}^* U_{s4} \phi \overline{\nu_4}\nu_i - \sum_{\alpha} \frac{g}{\sqrt{2}} U_{\alpha 4}  \overline{\nu_4} \slashed{W} \ell_{\alpha} + \text{ h.c.},
\end{equation}
where $U_{si}$ is the mixing between the sterile state $\nu_s$ and the massive eigenstate $\nu_i$, $g_s$ the parity-conserving, sterile neutrino coupling to the scalar $\phi$, and $g$ is the weak coupling constant.

\begin{figure}[ht]
    \centering
    \includegraphics[width=0.49\textwidth]{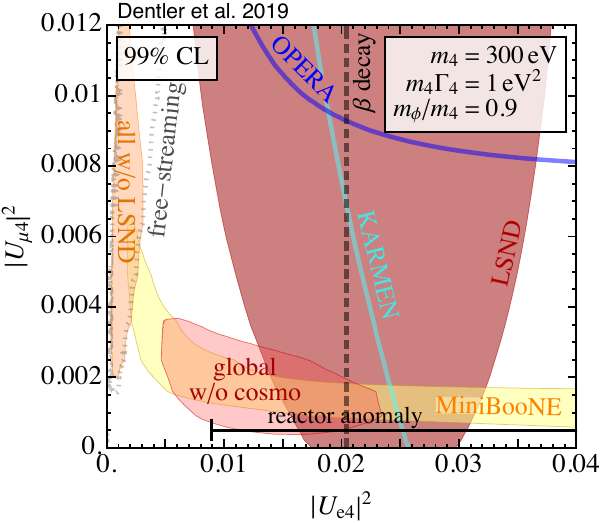}
    \caption{The preferred mixing of the light sterile neutrino with the muon and electron flavors, $|U_{e4}|^2$ and $|U_{\mu 4}|^2$, to explain the MiniBooNE and LSND anomalies in the decaying-sterile-neutrino model of \cref{eq:sterile_decay_vis}. The figure is taken from \cite{Dentler:2018sju}. 
    \label{fig:theory:visibledecay}}
\end{figure}

The authors in Ref.~\cite{Dentler:2019dhz} performed a fit to the MiniBooNE and LSND data. The results are shown in Fig.~\ref{fig:theory:visibledecay}. While most of the signal at MiniBooNE comes from $\pi^+\to \mu \nu_4$ production, at LSND, both $\pi^+\to\mu^+$, $\mu^+\to e^+ \nu_4 \overline{\nu_\mu}$ as well as $\mu^+\to e^+ \nu_e \overline{\nu_4}$ contribute to the rate of inverse-beta-decays. This is because of the subsequent $\phi \to \nu \overline{\nu}$ decays, which generate an apparent $\nu\to \overline{\nu}$ transition. This effective transition is strongly constrained by solar antineutrino searches~\cite{Hostert:2020oui}.

The presence of the scalar degree of freedom in the theory also helps reconcile the model with cosmology~\cite{Dasgupta:2013zpn,Hannestad:2013ana,Chu:2018gxk,Yaguna:2007wi,Saviano:2013ktj,Giovannini:2002qw,Bezrukov:2017ike,Farzan:2019yvo,Cline:2019seo,Dentler:2019dhz,Archidiacono:2020yey,DiValentino:2021rjj}. The secret, self-interactions in the sterile sector provide a new matter potential for sterile neutrinos in the early universe that suppresses their production. Another possibility is that the steriles interact with an ultra-light dark matter background, which also suppresses production in the early Universe~\cite{Cline:2019seo,Farzan:2019yvo}. For more details, see Sec.~\ref{sec-5:cosmo:evade_cosmo}.


\subsubsection{Lepton Number Violating Muon Decays}
\label{sec:LNV-muon-decay}

In this section, we focus on the possibility of using Lepton Number Violation (LNV) muon decays in addition to neutrino oscillations as an explanation for the LSND experiment.
While not necessarily providing a full solution to the short-baseline puzzle, this scenario is worth considering because it is allowed by all data, it can be realized in explicit models, and is testable. 
Additionally, when considered in tandem with the 3+1 model, it opens up some parameter space in the 3+1 neutrino oscillations scenario by accounting for some of the LSND observation.

Lepton-flavor violating non-standard interactions (NSI) are very strongly constrained by theoretical consistency requirements and charged lepton flavor experiments.
It has been pointed out in~\cite{Bergmann:1998ft} that $\Delta L \neq 0$ interactions can evade these constraints.
In~\cite{Babu:2016fdt} it was shown that, while most $\Delta L \neq 0$  effective operators are strongly constrained by high-precision measurements of the Michel parameters in muon decays, two such operators retain the SM prediction of $\rho=\delta=3/4$ and are thus allowed.
In addition, theoretical models that led to these two self-consistent effective operators were also developed.
These effective operators are:
\begin{align}
{\cal L}_1&\rightarrow[(\bar{\mu}_{R}\nu_{eL})(\nu_{aL}^{T}Ce_{L})-(\bar{\mu}_{R}\nu_{eL})(\ell_{aL}^{T}C\nu_{eL})]\langle |H^{0}|\rangle \ ,
\\
{\cal L}_2&\rightarrow (e_{L}^{T}C\nu_{eL})(\mu_{R}^{T}C\nu_{R})^{*}\langle |H^{0}|\rangle^{2} \ .
\end{align}

This type of NSI would lead to $\mu^+\to e^++\bar\nu_\mu+\bar\nu_e$, which would directly contribute to the muon decay-at-rest (DAR) signal in LSND.
Accommodating the entire DAR signal through such NSI would conflict with the KARMEN experiment, which also used a muon DAR beam.
However, it is possible to achieve good agreement between the two experiments when one combines the LNV NSI and oscillations through a sterile neutrino, due to the differences in baselines.
The LNV NSI would clearly not affect the pion decay beams in the other short-baseline accelerator experiments.
The oscillation parameters obtained in~\cite{Babu:2016fdt} in the presence of the LNV NSI for LSND+KARMEN are compatible with those of global fits that include the MiniBoone and other data which rely on neutrinos from semileptonic pion decays.

It is thus interesting to consider how the presence of such LNV NSI can change the allowed sterile neutrino oscillation parameter space through the additional contribution to the LSND signal.
This scenario would be testable by the different effects it produces in muon decay versus hadronic decay beams.
In specific model realizations, it might also be possible to observe the effects of the new particles associated with the generation of the LNV operators in future collider experiments.


\subsubsection{Large Extra Dimensions and Altered Dispersion Relations}
\label{sec:extra-dimensions}

Scenarios with sterile neutrino altered dispersion relations (ADRs) adopt additional terms in the standard dispersion relation $E^2=|\vec{p}|^2 +m^2$.
These terms make the oscillation amplitude energy-dependent, thus offering more freedom to accommodate various constraints and anomalies arising in short-baseline neutrino experiments.
There exist various realizations of this scenario, including Lorentz violation and sterile neutrino shortcuts in warped extra dimensions~\cite{Pas:2005rb,Carena:2017qhd,Doring:2018ncz}.
The effect implied resembles standard matter effects but features a different energy dependence and typically applies for neutrinos and antineutrinos in the same way.

The basic idea is that active-sterile neutrino mixing is unaltered at low energies; however, a resonance conversion is present when the effect of the ADR minimizes the effective mass squared difference between active and sterile neutrinos.
This effect is suppressed significantly for energies above the resonance energy. 
This allows to decouple sterile neutrinos at high energy and to evade stringent constraints from atmospheric and long-baseline accelerator neutrino experiments while offering the possibility to make small active-sterile neutrino mixing, observed or constrained in solar and reactor neutrino experiments, compatible with anomalies in short-baseline experiments such as LSND and MiniBooNE.   

\begin{figure}[ht]
    \centering
    \includegraphics[width=0.6\textwidth]{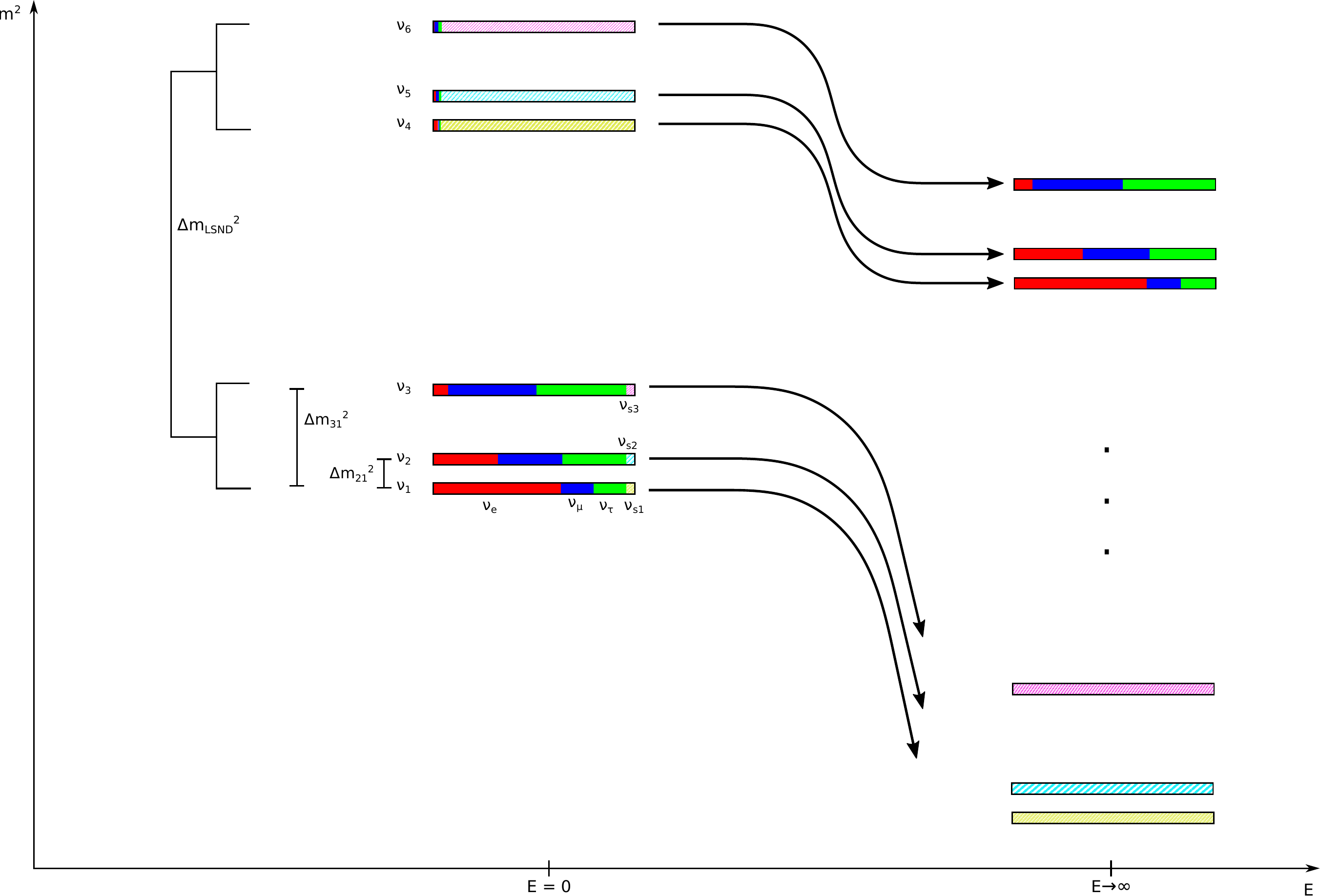}
    \caption{\label{fig:paes-0} Evolution of various $\Delta m^2$'s in ADR scenarios: symbolically. Figure from~\cite{Doring:2018cob}.}
\end{figure}

A challenge of such approaches is to find data sets compatible with all constraints in a complete framework with three active neutrinos.
As has been pointed out in~\cite{Doring:2018cob}, the requirement to obtain the same flavor structure of active neutrinos below and above the resonance requires three sterile neutrinos whose mass squared differences reflect the active neutrino mass spectrum and feature democratic active-sterile flavor mixing.
Moreover, different ADRs and, consequently, resonance energies are necessary for the three sterile flavors to avoid the cancellation of oscillation effects.
In Fig.~\ref{fig:paes-0}, the evolution of the various effective $\Delta m^2$'s is shown symbolically~\cite{Doring:2018cob}.  

\begin{figure}[ht]
    \centering
    \includegraphics[width=0.75\textwidth]{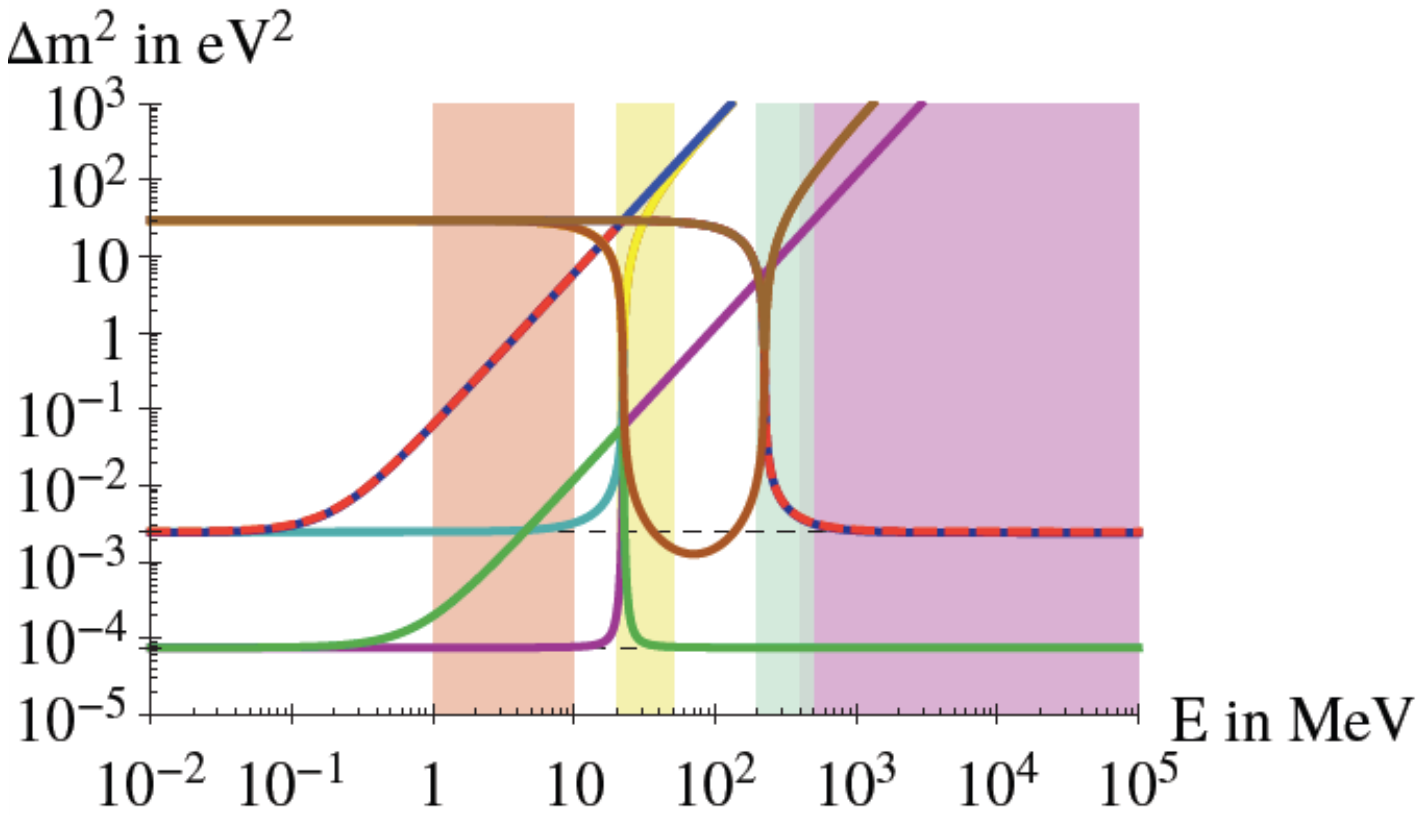}
    \caption{\label{fig:paes-1} Evolution of various $\Delta m^2$'s for the specific example of BMP4.
    The vertical colored regions correspond from left to right to the energy ranges probed by reactor and Gallium experiments, LSND, MiniBooNE, and long-baseline accelerator experiments, respectively.
    Figure  from~\cite{Doring:2018cob}.}
\end{figure}

\begin{figure}[ht]
    \centering
    \includegraphics[width=0.49\textwidth]{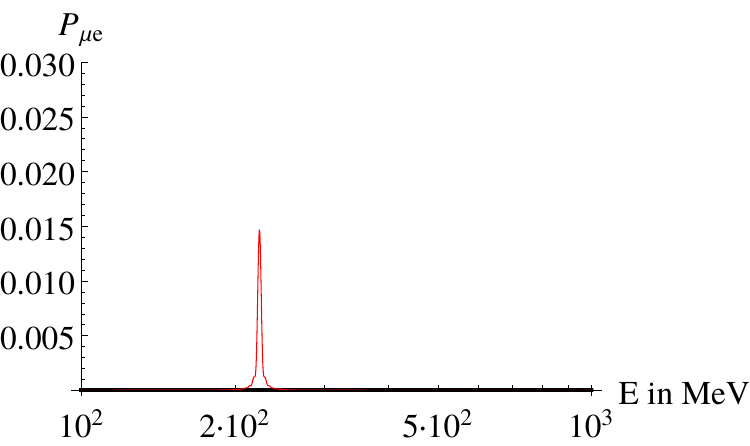}
    \includegraphics[width=0.49\textwidth]{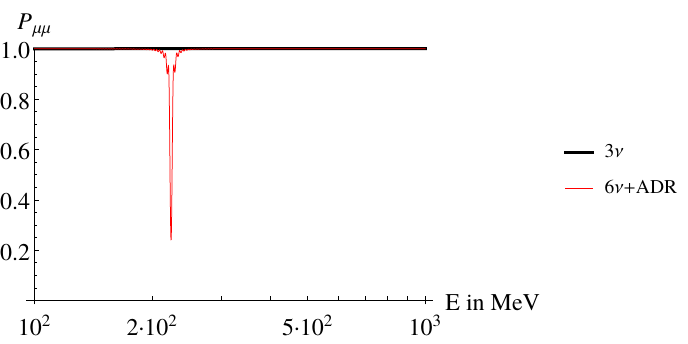}
    \caption{\label{fig:paes-2} $P(\nu_\mu \rightarrow \nu_e)$ and $P(\nu_\mu \rightarrow \nu_\mu)$ oscillation probabilities for MiniBooNE and the specific example of BMP4. Figures from Ref.~\cite{Doring:2018cob}).}
\end{figure}

The consequent parameter space has five parameters beyond that of the SM with three massive, active neutrinos; namely a universal active-sterile $\Delta m^2$ and mixing $\sin^2 \theta$ plus three ADR parameters or resonance energies for the sterile flavors, respectively.
As has been stressed in Ref.~\cite{Barenboim:2019hso}, where the phenomenology of two exemplary data points has been studied, it is not an easy task to find a combination of parameters that is compatible with all constraints, especially with both MiniBooNE and T2K that probe similar energy regions.
Obviously, a conclusive verdict on the potential of ADR models would require a thorough scan of the complete parameter space.
In Ref.~\cite{Doring:2018cob}, various promising Benchmark Mark Points (BMPs) have been analyzed.
In particular BMP4 (see Fig.~\ref{fig:paes-1} for the energy dependence of $\Delta m^2$'s for this concrete  data set) looks promising in this respect, as it leads to a muon-neutrino disappearance and electron-neutrino appearance probability that is sharply peaked around the resonance energy $E_{\rm Res}=223.6$~MeV (see Fig.~\ref{fig:paes-2}).
Moreover, BMP4 features a rather small active-sterile mixing angle $\sin^2\theta=10^{-4}$ that seems not to be excluded by MicroBooNE, according to the analysis in Ref.~\cite{Arguelles:2021meu}.
In fact, the large $\Delta m^2$ or order $\SI{30}\eV^2$ leads to fast oscillations that allow exploiting the difference in baselines between MiniBooNE and MicroBooNE ($\SI{541}\m$ versus $\SI{470}\m$) that amounts to roughly 8\% of the oscillation length 
(corresponding to an oscillation probability reduced by 25\% at resonance for MicroBooNE with respect to MiniBooNE) and 
that may be increased by finetuning the parameters. 

In summary, ADRs represent a well-motivated and interesting tool to alter standard neutrino oscillations that may prove useful to explain the intricate framework of anomalies and constraints characterizing short-baseline neutrino oscillations. 


\subsubsection{Lorentz Violation}

The Standard-Model Extension (SME) is an effective field theory framework to look for Lorentz symmetry violation (LV)~\cite{Kostelecky:2003cr}.
The main interest of LV as an explanation to short-baseline anomalies is the flexibility of the SME-based Hamiltonian.
One could design a suitable Hamiltonian using the SME to reproduce all neutrino and antineutrino oscillation data without the standard neutrino mass term.
For example, the bicycle model~\cite{Kostelecky:2003xn} has the seesaw-mechanism-like texture to reproduce $L/E$ oscillation behavior at high energy, even though Hamiltonian only has $CPT$-odd SME coefficient $a$ and $CPT$-even SME coefficient $c$ that make $\sim L$ or $\sim L\cdot E$ oscillation behaviors.
The effective Hamiltonian of the bicycle model in $3\times 3$ flavor basis matrix has following the texture, 
\begin{eqnarray}
  h_{eff}\sim\left(
  \begin{array}{ccc}
    aE&c&c\\
    c&0&0\\
    c&0&0
  \end{array}
  \right)~.
 \end{eqnarray}
This model demonstrates the possibility that LV can imitate the standard three massive neutrino oscillations.
The tandem model~\cite{Katori:2006mz} follows this, which can reproduce existing neutrino data at that time, including LSND. 

\begin{figure}[ht]
    \centering
    \includegraphics[width=0.49\textwidth]{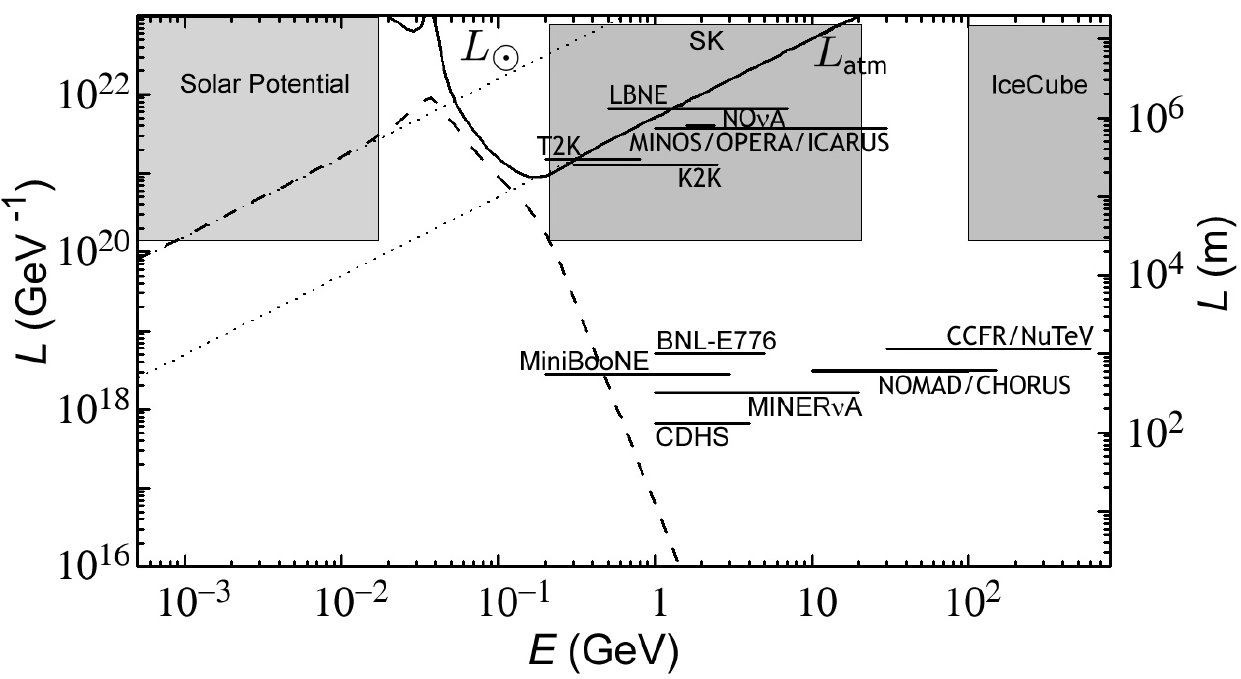}
    \includegraphics[width=0.49\textwidth]{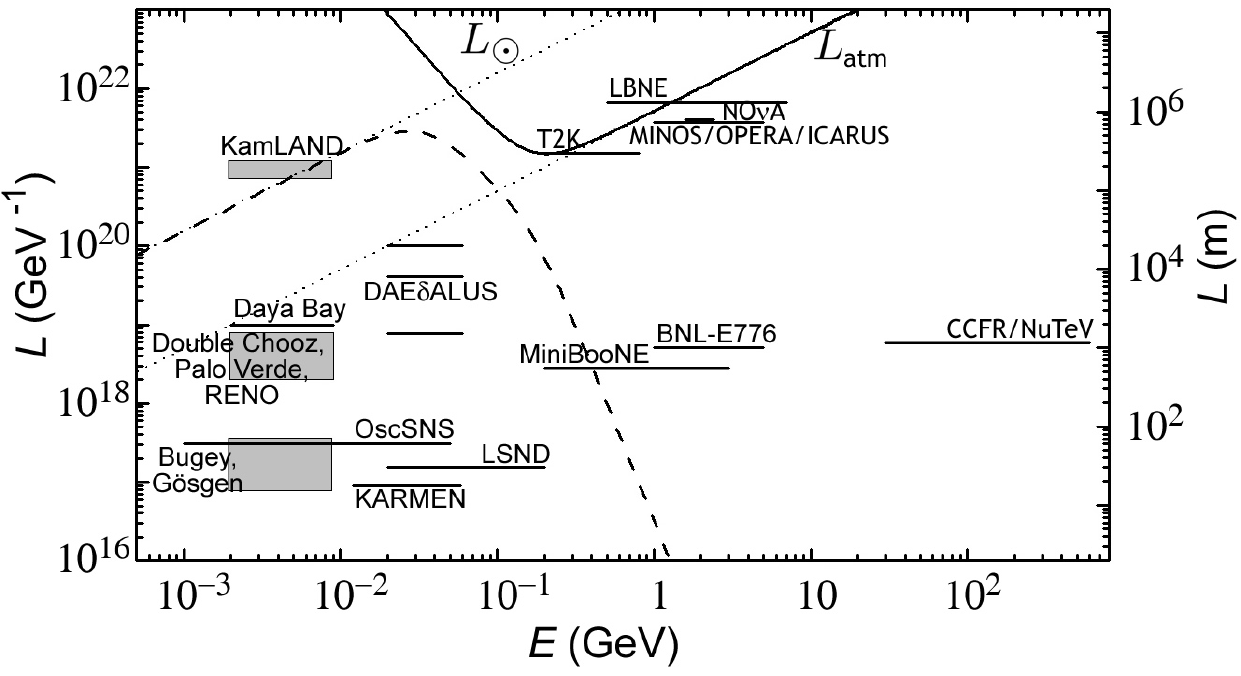}
    \caption{\label{fig:fig1} Solutions of the puma model~\cite{Diaz:2010ft,Diaz:2011ia} for neutrinos (left) and antineutrinos (right) are shown by solid and dashed curves.
    Dotted lines are solutions from the solar and atmospheric $\Delta m^2$.
    The horizontal axis is the energy and the vertical axis is the propagation length.
    Experimental regions are mapped by boxes or segments.}
\end{figure}

The SME Lagrangian can be extended to include higher-order terms~\cite{Kostelecky:2011gq}.
Since LV is related to a new space-time structure motivated by quantum gravity, it is natural to expect LV to show up in the non-renormalizable operators of the effective field theory.
This further increases the number of model-building possibilities. The puma model~\cite{Diaz:2010ft,Diaz:2011ia}, for instance, is based on higher-order SME terms.
The relevant Hamiltonian for this model is given by
\begin{eqnarray}
  h_{eff}\sim
  \frac{m^2}{2E}\left(
  \begin{array}{ccc}
    1&1&1\\
    1&1&1\\
    1&1&1
  \end{array}
  \right)
  +
    aE^2\left(
  \begin{array}{ccc}
    1&1&1\\
    1&0&0\\
    1&0&0
  \end{array}
  \right)
  +
   cE^5\left(
  \begin{array}{ccc}
    1&0&0\\
    0&0&0\\
    0&0&0
  \end{array}
  \right),
 \end{eqnarray}
where $m^2$, $a$, and $c$ are tunable parameters.
Suitable choices of these three parameters in this texture can reproduce all neutrino data at that time, including LSND and MiniBooNE.
However, at present time this model is in tension with measurements by MINOS.
The solution of this Hamiltonian provides energy-dependent oscillation lengths shown in Fig.~\ref{fig:fig1}.
One line is used to reproduce solar and reactor neutrino data at low energy, and another line is used to reproduce atmospheric and long-baseline muon-neutrino disappearance data.
Around $30$ to $300$~MeV, these two lines drastically change shapes, and this region is used to reproduce LSND and MiniBooNE data.
The model also uses a $CPT$-odd term which can make a difference between neutrino and antineutrino results.
However, this model does not produce short-baseline reactor neutrino disappearance data (Daya Bay, Double Chooz, Reno) and long-baseline electron neutrino appearance data (T2K, NOvA).
Thus, at present, this model is rejected as an explanation of the short-baseline puzzle.

\paragraph{Future of neutrino oscillation models based on Lorentz violation}

It may be possible to construct an LV-based neutrino oscillation model beyond the puma model to reproduce all existing data including LSND, MiniBooNE, and other short-baseline results.
However, such a model would have more fine-tuned parameters with an unusual texture or artificial cutoffs.

The difficulty to construct such a model is because LV-motivated terms have zero  ($\propto E^0$) to a higher power with energy ($\propto E^1, E^2,\ldots$), and they dominate neutrino oscillations at high energy.
So parameters introduced to explain short-baseline anomalies in general conflict with other oscillation data due to the lack of $L/E$ oscillation behavior, which requires $E^{-1}$ term in the Hamiltonian.
One possibility is to introduce an unusual texture discussed above because they can reproduce $L/E$ behavior with fine-tuned parameters.
Another possibility is to introduce fine-tuned cutoffs in LV terms so that LV terms are limited to only certain regions to explain short-baseline anomalies.
Therefore, LV-based models to explain short baseline anomalies would be unnatural, even if they exist.
This is true for any other similar approach based on effective field theory, regardless of whether they are Lorentz violating or not.

Finally, the LV-based neutrino oscillation models can also be tested by studying the time dependence of the anomalies.
This is the smoking-gun signature that differentiates this proposal from others.
Interestingly, the MiniBooNE antineutrino data set shows a preference for a non-zero time-dependent LV component; however, this is in tension with the constraint on LV from MINOS.


\subsection{Dark Sectors in Scattering and in the Beam}\label{sec:th_landscape:darksectors}

The difficulty of resolving the various short-baseline anomalies by invoking solely neutrino flavor transformations, as detailed throughout Sec.~\ref{sec:theory:flavorconv}, has led to more exotic proposals, where light dark sectors can be produced alongside neutrinos in the beam or inside neutrino detectors, and mimic the experimental signatures. These model scenarios typically explain, for instance, the MiniBooNE and LSND results without violating the null results from other experiments.

In this subsection, we highlight a few related classes of models that fit this description -- Sec.~\ref{subsubsec:TransitionMagneticMoment} discusses a model in which neutrinos are endowed with large transition magnetic moments, where upscattering from a light SM neutrino into a heavier neutral lepton $N$ can be mediated through photon exchange with the nucleus. Section~\ref{subsubsec:DarkNeutrinos} details a model in which $N$ interacts with a light-dark photon, where light-neutrino upscattering into this new state is mediated by the new force carrier, and $N$ subsequently decays into $e^+e^-$ pairs.

Models with long-lived particles produced at the neutrino source are also discussed. Section~\ref{subsubsec:LongLivedHNLs} discusses a model with long-lived HNLs that propagate to the detector and decay into $e^+e^-$ and single-$\gamma$ final states. In Sec.~\ref{subsubsec:LongLivedHNLs}, a dark matter model with light mediators is presented. The particles are produced in charged meson decays and upscatter inside the detector to produce electromagnetic showers.

\subsubsection{Transition Magnetic Moment}\label{subsubsec:TransitionMagneticMoment}

\newcommand\barparen[1]{\overset{(-)}{#1}}
\newcommand\barparena[1]{\overset{%
   \scriptscriptstyle(-)}{#1}}
\newcommand\bara[1]{\overset{%
   \scriptscriptstyle-}{#1}}
\newcommand\barparenb[1]{\overset{%
   \scalebox{0.4}{$(\mkern-1mu-\mkern-1mu)$}}{#1}}


Several extensions of the SM consider the existence of sterile neutrinos with Dirac or Majorana masses at or above the MeV scale. By convention, the mass eigenstates that are mostly in the direction of these sterile neutrino models are usually referred to as heavy neutral leptons (HNL), denoted here by $N$. Additional interactions notwithstanding, HNLs and light sterile neutrinos refer to the same class of particles which differ only in their mass value. Depending on the model, they can behave either as Majorana or (pseudo-)Dirac particles. 

Similar to the sterile neutrino models discussed above, $N$ can mix with SM neutrinos and interact with the $Z$ and $W$ bosons through mixing. However, it is also possible that this mixing is too small to be observed, and that these HNLs can be produced and decay via additional interactions. One interesting example is that of a transition magnetic moment between light SM neutrinos and $N$, described by the effective Lagrangian
\begin{equation} \label{eq:dipoleL}
\mathcal{L}_{N} = - d_\alpha \bar{\nu}_{\alpha L} \sigma_{\mu \nu} N F^{\mu \nu} - \frac{g}{\sqrt{2}} U_{\alpha N}^*  \overline{\nu}_\alpha \gamma_\mu N Z^\mu +  h.c.,
\end{equation}
where 
$d_\alpha$ is the transition magnetic moment between ${N}$ and SM neutrino weak eigenstate $\nu_\alpha$, $F^{\mu \nu}$ is the electromagnetic field strength tensor, $\sigma_{\mu \nu} = \frac{i}{2}(\gamma^\mu \gamma^\nu - \gamma^\nu \gamma^\mu)$, and $U_\alpha N$ is the mixing between $\nu_\alpha$ and $N$.
Note that in an effective theory language the operator that gives rise to the first term is found at dimension six, namely, $\frac{c_\alpha}{\Lambda^2} L_\alpha \tilde H \sigma_{\mu\nu}N B^{\mu\nu}$, and thus
$d_\alpha\propto c_\alpha v/\Lambda^2$, with $v=174$~GeV being the Higgs vacuum expectation value.
The second term above induces an effective vertex between an SM neutrino, a heavy neutrino, and the $Z$, which arrives due to mixing between $\nu_\alpha$ and $N$.

\begin{figure}[ht!]
    \centering
    \includegraphics[width=0.42\textwidth]{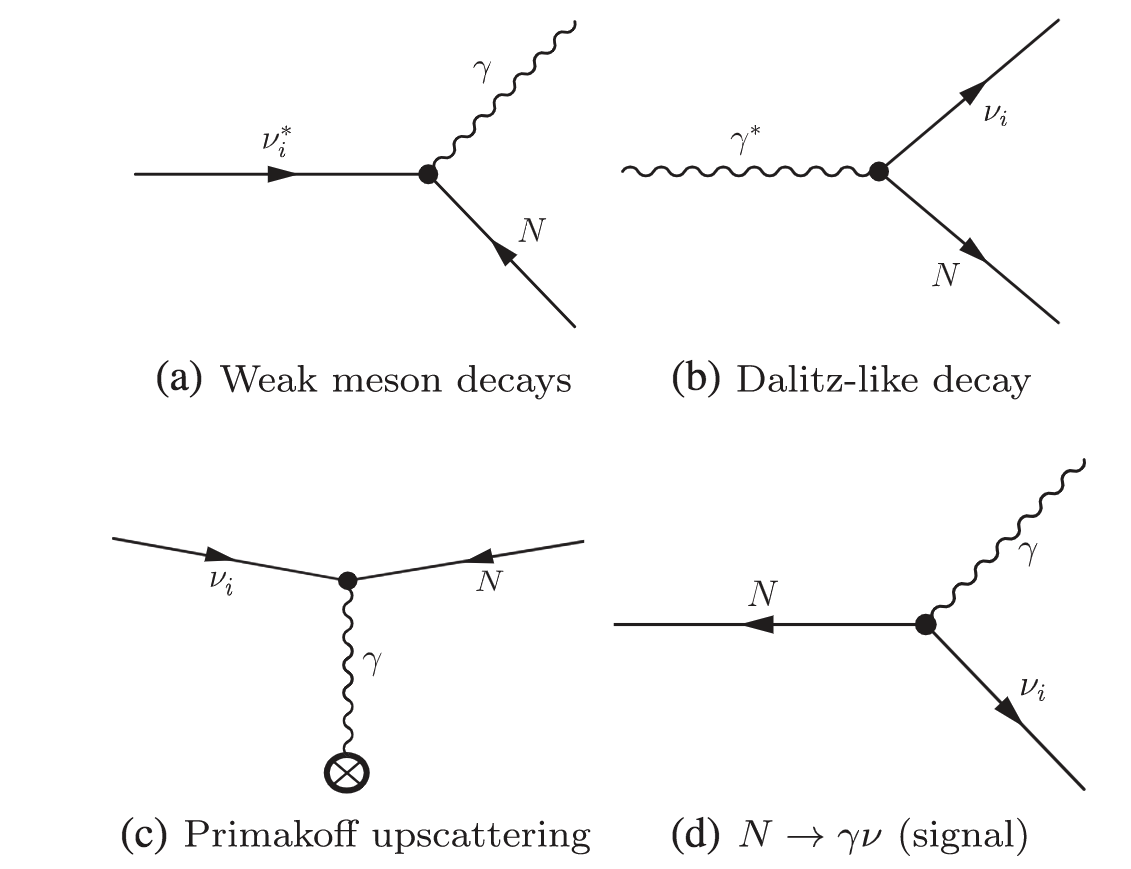}
    \includegraphics[width=0.47\textwidth]{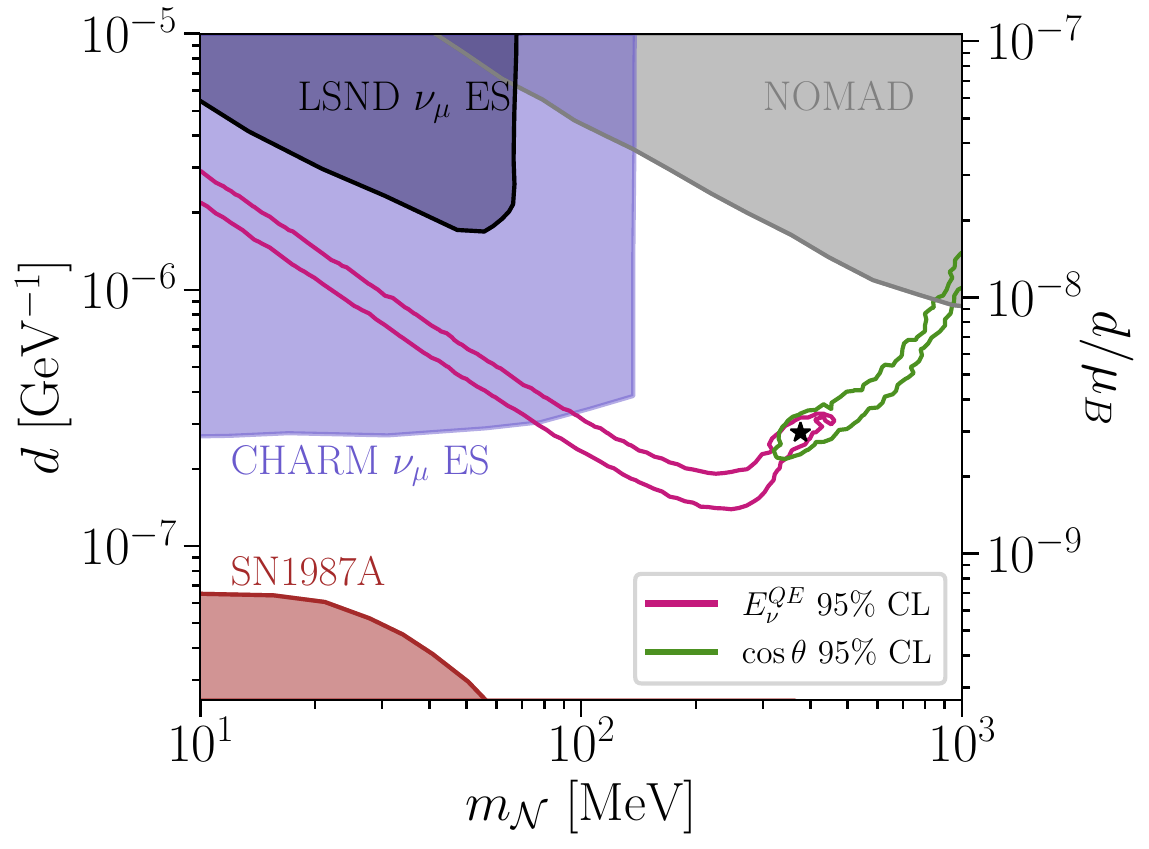}
    \caption{\label{fig:dipole_plots}
    Left: dipole model interactions channels relevant for neutrino experiments which are introduced by the effective vertex in Eq.(~\ref{eq:dipoleL}) (figure from Ref.~\cite{Magill:2018jla}); 
    Right: preferred regions in dipole model parameter space to explain short-baseline anomalies, along with constraints from existing experiments (figure from Ref.~\cite{Vergani:2021tgc}).}
\end{figure}

The effective vertex introduced in Eq.~(\ref{eq:dipoleL}) gives rise to new interactions relevant for existing and future neutrino experiments.
First, the dipole model opens up new decay modes for mesons through off-shell virtual mediators, which can provide a source of heavy neutrinos at beam-dump experiments.
For example, one introduces the weak mediated decay $\pi^+ \to \ell^+ \nu^* \to \ell^+ N \gamma$ and the Dalitz-like decay $\pi^0 \to \gamma \gamma^* \to \gamma N \nu$.
Additionally, heavy neutrinos can be produced by the Primakoff up-scattering of SM neutrinos off a nuclear target $A$ via the interaction $\nu A \to {N} X$.
Finally, the typical observable signal in the dipole model is the decay of a heavy neutrino to an SM neutrino and a photon via the process $N \to \nu \gamma$.
The relevant Feynman diagrams for these processes are shown in Fig.~\ref{fig:dipole_plots}.

It is also possible that HNLs have non-negligible mixing with SM neutrinos, in which case the upscattering of SM neutrinos to HNLs can be mediated by both the photon, referred to as electromagnetic (EM) production, or the SM $Z$ boson referred to as weak production. In this case, the mixing of $N$ with muon neutrinos is the most relevant since most neutrinos in accelerator experiments are $\nu_\mu$ and $\overline{\nu}_\mu$.

\textit{LSND ---}
The dipole model can explain the observed LSND excess via upscattering on carbon~\cite{Gninenko:2010pr},
\begin{equation} \label{eq:LSND}
\nu_\mu {}^{12}C \to {N} n X \to \nu \gamma n X,
\end{equation}
where the $\nu_\mu$ comes from $\pi^+ \to \mu^+ \nu_\mu$ decay-in-flight in the LANSCE beam-stop.
The Compton scattering and pair production of the photon from ${N}$ decay mimic the signal of the prompt $e^+$, which is detected in coincidence with the capture of the recoil neutron in Eq.~(\ref{eq:LSND}).
A heavy neutrino mass of $m_{N} \sim 50\;\si{\MeV}$ can reasonably explain the LSND anomaly while avoiding constraints from the KARMEN experiment~\cite{Gninenko:2010pr}. While Ref.~\cite{Gninenko:2010pr} considered weak production, subsequent studies showed that EM upscattering dominates in this region of parameter space. Nevertheless, the original solution can still be accommodated in non-minimal scenarios involving more than one heavy neutrino~\cite{Masip:2012ke}.

\textit{MiniBooNE ---}
Regarding MiniBooNE, since the detection of $e^\pm$ relies on reconstructing Cherenkov rings, this signature is indistinguishable from photons in the detector. Thus, the dipole model can provide an explanation of the MiniBooNE through the decay channel $N\to \nu \gamma$.
In this case, the dominant source of heavy neutrinos in MiniBooNE comes from the Primakoff up-scattering of SM neutrinos produced in the BNB.
Depending on the lifetime of the $N$, this can happen off of nuclei either in the dirt between the BNB and MiniBooNE or in the target material of the detector itself. We note that the decays of the heavy neutrino must be sufficiently prompt to be consistent with the timing distribution of the MiniBooNE excess~\cite{MiniBooNE:2020pnu}.

The upscattering to $N$ could proceed via EM or weak interactions with the protons of hydrogen or both with carbon nuclei in an incoherent or coherent fashion. The hadronic tensor in this case is the same as in elastic and quasi-elastic interactions of electrons and neutrinos on the corresponding targets (see Chapter 4 of Ref.~\cite{Edutesis} for details). As shown in Fig.~16 of Ref.~\cite{Alvarez-Ruso:2021dna}, the EM cross section on nuclei is dominated by the coherent mechanism; the incoherent one is suppressed by Pauli blocking at low four-momentum transfers, where the amplitude is enhanced by the photon propagator. On the contrary, the incoherent reaction is dominant in the weak part. 

\textit{Electromagnetic upscattering only ---}
The EM upscattering scenario with negligible $N$ mixing with SM neutrinos has been explored in a number of studies~\cite{Gninenko:2009ks,Gninenko:2009yf,Magill:2018jla,Masip:2012ke,Alvarez-Ruso:2017hdm}, which suggest that the energy distributions of the MiniBooNE excess can be accommodated by a dipole-coupled heavy neutrino with $m_N = \mathcal{O}(100)\;\si{\MeV}$ and $d_\mu = \mathcal{O}(10^{-7}{\text{\textemdash}}10^{-6})\;\si{\GeV}^{-1}$. The MiniBooNE angular distribution, however, can only be explained with EM upscattering when $m_N \gtrsim 300$~MeV. In that case, the produced HNL is less boosted and decays more isotropically.

One can also consider extending the model to include an $\mathcal{O}(1)\;\si{\eV}$ sterile neutrino in addition to the dipole-coupled HNL. Such a model can explain MiniBooNE through a combination of upscattering into $N$ and eV-sterile-neutrino oscillations. If $m_N > \mathcal{O}(100)$~MeV, other low-energy experiments like LSND would not be sensitive to it, but they would still be sensitive to oscillations. This combination of effects has been found to decrease the tension in the sterile neutrino global picture~\cite{Vergani:2021tgc}.

\begin{figure}[ht!]
  \centering
    \includegraphics[width=.44\textwidth]{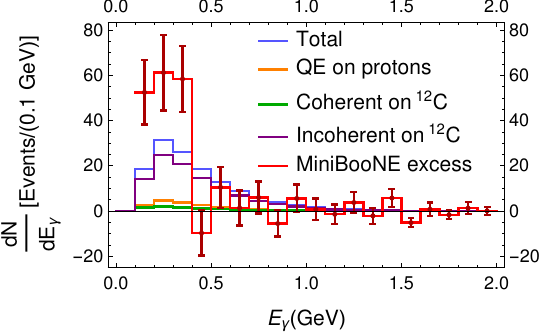}     
 $\qquad$
    \includegraphics[width=.44\textwidth]{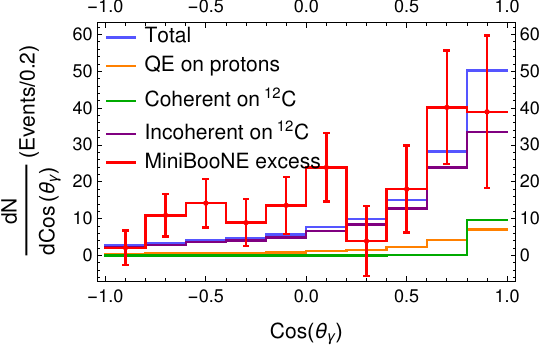}
  
    \includegraphics[width=.44\textwidth]{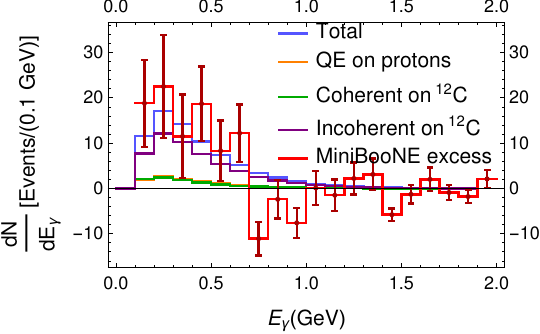}     
 $\qquad$
    \includegraphics[width=.44\textwidth]{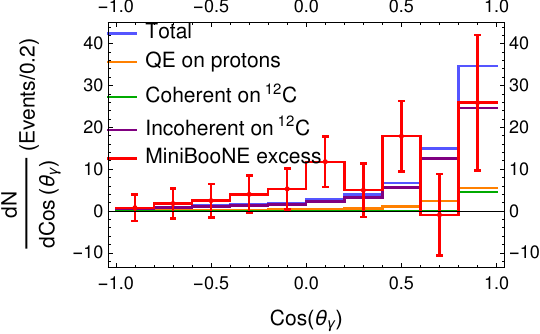}
  \caption{Photon events from radiative decay of $N$, $\overline{N}$ at the MiniBooNE detector in $\nu$-mode (top) and $\bar\nu$-mode (bottom) compared to the MiniBooNE excess. For details, see Refs.~\cite{Alvarez-Ruso:2017hdm} and \cite{Alvarez-Ruso:2021dna}. Figure from Ref.~\cite{Alvarez-Ruso:2021dna}.} 
  \label{fig:eventsMiniB}
\end{figure}

\textit{Transition magnetic moment with weak production ---} We now consider weak production of HNLs in MiniBooNE. With the parameters proposed in Ref.~\cite{Masip:2012ke}, the forward-peaked dominant coherent EM contribution leads to a very narrow angular distribution not observed in the experiment (see Fig.~2 of Ref.~\cite{Alvarez-Ruso:2017hdm}). The agreement can be improved by including production via the SM Z boson, which is less forward. The parameters can be constrained to the allowed range established in Ref.~\cite{Gninenko:2010pr}, but there are more stringent bounds for $U_{\mu N}$, in particular from radiative muon capture: $\mu^- \, p \rightarrow n \, \nu \, \gamma$, experimentally investigated at TRIUMF. Setting $m_N$ to the allowed minimum of 40~MeV to have the largest possible upper bound in the mixing: $|U_{\mu N}|^2 = 8.4 \times 10^{-3}$~\cite{McKeen:2010rx}, the best fit finds $\tau_N = 9.1^{+ 1.1}_{- 1.5} \times 10^{-10}$ seconds, BR$_\mu = 1.7^{+ 2.4}_{- 1.4} \times 10^{-5}$ with a $\chi^2/$DoF$=104/54$. The different contributions to the excess are singled out in Fig.~\ref{fig:eventsMiniB}. A reasonable description of the angular distribution requires a suppression of the EM strength, as reflected by the small BR$_\mu$ compared to the original proposal of BR$_\mu = 10^{-2}$, while increasing $U_{\mu N}$ as much as possible: its upper limit prevents from obtaining a more satisfactory description of the data. 

\begin{figure}[ht!]
  \centering
    \includegraphics[width=.42\textwidth]{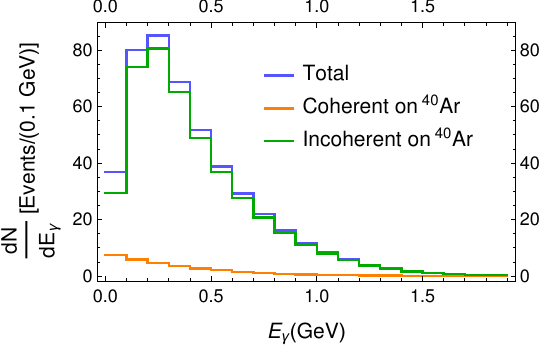}     
 $\qquad$
    \includegraphics[width=.42\textwidth]{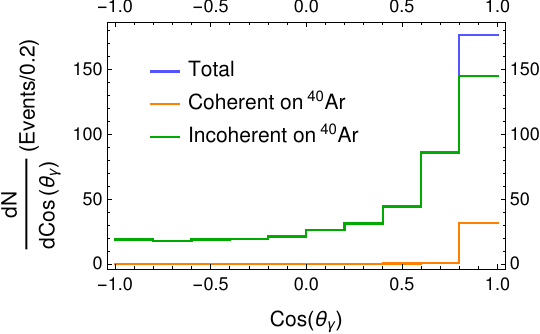}
    \caption{Photon events from $N$ radiative decay at MicroBooNE  for $6.6 \times 10^{20}$~POT in $\nu$-mode. For details, see Refs.~\cite{Alvarez-Ruso:2017hdm} and \cite{Alvarez-Ruso:2021dna}. Figure from Ref.~\cite{Alvarez-Ruso:2017hdm}.} 
  \label{fig:eventsMicroB}
\end{figure}

These results show that the heavy neutrino radiative decay hypothesis is not particularly successful in the simultaneous description of both the energy and the angular distribution of the excess, even with a degree of parameter fine-tuning. Nevertheless, based on MiniBooNE data alone, it cannot be fully excluded, at least as a partial source of the excess. Using the same number of POT as for Fig.~\ref{fig:ncgammaMicroBooNE} and the best-fit parameters, the photon events predicted at MicroBooNE are displayed in Fig.~\ref{fig:eventsMicroB}. With a total number of events of more than twice the SM ones and clearly more forward peaked, testing this possible explanation of the anomaly is within reach of the MicroBooNE experiment. It also warrants further studies for the new generation of experiments, SBND and ICARUS. 

\textit{Other constraints ---} We now briefly discuss existing and projected constraints on the heavy neutrino transition magnetic moments. 
For $m_N \lesssim 1\;\si{\GeV}$, neutrino-electron scattering cross section measurements from Borexino, CHARM-II, DONUT, and LSND can be translated into bounds on the transition magnetic moment~\cite{Coloma:2017ppo,Magill:2018jla,Auerbach:2001wg,Borexino:2017fbd,Schwienhorst:2001sj,CHARM-II:1991ydz} (see the discussion in Sec.~\ref{sec:minerva_and_charm}).
A single-photon search from the NOMAD experiment can be used to set a limit $\nu_\mu A \to (N \to \nu_\mu \gamma) A$~\cite{Gninenko:1998nn,Vannucci:2014wna}.
Future single-photon searches at Fermilab's short baseline program may be able to probe parameter space relevant for the MiniBooNE anomaly~\cite{Magill:2018jla}.

At higher heavy neutrino masses, collider experiments can provide bounds on transition magnetic moments by looking for single-photon events with missing energy~\cite{Magill:2018jla}.
One can also constrain heavy neutrino transition magnetic moments using measurements of the relic ${}^4$He abundance from Big Bang Nucleosynthesis and inferred Supernova 1987A cooling rates~\cite{Magill:2018jla}. 
The proposed SHiP detector at CERN may also be able to set leading limits on heavy neutrino transition magnetic moments, especially for the $d_{e,\tau}$ couplings~\cite{Magill:2018jla,Anelli:2015pba}.
Additionally, experiments such as Super-Kamiokande, IceCube, DUNE, and Hyper-Kamiokande may be able to place limits on the dipole model by taking advantage of the unique double-bang topology of $\nu-N$ up-scattering and subsequent $N$ decay~\cite{Coloma:2017ppo,Atkinson:2021rnp,Schwetz:2020xra}.
A subset of these limits, as well as the preferred regions to explain the short baseline anomalies, are shown in Fig.~\ref{fig:dipole_plots}.
We look forward to future experiments to shed light on heavy neutrino dipole portal explanations of short baseline anomalies.


\subsubsection{Dark Neutrinos}
\label{subsubsec:DarkNeutrinos}

Many extensions of the SM to accommodate neutrino masses involve HNLs that mix with the SM neutrinos. If these HNLs interact via additional mediators, e.g. a dark photon from a secluded $U(1)_X$, and their masses are in the ${\sim}$MeV-GeV, then the upscattering of SM neutrinos into these so-called ``dark neutrinos'', followed by dark neutrino decay, can explain the LEE observed by MiniBooNE~\cite{Bertuzzo:2018itn,Ballett:2018ynz}. Similar to the dipole model (Sec.~\ref{subsubsec:TransitionMagneticMoment}), since MiniBooNE cannot distinguish an electron from a photon, it also cannot distinguish either of these from a pair of overlapping $e^+$ and $e^-$. If the $e^+e^-$ pair are sufficiently collinear, or the energy of one of the particles falls below the detector energy threshold, the EM shower of the pair mimics that of a single electron or photon. 

\begin{figure}[ht]
    \centering
    \includegraphics[width=0.4\textwidth]{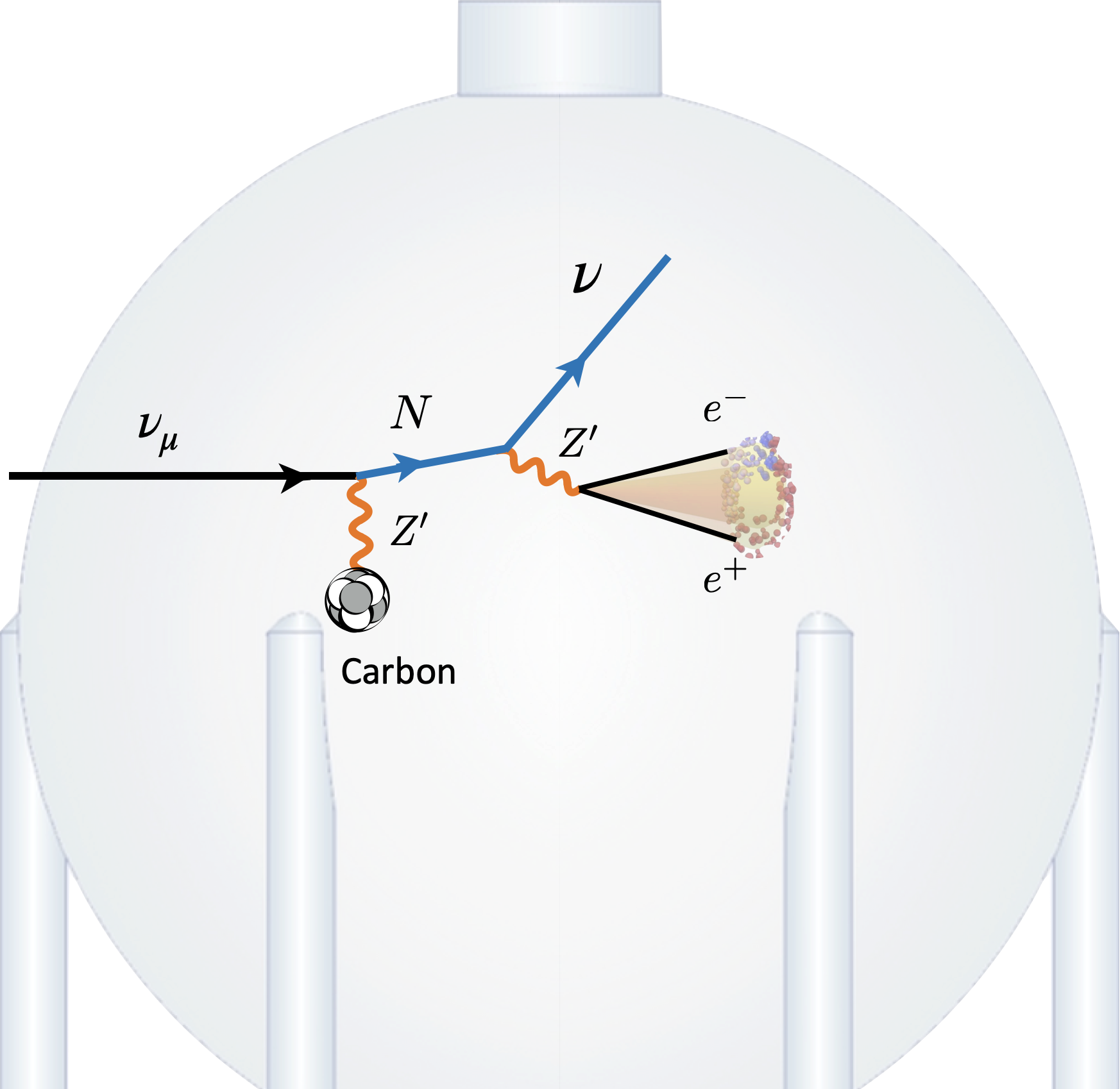}
    \caption{Dark neutrino production and decay inside the MiniBooNE detector.}
\end{figure}

HNLs can be produced via rare meson decays (e.g., from kaons), and then decay inside the detector~\cite{Fischer:2019fbw,Chang:2021myh}, however, the signals predicted in this scenario can be delayed with respect to the neutrino beam, and tend to lead to forward-peaked angular distributions. 
However, if the new particles are produced in neutrino upscattering inside the detector, provided their decays are sufficiently short-lived ($c\tau^{\rm LAB} < \mathcal{O}(10)$~ns at MiniBooNE), they can be registered in coincidence with the SM neutrinos. 
Scenarios where $e^+e^-$ pairs are produced in this fashion (not restricted to the dark neutrino model) have been studied in detail in Ref.~\cite{Bertuzzo:2018itn, Bertuzzo:2018ftf,Ballett:2018ynz, Ballett:2019pyw,Abdullahi:2020nyr,Datta:2020auq,Dutta:2020scq,Abdallah:2020biq,Abdallah:2020vgg,Hammad:2021mpl, Dutta:2021cip}. We first discuss the original proposals based on a dark photon and then move on to newer proposals involving scalar particles.

\textit{Dark photon models} 

In addition, these models face none of the problems of the popular 3+1 oscillation explanation of the MiniBooNE excess, as the phenomena observed at the different SBL experiments decouple in this framework. 
Finally, there is the possibility of connecting these models to other prominent questions of particle physics. 
The discovery of neutrinos with hidden interactions would be a very strong indication of the existence of dark sectors that could contain the theorized dark matter. 
With the SBN program currently underway, there is the opportunity to probe large regions of  the LEE model parameter space, but also more generic models of dark sector HNLs.

In Refs.~\cite{Bertuzzo:2018ftf,Bertuzzo:2018itn,Ballett:2018ynz,Ballett:2019cqp,Ballett:2019pyw,Hammad:2021mpl}, the mediator of upscattering is a dark photon. A simplified model can be used to understand the experimental signatures. In it, an electrically-neutral fermion $\nu_D$ is charged under dark U$(1)^\prime$ symmetry and is assumed to mix with SM neutrinos, $\nu_\alpha = \sum^{4}_{i=1} U_{\alpha i}\nu_{i}, \quad (\alpha=e,\mu,\tau,D)$ where $U$ is a $4\times4$ unitary matrix. The mediator of this new gauge group then kinetically mixes with the SM photon via $(\varepsilon/2) X^{\prime}_{\mu \nu} \hat{F}^{\mu \nu}$, which leads the electrically-charged SM fermions to acquire a small coupling to the dark photon, $Z^\prime$. The low-energy simplified Lagrangian reads
\begin{equation}
\mathcal{L}_{\rm int} \supset \;\;-g_D U_{D 4}^*U_{D i} \overline{\nu}_i \gamma_\mu N Z^{\prime \mu}
 - e \varepsilon Z^{\prime \mu}J^{\rm EM}_{\mu},
\end{equation}
where $g_D$ is the U$(1)^\prime$ gauge coupling, $J^{\rm EM}_{\mu}$ the SM electromagnetic current, and $\varepsilon$ is the kinetic mixing parameter. We define $N\equiv \nu_4$. In a full model, $N$ may be of Dirac or Majorana nature.

Since the dark photon interacts with the SM electric charge in an analogous way to the photon, the upscattering cross section can be calculated in an analogous way to the transition magnetic moment case, replacing the leptonic vertex and the propagator by a massive one. When the dark photon is light, coherent scattering with the nucleus is dominant and the rate is enhanced by the number of protons in the nucleus, $Z$, with respect to scattering on individual protons. For $m_{Z^\prime} \gtrsim 1$~GeV, the incoherent contributions start to dominate and the upscattering can kick a proton out of the Carbon nucleus in MiniBooNE, for example. In models where the HNL only mixes with muon neutrinos, the upscattering cross section can be shown to scale as $\sigma_{\nu_\mu A \to N A} \propto |U_{\mu 4} U_{D4}|^2$ by virtue of the unitarity of $U$. Therefore, $\sigma_{\nu_\mu A \to N A}$ can be readily constrained by existing limits on the active-heavy mixing $|U_{\mu 4}|^2$ for every choice of $|U_{D4}|^2$, which usually takes values close to unity.

The decay of the HNL is prompt when $m_{Z^\prime} < m_4$, where the dark photon is produced on shell and decays to $e^+e^-$. Decays to light neutrinos are suppressed by the mixing angle combination $|U_{Di} U_{Dj}|^2$, which is small for $i<4$. This regime produces very forward-going HNLs, which subsequently decay to boosted $e^+e^-$ pairs. While this produces more $e^+e^-$ events that mimic single photon or single electron showers, it also leads to very forward-peaked angular distributions, in contrast with MiniBooNE's observation~\cite{MiniBooNE:2020pnu}. The angular spectrum is less forward for $m_N \gtrsim 300$~MeV, however, at those masses constraints from high-energy experiments are severe~\cite{Arguelles:2018mtc}.

\begin{figure}[ht!]
  \centering
    
    \includegraphics[width=.4\textwidth]{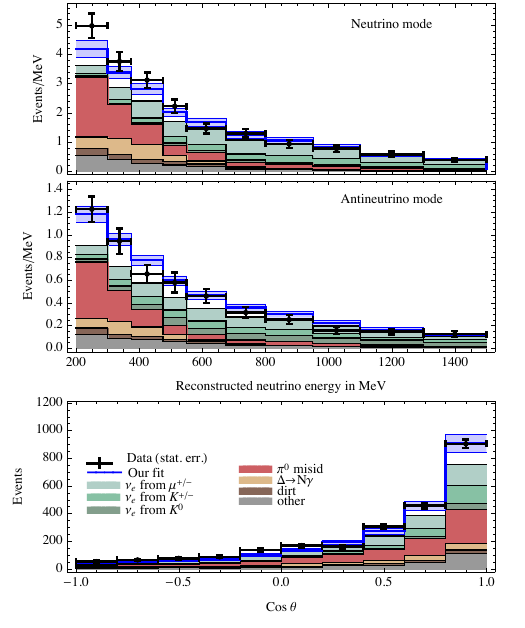}
     $\qquad$
    \includegraphics[width=.44\textwidth]{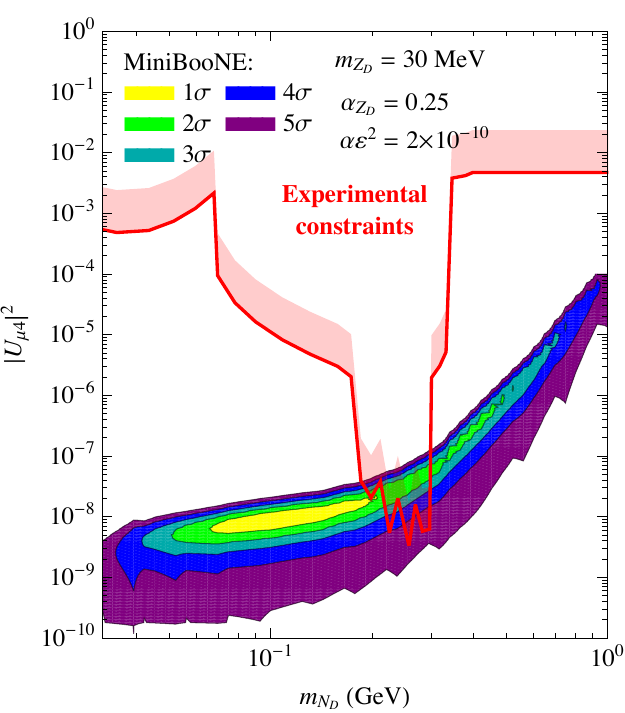}     
  \caption{Left: the spectrum of dark neutrino events in neutrino and antineutrino energy spectrum, as well as in the angular distribution at MiniBooNE. Right: fit to the neutrino-energy distribution at MiniBooNE in a dark neutrino model with a light dark photon ($m_{Z^\prime} < m_4$). Figure from \cite{Bertuzzo:2018itn}.} 
  \label{fig:bertuzzo_darknus}
\end{figure}

For heavy dark photons, $m_{Z^\prime} > m_4$, the decay is a three-body one and, therefore, the HNL is much longer-lived. Ref.~\cite{Ballett:2018ynz} proposed a model where $|U_{\tau 4}|^2 \gg |U_{\mu 4}|^2$, so that the decay process is effectively $N\to \nu_\tau e^+e^-$. A similar proposal was made in Ref.~\cite{Ballett:2019pyw,Abdullahi:2020nyr} where a model with two HNLs was used. In that case, the heaviest HNL decays to the intermediary state with a lifetime that is not suppressed by active-heavy mixing. Due to the heavy mediator, the upscattering happens with a larger $Q^2$, and the angular distribution can be less forward. The best agreement with the angular distribution is found when the dark photon interferes with the SM $Z$.

\begin{figure}[ht!]
  \centering
    \includegraphics[width=.42\textwidth]{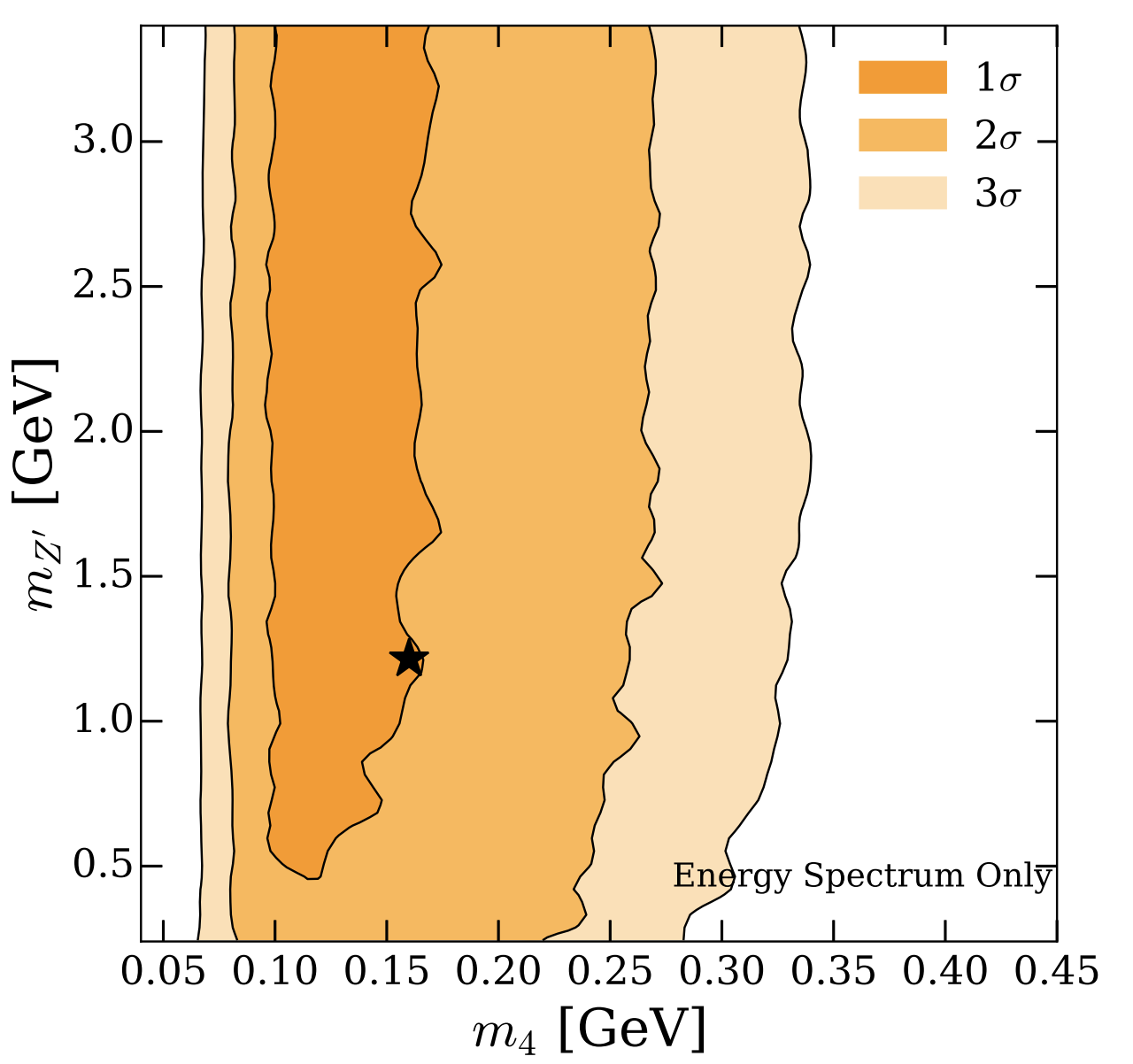}  
     $\qquad$
    \includegraphics[width=.4\textwidth]{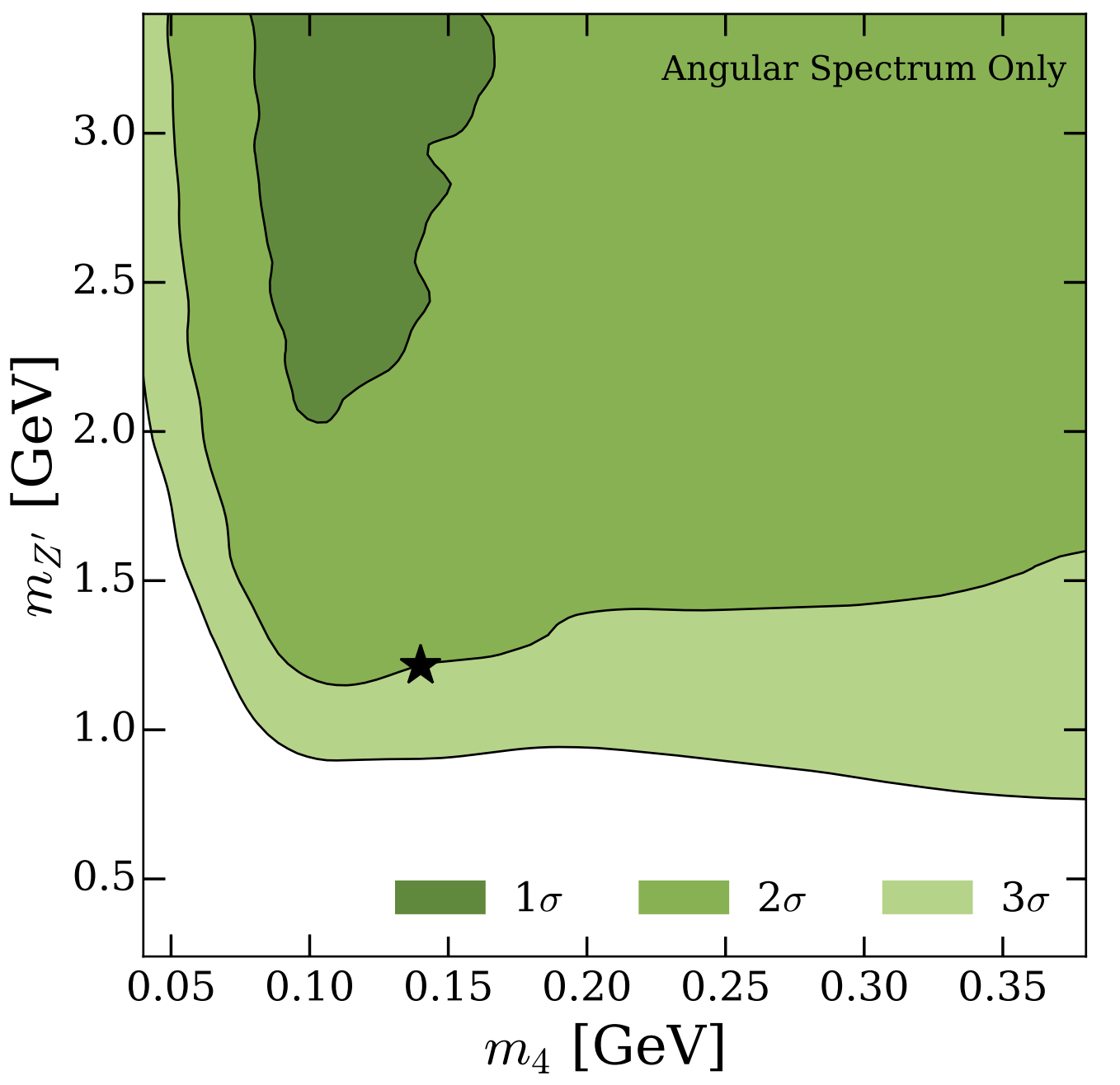}
  \caption{Left: Fit to the neutrino-energy distribution (left) and angular distribution (right) at MiniBooNE in a dark neutrino model with a heavy dark photon ($m_{Z^\prime} < m_4$). Figure from \cite{Ballett:2018ynz}.}
  \label{fig:ballett_darknus}
\end{figure}

\textit{Dark scalars}

HNLs can also interact with additional scalars that play the role of the dark photon discussed above. In this case, the upscattering cross section lacks the $t$-channel singularity and does not asymptote to a constant, but rather falls as $1/E_{\nu}^2$ at large energies. Therefore, if the dark sector particles have masses of $\mathcal{O}(100)$~MeV, the upscattering process is largest at the energies of LSND and MiniBooNE and may avoid constraints from high-energy experiments altogether. Models of this type have been discussed in Ref.~\cite{Datta:2020auq,Dutta:2020scq,Abdallah:2020biq,Abdallah:2020vgg}.

\textit{Dark scalars and neutrino polarizability}

As pointed out in Ref.~\cite{Datta:2020auq}, the scalar mediator can also lead to the production of photons pairs via the decay chain $N\to \nu (S \to \gamma \gamma)$. At MiniBooNE, if $S$ is lighter than the pion, it will be more boosted and its decays to overlapping photons would mimic the excess signal. It is also possible that the branching ratios for $S \to e^+e^-$ and $S\to \gamma \gamma$ are both sizeable, in which case the excess would display a non-trivial shape in energy and angle.

A scalar mediator coupling to both neutrinos and photons induces a parametrically enhanced neutrino polarizability, i.e., in low energy processes the scalar can be integrated out, resulting in dimension 7 Rayleigh operators of the form $(\bar\nu_i P_L \nu_j)F_{\mu\nu} F^{\mu\nu}$ (or $(\bar\nu_i P_L \nu_j)F_{\mu\nu}\tilde F^{\mu\nu}$ if the mediator is a pseudoscalar)~\cite{Bansal:2022zpi}. For light mediators, with masses below $\sim$MeV, there are stringent constraints on such neutrino polarizability models from cosmology and stellar cooling. For mediators in the MeV to few GeV regime, relevant for the MiniBooNE anomaly, these models can be probed at a multitude of neutrino facilities, from measurements of solar neutrino scattering (Borexino and Xenon-nT) to observations in high energy neutrino beams (DUNE ND), as well as using beam dump facilities and at precision $e^+e^-$ collider experiments such as Belle II~\cite{Bansal:2022zpi}. 


\textit{Explaining LSND}

Some of the models proposed in the literature above can, in principle, also explain the LSND anomaly. This sets a more stringent requirement on the theory since it requires the upscattering process to produce a neutron in the detector. The new signal should mimic the inverse-beta-decay signature with at LSND, with a prompt electromagnetic signal followed by delayed neutron capture. This already eliminates the dark photon as a potential solution, since the couplings with neutrons are much more suppressed than those with protons. In addition, depending on the mass of the heavy neutrino, the upscattering process may have too large of a threshold to be initiated by $\pi$ or $\mu$ decays at rest. Instead, for masses of $\mathcal{O}(100)$~MeV, one may take advantage of the number of pions that decay in flight. This requires a large cross section, since the at these energies is much smaller than the decay-at-rest one. Models of this type were proposed in Refs.~\cite{Abdallah:2020biq,Abdallah:2020vgg}, where the heavy neutrinos are produced via the exchange of a new scalar boson with large couplings to neutrons.

\textit{Broad requirements to explain MiniBooNE and LSND}

Such proposals must conform to strong  demands from  a) cross section requirements in order to yield a sufficient number of total events in both LSND and MB, b) the measured energy and angular distributions in both experiments, and finally, c) compatibility of the new physics model and its particle content with  bounds from an extensive swathe of particle physics experiments~\cite{Atre:2009rg,McKeen:2010rx,ISTRA:2011bgc,Drewes:2015iva,deGouvea:2015euy,Coloma:2017ppo,Magill:2018jla,MiniBooNEDM:2018cxm,Jordan:2018qiy,Arguelles:2018mtc,Bryman:2019ssi,Coloma:2019qqj,Bryman:2019bjg,Brdar:2020tle,Abdallah:2020vgg,Corsico:2014mpa,Diaz:2019kim,Studenikin:2021fmn,Radionov:2013mca}.

We compare how scalar and vector mediators perform in helping to achieve a simultaneous  understanding of both anomalies. Our treatment is necessarily brief, and for details  and  a full set of references on all that follows the reader is referred to~\cite{Abdallah:2022grs}.
 
\begin{figure*}[h!]
\begin{center}
\begin{tabular}{ccc}
\vspace*{-1cm}\hspace*{-0.5cm}\includegraphics*[width=0.3\textwidth]{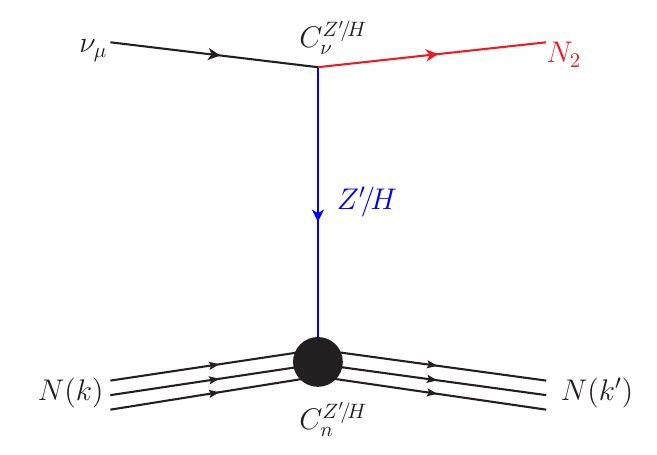} ~~ &
\hspace*{-0.0cm}\includegraphics*[width=0.3\textwidth]{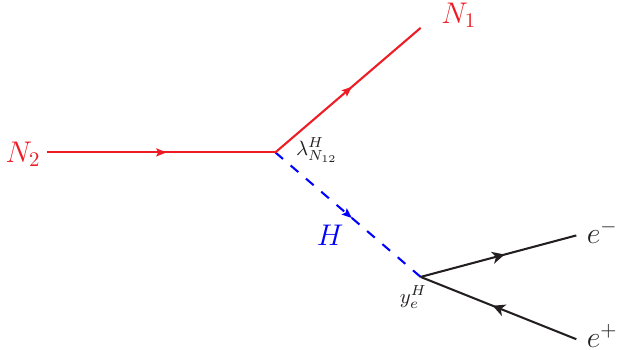} ~~ & 
\hspace*{-0.0cm}\includegraphics*[width=0.25\textwidth]{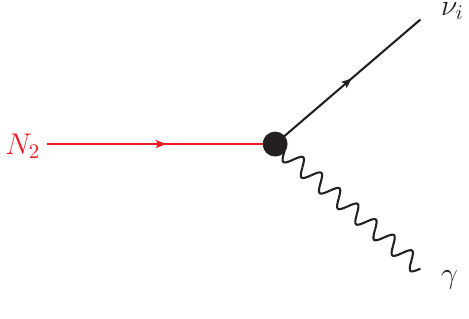} \\[1.2cm] 
 {\footnotesize (a)} & {\footnotesize (b)} & {\footnotesize (c)}
\end{tabular} 
\end{center}
\caption{\label{FD} Feynman diagrams for the production and the subsequent decay of the $N_2$ in LSND and MB.} 
\end{figure*}

We start by breaking up the interaction into two parts. We first consider the tree-level process leading to the up-scattering of an incoming muon neutrino, $\numu$, to a heavy neutral lepton ($N_2$) in the neutrino detector as shown in \cref{FD}(a), with the underlying assumption that it subsequently decays promptly in the detector into either of the final states shown in  \cref{FD}(b) and  \cref{FD}(c). The mediator for the up-scattering process could be either a light neutral vector boson $Z'$ or a light CP-even scalar $H$. The relevant interaction Lagrangian for the up-scattering process in each case  is given~by
\begin{eqnarray}
\mathcal{L}^{Z'}_{\rm{int}} \!\!&=&\!\! (C^{Z'}_{\nu} \bar{\nu}_{iL} \gamma^{\mu} N_{Lj}Z'_{\mu} +h.c.) +  C^{Z'}_{n} \bar{U}_{n} \gamma^{\mu}U_n Z'_{\mu},\\
\mathcal{L}^H_{\rm{int}} \!\!&=&\!\! (C^H_{\nu} \bar{\nu}_{iL} N_{Rj}H +h.c.) + C^H_{n} \bar{U}_{n} U_n H,
\end{eqnarray}
where $U_{n}$ is the nucleon field and $i, j=1,2,3$. For simplicity, we assume that for both mediators ($Z'$ and $H$), the coupling to a  proton is  the same as its  coupling to a neutron. The up-scattering cross section depends on the overall product of the coupling constants $C^{Z'}_{\nu} C^{Z'}_n$ ($C^{H}_{\nu} C^H_n$) for the vector (scalar) mediator. 

\begin{table}[h!]
\centering
 \begin{tabular}{|c|c|c|} 
 \hline
 $m_{Z'\!/\!H}$ & $C^{Z'}_{\nu} C^{Z'}_n$  &$C^{H}_{\nu} C^{H}_n$  \\   
 \hline
 50~MeV & 1.04$\times 10^{-8}$ & 9.3$\times 10^{-8}$ \\ 
 \hline
 1 GeV  & 8.5$\times 10^{-7}$ & 2.14$\times 10^{-6}$ \\
 \hline
  300~MeV & $-$ & 5.35$\times 10^{-7}$ \\
 \hline
\end{tabular}
\caption{\label{tab} The overall coupling values for the vector and scalar mediators for different values of mediator masses ($m_{Z'\!/\!H}$) to produce 560 $N_2$ in the MB final state. The mass of $N_2$ is 100~MeV.}
\end{table}

Based on the Lagrangian above and the benchmark parameter values in \cref{tab}, we compute the coherent and incoherent cross sections (shown in \cref{x1}) for both mediators,  required for generating the central value of the number of excess events in MB, which is $560$~\cite{MiniBooNE:2020pnu}. From \cref{x1}, we note that
the scalar and vector-mediated cross sections behave distinctly, and our representative calculations bring out the following qualitative points:
\begin{itemize}
 \item For all cases the cross section initially rises as the energy is increased from its lowest values, however, it subsequently drops for a scalar mediator whereas it remains approximately flat with increasing energy for a vector mediator. This is true for both the coherent and incoherent parts. As the neutrino energy rises, it is this relatively rapid drop in the cross section  for $H$  that allows solutions with a scalar mediator to comfortably
 skirt constraints~\cite{Arguelles:2018mtc} coming from CHARM~II~\cite{CHARM-II:1994dzw} and  MINER$\nu$A~\cite{MINERvA:2019hhc}, compared to the $Z'$ mediated process. 
 \item It can also be seen  from \cref{x1} that the coherent contribution dominates over the incoherent part for lighter mediator masses, whereas the reverse is true for the higher mass choice for mediators. 
\item For LSND, contributions to events come from the incoherent part of the cross section only, due to  the presence of a neutron in the final state. In the region in the left panel of  \cref{x1}, we note that the  energy drops from MB ($\sim 800$~MeV) to LSND DAR flux values ($\sim 150-200$~MeV). Note that  for $m_{Z'\!/\!H}=1$~GeV (solid curves), while the incoherent cross section drops  for both mediators,  the vector cross section has lower values to begin with compared to the scalar. It also drops  more rapidly.  For example, it can be seen that the cross section for the $Z'$ drops an order of magnitude over this energy range for $m_{Z'}= 1$~GeV.

\begin{figure*}[t!]
    \centering
\includegraphics[width=0.45\textwidth]{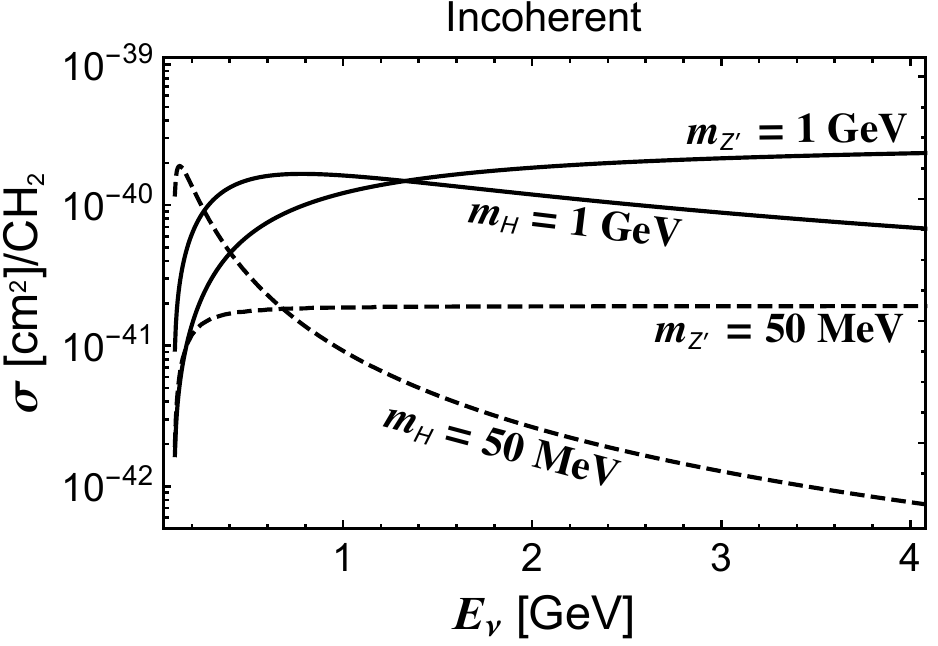}~~~~~\includegraphics[width=0.45\textwidth]{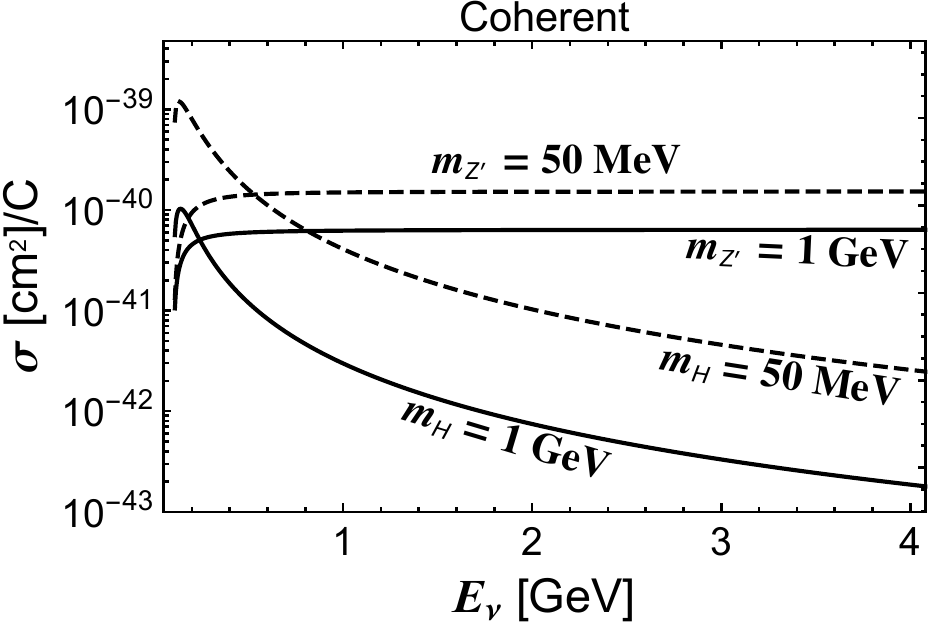}
\caption{\label{x1} The incoherent (coherent) cross section per CH$_2$ molecule (C atom) as a function of incoming neutrino energy. The overall constants for different kinds of mediator masses are taken from \cref{tab}.} 
\end{figure*}

 For $m_{Z'\!/\!H}=50$~MeV (dashed curves), over this energy range, the incoherent scalar cross section is significantly higher than  that for the vector. In fact,  it increases as the energy is lowered, unlike its vector counterpart. This reduction  in the incoherent vector cross section at   energies ($< 800$~MeV) makes it more difficult for models with a vector to give a sufficient number of electron-like excess events at LSND, even though by using a high enough $m_{Z'}$ one may  successfully evade the CHARM~II and  MINER$\nu$A bounds.  However, on the other hand, too low $m_{H}$ gives many more events than those observed in LSND,  
 both in the $20-60$~MeV visible energy range which recorded data, and beyond $60$ MeV, where only a limited number of events were seen.
 \item Finally, for  scalar mediators, especially those with low masses $m_H\simeq 100$~MeV, the cross section tends to rise  at low values of $E_\nu$. However, in such models, if  the primary decay modes of $N_2$ are to invisible daughters, as in~\cite{Dutta:2020scq}, the incoherent  interaction would mimic the neutral current interaction $\nu N \rightarrow\nu N$. This has  been measured at MB~\cite{Perevalov:2009mn} at these energies and found to be in agreement with the SM, providing an important restriction on such models.
 \end{itemize}

Overall, the cross section and mediator mass considerations for a common solution thus appear to favor scalar mediators over vectors.  Secondly, our representative calculations also point to a preference for  lighter (but not ultra-light)  mediators if both excesses are to have a simultaneous solution. 
  
An examination of the angular distribution of MB is also helpful from the point of view of  imposing requirements on proposed solutions. The  excess in MB  is distributed over all directions but is  moderately forward.   The  cross section responsible for the production of $N_2$  as a function of the cosine of the angle between the momentum direction of $N_2$ and the beam direction has been studied in a  bin-wise manner in~\cite{Abdallah:2022grs} for both the coherent and incoherent contributions.  It was found that when $m_{Z'\!/\!H}=50$~MeV, almost all the produced $N_2$ are in the  most forward bin for both mediators. However, as the mediator mass is increased to $1$~GeV,  there was a shift in the distribution,  and the other bins were also populated for both types of mediators, even though there were qualitative differences between the two.
From this, at first it appears that  using a single scalar mediator and adjusting its mass to an intermediate value, as well as the mass of $N_2$ will allow us to find a common solution to the two anomalies as well as match the angular distribution in MB. However, further examination based on considerations related to the energy distributions in LSND and MB (for details, see~\cite{Abdallah:2022grs}) reveals that this is not the case if good fits to both anomalies are desired. 
  
Overall, as detailed in~\cite{Abdallah:2022grs}, energy distributions in LSND and MB, the angular distribution in MB, when combined with the stringent constraints on light singlet scalars, all suggest the use of a scalar doublet, with one light and one moderately heavy partner.  We find this leads to a degree of angular isotropy while allowing a large number of events in the forward direction, consistent with observations. A combination of a moderately heavy and a light mediator complement each other  well when a common solution to the two anomalies is sought.   An example solution  to both  anomalies that incorporates  all the features that have been obtained in our study  has been provided in~\cite{Abdallah:2020vgg}.
    
Our insistence on a solution that addresses both anomalies simultaneously is, of course, a choice. It restricts  proposed solutions in ways that attempt to explain the anomalies individually do not.  However, it is noteworthy that  once we demand this, and adhere to the  dictates of the cross section,  the observed energy and angular distributions in both experiments  as well as the many constraints from various experiments~\cite{Abdallah:2022grs}, then we are  led to a simple extension of the SM that i) resolves both anomalies, ii) provides a portal to the dark sector, iii) accounts for the experimentally observed value of the muon $g-2$ and iv) addresses the issue of neutrino mass via a Type I seesaw, in conformity with the  global data on the observed values of neutrino mass-squared differences in oscillation experiments, as shown in~\cite{Abdallah:2020vgg}.

\subsubsection{Long-Lived Heavy Neutrinos}
\label{subsubsec:LongLivedHNLs}

\paragraph{Heavy neutrino decays to single photons} In Ref.~\cite{Fischer:2019fbw}, the authors propose a solution to the MiniBooNE excess with a heavy neutrino that interacts with muon-neutrinos through mixing as well as through a transition-magnetic moment. The model is the same as the one presented in \cref{eq:dipoleL}, but contrary to the solution in \cref{subsubsec:TransitionMagneticMoment}, the mixing angle $|U_{\mu N}|$ and transition magnetic moment $d_{\mu}$ are small, such that the heavy neutrino decays only in macroscopical distances. In particular, if the lifetime is larger than the distance between the target and the MiniBooNE detector, $c\tau^0 \gtrsim \mathcal{O}(500)$~m, then the heavy neutrino can be produced in pion and kaon decays via mixing, $K \to \mu N$, travel to the detector, and decay inside the active volume due to the transition magnetic moment, $N \to \nu \gamma$. 

While the model was shown to successfully reproduce the energy spectrum of the MiniBooNE excess, the decay process tends to produce very forward signatures. A better agreement with the angular spectrum is achieved for larger masses of the $N$. However, if the heavy neutrino is too massive, its arrival in the MiniBooNE detector is delayed with respect to the SM neutrinos. This delay is proportional to the $(m_N/E_N)^2$, and should not exceed values much larger than several nanoseconds, as the excess events have been observed to be in the same time window as the beam neutrinos~\cite{MiniBooNE:2020pnu}. This sets a strong constraint on the mass of the heavy neutrino. For masses below $150$~MeV, where agreement with the timing requirements can be achieved, production in pion decays should also be considered.

\paragraph{Heavy neutrino decays to axion-like particles}
In order to account for the MiniBooNE excess, Ref.~\cite{Chang:2021myh} proposes an extension of the SM with Dirac HNL that couples to a leptophilic axion-like particle ($\ell$ALP).
The HNL, denoted by $N_{D}$ to emphasize its Dirac nature, mixes with the three SM neutrino flavors, $\nu_{\beta}$. The flavor and mass eigenstates, $\nu_{jL}$ and $N_{D}$, can be transformed into each other by means of a unitary matrix $U$~\cite{Bertuzzo:2018itn}:
\begin{align}
    \nu_{\beta} = \sum^{3}_{j=1} U_{\beta j} \nu_{jL} + U_{\beta 4} N_{D}.
\end{align}
The relevant Lagrangian includes the interaction of $\ell$ALPs with sterile neutrinos and electrons~\cite{Alves:2019xpc}:
\begin{align}
    \mathcal{L}_{a\ell} = -\frac{\partial_{\mu}a}{2f_{a}} \left(c_{N}\overline{\nu}_{D}\gamma^{\mu}\gamma^{5}\nu_{D} + c_{e}\overline{e}\gamma^{\mu}\gamma^{5}e\right),
\end{align}
where $f_{a}$ is the $\ell$ALP decay constant, $c_{N}$ and $c_{e}$ are the dimensionless parameters for the $\ell$ALP-sterile neutrino and $\ell$ALP-electron couplings, respectively. 

\begin{figure}[ht]
    \centering
    \includegraphics[width=0.6\textwidth]{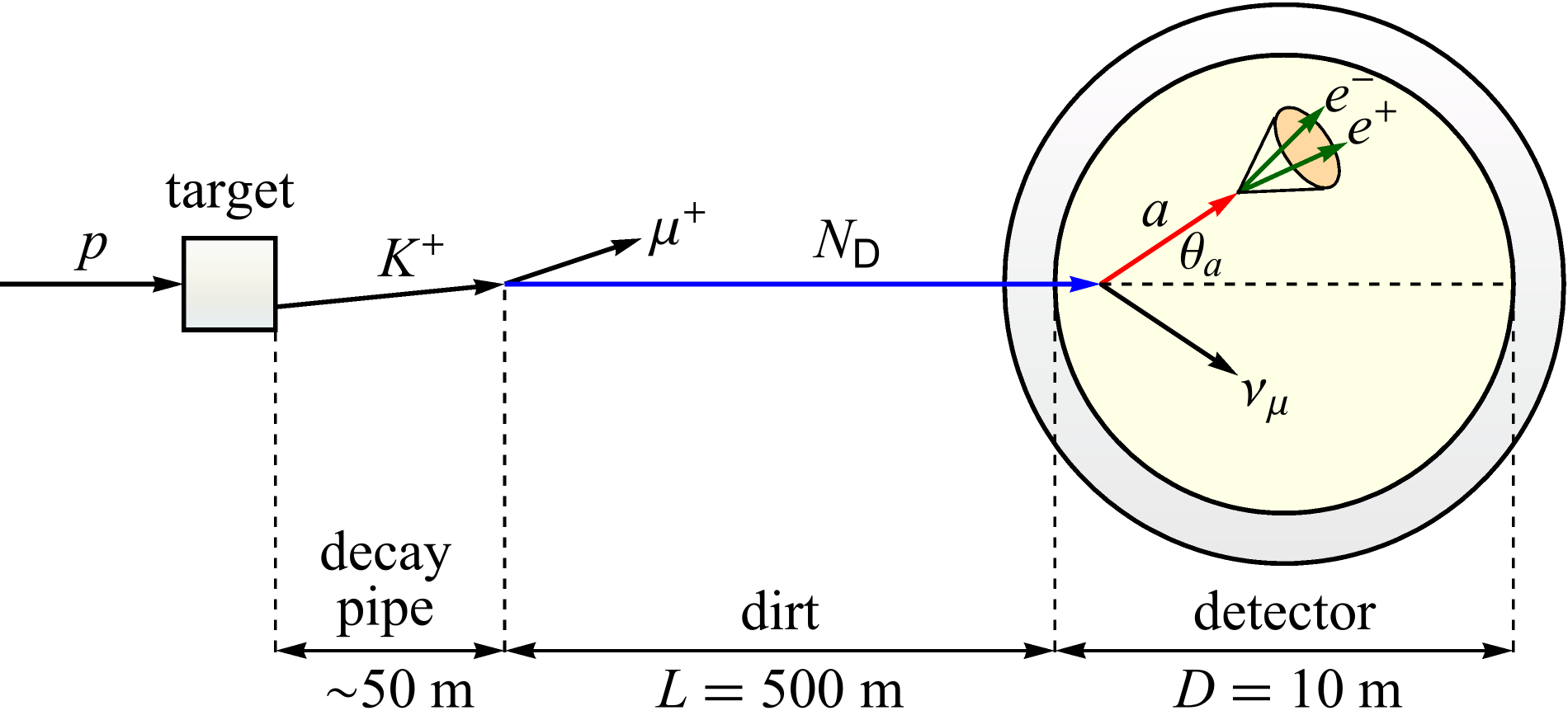}
    \caption{\label{fig:setup} $\ell$ALP scenario for the explanation of the MiniBooNE excess. $L$ is the travel distance of the sterile neutrino, $D$ is the diameter of the MiniNooBE detector, and $\theta_a$ is the scattering angle of the $\ell$ALP. Figure from \cite{Chang:2021myh}.}
\end{figure}
Our $\ell$ALP scenario is sketched in Fig.~\ref{fig:setup}. In this framework, the sterile neutrino with a mass $m_N$ of $\mathcal{O} \left(100\right)$~MeV, the $\ell$ALP with a mass $m_{a}$ of $\mathcal{O} \left(10\right)$~MeV and an inverse decay constant $c_{e}/f_{a} \simeq \mathcal{O} \left(10^{-2}\right)~\mathrm{GeV}^{-1}$ are considered. The Dirac-type sterile neutrino $N_{D}$, produced from charged kaon decays through its mixing with the muon neutrino, travels a distance of 500m and decays into a $\ell$ALP and a muon neutrino inside the MiniBooNE detector.
The electron-positron pairs produced from the $\ell$ALP decays can be interpreted as electron-like events provided that their opening angle is sufficiently small. We verify that with appropriate choices of the parameters, the sterile neutrino and $\ell$ALP have the proper mean decay lengths consistent with the setup of the MiniBooNE experiment. We also make sure that the values of the parameters adopted in our model are allowed by the astrophysical and experimental constraints.

\begin{figure}[ht]
    \centering
    \includegraphics[width=0.4\textwidth]{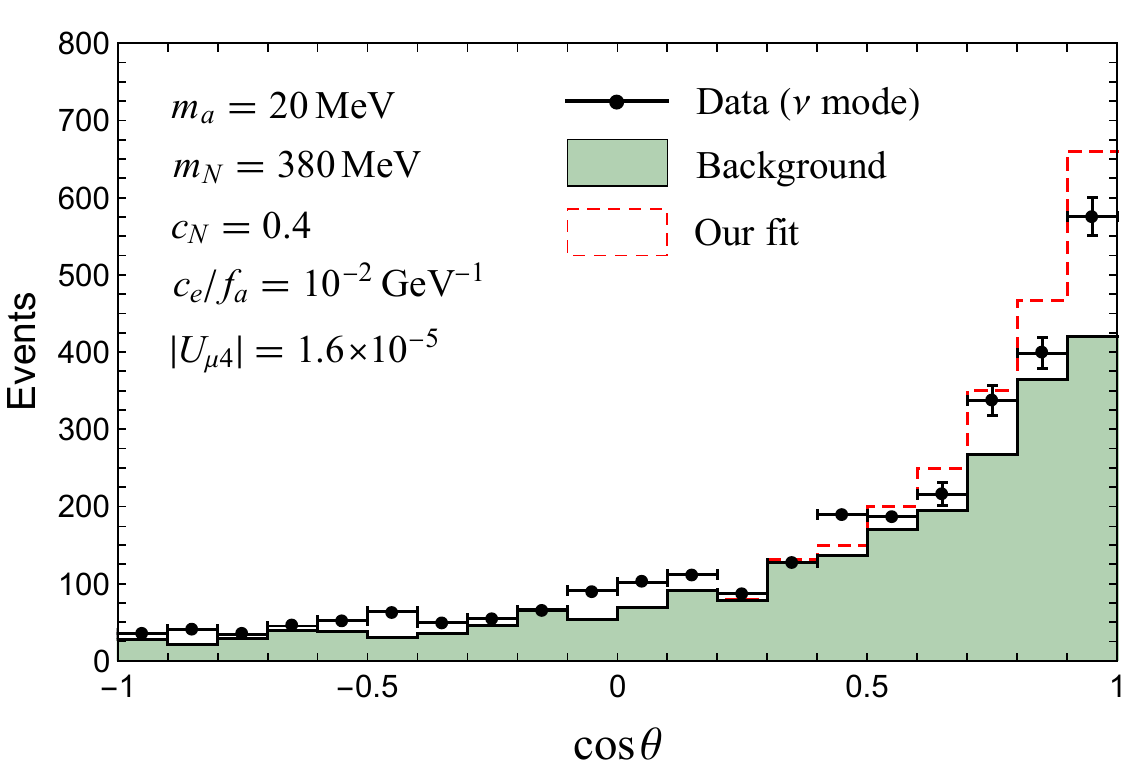}
    $\qquad$
    \includegraphics[width=0.4\textwidth]{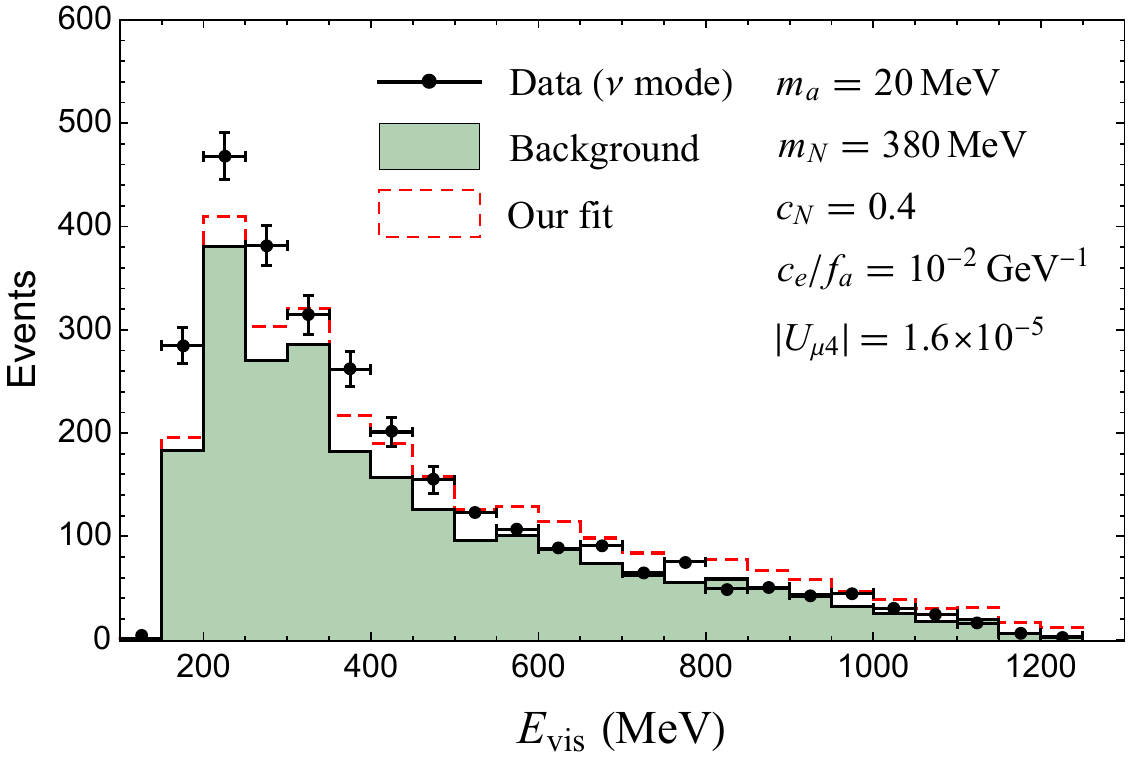}
    \caption{\label{fig:alp_spectra} The comparison of the numerical results of the angular and visible energy spectra in the $\ell$ALP scenario and the data of the MiniBooNE experiment in the neutrino mode. Figure from \cite{Chang:2021myh}.}
\end{figure}

Following the approach applied in~\cite{Fischer:2019fbw}, the angular and visible energy spectra in the neutrino mode are computed and compared with the results of the MiniBooNE experiment, as shown at the top of Fig.~\ref{fig:alp_spectra}. At the bottom panel, the predicted total event numbers are obtained after summing over the spectra and shown as the contours consistent with the MiniBooNE excess at the $1\sigma$ to $3\sigma$ levels on the $\left(m_{N},\left\vert U_{\mu 4}\right\vert^{2}\right)$ and $\left(c_{e}/f_{a},\left\vert U_{\mu 4}\right\vert^{2}\right)$ planes, with the constraints obtained from other experiments.
We find that the scenario with the sterile neutrino mass in the range 150 MeV $\lesssim m_{N} \lesssim$ 380~MeV and the neutrino mixing parameter between $10^{-10} \lesssim \left\vert U_{\mu 4} \right\vert^{2} \lesssim 10^{-8}$ can explain the MiniBooNE excess.

\begin{figure}[ht]
    \centering
    \includegraphics[width=0.44\textwidth]{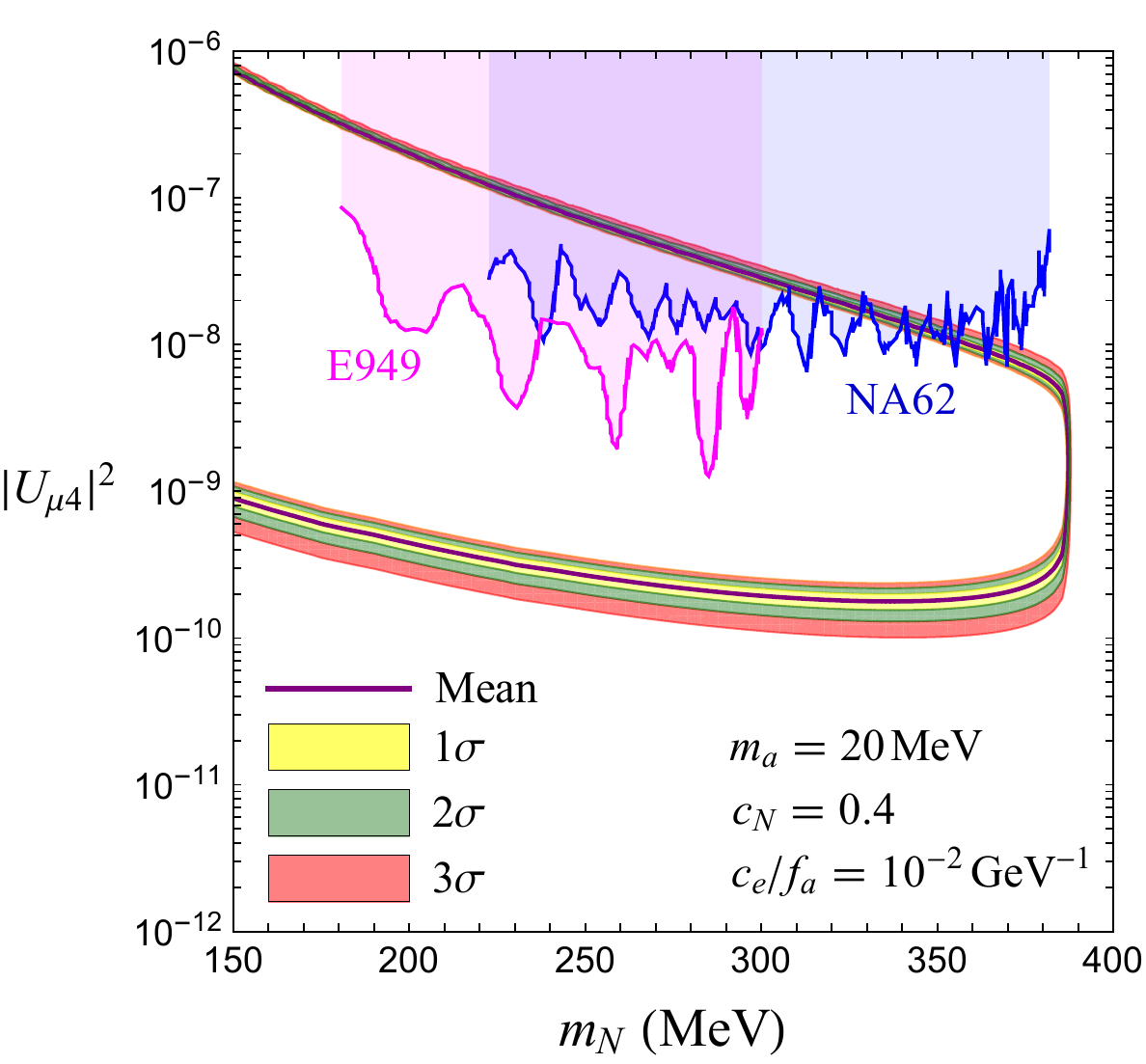}
    $\qquad$
    \includegraphics[width=0.44\textwidth]{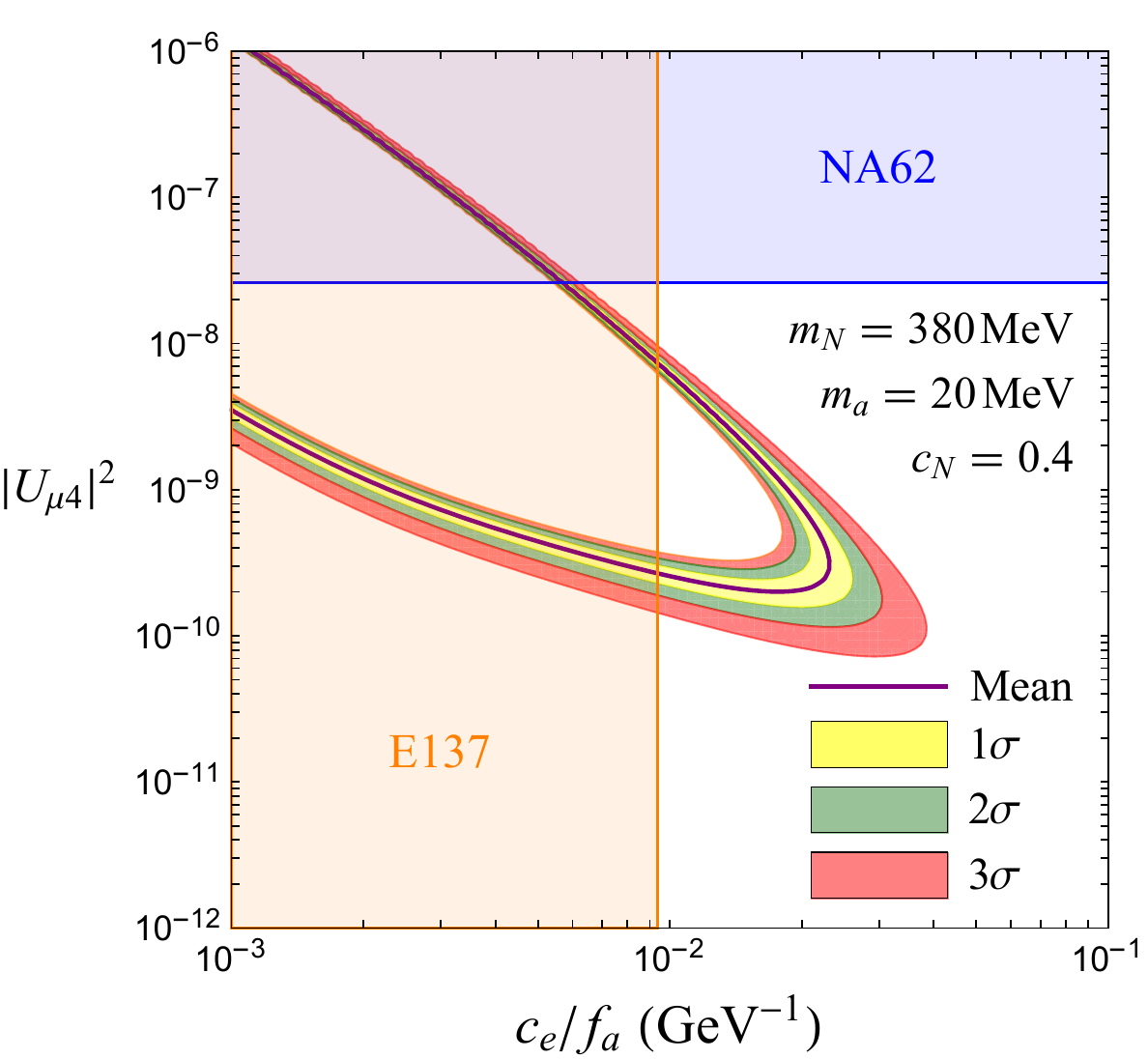}
    \caption{\label{fig:alp_result} Best-fit regions to the MiniBooNE excess in the HNL and $\ell$ALP parameter space. Limits and best-fit regions are shown on the $\left(m_{N},\left\vert U_{\mu 4}\right\vert^{2}\right)$ (left) and $\left(c_{e}/f_{a},\left\vert U_{\mu 4}\right\vert^{2}\right)$ (right) planes. Figure from \cite{Chang:2021myh}.}
\end{figure}




\subsubsection{Dark Matter Particles}
\label{subsub:darkmatter}

Since the neutrinos at MiniBooNE are produced primarily from charged meson decays and the decays of daughter muons of those charged mesons, neutrino-based solutions can accommodate the absence of any excess in the dump mode, in which the charged mesons are no longer focused by magnetic horns, unlike the neutrino and antineutrino modes. Essentially, the neutrino-based explanations work well because a key feature of the excess seems to be correlated to the focusing or suppression of charged mesons. 
\begin{figure}[ht]
    \centering
    \begin{tikzpicture}
        \begin{feynman}
            \vertex (a) {\(\pi^+, K^+\)};
            \vertex [right=1.6cm of a] (b);
            \vertex [above right=1.1cm of b] (f1) {\(\nu_\mu\)};
            \vertex [below right=1.0cm of b] (c);
            \vertex [above right=1.1cm of c] (f2) {\(\phi, a\)};
            \vertex [below right=1.1cm of c] (f3) {\(\mu^+\)};
            \diagram* {
            (a) -- [scalar] (b) -- [fermion] (f1),
            (b) -- [anti fermion] (c),
            (c) -- [scalar] (f2),
            (c) -- [anti fermion] (f3),
            };
        \end{feynman}
    \end{tikzpicture}
    \begin{tikzpicture}
        \begin{feynman}
            \vertex (a) {\(\pi^+, K^+\)};
            \vertex [right=1.6cm of a] (b);
            \vertex [above right=1.1cm of b] (f1) {\(\nu_\mu\)};
            \vertex [below right=1.0cm of b] (c);
            \vertex [above right=1.1cm of c] (f2) {\(A^\prime\)};
            \vertex [below right=1.1cm of c] (f3) {\(\mu^+\)};
            \diagram* {
            (a) -- [scalar] (b) -- [fermion] (f1),
            (b) -- [anti fermion] (c),
            (c) -- [boson] (f2),
            (c) -- [anti fermion] (f3),
            };
        \end{feynman}
    \end{tikzpicture}
    \caption{3-body charged meson decay into a scalar, pseudoscalar, or vector. Analogous processes exist for $\pi^-$ and $K^-$ decay. Figure from \cite{Dutta:2021cip}.}
    \label{fig:3body}
\end{figure}
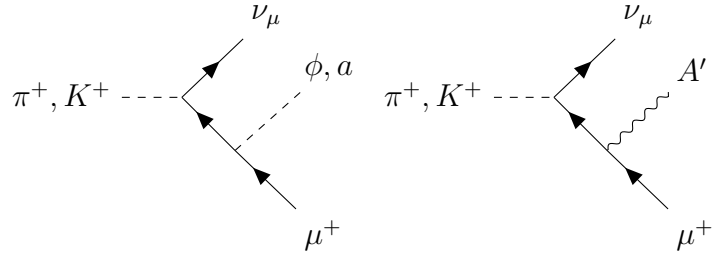

This poses a challenge to dark sector interpretations of the excess (e.g., using $\pi^0$ or dark bremsstrahlung production channels~\cite{MiniBooNEDM:2018cxm,Jordan:2018qiy}), which have been more constrained and less holistic in their explanation of the excess with respect to their counterparts in neutrino BSM physics thus far. One solution proposed in Ref.~\cite{Dutta:2021cip} opens up the possibility for dark sector explanations by means of connecting the dark sector to the physics of charged meson decays, something that had previously been overlooked. The authors in Ref.~\cite{Dutta:2021cip} considered various DM scenarios involving couplings to muons, namely those in Eq.~(\ref{eq:s}),(~\ref{eq:p}), and~(\ref{eq:v}):

\begin{eqnarray}
    \label{eq:s}
     \mathcal{L}_{S} &\supset&  g_\mu \phi \Bar{\mu} \mu + g_n Z^\prime_\alpha \Bar{u} \gamma^\alpha u + \frac{\lambda}{4} \phi F^\prime_{\mu\nu} F^{\mu\nu} + \textrm{h.c.}, \\
     \label{eq:p}
    \mathcal{L}_{P} &\supset& i g_\mu a \Bar{\mu} \gamma^5 \mu + g_n Z^\prime_\alpha \Bar{u} \gamma^\alpha u + \frac{\lambda}{4} a F^\prime_{\mu\nu} \Tilde{F}^{\mu\nu} + \textrm{h.c.}, \\
    \label{eq:v}
    \mathcal{L}_{V} &\supset& e (\epsilon_1 V_{1,\mu}+\epsilon_2 V_{2,\mu}) J_{\rm EM}^\mu + (g_1 V_{1,\mu}+g_2 V_{2,\mu}) J_D^\mu+(g^\prime_1 V_{1,\mu}+g^\prime_2 V_{2,\mu}) J^{\prime\mu}_D.
\end{eqnarray}

\begin{figure}[ht]
    \centering
    \includegraphics[width=\textwidth]{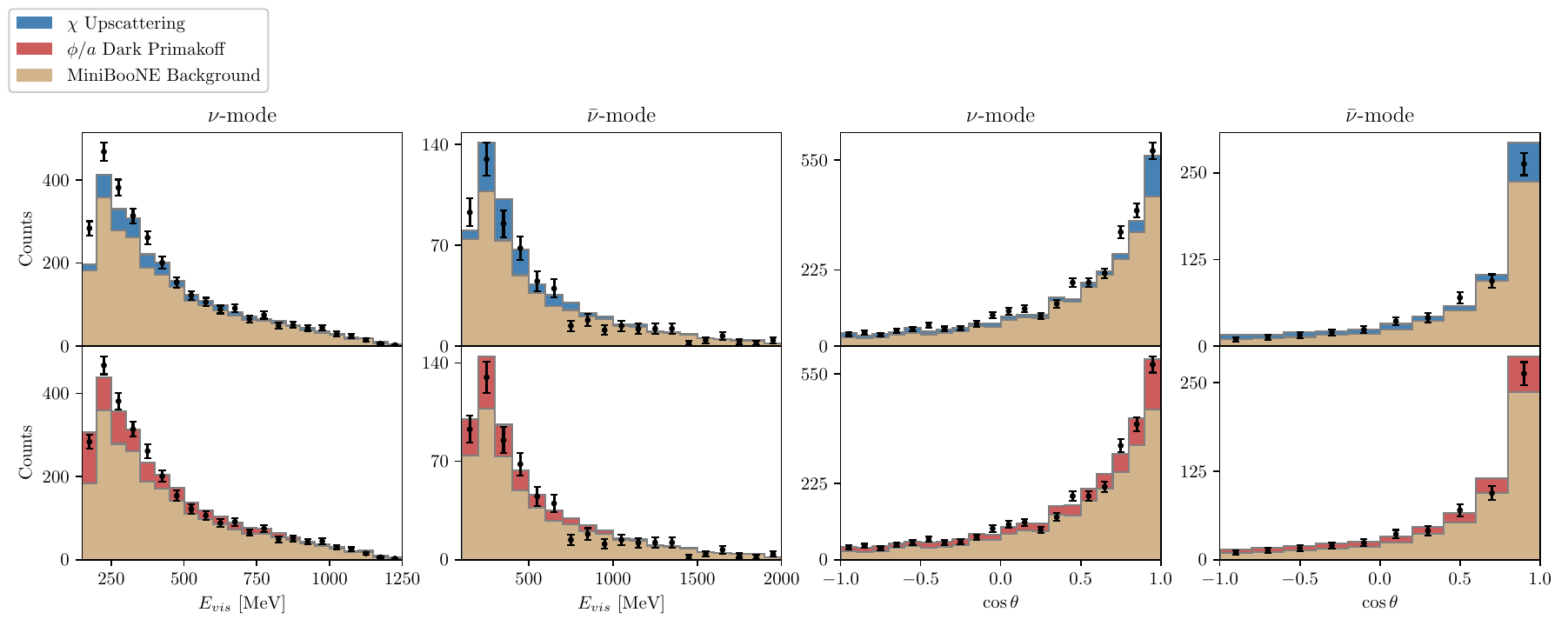}
    \caption{Branching ratios for the 3-body production of scalars, pseudoscalars, and vectors via a charged meson with mass $M = m_{\pi^+}, m_{K^+}$. Figure from \cite{Dutta:2021cip}.}
    \label{fig:br}
\end{figure}

\begin{figure}[ht]
    \centering
    \includegraphics[width=0.4\textwidth]{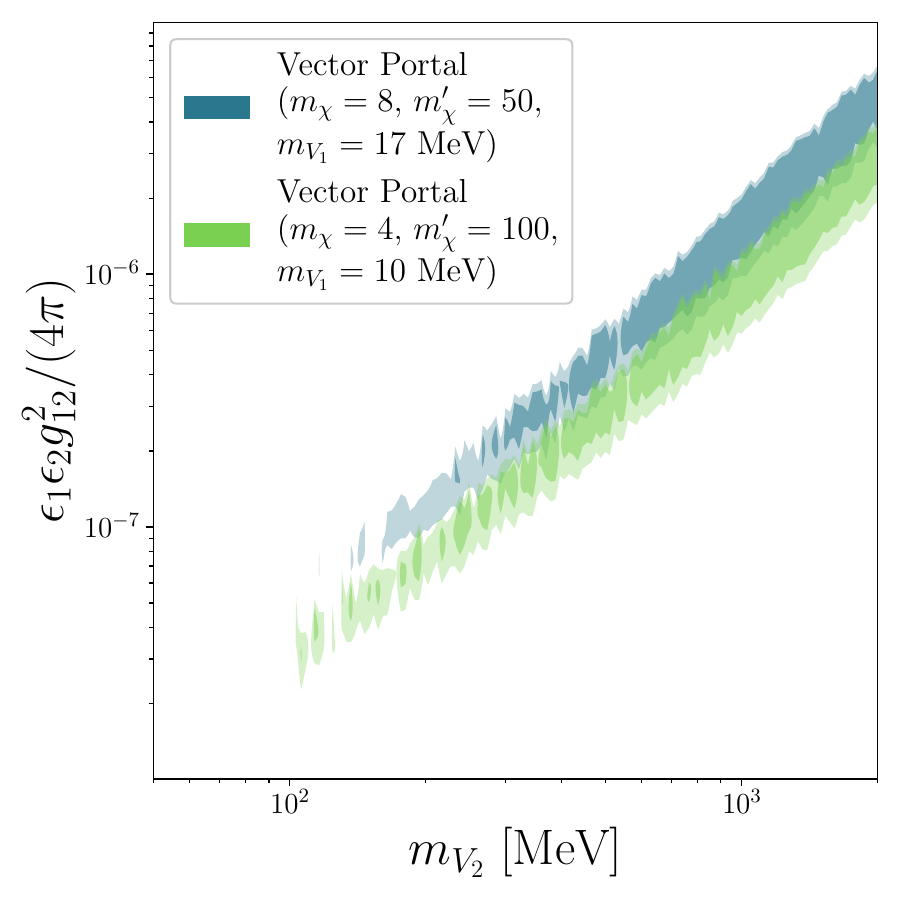}
    $\qquad$
    \includegraphics[width=0.4\textwidth]{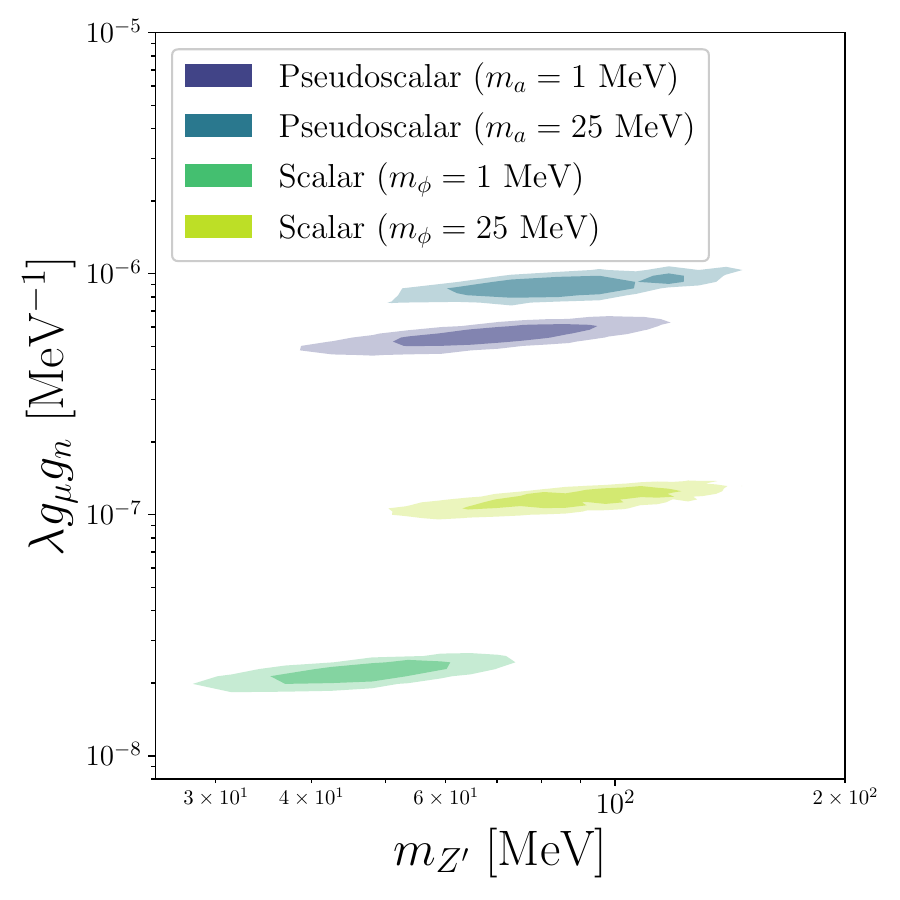}
    \caption{The credible regions for fits to the MiniBooNE cosine spectrum with $\mathcal{L}_V$ (left) and $\mathcal{L}_{S,P}$ (right) are shown at 68\% (dark-shaded) and 95\% (light-shaded). Figure from \cite{Dutta:2021cip}.
    }
    \label{fig:credible_regions}
\end{figure}

Here, three massive bosons have been introduced; a long-lived scalar $\phi$ and pseudoscalar $a$, and a short-lived vector $A^\prime_\alpha$ decaying to DM fermions $\chi$, $\chi^\prime$ with $F^\prime_{\mu\nu} \equiv \partial_\mu A^\prime_\nu - \partial_\nu A^\prime_\mu$. The muonic couplings allow for the 3-body decays of the form $M \to \mu \nu X$ ($M=\pi^\pm, K^\pm$) became possible, as shown in Fig.~\ref{fig:3body}. The 3-body nature of this decay mechanism is not phase-space suppressed in the contraction of $\nu^\dagger \mu_\uparrow$, as opposed to the ordinary 2-body decay which selects out only the combination $\nu_\downarrow \mu_\uparrow$. The branching ratios for scalar, pseudoscalar, and vector DM production are shown in Fig.~\ref{fig:br}. In the $\mathcal{L}_V$ scenario, the detector signature would then take place through DM upscattering, $\chi N \to \chi^\prime N (\chi^\prime \to \chi e^+ e^-$), while scenarios $\mathcal{L}_S$ and $\mathcal{L}_P$ consider long-lived $\phi$/$a$ scattering in the detector through a Primakoff-like process $\phi N \to \gamma N$ via a heavy mediator $Z^\prime$. The parameter space which fits the MiniBooNE excess for $\mathcal{L}_S$, $\mathcal{L}_P$, and $\mathcal{L}_V$ is shown in Fig.~\ref{fig:credible_regions}.

However, the muonic portal is not the only possibility, and DM production in the 3-body $\pi^\pm$ and $K^\pm$ decays from DM-quark couplings can also be treated. In this sense, the scope of the 3-body decay solution can cover several coupling schemes, and should be testable at other experiments with similar meson production capabilities.

This has broad implications for accelerator facilities, such as LBNF (DUNE), that also have magnetic focusing horns that should be sensitive to forward-produced DM from the meson decays. Constraints on the parameter space from accelerator-based searches at CHARM, MINER$\nu$A, and T2K can also be considered, but their smaller POT and exposures do not give them sensitivity to the MiniBooNE excess. Other neutrino experiments such as CCM, JSNS, and COHERENT that produce stopped mesons and lack magnetic focusing horns can also probe the parameter space relevant for the excess, since while the DM signal from meson decays will be isotropic, their detectors are situated much closer to the beam targets to be sensitive to the DM flux.

\subsection{Conventional Explanations}\label{sec:th_landscape:conventional}
While the majority of the explanations explored above rely on some new or BSM physics to give rise to the various anomalies, the possibility that the origin lies in more conventional explanations, such as an underestimated background, mis-modelling in simulation, or over-constrained cross-section uncertainties, must still be considered. While these explanations are generally difficult to test directly without access to (often) collaboration-internal experimental tools and data sets, there have been several attempts in recent years to test individual anomalies in this direction. In one such example \cite{Brdar:2021cgb}, it was shown that allowing a combination of theoretical uncertainties in different background channels to fluctuate in unison is not sufficient to resolve the MiniBooNE anomaly; however, it can reduce the significance of the MiniBooNE excess. In this section, we discuss several such possible ``conventional'' interpretations for the anomalies.

\subsubsection{Single-Photon Production}
\label{sec:single_gamma_luis}

The solution to the MiniBooNE puzzle may have important implications for our understanding of neutrinos and their interactions. In addition to the interpretation in terms of new physics, the MiniBooNE anomaly could be a manifestation of new forces of nature, while unaccounted or poorly modeled SM backgrounds cannot be entirely discarded. Once Cherenkov detectors like MiniBooNE misidentify single photon tracks as electrons, the excess of events could be due to their products through both SM and BSM mechanisms. 

\begin{figure}[ht!]
\centering
\includegraphics[width=0.23\textwidth]{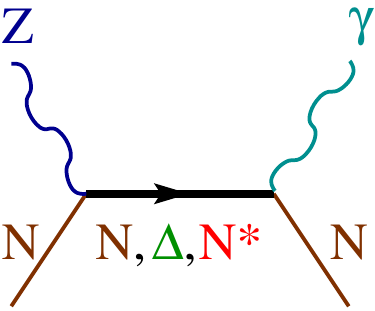}
\hspace{.05\textwidth} 
\includegraphics[width=0.23\textwidth]{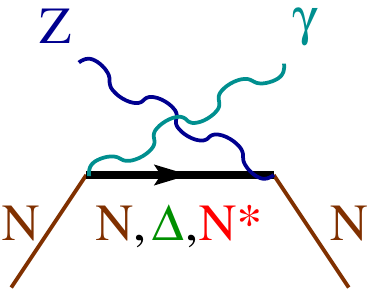}
\hspace{.05\textwidth} 
\includegraphics[width=0.2\textwidth]{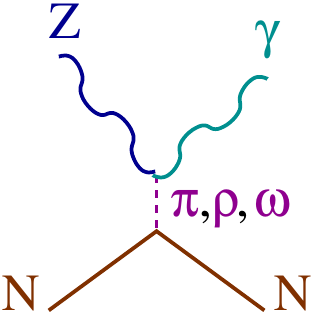}
\caption{\label{fig:NCgamma_diags}  Feynman diagrams for NC single photon emission considered in the literature. The first two diagrams stand for direct and crossed baryon pole terms with nucleons and baryon resonances in the intermediate state. The third diagram represents $t$-channel meson exchange contributions. Figure from Ref.~\cite{Alvarez-Ruso:2021dna}.
}
\end{figure}
In the SM, single photons can be emitted in NC interactions, NC$1\gamma$, on nucleons, $\nu (\bar{\nu})\, N \rightarrow \nu (\bar{\nu})\, \gamma \, N$, or on heavy nuclei, via incoherent or coherent (where the nucleus remains in its ground state) scattering. Theoretical models for the elementary NC$1\gamma$~\cite{Hill:2009ek,Serot:2012rd,Wang:2013wva} take into account $s$- and $u$-channel amplitudes with nucleons and $\Delta(1232)$, but also heavier baryon resonances, in the intermediate state, Fig.~\ref{fig:NCgamma_diags}. The structure of nucleon pole terms at threshold is determined by symmetries. The extension towards higher energy and momentum transfers, required to predict cross sections at MiniBooNE, is performed by the introduction of phenomenologically parametrized  weak and electromagnetic form factors. The same strategy has been adopted for resonance terms. The $\Delta(1232)$ excitation followed by radiative decay is the dominant mechanism, as correctly assumed by MiniBooNE, but the contribution of non-resonant terms is also sizable. The uncertainty in the elementary NC$1\gamma$ cross section is dominated by the leading $N-\Delta$ axial transition coupling, $C_5^A(q^2=0)$, which is related to the $\Delta N\pi$ coupling (known from $\pi N$ scattering) by a Goldberger-Treiman relation, but has also been found to be $C^A_5(0) = 1.18 \pm 0.07$~\cite{Hernandez:2016yfb} from data on $\pi$  production induced by neutrino scattering on hydrogen and deuterium. It is worth stressing that the $\Delta N \gamma$ couplings responsible for the resonance radiative decay are directly related to helicity amplitudes $A_{1/2,3/2}$ known with few-percent accuracy from photo-nucleon interactions~\cite{ParticleDataGroup:2020ssz}. Furthermore, owing to isospin symmetry, these quantities also constrain the vector part of the weak $\Delta$ production. This implies that large uncertainties in the $\Delta$ radiative decay couplings are at odds with hadron phenomenology but would also have an observable impact in weak pion production.  Among the non-resonant contributions, $t$-channel $\omega$-meson exchange, was proposed as a solution for the MiniBooNE anomaly~\cite{Harvey:2007rd} because of the rather large (although uncertain) couplings and the $\omega$ isoscalar nature, which enhances its impact on the coherent NC$1\gamma$ reaction. However, actual calculations found this contribution small compared to $\Delta(1232)$ excitation~\cite{Hill:2009ek,Rosner:2015fwa}. Nuclear effects, in particular the broadening of the $\Delta$ resonance in the nucleus, are important for single photon emission~\cite{Zhang:2012aka,Wang:2013wva}. On $^{12}$C they cause a reduction of about $30$\% in the cross section (see Fig.~9 of Ref.~\cite{Wang:2013wva}).

\begin{figure}[ht!]
  \centering
    \includegraphics[width=.44\textwidth]{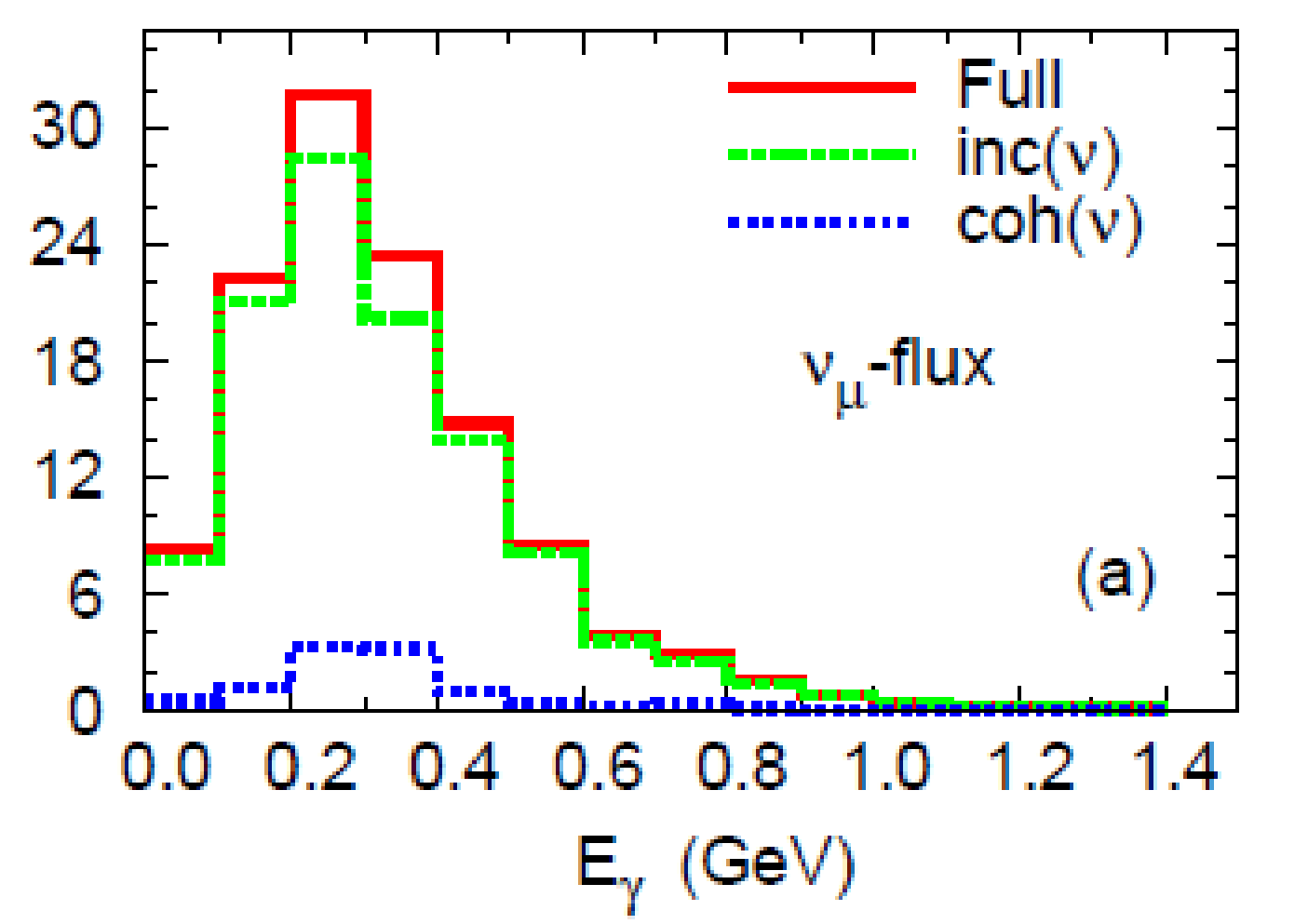}     
 $\qquad$
    \includegraphics[width=.44\textwidth]{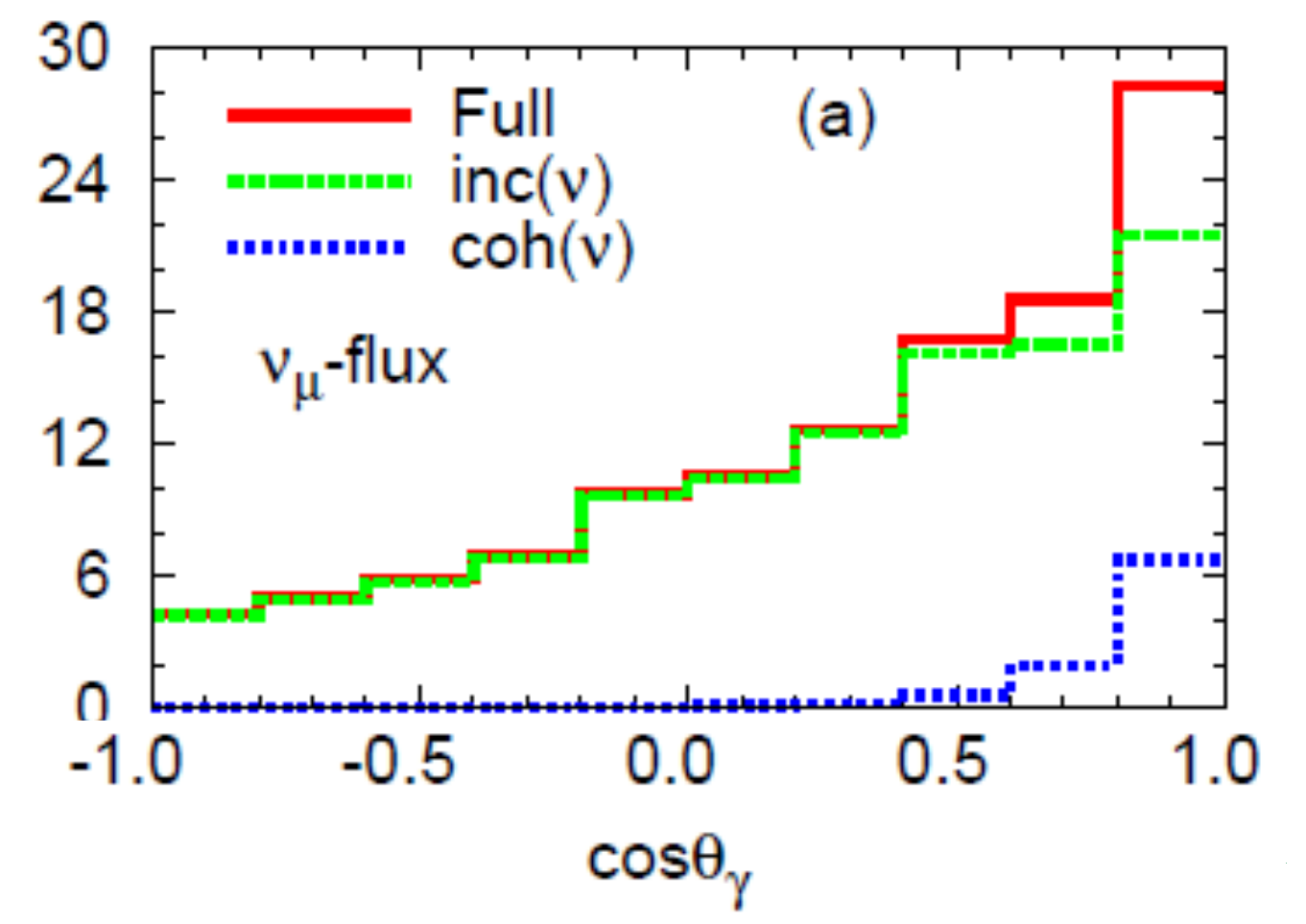}
  \caption{SM prediction for NC1$\gamma$ events at  MicroBooNE  for $6.6 \times 10^{20}$~POT in $\nu$-mode. Figure from Ref.~\cite{Alvarez-Ruso:2021dna}.} 
  \label{fig:ncgammaMicroBooNE}
\end{figure}

With the ingredients outlined above, the SM single-$\gamma$ contribution to  the number events in the MiniBooNE detector and their distributions have been calculated~\cite{Zhang:2012xn,Wang:2014nat} using the available information about the detector mass and composition (CH$_2$), neutrino flux and, quite significantly, photon detection efficiency. Results are to a large extent consistent with the (data based but relying on an improvable reaction model) MiniBooNE estimate (Figs. 4, 6 and 8 of Ref.~\cite{Wang:2014nat}) and, therefore, insufficient to explain the excess. The impact of two-nucleon meson-exchange reaction mechanisms has been recently investigated~\cite{Chanfray:2021wie} and found to be small (a factor of around 9 at $E_\nu=500$~MeV compared to single-nucleon mechanisms). In Ref.~\cite{Giunti:2019sag} it was estimated that the NC$1\gamma$ background should be enhanced by a factor between 1.52 and 1.62 over the MiniBooNE estimate, depending on the energy range and mode. Such an enhancement, shrinks the excess and significantly reduces the appearance-disappearance tension in global fits but is at odds with the earlier described theoretical calculations. An upper limit for the NC1$\gamma$ cross section on liquid argon has been  recently obtained by the MicroBooNE experiment~\cite{MicroBooNE:2021zai}, as discussed in the following section. It disfavors that the excess could be solely attributed to this reaction channel but new results with higher statistics are required for a firm conclusion. Assuming $6.6 \times 10^{20}$~POT from the experiment's run plan, the distributions of the NC $1\gamma$ events calculated with the model of Ref.~\cite{Wang:2013wva} are given in Fig.~\ref{fig:ncgammaMicroBooNE} (adapted from Ref.~\cite{Alvarez-Ruso:2018fdm}.). Comparison to future data shall offer valuable information about this process.


\subsubsection{Reactor Flux Modeling}
\label{sec:th_land_rx_models}

One conventional explanation for the Reactor Antineutrino Anomaly~(RAA) has been gaining momentum thanks to the significant experimental and theoretical progress made in the past decade.
Reactor neutrino experiments that can measure oscillations without any reliance on reactor neutrino models (discussed further in Sec.~\ref{sec:expt_landscape_reactors}) have been chipping away at RAA-suggested sterile neutrino oscillation parameter space.  
Additionally, neutrino, nuclear physics, and nuclear theory evidence have recently emerged suggesting that RAA may, at least in part, be caused by problems with reactor flux models.  
While experimental developments are discussed in Sec.~\ref{sec:expt_landscape_conventional_rates}, recent theoretical advancements in the flux prediction landscape that lend support to non-BSM origin to the RAA are discussed here.  


\begin{table}
\centering
\begin{tabular}{c|cccc}
    {} \bf Model & $\bm{\sigma_{235}}$ & $\bm{\sigma_{238}}$ & $\bm{\sigma_{239}}$ & $\bm{\sigma_{241}}$ \\
    \hline
    \bf HM
    &
    $6.74 \pm 0.17$
    &
    $10.19 \pm 0.83$
    &
    $4.40 \pm 0.13$
    &
    $6.10 \pm 0.16$
    \\
    \hline
    \bf EF
    &
    $6.29 \pm 0.31$
    &
    $10.16 \pm 1.02$
    &
    $4.42 \pm 0.22$
    &
    $6.23 \pm 0.31$
    \\
    \hline
    \bf HKSS
    &
    $6.82 \pm 0.18$
    &
    $10.28 \pm 0.84$
    &
    $4.45 \pm 0.13$
    &
    $6.17 \pm 0.16$
    \\
    \hline
    \bf KI
    &
    $6.41 \pm 0.14$
    &
    $ 9.53 \pm 0.48$
    &
    $4.40 \pm 0.13$
    &
    $6.10 \pm 0.16$  \\ \hline
\end{tabular}
\caption{\label{tab:model_2020}
Theoretical IBD yields of the four fissionable isotopes in units of $10^{-43} \text{cm}^{2}/\text{fission}$ predicted by different models~\cite{Giunti:2021kab}.}
\end{table}

\begin{table*}
\centering
\begin{tabular}{c@{\qquad}cc@{\qquad}cc@{\qquad}cc}
\bf Model & \multicolumn{2}{c}{\bf Rates} & \multicolumn{2}{c}{\bf Evolution} & \multicolumn{2}{c}{\bf Rates + Evolution}
\\
& $\bm{\overline{R}_{\text{mod}}}$ & \bf RAA & $\bm{\overline{R}_{\text{mod}}}$ & \bf RAA & $\bm{\overline{R}_{\text{mod}}}$ & \bf RAA
\\
\hline
\bf HM
&
$0.936 {}^{ + 0.024 }_{ - 0.023 }$
&
$2.5\,\sigma$
&
$0.933 {}^{ + 0.025 }_{ - 0.024 }$
&
$2.6\,\sigma$
&
$0.930 {}^{ + 0.024 }_{ - 0.023 }$
&
$2.8\,\sigma$
\\
\bf EF
&
$0.960 {}^{ + 0.033 }_{ - 0.031 }$
&
$1.2\,\sigma$
&
$0.975 {}^{ + 0.032 }_{ - 0.030 }$
&
$0.8\,\sigma$
&
$0.975 {}^{ + 0.032 }_{ - 0.030 }$
&
$0.8\,\sigma$
\\
\bf HKSS
&
$0.925 {}^{ + 0.025 }_{ - 0.023 }$
&
$2.9\,\sigma$
&
$0.925 {}^{ + 0.026 }_{ - 0.024 }$
&
$2.8\,\sigma$
&
$0.922 {}^{ + 0.024 }_{ - 0.023 }$
&
$3.0\,\sigma$
\\
\bf KI
&
$0.975 {}^{ + 0.022 }_{ - 0.021 }$
&
$1.1\,\sigma$
&
$0.973 {}^{ + 0.023 }_{ - 0.022 }$
&
$1.2\,\sigma$
&
$0.970 \pm 0.021$
&
$1.4\,\sigma$ \\ \hline
\end{tabular}
\caption{\label{tab:model_rates} Average ratio $\overline{R}_{\text{mod}}$ obtained in Ref.~\cite{Giunti:2021kab} from the least-squares analysis of the reactor rates in Tab.~\ref{tab:reactor_expts} and of the Daya Bay~\protect\cite{DayaBay:2017jkb} and RENO~\protect\cite{RENO:2018pwo} evolution data for the IBD yields of the models in Tab.~\ref{tab:model_2020}. The RAA columns give the corresponding statistical significance of the reactor antineutrino anomaly. The descriptions of all the models are given in text.}
\end{table*}

The statistical significance of the RAA depends not only on the magnitude of offset between reactor \anue data and the Huber-Mueller prediction but also on the size of the error bands applied to those predictions.  
In the years following the inception of the RAA in 2011, a variety of reactor modeling studies have argued that the 2-3\% error budget assigned to this prediction is likely underestimated.  
In particular, the role of forbidden beta transitions in altering the \anue flux and spectrum reported by conversion predictions is not considered in the formulation of Huber-Mueller model error bands~\cite{Huber:2011wv}.  
When naive treatments of forbidden decay contributions are included in the prediction, variations in the flux of 4\% or more are observed~\cite{Hayes:2013wra} and are reflected in the community-driven report in Ref.~\cite{bib:IAEA}.  
On the other hand, a more recent conversion calculation (termed the HKSS model) that attempts to account for forbidden transition contribution using nuclear shell model-based calculations shows strong deviations from the Huber-Mueller model in spectral shape.
However, it finds no major discrepancy in reported IBD yields~\cite{Hayen:2019eop}.  

The past ten years have also brought about substantial development of state-of-the-art summation calculations, thanks to improved nuclear data evaluations and new nuclear structure measurements for a range of high-Q, high-yield fission  daughters~\cite{fijalkowska2017impact,Guadilla:2019gws,Guadilla:2019zwz,IGISOL:2015ifm,rasco2016decays,Rice:2017kfj,Valencia:2016rlr}.  
The improved summations can be compared to conversion calculations to provide an assessment of the latter's robustness.  
As early as 2012, improved summation calculations were shown to generate reduced flux predictions with respect to earlier iterations~\cite{Fallot:2012jv}, indicating modest over-prediction of \uEight~fluxes in the Huber-Mueller prediction.  
When comparing to conversion predictions, modern summation calculations were shown to predict different fuel-dependent \anue flux variations~\cite{Hayes:2017res}.  
Most recently in 2019, Estienne, Fallot \textit{et al}~\cite{Estienne:2019ujo} used comprehensive improvements and updates in nuclear databases to generate a summation model (referred to as the EF Model) with a total predicted flux of a few percent lower than the measured global IBD yield average, but with a \uFive~prediction $\sim$6\% smaller than that predicted by the HM conversion model.  
These comparisons are suggestive of a possible issue with conversion-predicted fluxes for individual isotopes.  


\begin{figure}[ht]
\centering
\includegraphics[width=0.45\textwidth]{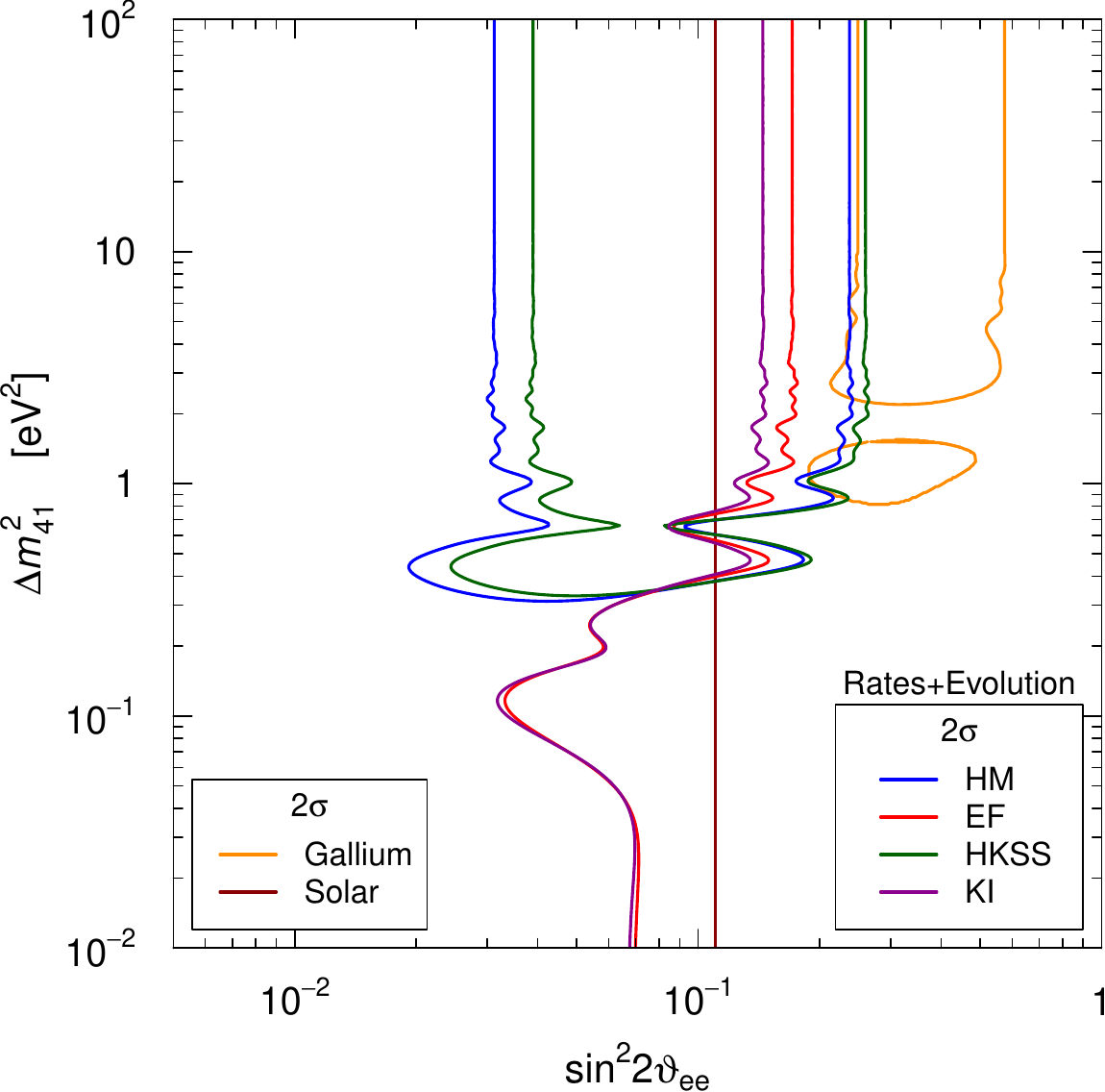}
\caption{\label{fig:osc-wp-rat+evo-2s} Contours of the $2\sigma$ allowed regions in the ($\sin^2\!2\vartheta_{ee},\Delta{m}^2_{41}$) plane obtained from the combined neutrino oscillation fit of the reactor rates in Tab.~\ref{tab:model_rates} and the Daya Bay~\protect\cite{DayaBay:2017jkb} and RENO~\protect\cite{RENO:2018pwo} evolution data. The blue, red, green, and magenta curves correspond, respectively, to the HM, EF, HKSS, and KI models in Tab.~\ref{tab:model_2020}. Also shown are the contour of the $2\sigma$ allowed regions of the Gallium anomaly obtained in Ref.~\cite{Barinov:2021asz} from the combined analysis of the GALLEX, SAGE and BEST data (orange curve), and the $2\sigma$ bound obtained from the analysis of solar neutrino data in Ref.~\cite{Goldhagen:2021kxe} (dark red vertical line). Figure from \cite{Giunti:2021kab}.}
\end{figure}

The IBD yields predicted by these various models can be compared to the global IBD yields listed in Tab.~\ref{tab:Rates_Giunti2012} and to specialized Daya Bay~\cite{DayaBay:2017jkb} and RENO~\cite{RENO:2018pwo} dataset reporting yields as a function of varying contents of the reactor cores; the latter datasets are described further in Sec.~\ref{sec:expt_landscape_conventional_rates}.  
In particular, if comparisons are made under the hypothesis that the RAA is generated by sterile neutrino oscillations, the different reactor flux models generate substantially differing interpretations.  
To illustrate, Fig.~\ref{fig:osc-wp-rat+evo-2s} shows the contours of the $2\sigma$ allowed regions in the ($\sin^2\!2\vartheta_{ee},\Delta{m}^2_{41}$) plane of 3+1 active-sterile neutrino mixing parameters~\cite{Giunti:2021kab}. 
One can see that an indication in favor of neutrino oscillations only for the HM and HKSS conversion models, which exhibit a significant reactor rate anomaly above $2\sigma$ (see Tab.~\ref{tab:model_rates}). 
Considering the EF model, for which the reactor rate anomaly is less than a few percent, the $2\sigma$ exclusion curves in Fig.~\ref{fig:osc-wp-rat+evo-2s} allow only small values of $\sin^2 2\vartheta_{ee}$, including $\sin^2 2\vartheta_{ee}=0$, which corresponds to a lack of any statistically significant indication of sterile neutrino oscillations.  
It should be noted here that error estimates for summation calculations are ill-defined, but are generally expected to be similar in magnitude to those provided by conversion predictions; efforts to provide more robust error envelopes are underway~\cite{MATTHEWS2021101441}.

For all the reactor flux models, it is also worth noting that upper bounds exist for the value of the mixing parameter $\sin^2 2\vartheta_{ee}$, with exact limits dependent on the value of $\Delta m_{41}^2$. 
For $\Delta m_{41}^2 \gtrsim 2 \, \text{eV}^2$ the upper bounds for $\sin^2 2\vartheta_{ee}$ are between $0.14$ and $0.25$.
Figure~\ref{fig:osc-wp-rat+evo-2s} shows that these bounds and the solar bound~\cite{Goldhagen:2021kxe} are in agreement, but in tension with the large mixing~\cite{Barinov:2021asz} required to explain the anomaly of the GALLEX~\cite{Kaether:2010ag}, SAGE~\cite{Abdurashitov:2005tb}, and BEST~\cite{Barinov:2021asz} gallium experiments with short baseline neutrino oscillations.
This is a puzzling recent development in the phenomenology of short-baseline neutrino oscillations that may require an explanation extending beyond both conventional explanations and the simplest possible model of 3+1 active-sterile neutrino mixing.

\subsubsection{The Gallium Anomaly and Interaction Cross-Section Uncertainties} \label{sec:GaAnomaly}

The first analysis establishing the existence of the Gallium Anomaly~\cite{Acero:2007su} did not consider the uncertainties on the cross section of the detection process in Eq.~(\ref{eq:nueGa}) and, as mentioned in previous sections, this has been a source of subsequent investigation as a possible avenue for resolving this anomaly.  

It is now clear that this quantity is of paramount relevance for the possible interpretation of the Gallium Anomaly. Different calculations of the cross section have been published after the seminal work by Bahcall \cite{Bahcall:1997eg}, and the values are summarized in Tab.~\ref{tab:nuGe_CrosSection}.
\begin{table}[ht]
    \centering
    \begin{tabular}{|c|c|}\hline
         $\sigma$ ($10^{-46}$ cm$^2$) & Ref.             \\\hline
         $58.1  ^{+0.21}_{-0.16}$ & \cite{Bahcall:1997eg} \\
         $59.3  \pm 0.14$  & \cite{Frekers:2015wga}      \\
         $59.10 \pm 0.114$ & \cite{Barinov:2017ymq}      \\
         $56.7  \pm 0.06$  & \cite{Kostensalo:2019vmv}   \\
         $59.38 \pm 0.116$ & \cite{Semenov:2020xea}      \\\hline
    \end{tabular}
    \caption{Summary of the cross section values recently published.}
    \label{tab:nuGe_CrosSection}
\end{table}

Also, it has been pointed \cite{Giunti:2010zu} that the rather large uncertainties come from the fact that only the cross section of the transition from the ground state of $^{71}\rm{Ga}$ to the ground state of $^{71}\rm{Ge}$ is known with precision from the measured rate of electron capture decay of $^{71}\rm{Ge}$ to $^{71}\rm{Ga}$. In fact, recent improvements on measurements of this transition \cite{Frekers:2011zz,Frekers:2013hsa,Frekers:2015wga} indicate that the Gallium Anomaly persists. However, electron neutrinos produced by processes in Eq.~(\ref{eq:CrAr_elCapture}), can also be absorbed through transitions from the ground state of $^{71}\rm{Ga}$ to two excited states of $^{71}\rm{Ge}$ at 175~keV and 500~keV.

\begin{figure}[hptb!]
    \centering
    \includegraphics[width=0.4\textwidth]{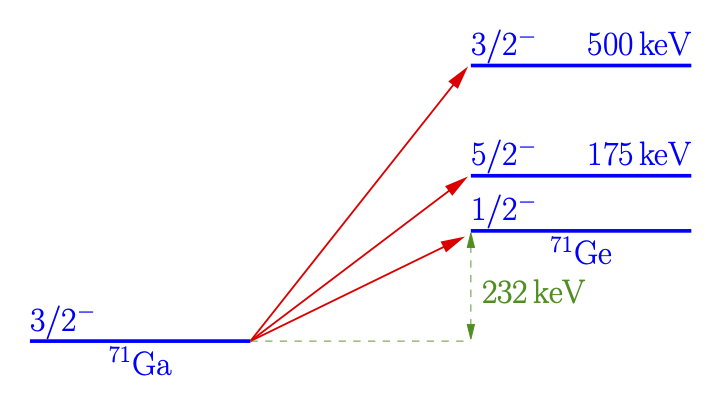}
    \caption{Nuclear levels for the $^{71}\rm{Ga}$ transitions to $^{71}\rm{Ge}$. Figure from \cite{Giunti:2012tn}.}
    \label{fig:GatoGe}
\end{figure}

When the aforementioned uncertainties are taken into account, different results for the total cross section for the radioactive sources are obtained, resulting in changes to the measured and expected $^{71}\rm{Ge}$ event rates. On the other hand, complete calculations of the cross sections of the interaction process, Eq.~(\ref{eq:nueGa}), for neutrinos produced by $^{51}\rm{Cr}$ and $^{37}\rm{Ar}$
sources are given in Ref.~\cite{Giunti:2012tn} as
\begin{equation}
\sigma = \sigma_{\rm{gs}} \left( 1 + \xi_{175}\frac{\rm{BGT}_{175}}{\rm{BGT}_{\rm{gs}}} + \xi_{500}\frac{\rm{BGT}_{500}}{\rm{BGT}_{\rm{gs}}} \right)
\,. 
\label{eq:CrossSec_BGT}
\end{equation}
Here, $\sigma_{\rm{gs}}$ is the cross section of the transitions from the ground state of $^{71}\rm{Ga}$ to the ground state of $^{71}\rm{Ge}$, $\rm{BGT}_{\rm{gs}}$ is the corresponding Gamow-Teller strength, and $\rm{BGT}_{175}$ and $\rm{BGT}_{500}$ are the Gamow-Teller strengths of the transitions from the ground state of ${}^{71}\rm{Ga}$ to the two excited states of ${}^{71}\rm{Ge}$ at about 175~keV and 500~keV as shown in Fig.~\ref{fig:GatoGe}. The coefficients of $\rm{BGT}_{175}/\rm{BGT}_{\rm{gs}}$ and $\rm{BGT}_{500}/\rm{BGT}_{\rm{gs}}$ are determined by phase space:
$\xi_{175}({}^{51}\rm{Cr}) = 0.669$,
$\xi_{500}({}^{51}\rm{Cr}) = 0.220$,
$\xi_{175}({}^{37}\rm{Ar}) = 0.695$,
$\xi_{500}({}^{37}\rm{Ar}) = 0.263$
\cite{Bahcall:1997eg}.

Table~\ref{tab:Rates_Giunti2012} shows the different values of the rate when four different approaches (other than the one by Baxton, $R_{\rm{B}}$) are used to compute the cross sections \cite{Giunti:2012tn}: $R_{\rm{HK}}$ uses information about the Gamow-Teller strengths from Haxton \cite{Haxton:1998uc} and Krofcheck \emph{et al.}~\cite{Krofcheck:1985fg}; for $R_{\rm{FF}}$, the corresponding numbers are taken from Frekers \emph{et al.}~\cite{Frekers:2011zz}; $R_{\rm{HF}}$ uses $\rm{BGT}_{175}$ from Haxton and $\rm{BGT}_{500}$ from Frekers \emph{et al.}; and $R_{\rm{JUN45}}$ uses calculations using nuclear shell-model wave functions obtained by exploiting recently developed two- nucleon interactions \cite{Honma:2009zz} (see also \cite{Kostensalo:2019vmv} for additional details).
\begin{table}[!hptb]
    \centering
    \begin{tabular}{|c|c|c|c|c|c|}\hline
	         & GALLEX$_1$ & GALLEX$_1$ & SAGE$_{\rm{Cr}}$ & SAGE$_{\rm{Ar}}$ & Avg.\\\hline
$R_{\rm{B}}$ & $0.95 \pm 0.11$ & $0.81 {}^{+0.10}_{-0.11}$ & $0.95 \pm 0.12$ & $0.79 \pm 0.08$ & $0.86 \pm 0.05$
\\
$R_{\rm{HK}}$	& $0.85 \pm 0.12$ & $0.71 \pm 0.11$ & $0.84{}^{+0.13}_{-0.12}$ & $0.71 \pm 0.09$ & $0.77 \pm 0.08$
\\
$R_{\rm{FF}}$	& $0.93 \pm 0.11$ & $0.79{}^{+0.10}_{-0.11}$ & $0.93{}^{+0.11}_{-0.12}$ & $0.77{}^{+0.09}_{-0.07}$ & $0.84 \pm 0.05$
\\
$R_{\rm{HF}}$	& $0.83{}^{+0.13}_{-0.11}	$ & $0.71 \pm 0.11$ & $0.83{}^{+0.13}_{-0.12}$ & $0.69{}^{+0.10}_{-0.09}$ & $0.75{}^{+0.09}_{-0.07}$ \\
$R_{\rm{JUN45}}$	& $0.97 \pm 0.11$ & $0.83 \pm 0.11$ & $0.97 \pm 0.12$ & $0.81 \pm 0.08$ & $0.88 \pm 0.05$ \\\hline
    \end{tabular}
    \caption{Ratios of measured and expected $^{71}\rm{Ge}$ event rates in the GALLEX and SAGE experiments. The last column corresponds to the weighted average \cite{Giunti:2012tn}. Information in the last row is from Ref.~\cite{Kostensalo:2019vmv}.}
    \label{tab:Rates_Giunti2012}
\end{table}

\begin{figure}[!hptb]
    \centering
    \includegraphics[width=0.5\textwidth]{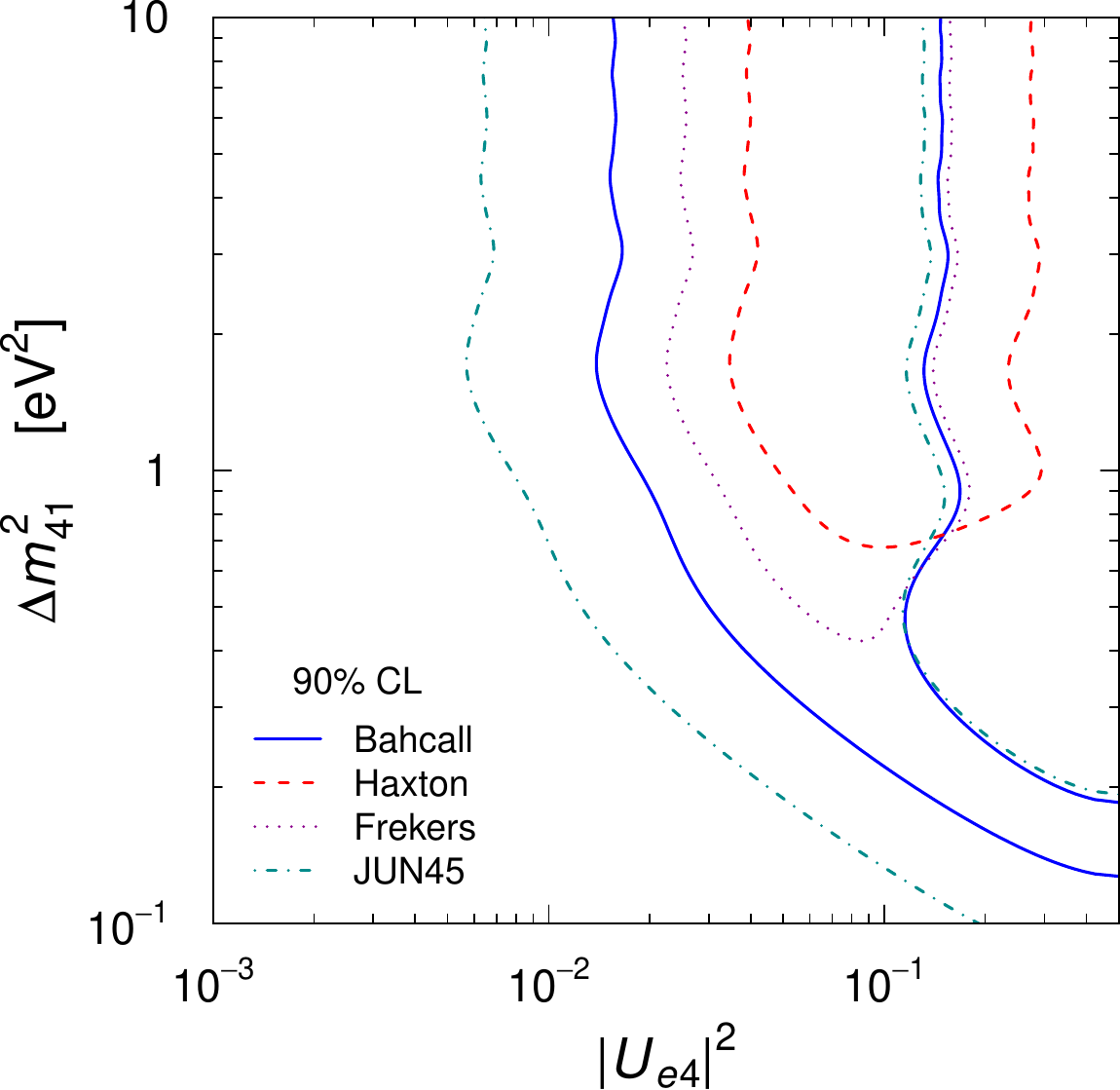}
    \caption{The 90\% allowed regions in the $|U_{e4}|^2$--$\Delta{m}^2_{41}$ plane
obtained from the analysis of the measured and expected $^{71}\rm{Ge}$ event rates \cite{Kostensalo:2019vmv}.}
    \label{fig:GaAnomaly_contours}
\end{figure}

The last column of Tab.~\ref{tab:Rates_Giunti2012} shows the corresponding weighted average for each case showing the Gallium Anomaly with a statistical significance of $3\sigma$, $2.9\sigma$, $3.1\sigma$, and $2.3\sigma$, respectively for the last four cases. This confirms the Gallium Anomaly and retains the indication in favor of a short-baseline disappearance of electron neutrinos, possibly due to neutrino oscillations. Figure~\ref{fig:GaAnomaly_contours} shows the 90\% contours in the $|U_{e4}|^2$--$\Delta{m}^2_{41}$ plane obtained from the analysis of the measured and expected $^{71}\rm{Ge}$ event rates, considering the neutrino survival probability Eq.~(\ref{eq:nueSurvP}), where one can see that the squared-mass difference $\Delta{m}^2_{41}$ allowed values are $\sim1$~eV$^2$ or larger, for the different approaches to compute the $\nu_e \,-\,{}^{71}\rm{Ga}$ cross section.

A revision of the statistical significance of the Gallium Anomaly considering different calculations of the neutrino detection cross section, has been performed to compare the 3+1 neutrino oscillation hypothesis with the Reactor Antineutrino Anomaly and with the inclusion of data from tritium experiments and from experiments measuring solar neutrinos  [Giunti:2022btk]. Remarkably, it was found that the Gallium Anomaly is in strong tension with bounds obtained from the other data sets. In addition, when all data are combined, the corresponding parameter goodness of fit is below 0.042\%, implying a tension of ~$\sim5\sigma$, leading to the conclusion that it should be necessary to seek for alternative solutions to the short-baseline oscillations for this anomaly.

\subsection{Summary of Interpretations}

A summary of the interpretations detailed in this section is provided in Tab.~\ref{tab:sec-3:big_picture}.
The columns, from left to right, are the following: broad classes of models; specific models that fall in each class; the experimental signature of each model; which anomalies each model can address (LSND, MiniBooNE, Reactor and Gallium anomalies, respectively), and the corresponding references.
For convenience, in the first column, we also indicate the sections of this document that are related to each class of models.

Finally, in Tab.~\ref{fig:sec-3:prospects_big_picture} we summarize which experimental efforts can probe which signatures of new physics that can address the anomalies.
For concreteness, we focus on experiments that are either recent, are expected to be upgraded, or are still under proposal.
The leftmost column shows broad classes of experiments grouped arbitrarily by their source and type of experiment.
The other columns present which specific experiments, in each of these classes, can probe a given experimental signature resulting from different interpretations of the anomalies.
For clarity, the ``Decays in flight'' column title refers to particles produced by the decay of mesons, muons, or taus in flight.



\renewcommand{\arraystretch}{1}
\newcommand{\cmark}{\ding{51}}%
\newcommand{\xmark}{\ding{55}}%
\newcommand{\nocheck}{\textcolor{Red}{\xmark}}
\newcommand{\semicheck}{\textcolor{Orange}{\cmark}}
\newcommand{\fullcheck}{\textcolor{Green}{\cmark}}
\newcommand{\newhline}{\cline{2-8}}

\begin{table*}[!htp]
    \centering
    \begin{sideways}
\resizebox{20cm}{!}{%
    \begin{tabular}{|>{\centering\arraybackslash}p{3 cm}|>{\centering\arraybackslash}p{4 cm}|>{\centering\arraybackslash}p{3.2 cm}|>{\centering\arraybackslash}p{2 cm}|>{\centering\arraybackslash}p{1.9 cm}|>{\centering\arraybackslash}p{1.4 cm}|>{\centering\arraybackslash}p{1.3 cm}|>{\centering\arraybackslash}p{2.5 cm}|}
\hline\hline
    \multirow{2}{*}{Category} & \multirow{2}{*}{Model}& \multirow{2}{*}{Signature} & \multicolumn{4}{c|}{Anomalies} & \multirow{2}{*}{References} \\ \cline{4-7} & & & LSND & MiniBooNE & Reactors & Sources & \\
\hline\hline
\multirow[c]{3}{*}[-3em]{ \parbox[t]{3 cm}{\vspace{-0.2cm}\centering Flavor transitions Secs.~\ref{sec:3+1}-\ref{subsubsec:3plusN}, \ref{sec:light-sterile-decay} }}
        & (3+1) oscillations & oscillations & \fullcheck & \fullcheck & \fullcheck & \fullcheck & Reviews and global fits \cite{Dentler:2018sju,Diaz:2019fwt,Boser:2019rta,Dasgupta:2021ies}
        \\
        \newhline
        & (3+1) w/ invisible sterile decay & oscillations w/ $\nu_4$ invisible decay &  \fullcheck &  \fullcheck & \fullcheck & \fullcheck & \cite{Moss:2017pur,Moulai:2019gpi}
        \\
        \newhline
        & (3+1) w/ sterile decay & $\nu_4 \to \phi \nu_e$ &  \fullcheck &  \fullcheck & \semicheck & \semicheck & \cite{Palomares-Ruiz:2005zbh,Bai:2015ztj,deGouvea:2019qre,Dentler:2019dhz,Hostert:2020oui}
        \\
\hline\hline
\multirow[c]{2}{*}[-2em]{ \parbox[t]{3 cm}{\vspace{-0.2cm}\centering Matter effects  Secs.~\ref{sec:steriles-and-NSI}, \ref{sec:extra-dimensions}}}
        & (3+1) w/ anomalous matter effects  & $\nu_\mu \to \nu_e$ via matter effects & \fullcheck  & \fullcheck & \nocheck & \nocheck & \cite{Akhmedov:2011zza,Bramante:2011uu,Karagiorgi:2012kw,Asaadi:2017bhx,Smirnov:2021zgn}
        \\ 
        \newhline
        & (3+1) w/ quasi-sterile neutrinos  & $\nu_\mu \to \nu_e$ w/ resonant $\nu_s$ matter effects & \fullcheck & \fullcheck & \semicheck & \semicheck & \cite{Alves:2022vgn}
        \\
\hline\hline
\multirow[c]{2}{*}[-2em]{ \parbox[t]{3 cm}{\vspace{-0.2cm}\centering Flavor violation  Sec.~\ref{sec:LNV-muon-decay}}} 
        & Lepton-flavor-violating $\mu$ decays & $\mu^+\to e^+ \nu_{\alpha}\overline{\nu_e}$ & \fullcheck & \nocheck & \nocheck & \nocheck & \cite{Bergmann:1998ft,Babu:2016fdt,Jones:2019tow}
        \\ 
        \newhline
        & neutrino-flavor-changing bremsstrahlung & $\nu_\mu A \to e \phi A$ & \fullcheck & \fullcheck & \nocheck & \nocheck  & \cite{Berryman:2018ogk} 
        \\
\hline\hline
\multirow[c]{2}{*}[-2em]{ \parbox[t]{3 cm}{\vspace{-0.9cm}\centering Decays in flight Sec.~\ref{subsubsec:LongLivedHNLs}}}
        & Transition magnetic mom., heavy $\nu$ decay & $N\to \nu \gamma$ &  \nocheck & \fullcheck & \nocheck & \nocheck & \cite{Fischer:2019fbw}
        \\
        \newhline
        & Dark sector heavy neutrino decay & $N\to \nu (X \to e^+e^-)$ or $N\to \nu (X\to \gamma \gamma)$ & \nocheck & \fullcheck & \nocheck & \nocheck & \cite{Chang:2021myh}
        \\
        \newhline
\hline\hline
\multirow[c]{2}{*}[-2em]{\parbox[t]{3 cm}{\vspace{-0.7cm}\centering Neutrino Scattering Secs.~\ref{subsubsec:TransitionMagneticMoment}, \ref{subsubsec:DarkNeutrinos}}} 
    & neutrino-induced upscattering & $\nu A \to N A$,  $N\to \nu e^+e^-$ or $N \to \nu \gamma \gamma$ & \semicheck & \fullcheck & \nocheck  & \nocheck & \cite{Bertuzzo:2018itn,Bertuzzo:2018ftf,Ballett:2018ynz,Ballett:2019pyw,Datta:2020auq,Dutta:2020scq,Abdullahi:2020nyr,Abdallah:2020biq,Abdallah:2020vgg,Hammad:2021mpl}
    \\
    \newhline
    & Transition magnetic mom. or polarizability photons & $\nu A \to N A$,  $N\to \nu \gamma $ or $\nu A \to \nu \gamma A$  & \semicheck & \fullcheck & \nocheck & \nocheck  & \cite{Gninenko:2009ks,Gninenko:2010pr,Gninenko:2012rw,Masip:2012ke,Radionov:2013mca,Magill:2018jla,Vergani:2021tgc,Alvarez-Ruso:2021dna,Bansal:2022zpi}
    \\
\hline\hline
\multirow[c]{2}{*}[-2em]{ \parbox[t]{3 cm}{\vspace{-0.9cm}\centering Dark Matter Scattering Sec.~\ref{subsub:darkmatter}}}
    & dark particle-induced upscattering  & $\gamma$ or $e^+e^-$ & \nocheck & \fullcheck & \nocheck & \nocheck & \cite{Dutta:2021cip}
    \\
    \newhline
    & dark particle-induced inverse Primakoff & $\gamma$ & \fullcheck & \fullcheck & \nocheck & \nocheck & \cite{Dutta:2021cip}
    \\
\hline\hline
\end{tabular}
}
 \end{sideways}
    \caption{New physics explanations of the short-baseline anomalies categorized by their signature. Notation: \fullcheck -- the model can naturally explain the anomaly,  \semicheck -- the model can partially explain the anomaly,   \nocheck -- the model cannot explain the anomaly.~\label{tab:sec-3:big_picture}}
\end{table*}



\renewcommand{\arraystretch}{1.2}
\newcommand{\fixedcol}{{\centering\arraybackslash}p{2.6 cm}}
\begin{table}[!htp]
\centering
\begin{sideways}
\resizebox{22cm}{!}{%
\begin{tabular}{
|>{\centering\arraybackslash}p{2.4 cm}
|>{\centering\arraybackslash}p{4.8 cm}
|>{\centering\arraybackslash}p{2.4 cm}
|>{\centering\arraybackslash}p{2.4 cm}
|>{\centering\arraybackslash}p{2.4 cm}
|>{\centering\arraybackslash}p{2.4 cm}
|>{\centering\arraybackslash}p{2.4 cm}|
            }
\hline \hline
Source 
& 3+1 Oscillations 
& Anomalous matter effects 
& Lepton flavor violation 
& Decays in flight 
& Neutrino-induced upscattering 
& Dark-particle-induced upscattering 
\\  \hline \hline
Reactor  
        & DANSS upgrade, JUNO-TAO, NEOS II, Neutrino-4 upgrade, PROSPECT-II & & & & & \\ 
\hline   
Radioactive Source
      & BEST-2, IsoDAR, THEIA, Jinping & & & & & \\  
\hline
Atmospheric
        & \multicolumn{2}{|>{\centering\arraybackslash}p{7.2 cm}|}{IceCube upgrade, KM3NET, ORCA and ARCA, DUNE, Hyper-K, THEIA} & & &  \multicolumn{2}{|>{\centering\arraybackslash}p{4.8 cm}|}{IceCube upgrade, KM3NET, ORCA and ARCA, DUNE, Hyper-K, THEIA}\\ 
\hline
Pion/Kaon decay-at-rest
        & JSNS$^2$, COHERENT, CAPTAIN-Mills, IsoDAR, KPIPE & & JSNS$^2$, COHERENT, CAPTAIN-Mills, IsoDAR, KPIPE, PIP2-BD & & & COHERENT, CAPTAIN-Mills, KPIPE, PIP2-BD \\ 
\hline  
Beam Short Baseline
        & SBN & & & \multicolumn{3}{|>{\centering\arraybackslash}p{7.2 cm}|}{SBN} \\  
\hline
Beam Long Baseline
        & \multicolumn{3}{|>{\centering\arraybackslash}p{9.6 cm}|}{DUNE, Hyper-K, ESSnuSB} &  \multicolumn{3}{|>{\centering\arraybackslash}p{7.2 cm}|}{DUNE, Hyper-K, ESSnuSB, FASER$\nu$, FLArE} \\  
\hline
Muon decay-in-flight 
      &  \multicolumn{3}{|>{\centering\arraybackslash}p{9.6 cm}|}{$\nu$STORM} & & $\nu$STORM &\\  
\hline
Beta Decay and Electron Capture  & KATRIN/TRISTAN, Project-8, HUNTER, BeEST, DUNE-$^{39}$Ar, PTOLEMY, $2\nu\beta\beta$ &  & & & & \\  
\hline\hline
\end{tabular}
}
\end{sideways}
\caption{Summary of future experimental prospects to probe new physics explanations of the anomalies. We emphasize that all experiments can constrain the new physics models discussed in this paper in one way or another, but we list those that can provide a direct test of the respective model.\label{fig:sec-3:prospects_big_picture}}
\footnotesize
\vspace{0.2in}
\end{table}

\clearpage

\section{Broader Experimental Landscape}
\label{sec:null}

In this section, we review the broader landscape of existing experimental results with relevant sensitivity to interpretations of the LSND, MiniBooNE, Reactor, and Gallium anomalies.

\subsection{Flavor Conversion}

Other short-baseline, as well as long-baseline, neutrino experiments can look for anomalous flavor conversions. In the following sections, we discuss direct tests of MiniBooNE and LSND in $\nu_\mu \to \nu_e$ and $\overline{\nu}_\mu \to \overline{\nu}_e$ searches, as well as in $\nu_\mu$ and $\nu_e$ disappearance. We also discuss the $\nu_e$ and $\overline{\nu}_e$ disappearance in the context of the Gallium and reactor anomalies, where both null results and hints are observed.

\subsubsection{Pion Decay-at-Rest Accelerator Experiments}

The KARMEN (KArlsruhe Rutherford Medium Energy Neutrino) experiment was located at the highly pulsed
spallation neutron source ISIS of the Rutherford Laboratory (UK).  ISIS protons had an energy of 800~MeV and were delivered to the water-cooled Ta-D$_2$O target with a repetition rate of 50 Hz.  The time
structure of the ISIS protons (double pulses with a width of 100 ns separated by 325 ns) allowed a clear separation of $\nu_\mu$ events due to $\pi^+$ decay from $\bar\nu_\mu$ and $\nu_e$ events due to $\mu^+$ decay.

\begin{figure}[!htbp]
    \centering
    \includegraphics[width=0.5 \textwidth]{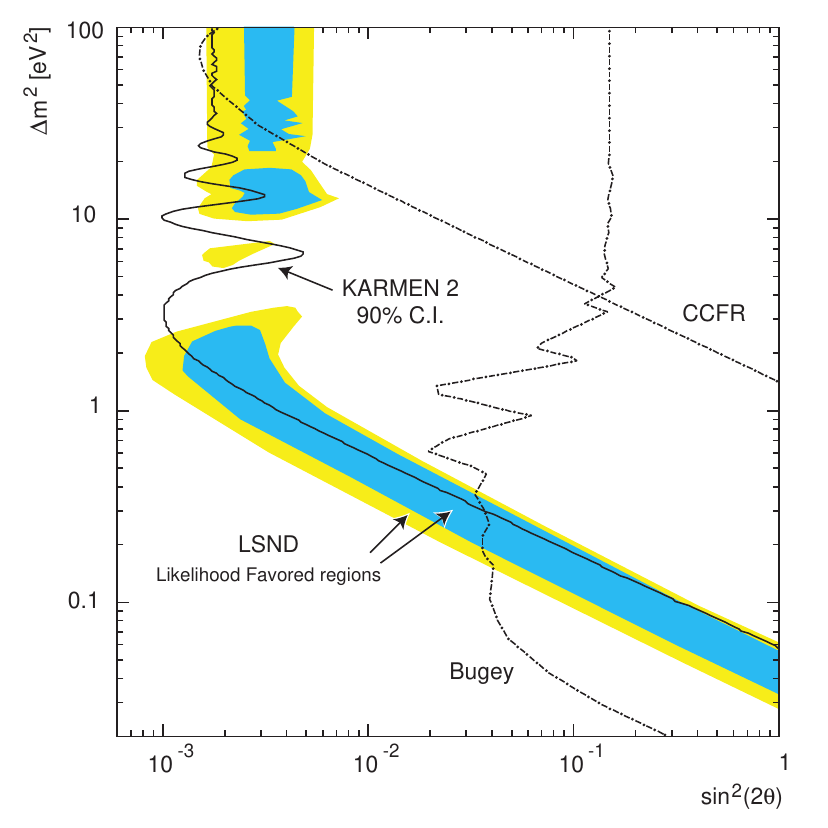}
    \caption{
      The KARMEN sterile neutrino 90\% C.I.~compared to other then contemporary experiments. Figure from~\cite{KARMEN:2002zcm}.
    }
    \label{FIG:Karmen}
  \end{figure}

The KARMEN detector~\cite{Gemmeke:1990ix} was a segmented liquid scintillator calorimeter, located 17.7 m from the ISIS target at an angle 100 degrees relative to the proton beam. The active target consisted of 65 $\mbox{m}^3$ of liquid scintillator segmented into 608 modules with gadolinium-coated paper placed between modules for efficient detection of thermal neutrons.  KARMEN performed a search for $\bar\nu_\mu \rightarrow \bar\nu_e$ oscillations, analogous to the LSND search, using $p(\bar\nu_e,e^+)n$ and found measured rates agreed with background expectations~\cite{KARMEN:2002zcm}. The sterile neutrino 90\% Confidence Interval (C.I.) obtained by the KARMEN measurement is shown in Fig.~\ref{FIG:Karmen} in relation to the LSND allowed regions. Although KARMEN did not see the LSND-like signal, it did not exclude the entirety of the LSND 99\% allowed regions although it did strongly disfavour the larger $\Delta m^2 > 10 \text{ eV}^2$ solutions. 

\subsubsection{Pion Decay-in-Flight Accelerator Experiments}

\subsubsubsection{Short-Baseline Experiments}\label{dif_sbl}

\begin{figure}[!htbp]
    \centering
    \includegraphics[width=0.5\textwidth]{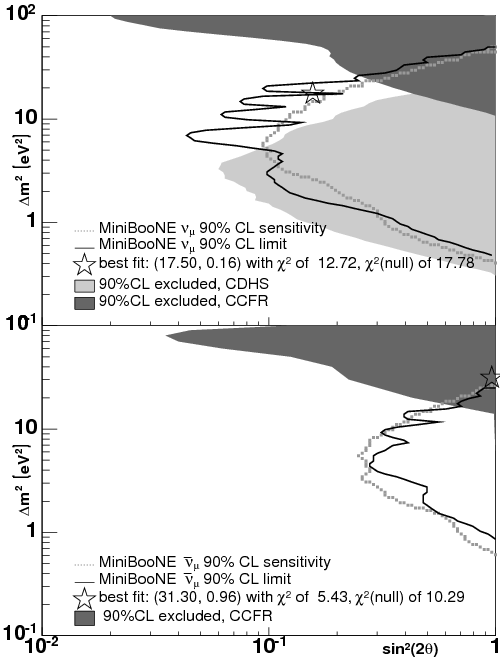}
    \caption{MiniBooNE's 90\% CL sensitivity (dashed line) and limit (solid line) for $\nu_{\mu}$ (top) and $\bar\nu_{\mu}$ (bottom) disappearance under a 3+1 scenario. Previous limits by CCFR (dark grey) and CDHS (light grey) are also shown. The figure is from~\cite{MiniBooNE:2009ozf}.}
    \label{fig:mbdis}
\end{figure}

{\bf MiniBooNE}
While MiniBooNE's $\nu_e$ and $\bar{\nu}_e$ appearance searches in the Fermilab BNB have resulted in the observation of anomalous excesses, MiniBooNE BNB $\nu_\mu$ and $\bar{\nu}_\mu$ CC measurements have been relatively well understood, including a disagreement between data and the Monte Carlo prediction that has been attributed to cross-section effects and uncertainties \cite{MiniBooNE:2007iti}. As a result, MiniBooNE has been able to perform searches for $\nu_\mu$ and $\bar{\nu}_\mu$ disappearance, both exclusively, and inclusively, resulting in limits (using approximately the first half of data collected by MiniBooNE) as shown in Fig.~\ref{fig:mbdis} \cite{MiniBooNE:2009ozf}. The results complement those from prior short-baseline $\nu_\mu$ and $\bar{\nu}_\mu$ disappearance searches, from the CCFR \cite{CCFRNuTeV:1998gjj} and CDHS \cite{Stockdale:1984ce} experiments.

It should be noted that, in these searches, due to the lack of a near detector (ND), large flux and cross-section uncertainties limited MiniBooNE's sensitivity particularly in the high $\Delta m^2$ range, where (fast) oscillations are expected to lead to an overall normalization deficit, and are thus masked by systematic uncertainties on the overall $\nu_\mu$ and $\bar{\nu}_\mu$ CC rate normalization. More powerful searches were performed by MiniBooNE in combination with measurements by the SciBooNE experiment, as described next.

\begin{figure}[!htbp]
    \centering
    \includegraphics[width=0.45 \textwidth]{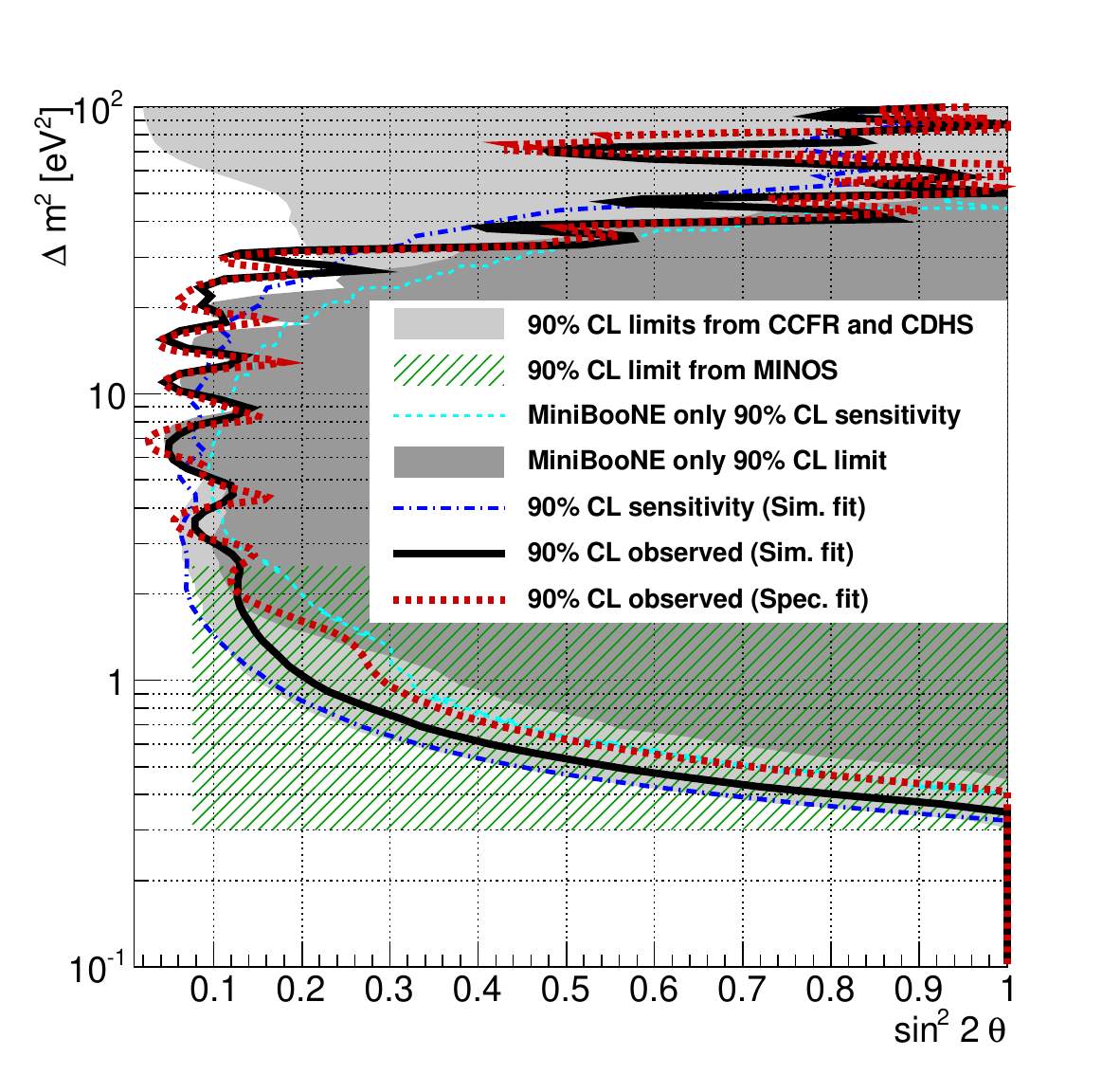}
    \includegraphics[width=0.45 \textwidth]{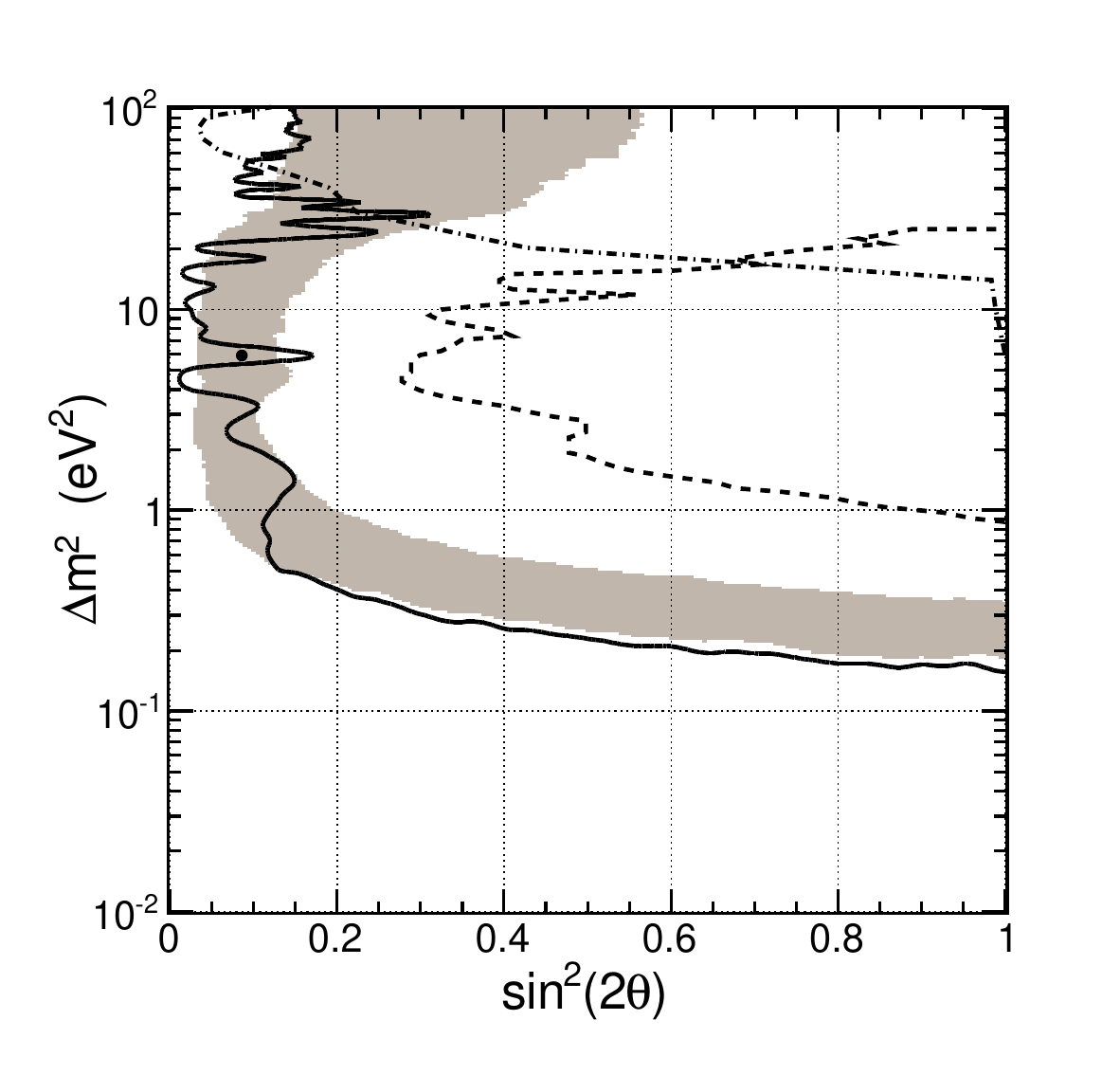}
    \caption{
      MiniBooNE-SciBooNE 90\% CL limits from a joint $\nu_\mu$ disappearance search {\it (left)} \cite{SciBooNE:2011qyf} and a joint $\bar\nu_\mu$ disappearance search {\it (right)} \cite{MiniBooNE:2012meu}.
    }
    \label{FIG:MB-SB}
  \end{figure}

{\bf MiniBooNE/SciBooNE} The SciBooNE experiment was located 100~m downstream from the BNB target and ran simultaneously with MiniBooNE from 2007-2008.  A simultaneous $\nu_\mu$ disappearance search under a (3+1) scenario was performed in the SciBooNE and MiniBooNE detectors with the BNB operating in forward horn current mode~\cite{SciBooNE:2011qyf}.  A separate $\bar\nu_\mu$ disappearance search was performed in the SciBooNE and MiniBooNE detectors with the BNB operating in reverse horn current mode~\cite{MiniBooNE:2012meu}. Exclusion (at 90\% CL) regions for these two searches are shown in Fig.~\ref{FIG:MB-SB}, and are consistent with exclusion limits from other accelerator-based neutrino experiments, including the short-baseline CCFR and CDHS experiments, and the long-baseline MINOS experiment \cite{MINOS:2011ysd}, as well as a MiniBooNE-only search limit.

\begin{figure}[!htbp]
    \centering
      \includegraphics[width=0.48\textwidth]{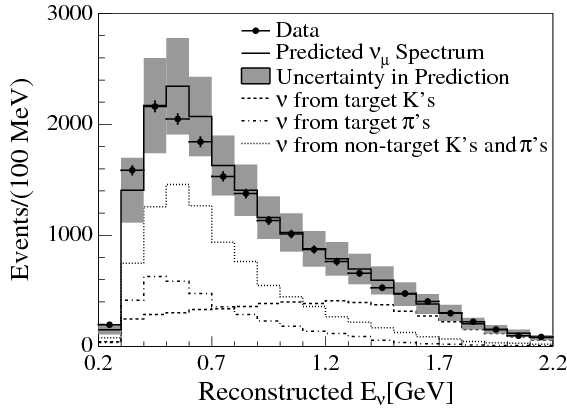}
      \includegraphics[width=0.48\textwidth]{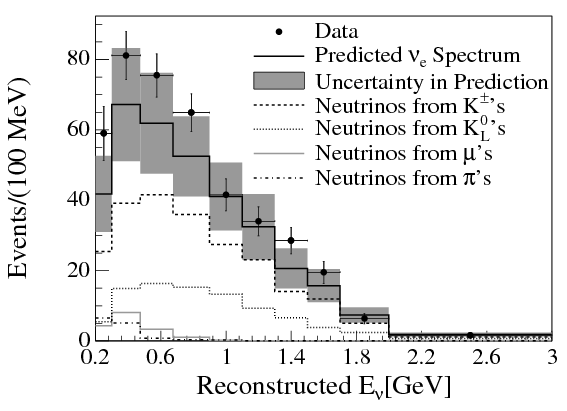}
    \includegraphics[width=0.48\textwidth]{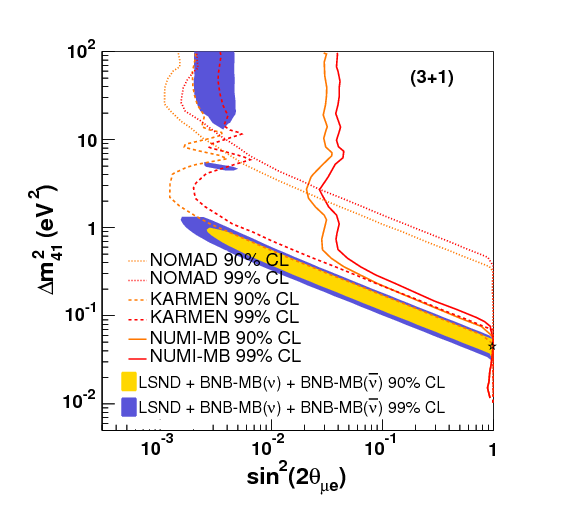}
    \caption{Top: Reconstructed energy distributions of $\nu_{\mu}$ (left) and $\nu_e$ (right) events in MiniBooNE from the NuMI off-axis beam. The figure is
  from \cite{MiniBooNE:2008hnl}. Bottom: Limit extracted from a fit to the $\nu_e$ distribution under a 3+1 hypothesis, from Ref.~\cite{Karagiorgi:2009nb}. }
    \label{fig:numi}
\end{figure}

{\bf MiniBooNE-NuMI} In addition to neutrinos from the BNB, MiniBooNE has also been able to study neutrinos from the Fermilab-based NuMI beam, viewed by MiniBooNE at an off-axis angle of 6.3$\degree$ \cite{MiniBooNE:2008hnl}.
MiniBooNE measured both $\nu_e$ and $\nu_{\mu}$ CCQE events from the NuMI off-axis beam (with no wrong-sign discrimination), as shown in Fig.~\ref{fig:numi} (top).
While an appearance or disappearance search with these data sets was not performed by MiniBooNE\footnote{The MiniBooNE-NuMI $\nu_e$ data set has been analyzed under a 3+1 appearance scenario in Ref.~\cite{Karagiorgi:2009nb}, where it was shown that, despite the small observed data excess, the data showed no significant preference for oscillations over the null hypothesis.}, the data was found to agree with expectation. The NuMI off-axis beam at the MiniBooNE location is much higher in $\nu_e$ content than the BNB, as well as in neutrinos contributed from kaon decays in the beamline.
Because of a higher intrinsic $\nu_e$ background, this data set was particularly limited in sensitivity to light sterile neutrino oscillations.

\paragraph{MicroBooNE}

\begin{figure}[!htbp]
    \centering
      \includegraphics[width=0.48\textwidth]{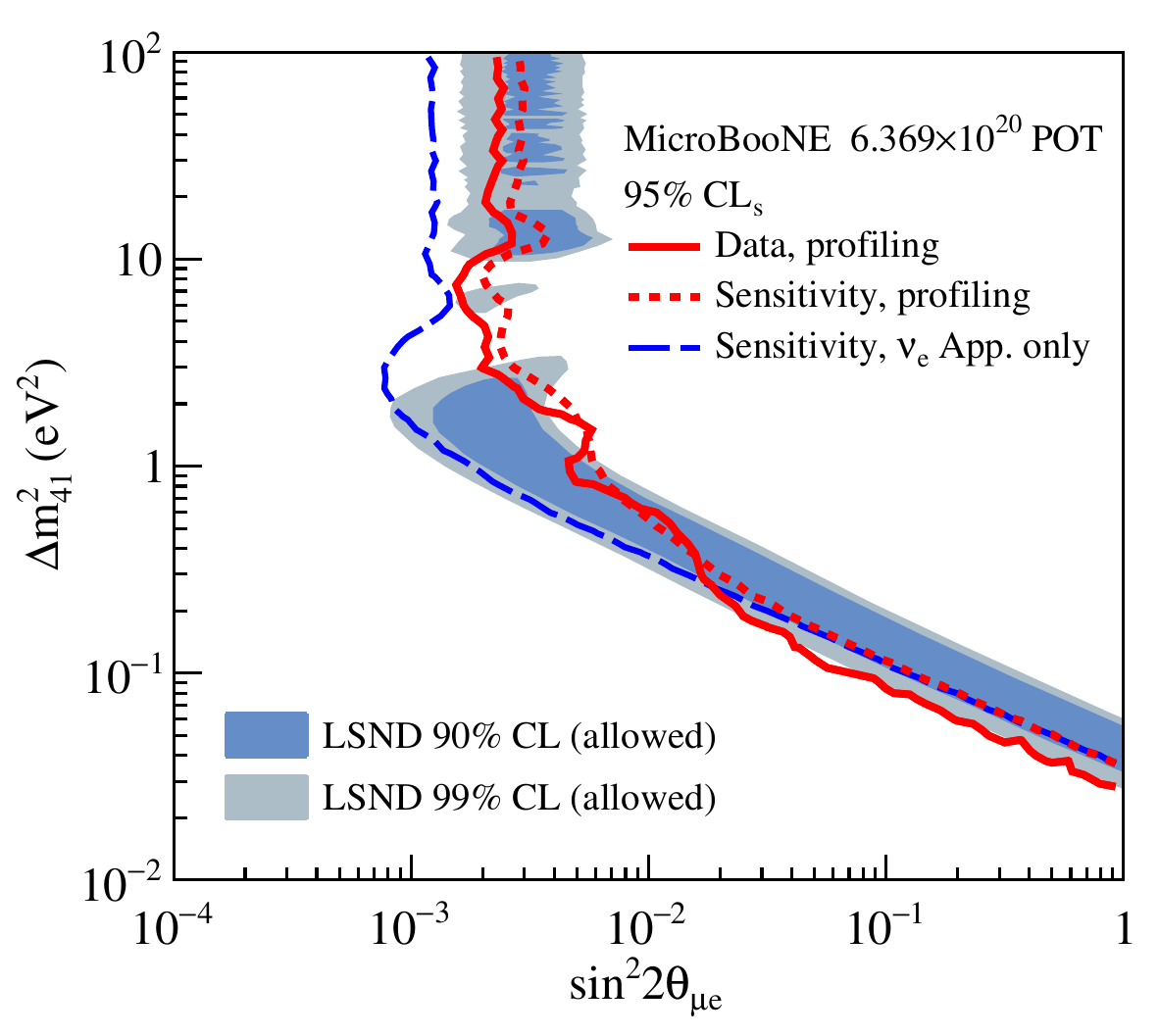}
      \includegraphics[width=0.48\textwidth]{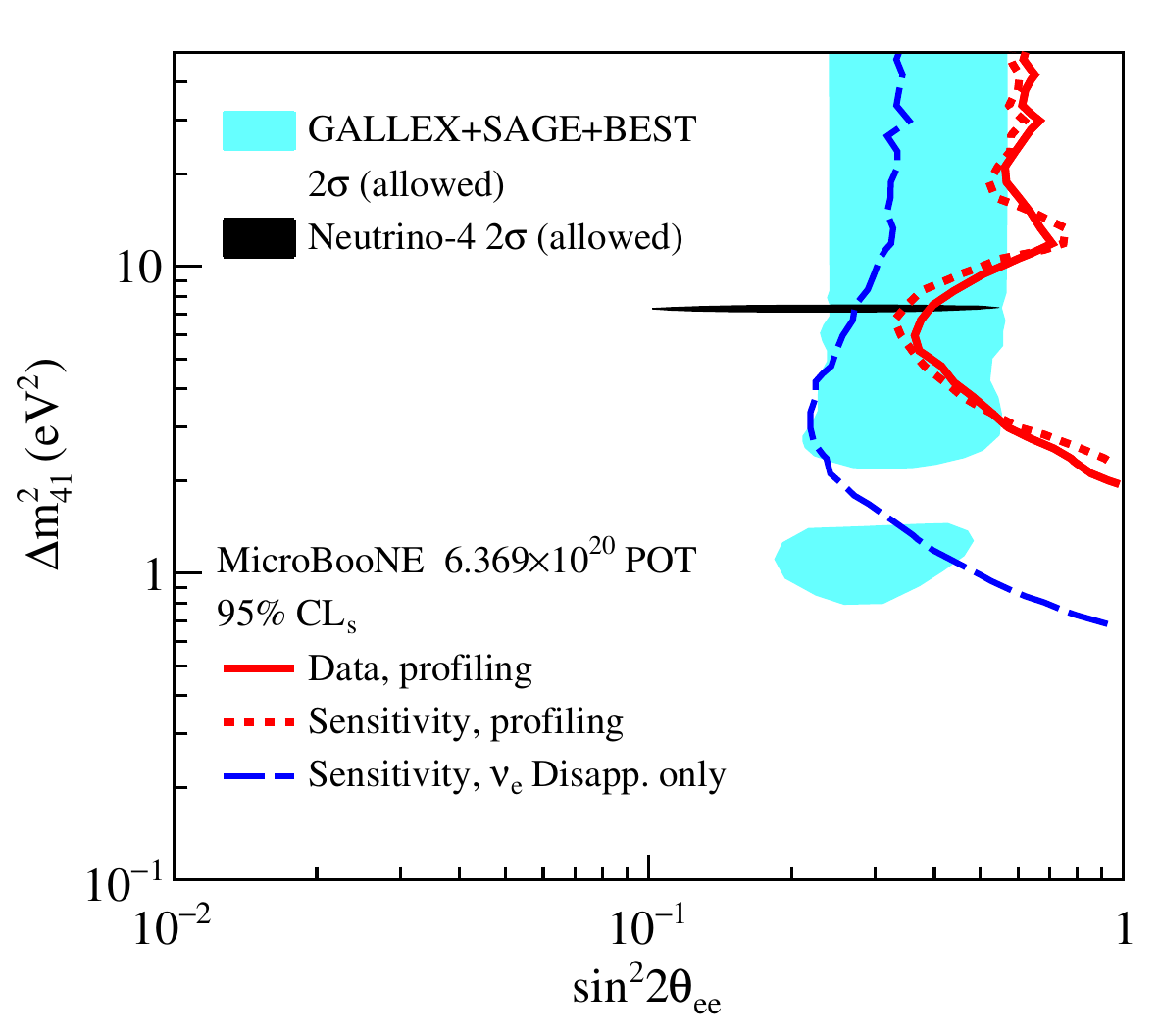}
    \caption{The MicroBooNE 3+1 CLs exclusion contours at the 95\% CL in the plane of $\Delta m_{41}^2$ versus (left) $\sin^2 2 \theta_{\mu e}$ and (right) $\sin^2 2 \theta_{ee}$. A full 3+1 treatment was implemented, with parameters not shown profiled over for the sensitivity and resulting data exclusions. Shown also is the (unphysical) $\nu_e$ appearance-only and $\nu_e$ disappearance-only sensitivities for comparison. Figures from \cite{MicroBooNE:2022wdf}. }
    \label{fig:microboone_osc_2022}
\end{figure}

Recently MicroBooNE has published its first results on direct oscillation searches due to light sterile neutrinos \cite{MicroBooNE:2022wdf}. The analysis uses high-statistics CC $\nu_\mu$ and CC $\nu_e$ selections, developed as part of the inclusive $\nu_e$ search ~\cite{MicroBooNE:2021nxr}, with data collected over a three year period with a total exposure of $6.369\times 10^{20}$ protons-on-target, representing approximately half of the total MicroBooNE data-set. A full 3+1 neutrino model was studied, capitalizing on the seven channels of $\nu_e$ and $\nu_\mu$ selections. Although a small deficit of $\nu_e$ was observed, the data was found to agree with the $3\nu$ (no-sterile) hypothesis within the $1\sigma$ level. Since the data was found to be consistent with the no-sterile hypothesis, exclusion contours were calculated and can be found in Fig.~\ref{fig:microboone_osc_2022}, where regions of the LSND anomaly at both high $\Delta m^2$ and low $\Delta m^2$ were excluded. As the analysis makes use of a a very pure $\nu_e$ selection, exclusions are shown both for $\nu_e$ appearance ($\sin^2 2\theta \mu e)$ and $\nu_e$ disappearance ($\sin^2 2\theta e e)$ profiling over the other parameters of the full 3+1 fit. 

This work highlighted an important aspect of studying full 3+1 neutrino oscillations in the BNB. As the BNB is a $\nu_\mu$ dominated beam with a non-negligible intrinsic $\nu_e$ component, cancellation can occur in any $\nu_\mu\rightarrow \nu_e$ appearance signal due to $\nu_e \rightarrow \nu_e$ disappearance occurring in parallel. This cancellation can lead to a diminished oscillation effect in comparison to when one studies the (un-physical) $2\nu$ approximation. The use of the NuMI beam, which has a different ratio of intrinsic $\nu_e$ to $\nu_\mu$ could help break this cancelletion effect and includion of the NuMI beam data in the 3+1 result is an ongoing effort for the MicroBooNE collaboration.   

Prior to this collaboration result, the publicly available MicroBooNE $\nu_e$ CC data sets~\cite{MicroBooNE:2021rmx,MicroBooNE:2021jwr,MicroBooNE:2021sne,MicroBooNE:2021nxr} have also been studied by the phenomenology community~\cite{Denton:2021czb, Arguelles:2021meu} and in combination with the MiniBooNE data by the MiniBooNE collaboration~\cite{MiniBooNE:2022emn}.  Although these studies involve several assumptions and approximations that are not inherent in the official results, the qualitative conclusions are largely the same, with the exception of the analysis~\cite{Denton:2021czb} which founds a preference for $\nu_e$ disappearance at the ${\sim}2\sigma$ level but notably did not use the full systematic uncertainty accounting of the collaboration's data release. However, a more recent analysis~\cite{Arguelles:2021meu}, as well as the official collaboration release, incorporated systematic uncertainties from the collaboration's data release and accounting for neutrino energy reconstruction smearing found this preference for $\nu_e$ disappearance to be statistically insignificant.

\begin{figure}
    \centering
    \includegraphics[width=0.49\textwidth]{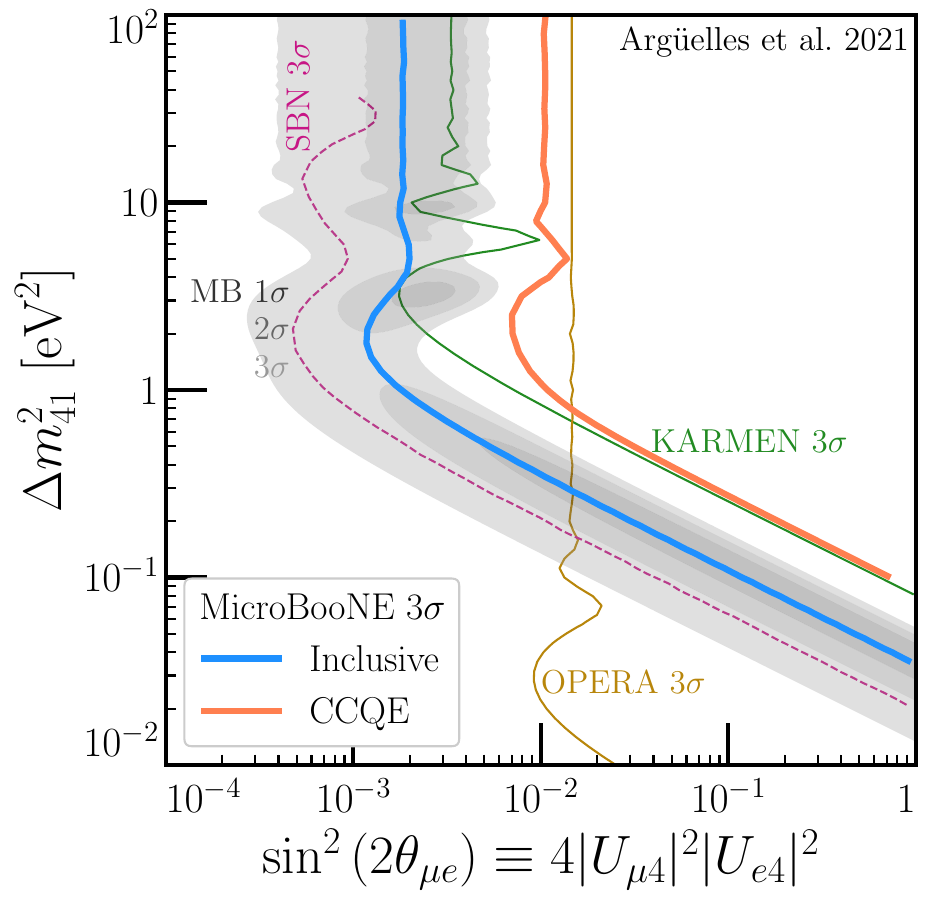}
    \includegraphics[width=0.49\textwidth]{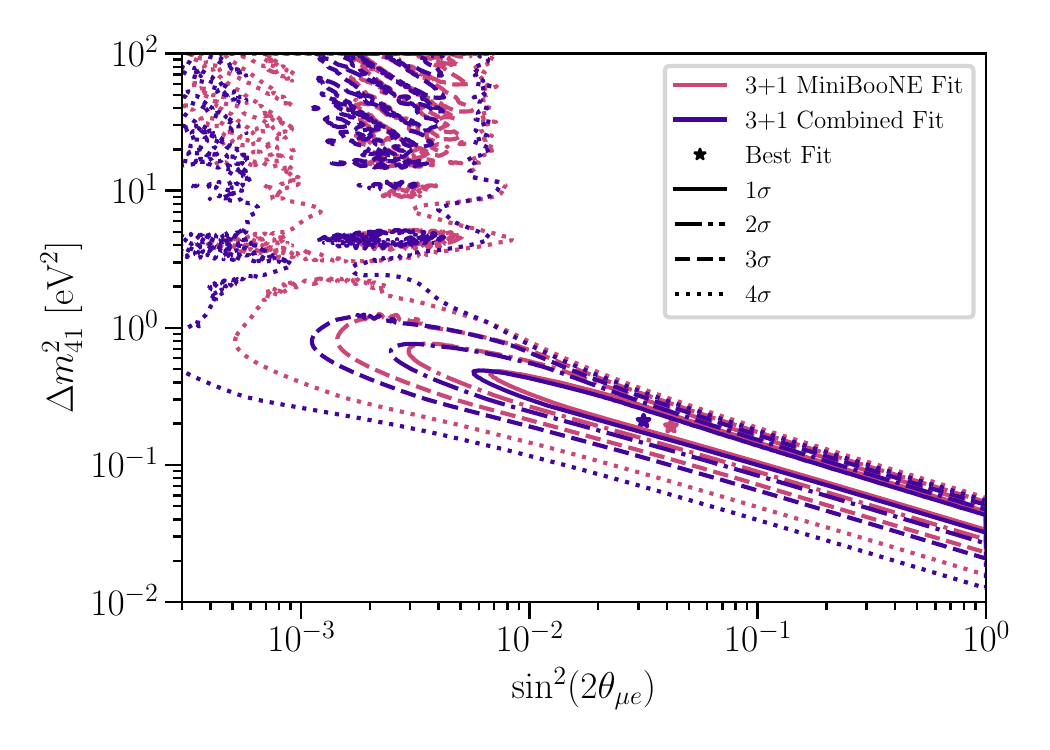}
    \caption{Left: Preferred regions in the simplified two-flavor model for MiniBooNE (gray), MicroBooNE inclusive analysis (blue), MicroBooNE CCQE analysis (red), KARMEN (green), and OPERA (brown) as indicated. For reference, the predicted sensitivity of SBN is also shown. Figure taken from Ref.~\cite{Arguelles:2021meu}. Right: Preferred regions in the 3+1 model parameter space for a MiniBooNE-only fit and a joint fit including the MicroBooNE $\nu_e$ CCQE result. 
    Figure adapted from Ref.~\cite{MiniBooNE:2022emn}.}
    \label{fig:contours}
\end{figure}

Ref.~\cite{Arguelles:2021meu} has also provided the first analysis of the effect of MicroBooNE's $\nu_e$ results in the context of a light sterile neutrino explanation of MiniBooNE. This includes both an analysis in the ``simplified two-flavor'' model, in which only $\nu_\mu \to \nu_e$ oscillations are taken into account, without significant $\nu_\mu \to \nu_\mu$ or $\nu_e \to \nu_e$ effects, consistent with MiniBooNE analyses prior to 2022 as well as the complete 3+1 model. It is found that, while MicroBooNE prefers the null hypothesis, i.e.~no new sterile neutrino, the allowed regions from MiniBooNE at ${\sim}3\sigma$ are still allowed at the same confidence level by the MicroBooNE results, see left panel of Fig.~\ref{fig:contours}. 
Given its large dataset, the results of MicroBooNE's inclusive analysis~\cite{MicroBooNE:2021nxr} are found to be the most statistically powerful, while the CCQE sample provides a considerably weaker constraint.
Besides that, the MiniBooNE collaboration has performed a fit of both MiniBooNE and MicroBooNE CCQE data~\cite{MicroBooNE:2021jwr} to a 3+1 sterile neutrino model~\cite{MiniBooNE:2022emn}, properly accounting for oscillations in the backgrounds.
As expected, the impact of adding MicroBooNE's CCQE sample to the fit has a marginal effect on the preferred regions for the 3+1 model, see right panel of Fig.~\ref{fig:contours}.
While the two experiments' results are combined into a joint likelihood, the systematic uncertainties that the two experiments have in common, particularly those associated with their common neutrino flux from the Booster Neutrino Beam and the neutrino-nucleus cross section model, have not been combined.

{\bf NOMAD} The NOMAD experiment \cite{NOMAD:2003mqg}, which ran at CERN using protons from the 450~GeV SPS accelerator, employed a conventional neutrino beamline to create a wideband 2.5 to 40~GeV neutrino energy
source. These neutrinos were created with a carbon-based, low-mass, tracking detector located 600 m
downstream of the target. This detector had fine spatial resolution and could search for muon-to-electron and muon-to-tau oscillations. No signal was observed in either mode, and this experiment
set a limit to $\nu_\mu\rightarrow\nu_e$ appearance that excluded the majority of the 99\% CL-allowed LSND region above 10~eV$^2$ at 90\% CL, but had a significantly worse limit than KARMEN below 10~eV$^2$.

{\bf CCFR} The CCFR experiment was carried out at Fermilab in 1984 \cite{Stockdale:1984cg}. The experiment made use of a narrow band beamline, with meson energies set to 100, 140, 165, 200, and 250~GeV, yielding $\nu_\mu$ and $\bar\nu_\mu$ beams that ranged from 40 to
230~GeV in energy. A two-detector experimental setup carried out a disappearance search, with the ND at 715~m and
the far detector (FD) at 1116~m from the center of the 352~m long decay pipe. The calorimetric detectors
were constructed of segmented iron with a scintillator and spark chambers, and each had a downstream
toroid to measure the muon momentum. The data showed no evidence for a distance-dependent modification of the neutrino flux and ruled out oscillations of $\nu_\mu$ into any other single type of neutrino for $30<\Delta m^2<1000$~eV$^2$ and $\sin^2(2\theta)>0.02-0.20$.

{\bf CDHS} The CDHS experiment \cite{Dydak:1983zq} at CERN searched for $\nu_\mu$ disappearance with a two-detector design
of segmented calorimeters with iron and scintillator. The experiment used 19.2~GeV protons on a
beryllium target to produce mesons that were subsequently focused into a 52 m decay channel. The
detectors were located 130 m and 885 m downstream of the target. The experiment set a limit at 95\% CL and set constraints that are comparable to the MiniBooNE $\nu_\mu$ disappearance limit described above, but extending to slightly lower $\Delta m^2$.

\subsubsubsection{Long-Baseline Experiments}\label{dif_lbl}
The phenomenology of sterile neutrino-driven oscillations at long baselines
consists of interference phenomena arising from at least two
distinct oscillation frequencies and several scale-determining mixing angles and phases.
In the context of a 3+1 model with the addition of one new flavor state $\nu_s$ and a new mass state $\nu_4$ extending the three-flavor PMNS matrix to a $4\times4$ unitary matrix, and assuming the parametrization $U=R_{34}R_{24}R_{14}R_{23}R_{13}R_{12}$, the extended matrix is written as:
\begin{equation}
   U = 
    \begin{bmatrix}
        U_{e1}    &  U_{e2}   &  e^{-i\delta_{13}}s_{13}c_{14}                                                                                             & e^{-\delta_{14}}s_{14} \\
        U_{\mu1}  & U_{\mu2}  & -e^{-i(\delta_{13}-\delta_{14}+\delta_{24})}s_{13}s_{14}s_{24} + c_{13}s_{23}c_{24}                                        & e^{-i\delta_{24}}c_{14}s_{24} \\
        U_{\tau1} & U_{\tau2} & -e^{i\delta_{24}}c_{13}s_{23}s_{24}s_{34} + c_{13}c_{23}c_{34} - e^{-i(\delta_{13}-\delta_{14})}s_{13}s_{14}c_{24}s_{34}   & c_{14}c_{24}s_{34} \\
        U_{s1}    & U_{s2}    & -e^{i\delta_{24}}c_{13}s_{23}s_{24}c_{34} - c_{13}c_{23}s_{34} - e^{-i(\delta_{13}-\delta_{14})}s_{13}s_{14}c_{24}c_{34}   & c_{14}c_{24}c_{34} \\
    \end{bmatrix},
\end{equation}
\noindent which introduces three new mixing angles, $\theta_{14}$, $\theta_{24}$, and $\theta_{34}$, two new CP violating phases, $\delta_{14}$ and $\delta_{24}$, and one new linearly independent mass-splitting that is chosen to be $\Delta m^2_{41}$ in this case.  
Neglecting $\Delta_{21}$ and using unitarity to rewrite any terms containing $U_{\alpha 1}$ or $U_{\alpha 2}$, the probability of $\nu_{\mu}$ survival in a 3+1 model can be written as
\begin{equation}\label{eq:SterileDisFull}
\begin{split}
    P(\nu_{\mu} \rightarrow \nu_{\mu}) \approx 1 & - 4\left|U_{\mu 3}\right|^2\left(1 - \left|U_{\mu 3}\right|^2 - \left|U_{\mu 4}\right|^2\right)\sin^2\Delta_{31} \\
    & - 4\left|U_{\mu 3}\right|^2\left|U_{\mu 4}\right|^2\sin^2\Delta_{43} -4\left|U_{\mu 4}\right|^2\left(1 - \left|U_{\mu 3}\right|^2 - \left|U_{\mu 4}\right|^2\right)\sin^2\Delta_{41}.
\end{split}
\end{equation}

Of particular interest are NC neutrino interactions, since as the three active flavors participate in the NC interaction at the same rate, the NC sample is insensitive to three-flavor oscillations between ND and FD. However, if a sterile neutrino exists, sterile-neutrino appearance would cause a depletion in the NC channel. The NC survival probability is thus defined as $1 - P(\nu_{\mu} \rightarrow \nu_{s})$. Similar to Eq.~(\ref{eq:SterileDisFull}), this can be written as
\begin{equation}\label{eq:SterileAppFull}
  \begin{split}
    1 - P(\nu_{\mu} \rightarrow \nu_{s}) \approx 1 & - 4\left|U_{\mu 3}\right|^2 \left|U_{s 3}\right|^2 \sin^2\Delta_{31} \\
                                             & - 4\left|U_{\mu 4}\right|^2 \left|U_{s 4}\right|^2 \sin^2\Delta_{41} \\
                                             & - 4\operatorname{Re}\left(Z\right)\left( \sin^2\Delta_{31} - \sin^2\Delta_{43} + \sin^2\Delta_{41} \right) \\
                                             & - 2\operatorname{Im}\left(Z\right)\left( \sin2\Delta_{31} + \sin2\Delta_{43} - \sin2\Delta_{41} \right),
  \end{split}
\end{equation}
\noindent where $Z = U^{*}_{\mu 4} U_{s 4} U_{\mu 3} U^{*}_{s 3}$.
The phenomenology of $\nu_{\mu}$-CC and NC disappearance driven by sterile neutrino oscillations is complicated at long baselines due to the interference of three-flavor oscillations and sterile oscillations, which do not occur at short baselines. To perform two-detector analyses typical of long-baseline accelerator neutrino experiments, effects at both short and long baselines must thus be understood.

For instance, in the case of the short-baseline oscillation, only probed by the ND, the oscillation probability for NC disappearance is approximately given by
\begin{equation}
1 - P(\nu_{\mu} \rightarrow \nu_s) \approx 1 - \cos^4\theta_{14}\cos^2\theta_{34}\sin^{2}2\theta_{24}\sin^2\Delta_{41},
\label{eqn:sblNC}
\end{equation}
\noindent where $\Delta_{ji} = \frac{\Delta m^2_{ji}L}{4E}$. For the typical beam neutrino energies and ND baselines, when $\Delta m^2_{41} < 0.05$, oscillations are not visible in the ND. Starting at $\Delta m^2_{41}$ $\sim$ 0.5~eV$^2$, oscillations begin to be visible at low energies in the ND, and as $\Delta m^2_{41}$ increases, the first oscillation maximum moves to higher energies. At sufficiently high $\Delta m^2_{41}$ values, the entire ND sees rapid oscillations that can no longer be resolved and are seen as a constant normalization shift described by
\begin{equation}
1 - P(\nu_{\mu} \rightarrow \nu_s) \approx 1 - \frac{1}{2}\cos^4\theta_{14}\cos^2\theta_{34}\sin^{2}2\theta_{24}.
\end{equation}

For $\nu_{\mu}$-CC at the ND, the oscillation probability can be approximated as
\begin{equation}
P(\nu_{\mu} \rightarrow \nu_{\mu}) \approx 1 - \sin^22\theta_{24}\sin^2\Delta_{41},
\label{eqn:sblCC}
\end{equation}
\noindent which behaves similarly to NC disappearance except it depends only on $\theta_{24}$, and in the rapid oscillation case the normalization shift is given by $(1/2)\sin^22\theta_{24}$.

However, when considering long baselines, terms oscillating at the atmospheric frequency cannot be neglected. Approximating the NC disappearance probability to first order in small mixing angles gives
\begin{equation}
\begin{aligned}
1 - P(\nu_{\mu} \rightarrow \nu_s) & \approx 1 - \cos^4\theta_{14}\cos^2\theta_{34}\sin^{2}2\theta_{24}\sin^2\Delta_{41} \\
& - \sin^2\theta_{34}\sin^22\theta_{23}\sin^2\Delta_{31} \\
& + \frac{1}{2}\sin\delta_{24}\sin\theta_{24}\sin2\theta_{23}\sin\Delta_{31}.
\end{aligned}
\end{equation}
\noindent In this expression, the first term is identical to the short-baseline approximation. The second and third terms both oscillate at the atmospheric frequency. If $\theta_{34}$ $>$ 0, the second term is non-zero, and if $\sin\delta_{24}$ and $\theta_{24}$ are non-zero, the third term will not be zero. In either case, this creates an oscillation dip visible at the FD regardless of the value of $\Delta m^2_{41}$. It is notable that this will happen even though NC disappearance cannot occur in the standard three-flavor paradigm. It should also be noted that the third term is CP-odd, since NC disappearance is effectively sterile neutrino appearance, so $\bar{\nu}$ data will also add to the sensitivity of long-baseline experiments to sterile neutrinos.

For $\nu_{\mu}$-CC disappearance, expanding to second order in small mixing angles one finds
\begin{equation}
\begin{aligned}
P(\nu_{\mu} \rightarrow \nu_{\mu}) &\approx 1 - \sin^22\theta_{23}\cos2\theta_{24}\sin^2\Delta_{31}\\ 
&- \sin^22\theta_{24}\sin^2\Delta_{41},
\end{aligned}
\end{equation}
which can be rewritten as
\begin{equation}
\begin{aligned}
P(\nu_{\mu} \rightarrow \nu_{\mu}) &\approx 1 - \sin^22\theta_{23}\sin^2\Delta_{31} \\
& + 2\sin^22\theta_{23}\sin^2\theta_{24}\sin^2\Delta_{31} \\ 
& - \sin^22\theta_{24}\sin^2\Delta_{41}.
\label{eq:NuMuDisFull}
\end{aligned}
\end{equation}
\noindent The first term is the standard approximation for three-flavor $\nu_{\mu}$-CC disappearance. The second term also oscillates as a function of $\Delta m^2_{31}$, the atmospheric frequency, but it is driven by sterile mixing. Even at large $\Delta m^2_{41}$ values, this term does not enter into rapid oscillations. Thus, even for large mass splittings where the ND is well inside the rapid oscillation regime, the FD will still show shape variations in addition to normalization changes due to the terms oscillating at the atmospheric frequency with a magnitude that scales with $\sin^2\theta_{24}$. Furthermore, when considering $\nu_{\mu}$ disappearance alone, one could define an effective atmospheric mixing angle to account for both sterile and standard oscillations by combining the first two terms in Eq.~(\ref{eq:NuMuDisFull}):
\begin{equation}
\sin^22\theta_{23}^{\mathrm{eff}} = \sin^22\theta_{23}\cos2\theta_{24}.
\end{equation}
\noindent While it may seem this would imply that the depth of the atmospheric dip would be insensitive to the large $\Delta m^2_{41}$ regime, it actually provides a constraint due to $\theta_{23}^{\mathrm{eff}}$ having been measured to be close to maximal. Due to the $\cos2\theta_{24}$ factor, a non-zero $\theta_{24}$ can only drive $\theta_{23}^{\mathrm{eff}}$ away from maximal. 

In addition, experiments that can make precise measurements of \nue~CC interactions can look for $\overset{(-)}\nu\!\!_e$-CC disappearance at the far detector following
\begin{equation}
\begin{aligned}
P(\overset{(-)}\nu\!\!_e \rightarrow \overset{(-)}\nu\!\!_e) &\approx 1 - \sin^22\theta_{13}\sin^2\Delta_{31} \\
& - \sin^22\theta_{14}\sin^2\Delta_{41},
\label{eq:NueDisFull}
\end{aligned}
\end{equation}
valid only if one assumes sterile-driven muon neutrino disappearance and electron neutrino appearance do not occur. 

Finally, it is worth noting the near detectors of long-baseline experiments can be used to conduct short-baseline searches following the same methodologies described in Sec.~\ref{dif_sbl}

\begin{figure}[!ht]
	\centering
  		\includegraphics[width=0.48\textwidth]{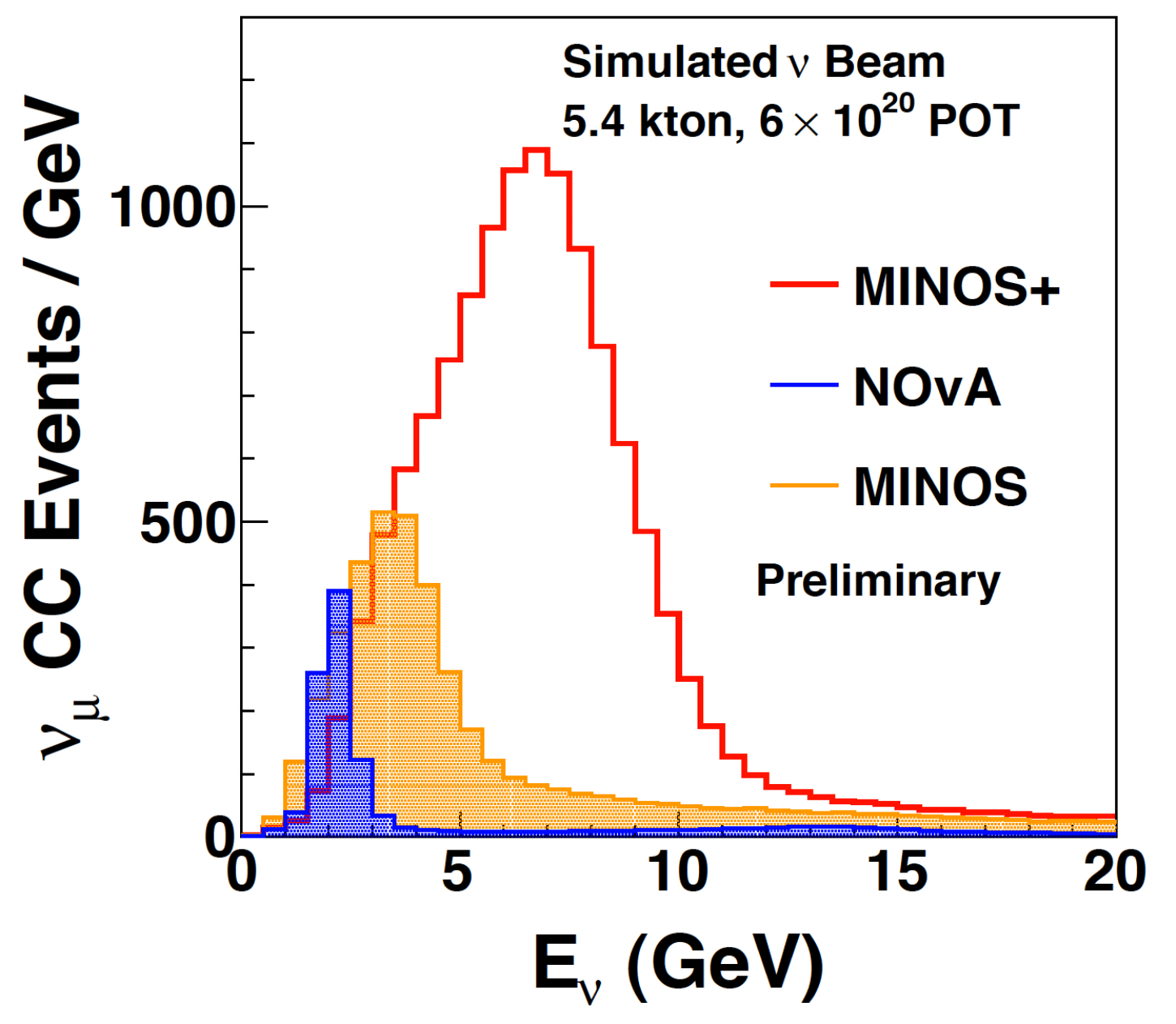}
\caption{The NuMI neutrino energy spectrum for the MINOS+ medium-energy tune, shown as the red solid line. The \nova~spectrum shown in blue is obtained with the same tune at a 14\,mrad offset from the beam axis (MINOS+ is on-axis). For comparison, the spectrum corresponding to the NuMI low-energy tune used by MINOS is shown as the gold histogram. Figure from~\cite{Sousa:2015bxa}.}
  \label{fig:numispectrum}
\end{figure}

\begin{figure}[!ht]
	\centering
  	\includegraphics[width=0.48\textwidth]{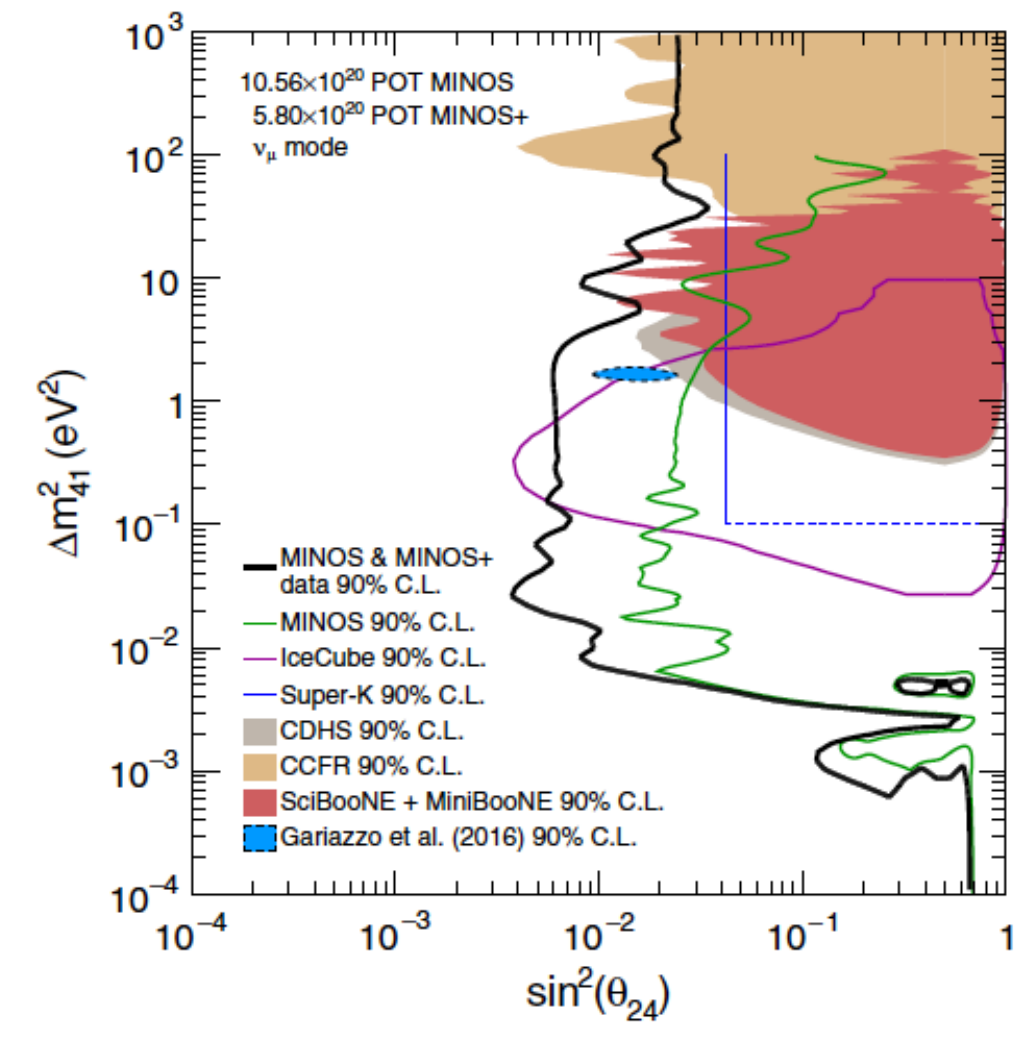}
	\caption{The MINOS and MINOS+ 90\% Feldman-Cousins exclusion limit compared to the previous MINOS result~\cite{MINOS:2016viw} and results from other experiments. The {\it Gariazzo et al.} region is the result of a global fit to neutrino oscillation data~\cite{Gariazzo:2015rra}. Figure from~\cite{MINOS:2017cae}.}
  \label{fig:minosplus_result}
\end{figure}

{\bf MINOS/MINOS+} The Main Injector Neutrino Oscillation Search (MINOS) experiment was a long-baseline neutrino oscillation experiment using the Neutrinos at the Main Injector (NuMI) neutrino beam and two detectors placed within a $735\,$km baseline. The NuMI beam is produced by collisions with a graphite target of protons accelerated to 120~GeV at Fermilab's Main Injector. The secondary products of these collisions, pions, and kaons, are focused by two parabolic magnetic horns and eventually decay into muons and neutrinos inside a 675\,m-long decay pipe filled with helium. The muons are absorbed in the rock and neutrinos continue towards the 1\,kton ND, 1\,km downstream of the target, and beyond, towards the 5.4\,kton FD. The relative positions of target and horns can be changed to tune the beam spectrum to lower or higher neutrino energies. The MINOS run concluded in 2012, with a total exposure of over $15\times10^{20}\,$protons-on-target (POT) in neutrino and antineutrino mode since the start of data taking in 2005. The MINOS+ experiment operated the MINOS detectors using the NuMI beam upgraded from 320\,kW to 700\,kW of beam power, part of the NOvA experimental setup. Instead of the low-energy configuration used for the MINOS run, MINOS+ used the NOvA medium-energy configuration, which for MINOS+ corresponds to a neutrino energy spectrum peaked around 7~GeV, as shown in Fig.~\ref{fig:numispectrum}.
Exposure of the MINOS+ detectors to a beam peaked above the three-flavor oscillation maximum provided excellent sensitivity to new physics through precise measurements of muon neutrino disappearance between the ND and FD. MINOS+ operated from 2014 through 2016 and accumulated only neutrino mode data. A combined analysis of the $10.6\times10^{20}$\,POT of MINOS neutrino data and $5.8\times10^{20}$\,POT of MINOS+ neutrino data using a two-detector fitting technique placed stringent limits on sterile driven muon neutrino disappearance within a 3+1 model, as shown in Fig.~\ref{fig:minosplus_result}~\cite{MINOS:2017cae}.
The null results from MINOS/MINOS+ are one of the primary drivers, along with IceCube results, of the large tension between appearance and disappearance data when attempting to explain observations purely through sterile neutrino mixing. This is further evidenced in combinations with reactor data looking for electron (anti)neutrino disappearance, as described in Sec.~\ref{lbl_reactor_combo}

\begin{figure}[!ht]
	\vspace{0pt}
	\begin{center}
  		\includegraphics[width=0.75\textwidth]{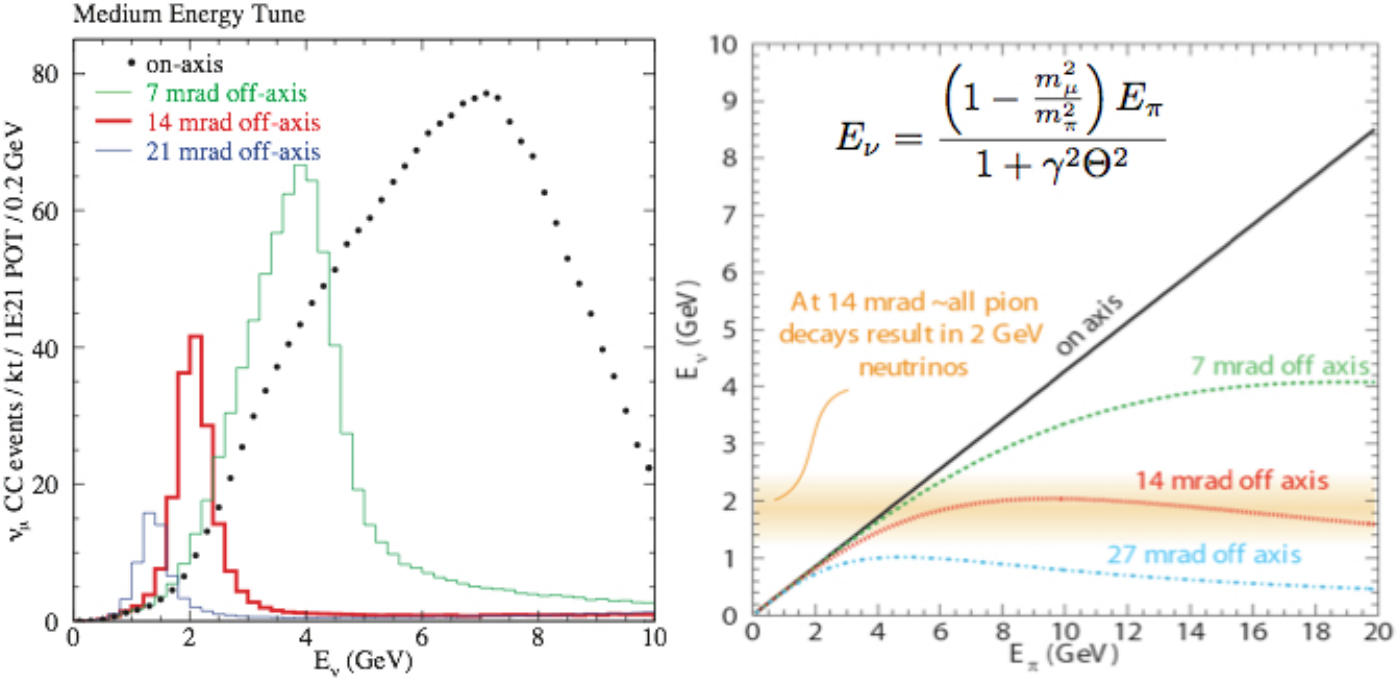}
	\end{center}
	\vspace{0pt}
	\caption{The plot on the left displays the predicted neutrino energy spectra of the NuMI beam  during the \nova run, including the on-axis spectrum sampled by MINOS+ as the dotted line, as well as the spectrum at different off-axis positions. The 14 mrad off-axis position of the NOvA detectors is identified by the red line. The plot on the right shows neutrino energy as a function of the parent pion energy and the angle between the neutrino and the decaying pion. Figure adapted from~\cite{Sousa:2011zz}.}
  \label{fig:nova_offaxis}
  \vspace{0pt}
\end{figure}

{\bf NOvA} The NuMI Off-Axis $\nu_e$ Appearance (NOvA) experiment is a long-baseline accelerator neutrino experiment based at Fermilab and the Far Detector Laboratory at Ash River, Minnesota. NOvA has as its primary goal to measure three-neutrino mixing parameters, including the determination of the neutrino mass ordering, by looking for the appearance of electron neutrinos or antineutrinos, and the disappearance of muon neutrinos or antineutrinos, using the NuMI neutrino beam produced at Fermilab. This is accomplished by using two detectors separated by 810~km, placed 14~mrad off the NuMI beam axis. Due to the off-axis placement, the detectors sample a narrow range of neutrino energies between 1 and 4~GeV, peaking at 2~GeV as shown in Fig.~\ref{fig:nova_offaxis}. This configuration is chosen to drastically reduce the feed-down of NC interactions of higher-energy neutrinos, which typically represent the dominant background to the measurement of \nue~CC interactions in on-axis experiments. 
The 0.33~kton ND is located underground next to the MINOS ND hall at Fermilab, while the 14~kton FD is positioned at the surface in Ash River, Minnesota. Both detectors are composed of extruded 32-cell PVC modules filled with liquid scintillators. The cells are read out by 32-pixel avalanche photodiodes (APDs).
NOvA began collecting data in 2014 and has so far accumulated large samples in both neutrino-dominated and antineutrino-dominated modes. 

\begin{figure}[!ht]
    \centering
    \includegraphics[width=0.49\textwidth]{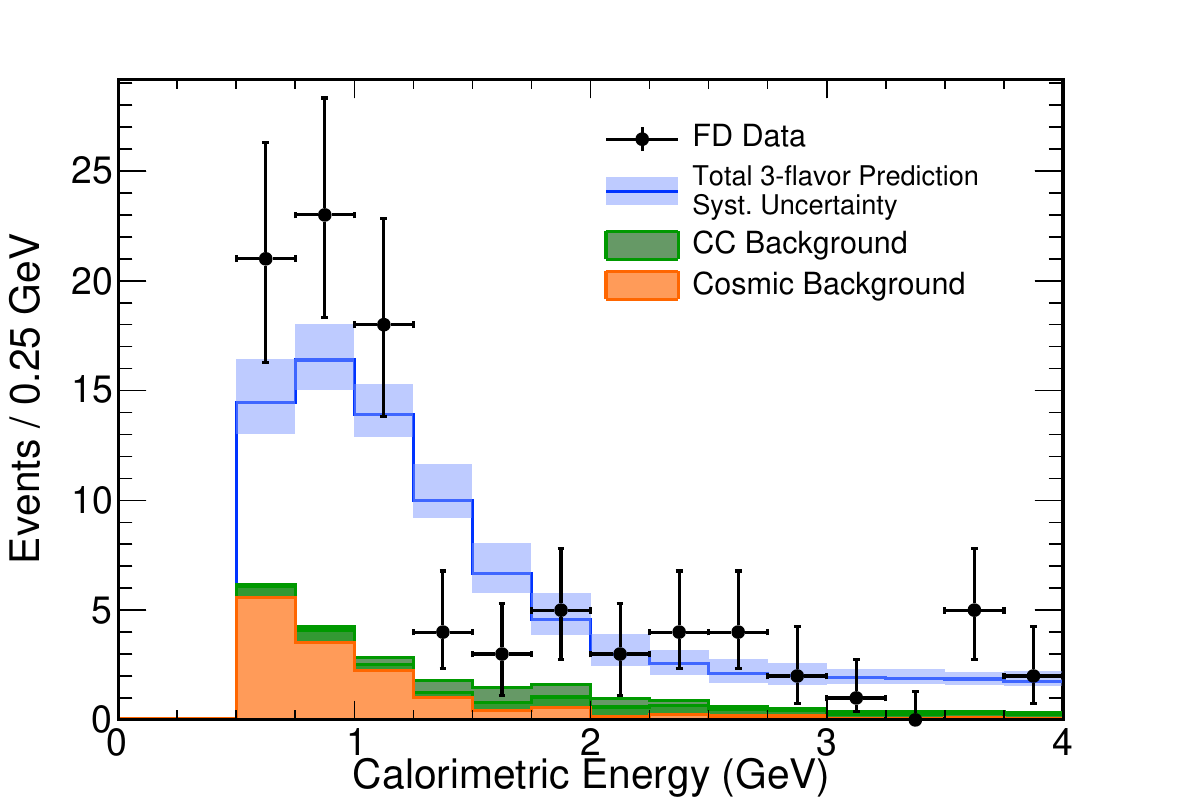}
    \includegraphics[width=0.49\textwidth]{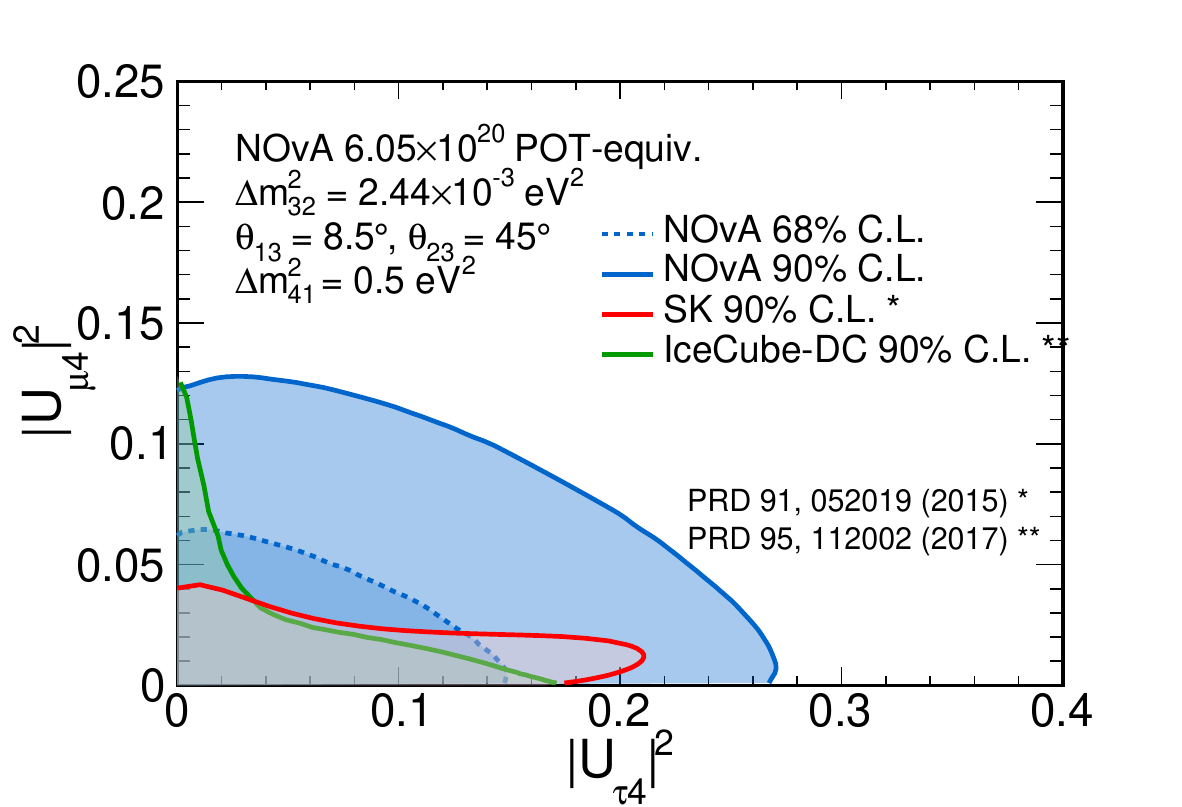}
    \caption{Left: Calorimetric energy spectrum for NC neutrino candidates in the NOvA FD data compared with three-flavor prediction; Right: The NOvA 68\% (dashed) and 90\% (solid) Feldman-Cousins non-excluded regions (shaded) in terms of $|U_{\mu4}|^{2}$ and $|U_{\tau4}|^{2} $, where it is assumed that $\cos^2\theta_{14}=1$ in both cases, compared to SuperK and IceCube-DeepCore constraints. Figures from~\cite{NOvA:2017geg}.}
    \label{fig:nova_neutrino}
\end{figure}

NOvA placed constraints on sterile neutrinos via searches for differences in the rate of NC neutrino interactions between the Near and Far detectors. The analysis was based on  $6.05\times10^{20}$ protons-on-target taken in neutrino-dominated mode, and 95 NC candidates were selected at the Far Detector compared with $83.5\pm9.7(\rm{stat.})\pm9.4(\rm{syst.})$ events predicted assuming mixing only occurs between active neutrino species. Therefore, NOvA found no evidence of active-sterile neutrino mixing. Interpreting these results within a 3+1 model results in constraints on the sterile mixing angles of $\theta_{24}$\,$<$\,$20.8^{\circ}$ and $\theta_{34}$\,$<$\,$31.2^{\circ}$ at the 90\% C.L. for $0.05$~eV$^2\leq \Delta m^2_{41}\leq 0.5$~eV$^2$, the range of mass splittings for which no significant oscillations over the ND baseline are expected~\cite{NOvA:2017geg}. The  energy spectrum of the NC selected candidates at the Far detector is shown along with a comparison of the allowed regions for the sterile matrix elements obtained by NOvA with similar constraints by SuperK and IceCube-DeepCore in Fig.~\ref{fig:nova_neutrino}.

\begin{figure}[!ht]
    \centering
     \includegraphics[width=0.49\textwidth]{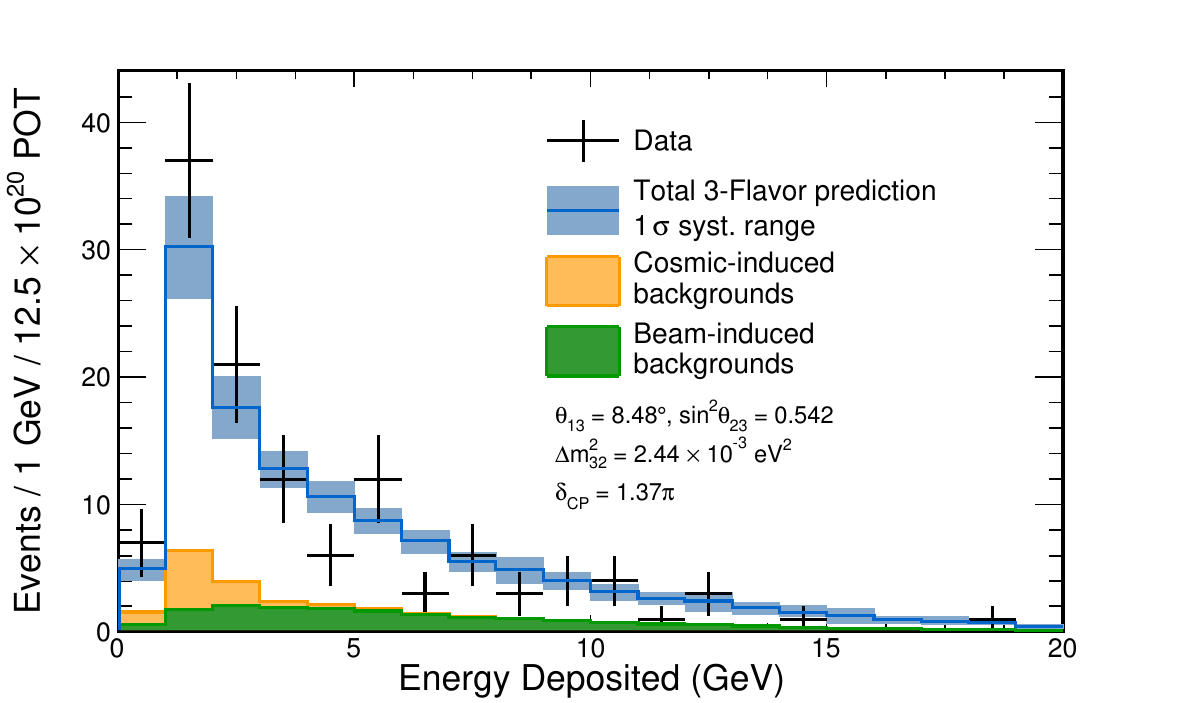}
    \includegraphics[width=0.49\textwidth]{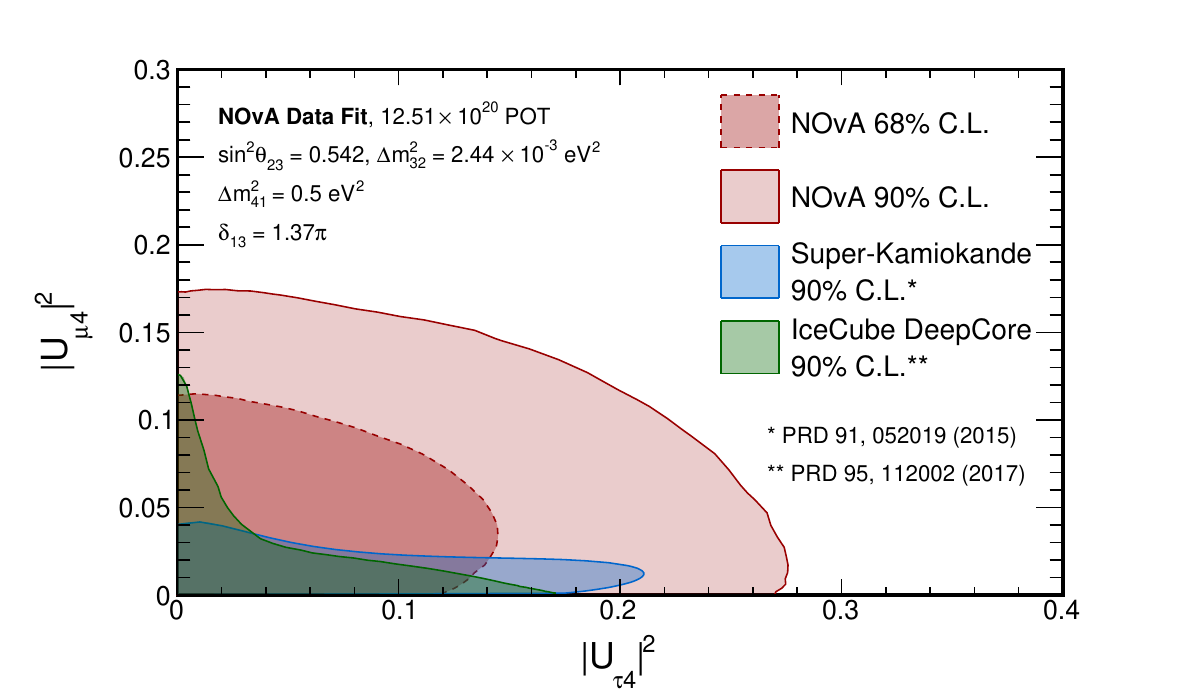}
    \caption{Left: Calorimetric energy spectrum for NC antineutrino candidates in the NOvA FD data compared with three-flavor prediction; Right: The NOvA 68\% (dashed) and 90\% (solid) Feldman-Cousins non-excluded regions (shaded) in terms of $|U_{\mu4}|^{2}$ and $|U_{\tau4}|^{2} $, where it is assumed that $\cos^2\theta_{14}=1$ in both cases, compared to SuperK and IceCube-DeepCore constraints. Figure from~\cite{NOvA:2021smv}.}
    \label{fig:nova_antineutrino}
\end{figure}

\begin{figure}[!ht]
    \centering
     \includegraphics[width=0.4\textwidth]{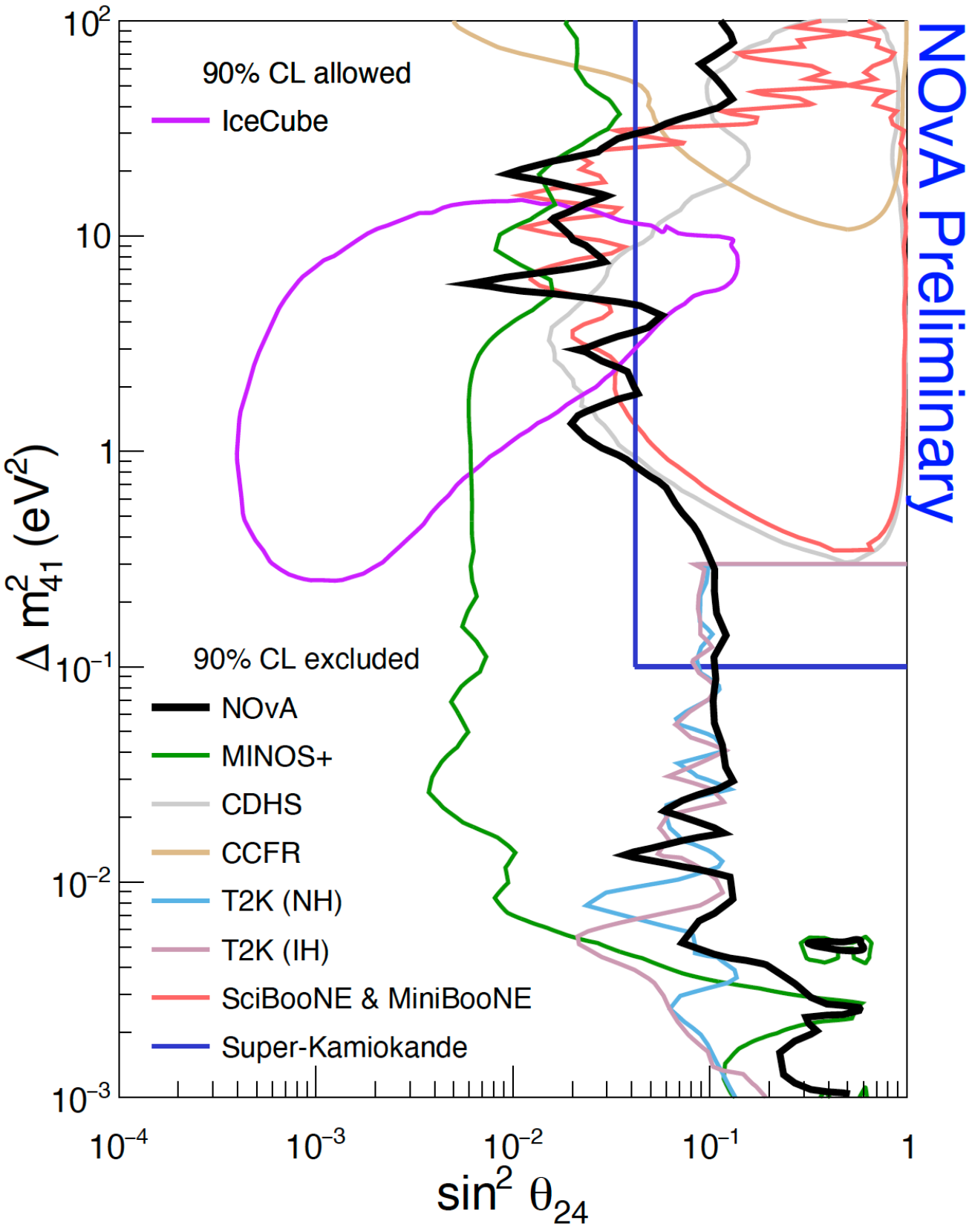}
    \includegraphics[width=0.4\textwidth]{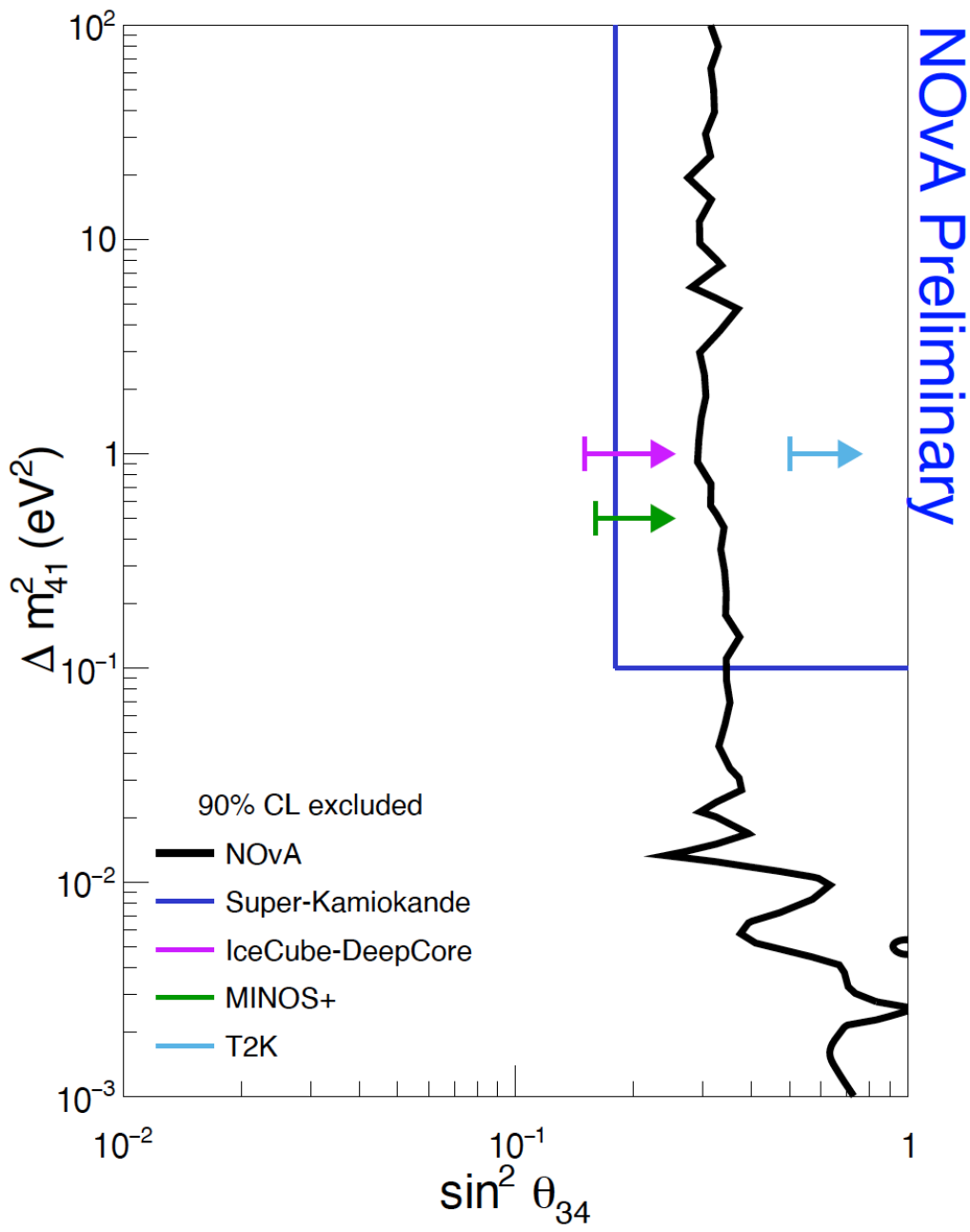}
    \caption{Left: NOvA 90\% Feldman-Cousins excluded region in $\sin^2\theta_{24}$ vs $\Delta m_{41}^2$, obtained from a two-detector fitting method, compared to allowed and exclude regions from other experiments. Right: NOvA 90\% Feldman-Cousins excluded region in $\sin^2\theta_{34}$ vs $\Delta m_{41}^2$, compared to limits reported by other experiments. Figure from~\cite{jeff_hartnell_2022_6683827}.}
    \label{fig:nova_nu2022}
\end{figure}

NOvA has also reported results on the first search for sterile antineutrino mixing in an antineutrino beam, using an exposure of $12.51\times10^{20}$~protons-on-target from the NuMI beam at Fermilab running in antineutrino-dominated mode. NOvA observed 121 NC antineutrino candidates at the FD, compared to a prediction of $122\pm11\rm{(stat.)}\pm15\rm{(syst.)}$ assuming mixing between only three active flavors. Therefore, no evidence for $\bar{\nu}_{\mu}\rightarrow \bar{\nu}_s$ oscillations is observed.  In this case the 3+1 model constraints on the mixing angles are found to be $\theta_{24} < 25^{\circ}$ and $\theta_{34} < 32^{\circ}$ at the 90\% C.L. for $0.05~\rm{eV}^2 \le \Delta m_{41}^2 \le 0.5~\rm{eV}^2$~\cite{NOvA:2021smv}. The antineutrino energy spectrum at the FD and obtained allowed regions are shown in Fig.~\ref{fig:nova_antineutrino}.
Finally, NOvA has recently presented results from a two-detector fit at Neutrino 2022, using similar techniques to the MINOS/MINOS+ experiment, extending limits on the $\theta_{24}$ and $\theta_{34}$ angles over a large range of $\Delta m_{41}^2$ values. These results are presented in Fig.~\ref{fig:nova_nu2022}.

{\bf T2K} The Tokai-to-Kamioka experiment~\cite{Abe:2011ks} is a neutrino oscillation experiment based in Japan. T2K's primary goal is to measure three flavor neutrino oscillation parameters. An intense source of (anti)neutrinos is produced at J-PARC and is directed toward a series of detectors placed \SI{280}{\meter} from the target, and a massive detector \SI{295}{\kilo\meter} away (Super-Kamiokande, SK~\cite{Super-Kamiokande:2002weg}). The neutrino beam composition is predominantly muon neutrino flavor, with a small (0.5\%) fraction of electron neutrinos~\cite{T2K:2012bge}; the beamline elements can be configured to produce a predominantly antineutrino beam as well. The ``near detectors'' include the INGRID detector~\cite{Abe:2011xv}, WAGASCI detectors~\cite{Koga:2015iqa,Hallsjo:2018mmo}, and ND280 detector suite. These detectors observe interactions from slightly different neutrino energy spectrums, with INGRID's peak neutrino energy approximately at \SI{1}{\giga\electronvolt} (on-axis), ND280 at \SI{0.6}{\giga\electronvolt} (2.5 degrees off-axis), and WAGASCI at \SI{0.8}{\giga\electronvolt} (1.5 degrees off-axis).
T2K has operated since 2009 and has produced a series of results on light sterile neutrinos using the ND280 detectors and SK with subsets of the data taken. T2K plans to install improvements to the ND280 detector~\cite{T2K:2019bbb}, and, with improvements to the beamline, T2K will have further opportunities for updated or expanded analyses between detectors.


CC electron neutrino interactions were selected in ND280 to test for 3+1 \nue disappearance, motivated by radiochemical experiments~\cite{SAGE:1998fvr,SAGE:2009eeu,Hampel:1997fc} and discrepancies in reactor experiments energy spectra~\cite{Mention:2011rk}. Isolating CC \nue candidates was challenging, with significant backgrounds to the \nue selection from  photons (from, for example, the inactive material surrounding the active scintillator target). These backgrounds were controlled by dedicated photon selections and with additional systematic uncertainty estimation on the production of $\pi^0$ from neutrino interactions on non-scintillator materials. The analysis also assumes no \numu disappearance, and systematic uncertainty from the flux and cross section models were reduced using dedicated \numu selections as is done for T2K three flavor oscillation analyses. T2K excludes regions of interest at 95\% CL with \num{5.9e20} POT between approximately $\sinstt{ee}>0.3$ and $\dm{\rm eff} > \SI{7}{\square\electronvolt}$~\cite{T2K:2014xvp}. An updated analysis of large mixing regions would be useful in light of recent MicroBooNE results~\cite{MicroBooNE:2021rmx}.


A search for light sterile neutrinos using SK has been performed in 2019~\cite{T2K:2019efw}, using \num{14.7e20} (\num{7.6e20}) POT in neutrino (antineutrino) mode. It is focusing on a 3+1 model where a single sterile neutrino is added and assumed to be mixing with $\nu_2$ and $\nu_3$ mass eigenstates through new parameters $\theta_{24}$ and $\theta_{34}$ (T2K has limited sensitivity to other parameters, that are thus neglected $\theta_{14}=\delta_{14}=\delta_{24}=\SI{0}{\degree}$).

The analysis consists of a simultaneous fit of the CC muon, electron and NC neutrino samples. While CC channels are mainly sensitive to the new mass splitting $\Delta m^2_{41}$ and to $\theta_{24}$, the NC channel measures the active neutrino survival probability and is also sensitive to $\theta_{34}$. The five CC analysis samples are the same as the one used in standard oscillation analyses~\cite{T2K:2021xwb}, but it has been the first time NC$\pi^0$ (single $\pi^0$ production where the pion decays and produces two Cherenkov rings) and NC $\gamma$-deexcitation samples are used in oscillation measurements.

\begin{figure}[ht!]
    \centering
    \includegraphics[width=0.48\textwidth]{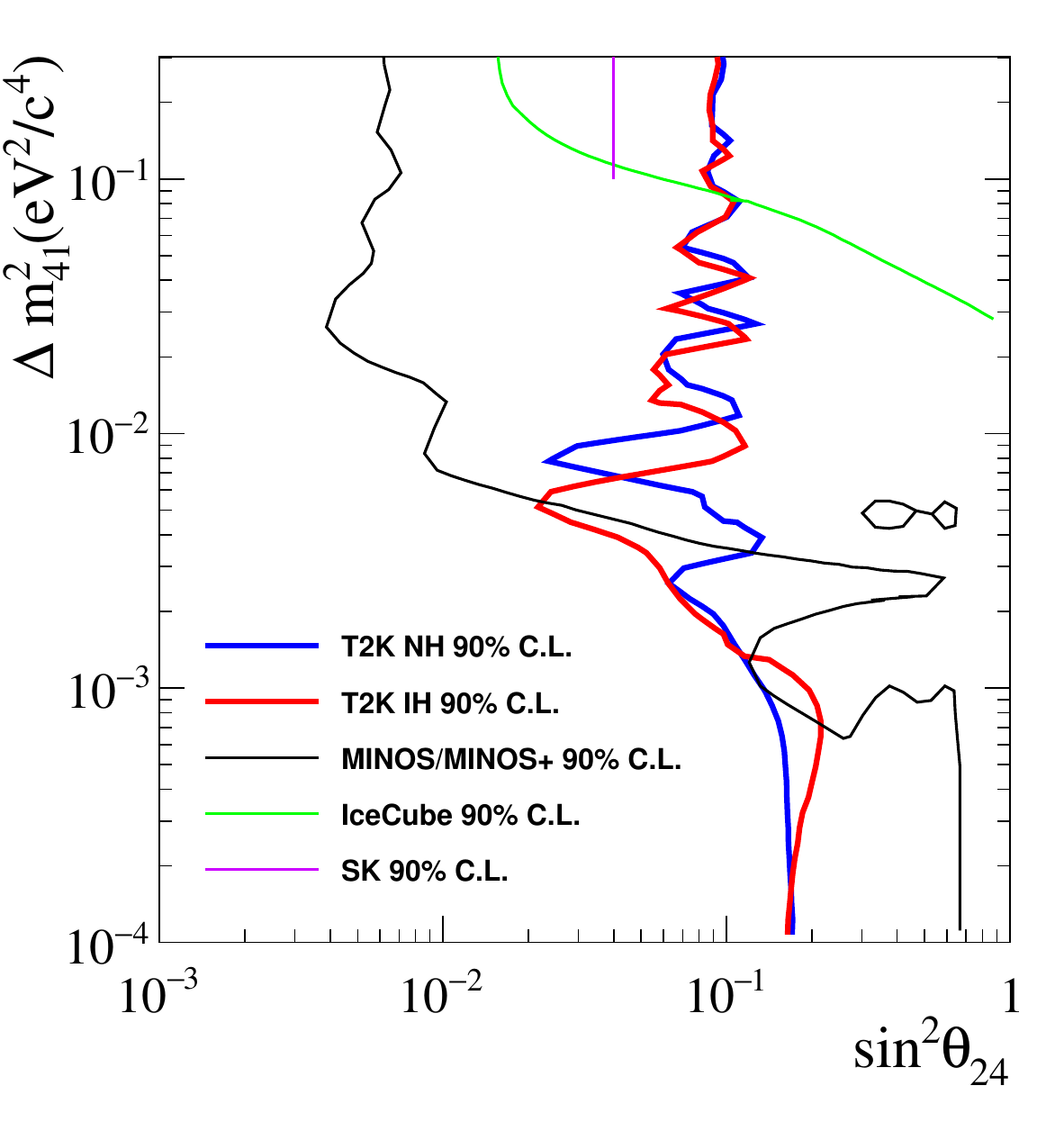}
    \caption{The T2K 90\% CL. exclusion limits on $\sinst{24}$ as a function of \dm{41} compared to other experiments. Areas on the right are excluded. Figure from~\cite{T2K:2019efw}.}
    \label{fig:t2k_overlay}
\end{figure}

Most of the systematic parameters are constrained in the same way as in the 3 flavor analysis (using e.g.~ND constraints), but additional 30\% normalization uncertainties are added for the new NC samples. The joint maximum-likelihood fits allows to draw exclusion limits in the $(\sinst{24}, \dm{41})$ plane: the most stringent limits on $\theta_{24}$ are obtained for $\dm{41} < \SI{3e-3}{\square\electronvolt}$. Similarly, limits in the $(\sinst{24}, \cos^2 \theta_{24} \sinst{34})$ plane have been obtained although more statistics in the NC samples and additional systematic studies are needed to further improve the measurement.
A comparison of the exclusion regions obtained with other existing constraints is shown in Fig.~\ref{fig:t2k_overlay}.


One limitation of work to date on T2K is the completeness of the assessment of interaction model uncertainties as applied to short-baseline analyses. T2K analyses so far assume no \numu disappearance, however the interaction model systematic uncertainties are assessed based on external and ND280 measurements. Those measurements are placed close to production and therefore could be sensitive to a \numu disappearance signal, potentially biasing a dedicated \numu disappearance search.  T2K studied the possible impact of a subset of interaction model uncertainties on a ND280 \numu disappearance result~\cite{Bordoni:2017zwi} and found it to be robust, but this does not consider a full re-assessment of where external data is used to inform the model. Current efforts in T2K cross-section measurements and the implementation of ab initio computations in the context of three-flavor analysis would greatly benefit such studies as well. Certain event selections, like coherent neutrino interactions or low $\nu$ selections~\cite{Bodek:2012uu} may have better theoretical understanding for a single detector analysis, but can be challenging to use due to statistics or acceptance for T2K.

In the future, joint analyses, including \numu disappearance, \nue  appearance/disappearance may also be performed either between several near detectors (with different technologies and/or neutrino energy spectra) or benefiting from the complementary coverage of near versus far facilities. The upgraded near detector, for example, will have improved detection thresholds and sensitivity to neutrons, possibly enabling the selection of NC events at the near site. 

{\bf OPERA and ICARUS}
The Oscillation Project with Emulsion-tRacking Apparatus (OPERA) was a long-baseline accelerator neutrino experiment that sampled the CERN Neutrinos to Gran Sasso (CNGS) beam with a detector placed at the Gran Sasso laboratory (LNGS) 730\,km from the neutrino production source, having as primary physics goal to observe $\numu\rightarrow\nutau$ appearance. This goal required the use of emulsion detection technology, such that the detector was a hybrid apparatus made of a nuclear emulsion/lead target complemented by electronic detectors.
The detector ``target'' region had a total mass of about 1.25 kt and was composed of two identical sections, each with 31 walls of emulsion cloud chamber bricks, interleaved by planes of horizontal and vertical scintillator strips used to select bricks in which a neutrino interaction had occurred. The exquisite spatial resolution afforded by the emulsion layers enabled the identification of the characteristic ``kink'' due to the decay of the final-state $\tau$ particle from a $\nutau$ interaction. The target region was complemented by a magnetized muon spectrometer. 

\begin{figure}[ht!]
    \centering
    \includegraphics[width=0.5\textwidth]{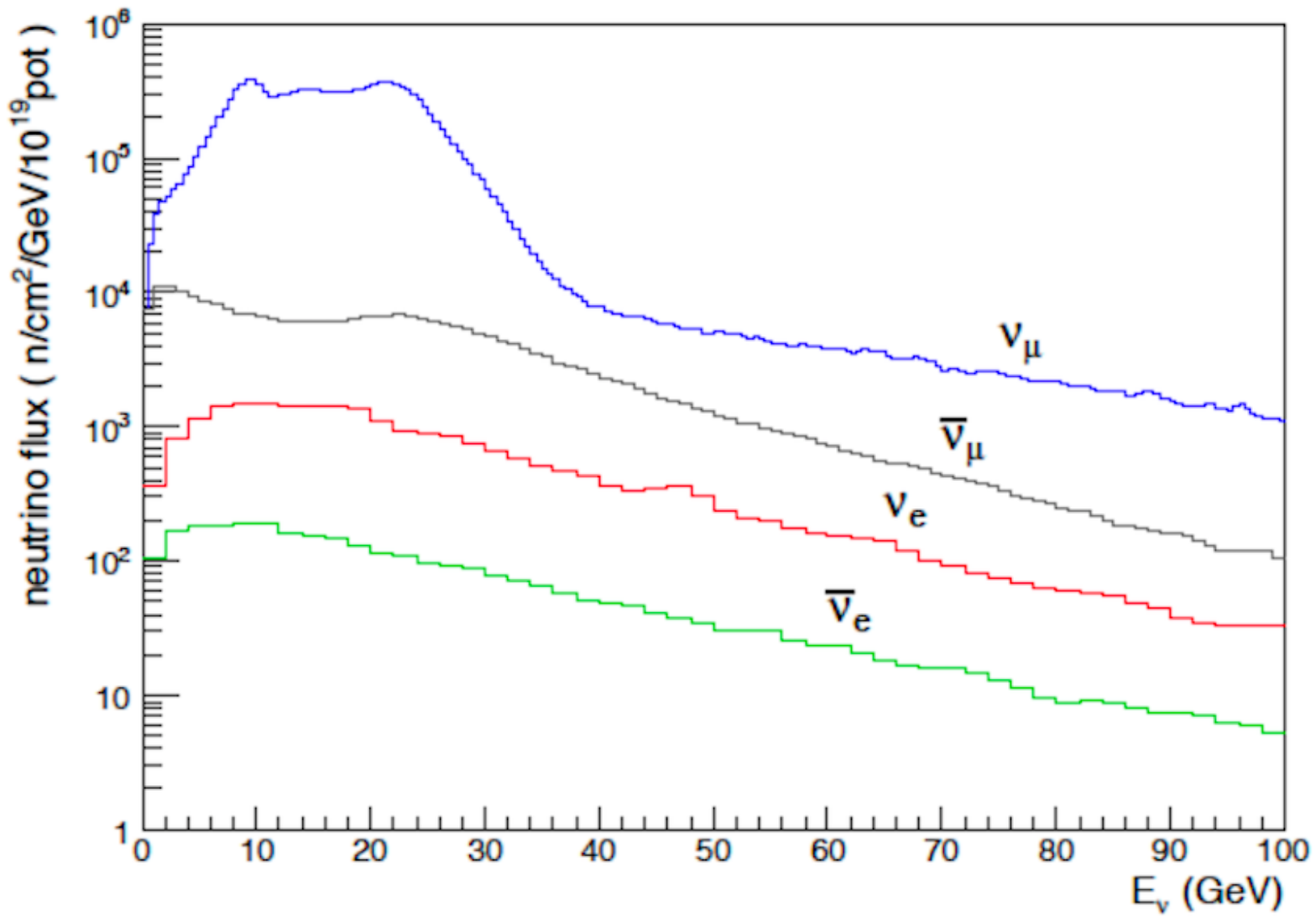}
    \caption{Fluxes of the different CNGS beam neutrino components at LNGS. Figure from~\cite{Acquafredda:2009zz}.}
    \label{fig:cngs}
\end{figure}

Reconstruction of neutrinos interacting within the emulsion layers was conducted by automated scanning robots. Due to the high-energy tau production threshold ($E_{\nu_\tau} \gtrsim 3.5$\,GeV), the CNGS neutrino flux was distributed at higher energies than other LBL experiments, as shown in Fig.~\ref{fig:cngs}~\cite{Acquafredda:2009zz}. OPERA collected CNGS beam data from 2008 to 2012, with an integrated exposure of $17.97\times10^{19}$ protons-on-target. A total of 19505 neutrino interaction events in the detector target were recorded by the electronic detectors, of which 5603 were fully reconstructed in the emulsion layers.

\begin{figure}[!ht] 
\centering
\includegraphics[width=0.48\textwidth]{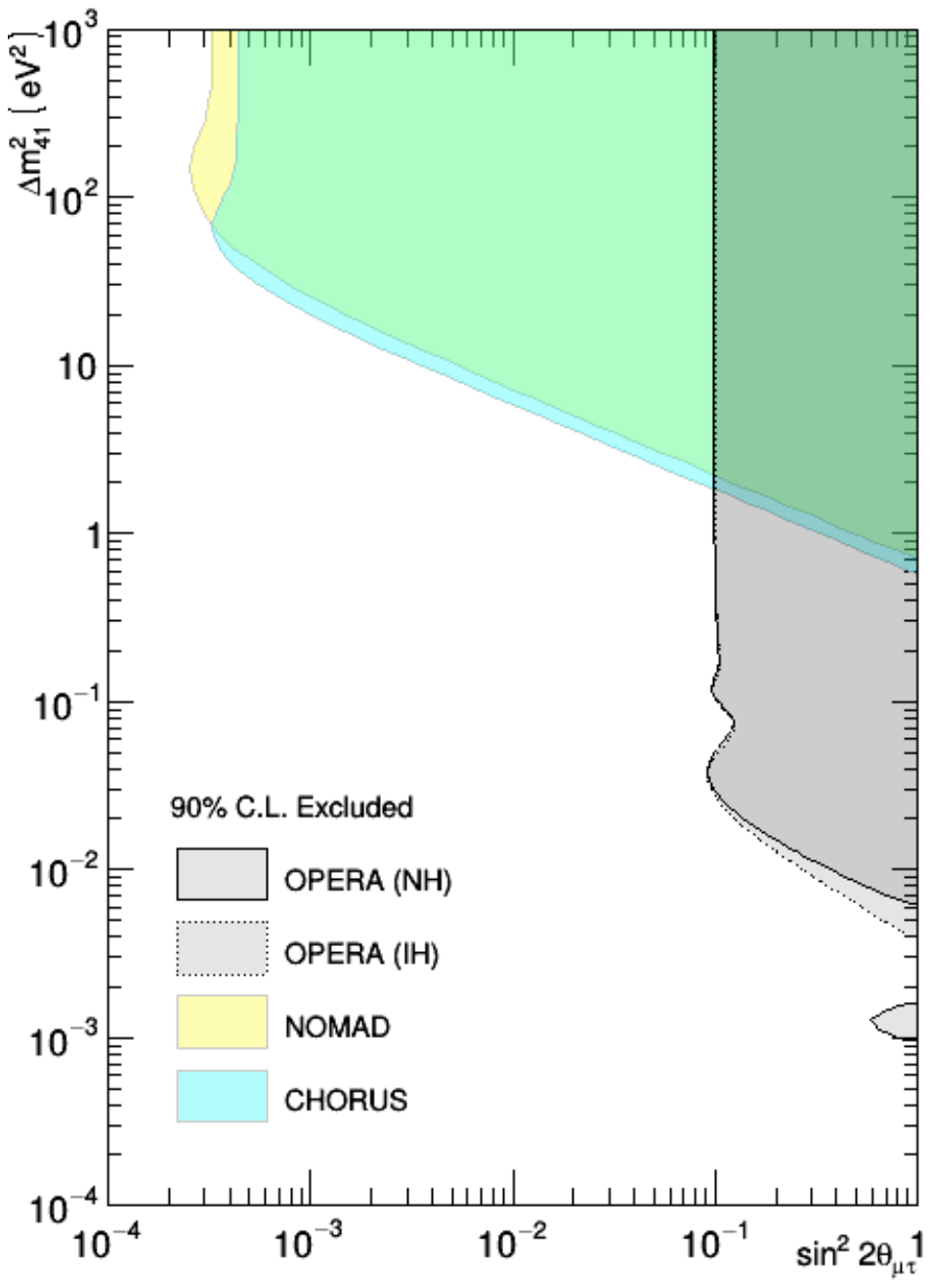}
\includegraphics[width=0.48\textwidth]{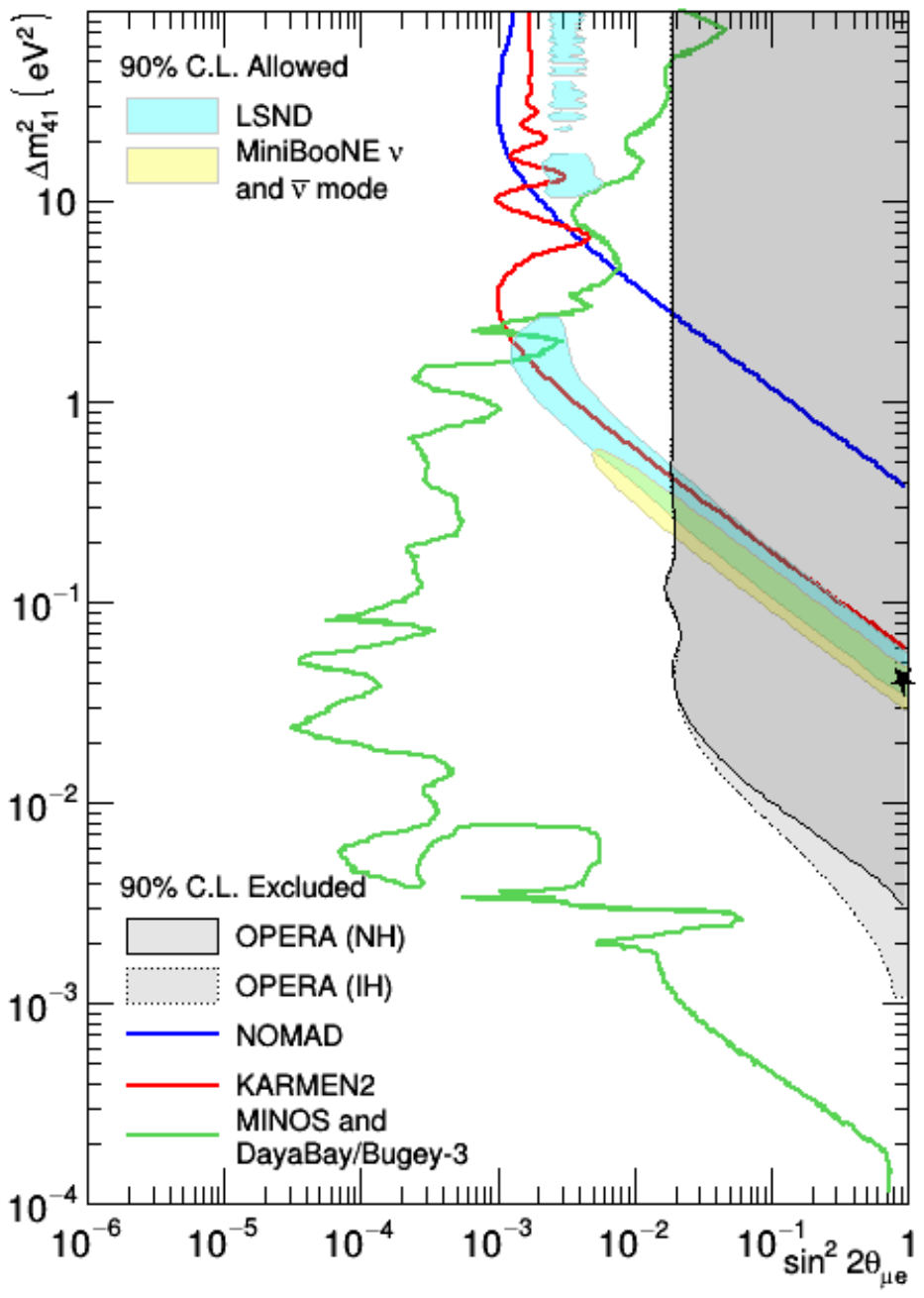}
\caption{Left: OPERA 90\% C.L. exclusion region in the $\Delta m^{2}_{41}$ and $\sin^{2} 2\theta_{\mu\tau}$ parameter space for the normal (NH, solid line) and inverted (IH, dashed line) hierarchy of the three standard neutrino masses. The exclusion regions by NOMAD \cite{NOMAD:2001xxt} and CHORUS \cite{CHORUS:2007wlo} are also shown. 
Right: OPERA 90\% C.L. exclusion region in the $\Delta m_{41}^{2}$ and $\sin^{2}2\theta_{\mu e}$ plane is shown for the normal (NH, solid line) and inverted (IH, dashed line) hierarchy of the three standard neutrino masses. The plot also reports the 90\% C.L. allowed region obtained by LSND \cite{Aguilar:2001ty} (cyan) and MiniBooNE combining $\nu$ and $\bar{\nu}$ mode \cite{Aguilar-Arevalo:2018gpe} (yellow). The blue and red lines represent the 90\% C.L. exclusion regions obtained in appearance mode by NOMAD \cite{NOMAD:2003mqg} and KARMEN2 \cite{KARMEN:2002zcm}, respectively. The 90\% C.L. exclusion region obtained in disappearance mode by the MINOS and DayaBay/Bugey-3 joint analysis \cite{DayaBay:2016lkk} is shown as green line. The black star ($\star$) corresponds to the MiniBooNE 2018 best-fit values for the combined analysis of $\nu$ and $\bar{\nu}$ data. figure from~\cite{OPERA:2019kzo}.}
\label{fig:opera_results}
\end{figure}

OPERA used its full data sample to conduct a search for sterile neutrino mixing within the context of a 4-flavor (3 active + 1 sterile) model based on both the $\nutau$ and $\nue$ appearance channels, with the $\numu$ CC disappearance channel not being considered given the low sensitivity to sterile mixing for that channel due to ambiguities with potential NC disappearance~\cite{OPERA:2019kzo}. 
Defining $\sin^{2}2\theta_{\mu\tau} = 4|U_{\tau4}|^2|U_{\mu4}|^2$, $\sin^{2}2\theta_{\mu e} = 4|U_{e4}|^2|U_{\mu4}|^2$, exclusion regions of $\Delta m^{2}_{41}$ versus $\sin^22\theta_{\mu\tau}$ and $\sin^22\theta_{\mu e}$ were computed and are shown in Fig.~\ref{fig:opera_results}. The result is restricted to positive $\Delta m^{2}_{41}$ values since negative values are disfavored by results on the sum of neutrino masses from cosmological surveys \cite{Planck:2015fie}. 
The results are consistent with no active-sterile neutrino mixing. For $\Delta m^{2}_{41} > 0.1 \textrm{ eV}^{2}$, the upper limits on $\sin^{2}2\theta_{\mu\tau}$ and $\sin^{2}2\theta_{\mu e}$ are set to 0.10 and 0.019 both for the case of Normal Ordering and Inverted Ordering. The values of the oscillation parameters $\left(\Delta m^{2}_{41} = 0.041 \textrm{ eV}^{2},\, \sin^{2}2\theta_{\mu e} = 0.92\right)$ corresponding to the MiniBooNE combined neutrino and antineutrino best-fit~\cite{Aguilar-Arevalo:2018gpe} are excluded with a $p$-value of $8.9\times10^{-4}$, corresponding to a significance of 3.3 $\sigma$. 

The ICARUS (Imaging Cosmic And Rare Underground Signals) experiment operated the T600 liquid argon time projection chamber detector, with a fiducial mass of 447 tons, exposed to the same CNGS beam as OPERA, and located at the same 730\,km from the neutrino production target in the LNGS. The baseline and CNGS beam typical energy determine that ICARUS is primarily sensitive to sterile mixing in the region where $L/E \sim 36.5$~km/GeV. ICARUS conducted a search for sterile neutrino mixing by looking for excess $\nue$ appearance in the data sample collected in 2010 and 2011. This search assumed a simplified two-flavor model (one active, one sterile) and found no evidence of sterile neutrino oscillations, as detailed in Ref.~\cite{Antonello:2012pq}. However, it was pointed out soon after the publication of the ICARUS results that the limits obtained on $\sin^{2}2\theta_{\mu e}$ by using a two-flavor approximation (which works well for a short-baseline experiment) in a long-baseline setup neglects sizable four-flavor effects, induced by the interference of the new large squared-mass splitting $\Delta m^2_{41}$ with the atmospheric one. The analysis also neglected contributions to the four-flavor oscillation probabilities arising from the intrinsic $\nue$ component of the CNGS beam. It is estimated that these four-flavor effects weaken the reported ICARUS constraints by up to a factor of 3~\cite{Palazzo:2015wea}.
The ICARUS T600 detector has since been moved to Fermilab, where it will be operated as the ``Far'' detector for Fermilab's Short-Baseline Neutrino program. ICARUS completed installation at Fermilab in Spring 2022 (the last section of the cosmic ray tagger and overburden) and is beginning its first physics run. The prospective ICARUS contributions to the Fermilab SBN program are detailed in Section~\ref{fermilab_sbn}.

\subsubsection{Reactor Neutrino Experiments}
\label{sec:expt_landscape_reactors}
\hfill\\

A conclusive way to test whether RAA is due to mixing between active and sterile neutrinos is by searching for baseline-dependent sterile neutrino-induced spectral variations.
In this section, we summarize the results from the experiments that performed a relative spectral search for 3+1 \anue{} oscillations induced by active to sterile mixing.
Relative oscillation searches curtail the dependence on the reactor \anue~spectrum model and detection efficiency by comparing the ratio of energy spectrum at different baselines to the corresponding predicted ratio under the oscillation hypothesis.
Baseline-dependent spectral measurement is done either by using a segmented detector, a single-volume movable detector, or by the use of multiple detectors placed at different baselines.
At reactor \anue~energies, since eV-scale neutrinos induce oscillations at meter scales, purpose-built experiments place their detectors as close to the reactor as possible, ideally $<$10~m.  
Furthermore, experiments built to search for $\theta_{13}$ are also sensitive to  sterile neutrino oscillations but at lower oscillation frequencies, typically for $\dm{13}<0.1 \mathrm{eV}^2$.

All the experiments discussed in this section use scintillator detectors and the inverse beta decay interaction process, where a characteristic signature is provided by the timing and spatial coincidence between prompt positron annihilation signal and delayed neutron capture signal producing two separate flashes ($\Delta t \sim \mu$s).
Refer to Sec.~\ref{sec:anomaly_reactor} for more details on the IBD mechanism.
Experiments also typically use an external neutron capture agent~(other than hydrogen), either Gadolinium or $^6$Li, to increase IBD efficiency by increasing the neutron capture cross-section.
$^6$Li has the added benefit of constraining the spatial extent of the delayed signal since the capture products have high $dE/dx$ in scintillators.
In addition to the delayed coincidence, experiments also use various combinations of active and passive shielding, detector segmentation, and pulse shape discrimination~(PSD) for background reduction.
PSD-capable scintillators generate pulse shapes that are particle-dependent and could in principle be used for both prompt and delayed selection.
Whenever possible, experiments also measure reactor-off data to constrain reactor-uncorrelated IBD-mimicking backgrounds.
Following is a discussion of the individual reactor experiments that searched for sterile neutrino-induced oscillations.
A list of purpose-built experiments and their features pertinent to the eV-scale oscillation search can be found in Tab.~\ref{tab:current_osc}.  

\begin{table*}[thbp!]
\begin{adjustbox}{width=\columnwidth,center}
\begin{tabular}{l||c|c|c|c|c|c}
\hline
Experiment & Baseline (m) & Reactor & Reactor & Detector & Target & Sterile $\nu$ \\
 &  & Type & Power~(MW\emph{th}) & Size & & Search Strategy \\\hline \hline 
DANSS~\cite{DANSS:2018fnn}  & 11--13 & LEU & 3000 & 1 m$^3$ & Segmented PS with Gd coating & Multi-Site \\ \hline
NEOS~\cite{NEOS:2016wee}   & 24 & LEU & 2800 &  1 m$^3$ & Single-volume GdLS + PSD & Single-Site \\ \hline
Neutrino-4~\cite{Serebrov:2020kmd} & 6--12 & HEU & 90 &  2 m$^3$ & Segmented GdLS & Multi-Site/Zone \\ \hline
PROSPECT~\cite{PROSPECT:2018dtt}  & 7--9 & HEU & 85 & 4 m$^3$  & Segmented $^6$LiLS + PSD & Multi-Zone \\ \hline
STEREO~\cite{STEREO:2019ztb}  & 9--11 & HEU & 57 & 2 m$^3$ & Segmented GdLS + PSD & Multi-Zone \\ \hline \hline
\end{tabular}
\end{adjustbox}
\caption{Details of the short-baseline experiments that have been specifically built to search for eV-scale sterile neutrinos. Short baselines are preferable for oscillation searches involving eV-scale sterile neutrinos. HEU reactors are typically $\sim 10x$ smaller than commercial LEU power reactors and are preferred for eV-scale oscillation searches since they have smaller source positions oscillation washout. Detector segmentation, capture agent, and pulse shape discrimination capabilities are highly beneficial for background reduction.} 
\label{tab:current_osc}
\end{table*}

\subsubsubsection{Short-Baseline Experiments}

{\bf DANSS} The Detector of AntiNeutrinos based on Solid State Scintillator~(DANSS) experiment \cite{danilov:2020danss} consists of 1 m\textsuperscript{3} highly segmented, plastic scintillator detector with each scintillator bar wrapped in Gd-loaded reflector.  
Light and signal readout is performed using wavelength-shifting fibers and a combination of pixel photon detectors and PMTs operated at room temperature.  
The experiment samples \anue from a 3 GW$_{th}$ low enriched uranium~(LEU) reactor at Kalinin Nuclear Power Plant in Russia.  
As opposed to most other reactor \anue experiments, the DANSS detector is located below the reactor core, benefiting from its 50-meter water equivalent~(m.w.e) overburden.
The detector is placed on a lifting platform which enables measurements at the baselines of 10.9 m, 11.9 m, and 12.9 m.
Due to its close proximity to the reactor core, the detector measures a high \anue flux of around 4000 events per day.

\begin{figure}[ht!]
    \centering
    \includegraphics[height=3.5in,width=0.47\textwidth]{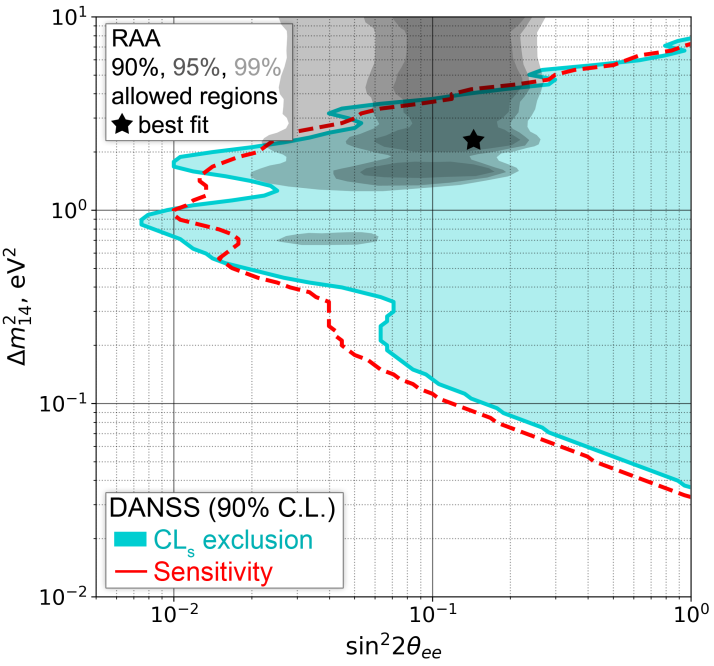}
    \includegraphics[height=3.5in,width=0.47\textwidth]{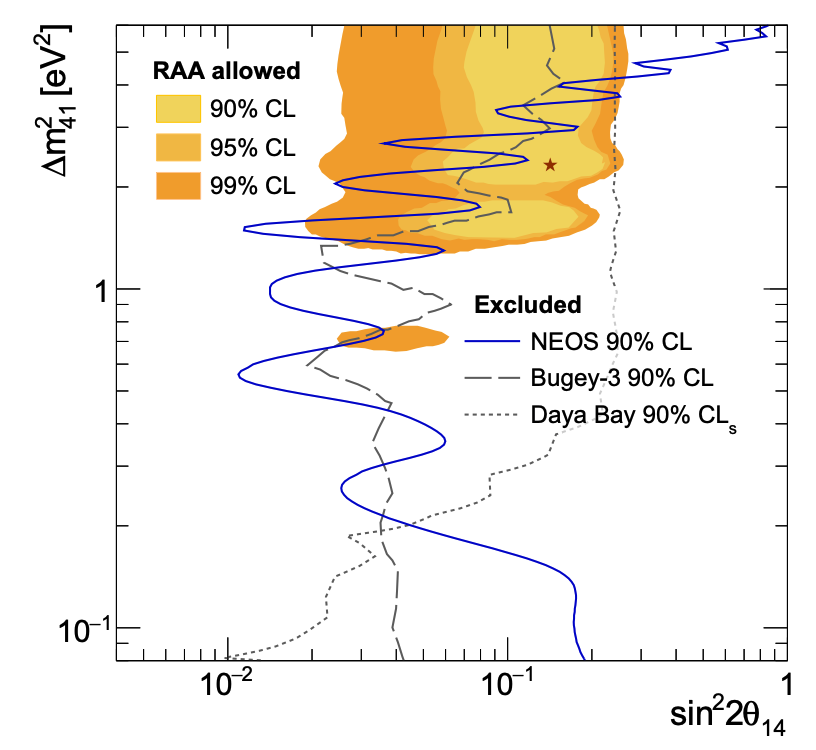}
    \caption{Left panel: Exclusion curves at 90\% C.L. (filled area) and 90\% C.L. sensitivity contours (dashed line) are shown. Expected regions from
     RAA and GA are also shown. Figure from~\cite{Danilov:2022str}. 
    Right panel: Exclusion curves for 3+1 neutrino oscillations in the  $\sin^{2}{2\theta_{14}}$, $\Delta m^{2}_{14}$ parameter space obtained by NEOS. Figure from \cite{NEOS:2016wee}.}
    \label{fig:dans_neos}
\end{figure}

The DANSS detector was commissioned in  2016 and started taking data in October 2016. The experiment has collected around 4 million signal events over three years at three different positions.  
The biggest source of background that constitutes 13.8\% of the IBD signal events is the accidental coincidence background, along with non-negligible cosmogenic backgrounds and IBD backgrounds from neighboring reactors. The data excludes a large area in sterile neutrino parameter space ($\Delta m^{2}$, $\sin^{2}{2\theta_{ee}}$) and most interestingly excludes the best-fit point of RAA and Ga experiments at more than $5\sigma$ level as shown in Fig.~\ref{fig:dans_neos}. Additionally, the IBD positron spectrum was measured and compared to the simulated Huber-Mueller spectrum.  Although the measured spectrum disagrees with DANSS's Huber-Mueller-derived prediction, final oscillation measurements are largely independent of the choice of models due to the ratios taken at different baselines \cite{danilov:2020danss}.
\\  

\noindent {\bf NEOS} The Neutrino Experiment for Oscillation at Short baseline (NEOS) aims to search for light sterile neutrinos by detecting electron antineutrinos ($\bar{\nu_{e}}$) from a reactor at a very short baseline. The NEOS detector consists of a liquid scintillator (LS) target ($\sim$1000 liter with $\sim$0.5\% Gd loading by weight), two buffer zones filled with mineral oil where 19 8-inch PMTs per a zone are attached, and a muon veto system. NEOS had been installed in the tendon gallery of the fifth reactor of the Hanbit Nuclear Power Plant in Korea for two periods, from Aug.~2015 to May 2016 (NEOS-I) and from Sept.~2018 to Oct.~2020 (NEOS-II). For the NEOS-II, the modifications include newly produced target GdLS and a minor modification in the muon veto system. With a 24~m baseline from the core of the reactor (2.8~GW$_{th}$ in 100\%) about 2\,000 inverse beta decay (IBD), $\bar{\nu_{e}} + p \rightarrow e^+ + n$, events per day are observed. The signal-to-background ratio is $~$22 thanks to relatively good overburden ($\sim$20 m.w.e.). Background contribution from the nearest neighboring reactor ($d = 256$~m) is found to be less than 1\% of the total $\bar{\nu_e}$ flux from the fifth reactor. Calibration data using radioactive sources, $^{22}$Na, $^{137}$Cs, $^{60}$Co, PoBe, and $^{252}$Cf, had been taken regularly. 


NEOS-I~\cite{NEOS:2016wee} using 180 (46) live days of reactor-ON (-OFF) data excluded the RAA best fit at 90\% C.L.~by comparing the prompt energy spectrum of Daya Bay where the sterile neutrino oscillation effect is averaged out. The best-fit was found at (sin$^2 2\theta, \Delta m^2)$ = (0.05, 1.70~eV$^2$) with $\chi^2$/NDF for 3$\nu$ and 4$\nu$ are 64.0/61, and 57.5/59, respectively. The corresponding p-value is estimated to be 22\%. The well-known ``5~MeV excess'' was clearly observed as well in NEOS-I. Recently a joint analysis~\cite{Atif:2020glb} between NEOS-I and RENO was performed, yielding a slightly improved result beyond its previous result using early Daya Bay data.  More details on the NEOS-I and RENO joint analysis is discussed in the following sub-section (Sec.~\ref{sec:null_results_joint_rx_fits}).  
\\

\noindent {\bf Neutrino-4} Neutrino-4 is the only reactor neutrino experiment performing relative spectral comparison that has reported evidence of oscillations.
The detector is a 1.8\,m$^3$ Gd-doped liquid scintillator detector divided into 10 rows each row consisting of 5 sections each of size 0.225\,m~$\times$ 0.225\,m~$\times$ 0.85\,m.
The detector samples \anue from the 57 MW$_{th}$ SM-3 reactor, and HEU in Dimitrovgrad, Russia.
A movable platform enables the detector to sample baselines from 6\,m--12\,m.
The short reactor on~(off) cycles of 8--10~(2--5) days enable the experiment to perform rapid signal and background measurements. 

\begin{figure}[ht!]
    \centering
    \includegraphics[width=0.44\textwidth]{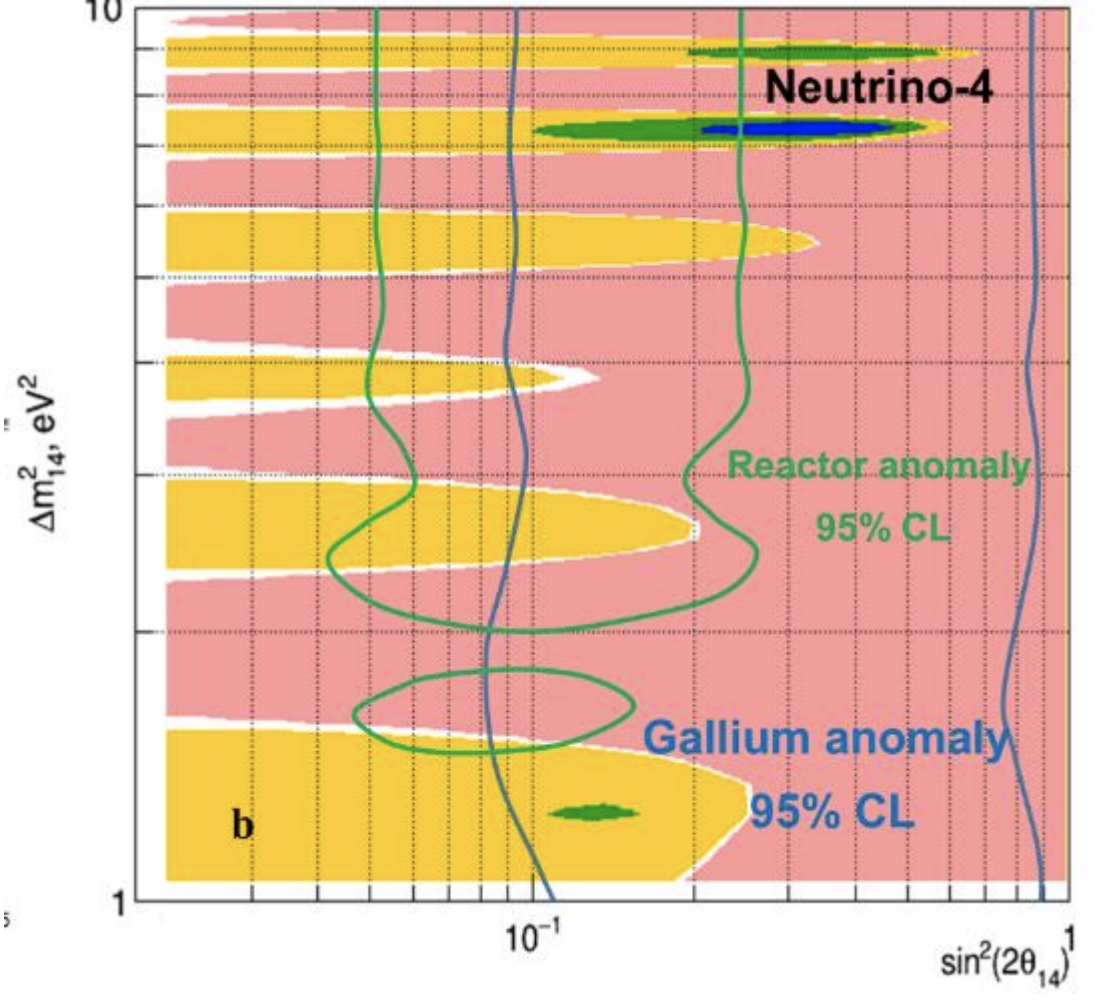}
    \includegraphics[width=0.55\textwidth]{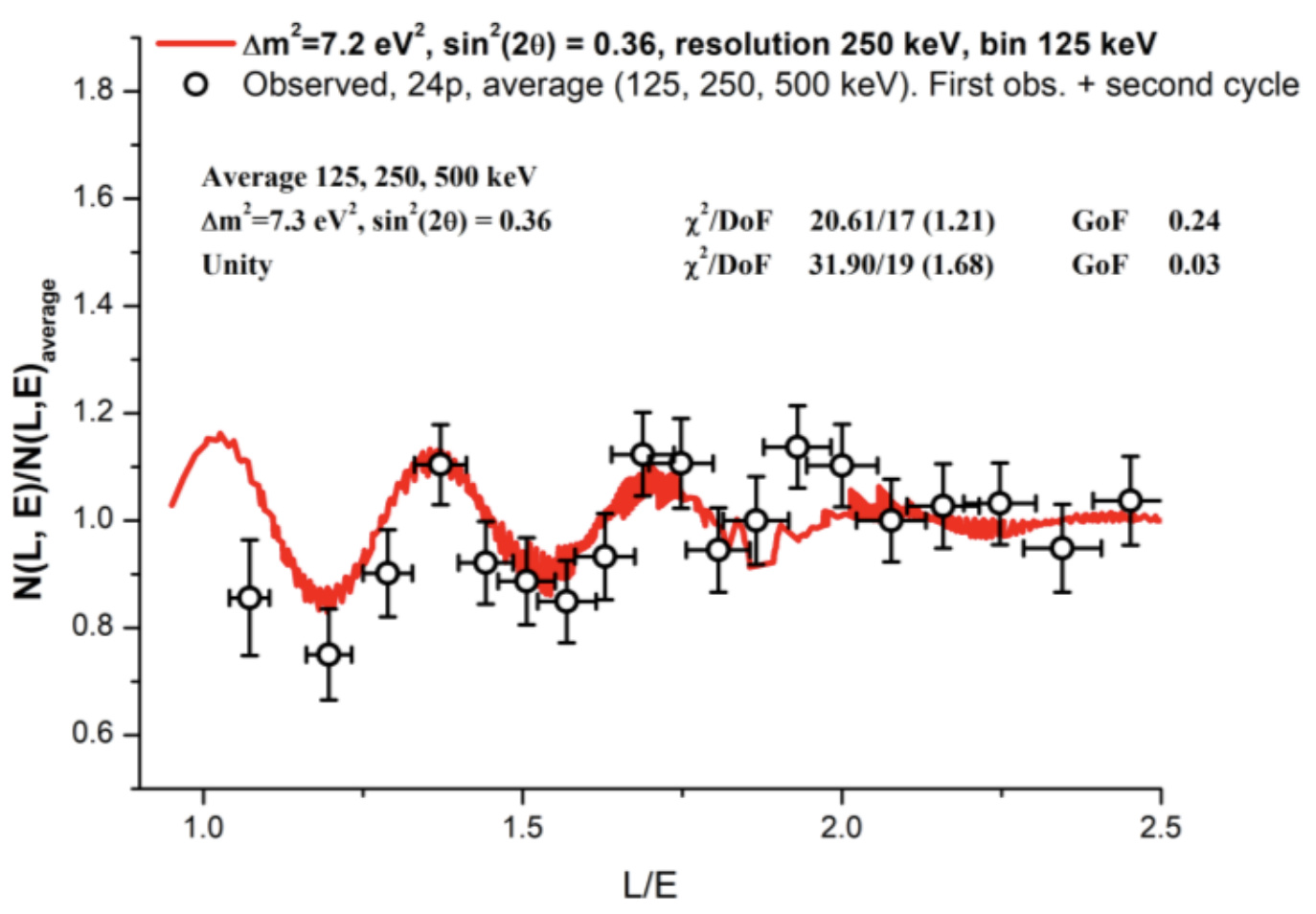}
    \caption{Left: 1$\sigma$ (blue), 2$\sigma$ (green), and 3$\sigma$ (yellow) suggested regions from the Neutrino-4 sterile neutrino oscillation search.  Right: $L/E$ distribution of background-subtracted IBD rates reported by the Neutrino-4 Experiment.  Plots from~\cite{Serebrov:2020kmd}}
    \label{fig:N4}
\end{figure}

The detector collected data for five years from 2016--2021 with $\sim$300 events/day.
The latest analyzed dataset consists of 720~(860) calendar days of reactor on~(off) data with a signal-to-background ratio of $\sim$0.5.
The experiment performed a sterile neutrino search in the \textit{L/E} space and observed evidence for 2.9$\sigma$ neutrino oscillation effect with the best-fit at \dm{14}=7.3$\pm$0.13(stat) $\pm$1.16(syst) and $\sin^2\theta_{14}$=0.36$\pm$0.12(stat) (Fig.~\ref{fig:N4}).
It is important to point out that these results are widely debated by several groups~\cite{Coloma:2020ajw,Danilov:2020rax,Giunti:2020uhv,PROSPECT:2020raz}.
Additionally, PROSPECT and STEREO experiments~\cite{PROSPECT:2018hzo,STEREO:2022nzk} also individually disfavor the best-fit point and a significant portion of the Neutrino-4 suggested parameter space.
\\

\begin{figure}[htb]
    \centering
    \includegraphics[width=0.8\textwidth]{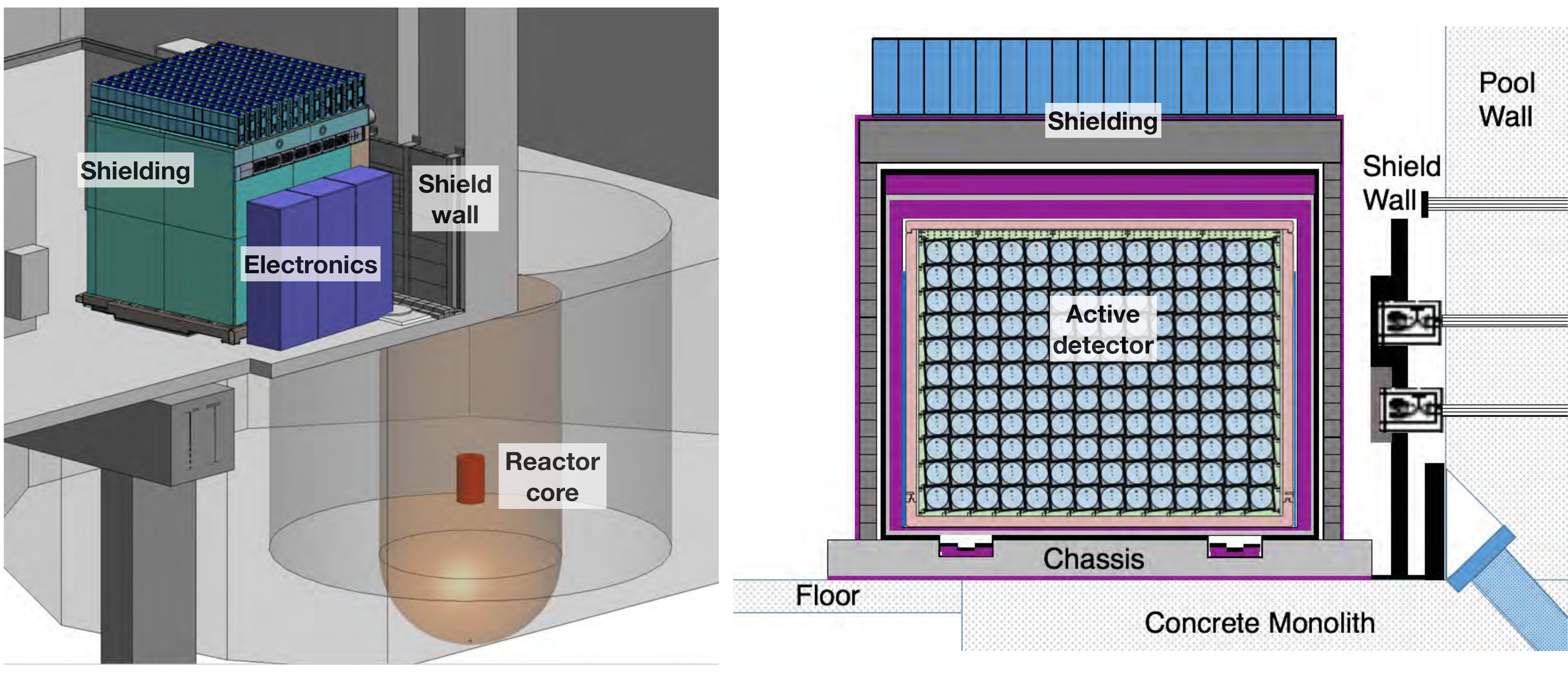}
    \caption{Left: Schematic of the PROSPECT detector, shielding, and electronics in the HFIR Experiment Room. The center of the detector is located at a distance of 7.84 m from the center of the reactor core~(red). Right: Cutout of the PROSPECT active segmented detector enclosed in the containment and shielding. Also shown is the vertical local lead shield adjoining the reactor pool wall that was installed to mitigate the reactor-related backgrounds. The figure was a modified version of the figure from \cite{PROSPECT:2021jey}}
    \label{fig:PROSPECT_1_Detector}
\end{figure}

\noindent {\bf PROSPECT}\label{sec:PROSPECTI} The Precision Reactor Oscillation and SPECTrum~(PROSPECT) experiment~\cite{PROSPECT:2015iqr,PROSPECT:2018dnc} is a U.S.-based reactor neutrino experiment installed at short baselines of 6.7--9.2 m from the 85 MW High Flux Isotope Reactor~(HFIR) at Oak Ridge National Laboratory.
The 4-ton~\textsuperscript{6}Li-loaded PSD-capable liquid scintillator~\cite{PROSPECT:2019gwi} detector is composed of a two-dimensional grid of 
optically isolated segments~\cite{PROSPECT:2015rce, PROSPECT:2018hzo} as shown in Fig.~\ref{fig:PROSPECT_1_Detector}.
Low mass, highly reflective, rigid separators~\cite{PROSPECT:2019enz} were used to achieve segmentation and photomultiplier tubes~(PMTs) enclosed in mineral oil-filled acrylic housings were installed at either ends of the segments for signal readout.

The detector shielding was optimized based on the neutron and $\gamma$ ray background measurements performed at HFIR~\cite{PROSPECT:2015eri} and consisted of top-heavy hydrogenous shielding to reduce cosmogenic backgrounds and a fixed local lead shield to mitigate reactor-specific backgrounds.
PROSPECT was the first on-surface reactor neutrino experiment to achieve a signal-to-background~(S:B) > 1, thanks to the high background suppression enabled by the detector design.

PROSPECT was commissioned in February 2018 and started collecting data in March.
The first oscillation search result was published in 2018~\cite{PROSPECT:2018dtt} with a relatively small dataset of 33~(28) reactors on~(off) live days which was followed by a result with a longer dataset consisting of 96~(73) reactor on~(off) live days composed of $>$ 50k signal events in 2020.
The oscillation search was done by performing a relative comparison of baseline-dependent spectra minimizing the dependence on the reactor neutrino model \cite{PROSPECT:2020sxr}.

\begin{figure}[ht!]
    \centering
    \includegraphics[width=0.5\textwidth]{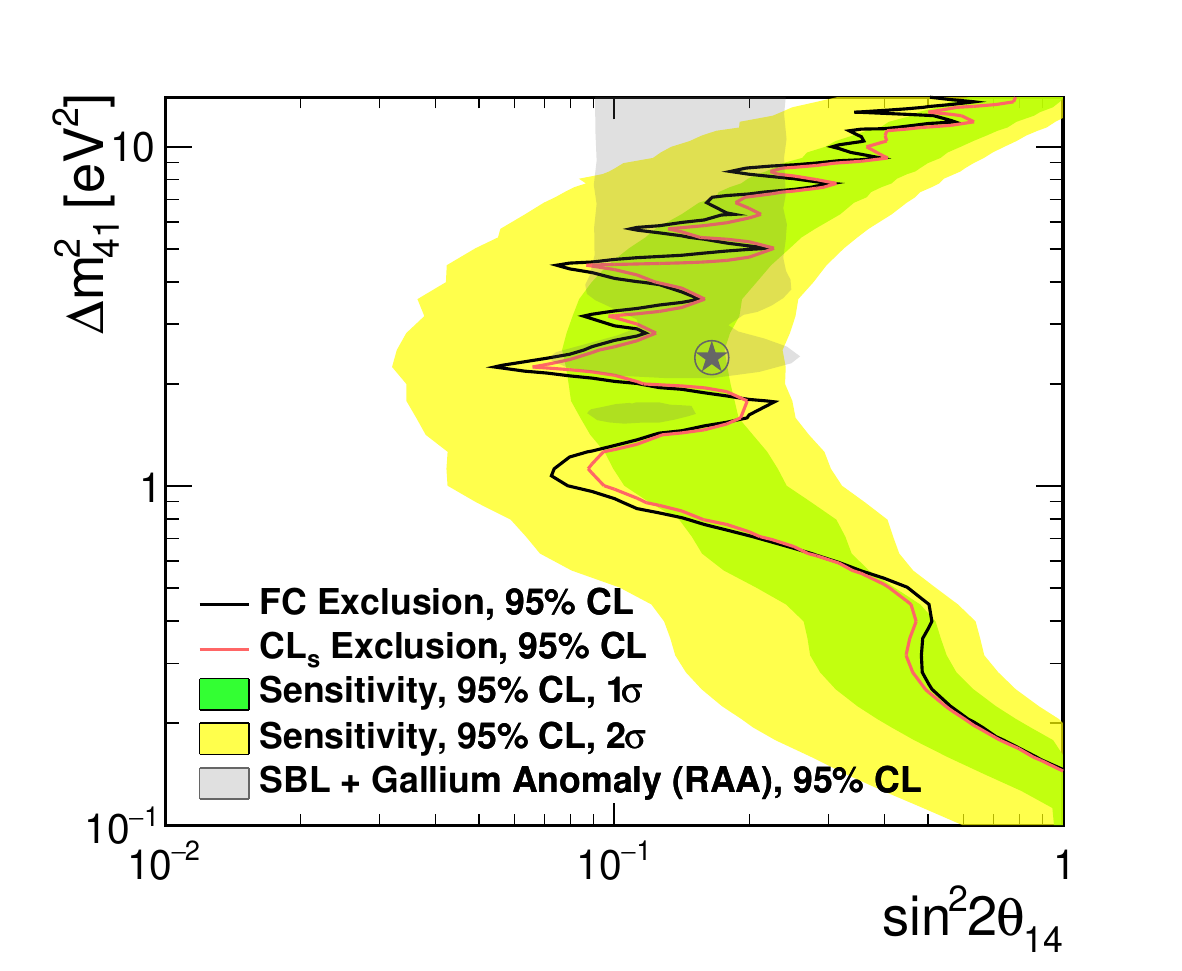}
    \caption{Results from PROSPECT's search for $\anue$ to sterile neutrino oscillations. Exclusion contours were drawn using Feldman-Cousins~(black) and Gaussian CLs~(red) methods. Green and yellow bands show the 1~$\sigma$ and 2~$\sigma$ PROSPECT sensitivities to the sterile neutrino oscillations. Also shown for comparison are the RAA-suggested parameter space and the best-fit point. Figure from \cite{PROSPECT:2020sxr}.}
    \label{fig:PROSPECT_1_Results}
\end{figure}

A combination of the compact HFIR reactor core~(a cylinder of diameter 0.435 m and height of 0.508 m) with the fine detector segmentation~(14.5 cm $\times$ 14.5 cm cross-section) enabled high sensitivity to oscillation frequencies of $\Delta m^2>1$~eV$^2$.
PROSPECT observed no statistically significant indication of \anue to sterile neutrino oscillations. 
Using Feldman-Cousins technique~\cite{Feldman:1997qc}, PROSPECT excluded a significant portion of the RAA suggested parameter space at 95\% CL and the best-fit point at 2.5 $\sigma$ as shown in Fig.~\ref{fig:PROSPECT_1_Results}.
A complementary technique using the Gaussian CLs method~\cite{Qian:2014nha} also provides similar exclusion.

The PROSPECT detector was decommissioned in 2020 after an unexpected HFIR downtime.
During the course of data taking, it was also observed that a number of segments have slowly lost functionality due to failures of PMT electronics induced by the ingress of the liquid scintillator into the PMT housings.
Due to these reasons, the results shown above--corresponding to the full PROSPECT dataset--are primarily dominated by statistics.
However, the collaboration is carrying out two major analysis modifications by leveraging distinctive detector features.
a) Splitting the data-taking period into multiple time-frames and b) allowing the use of segments with single PMTs. 
The former would allow for an increase in statistics by allowing segments with non-functioning PMTs to be used for part of the dataset and the latter increases the S:B by improving particle identification and consequently reduces the backgrounds. 
Each of the analysis improvements is individually expected to increase the effective statistics by $\sim$50\% and enable considerable improvement in the statistical power of the PROSPECT's oscillation search.
\\

\begin{figure}[ht!]
    \centering
    \includegraphics[height=3in,width=0.47\textwidth]{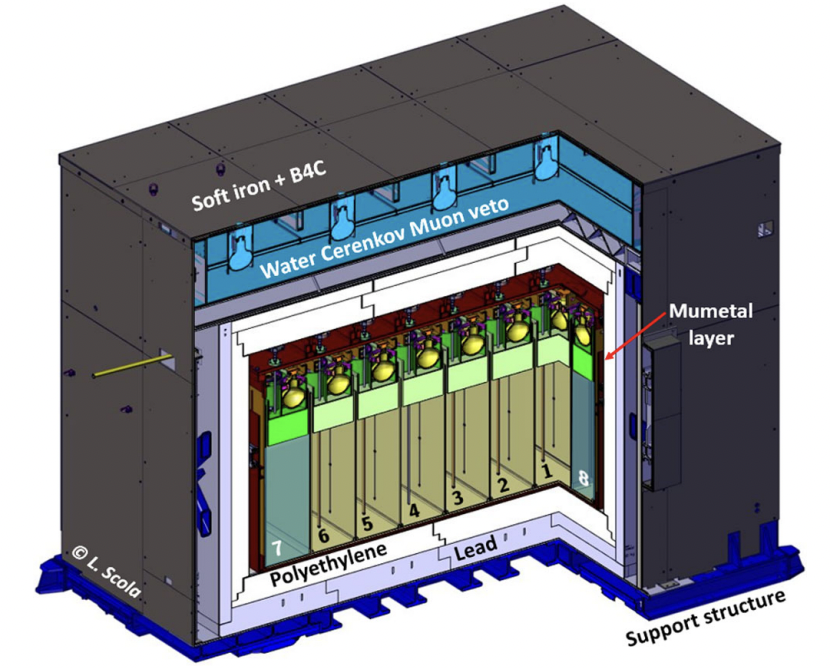}
    \includegraphics[height=3in,width=0.47\textwidth]{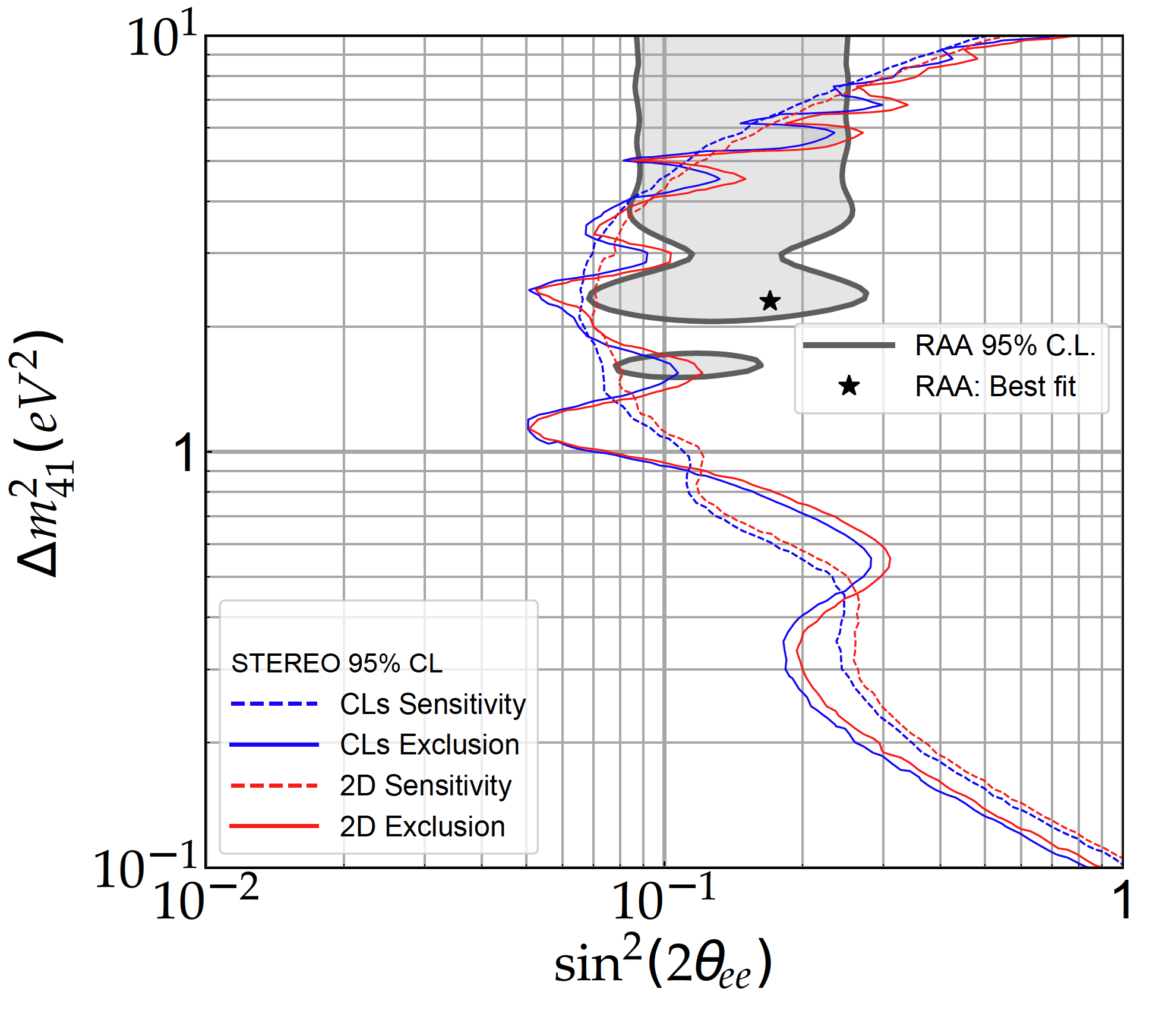}
    \caption{(Left) STEREO setup. 1–6: target cells (baselines from core: 9.4–11.1~m). Figure from \cite{Almazan:2018stereo}. (Right) Exclusion contour (red) and exclusion sensitivity (blue) at 95\% C.L. Overlaid are the allowed regions of the RAA (grey) and its best fit point (star). Figure from \cite{STEREO:2022nzk}.}
    \label{fig:stereo}
\end{figure}

\noindent {\bf STEREO} The STEREO experiment \cite{Almazan:2018stereo} is a Gd-doped liquid scintillator detector located at the ILL in Grenoble, France. The detector uses a high-flux 58 MW$_{th}$ research reactor that consists of highly enriched Uranium. The detector is located 9~m away from the reactor core and is a segmented detector where the position of different segments of the detector serves as different baselines varied between 9-11~m as shown in Fig.~\ref{fig:stereo} (Left).

The IBD signal events are selected using a cut-based approach where selection cuts on energy and time variables are optimized by trading off between detection efficiency and background rejection. Furthermore, antineutrino signal rates separated from the remaining background using Pulse Shape Discrimination (PSD) where pulses generated from neutrons have longer tails compared to that of gamma. Hence, a ratio of the pulse tail to total charge is used to mitigate neutron-related background.\\
The test of sterile neutrino oscillations is performed using ratios of energy spectrum at six different segments to that of the first segment and therefore making the measurement independent of absolute normalization and of the prediction of the reactor spectrum. With 273 (520) days of reactor-on (-off) data, STEREO found no evidence for sterile neutrino oscillations, and the results are compatible with the null oscillation hypothesis. The data excluded RAA best-fit point at p-value $<10^{-4}$ and exclusion curve with 3+1 neutrino oscillations scenario is shown in Fig.~\ref{fig:stereo} (Right).\\



\subsubsubsection{Medium-Baseline Experiments}

\noindent {\bf Daya Bay} Daya Bay's unique configuration makes it an excellent experiment to search for sterile-active neutrino mixing~\cite{DayaBay:2014fct,DayaBay:2016qvc,DayaBay:2016lkk,MINOS:2020iqj}. In this experiment, electron antineutrinos emitted from six $2.9$~GW$_\mathrm{th}$ nuclear reactors are detected in eight identically-designed antineutrino detectors (ADs) placed underground in two near experimental halls (EHs) and one far hall. 
The two near halls, EH1, and EH2, are located $\sim$350-600~m away from the reactors, whereas the far hall, EH3, is located $\sim$1500-1950~m from the reactors. 

\begin{figure}[htbp!]
    \centering
    \includegraphics[width=0.51\textwidth]{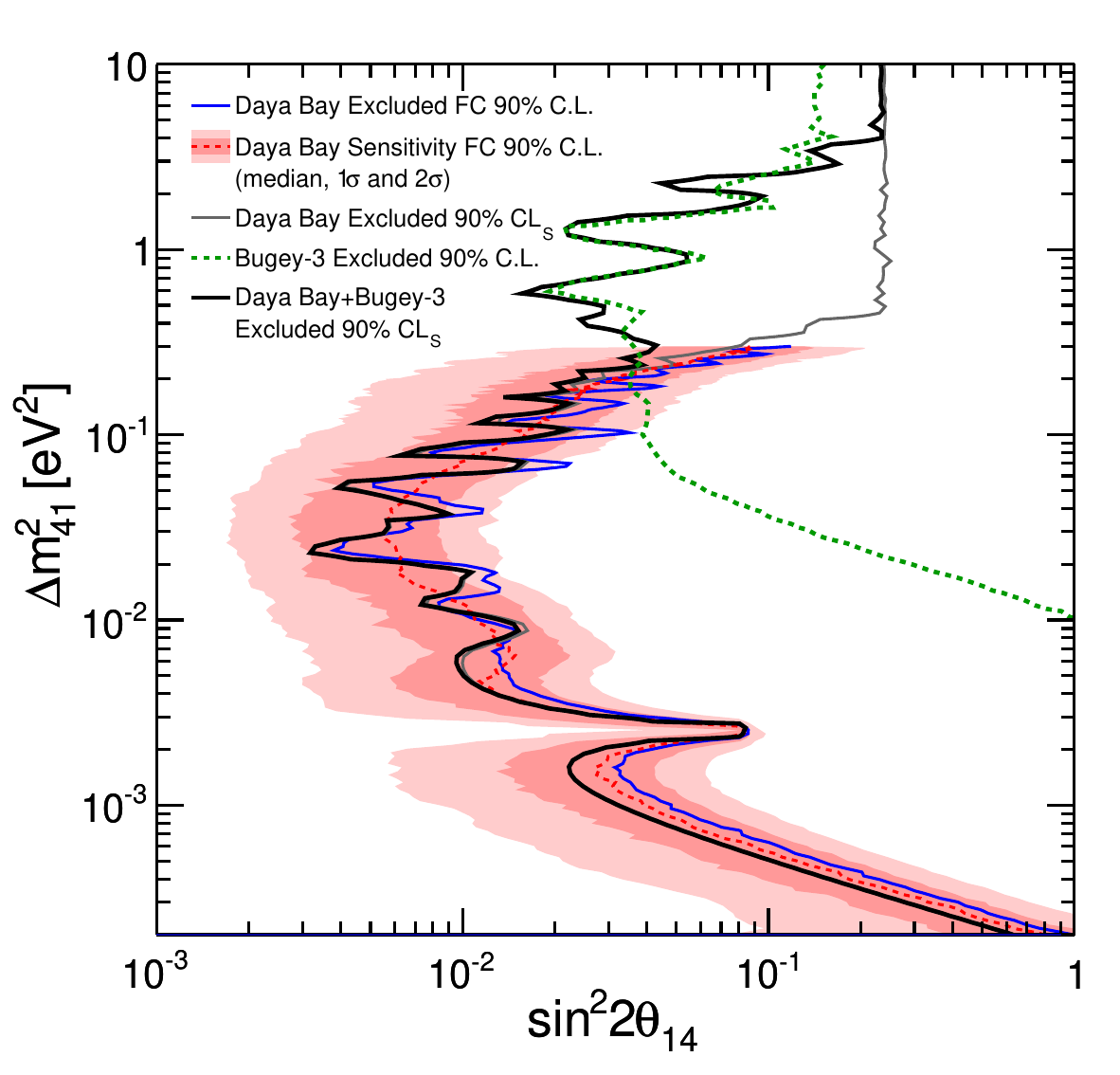}
    \includegraphics[height=0.48\linewidth]{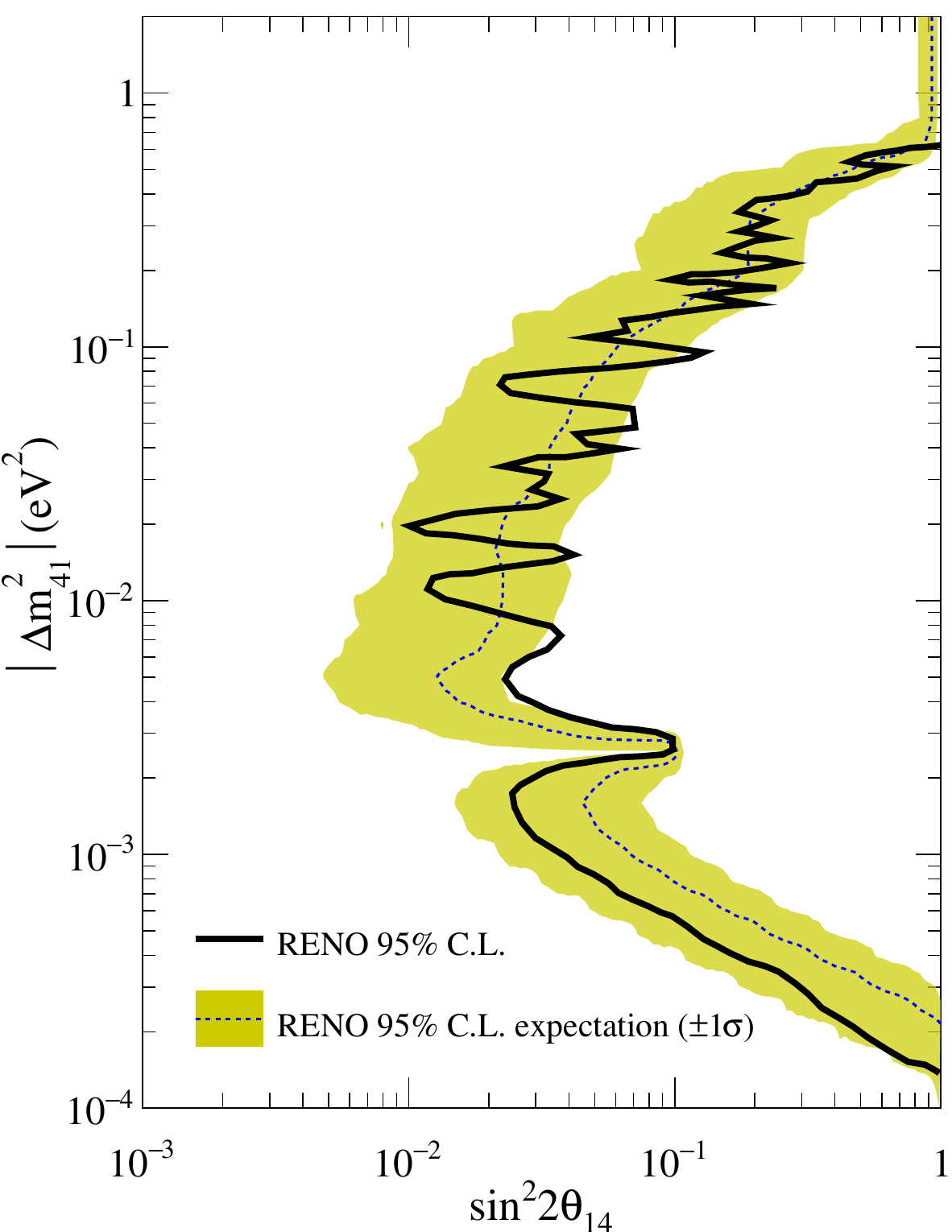}
    \caption{\label{fig:dyb} Left: Feldman-Cousins (FC) 90\%~C.L. and 90\%~CL$_s$ exclusion regions from an oscillation analysis of 1230 days of Daya Bay data. The dashed red line shows the 90\%~C.L. median sensitivity along $1\sigma$ and $2\sigma$ bands. The excluded region for the original Bugey-3 limit~\cite{Declais:1994su} is shown in green, while the resulting CL$_s$ contour from Daya Bay and its combination with the reproduced Bugey-3 results are shown in grey and black, respectively. From \cite{MINOS:2020iqj}.  Right: RENO's 95\,\% C.L. exclusion contour for the oscillation parameters $\sin^2 2\theta_{14}$ and $|\Delta m^2_{41}|$. The black solid contour represents an excluded region obtained from spectral distortion between near and far detectors. The green shaded band represents expected 1$\sigma$ exclusion contours due to a statistical fluctuation. The blue dotted contour represents its median. From \cite{RENO:2020uip}.
    }
\end{figure}

Daya Bay's latest constraints in the $(\sin^2 2\theta_{14},\Delta m^2_{41})$ parameter space, obtained with a 1230-day data set, are shown in Fig.~\ref{fig:dyb}~\cite{MINOS:2020iqj}. Two complementary analysis methods are used to set the exclusion contours, one relying on the Feldman-Cousins (FC) frequentist approach and the other on the CL$_s$ approach. Daya Bay is most sensitive to $\sin^2 2\theta_{14}$ in the $10^{-3}$~eV$^2$ $\lesssim |\Delta m^2_{41}| \lesssim 0.3$~eV$^2$ region, where a distortion from the standard three-neutrino oscillation framework would be visible through a relative comparison of the rate and energy spectrum of reactor antineutrinos in the different EHs. For $|\Delta m^2_{41}| \gtrsim 0.3$~eV$^2$, the oscillations are too fast to be resolved, and the sensitivity arises primarily from comparing the measured rate with the expectation, resulting in less stringent limits. 

Daya Bay ceased operations in December 2020 after collecting data for over 3000 days. Throughout this time it accumulated the largest sample of reactor antineutrinos to date, consisting of more than six million events. This sample is still being analyzed and final results are expected to be released by early 2023. The sizable increase in statistics, combined with potential reductions in systematic uncertainties, implies that significant improvements over the existing limits are expected. The new constraints will likely remain the best in the world for the foreseeable future in the $|\Delta m^2_{41}| \lesssim 0.3$~eV$^2$ region, which no experiment in the horizon is expected to cover at the time of writing.  

\noindent {\bf Double Chooz} Double Chooz \cite{Abrahao:2021DC} consists of two nearly identical gadolinium-doped liquid scintillator detectors located close to the nuclear power plant comprising two 4.25 GW nuclear reactors.
The near~(far) detectors are located underground at an overburden of 120~m~(300 m) at a distance of 469~m and 355~m~(1115 m and 998 m) from the two reactor cores.
The detector-reactor locations are such that the relative contributions to both the detectors from the reactors are very similar which helps reduce reactor-related uncertainties.

This analysis includes three datasets amounting to a total 5-year long dataset. 
The first~(FD-I) dataset consists of 455.21 days of livetime collected with the far detector before the commissioning of the near detector. 
The second~(FD-II) and third~(ND) datasets are collected during the same period of time and consists of 362.97 days and 257.96 days of livetime respectively.
The livetime for ND is lower than FD-II because of the higher muon rate causing larger deadtime in the near detector. 
In order to obtain a measurement independent of absolute flux predictions, the experiment directly compares the event rates measured in the two identical detectors which helps in canceling most of the reactor flux and detection efficiency-related uncertainties.
The experiment does not see any indications of sterile neutrino oscillations and set exclusion limits in similar regions of \dm{14} as the other $\theta_{13}$ experiments.  
\\

\noindent {\bf RENO} The RENO collaboration has reported a search result for light sterile neutrino oscillations.
The search is performed using six 2.8\,GW$_{\text{th}}$ reactors
and two identical detectors located at 294\,m (near) and 1383 \,m (far), respectively, from the center of six reactor cores at the Hanbit Nuclear Power Plant Complex in Yonggwang.
The reactor flux-weighted baseline is 410.6\,m for the near detector and 1445.7\,m for the far detector, respectively.
The near (far) underground detector has 120\,m (450\,m) of water equivalent overburden.
A spectral comparison between near and far detectors was performed to search for reactor $3+1$ light sterile neutrino oscillations
\cite{RENO:2015ksa,Park:2012dv,Park:2013nsa,Ma:2009aw,Bak:2018ydk}.  




The RENO sterile analysis uses roughly 2200 live days of data taken in the period between August 2011 and February 2018 amounting to 850\,666 (103\,212) $\overline{\nu}_e$ candidate events
in the near (far) detector.
The details of pull terms and systematic uncertainties are described in Ref.~\cite{Bak:2018ydk}.
Exclusion regions at 95\,\% confidence level are set for $\Delta \chi^2 > 5.99$ and are shown in Fig.~\ref{fig:dyb}.
Exclusion contours obtained using the Gaussian CL$_s$ method~\cite{Read:2002hq, Qian:2014nha} show negligible difference with the pictured $\Delta\chi^2$ method.
Fig.~\ref{fig:dyb} shows the 95\,\% C.L. exclusion contour and median sensitivity including 1$\sigma$ band due to statistical fluctuations.

The limit of $\sin^2 2\theta_{14}$ is mostly determined by a statistical uncertainty, while the systematic uncertainties dominate in the $|\Delta m^2_{41}| \lesssim 0.06$\,~eV$^2$.
The uncertainty of background is a dominant systematic source in the $0.003 \lesssim |\Delta m^2_{41}| \lesssim 0.06$\,~eV$^2$ region, and the energy-scale uncertainty  is a major limiting factor in the $|\Delta m^2_{41}| \lesssim 0.008$\,~eV$^2$ region.
The RENO result provides the most stringent limits on sterile neutrino mixing at $|\Delta m^2_{41}| \lesssim 0.002$\,~eV$^2$ using the $\overline{\nu}_e$ disappearance channel.
Adding data taken since 2018 and reducing the above systematic uncertainties will improve the results significantly.

\begin{figure}[bt!]
\centering
\includegraphics[height=0.5\linewidth]{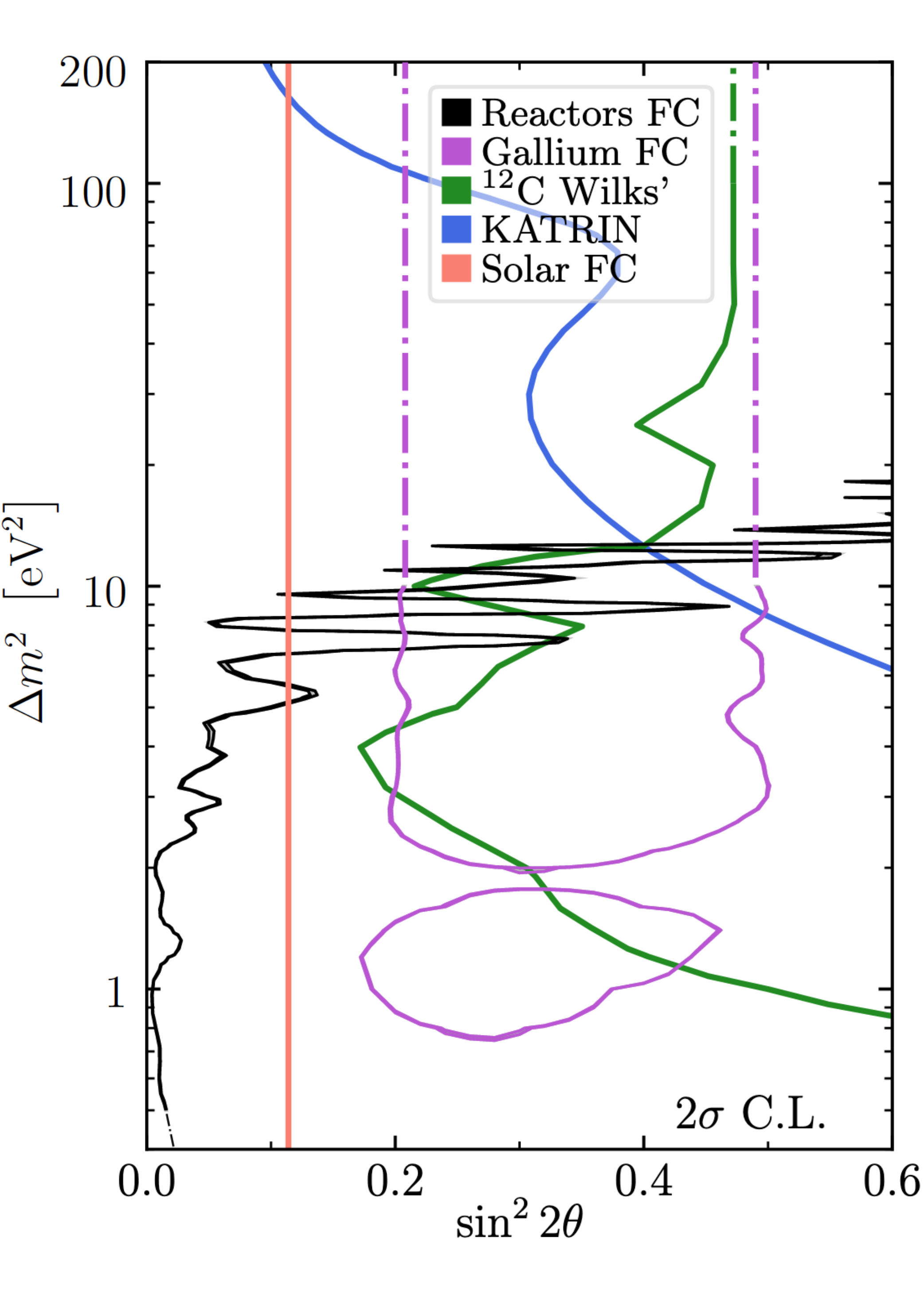}
\includegraphics[height=0.5\linewidth]{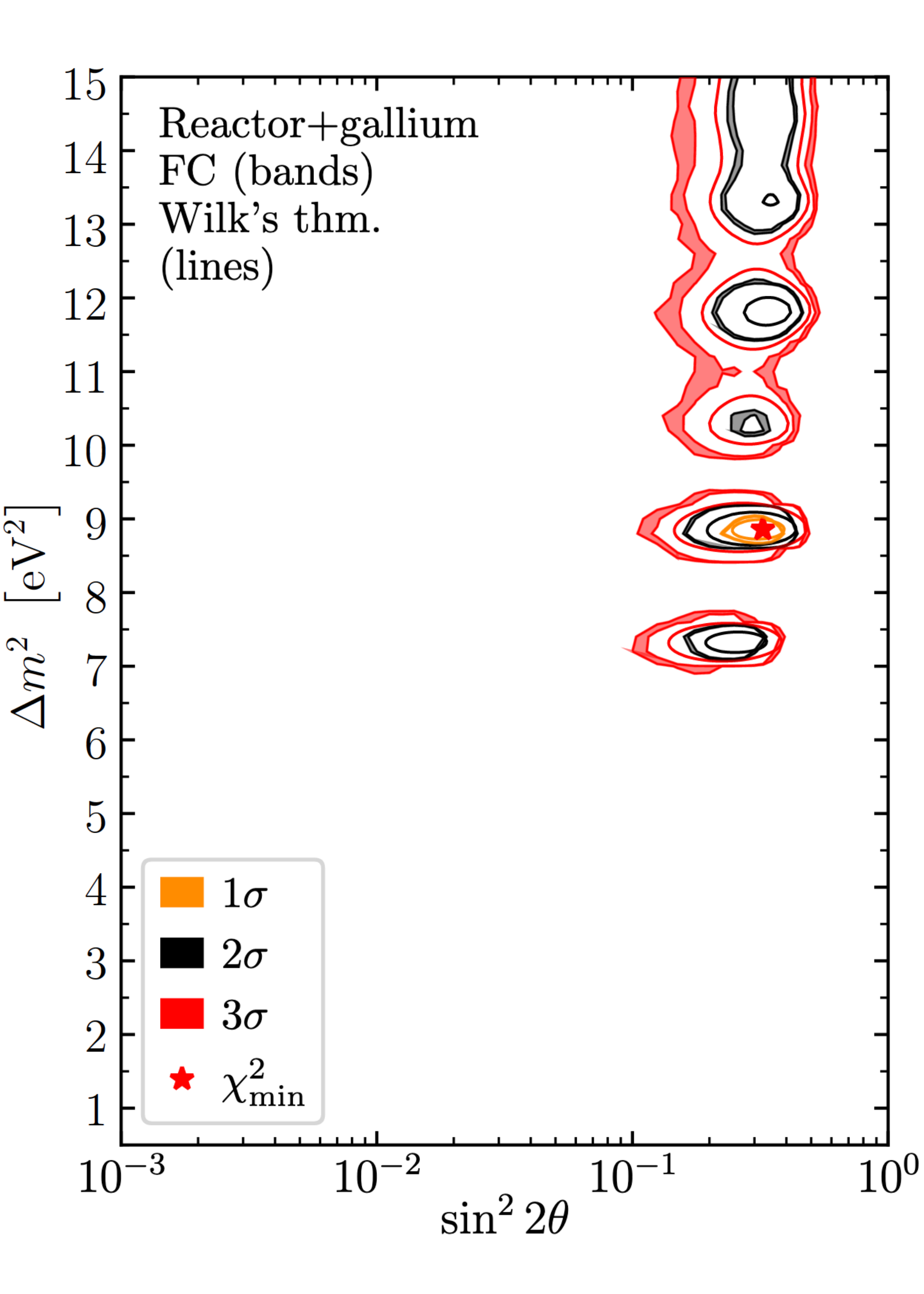}
\caption{Left: 2$\sigma$ CL Feldman-Cousins~(FC) exclusion curve of the combined reactor neutrino data. Also overlayed are the 2$\sigma$ exclusion FC curves from solar data, $\nu_{e}-^{12}$C scattering data from LSND and KARMEN, 95\% C.L. exclusion from the KATRIN experiment, and 2$\sigma$ gallium suggested region. Right: 1$\sigma$ , 2$\sigma$, and 3$\sigma$ CL FC suggested regions for a combination of gallium anomaly and relative reactor measurements. The suggested regions are at higher \dm{} mainly driven by the BEST and Neutrino-4 experiments. Figures adapted from Ref.\cite{Berryman:2021yan}.}
\label{fig:reactor_global_fits}
\end{figure}

\subsubsubsection{Joint Fits of Reactor Neutrino Experiments}
\label{sec:null_results_joint_rx_fits}
As discussed above, the conclusive way to test whether RAA is due to active-sterile mixing is by searching for sterile neutrino-induced spectral variations as a function of the baseline. 
With the exception of the Neutrino-4 experiment, no experiment has claimed to observe statistically significant hints of oscillations. 
Nonetheless, modest hints of oscillation in the other experiments have been reported, and it is worth considering the global fits of these datasets to determine and build a broader context from their individual results.  
A joint analysis~\cite{Berryman:2021yan} using Monte Carlo simulations with data from a combination of the short-baseline reactor neutrino experiments DANSS, NEOS, Neutrino-4, PROSPECT, and STEREO shows that the combination of these datasets is statistically compatible with the three neutrino model.
The exclusion curve from this joint fit is shown in Fig.~\ref{fig:reactor_global_fits}.

Although a combination of relative reactor neutrino datasets is incompatible with sterile neutrino-induced oscillations, they are in good agreement with the Gallium Anomaly.
This is primarily driven by the data from BEST and Neutrino-4 experiments which prefer oscillations with \dm{14}$> 5 \textrm{ eV}^2$--a region where other reactor experiments have minimal sensitivity.
Experiments with sensitivity in the \dm{14}$ > 5 \textrm{ eV}^2$ are needed to fully address these suggested regions. 
This could be achieved by a combination of upcoming reactor experiments~(see Sec.~\ref{sec:future_rx}) and $\beta$-decay experiments~(see Sec.~\ref{sec:null:numass}, \ref{sec:future_direct}).

\begin{figure}[!ht]
	\centering
  	\includegraphics[width=0.48\textwidth]{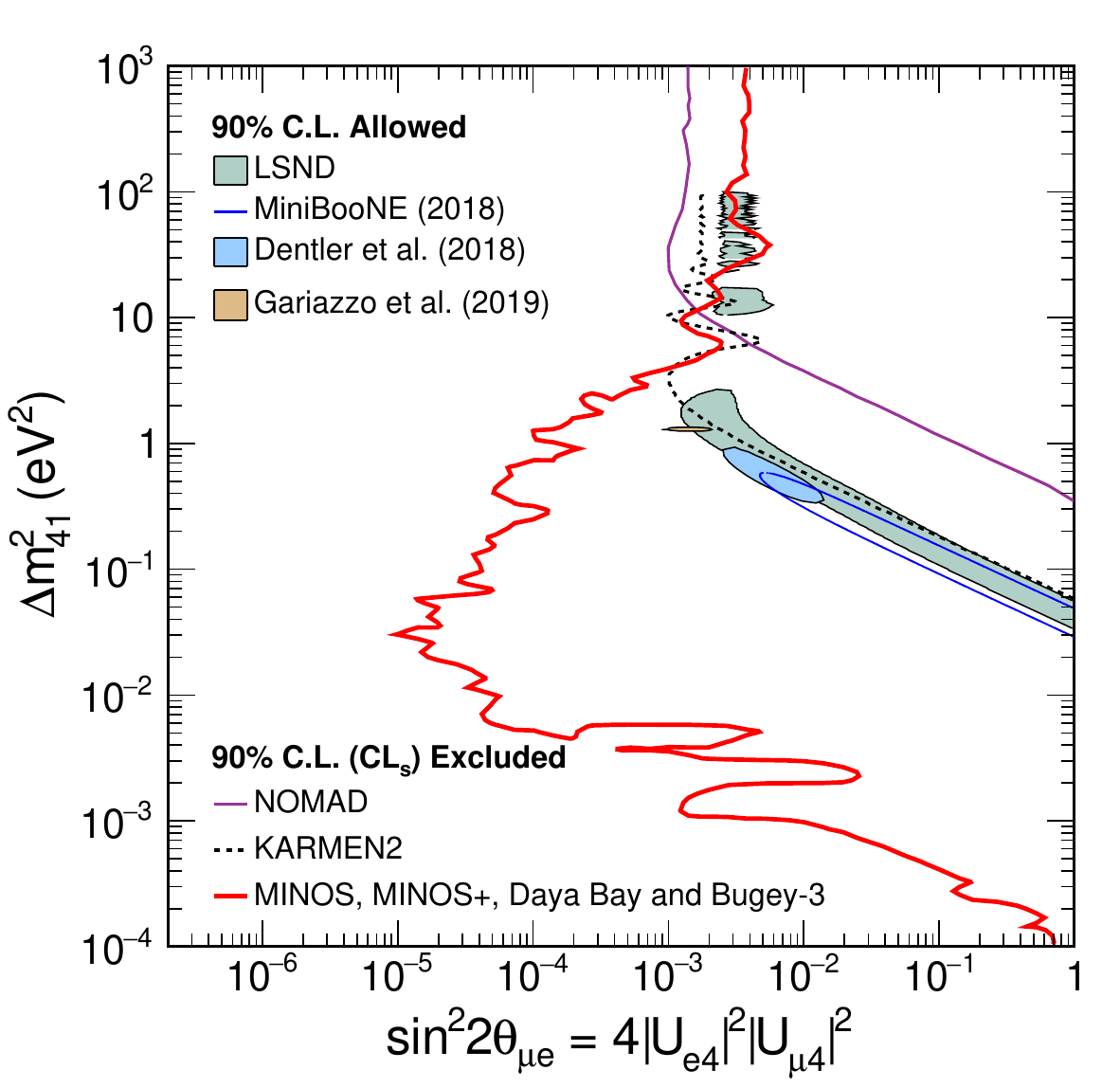}
    \includegraphics[width=0.48\textwidth]{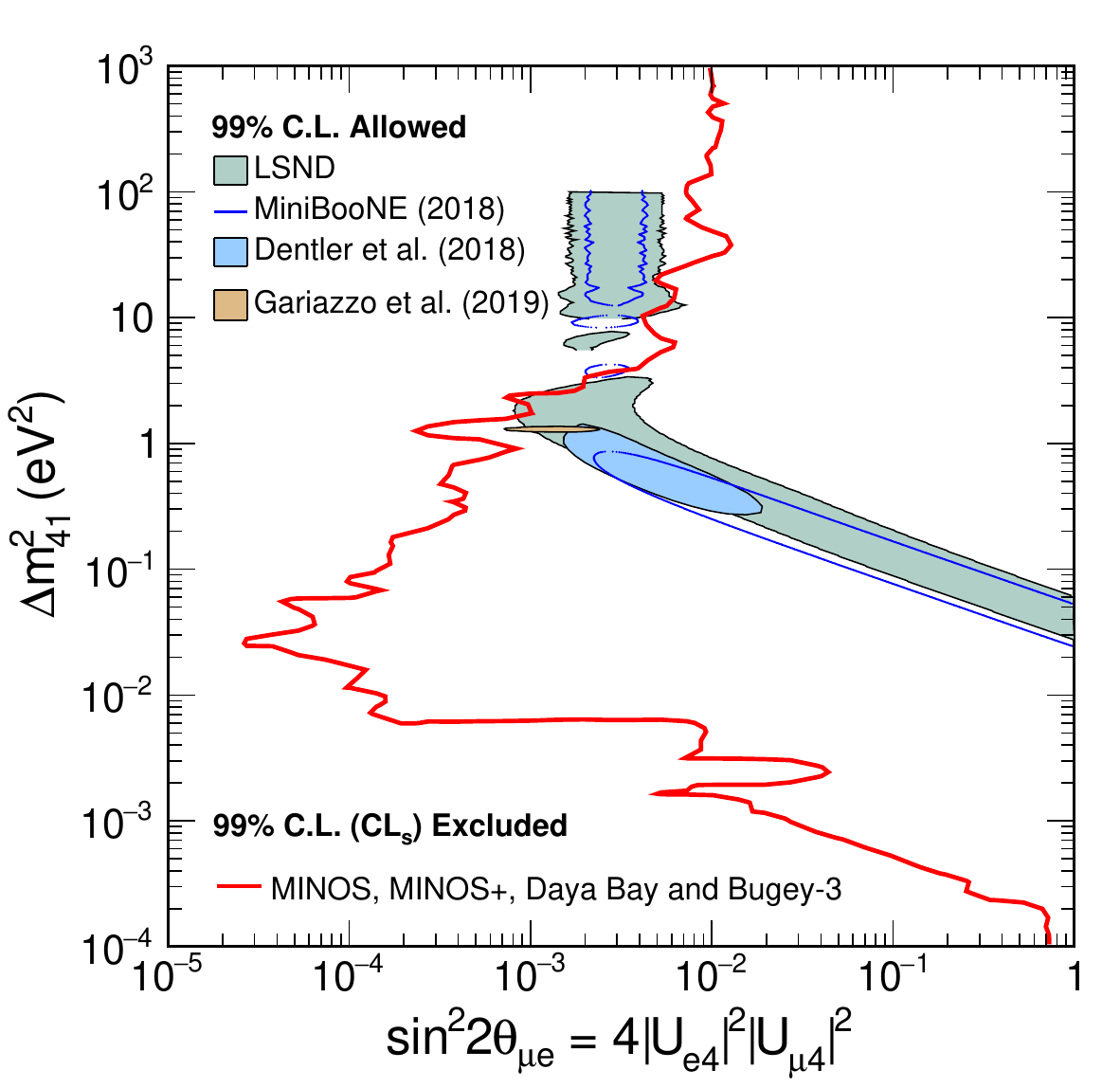}
	\caption{Comparison of the MINOS, MINOS+, Daya Bay, and Bugey-3 combined 90\% $\mathrm{CL_s}$ (left) and 99\%~$\mathrm{CL_s}$ (right) limits on ${\sin}^22{\theta}_{\mu e}$ to the LSND and MiniBooNE 90\% and 99\%~C.L. allowed regions, respectively. The limit also excludes the 90\% and 99\%~C.L. regions allowed by a fit to global data by Gariazzo {\it et al.} where MINOS, MINOS+, Daya Bay, and Bugey-3 are not included~\cite{Gariazzo:2017fdh,Gariazzo:2018mwd}, and the 90\% and 99\%~C.L. regions allowed by a fit to all available appearance data by Dentler {\it et al.}~\cite{Dentler:2018sju} updated with the 2018 MiniBooNE appearance results~\cite{Aguilar-Arevalo:2018gpe}. Figures from\cite{MINOS:2020iqj}.}
  \label{fig:dyb_minosplus}
\end{figure}

\subsubsubsection{Joint Analysis with Accelerator Experiments}\label{lbl_reactor_combo}
It is impossible for a single experiment to cover all the parameter space of interest to experimentalists and phenomenologists, which motivates the undertaking of joint analyses carried out by the members of the experimental collaborations that properly treat systematic uncertainties and their correlations. Of particular interest is the combination of Daya Bay's data with those of other reactor experiments operating at shorter baselines to cover a wide range of $|\Delta m^2_{41}|$ values. A case in point is the joint fit of the Daya Bay and Bugey-3 data presented in Ref.~\cite{MINOS:2020iqj} that results in the black contour of Fig.~\ref{fig:dyb}. Furthermore, powerful constraints on sterile-driven neutrino oscillations can also be extracted from combining data from reactor experiments with data from long-baseline accelerator experiments. In a 3+1 scenario, reactor experiments primarily measure $|U_{e4}|^2 = \,\,\sin^2\theta_{14}$ through electron antineutrino disappearance, while long-baseline accelerator experiments are typically most sensitive to $|U_{\mu4}|^2 = \,\,\sin^2\theta_{24}\cos^2\theta_{14}$ through measurements of muon (anti)neutrino disappearance. The product of the two matrix elements provides the amplitude of short-baseline electron neutrino appearance in a primary muon neutrino source:
\begin{equation}
P_{\overset{(-)}\nu\!\!_\mu\rightarrow\overset{(-)}\nu\!\!_e}^{SBL} = \,4|U_{e4}|^2|U_{\mu 4}|^2\sin^2\left(\frac{\Delta m^2_{41}L}{4E}\right), \label{eq:PmueU}
\end{equation}
where $4|U_{e4}|^2|U_{\mu4}|^2 = \sin^22\theta_{14}\sin^2\theta_{24} \equiv \sin^22\theta_{\mu e}$. 
Possibly the most representative examples of this type of combination are found in Refs.~\cite{DayaBay:2016lkk, MINOS:2020iqj}, where constraints on $\theta_{14}$ from the Daya Bay and Bugey-3 experiments are combined with constraints on $\theta_{24}$ from the MINOS/MINOS+ experiments to constrain the effective mixing parameter $\sin^{2} 2\theta_{\mu e}$. This work culminated in the most stringent constraints to date from disappearance searches on active to sterile neutrino oscillations and probed the parameter space allowed by the LSND and MiniBooNE anomalies, as shown in Fig.~\ref{fig:dyb_minosplus}. 

As highlighted in Ref.~\cite{bib:jointanaloi}, there are attractive opportunities in combining Daya Bay's data with other current and future experiments, such as PROSPECT, STEREO, NEOS, and JUNO-TAO. It is worth noting that the Daya Bay collaboration plans to publicly release its full data set once all final results have been released~\cite{bib:dybloisnowmass}, allowing such combinations to occur even well after the collaboration has dissolved. Similarly, following end of data taking in 2016, the MINOS/MINOS+ $\mathrm{CL}_s$ surfaces remain available for use in future combinations.

\subsubsection{Atmospheric Neutrino Experiments} 

\paragraph{IceCube} Neutrino telescopes, such as the IceCube Neutrino Observatory in the South Pole, play a unique role in searches for new physics associated with the short-baseline anomalies~\cite{IceCube:2020phf,IceCube:2020tka,IceCube:2017ivd,Aartsen:2017mnf}.
For atmospheric neutrino oscillation baselines, the $L/E$ range of the short-baseline anomalies corresponds to neutrino energies of order 100~GeV-1~TeV.
Because this closely matches the peak of the detected neutrino flux at IceCube, it is natural to expect any Lorentz invariant (LI, implying scaling as $L/E$) phenomenon connected with the short-baseline anomalies~\cite{Abazajian:2012ys} may be meaningfully testable at IceCube.  

Sensitivity to certain models is even further enhanced by fortuitous features of this energy range, notably: 1)  For sterile neutrinos with 0.1-10~eV$^2$ mass splittings, a resonant matter effect leads to dramatic enhancement of oscillations at $\sim$1~TeV, leading to sensitivity far exceeding that for vacuum-like oscillations~\cite{Esmaili:2013vza,Nunokawa:2003ep,Petcov:2016iiu,Barger:2011rc}; 2) Any new physics model that invokes non-LI effects scaling positively with $E$ will be enhanced at IceCube relative to all other experiments at lower $L/E$ - these include anomalous decoherence~\cite{Coloma:2018idr,Stuttard:2020qfv,Stuttard:2021uyw} and Lorentz violation~\cite{Abbasi:2010kx,Aartsen:2017ibm,Katori:2019xpc} models; 3) This energy range offers unique access to the $\nu_\tau$ appearance sector~\cite{Smithers:2021orb,Esmaili:2013vza}.
Other notable features of the IceCube event sample are its ``broadband'' nature, covering five decades of energy of atmospheric neutrinos (10~GeV-1~PeV) and 3.5 decades of baselines (20-12,750~km);  high statistical precision, owing to the high total exposure of a billion-ton detector operating stably for ten years collecting 70,000 atmospheric neutrino events per year, and well-controlled cross section uncertainties due to the predominance of deep-inelastic scattering interactions.
These features have enabled world-leading sensitivity to the parameters governing three-flavor oscillations as well as non-standard oscillation models including sterile neutrinos~\cite{IceCube:2020tka,IceCube:2020phf,IceCube:2017ivd,IceCube:2016rnb}, tests of low-energy manifestations of quantum gravity~\cite{IceCube:2017qyp,IceCube:2021tdn}, neutrino decay~\cite{Moss:2017pur}, and non-standard interactions~\cite{Salvado:2016uqu,IceCubeCollaboration:2021euf,IceCube:2022ubv}.

\begin{figure}[t]
    \centering
    \includegraphics[width=0.99\textwidth]{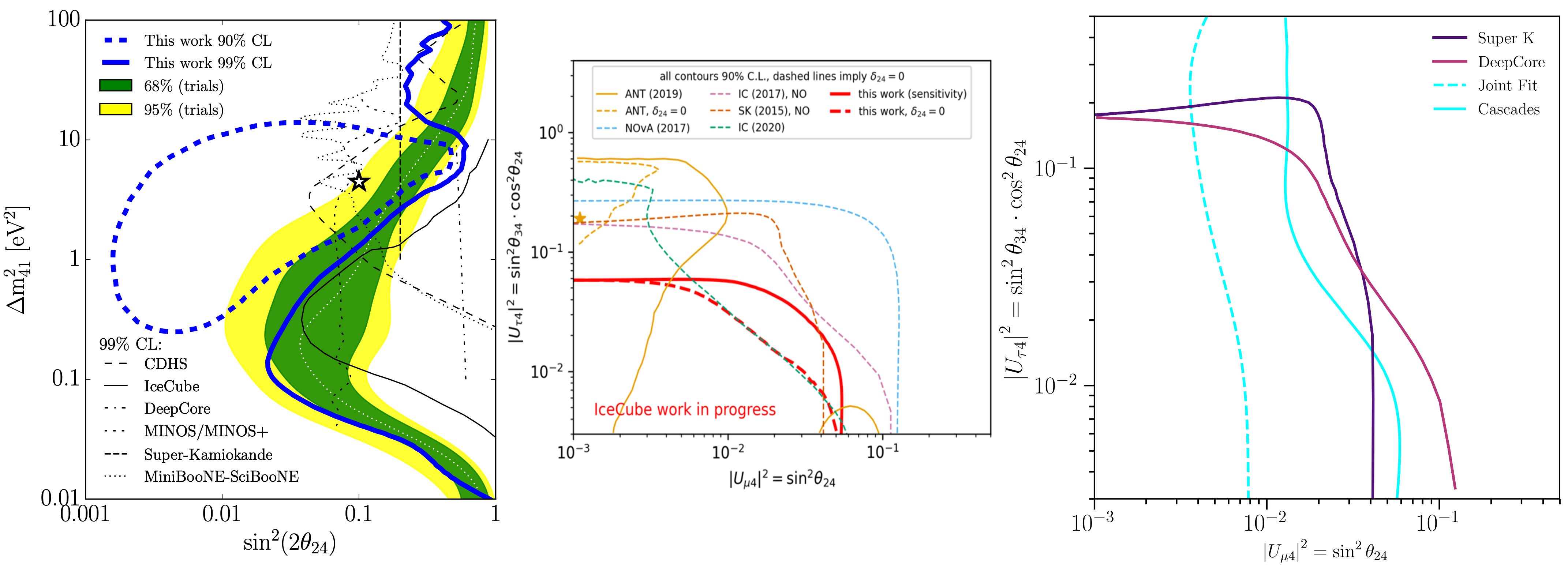}
    \caption{\label{fig:ICSterile} Left: Results from IceCube's high-energy muon-neutrino disappearance search~\cite{IceCube:2020tka}; Middle: results from IceCube low- and high-energy analyses~\cite{Trettin:2021bmn}; Right: expected sensitivity with cascades~\cite{Smithers:2021orb}.}
\end{figure}

\paragraph{Present generation light sterile neutrino searches at IceCube}
IceCube has made powerful sterile neutrino searches in both high ($\geq 400$ GeV) and  low ($\leq$ 60 GeV) energy ranges.
The former targets the matter resonance~\cite{Nunokawa:2003ep,Esmaili:2013vza} expected for $\Delta m^2\sim$ ${\cal O}$(1 eV$^2$) splittings, and is one of the world's most sensitive in the $\nu_\mu$ disappearance channel at eV$^2$-scale mass splittings.
The latest generation analysis~\cite{IceCube:2020tka,IceCube:2020phf} uses a sample of 305735 reconstructed $\nu_\mu$ events and excludes mixing angles down to $\sin^2 2\theta_{24} \sim$ 0.02 at $\Delta m^2\sim 0.2$ at 99\% CL.
At 90\% CL the analysis yields a closed contour that may be interpreted as a statically weak hint of a signal, with a best-fit point at $\sin^2 2\theta_{24}$ = 0.10 and $\Delta m^2_{41}$ = 4.5~eV$^2$.
This result is consistent with the no sterile neutrino hypothesis with a p-value of 8.0\%.  The 90\%  CL contour is shown in Fig.~\ref{fig:ICSterile}, left.

At low energy, sterile neutrino mixing within an extended neutral lepton mixing matrix enhances the standard oscillation probability proportionally to the matter column density traversed~\cite{Razzaque:2012tp}.
The effect is approximately independent of $\Delta m^2$, as oscillation cycles are irresolvable within detector energy resolution.
IceCube has tested for this effect using a multi-flavor sample over an energy range of 10-60~GeV~\cite{IceCube:2017ivd}, with the strongest effect expected at an energy of $\sim$20~GeV for upgoing muons.
The analysis yielded no evidence of anomalous oscillations, setting a limit on the extended PMNS matrix elements $|U_{\mu 4}|^2=\sin^2 \theta_{24}$ and $|U_{\tau 4}|^2=\sin^2 \theta_{34}\cos^2 \theta_{34}$, marked {\tt IC2017(NO)} in Fig.~\ref{fig:ICSterile}, center.

\paragraph{Next generation sterile neutrino searches at IceCube}

The sterile neutrino sensitivity at IceCube has yet to be exhausted, with near-term improvements expected from event samples already in hand.
At low energies, a sterile neutrino search using the full ten year dataset of $\geq$300,000 events with $E_{\nu}\leq 100$~GeV spanning all flavors is underway~\cite{Trettin:2021bmn}, promising unique sensitivity in the  $|U_{\mu 4}|^2$, $|U_{\tau 4}|^2$ plane (Fig.~\ref{fig:ICSterile}, center).
At higher energies ($\geq$ 400~GeV), attention to date has been focused on searches for $\nu_\mu$ disappearance through non-zero $\theta_{24}$, motivated by the necessity of its finite value if sterile neutrino osculations were responsible for the short-baseline ($\nu_\mu\rightarrow\nu_e) / (\bar{\nu}_\mu\rightarrow\bar{\nu}_e $) anomalies ($\theta_{14}= \theta_{34}=0$ leads to the most conservative limits on $\theta_{24}$~\cite{Lindner:2015iaa}).
Efforts are now underway to incorporate the high energy cascade event sample into these analyses, which includes topologies associated with $\nu_e$ and $\nu_\tau$ CC and all flavor NC interactions.
This extension promises sensitivity to both $\nu_e$ and $\nu_\tau$ appearance signatures associated with non-zero $\theta_{14}$ and $\theta_{34}$, respectively.
Preliminary studies~\cite{Smithers:2021orb} of the sensitivity in this channel suggest that  $\nu_\tau$ appearance signatures are discoverable for values of  $\theta_{34}$ consistent with  world data and IceCube's existing $\theta_{24}$ limits; and that values of $\theta_{14}$ consistent with reactor~\cite{Mention:2011rk} and gallium/BEST~\cite{Barinov:2021asz,Kostensalo:2019vmv,Gavrin:2010qj} anomalies may yield observable $\nu_e$ appearance signatures.
The expected sensitivity of the combined high energy $\nu_\mu$ disappearance and cascade appearance signatures is shown in Fig.~\ref{fig:ICSterile}, right.
Augmentations of the sterile neutrino searches using machine learning techniques, starting-event topologies, and improved reconstruction methods are also underway, expected to provide continuing improvements over the coming Snowmass period.

\paragraph{Super-K} Super-Kamiokande (Super-K, SK) has performed a search for light sterile neutrinos using approximately 4,000 live-days of atmospheric neutrino data~\cite{Super-Kamiokande:2014ndf}. 
The SK analysis focuses on light sterile neutrinos with mass-squared differences greater than $\SI{0.1}\eV^2$.
In the energy range of the Super-K analysis, predominantly below 10~GeV, such a large mass-square difference implies that the oscillation effects due to light sterile neutrinos are averaged out.
Thus, Super-K analysis is insensitive to the mass-square difference, but only to the mixing elements.
In the Super-K analysis, they choose to constraint the mixing elements: $|U_{\tau 4}|$ and $|U_{\mu 4}|$.
In principle, Super-K has also sensitivity to $|U_{e 4}|$, but this is a subleading effect.
Additionally, the Super-K also is sensitive to the additional CP-violating phases that appear in the presence of a sterile neutrino.
The effect of these CP-violating phases is subleading but does not always yield conservative results on the mixings and thus the results should be taken with this caveat~\cite{Esmaili:2013vza}.

\begin{figure}[t]
    \centering
    \includegraphics[width=0.5\textwidth]{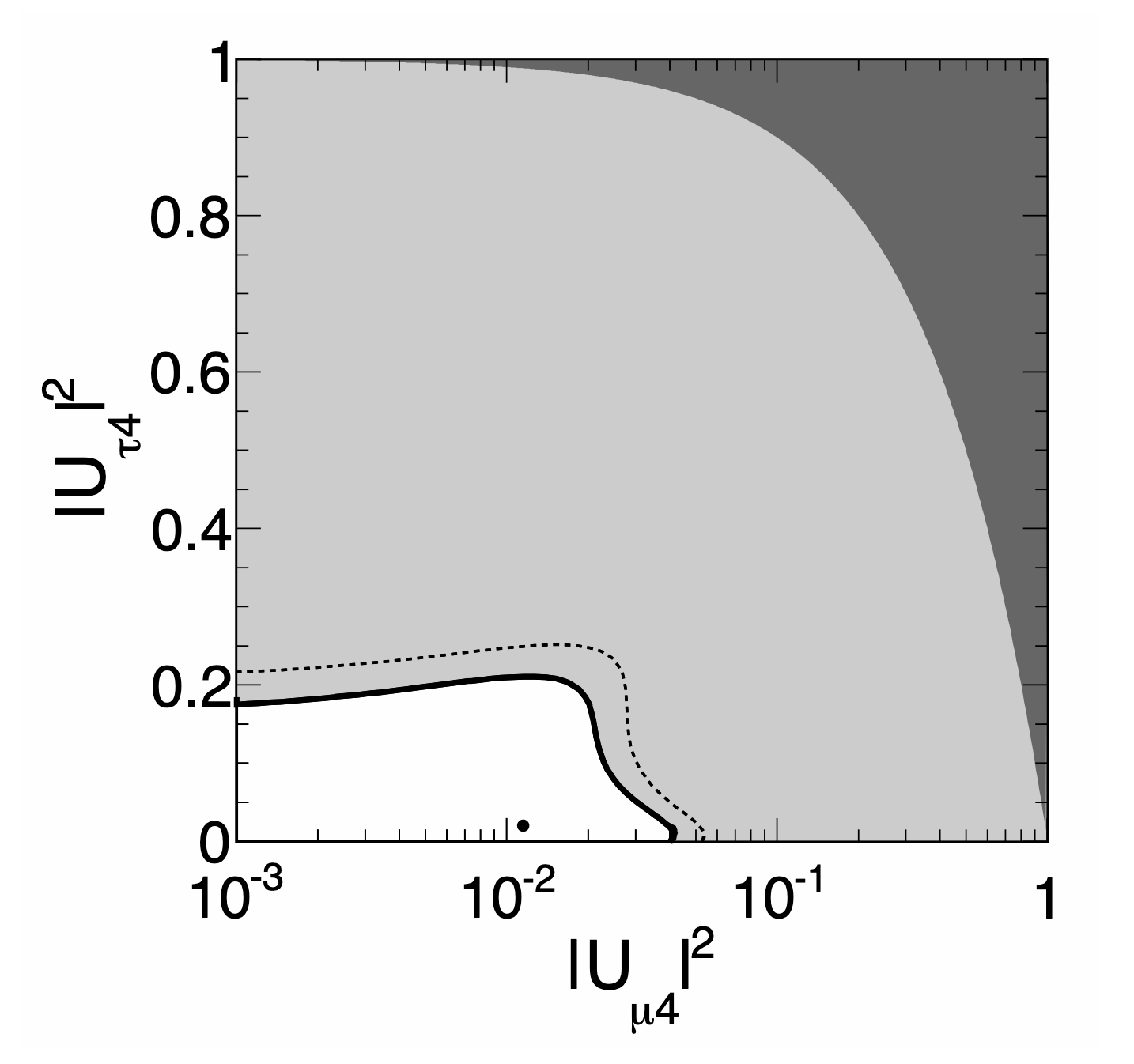}
    \caption{\label{fig:sk_atm_results}Constraint on $|U_{\tau 4}|$ and $|U_{\mu 4}|$ for $\Delta m^2_{41} > \SI{0.1}\eV^2$ obtained by Super-K. Figure from~\cite{Super-Kamiokande:2014ndf}.}
\end{figure}

The Super-K analysis found no significant evidence of a light sterile neutrino and reported constraints on $|U_{\tau 4}|$ and $|U_{\mu 4}|$ for $\Delta m^2_{41} > \SI{0.1}\eV^2$; concretely they limit $|U_{\mu 4}|$ to less than 0.041 and $|U_{\tau 4}|$ to be less than 0.18 at 90\% C.L.
These results are shown in Fig.~\ref{fig:sk_atm_results} and have been superseded by constraints from ANTARES and IceCube; compare to Fig.~\ref{fig:ICSterile}.

\paragraph{ANTARES} 
The ANTARES neutrino telescope~\cite{ANTARES:2011hfw}, a sub-gigaton-scale neutrino telescope, had been designed and optimized for the exploration of the high-energy universe by using neutrinos as cosmic probes. However, its energy threshold of about 20\,GeV was sufficiently low to be sensitive to the first atmospheric oscillation minimum. The majority of the neutrino events have been recorded with energies between 100~GeV and few TeV, an energy range rich in signatures of eV-scale sterile neutrinos. 
\begin{figure}[htbp]
\centering
\includegraphics[width=0.8\linewidth]{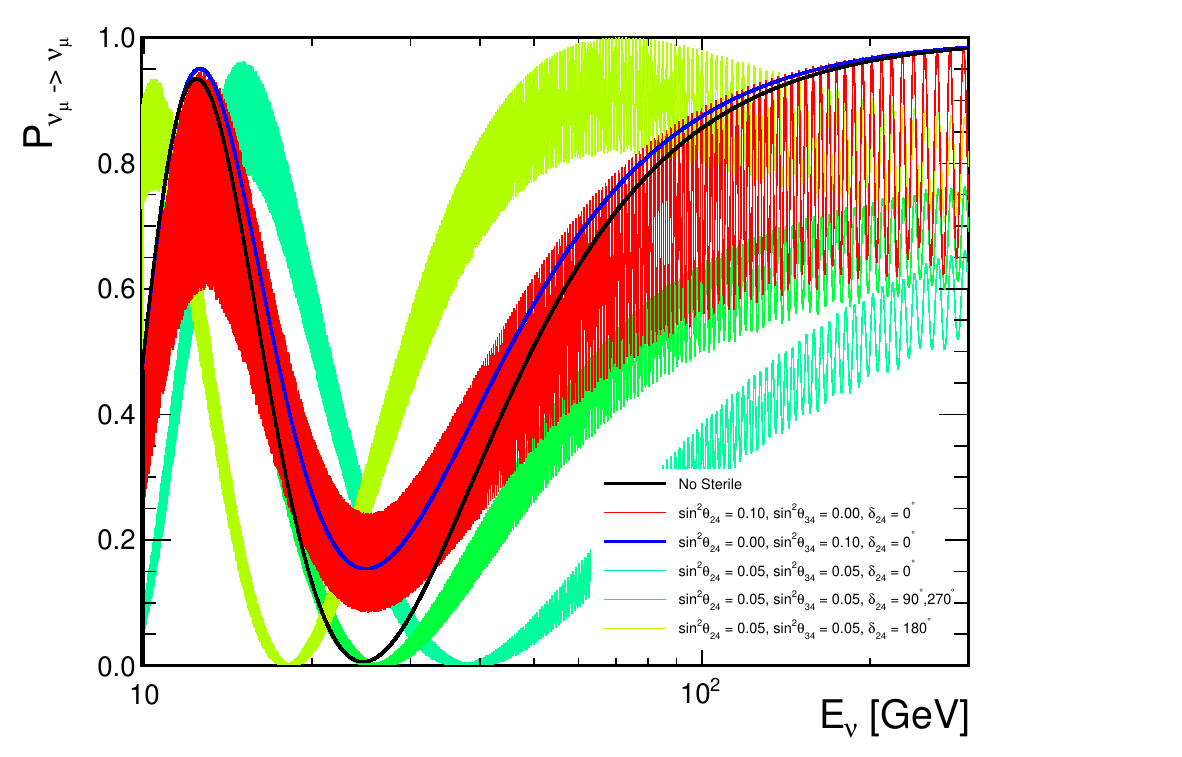}
\caption{Survival probability of vertically up-going $\nu_\mu$ at ANTARES as a function of neutrino energy for different values of mixing angles $\theta_{24}, \theta_{34}$ and $\delta_{24}$ with $\Delta m^2_{41}=0.5$~eV$^2$, \textcolor{black}{$\Delta m^2_{31}=2.5\cdot 10^{-3}$~eV$^2$ and $\sin^22\theta_{23}=1$.} Figure taken from~\cite{ANTARES:2018rtf}.}
\label{fig:Ant_Sterile}
\end{figure}
This is illustrated in \cref{fig:Ant_Sterile} which shows the $\nu_{\mu}$ survival probability for maximal mixing of $\theta_{23}$ and different combinations of the mixing parameters $\theta_{24},\theta_{34}$ and $\delta_{24}$ with
\begin{eqnarray}
U_{\mu 4}  &=& e^{-i\delta_{24}}\sin\theta_{24},\\ 
U_{\tau 4} &=& \sin\theta_{34}\cos\theta_{24}. 
\label{Umu4}
\end{eqnarray}
The fast oscillations due to $\Delta m^2_{41} \gtrsim 0.5$\,eV$^2$ are unobservable due to the limited energy resolution of the detector, making $\Delta m^2_{41}$ not measurable. 

\begin{figure}[htbp]
\centering
\includegraphics[width=\linewidth]{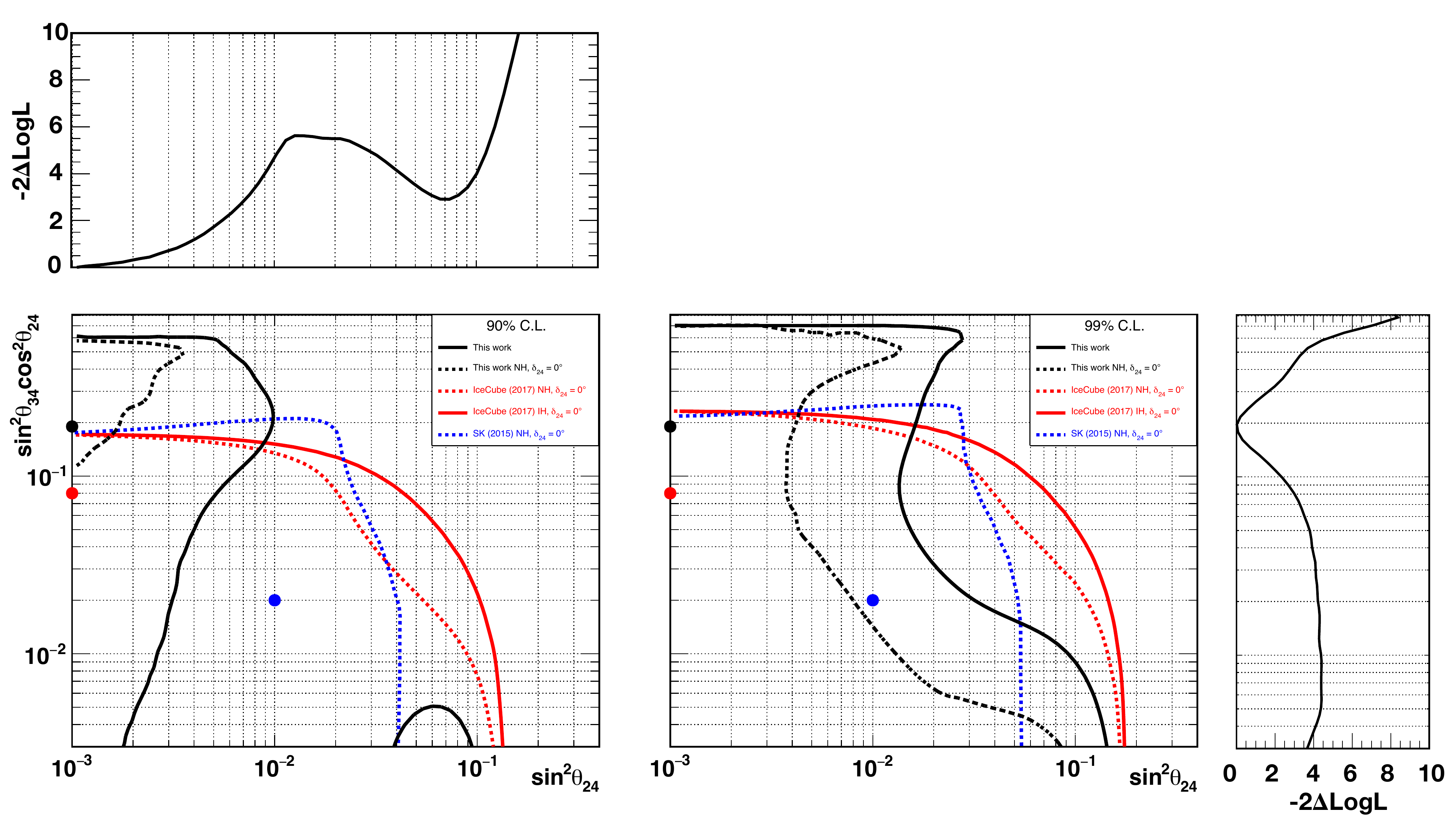}
\caption{90\% (left) and 99\% (right) CL limits for the 3+1 neutrino model in the parameter plane of $|U_{\mu 4}|^2=\sin^2\theta_{24}$ and $|U_{\tau 4}|^2=\sin^2\theta_{34}\cos^2\theta_{24}$ obtained by ANTARES (black lines), and compared to the ones published by IceCube/DeepCore~\cite{IceCube:2017ivd} (red) and Super-Kamiokande~\cite{Abe:2014gda} (blue). \textcolor{black}{The dashed lines are obtained for NH and $\delta_{24}=0^\circ$ while the solid lines are for an unconstrained $\delta_{24}$ (ANTARES) or for IH and $\delta_{24}=0^\circ$ (IceCube/Deepcore) respectively.} The colored markers indicate the best-fit values for each experiment. 
The 1D projections after profiling over the other variable are also shown for the result of this work. Figure taken from~\cite{ANTARES:2018rtf}.}
\label{fig:Ant_Results}
\end{figure}

The ANTARES neutrino telescope was located in the Mediterranean Sea, 40\,km off the coast of Toulon, France, at a mooring depth of about 2475\,m. The detector was completed in 2008 and took data until February 2022. ANTARES was composed of 12 detection lines, instrumenting a water volume of about 15 Mtons. ANTARES data collected from 2007 to 2016 with a total detector lifetime of 2830 days have been used to constrain $U_{\mu 4}$ and $U_{\tau 4}$~\cite{ANTARES:2018rtf}. A total of 7710 low-energetic atmospheric neutrino candidate events have been selected, largely dominated by muon-(anti)-neutrino charge-current events, identified thanks to a long-range up-going muon track. Particular attention was paid to consistent handling of the complex phase $\delta_{24}$ in conjunction with the neutrino mass ordering. The limits from the ANTARES analysis are shown in \cref{fig:Ant_Results}. As expected from Fig.~\ref{fig:Ant_Sterile} ANTARES is particularly sensitive if both $U_{\mu 4}$ and $U_{\tau 4}$ are non-zero and improves existing  limits from \cite{IceCube:2017ivd,Abe:2014gda} substantially.

\textcolor{black}{After profiling over the other variable,} the following limits on the two matrix elements can be derived:
\begin{eqnarray}
|U_{\mu 4}|^2  &<& 0.007~(0.13)~ \mathrm{at}~ 90\%~(99\%)~\mathrm{CL},\\ 
|U_{\tau 4}|^2 &<& 0.40~(0.68)~ \mathrm{at}~ 90\%~(99\%)~\mathrm{CL}. 
\end{eqnarray}

\subsubsection{Radioactive Source Experiments}
\label{sec:BExpST}
\hfill\\

{\bf BEST} The Baksan Experiment on Sterile Transitions (BEST) \cite{Gavrin:2010qj} was proposed to probe the possibility of the short-baseline electron neutrino disappearance using the same process studied by GALLEX and SAGE, Eq.~(\ref{eq:nueGa}). The disappearance of neutrinos was suggested as an explanation of the deficit in observed events measured by the previous radioactive source experiments (see Sec.~\ref{sec:GaAnomaly}).

\begin{figure}[ht!]
    \centering
    \includegraphics[width=0.35\textwidth]{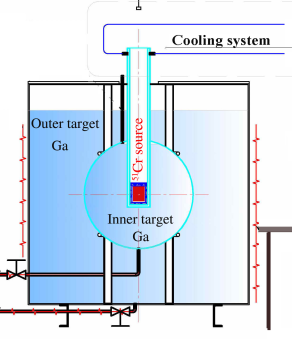}
    \caption{BEST detector configuration.  Two Ga target volumes detect neutrino interactions from a $^{51}$Cr source.  Figure adapted from~\cite{Barinov:2021asz}.}
    \label{fig:BEST_det}
\end{figure}

Although the main experimental idea is the same, the BEST configuration provides for simultaneous measurements at two different baselines. In this case, a spherical vessel (\textit{inner target}) is located inside a cylindrical container (\textit{outer target}), both filled with liquid gallium (the detector). The radioactive source ($^{51}\rm{Cr}$) is placed inside the sphere, as shown in Fig.~\ref{fig:BEST_det}.

Here again, the emitted neutrinos interact with the detector through reaction Eq.~(\ref{eq:nueGa}), and the produced $^{71}\rm{Ge}$ atoms are extracted and counted for each vessel separately \cite{Barinov:2016znv}. The ratio of the measured rate of $^{71}\rm{Ge}$ production at each distance to the expected one considering the cross section and experimental efficiencies are presented in Tab.~\ref{tab:BEST_Rate} \cite{Barinov:2022wfh}.
\begin{table}[t]
    \centering
    \begin{tabular}{|c|c|}\hline
    Inner                  & Outer                  \\\hline
    $0.79\pm 0.05$ & $0.77\pm 0.05$\\\hline
    \end{tabular}
    \caption{Ratio of observed to predicted $^{71}\rm{Ge}$ event rates as measured by BEST using $^{51}\rm{Cr}$ with its inner and outer targets.}
    \label{tab:BEST_Rate}
\end{table}
Recall that the cross section of the reaction in Eq.~(\ref{eq:nueGa}) is used in the calculation of these ratios, and BEST used the Bahcall results~\cite{Bahcall:1997eg}, including conservative uncertainties~\cite{Barinov:2021asz}. The ratios are $4.2\sigma$ and $4.8\sigma$ less than unity, respectively, supporting the gallium anomaly observed by other experiments.

\begin{figure}[ht]
    \centering
    \includegraphics[width=0.4\textwidth]{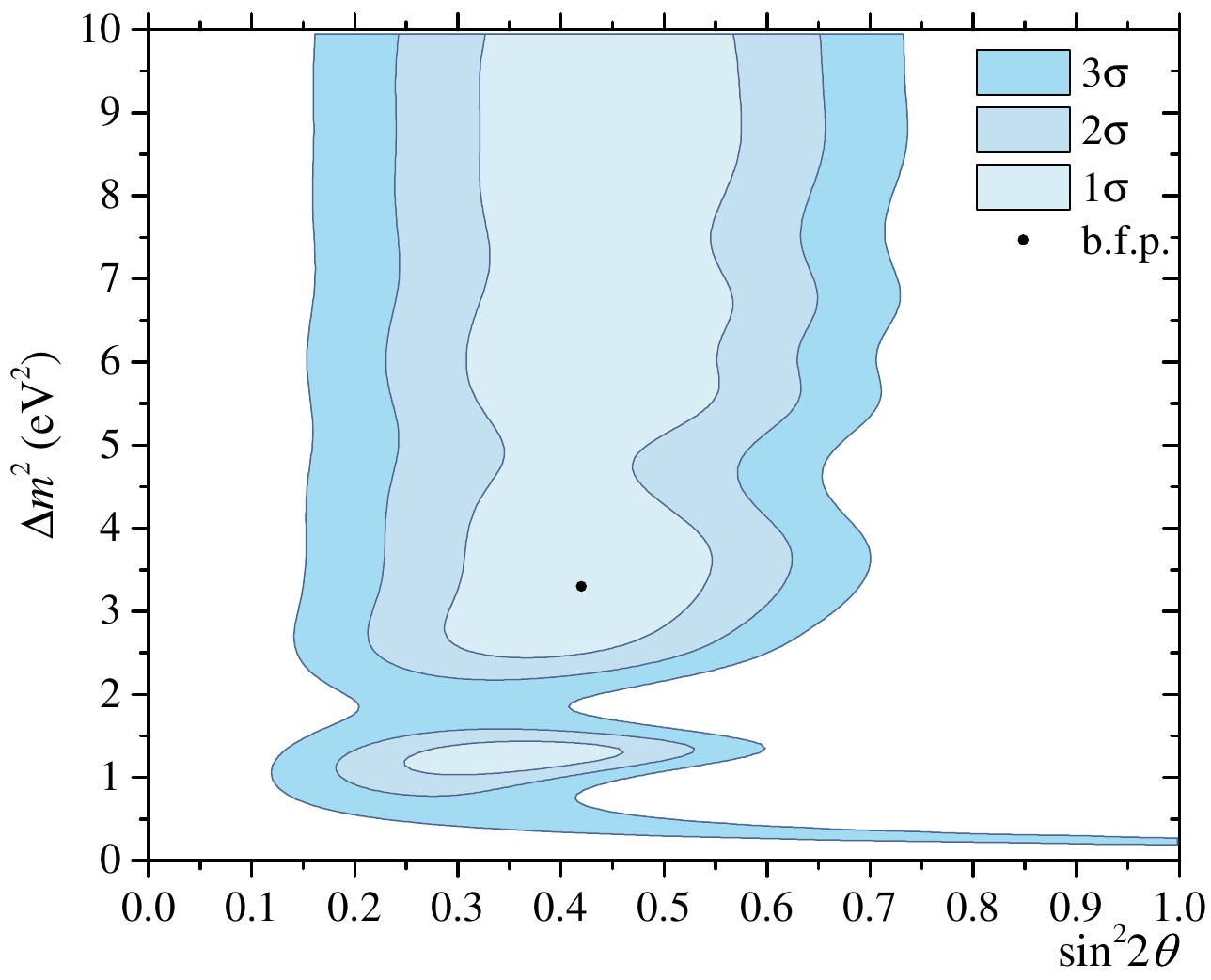}
    \includegraphics[width=0.4\textwidth]{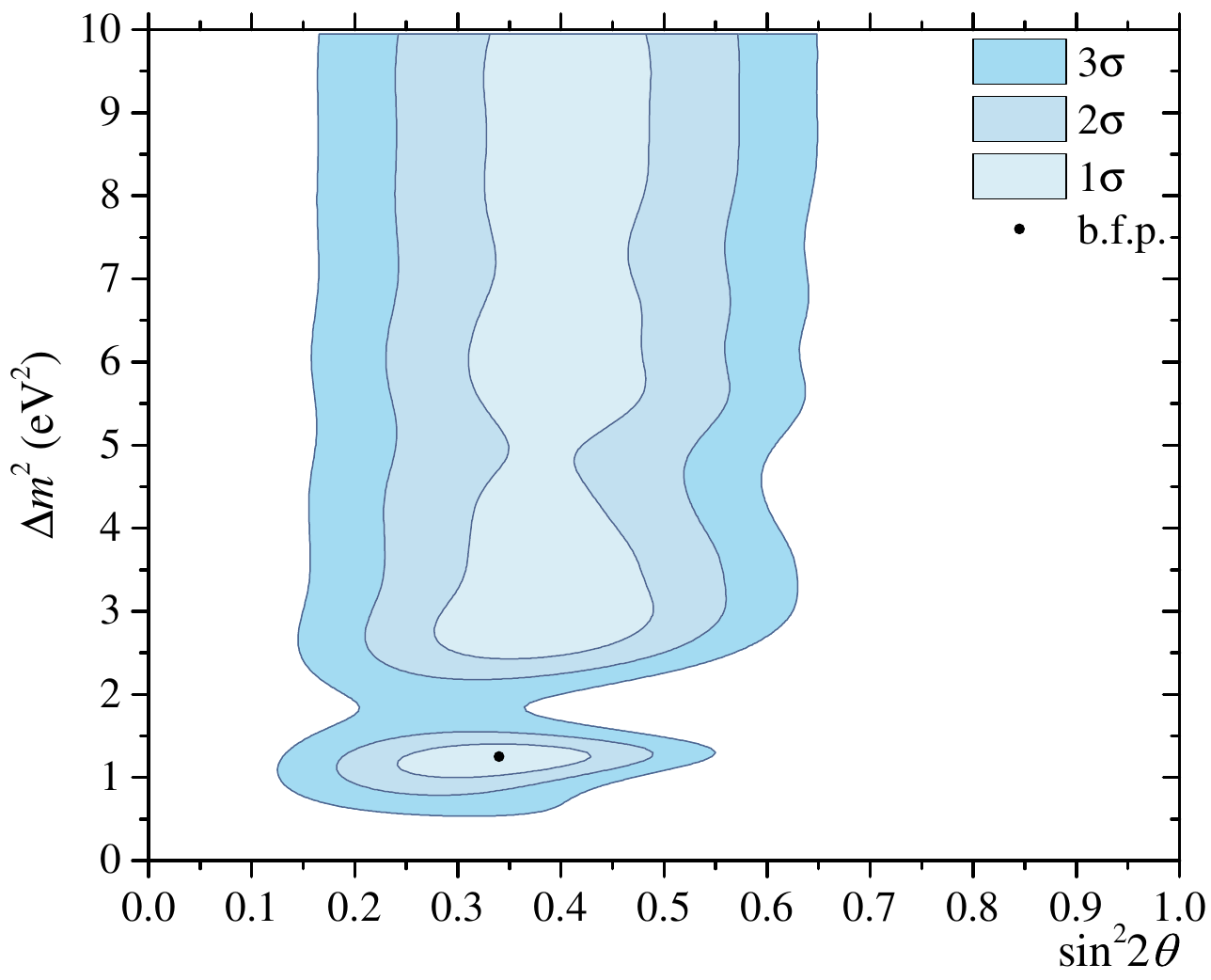}
    \caption{Allowed regions obtained in the $\sin^22\theta$--$\Delta{m}^2_{41}$ parameter space from the analysis of the BEST results only (left) and from the BEST results combined with results from GALLEX and SAGE (right). Figures taken from~\cite{Barinov:2022wfh}. Note that here $\sin^22\theta  = 4\left|U_{e4}\right|^2\left(1 - \left|U_{e4}\right|^2\right)$.}
    \label{fig:GaAnomaly_contours_BEST}
\end{figure}

BEST performed an analysis of these results in the framework of short-baseline electron neutrinos disappearance due to neutrino oscillation governed by the survival probability in Eq.~(\ref{eq:nueSurvP}), using $\sin^22\theta  = 4\left|U_{e4}\right|^2\left(1 - \left|U_{e4}\right|^2\right)$. The study leads to the allowed regions shown in Fig.~\ref{fig:GaAnomaly_contours_BEST}, corresponding to $1\sigma, \,2\sigma$, and $3\sigma$ confidence levels (left plot), with the best-fit at ($\sin^22\theta $,  $\Delta m^2) = (0.42, 3.3$~eV$^2$).
When combined with the results from GALLEX and SAGE (Tab.~\ref{tab:Ge71_Rate}), the allowed regions are the ones illustrated in the right plot of Fig.~\ref{fig:GaAnomaly_contours_BEST}, where correlated cross section uncertainties were considered. In this case, the best fit is located at $(\sin^22\theta, \Delta m^2) = (0.34, 1.25$~eV$^2)$ \cite{Barinov:2022wfh}.

Some of the calculations of the cross section of the process Eq.~(\ref{eq:nueGa}) appear to decrease the significance of the gallium anomaly related to the GALLEX and SAGE experiments (Sec.~\ref{sec:GaAnomaly}); nonetheless, the results are still consistent with a possible short-baseline electron neutrino disappearance produced by active to sterile neutrino oscillations. On the other hand, the measurements by BEST are in agreement with the former source experiments and confirm the gallium anomaly with larger significance.

\subsubsection{Beta Spectrum Searches}
\label{sec:null:numass}

High precision beta spectroscopy enables searches for sterile neutrinos from the sub-eV- up to the MeV-scale.
Exploiting the kinematics of the weak process, these experiments offer a clean probe of the coupling of the electron flavor neutrino to the different mass states.
The electron flavor neutrino emitted in $\beta$-decay does not have a well-defined mass but is rather an admixture of the neutrino mass eigenstates.
The existence of a hypothetical sterile neutrino(s) implies the electron flavor neutrino may also contain a small admixture of (at least) a new fourth neutrino mass eigenstate, $m_4$.

The $\beta$-decay spectrum, $R_\beta(E)$ will be altered:
\begin{equation}
R_\beta(E) = \cos^2\theta_s \, R_{\beta}(E, m_{\beta}^2) + \sin^2\theta_s \, R_\beta(E,  m_4^2) \,
\label{eq:sterile}
\end{equation}
composed of both the spectrum corresponding to the traditional electron-weighted neutrino mass $m_\beta$ (from the three active neutrinos) and a spectrum associated with $m_4$.
The maximal energy of each spectrum contribution is $E_0 - m_i$, with the largest neutrino mass resulting in the lowest endpoint energy.
The amplitudes of the two decay branches are given by $\cos^2\theta_s$ and $\sin^2\theta_s$, respectively, where $\theta_s$ is the active-to-sterile mixing angle.
The resulting signature of a sterile neutrino is thus a kink-like distortion of the measured spectrum at an energy of $E_0 - m_4$.

Spectral measurements across a broad energy range therefore enable searches for kink features and thus sterile neutrinos.
Although not an L/E oscillation signature, detection would be an unambiguous detection of sterile neutrinos subject to drastically different systematics.
As a complementary probe, these searches exclusively probe the sterile neutrino hypothesis, and not other physics explanations of the existing anomalies.
For eV-scale sterile neutrinos, this signature naturally appears in the region of interest for direct neutrino mass measurements studying the beta endpoint.

\subsubsubsection{Historical Context}

\begin{figure}[t]
\begin{center}
\includegraphics[width=0.6\textwidth]{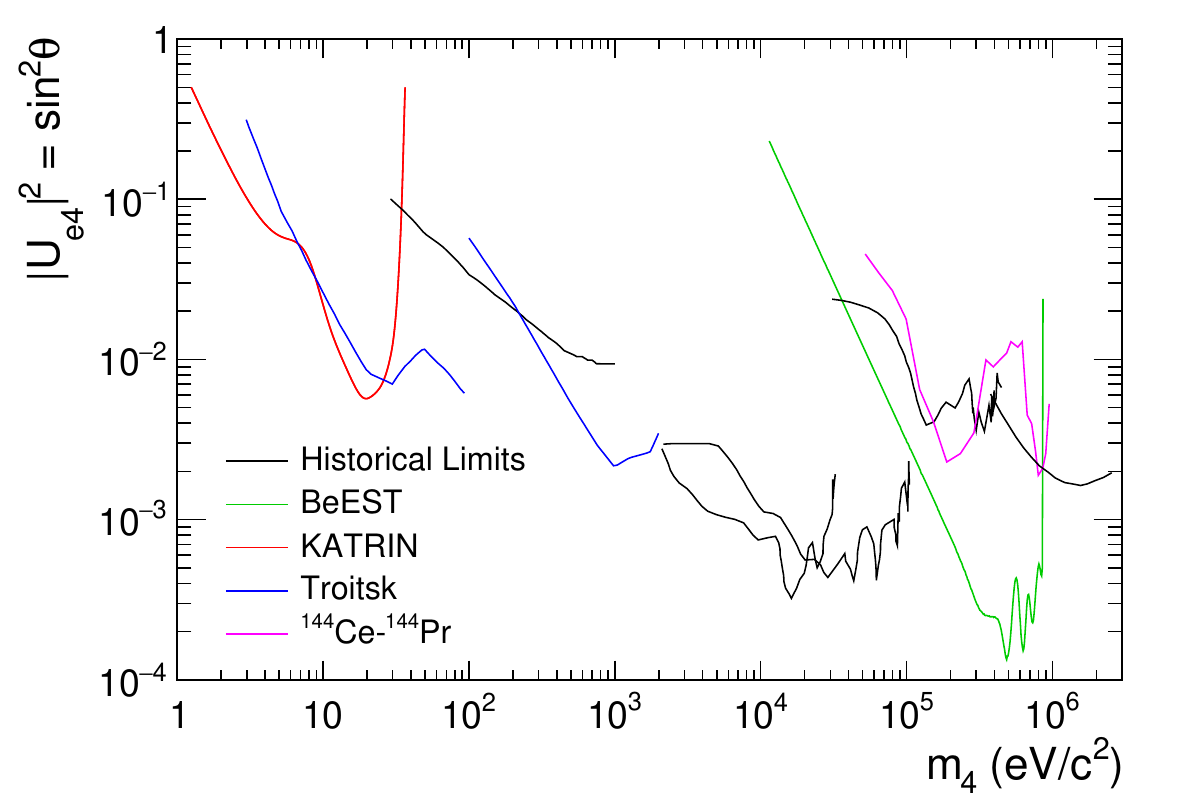}
\caption{\label{fig:null:beta_history} Landscape of historical limits on sterile neutrino from beta spectrum searches.  Results from the last decade are highlighted in color and labeled.}
\end{center}
\end{figure}

The flexibility of the beta spectrum method has led to its use in placing strong constraints on sterile neutrinos.
These limits span more than six decades in $m_4$ -- from 1.5\,eV up to 2.5\,MeV -- and down to $\sin^2 \theta \sim 10^{-4}$ (see Fig.~\ref{fig:null:beta_history}).
Isotopes placing the strongest constraints across this energy range include
$^{3}$H~\cite{Belesev:2013cba,Abdurashitov:2017kka,KATRIN:2022ith,KATRIN:2022spi}, $^{187}$Re~\cite{Galeazzi:2001py}, $^{63}$Ni~\cite{Holzschuh:1999vy}, $^{35}$S~\cite{Holzschuh:2000nj}, $^{64}$Cu~\cite{Schreckenbach:1983cg}, $^{144}$Ce-$^{144}$Pr~\cite{Derbin:2018dbu}, $^{7}$Be~\cite{Friedrich:2020nze}, and $^{20}$F~\cite{Deutsch:1990ut}.

Tritium endpoint searches for direct neutrino mass study have enjoyed great success, leveraging the sophisticated detector and source development for that science.
Both KATRIN~\cite{KATRIN:2022ith,KATRIN:2022spi} and Troitsk~\cite{Belesev:2013cba,Abdurashitov:2017kka} have derived eV-scale limits from the primary neutrino mass physics data.
Additionally, dedicated searches away from the endpoint open up a wider range to keV masses.
The progress of the ongoing KATRIN experiment for eV steriles is described in detail below (Sec.~\ref{sec:null:katrin}), with discussion of sensitivity to keV steriles with an upgraded detector later (Sec.~\ref{sec:future:TRISTAN}).
At higher mass, the BeEST experiment has obtained the strongest limits to steriles up to 0.85\,MeV based on their Phase II prototype~\cite{Friedrich:2020nze}, heralding the advent of new technologies with significant improvements in sensitivity; BeEST is discussed later in the context of their full sensitivity (Sec.~\ref{sec:future:BeEST}).

\begin{figure}[ht!]
\begin{center}
\includegraphics[width=0.5\textwidth]{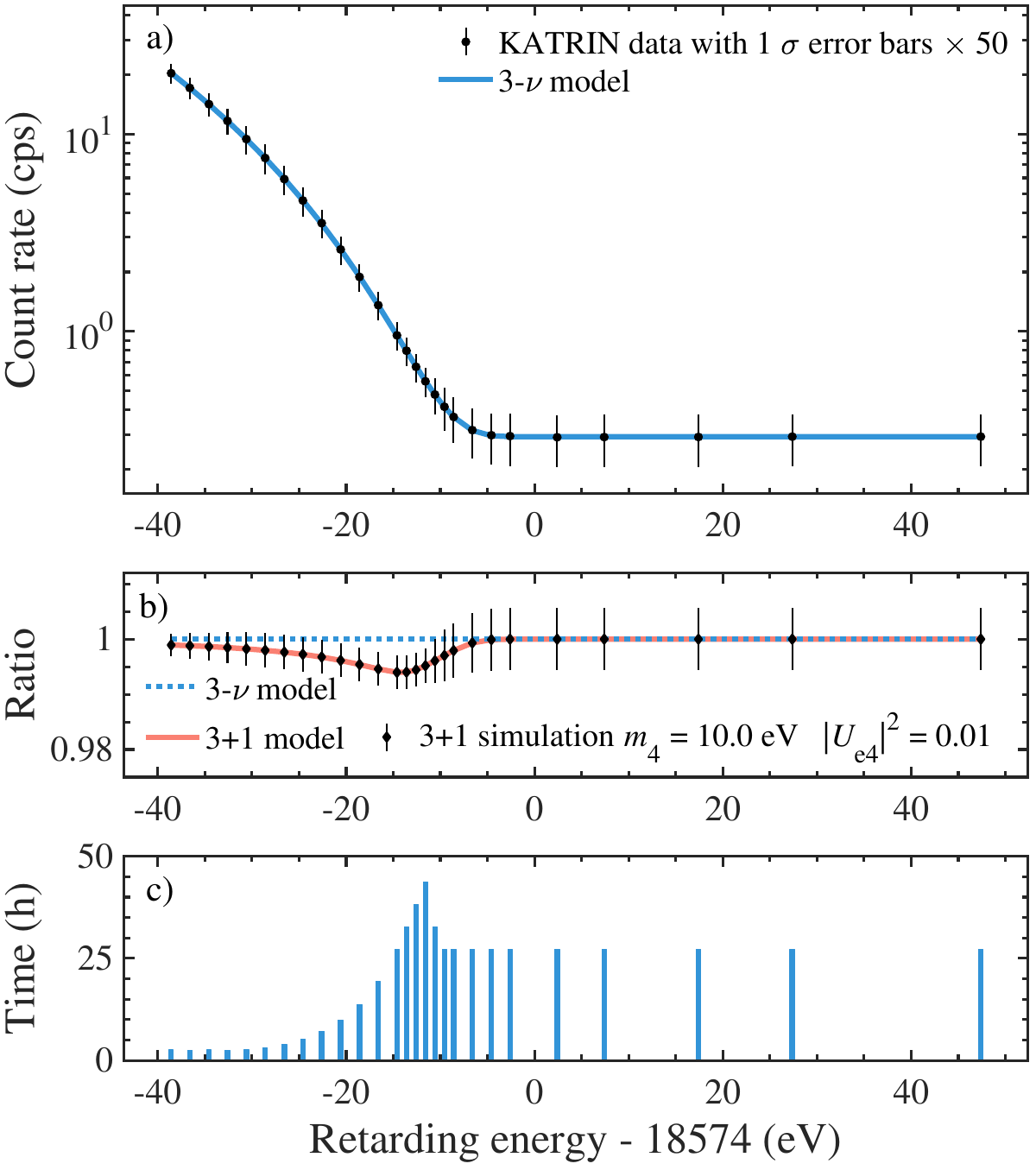}
\caption{\label{fig:null:beta_spectrum} a) Electron spectrum of KATRIN data $R(\langle qU \rangle)$ over the interval $\lbrack E_0 -40$~eV, $E_0 +50$~eV$\rbrack$ from all 274 campaign 1 tritium scans and the three-neutrino mixing best-fit model $R_\mathrm{calc} (\langle qU \rangle)$ (line). The integral $\beta$-decay spectrum extends to $E_0$ on top of a flat background $R_\mathrm{bg}$. 1-$\sigma$ statistical errors are enlarged by a factor 50. b) Simulation of an arbitrary sterile neutrino imprint on the electron spectrum. The ratio of the simulated data without fluctuation, including a fourth neutrino of mass $m_4=10$~eV and mixing $|U_{e4}|^2=0.01$, to the three-neutrino mixing model is shown (red solid line). c) Integral measurement time distribution. Figure from~\cite{KATRIN:2020dpx}.}
\end{center}
\end{figure}

\subsubsubsection{KATRIN}
\label{sec:null:katrin}
The Karlsruhe Tritium Neutrino experiment (KATRIN)~\cite{KATRIN:2001ttj,KATRIN:2005fny,KATRIN:2018sds} provides a high-precision electron spectrum measurement of tritium $\beta$-decay, $^3\rm{H} \rightarrow {^3\rm{He}^+} + \rm{e}^- + \bar{\nu}_\mathrm{e}$
(endpoint $E_0$ = 18.57~keV, half-life $t_{1/2}$ = 12.32~yr). KATRIN is designed to improve the sensitivity on the effective neutrino mass, $m_\beta$, to 0.2-0.3~eV (90$\%$~CL). Based on the science measurement campaigns taken in 2019~\cite{KATRIN:2019yun,KATRIN:2021uub}, KATRIN can constrain the mass and mixture of a sterile neutrino that would manifest itself as a distortion of the $\beta-$electron spectrum.
The signature is a kink-like feature, as shown in a simulation presented in Fig.~\ref{fig:null:beta_spectrum}.
The first light sterile neutrino result is based on data from KATRIN's first high-purity tritium campaign, which ran from April to May, 2019, at an average source activity of $2.45\cdot10^{10}$ Bq~\cite{KATRIN:2020dpx}.
An updated result is based on KATRIN's second campaign, which ran from October to November, 2019, achieving a source activity of $9.5\cdot10^{10}$ Bq~\cite{KATRIN:2022ith}.

The integral $\beta$-electron spectrum is scanned repeatedly in the range of $\lbrack E_0 -$90~eV, $E_0 +$135~eV$\rbrack$ by applying non-equidistant HV settings to the spectrometer electrode system.
Each scan lasted 2~h. At each HV set point, the transmitted electrons are counted.
Fig.~\ref{fig:null:beta_spectrum} shows the measurement time distribution. Stable scans are selected with an overall scanning time of 522~hr (campaign one) and 744~hr (campaign two).
Detector variations are minimized by gathering events from the 117 most similar pixels and combining them into a single effective pixel analysis.
The resulting spectra, \emph{R}($\langle qU \rangle$), include a combined $5.2\cdot10^6$~expected tritium events on a flat background.
Even for the campaign one search, which exhibited lower source rates and higher backgrounds, a high signal-to-background ratio was achieved, rapidly increasing from~1~at~$\langle qU \rangle=E_0-$12~eV to $>$70 at~$\langle qU \rangle=E_0-$40~eV.
In the campaign two search, the highest signal-to-background ratio improves to 235 at~$\langle qU \rangle=E_0-$40~eV.
The modeled experimental spectrum $R_{\mathrm{calc}}(\langle qU \rangle)$ is the convolution of the differential $\beta$-spectrum $R_\beta(E)$  with the response function $f(E - \langle qU \rangle)$, and an energy-independent background rate $R_\mathrm{bg}$:
\begin{equation}  
R_\mathrm{calc} (\langle qU \rangle) = A_\mathrm{s}\,\cdot\,N_\mathrm{T}\int R_{\mathrm{\beta}}(E) \cdot  f(E - \langle qU \rangle)~ dE + R_\mathrm{bg},
\end{equation}
where $A_{\mathrm{s}}$ is the tritium signal amplitude.
$N_\mathrm{T}$ denotes the number of tritium atoms in the source multiplied with the accepted solid angle of the setup $\Delta \Omega/4\pi = (1- \cos{ \theta_\mathrm{max}})/2$, with $\theta_{\mathrm{max}} = 50.4^{\circ}$, and the detector efficiency (0.95).
The function $f(E - \langle qU \rangle)$ describes the transmission probability of an electron as a function of its surplus energy $E-\langle qU \rangle$.
KATRIN extends the experimental modeling and statistical analysis to constrain both the sterile neutrino mass squared $m^2_{4}$ and its mixing amplitude $|U_{e4}|^2$, following the same strategy as for the $m_\beta$ analysis~\cite{KATRIN:2021fgc}.
In the 3+1 active-sterile neutrino model extension the electron spectrum, $R_\beta$, is replaced by $R_\beta(E,m_\beta,m_4) = (1~-~|U_{e4}|^2) R_\beta(E,m^2_\beta) + |U_{e4}|^2 R_\beta(E,m^2_{4})$, where $U$ is the extended $4\times4$ unitary matrix, $R_\beta(E,m^2_\beta)$ is the differential electron spectrum associated with decays the include active neutrinos in the final state, and $R_\beta(E,m^2_{4})$ describes the additional spectrum associated to decays involving a sterile neutrino of mass $m_{4}$.

The observable integral spectrum $R_\mathrm{calc}$ is henceforth modeled with six free parameters: the four original parameters ($A_\mathrm{s}$, $E_0$, $R_\mathrm{bg}$, $m_\nu^2$)~\cite{KATRIN:2019yun}, $m_4^2$ and $|U_{e4}|^2$.
This extended model $R_{\mathrm{calc}}(\langle qU \rangle)$ is then fitted to the experimental data $R(\langle qU \rangle)$. 

\begin{figure}[ht!]
  \begin{center}
  \includegraphics[width=0.8\textwidth]{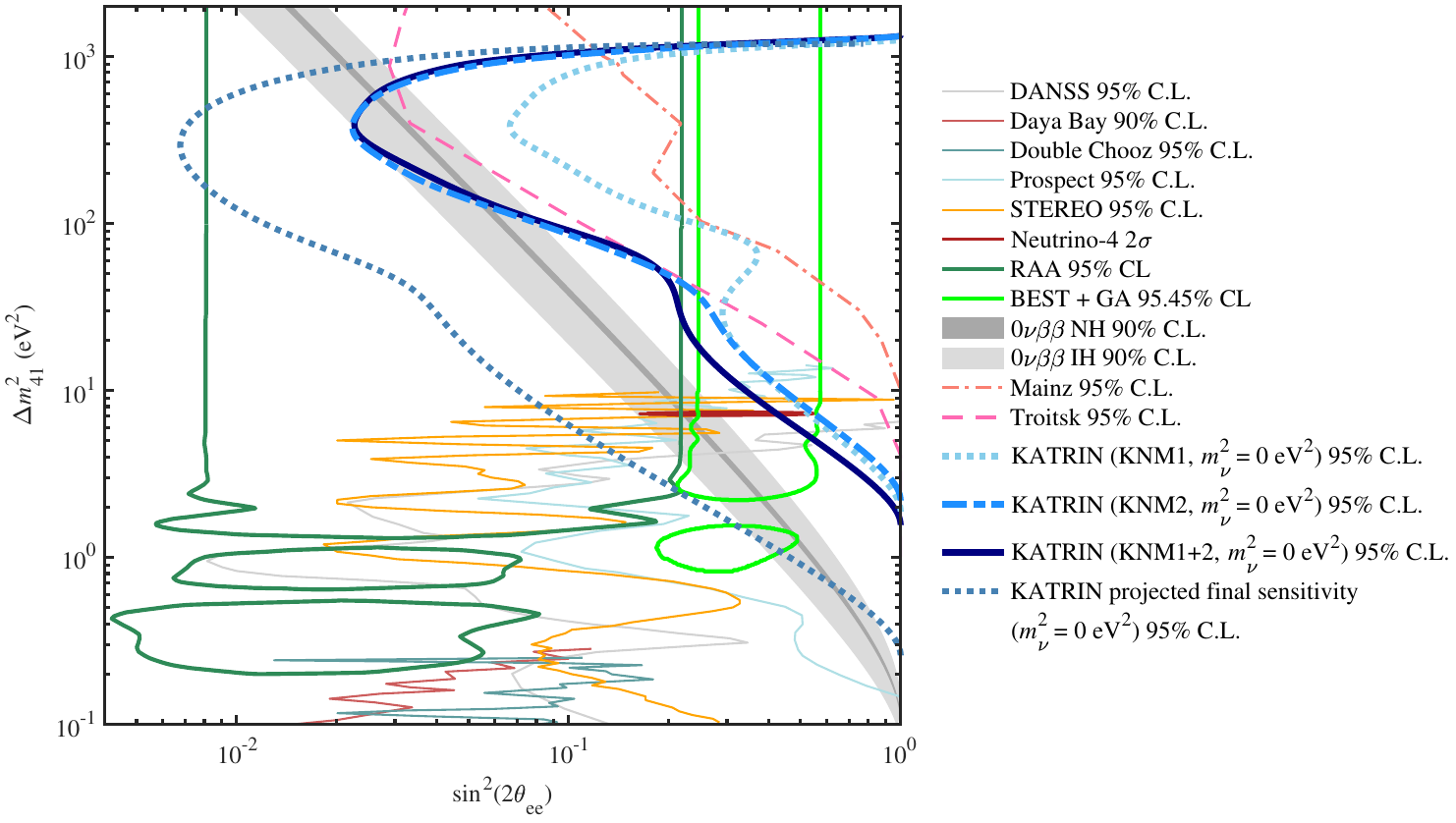}
  \caption{\label{fig:null:contourfinal} 
  95\%~confidence level exclusion curves in the ($\sin^2(2\theta_{ee}),\Delta{m}^2_{41}$) plane obtained from the analysis of KATRIN data with fixed $m_\nu (m_\beta)=0$. The green contours denote the 3+1 neutrino oscillations allowed at 95\%~confidence level by the reactor and gallium anomalies~\cite{Mention:2011rk}. KATRIN data improve the exclusion of the high $\Delta{m}^2_{41}$ values with respect to DANSS, PROSPECT, and STEREO reactor measurements~\cite{Danilov:2019aef, PROSPECT:2018dtt,STEREO:2019ztb}. Mainz~\cite{Kraus:2012he} and Troitsk~\cite{Belesev:2012hx} exclusion curves~\cite{Giunti:2012bc} are also displayed for comparison. An estimation of KATRIN's final sensitivity  is represented by the dotted line. Figure from~\cite{KATRIN:2022ith}.}
  \end{center}
\end{figure}

In order to mitigate bias, the full analysis is first conducted on a Monte Carlo (MC) data set before turning to the actual data. The fit of $R(\langle qU \rangle)$ with $R_\mathrm{calc} (\langle qU \rangle)$ is performed by minimizing the standard $\chi^2$-estimator.
In a ``shape-only'' fit, both $E_0$ and $A_{\mathrm{s}}$ are left unconstrained. To propagate systematic uncertainties, a covariance matrix is computed after performing ${\cal O}(10^4)$ simulations.
The sum of all matrices encodes the total uncertainties of $R_\mathrm{calc} (\langle qU \rangle)$, including HV set-point-dependent correlations.
The $\chi^2$-estimator is then minimized to determine the best-fit parameters, and the shape of the $\chi^2$-function is used to infer the uncertainties. 
The fit range $\lbrack E_0 -$40~eV, $E_0 +$50~eV$\rbrack$ is chosen such that statistical uncertainties on $ |U_{e4}|^2 $ dominate over systematic uncertainties in the whole range of $m_{4}^2$ considered~\cite{KATRIN:2019yun,KATRIN:2021fgc}.
The experimental result agrees well with the sensitivity estimates and are displayed in Fig.~\ref{fig:null:contourfinal}.
They are showing no evidence for a sterile neutrino signal and are compared with short-baseline neutrino oscillation experiments measuring the electron (anti-)neutrino survival probability $P(\Delta{m}^2_{41},\sin^2(2\theta_{ee}))$.
An estimation of KATRIN's five-year (1000 live-day) sensitivity is also presented.

\subsection{Dark Sectors in Scattering and in the Beam}

The MiniBooNE low energy excess has traditionally been interpreted
as a potential hint of neutrino oscillations at short baseline driven
by a new sterile state. However, simple interpretations (e.g.~$3+1$ models) 
are in tension with other oscillation measurements (see the discussion in the sections above), in
particular searches for muon flavor disappearance driven by the same
$\Delta m^2$. Given this tension, it is natural to consider other
possible new physics explanations for the LEE. These new physics
explanations must have detector signatures and production mechanisms
which are consistent with the existing MiniBooNE measurements.

The measured angular distribution of the events in the LEE constrains
the allowed interaction channels in the detector. For concreteness,
let’s suppose that the LEE is due to some new particle $X$. Simple
signatures involving elastic scattering on electrons (e.g.~$X e^-
\rightarrow X e^-$) or fully visible decays (e.g.~$X \rightarrow e^+
e^-$) produce events which are very forward, inconsistent with the
measurement which features a broad angular distribution. While
there is a highly significant forward component to the excess,
especially apparent in the 2021 MiniBooNE LEE result, the excess
remains significant across a broad range of angles, and one cannot
simply presume that the forward excess is due to new physics while the
rest of the excess is due to background underestimation. Considering
semi-visible decays (e.g.~$X \rightarrow X' + \gamma$) lessens the
disagreement somewhat, but such models are still strongly disfavored,
with either too many forward events or too many backward events
depending on the $X$ mass. In light of these constraints, the only
class of models which can adequately reproduce the LEE angular
distribution are those involving inelastic scattering, similar to the
neutrino-nucleus scattering in the standard sterile neutrino
interpretation.

\subsubsection{Beam Dump Searches}

\label{sec:miniboone_beam_dump} 

Neutrino experiments as well as dedicated electron or proton beam dump experiments can search for dark sector particles by looking for the decays or scatterings of states produced at the target station. This technique has been extensively explored in the literature in the search for light dark matter and mediators~\cite{Batell:2009di,Bjorken:2009mm,deNiverville:2011it,deNiverville:2012ij}. Measurements at LSND and MiniBooNE provide competitive limits in several light dark sectors.

Of particular relevance are the results of the MiniBooNE-DM run. MiniBooNE ran in a beam dump configuration where the proton beam was
aimed directly at the downstream beam dump instead of onto the
neutrino production target. In this mode, charged meson decay in
flight was suppressed, reducing the ``background'' neutrino flux and
enhancing sensitivity to new physics. No excess of events was observed
in beam dump mode, implying that any new physics production modes that
have a simple scaling with the number of protons on target like
neutral meson decays or continuum processes (i.e. bremsstrahlung) are
ruled out because they should produce a signature in beam dump mode.
This leaves production from charged meson decay in flight as the only
viable production mode for new particles that can explain the LEE.

New physics explanations of the MiniBooNE LEE based on inelastic scattering signatures in the detector are favored due to the
compatibility with the measured angular distribution of the excess. The MiniBooNE-DM results imply that such processes should be initiated by either neutrinos or new particles produced in charged meson decays at the target~\cite{Jordan:2018qiy}.

\subsubsection{Neutrino-Electron Scattering Measurement}
\label{sec:minerva_and_charm}

\begin{figure}[ht]
    \centering
    \begin{center}
    \includegraphics[width=0.55\textwidth]{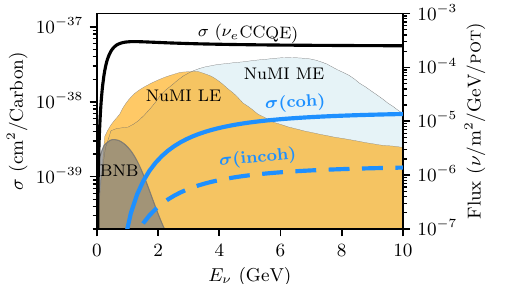}
    \includegraphics[width=0.44\textwidth]{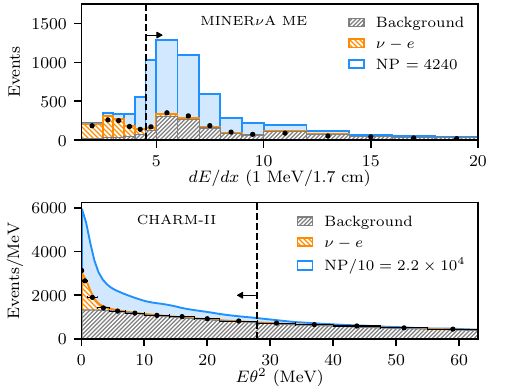}
    \end{center}
    \caption{\label{fig:minervacharm_1} Left: the neutrino flux at MiniBooNE, MINER$\nu$A, and CHARM-II are shown as shaded regions. In lines the upscattering cross section for producing heavy neutrinos through coherent and incoherent $Z^\prime$ scattering is shown. Right: The single shower event spectra as a function of $dE/dx$ at MINER$\nu$A (top) and of $E\theta^2$ at CHARM-II (bottom). Dashed lines indicate analysis cuts. Figures taken from Ref.~\cite{Arguelles:2018mtc}.}
\end{figure}

Models that can explain the LSND or MiniBooNE anomaly through the production of new particles that decay to electromagnetic showers can be constrained by measurements of neutrino-electron ($\nu-e$) scattering. Since single photons and collimated $e^+e^-$ pairs appear as single showers, they can be searched for by looking for in the photon-like sidebands of this measurement. In Ref.~\cite{Arguelles:2018mtc}, the authors propose a technique to constrain $e^+e^-$ explanations of LSND and MiniBooNE using data from MINER$\nu$A and CHARM-II. These experiments were located in the NuMI and CERN SPS beams, respectively, and therefore cover a much broader and higher-energy neutrino flux than LSND and MiniBooNE (see left panel of \cref{fig:minervacharm_1}).

In the dark photon model discussed in Sec.~\ref{sec:th_landscape:darksectors}, the HNL decays to overlapping $e^+e^-$ predict new signals in the large-$dE/dx$ sideband of $\nu-e$ scattering, with moderate values of $E \theta^2$. The energy deposition $dE/dx$, defined as the deposited energy in the first centimeters of a electromagnetic shower, helps discriminate between electron-like showers from photon-like showers, which have typically twice the energy deposited per cm than the former. 
The variable $E \theta^2$, defined as the shower energy times the square of the shower angle with respect to the neutrino beam, is also used to reduce backgrounds since in the boosted electron in $\nu-e$ scattering obeys $E \theta^2 < 2m_e$, while backgrounds, mostly from $\pi^0$ decay and $\nu_e$ CCQE, can have much broader angular distributions. For the dark sector signatures, the resulting overlapping $e^+e^-$ pairs can be quite forward, especially when the dark photon is light. Examples of the dark sector predictions at MINER$\nu$A as a function of $dE/dx$ and at CHARM-II as a function of $E\theta^2$ are shown in \cref{fig:minervacharm_1}.

\begin{figure}[ht]
    \centering
    \begin{center}
    \includegraphics[width=0.49\textwidth]{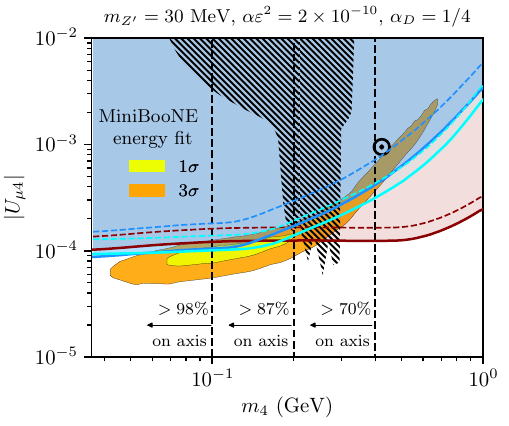}
    \end{center}
    \caption{\label{fig:minervacharm_2} The MINER$\nu$A and CHARM-II constraints in the parameter space of the HNL mixing with muons as a function of its mass, at $90$\% C.L. Solid lines show the nominal constraints, while dashed ones show constraints with inflated systematic errors on the background. The arrows on the vertical lines indicate where more than a given percentage of the total excess events at MiniBooNE are predicted in the forward-most angular bin. Figure taken from Ref.~\cite{Arguelles:2018mtc}.}
\end{figure}

Since the backgrounds have large uncertainties and the search is not tailored to the dark sector signals, the sensitivity is not enough to rule out all explanations of the MiniBooNE excess. However, due to the higher-energies at MINER$\nu$A and CHARM-II, explanations with large HNL masses, often preferred due to the broader angular distributions at MiniBooNE, can be robustly excluded. In \cref{fig:minervacharm_2}, the limits are shown in the mixing of the HNL with muon neutrinos, $|U_{\mu 4}|^2$, and their mass $m_4$. Other parameters like kinetic mixing  $\epsilon$ and the dark sector coupling $\alpha_D$, as well as the dark photon mass $m_{Z^\prime}$ are fixed. Two curves are shown for each experiment: a solid curve for the nominal choice of background uncertainty, and a dashed curve corresponding to the case where uncertainties were inflated by a factor of a few (see caption). For MINER$\nu$A LE (ME), this corresponds to $30\%$ ($40\%$) background normalization uncertainty in the nominal case and $100\%$ uncertainty in the conservative case. For CHARM-II these correspond to $3\%$ and $10\%$, respectively. The uncertainties at CHARM-II are constrained by the sideband at large $E \theta^2$.

Future measurements, including antineutrino-electron scattering at MINER$\nu$A can improve on the limits discussed above. The limits can also be recasted onto models with photon final states, like those discussed in \cref{subsubsec:TransitionMagneticMoment}. Finally, HNLs produced via scalar mediators are less constrained by this technique since the upscattering cross section decreases at higher energies. 

\subsubsection{Searches for Long-Lived Particles}

\begin{figure}[ht]
    \centering
    \includegraphics[width=0.49\textwidth]{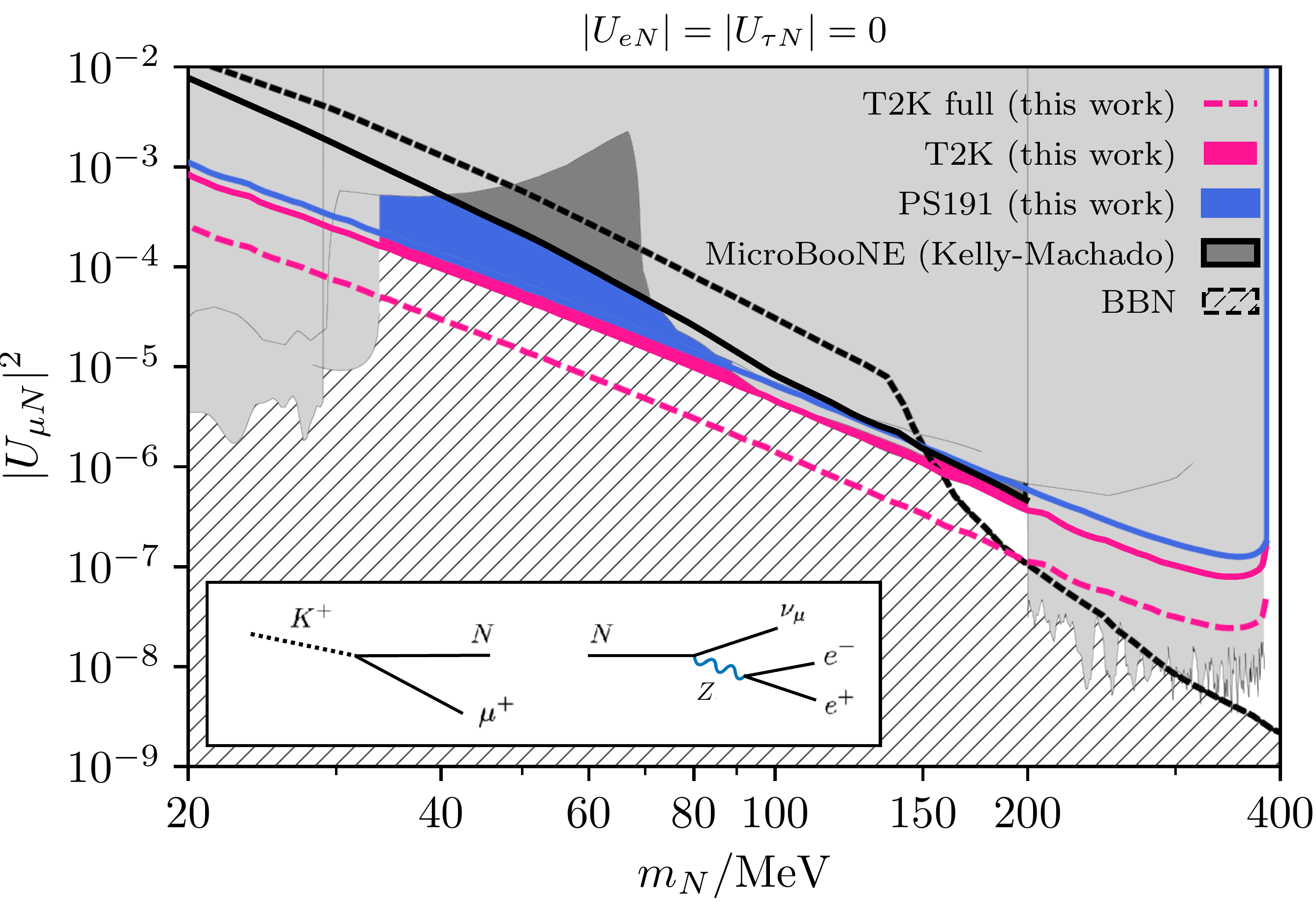}
    \includegraphics[width=0.49\textwidth]{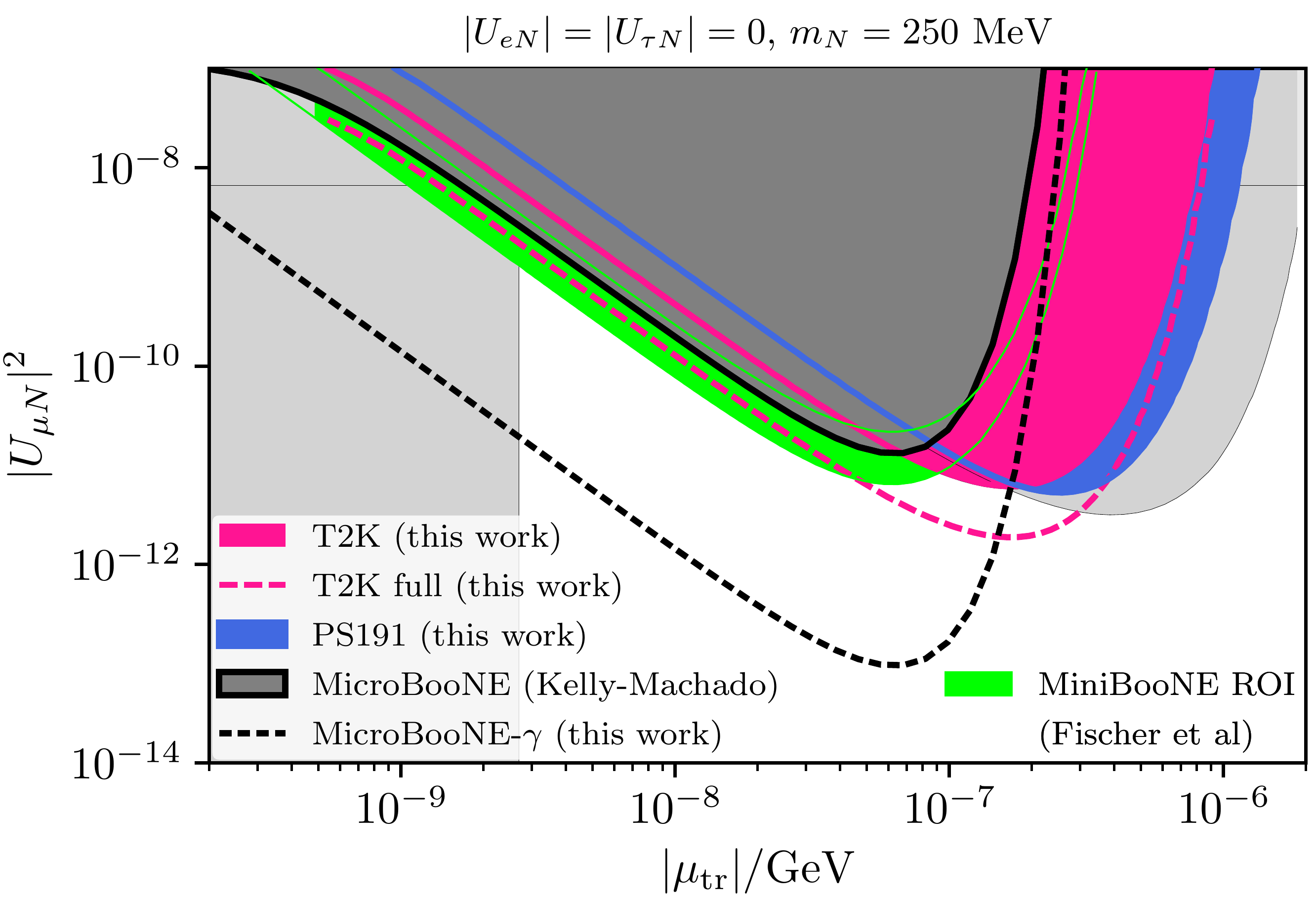}
    \caption{\label{fig:exclusion_plots}
    Left: Exclusion plot for the minimal scenario, showing the mixing with the muon flavour versus the HNL mass.
    Right: Exclusion plot for the non-minimal scenario where HNLs possess a magnetic moment.
    It shows the mixing with the muon flavour versus the magnetic dipole moment for an HNL mass of 250~MeV.
    The region of interest to explain the MiniBooNE excess is shown in green~\cite{Fischer:2019fbw}.
    The dark grey region is taken from or extrapolated by~\cite{Kelly:2021xbv}.
    Both plots are taken from~\cite{Arguelles:2021dqn}.
    }
\end{figure}

The possibility that meson decays produce long-lived HNLs in neutrino beams is strongly constrained by direct searches at T2K~\cite{T2K:2019jwa}, MicroBooNE~\cite{MicroBooNE:2019izn}, and ArgoNeuT~\cite{ArgoNeuT:2021clc}, as well as by other past-generation experiments such as PS-191~\cite{Bernardi:1987ek} (recently re-evaluated in Ref.~\cite{Arguelles:2021dqn} and ~\cite{Gorbunov:2021wua}), CHARM~\cite{CHARM:1985nku}, CHARM-II~\cite{CHARMII:1994jjr}, and NuTeV~\cite{NuTeV:1999kej}. The constraints are placed on minimal models with a single HNL $N$ that interacts with the weak bosons only, with interaction strength suppressed by small mixing angles $U_{\alpha N}$. At first sight, this model differs from the light sterile neutrino only for the mass of the additional neutrino, which is supposed to be in the $10 - 500$~MeV range.
However, this difference results in a radically different phenomenology.
Because these new particles are so much heavier than the standard neutrinos, no oscillation is possible: the HNL mixes with the other neutrinos but loses coherence immediately with the rest of the wave packet.
As a result, in the minimal scenario, in which no other particle or interaction is present aside from one or more HNLs, the HNL is produced in the beam through mixing and decays in the detector.
The available decay channels depend on the mass and include $e^+e^- \nu$, $e \mu \nu$, $e \pi$, $\mu^+ \mu^- \nu$, $\mu \pi$.
However, this minimal model is ruled out by a combination of Big Bang Nucleosynthesis, which provides a lower bound on the mixing parameters~\cite{Boyarsky:2020dzc}, and experiments with neutrino beams, which lead to upper bounds, with no available parameter space in between these bounds (left plot in Fig.~\ref{fig:exclusion_plots}).

\begin{figure}[ht!]
    \centering
    \includegraphics[width=0.9\textwidth]{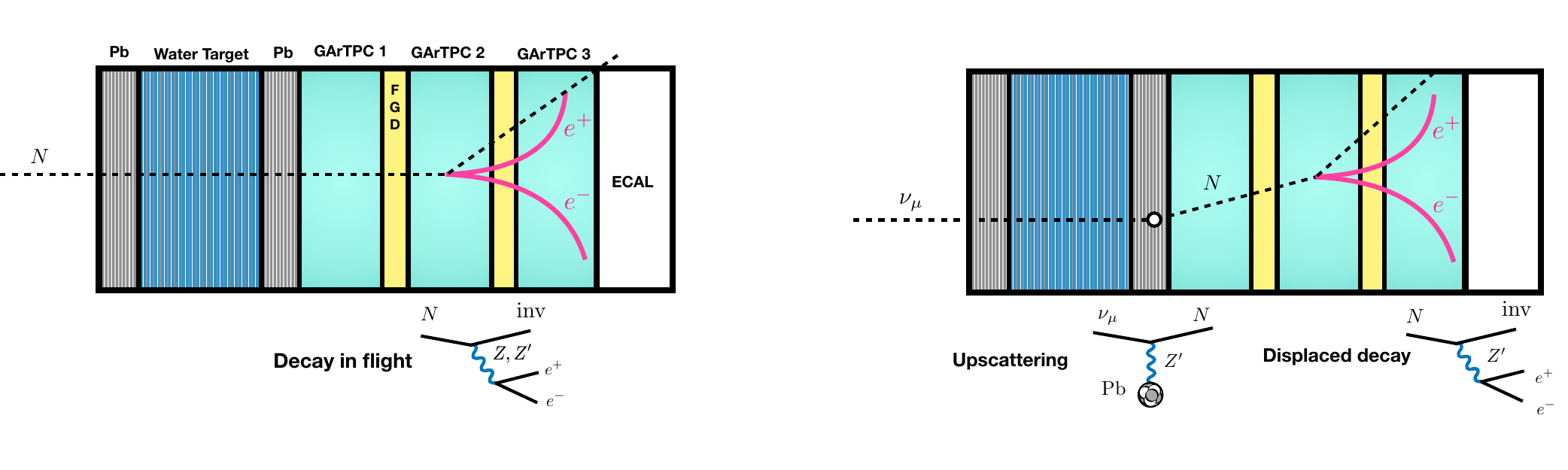}
    \caption{\label{fig:nd280_schematics}
    Left: Schematic representations of an HNL decaying in flight in one of the ND280 GArTPCs, as considered in~\cite{Abe:2019kgx,Arguelles:2021dqn}. 
    Right: The HNL is produced in the dense layers of lead through upscattering and subsequently decays in one of the GArTPCs as considered in~\cite{Arguelles:2022lzs}. }
\end{figure}

In particular, the T2K Near Detector ND280, a modular detector able to resolve details of interactions track and identify individual final state particles, sets the best limits on these models.
For this analysis, the three TPCs filled with gaseous argon provide a low-density decay volume for the HNLs, with zero background from neutrino interactions, as illustrated in the left plot in Fig.~\ref{fig:nd280_schematics}.
The limits have been derived through a search with little expected background and zero observed events in every analysis channel~\cite{Abe:2019kgx}, later extended and combined with BBN constraints~\cite{Arguelles:2021dqn}.

This model becomes more interesting if new interactions are present.
These new interactions allow new decay modes, hence shortening the lifetime and relaxing BBN limits.

For example, if the HNL possesses a magnetic moment, it could decay electromagnetically into $\nu \gamma$.
For magnetic moments of the order of PeV, this model could explain the MiniBooNE anomaly~\cite{Fischer:2019fbw}.
However, this explanation is constrained by short-baseline experiments, like MicroBooNE and ND280 (right plot in Fig.~\ref{fig:exclusion_plots}).
Thanks to the high density of the liquid argon, MicroBooNE identifies single photons through conversion to $e^+e^-$ pairs.
ND280 is instead sensitive to the branching ratio into off-shell photons, which results in a genuine $e^+e^-$ pair.
In this case, the rate is lower by a factor of ~100 but benefits from a zero background search.

\paragraph{Upscattering Recasts}

As discussed in Secs.~\ref{sec:th_landscape:darksectors}, HNLs can be short-lived if they interact via additional forces via a dark photon or scalar, for instance. 
In these models, the lifetime of the HNL can range from tens of meters to sub-mm, depending on the choice of parameters.
In this regime, standard neutrinos scattering on nuclei can produce a HNL, as illustrated in the right panel of Fig.~\ref{fig:nd280_schematics}.
This particle propagates inside the detector and subsequently decays into an $e^+e^-$ pair plus another neutrino. Searches for $e^+e^-$ final states at neutrino detectors can be used to constraint this possibility. In particular, the near detector of T2K, ND280, and the MicroBooNE detector have both been used to search for the decay of long-lived particles into $e^+e^-$

In the scenario of a dark photon heavier than the HNL, lifetimes range between a few cm to several meters.
In this case, a dark neutrino would be produced in the Pi0 detector (POD), a very dense detector composed of high-Z materials such as lead, which lies upstream with respect to the TPCs.
The production is particularly copious as the process is coherent, thus scaling with $Z^2$ and benefiting the high Z materials.
In the scenario of a dark photon lighter than the HNL, the decay proceeds through an on-shell dark photon, resulting in a larger decay width and shorter lifetimes, sub-millimetre.
In this case, the HNL is produced and decays at the same point in the detector.
The Fine-Grained Detectors (FGD) of ND280 come into play for this study.
This plastic scintillator is dense enough to make the rate for production through upscattering significant while allowing precise tracking and identification of the $e^+e^-$ pair.
Because of the larger density, backgrounds from the beam or photons that convert inside the FGD are present.
On the other hand, the invariant mass of the $e^+e^-$ pair is an excellent quantity to discriminate this background as it peaks around the dark photon mass because of the on-shell decay - making this analysis a peak search.
Explanations of MiniBooNE under this model are not entirely ruled out yet~\cite{Arguelles:2022lzs}, but the next generation of analyses, together with an upgrade to the detector~\cite{T2K:2019bbb} and a much larger dataset~\cite{Abe:2016tii} that will be collected by T2K, are expected to probe the entire parameter space of interest.

\subsection{Conventional Explanations and Other Searches}
\label{sec:conventional}

Partly motivated by the lack of a single interpretation that can simultaneously explain all observed experimental anomalies, the possibility that the anomalies represent a collection of conventional explanations (or that in combination with new physics) has also been discussed; see, e.g.~Ref.~\cite{Brdar:2021cgb}, in which the possibility that the MiniBooNE anomaly is a combination of underestimated backgrounds is explored. In this case it was shown that no combination of varying the  known backgrounds can completely explain the MiniBooNE LEE, it can potentially reduce the significance to $3\sigma$ if several less well-known backgrounds saturate their prescribed errors. 

Key among the anomalies is the MiniBooNE low-energy excess, where it has been shown repeatedly that the excess shape at low energy is incompatible with 3+$N$ light sterile neutrino oscillations, raising the question whether this excess is due to another effect. Furthermore, as this energy region is dominated by non-$\nu_e$ backgrounds, conventional interpretations for the excess that do not rely on new physics or introduce electrons into the MiniBooNE detector have been raised. The majority of these rest on the known fact that MiniBooNE---as a Cherenkov detector---had no ability to distinguish a single photon from a single electron, and in particular target the various photon backgrounds irreducibly contributing to and dominating the MiniBooNE observed $\nu_e$ CCQE rate at low energy.

In the reactor sector, the experimental search for sterile neutrinos were primarily motivated by the anomalously low rates of fluxes measured by the past reactor neutrino experiments.
Overestimated \anue~rates based on mismodeled reactor models, against which the fluxes were compared, could also explain the observed discrepancy without invoking new BSM particle physics.

\subsubsection{Constraints on Single-Photon Production} 

The majority of conventional interpretations to the MiniBooNE excess propose that the excess is due to mis-identified photons from various neutrino interactions that produce photons and no charged lepton in the final state. In MiniBooNE, these mis-identified backgrounds are primarily contributed by $\pi^0$ production, due to reconstruction inefficiencies (where one photon from the $\pi^0$ decay may be missed) or more rare processes such as NC single-photon production through $\Delta(1232)$ resonance production and subsequent radiative decay. Many of these processes can and have been constrained in situ at MiniBooNE. This was the case for the rate and momentum-dependence of misidentified NC $\pi^0$ decays, which were constrained by a high-statistics measurement of events with two reconstructed electromagnetic shower Cerenkov rings, representing the two photons from NC $\pi^0$ decay. Similarly, the ``dirt'' background component in MiniBooNE, which was NC $\pi^0$-dominated, was directly constrained by a high-statistics measurements of events close to the detector boundary.  These processes were further studied and disfavored as the source of the MiniBooNE anomaly by studying both the radial distribution of the excess as well as the timing of the events relative to the known beam bunch timing \cite{MiniBooNE:2020pnu}. On the other hand, photons from the radiative decay of the $\Delta(1232)$ baryon produced in neutrino NC interactions (NC $\Delta\rightarrow N\gamma$, where $N=p,n$) were an irreducible background in MiniBooNE. This background was not constrained in situ, but rather the rate was constrained indirectly through its correlation with the NC $\pi^0$ measured rate (which proceeds predominantly via $\Delta\rightarrow N\pi^0$ decay).

The MicroBooNE collaboration has recently performed a direct search for single-photon events coming from neutrino-induced NC production of the $\Delta$(1232) baryon resonance with subsequent $\Delta$ radiative decay~\cite{MicroBooNE:2021zai}. As discussed in Sec.~\ref{sec:single_gamma_luis}, $\Delta$ decay is expected to be the dominant source of single-photon events in neutrino-argon interactions below 1~GeV. Although $\Delta$ radiative decay is predicted in the SM, and measurements of photoproduction~\cite{Blanpied:1997zz} and virtual Compton scattering~\cite{Sparveris:2008jx} are well described by theory, this process has never been directly observed in neutrino scattering. In a fit to the radial distribution of the MiniBooNE data with statistical errors only, an enhancement of NC $\Delta\rightarrow N \gamma$ by a normalization factor of $3.18$ (quoted with no uncertainty) was found to provide the best fit for the observed excess, and in good spectral agreement with the observed excess across a number of reconstructed kinematic variables \cite{MiniBooNE:2020pnu}. The MicroBooNE collaboration searched directly for an excess of this magnitude, interpreted as an overall enhancement to the theoretically-predicted NC $\Delta\rightarrow N\gamma$ rate. 

MicroBooNE utilized the strength of LArTPC neutrino detectors to search for this excess of single-photons from the $\Delta$ decay, with and without a proton track present in the interaction (referred to as $1\gamma1p$ and $1\gamma0p$ final states). The presence of the reconstructed track in the $1\gamma1p$ selection allowed for the tagging of the neutrino interaction vertex, and subsequent reconstruction of the photon conversion distance in argon, which led to improved background rejection compared to the $1\gamma0p$ selection, where the lack of associated hadronic activity prevented neutrino vertex tagging.

The analysis used approximately half of the total collected MicroBooNE data to date ($6.9\times 10^{20}$~POT).  It selected and simultaneously fitted the $1\gamma1p$ and $1\gamma0p$ samples together with two additional, mutually exclusive but highly correlated samples with high NC $\pi^0$ purity. These high-statistics NC $\pi^0$ samples effectively constrained the rate and systematic uncertainty of NC $\pi^0$ production in argon, which was the dominant background to the $1\gamma1p$ and $1\gamma0p$ selections. 

\begin{figure}
    \centering
    \includegraphics[width=0.45\textwidth]{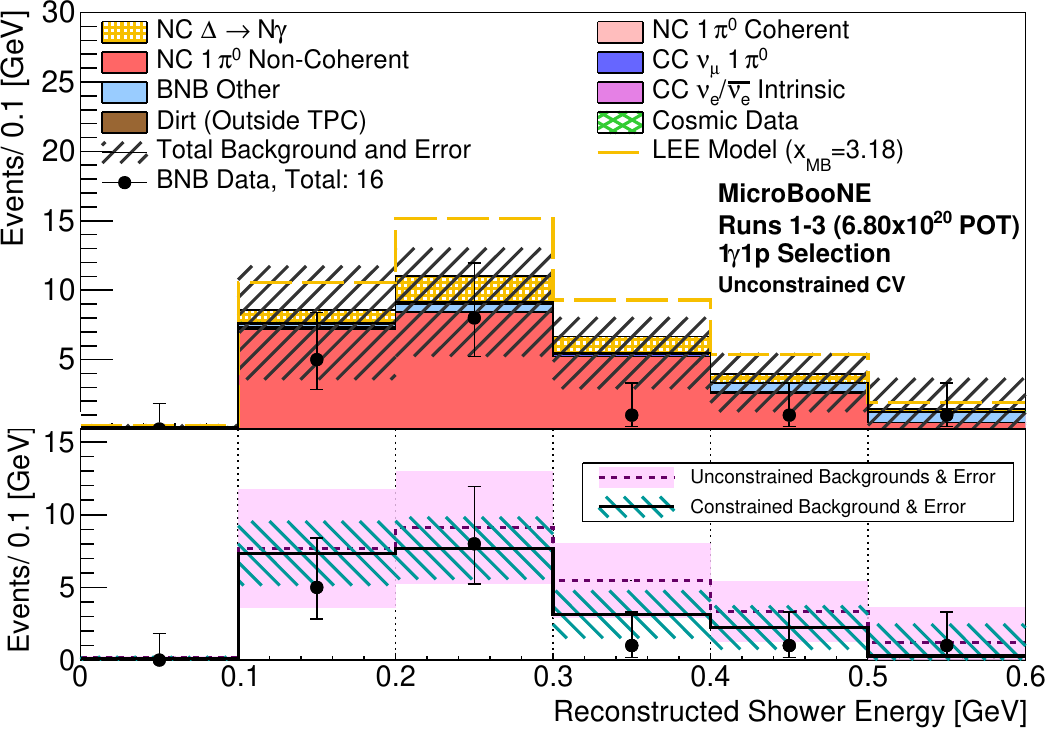}
    %
    \includegraphics[width=0.45\textwidth]{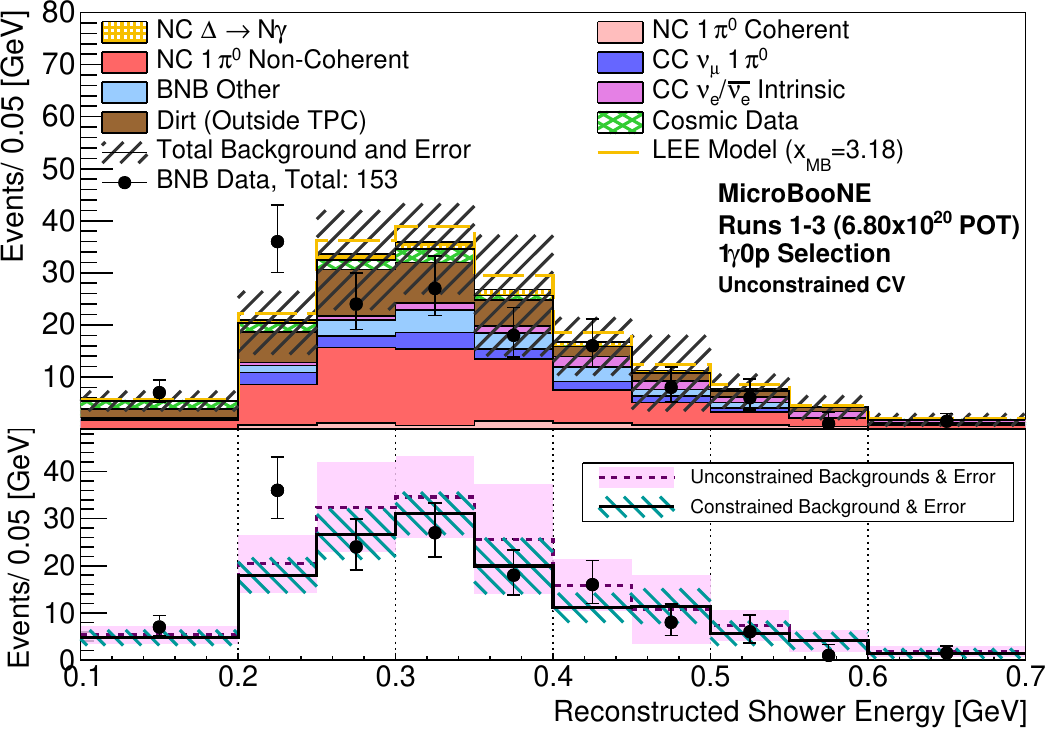}
    \caption{Summary of MicroBooNE's NC $\Delta \rightarrow N \gamma$ single-photon searches for both with (left) and without (right) an associated proton. Figures from \cite{MicroBooNE:2021zai}. No evidence of an enhanced NC $\Delta\rightarrow N\gamma$ rate is observed. } 
    \label{fig:ub_gamma_results}
\end{figure}

MicroBooNE observed no evidence for an excess of NC $\Delta$ radiative decay, as shown in Fig.~\ref{fig:ub_gamma_results}. The measurement ruled out the normalization enhancement factor of 3.18 as an explanation to the MiniBooNE low-energy excess at 94.8\% CL (1.9$\sigma$), in favor of the nominal prediction for NC $\Delta\rightarrow N\gamma$. Note that this was a model-dependent test of the MiniBooNE excess, and therefore does not necessarily apply universally to any photon-like interpretation. Those include, e.g. coherent single-photon production, which is expected to be a rarer process in MicroBooNE than NC $\Delta\rightarrow N\gamma$, or BSM processes that manifest as single-photon events, such as co-linear $e^+e^-$ from the decay of new particles. Those will be the target of dedicated follow-up MicroBooNE analyses, as well as model-independent single-photon searches.

\subsubsection{Reactor Flux Models}
\label{sec:expt_landscape_conventional_rates}

As discussed in Sec.~\ref{sec:anomaly_reactor}, modeling the reactor \anue~flux is quite challenging. 
State-of-the art models used to compare against measured IBD yields employed the conversion approach and relied on $\beta$-decay measurements performed at ILL in the 1980s. 
The presence of mistakes in these beta-decay measurements, as well as incorrect assumptions present in the conversion process, could lead to mis-modeled reactor \anue{} fluxes.  
Recent \anue and $\beta$-decay measurements have played a crucial role in further exploring  flux prediction issues as a possible source of the RAA.

\begin{figure}
    \centering
    \includegraphics[width=0.55\textwidth]{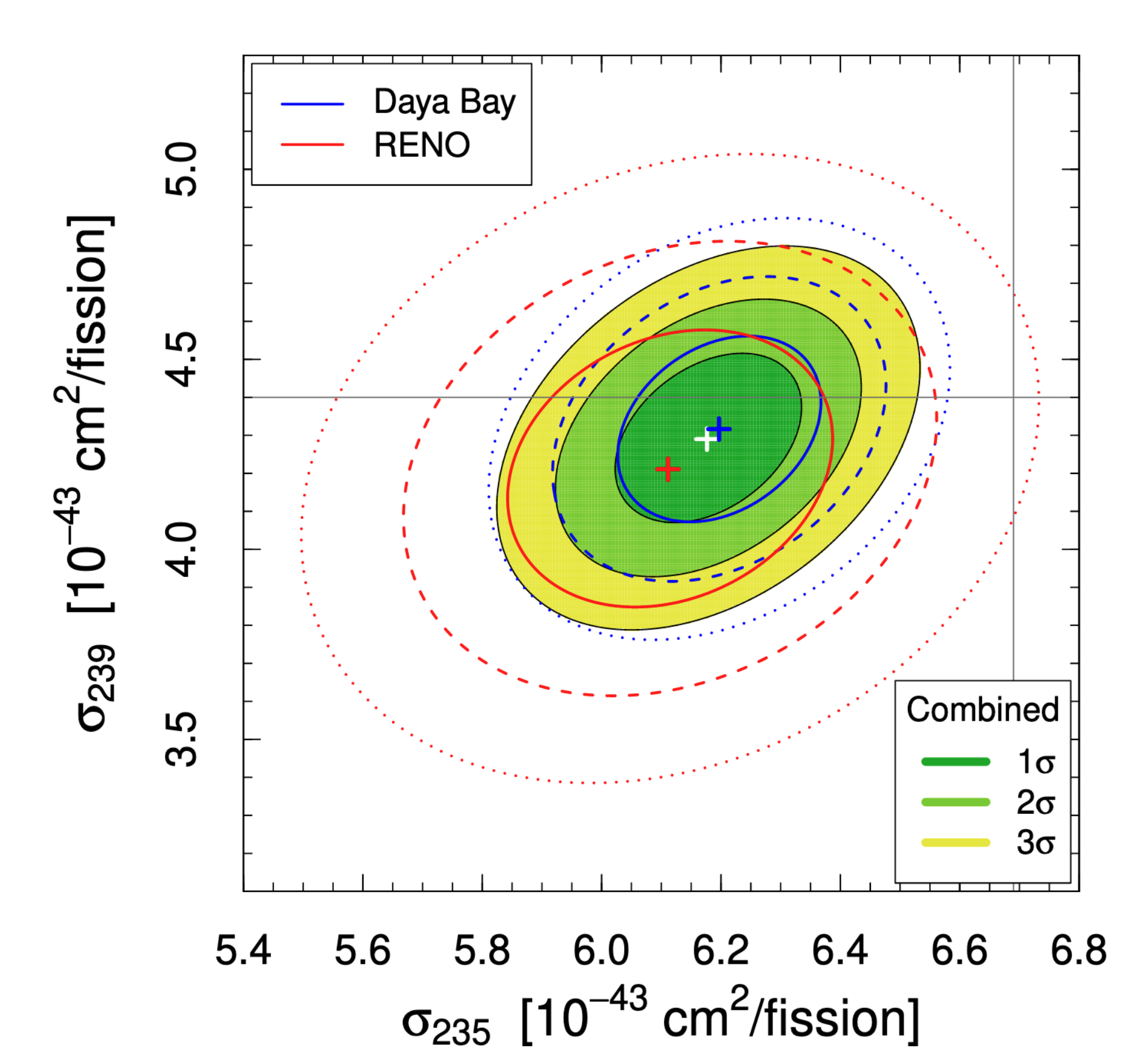}
    \caption{IBD yields of \uFive{} and \pNine{} from Daya Bay~(blue), RENO~(red), and a combined fit to Daya Bay and RENO. Horizontal and vertical black lines show the predicted IBD yields for \pNine{} and \pNine{} respectively based on the Huber-Mueller (HM) model. While \pNine{} yields agree with the model within $1\sigma$, a $>3 \sigma$ discrepancy can be noticed in \uFive~yields. Figure from Ref.~\cite{Giunti:2019qlt}.}
    \label{fig:null_results_DYB_RENO}
\end{figure}

Daya Bay~\cite{DayaBay:2017jkb} and RENO~\cite{RENO:2018pwo} experiments -- leveraging their IBD yield measurements as a function of evolving fission fractions in the reactors -- were able to perform simultaneous IBD yield measurements of \uFive~and \pNine.  
While continuing to observe the same time-integrated IBD yield deficit that defines of the RAA,  these results also showed that while \pNine~yields are in good agreement with the models, \uFive~is at 0.925 $\pm$ 0.015~($>3\sigma$) of the modeled yield as shown in Fig.~\ref{fig:null_results_DYB_RENO}.  
While these results are in slight tension~($\sim 1 \sigma$) with the average \uFive~IBD yields from the pure \uFive~measurements done using HEU reactors~\cite{Gebre:2017vmm}, they are in approximate agreement~($\sim 0.5 \sigma$)  with the modern \uFive~yield measured by the STEREO experiment~\cite{STEREO:2020fvd}.  
It is also worth noting that summation-predicted IBD yields are in good agreement with the Daya Bay's flux evolution measurements~\cite{Hayes:2017res,Estienne:2019ujo}.  

\begin{figure}
    \centering
    \includegraphics[width=0.55\textwidth]{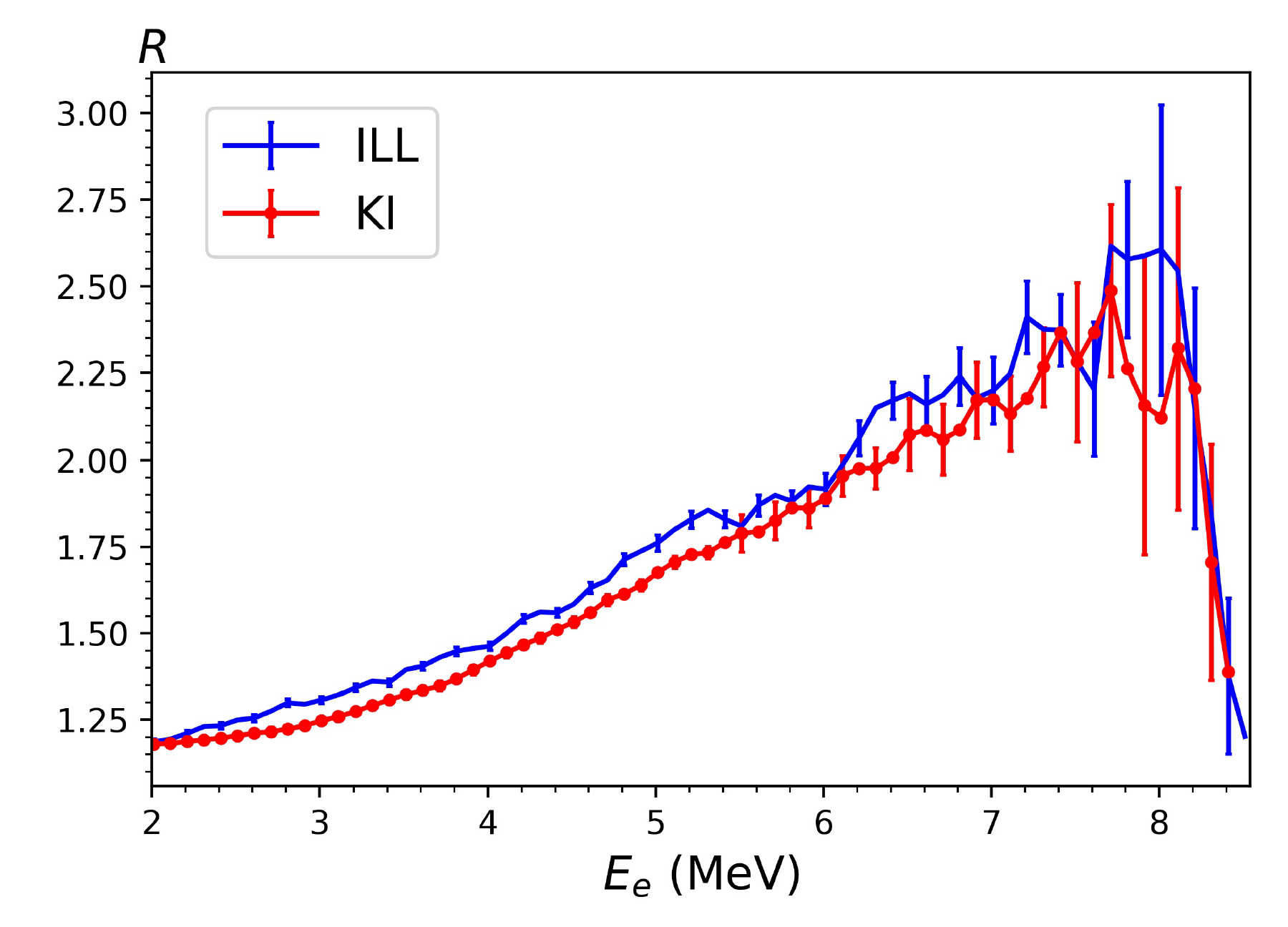}
    \caption{Ratio of the cumulative $\beta$ spectra for \uFive~and~\pNine~as measured at ILL~\cite{Haag:2014kia} (blue) and at KI~\cite{Kopeikin:2021rnb} (red). KI sees consistently lower ratios than ILL. Figure from ~\cite{Kopeikin:2021ugh}.}
    \label{fig:null_results_KI}
\end{figure}

To test the possibility of mis-modeled IBD yields in the conversion model, Kopeikin \textit{et al.}~\cite{Kopeikin:2021ugh} performed simultaneous $\beta$-decay spectrum measurements of \uFive~and \pNine~at the Kurchatov Institute, generating a conversion prediction, referred to as the KI Model that uses completely different primary inputs than those used by the Huber model.  
While the measurement was designed to achieve an extremely high degree of correlation between \uFive~and \pNine~measurements, it exhibited lower statistical precision than the ILL measurements of the 1980s.  
The KI model found a consistently higher ratio of \uFive~to \pNine~$\beta$-decay rates over the full energy range compared to the ILL measurements as shown in Fig.~\ref{fig:null_results_KI}.  
Potential bias in $\beta$ decay measurements have also been pointed out in a separate study~\cite{Bergevin:2019tcg}.  

\begin{figure}[!ht]
    \centering
\includegraphics[trim = 0.0cm 7cm 37.5cm 0.0cm, clip=true, width=0.5\textwidth]{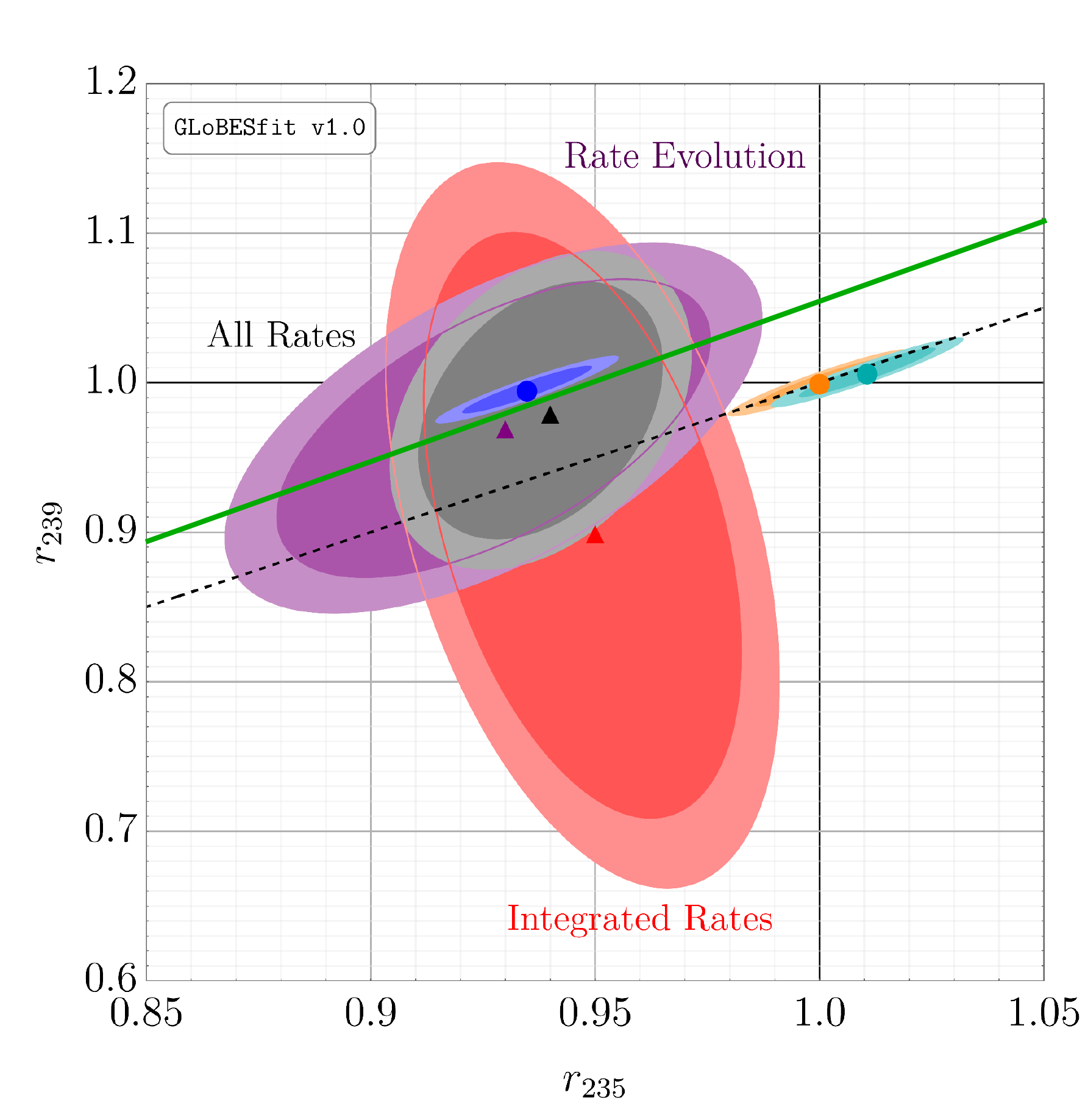}
    \caption{Best-fit points and 95 \% CL, and 99 \% CL contours of \uFive{} and \pNine{} IBD yields obtained using integrated rates (red), evolution measurements by Daya Bay and RENO (purple), and combined integrated rates and evolution measurements (gray). The axes correspond to $r_{i}$~($\overline{R}_{\text{HKSS}}$ as described above) for \uFive{} and \pNine{} along x and y direction respectively. Predictions from HM, EF, and HKSS models are shown in cyan, orange, and blue contours respectively. Black dashed line represents $r_{235} = r_{239}$. It is clear from the plot that while the measured IBD yields agree with the predictions from the EF (\emph{ab inito}) model for both isotopes, only \pNine{} agrees with the HM and HKSS models--both of which are conversion models relying on ILL $\beta$ spectra measurements--pointing towards an issue with the conversion approach-based \uFive{} predictions. Figure from Ref.~\cite{Berryman:2020agd}.}
    \label{fig:IBD_Yield_Comparison}
\end{figure}

These new conversion prediction results, together with the \anue flux evolution measurements from Daya Bay and RENO and the summation-conversion flux evolution mis-matches mentioned in Sec.~\ref{sec:th_land_rx_models} collectively suggest mis-modeled \uFive{}~IBD yields as a major contributor to the RAA.   
This collective picture is well-illustrated in Fig.~\ref{fig:IBD_Yield_Comparison}, which shows predictions and \anue-based measurements of IBD yields for the dominant fission isotopes \uFive~and \pNine, given in terms of a ratio with respect to the Huber-Mueller conversion prediction.  
Conversion predictions based on the ILL beta measurements (Huber-Mueller and the HKSS model discussed in Sec.~\ref{sec:th_land_rx_models}) appear to deviate from all other measurements and predictions.  
Huber-Mueller models show substantial deficits with respect to direct IBD yield measurements, whereas discrepancies between EF and KI models with respect to the data are not large enough to claim the existence of any anomaly whatsoever.  

It is important to stress that while these flux-based indications are fairly clear in their suggestion of flux prediction issues in the Huber-Mueller model and cast doubt on a pure sterile neutrino interpretation of the RAA, this global flux picture still leaves substantial room for short-baseline sterile oscillation phenomena at reactors.  
As previously mentioned in Sec.~\ref{sec:th_land_rx_models}, it is likely that the error bars assigned to most, if not all, predictions in Fig.~\ref{fig:IBD_Yield_Comparison} are under-estimated, leaving ample room for sterile oscillation amplitudes around the 20\% level or lower.  
Moreover, multiple studies show that hybrid models containing both incorrect flux predictions and sterile neutrinos also provide a good fit to global IBD yield datasets~\cite{Giunti:2017yid,Giunti:2019qlt}; some of these scenarios produce best-fit oscillation-induced deficits well in excess of 10\%.  
In light of this degeneracy between incorrect flux predictions and constant oscillation-induced deficits, short-baseline reactor measurements capable of directly probing the deficit's $L/E$ character are the better bet for cleanly elucidating the the role played by short-baseline sterile oscillations in the reactor sector.  

\subsubsection{MicroBooNE \nue CC Measurement} 

MicroBooNE has additionally performed a search that explicitly tests the nature of the MiniBooNE excess in a physics-model-agnostic way. Specifically, MicroBooNE has carried out three independent analyses to investigate whether the MiniBooNE observed low-energy excess can be described by an effective enhancement of $\nu_e$ CC scattering at low energy given by unfolding the MiniBooNE observed excess distribution. This unfolding predicts a factor of 5--7 enhancement of $\nu_e$ CC interactions below 500~MeV in true neutrino energy. 

\begin{figure}[!htbp]
    \centering
    \includegraphics[width=0.4\textwidth]{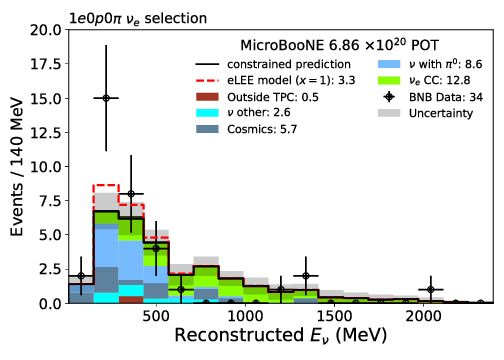}
    \includegraphics[width=0.4\textwidth]{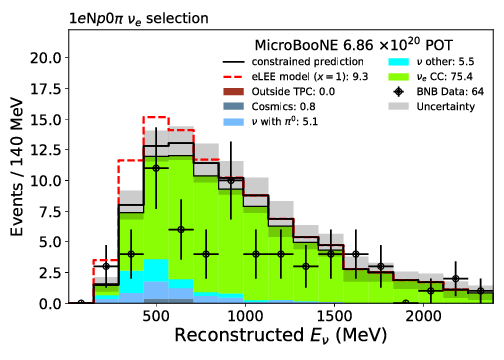}\\
    \includegraphics[width=0.4\textwidth]{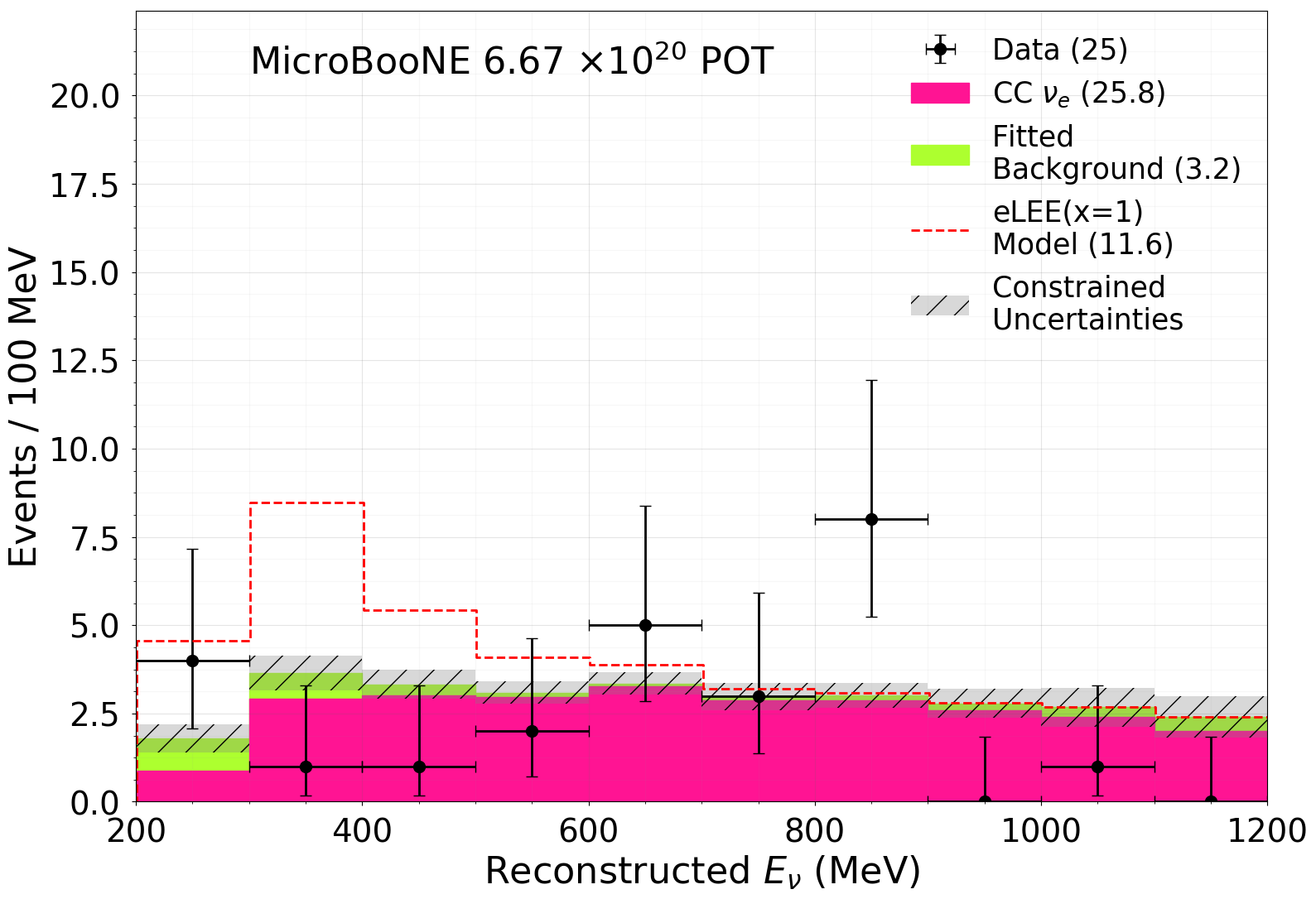}
    \includegraphics[width=0.4\textwidth, trim=150 100 0 0]{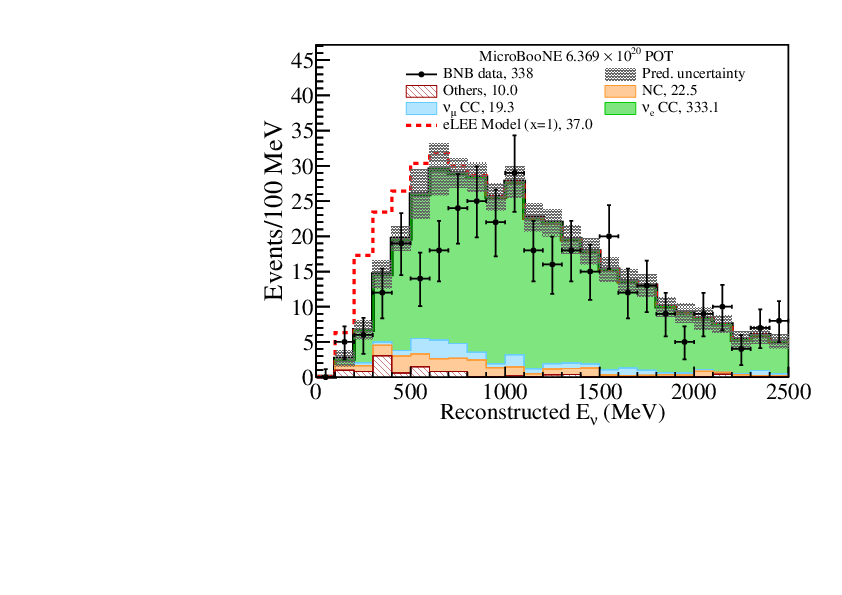}
    \caption{MicroBooNE's four targeted $\nu_e$ CC spectra. Figures taken from \cite{MicroBooNE:2021rmx}.}
    \label{fig:ub_nue_cc_distributions}
\end{figure}

\begin{figure}
    \centering
    \includegraphics[width=7cm]{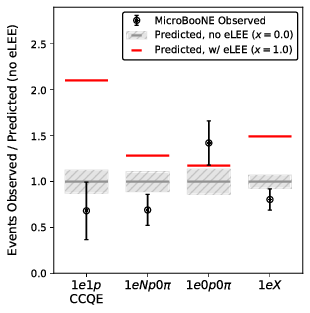}
    \caption{Summary of MicroBooNE $\nu_e$ CC measurements, from \cite{MicroBooNE:2021rmx}.}
    \label{fig:ub_nue_results}
\end{figure}

The data used in this search 
correspond to approximately half the data set collected by MicroBooNE during its entire operational run time in the Fermilab BNB. The search was carried out by three separate analyses, each targeting different exclusive and inclusive $\nu_e$ CC final states, and using separate reconstruction paradigms and signal selections. The $\nu_e$ final states explicitly targeted by MicroBooNE include a pion-less final state topology with one electron and with 0 or $N>1$ protons as part of the visible hadronic final state; a CCQE-like final state topology with one electron and with only 1 proton as part of the visible hadronic final state; and a CC-inclusive final state topology with one electron and with any number of charged pions or protons as part of the visible hadronic final state. While probing different event topologies with distinct event reconstruction and selection methods, the three independent analyses still share several common aspects, including the  signal model, a Geant4-based simulation of the neutrino beam, the detector response model, and a tuned variation of the GENIE v3 event generator incorporating the most up to date knowledge of neutrino scattering in the $<1$~GeV energy range. The results from each of the three analyses are presented in detail in \cite{MicroBooNE:2021sne,MicroBooNE:2021jwr,MicroBooNE:2021nxr}, and are summarized in \cite{MicroBooNE:2021rmx}. The final observed and predicted distributions from each search are reproduced in Fig.~\ref{fig:ub_nue_cc_distributions}, with the observed data to Monte Carlo ratios shown in Fig.~\ref{fig:ub_nue_results}. 

While statistics-limited, the three mutually-compatible analyses collectively reported no excess of low-energy $\nu_e$ candidates, and were found to be either consistent with or modestly lower than the predictions for all $\nu_e$ event classes, including inclusive and exclusive hadronic final-states, and across all energies. With the exception of the pion-less, zero-proton 
selection, 
which was the least sensitive to a simple model of the MiniBooNE low-energy excess, MicroBooNE rejected the hypothesis that an enhancement of $\nu_e$ CC interactions at low energy is fully responsible for
the MiniBooNE low-energy excess at $>97$\% CL for both exclusive and inclusive event classes. Additionally, MicroBooNE disfavored generic $\nu_e$ interactions as the primary contributor to the excess, with a $1\sigma$ ($2\sigma$) upper limit on the inclusive $\nu_e$ CC contribution to the excess of 22\% (51\%). 

It should further be noted that while these searches do not explicitly test the possibility of light sterile neutrino oscillations, a recent MicroBooNE publication \cite{MicroBooNE:2022wdf}, as well as independent phenomenological studies, have further analyzed the observed data within that context, including $\nu_e$ appearance and/or $\nu_e$ disappearance. The results of those are summarized in Sec.~\ref{dif_sbl}.


\clearpage

\section{Astrophysical and Cosmological Indirect Probes}
\label{sec:astro_cosmo}

 It has been known for a long time that astrophysical observations can provide powerful constraints on BSM physics.  Stellar evolution arguments have long been invoked to constrain couplings between light BSM particles and SM particles~\cite{Raffelt:1996wa}.  In particular, the success of a core-collapse supernova and the synthesis of heavy elements via the $r$-process therein maybe turned into an argument for or against sterile neutrino states (see, e.g., Ref.~\cite{Abazajian:2012ys}).  In the past two decades, however, the strongest astrophysical statements on light sterile neutrinos have come from cosmological measurements, i.e., those observations that probe the universe on the largest length scales.  We discuss in this section the relevant theoretical arguments and the most recent observational constraints.

\subsection{Cosmology}

The standard hot big bang model predicts that the three generations of SM neutrinos are held in a state of thermodynamic equilibrium with other SM particles via the weak interaction in the first second post-big bang.  At temperatures below about 20~MeV, the dominant equilibrating interactions are
\begin{equation}
\begin{aligned}
 \nu e & \to \nu e, \\
 \nu \bar{\nu} & \to e^+ e^-, \\
 \nu \nu & \to \nu \nu.
\end{aligned}    
\end{equation}
As the universe expands and cools, these interactions become less frequent.  When the universe cools to a temperature of ${\cal O}(1)$~MeV, the interaction rate per neutrino $\Gamma$ drops below the Hubble expansion rate $H$.  From this point onwards, the neutrinos are said to be ``decoupled'' from the thermal plasma.  The typical energy of the neutrino ensemble at decoupling is $E \sim 3 T \gg m_\nu$, i.e., the ensemble is largely ultra-relativistic at decoupling. Because of this, the neutrinos retain to a high degree of accuracy their relativistic Fermi-Dirac phase space distribution parametrized by a temperature and possibly a nonzero chemical potential.

Shortly after neutrino decoupling, the $e^+e^-$ plasma becomes non-relativistic at $T \sim 0.5$~MeV.  Here, kinematics favor the annihilation of $e^+e^-$ pairs.  The energy released in this process ``reheats'' the photon population.  The neutrino population, however, does not feel this reheating, because they have already decoupled: in other words, the weak interaction processes are no longer efficient at transporting the energy released from the annihilation to the neutrino sector.  The net effect is that the neutrinos emerge from the annihilation event a little colder than the photons. Assuming (i)~ideal gases, (ii)~instantaneous neutrino decoupling, and (iii)~ultra-relativistic electrons/positrons at neutrino decoupling, one can show using entropy conservation arguments
\begin{equation}
\label{eq:nutemp}
T_\nu = \left(\frac{4}{11} \right)^{1/3} T_\gamma,
\end{equation}
where $T_\nu$ and $T_\gamma$ are the neutrino and the photon temperatures, respectively, after $e^+e^-$ annihilation.

Eq.~\ref{eq:nutemp} is generally taken to define the standard neutrino temperature
after $e^+e^-$ annihilation.  In reality, however, dropping any one of the three aforementioned assumptions can result in percent-level corrections to the relation.  These corrections are commonly absorbed into the definition of the effective number of neutrinos $N_{\rm eff}$ via
\begin{equation}
\label{eq:neff}
\rho_\nu = N_{\rm eff} \, \frac{7 \pi^2}{120} \, T_\nu^4 = N_{\rm eff} \, \frac{7}{8} \left(\frac{4}{11} \right)^{4/3} \rho_\gamma.    
\end{equation}
Here, $\rho_\nu$ is the actual total neutrino energy density after $e^+e^-$ annihilation (but still deep in the radiation domination era when the neutrinos are ultra-relativistic), the quantity $(7 \pi^2/120)T_\nu^4$ denotes the energy density in one family of thermally-distributed relativistic neutrinos with a temperature $T_\nu$ defined in Eq.~\ref{eq:nutemp}, and $\rho_\gamma$ is the actual energy density in the photon population, i.e., what eventually becomes the cosmic microwave background (CMB) radiation.  Precision calculations of the SM prediction of $N_{\rm eff}$ find $N_{\rm eff}^{\rm SM} = 3.0440 \pm 0.0002$~\cite{Froustey:2020mcq,Bennett:2020zkv}, including the effects summarized in Tab.~\ref{tab:Split}.\footnote{Note that because the parameter $N_{\rm eff}$ is defined in relation to the neutrino energy density in an epoch when the neutrinos are ultra-relativistic, there is no ambiguity in the definition (Eq.~\ref{eq:neff}) even if the neutrinos
should become non-relativistic at a later time because of their nonzero masses, provided the masses do not exceed the eV scale.}

\begin{table}[t]
\centering
\begin{tabular}{|l|c|}
\hline
Standard-model corrections to $N_{\rm eff}^{\rm SM}$ & Leading-digit contribution \\
\hline
$m_e/T_d$ correction& $+0.04$ \\
$\mathcal{O}(e^2)$ FTQED correction to the QED EoS& $+0.01$\\
Non-instantaneous decoupling+spectral distortion & $-0.005$\\
$\mathcal{O}(e^3)$ FTQED correction to the QED EoS& $-0.001$\\
Neutrino flavor oscillations & $+0.0005$\\
Type (a) FTQED corrections to the weak rates & $\lesssim 10^{-4}$\\
\hline
\end{tabular}
\caption{Leading-digit contributions from various SM corrections, in order of importance, that make up the final $N_{\rm eff}^{\rm SM}-3$ (adapted from Ref.~\cite{Bennett:2020zkv}). The largest, $m_e/T_d$ correction results from dropping the assumption of an ultra-relativistic electron-positron population; finite-temperature quantum electrodynamics corrections (FTQED) to the QED equation of state (EoS) enter at ${\cal O}(e^2)$ and ${\cal O}(e^3)$, where $e$ is the elementary electric charge; the non-instantaneous decoupling+spectral distortion correction is defined relative to an estimate of $N_{\rm eff}^{\rm SM}$ in the limit of instantaneous decoupling assuming $T_d = 1.3453$~MeV~\cite{Bennett:2019ewm}; and Type (a) FTQED corrections to the weak rates refer to neutrino-electron scattering rates corrected with thermal masses.}
\label{tab:Split}
\end{table}

The effective number of neutrinos $N_{\rm eff}$ is of interest to cosmology primarily because energy density in relativistic particles affect directly the Hubble expansion expansion rate during the radiation domination era. In the epoch after $e^+e^-$ annihilation, the expansion rate is given by
\begin{equation}
\label{eq:hubble}
H^2(t)  = \frac{8 \pi G}{3} \left(\rho_\gamma + \rho_\nu \right),
\end{equation}
where $G$ is the gravitational constant.  With the photon energy density $\rho_\gamma$ having been measured to better than $0.1$\% accuracy by the FIRAS instrument on board COBE~\cite{Fixsen:1996nj,Fixsen:2009ug}, 
constraints on $H(t)$ in the early universe can be interpreted as bounds on the ratio $\rho_\nu/\rho_\gamma$ and hence $N_{\rm eff}$.

From a particle physics standpoint, a thermal
population of light sterile neutrinos is one possible cause of a $N_{\rm eff}$ that differs from the standard $N_{\rm eff}^{\rm SM}$ value.   However, it is
important to emphasize that, as far as the Hubble expansion rate $H(t)$ is concerned, any thermal background or non-thermal population (e.g., from decays) of non-photon light particles such as axions,  majorons, or even gravitons will contribute to $N_{\rm eff}$. Such scenarios have been considered by many authors, including Refs.~\cite{Hasenkamp:2011em,Hasenkamp:2014hma,Chen:2015dka,Roland:2016gli,Shakya:2015xnx,Shakya:2016oxf,Shakya:2018qzg}.
   Likewise, any process that alters the thermal abundance
of neutrinos (e.g., a low reheating temperature) or affects directly the expansion rate itself (e.g., a time-dependent gravitational constant $G$) can mimic a non-standard $N_{\rm eff}$ value. Yet another way to change $N_{\rm eff}$ is to tinker with the photon energy density itself (via, e.g., interaction with millicharged particles~\cite{Davidson:2000hf} or late kinetic decoupling of the dark matter~\cite{Diacoumis:2018nbq}), 
while preserving the neutrino energy density.

In the case of a non-standard $N_{\rm eff}$ due to a BSM light particle, the mass of the new particle can also impact on the evolution of cosmological density perturbations via its role as a ``hot dark matter''; the mathematical description of this effect goes beyond the Hubble expansion rate (Eq.~\ref{eq:hubble}).

In the following, we describe first how a thermal population of light sterile neutrinos can arise in the early universe through a combination of neutrino flavor oscillations and scattering with other SM particles.
We then discuss the various signatures of light sterile neutrinos in cosmological observables such as the light elemental abundances from big bang nucleosynthesis (BBN), the CMB anisotropies and the large-scale structure (LSS) distribution, constraints from current observations, as well as various proposals on how to get around them.

\subsubsection{Light Sterile Neutrino Thermalization}
\label{sec:thermalise}

If a sterile neutrino state mixes sufficiently strongly with any of the active  neutrino states, a thermal population of light sterile neutrinos that adds to $N_{\rm eff}$ can be produced prior to neutrino decoupling via a combination of active-sterile neutrino oscillations and collisions (i.e., elastic and inelastic scattering) with the primordial plasma of SM particles.  Roughly speaking, as the universe cools, an initial population comprising only active neutrinos can begin to oscillate into sterile neutrinos once the oscillation frequency, given by $\Delta m^2/(2 E)$, becomes larger than the Hubble expansion rate $H(t)$.  The role of collisions is then to force a neutrino into a flavor eigenstate and hence ``measure'' the flavour content of the ensemble.  Since the probability of measuring a sterile flavor is nonzero, signifying that an active neutrino has turned into a sterile state, collisions also play the role of refilling any gap in the active neutrino distribution vacated by the oscillation process.
This effect and the region of parameter
space leading to thermalization of the sterile neutrino was found by the early works of, e.g., \cite{Kainulainen:1990ds,Enqvist:1991qj,Cline:1991zb,Shi:1993hm}
If sterile neutrinos do become thermalized, then we expect them to have the same temperature as the
active neutrinos.

Nowadays, light sterile neutrino thermalization in the early universe can be computed precisely, using a generalized Boltzmann formalism developed in~\cite{Sigl:1993ctk,McKellar:1992ja} which tracks the flavor evolution of a neutrino ensemble under the influence of neutrino oscillations and scattering.  Schematically, the generalized Boltzmann equation for the one-particle reduced density matrix of the neutrino ensemble, $\varrho(t,p)$, in the Friedmann-Lema\^{\i}tre-Robertson-Walker (FLRW) universe of standard hot big bang cosmology is given by
\begin{equation}
\partial_t \varrho - p H \partial_p \varrho = -{\rm i} [\mathbb{H},\varrho] + {\cal I}[\varrho],
\end{equation}
where $\partial_t$ and $\partial_p$ are partial derivatives with respect to the cosmic time~$t$ and physical momentum~$p$ respectively, $H$ is the Hubble expansion rate, $[\mathbb{H},\varrho] \equiv \mathbb{H} \varrho-\varrho \mathbb{H}$ denotes a commutator between the flavor oscillations Hamiltonian~$\mathbb{H}$ and~$\varrho$, and the collision integrals ${\cal I}[\varrho]$ encapsulate all non-unitary (scattering) effects on $\varrho$.  In the fully CP symmetric case, one set of density matrix $\varrho(t,p)$ suffices to describe the evolution of the whole neutrino ensemble including antineutrinos. If however the system is CP asymmetric, we will need to introduce a 
 separate one-particle reduced density matrix $\bar{\varrho}(t,p)$ for the antineutrino ensemble.  Here, we follow the convention of Ref.~\cite{Sigl:1993ctk}, and define the density matrices using the ``transposed notation, e.g., for a 3 active+1 sterile system one would have 
 \begin{equation}
\varrho(t,p) = \left(\begin{array}{cccc}
\varrho_{ee} & \varrho_{e \mu} & \varrho_{e \tau} & \varrho_{es} \\
\varrho_{\mu e} & \varrho_{\mu \mu} & \varrho_{\mu \tau} & \varrho_{\mu s}\\
\varrho_{\tau e} & \varrho_{\tau \mu} & \varrho_{\tau \tau} & \varrho_{\tau s} \\
\varrho_{s e} & \varrho_{s \mu} & \varrho_{s \tau} & \varrho_{s s} \\
\end{array}
\right), \qquad
\bar{\varrho}(t,p) = \left(\begin{array}{cccc}
\bar{\varrho}_{ee} & \bar{\varrho}_{\mu e} & \bar{\varrho}_{\tau e} & \bar{\varrho}_{se} \\
\bar{\varrho}_{e\mu} & \bar{\varrho}_{\mu \mu} & \bar{\varrho}_{\tau\mu} & \bar{\varrho}_{s \mu}\\
\bar{\varrho}_{e \tau} & \bar{\varrho}_{\mu \tau} & \bar{\varrho}_{\tau \tau} & \bar{\varrho}_{s \tau} \\
\bar{\varrho}_{e s} & \bar{\varrho}_{\mu s} & \bar{\varrho}_{\tau s} & \bar{\varrho}_{s s} \\
\end{array}
\right) .
 \end{equation}
 This convention enables the equations of motion to be expressed in a more compact form.

Working in the flavor basis, the oscillations Hamiltonian under this convention is
\begin{equation}
\begin{aligned}
\mathbb{H} (p,t,T(t)) =\, &   \pm \frac{U\mathbb{M}U^\dagger}{2p} + \sqrt{2} G_F \left[\mathbb{N}_\ell(T(t)) - \mathbb{N}_{\bar{\ell}}(T(t))+ \mathbb{N}_\nu(t)-\mathbb{N}_{\bar\nu}(t) \right] \\
& \mp 2\sqrt{2}G_F p
\left[
\frac{{\mathbb{E}}_\ell(T(t))+{\mathbb{E}}_{\bar\ell}(T(t))+{\mathbb{P}}_\ell(T(t))+{\mathbb{P}}_{\bar\ell}(T(t))}{m_W^2}
+
\frac{4}{3} \frac{{\mathbb{E}}_\nu(t)+{\mathbb{E}}_{\bar\nu}(t)}{m_Z^2}
\right],
\label{eq:hamiltonian}
\end{aligned}
\end{equation}
which contains a vacuum and an in-medium part.
The vacuum part consists of the neutrino squared-mass difference matrix $\mathbb{M}$ and the vacuum mixing matrix $U$, with the ``$+$'' sign pertaining to neutrinos and the ``$-$'' sign to the antineutrinos.
The in-medium part proportional to the Fermi constant $G_F$, i.e., the ``matter potential'', contains a CP symmetric and a CP asymmetric correction to the neutrino dispersion relation.  The CP asymmetric part is similar to the usual matter potential found in, e.g., the Sun, with the matrix $\mathbb{N}_\ell-\mathbb{N}_{\bar{\ell}}$ denoting the asymmetry in the number density of charged leptons; in a 3+1 system $\mathbb{N}_\ell$ would take the form
\begin{equation}
\mathbb{N}_\ell \equiv \frac{1}{2 \pi^2} \int {\rm d} p \, p^2 \, {\rm diag}\Big(f_e,f_\mu, f_\tau,0  \Big) = {\rm diag}(n_e,n_\mu,n_\tau,0),
\end{equation}
where $f_\ell(p,T)$ is the (equilibrium) occupation number of the charged lepton $\ell$.  A similar expression exists also for $\mathbb{N}_{\bar{\ell}}$.  Typically, the charged-lepton asymmetries are of order $10^{-10}$, so this $\mathbb{N}_\ell-\mathbb{N}_{\bar{\ell}}$ term is not strictly a necessary ingredient.
Observe however the additional term proportional to ${\mathbb{N}}_\nu-{\mathbb{N}}_{\bar\nu}$, where
\begin{equation}
\mathbb{N}_\nu\equiv\frac{1}{2 \pi^2} \int \mathrm{d}p\,p^3 \, S_a \varrho S_a,
\end{equation}
and similarly for $\mathbb{N}_{\bar\nu}$.
This CP asymmetric term describes neutrino self-interaction in the presence of a large ($>10^{-5}$) neutrino asymmetry.  Current cosmological data constrain neutrino asymmetries only to  ${\cal O}(10^{-2})$~\cite{Oldengott:2017tzj}, so the presence of a sizeable $\mathbb{N}_\nu-\mathbb{N}_{\bar{\nu}}$   term is an interesting possibility.  In standard calculations of light sterile neutrino thermalization, however, this asymmetry is set to zero.
The quantity $S_a$ is a diagonal matrix that projects out only the active neutrino states, since sterile states are by definition interaction-less.  In a 3+1 system it takes the form $S_a = {\rm diag}(1,1,1,0)$.

In the CP symmetric portion of the in-medium terms,
$m_W$ and $m_Z$ denote respectively the $W$ and $Z$ boson mass, while the terms $\mathbb{E}_\ell$ and $\mathbb{P}_\ell$ are momentum-integrals of some combinations of the charged-lepton energy $E_\ell(p)\equiv (p^2+ m_\ell^2)^{1/2}$ and occupation number~$f_\ell(p,T)$, and $\mathbb{E}_{\bar\ell}$, $\mathbb{P}_{\bar\ell}$ are their antiparticle counterparts.  For a 3+1 system, we have
\begin{eqnarray}
\mathbb{E}_\ell & \equiv& \frac{1}{2 \pi^2} \int {\rm d} p \, p^2 \, {\rm diag}\Big(E_e f_e,E_\mu f_\mu, E_\tau f_\tau,0  \Big)
=
{\rm diag}(\rho_e,\rho_\mu,\rho_\tau,0),\label{eq:eell}\\
\mathbb{P}_\ell &\equiv& \frac{1}{6 \pi^2} \int {\rm d} p \, p^2 \, {\rm diag}\Bigg(\frac{p^2}{E_e} f_e,\frac{p^2}{E_\mu} f_\mu,\frac{p^2}{E_\tau} f_\tau,0\Bigg)
={\rm diag}(P_e,P_\mu,P_\tau, 0), \label{eq:pell}
\end{eqnarray}
where $\rho_\ell$ and $P_\ell$ are, respectively, the energy density and pressure of the charged lepton $\ell$.  Lastly, $\mathbb{E}_\nu$ is the equivalent of Eq.~\ref{eq:eell} for an ultra-relativistic neutrino gas, 
\begin{equation}
\label{eq:nurhosum}
\mathbb{E}_\nu\equiv\frac{1}{2 \pi^2} \int \mathrm{d}p\,p^3 \, S_a\varrho S_a,
\end{equation}
where $S_a$ is again a diagonal matrix that projects out only the active neutrino states.

Note that the $\mathbb{E}_\ell +\mathbb{P}_\ell$ term in Eq.~\ref{eq:hamiltonian} differs from its usual presentation found in, e.g., equation~(2.2) of~\cite{Gariazzo:2019gyi}, which has $\mathbb{E}_\ell +\mathbb{P}_\ell$ replaced with $(4/3)\, \mathbb{E}_\ell$.
First reported in~\cite{Notzold:1987ik}, the former is in fact the more general result, while $(4/3)\, \mathbb{E}_\ell$ applies strictly only when the charged leptons are ultra-relativistic.

The collision integral ${\cal I}[\varrho]$ incorporates in principle all weak scattering processes wherein at least one neutrino appears in either the initial or final state.
All published calculations to date, however, account only for $2 \to 2$ processes involving
(i) two neutrinos and two charged leptons anyway distributed in the initial and final states, and
(ii) neutrino-neutrino scattering.
Then, schematically, ${\cal I}[\varrho]$ comprises 9D momentum-integrals that, at tree level, can be systematically reduced to 2D integrals of the form
\begin{equation}
{\cal I}[\varrho(p,t)]
\propto G_F^2 \int {\rm d}p_2\, {\rm d}p_3\, \Pi(p,p_2,p_3;t) \, F(\varrho), \label{eq:collision1}
\end{equation}
where $\Pi$ is a scalar function representing the scattering kernel, and $F$ is a phase space matrix including quantum statistics.

The vast majority of existing works on light sterile neutrino thermalization solve the above set of equations of motion for a 1 active+1 sterile system. Because of the complexity of the collision integrals~${\cal I}[\varrho]$, a variety of approximations have been introduced to simplify the integrals
and hence speed up the computation~\cite{Hannestad:2012ky,McKellar:1992ja,Bell:1998ds,Kainulainen:2001cb,Chu:2006ua,Hannestad:2015tea}, although of course it is also possible to solve the collision integrals exactly if percent-level precision is required~\cite{Hannestad:2015tea}.
The publicly available code {\tt LASAGNA}~\cite{Hannestad:2013pha} solves a 1+1 system and also allows for the possibility of a CP asymmetry.  On the other hand, 
multi-flavor (i.e., 3+1) effects have been  considered in Ref.~\cite{Melchiorri:2008gq} using the momentum-averaged approximation and in Ref.~\cite{Saviano:2013ktj} using a multi-momentum approach.  Most recently, a publicly available multi-flavor code~{\tt FortEPiaNO}~\cite{Gariazzo:2019gyi} has also become available, which is capable of solving a system with up to 3 active+3 sterile species, although it does not provide the CP asymmetric case.

\begin{figure}[t]
    \centering
    \includegraphics[width=0.6\textwidth]{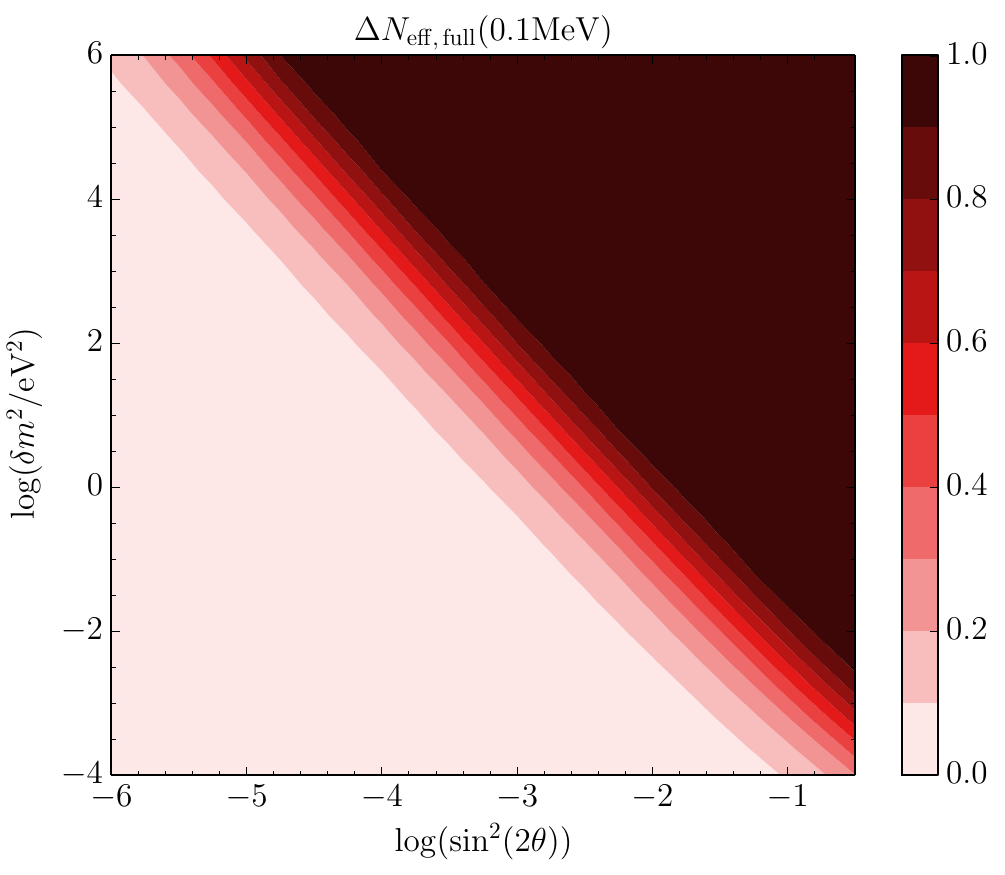}
    \caption{\label{fig:thermal1+1} The change in the effective number of neutrinos, $\Delta N_{\rm eff}  \equiv N_{\rm eff} - N_{\rm eff}^{\rm SM}$, at a temperature $T = 0.1$~MeV from active-sterile neutrino oscillations in a 1+1 system, where $\delta m^2$ is the squared mass difference and $\theta$ is the effective mixing angle between the active and the sterile state.  Figure taken from Ref.~\cite{Hannestad:2015tea}.}
\end{figure}

\begin{figure}[p]
    \centering
    \includegraphics[width=0.6\textwidth]{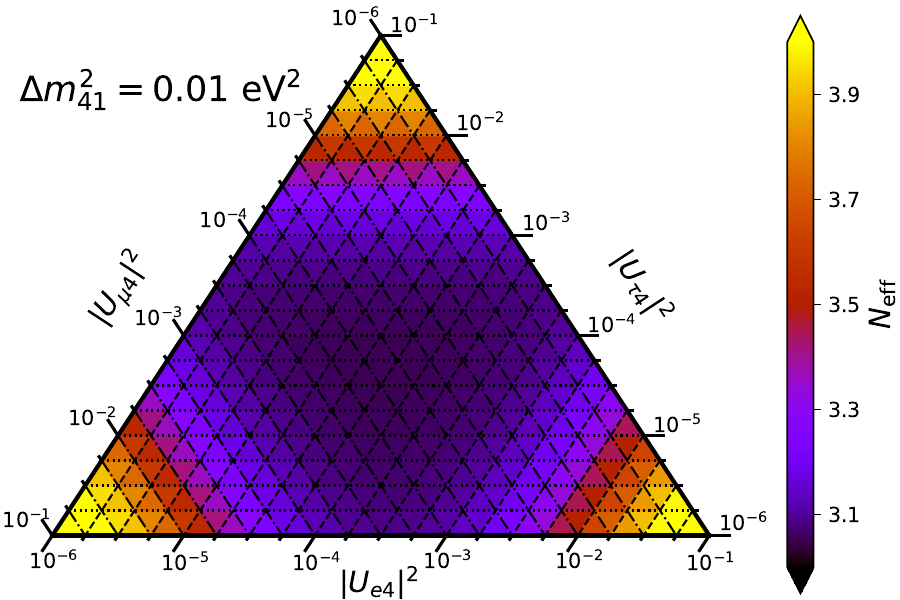}
    \includegraphics[width=0.6\textwidth]{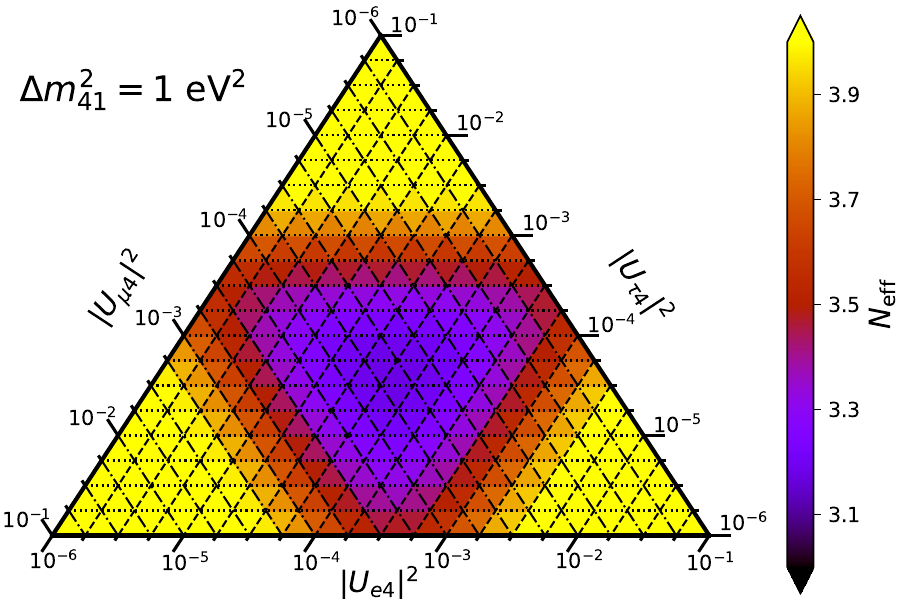}
    \includegraphics[width=0.6\textwidth]{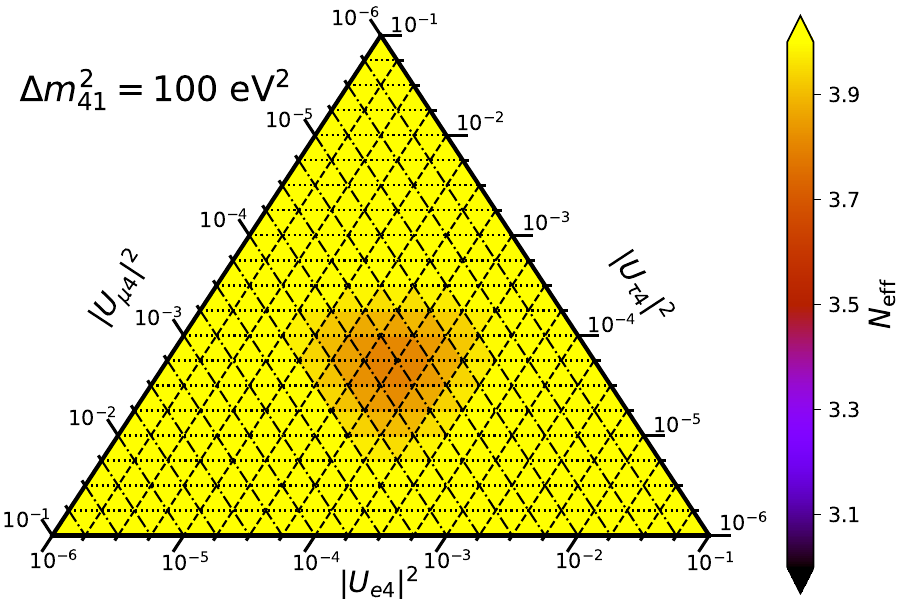}
    \caption{\label{fig:thermal3+1} The final effective number of neutrinos from active-sterile neutrino oscillations in a 3+1 scheme, under the constraint $\sum \log_{10} |U_{i4}|^2 = -13$.  Figure taken from Ref.~\cite{Gariazzo:2019gyi}.}
\end{figure}

Irrespective of how exactly the system is solved (3+1, 1+1, approximate or exact collision integrals, etc.), however, the general conclusion is that the active-sterile neutrino squared mass difference  and mixing required to explain the short-baseline anomalies will inevitably lead to a fully thermalized light sterile neutrino population with a temperature similar to that of the active neutrinos.  In other words, in terms of the $N_{\rm eff}$ parameter, we expect the canonical light sterile neutrino solution to the short-baseline anomalies to lead to $N_{\rm eff} \sim 4$. Fig.~\ref{fig:thermal1+1} and ~\ref{fig:thermal3+1} show the prediction for $N_{\rm eff}$ as a function of the mixing parameters in a 1+1 and 3+1 scheme respectively.  We discuss in the next section the observable consequences of this light sterile neutrino population.

\subsubsection{Observable Consequences}

Three standard cosmological probes are sensitive to the presence of a light sterile neutrino population primarily through its contribution to increasing $N_{\rm eff}$ and hence the Hubble expansion rate, and/or through its nonzero mass and hence its role as a ``hot dark matter''.  We describe these probes below.

\subsubsubsection{Big Bang Nucleosynthesis}  

The discussion of light element formation from protons and neutrons via the process of big bang nucleosynthesis (BBN) (see Ref.~\cite{Pitrou:2018cgg} for a recent review) usually begins at a temperature around $T \sim 0.7$~MeV, when the weak processes 
\begin{equation}
    \begin{aligned}
    \label{eq:betaprocess}
         \nu_e + n & \leftrightarrow e^- + p,\\
        \bar{\nu}_e + p &\leftrightarrow e^+ + n 
    \end{aligned}
\end{equation}
become inefficient compared with the Hubble expansion rate and equilibrium can no longer be maintained.  When these process go out of equilibrium, the ratio of neutron-to-proton number density also freezes out, to a value given by
\begin{equation}
\label{eq:np}
\left. \frac{n_n}{n_p} \right|_{\rm fr}   = \exp \left(- \frac{m_n-m_p}{T_{\rm fr}} \right),
\end{equation}
where $T_{\rm fr}$ is the freeze-out temperature, and $m_{n,p}$ are the neutron and proton masses.  For $T_{\rm fr} \simeq 0.7$~MeV, this ratio evaluates approximately to $1/6$. Free neutron decay over a lifetime of about 880~s, however, will reduce it to a smaller number by the end of BBN.

In standard BBN, the formation of elements commences at a temperature controlled by the baryon-to-photon ratio $\eta$, when the energy in the photon bath per baryon has become sufficiently low such that newly formed nuclei are no longer immediately broken apart.  For $\eta \sim 6 \times 10^{-10}$, this temperature is around $T \sim 0.1$~MeV.
The first element to be formed is Deuterium, followed by the production of heavier nuclei. Of particular note is Helium-4, whose mass fraction is defined as
\begin{equation}
\label{eq:yhe}
Y_p \equiv \frac{4 n_{\rm He 4}}{n_n+n_p},    
\end{equation}
where $n_{\rm He4}$ is the number density of $^4$He.  Because $^4$He has the largest binding energy amongst the light elements, the bulk of all initially available neutrons will eventually end up bound in Helium-4 nuclei,
i.e., we expect $n_{\rm He 4} \simeq n_n/2$.  Then, to estimate $Y_p$ from Eq.~\ref{eq:yhe} we simply need to note that the neutron-to-proton ratio typically drops to about $1/7$ at the end of BBN via neutron decay.  From this we find $Y_p \simeq 0.25$.  

Besides Helium-4 mass fraction, small amounts of Deuterium and $^3$He (D/H $\sim ^3$He/H $\sim O(10^{-5})$), 
as well as traces of $^6$Li and $^7$Li, are expected to remain.  Unlike for $^4$He, however, there are no simple ways to estimate their abundances, and we must rely on solving a set of Boltzmann equations to track their number densities.  Several publicly available codes can perform this task, including {\tt AlterBBN}~\cite{Arbey:2011nf,Arbey:2018zfh}, {\tt PArthENoPE}~\cite{Pisanti:2007hk,Consiglio:2017pot}, and {\tt PRIMAT}~\cite{Pitrou:2018cgg}.  In standard BBN, barring experimental uncertainties in the nuclear reaction rates and the free neutron lifetime, the baryon-to-photon ratio $\eta$ alone enters these Boltzmann equations and is hence the sole free parameter in the determination of the elemental abundances.

A non-standard neutrino sector can alter this picture in two different ways.  Firstly, as can be seen in Eq.~\ref{eq:betaprocess}, electron neutrinos participate directly in the CC weak interactions that determine the neutron-to-proton ratio.  If because of non-standard physics these neutrinos should end up at $T \lesssim 1$~MeV with an energy spectrum that departs strongly from an equilibrium relativistic Fermi-Dirac distribution with zero chemical potential, then the equilibrium of the processes (Eq.~\ref{eq:betaprocess}) could shift to a different point and in so doing alter the neutron-to-proton ratio.  A particularly well studied example in this regard is the case of a nonzero electron neutrino chemical potential $\mu_e$, which shifts the neutron-to-proton ratio at weak freeze-out in a manner well described by
\begin{equation}
\left. \frac{n_n}{n_p} \right|_{\rm fr}   = \exp \left(- \frac{m_n-m_p}{T_{\rm fr}}-\frac{\mu_e}{T_{\rm fr}} \right).
\end{equation}
The effects of more general distortions to the $\nu_e$ and/or $\bar{\nu}_e$ energy spectra on $n_n/n_p$ need to be computed numerically  using a Boltzmann code.

\begin{figure}[t]
    \centering
      \includegraphics[angle=-90,width=0.49\textwidth]{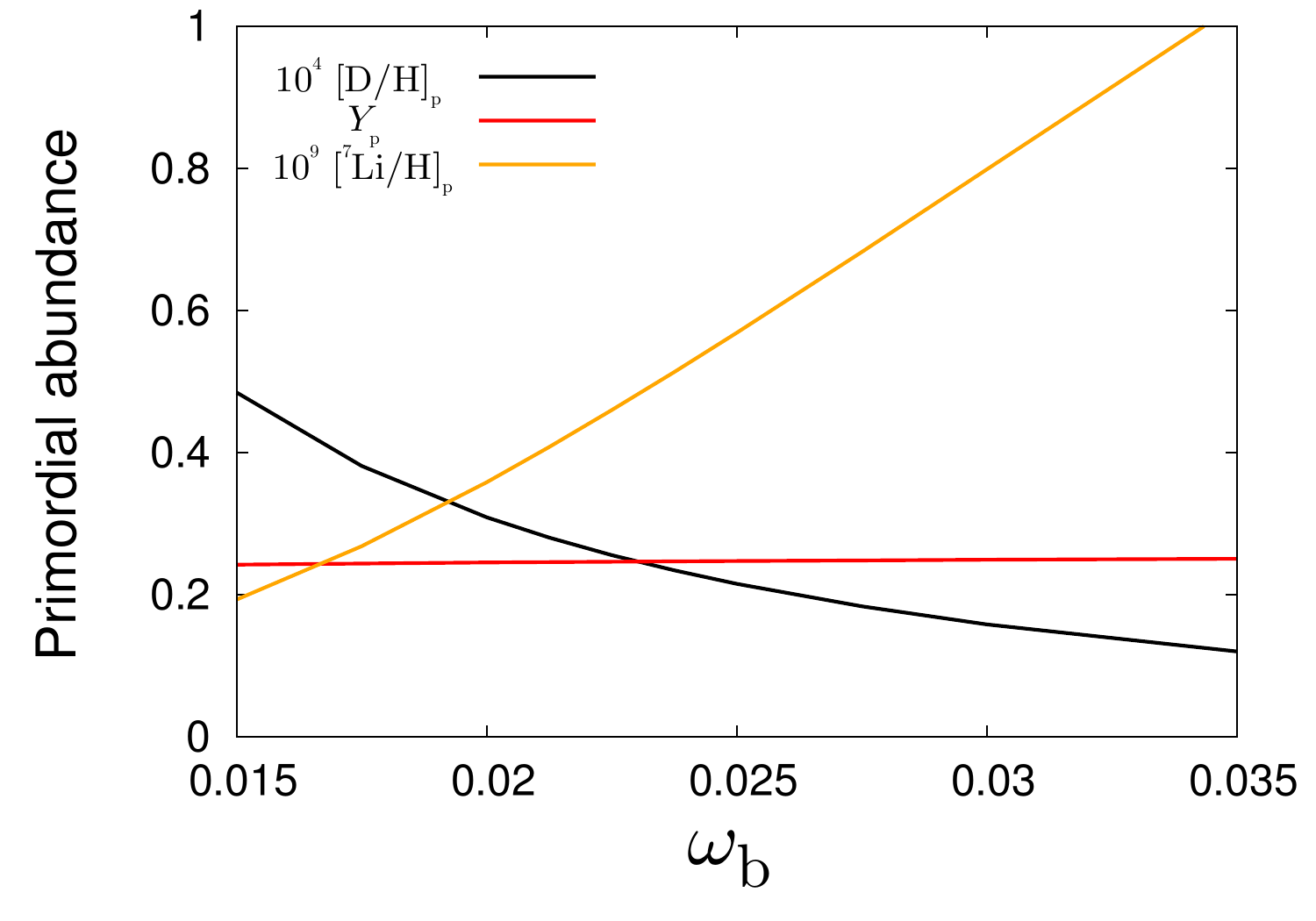}
    \includegraphics[angle=-90,width=0.49\textwidth]{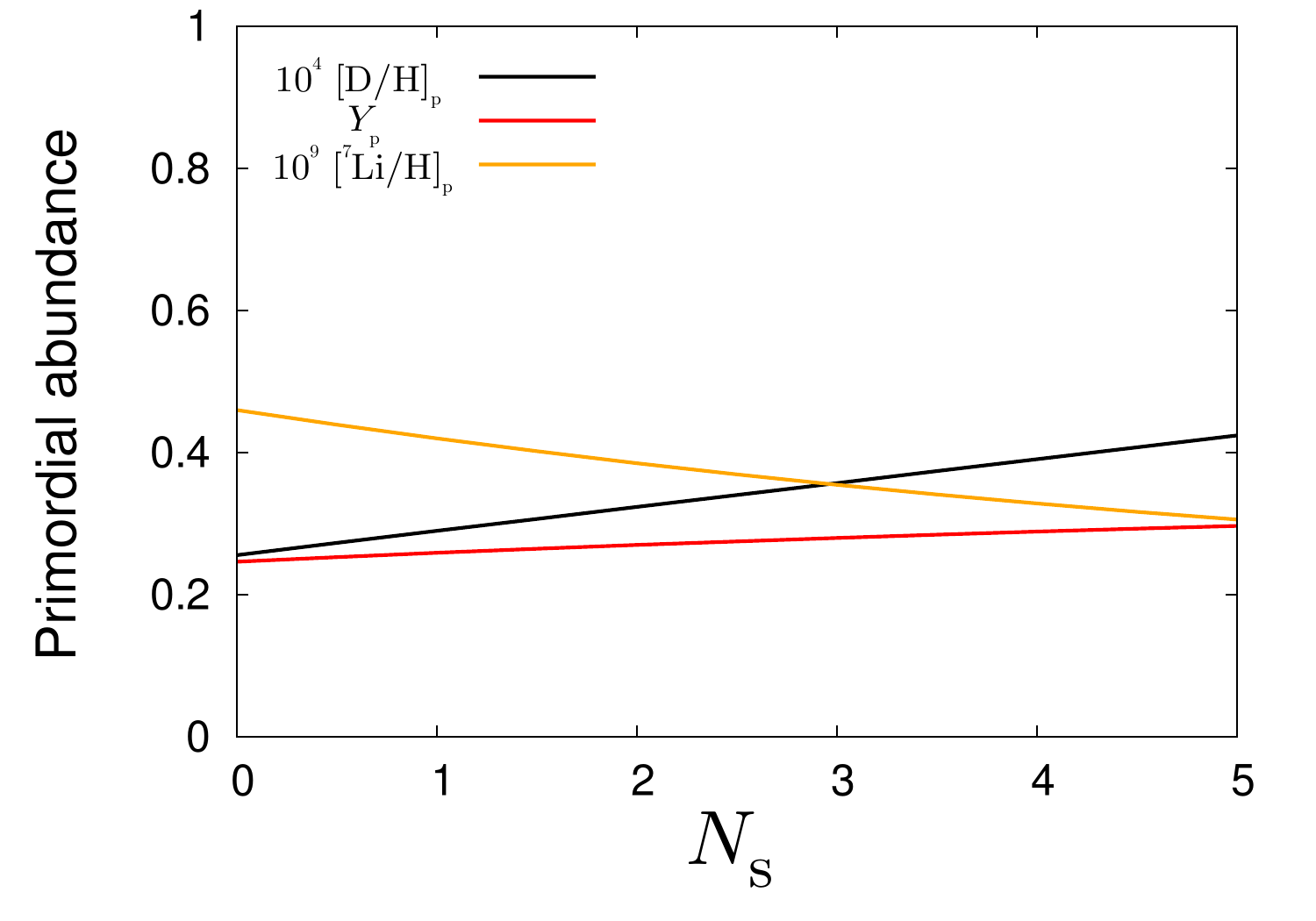}
      \includegraphics[angle=-90,width=0.49\textwidth]{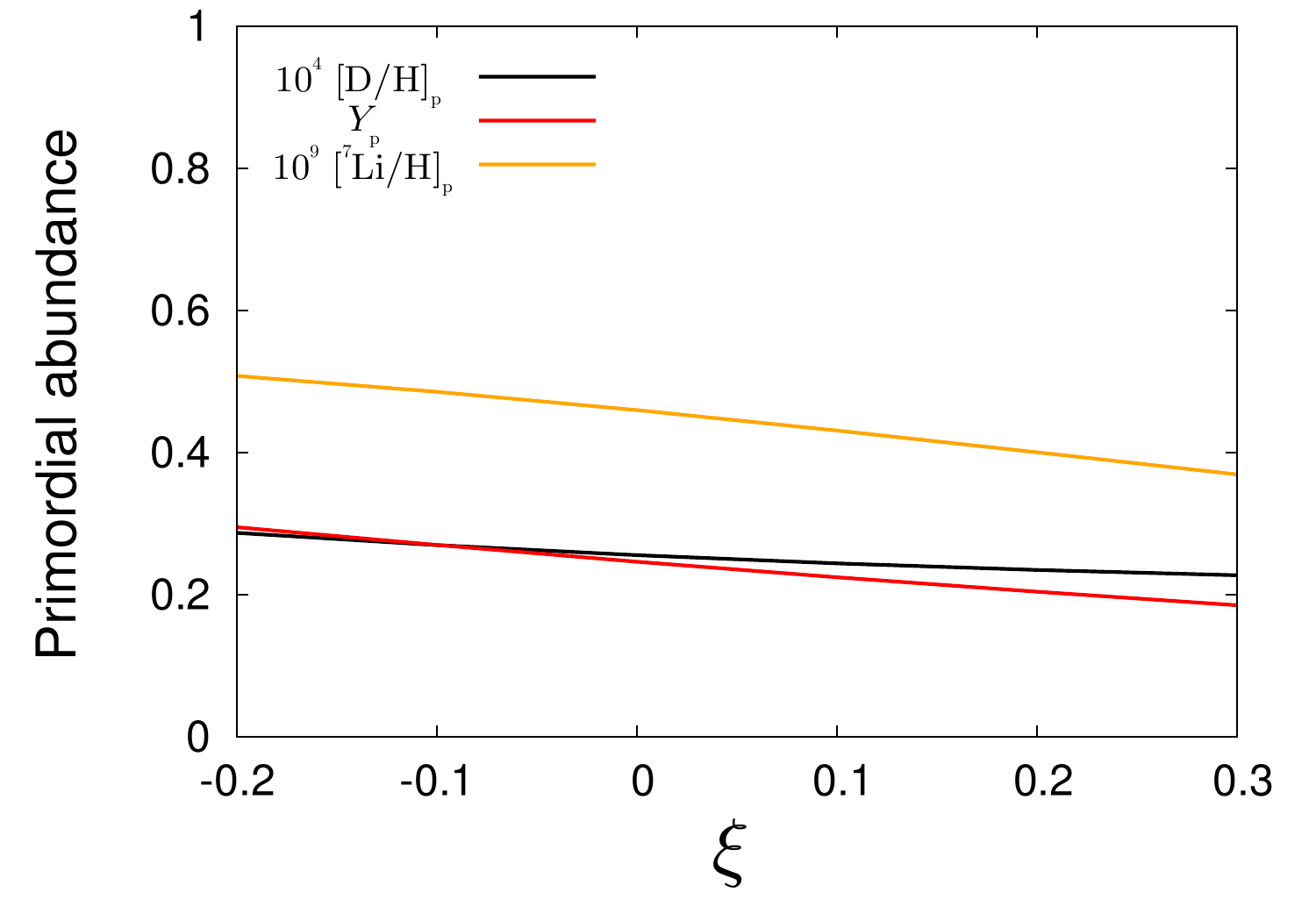}
    \caption{\label{fig:bbn} Deuterium, Helium-4 and Lithium-7 abundances, as computed using {\tt PArthENoPE}~\cite{Pisanti:2007hk,Consiglio:2017pot}, as a function of cosmological parameters: $\omega_b$ is the baryon density, $N_s$ is our $\Delta N_{\rm eff}  \equiv N_{\rm eff} - N_{\rm eff}^{\rm SM}$, and $\xi \equiv \mu_e/T$ is the electron neutrino degeneracy parameter.
    Figure taken from Ref.~\cite{Hamann:2011ge}.}
\end{figure}

Secondly, neutrinos of all flavors influence the expansion rate of the universe prior to and during BBN through their energy densities via Eq.~\ref{eq:hubble}.  Therefore, if because of new physics the total neutrino energy should be larger than the standard $N_{\rm eff}^{\rm SM}$, the freeze-out of the processes (Eq.~\ref{eq:betaprocess}) would occur at a higher temperature.  This in turn pushes up the neutron-to-proton ratio via Eq.~\ref{eq:np} and hence the Helium-4 mass fraction as well via Eq.~\ref{eq:yhe}.
Fig.~\ref{fig:bbn} shows the Deuterium, Helium-4 and Lithium-7 abundances computed using {\tt  PArthENoPE}~\cite{Pisanti:2007hk,Consiglio:2017pot}, as a function of the baryon density $\omega_b$, the excess number of relativistic degrees of freedom $N_s \equiv \Delta N_{\rm eff}  \equiv N_{\rm eff} - N_{\rm eff}^{\rm SM}$, and the electron neutrino degeneracy parameter  $\xi \equiv \mu_e/T$.

For the specific problem of a light sterile neutrino with  mixing parameters compatible with hints from terrestrial experiments, only the second effect is relevant. This is because the relatively large mass-squared difference and mixing between the active and the sterile neutrino states essentially guarantee full thermalization of the sterile species prior to the decoupling of the active neutrinos (see Fig.~\ref{fig:thermal1+1} and ~\ref{fig:thermal3+1}).   In other words, $\Delta N_{\rm eff} \simeq 1$ and the phase space distribution of the sterile states follows closely the equilibrium relativistic Fermi-Dirac distribution of the active neutrinos. This also implies equipartition amongst the four neutrino flavors, such that any further active-sterile flavor oscillations after neutrino decoupling will not cause the $\nu_e$ phase space distribution to deviate from a thermal distribution.  Thus, to constrain such light sterile neutrino scenarios, we only need to extend standard BBN with one extra free parameter, namely, $N_{\rm eff}$.  We emphasize, however, that more general cases of active-sterile neutrino mixing would require that we determine the $\nu_e$ energy spectrum as well, in order to determine the full effect of light sterile states on the light elemental abundances.

\begin{figure}[t]
    \centering
      \includegraphics[angle=0,width=0.55\textwidth]{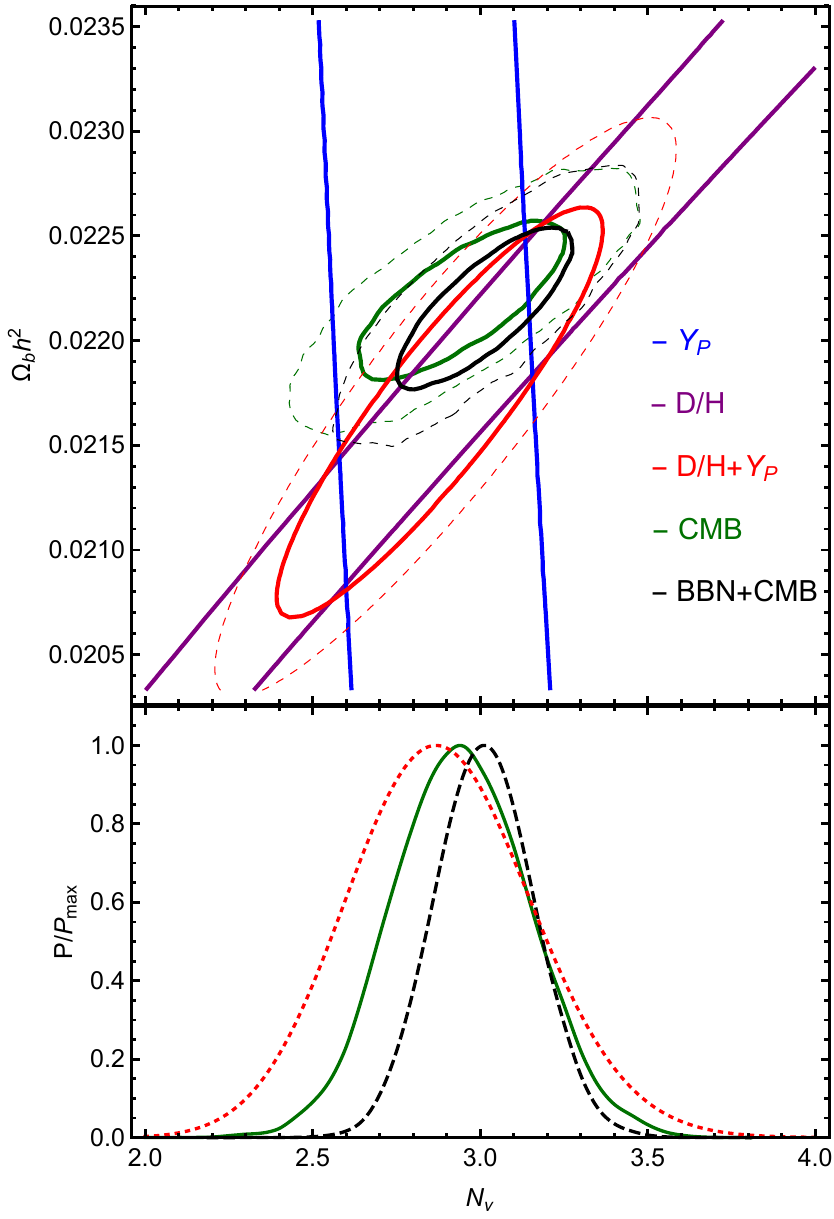}
    \caption{\label{fig:bbnconstraints} {\it Top}: 68.27\% and 95.45\% contours in the $(\Omega_b h^2, N_{\rm eff})$-plane obtained from various data combinations.  {\it Bottom}: 1D marginalized posterior for $N_{\rm eff}$.
    Figure taken from Ref.~\cite{Pitrou:2018cgg}.}
\end{figure}

\paragraph{Current BBN constraints} Current astrophysical observations put the Helium-4 mass fraction at 
$Y_p = 0.2449 \pm 0.0040$~\cite{Aver:2015iza}, the Deuterium abundance at ${\rm D}/{\rm H} = (2.527 \pm 0.030) \times 10^{-5}$~\cite{Cooke:2017cwo}, and the Lithium-7 abundance at $^7{\rm Li}/{\rm H} = (1.58 \pm 0.3) \times 10^{-10}$~\cite{Sbordone:2010zi}, while  the Helium-3 abundance is constrained to $^3{\rm He}/{\rm H}  < (1.1 \pm 0.2) \times 10^{-5}$~\cite{Bania:2002yj}.  Of these, only the D/H and $Y_p$ measurements have sufficient precision to probe $N_{\rm eff}$ during the BBN epoch.  How D/H and $Y_p$ probe $N_{\rm eff}$ can be seen in Fig.~\ref{fig:bbnconstraints}, which shows in the top panel the 68.27\% and 95.45\% contours in the $(\Omega_b h^2, N_{\rm eff})$-plane, where $\Omega_b h^2 \equiv \omega_b$ is the baryon density, obtained from 
various data combinations, and in the bottom panel the corresponding 1D marginalized posterior for $N_{\rm eff}$.

As can be seen, the Helium-4 mass fraction alone is already quite sensitive to $N_{\rm eff}$, although this measurement is not particularly useful for pinning down the baryon-to-photon ratio $\eta$ and hence the baryon density $\Omega_b h^2$.  This is because $\eta$ has no strong influence on the rate at which $^4$He is formed, other than setting the initial time of BBN.
In contrast, the Deuterium abundance is strongly sensitive to both $N_{\rm eff}$ and $\eta$, which do directly affect the formation rate. However, because these two measurements have opposite degeneracy directions on the $(\Omega_b h^2, N_{\rm eff})$-plane, in combination they provide a good measurement of both $\Omega_b h^2$ and $N_{\rm eff}$.  For the latter, Ref.~\cite{Pitrou:2018cgg} finds
\begin{equation}
 N_{\rm eff} = 2.88 \pm 0.27, \quad   (68\%~{\rm CL})
\end{equation}
using the code {\tt PRIMAT}~\cite{Pitrou:2018cgg}. Using {\tt PArthENoPE}~\cite{Pisanti:2007hk,Consiglio:2017pot} instead would have yielded a central value about 2\% smaller~\cite{Pitrou:2018cgg}.  
Thus, the conclusion here is that current measurements of primordial elemental abundances are completely consistent with the $N_{\rm eff}^{\rm SM}=3.0440$, and shows no evidence of any extra relativistic degrees of freedom.

\subsubsubsection{Cosmic Microwave Background and Large-Scale Structure} 

Unlike BBN, probes of the universe’s inhomogeneities such as the CMB temperature and polarization anisotropies
and the large-scale matter distribution are not sensitive to the flavor content of the neutrino
sector, only to its contribution to the stress-energy tensor. If neutrinos are massless, then the $N_{\rm eff}$
parameter as defined in Eq.~\ref{eq:neff} alone characterizes their effects on the universe’s evolution.
If neutrinos are massive, then, in addition to $N_{\rm eff}$ and the neutrino masses, in principle it is also necessary to know the exact form of the neutrino
momentum  distribution in order to solve the evolution equations for the inhomogeneities exactly.
However, unless the deviations from a thermal relativistic Fermi-Dirac spectrum is of order unity, it suffices to specify only the temperature of the distribution, as late-time cosmological probes are currently not very sensitive to spectral distortions in the neutrino sector.  For this reason, most existing analyses characterize the neutrino sector, including light sterile neutrinos produced as described in Sec.~\ref{sec:thermalise}, only in terms of the neutrino mass spectrum and the $N_{\rm eff}$ parameter.  The light sterile states are assumed to share the same temperature as the standard active neutrinos.

One interesting variation to the above is the case in which an excess $N_{\rm eff} > N_{\rm eff}^{\rm SM}$ is due to a thermalized particle species that has temperature different from the standard neutrino temperature and/or has a different spin statistics, e.g., a thermalized bosonic particle species such as an axion that follows the Bose-Einstein distribution.  If this new particle is massless, then again it suffices to describe its phenomenology in terms of its contribution to $N_{\rm eff}$ alone.  If however the particle species is massive, then in addition to its mass, its temperature also plays a role in determining the ``hotness'' of the resulting hot dark matter, where the relation between the hot dark matter's temperature and abundance is fixed by the particle's spin statistics.

\paragraph{Cosmic microwave background primary anisotropies}

The CMB primary anisotropies refer to the temperature and polarization fluctuations imprinted on the last scattering surface.  These are sensitive to the physics of the early universe up to the time of photon decoupling ($T \sim 0.1$~eV or redshift $z \sim 1000$), and differ from the secondary anisotropies which are additional spatial fluctuations gathered by the CMB photons as they free-stream from the last scattering surface to the observer and hence sensitive to late-time/low-redshift physics.

\begin{figure}[t]
    \centering
      \includegraphics[angle=0,width=0.52\textwidth]{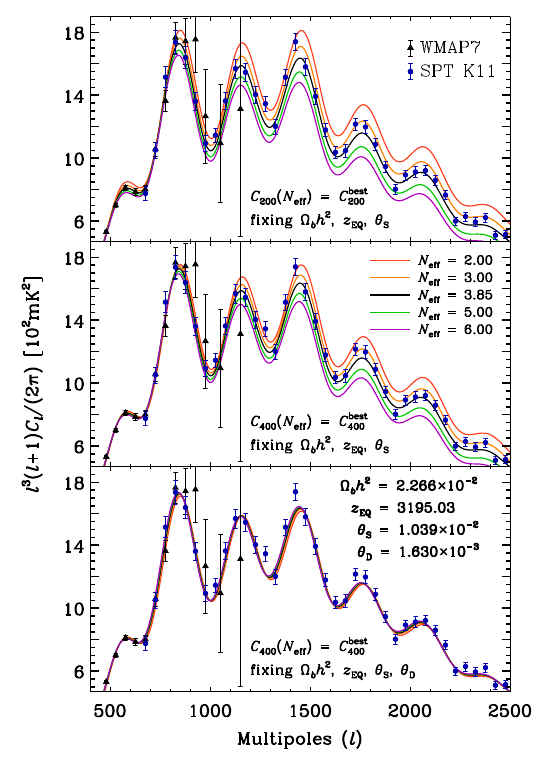}
    \caption{\label{fig:neffcmb} Signature of $N_{\rm eff}$ in the damping tail of the CMB TT power spectrum.  {\it Top}: Here, the cosmological parameters have been adjusted such that the baryon density $\Omega_b h^2$, the redshift of matter-radiation equality $z_{\rm eq}$, the angular sound horizon $\theta_s$, and the normalisation at $\ell=200$ are fixed for all $N_{\rm eff}$ cases shown.  {\it Middle}: Like the top panel, but with the spectrum normalisation fixed at $\ell=400$ instead.  {\it Bottom}: Like the middle panel, but here the Helium-4 fraction $Y_p$ is also allowed to vary such that all $N_{\rm eff}$ cases end up with roughly the same angular diffusion scale $\theta_d$.
    Figure taken from Ref.~\cite{Hou:2011ec}.}
\end{figure}

The effects of a non-standard $N_{\rm eff}$ value on the CMB primary anisotropies and its associated parameter degeneracies with the present-day Hubble expansion rate $H_0$ and physical matter density $\omega_{\rm m}$ in the context of $\Lambda$CDM cosmology
have been discussed extensively in, e.g., Ref.~\cite{Hou:2011ec,Abazajian:2012ys}.  Broadly speaking, if the particles that make up $N_{\rm eff}$ are ultra-relativistic at the time of CMB formation, then a non-standard $N_{\rm eff}$ can manifest itself in the following ways:
\begin{enumerate}
    \item The redshift of matter-radiation equality $z_{\rm eq}$ controls the ratio of radiation to matter at the time of photon decoupling and hence the evolution of the potential wells.  This in turn affects the peak height ratios of the CMB temperature anisotropy spectrum.  With seven acoustic peaks measured by the Planck mission~\cite{Planck:2018vyg}, the equality redshift $z_{\rm eq}$ has been measured to percent level precision in $\Lambda$CDM-type cosmologies.  As a probe of $N_{\rm eff}$, however, we note that
    \begin{equation}
    \label{eq:zeq}
       z_{\rm eq} =  \frac{\omega_{\rm m}}{\omega_{\gamma}} \frac{1}{+0.227 N_{\rm eff}}-1,
    \end{equation}
    where $\omega_\gamma$ is the present-day photon energy density.  In other words, $N_{\rm eff}$ is exactly degenerate with the physical matter density $\omega_{\rm m}$, and measuring $z_{\rm eq}$ alone does not determine $N_{\rm eff}$.
    
    \item The angular sound horizon $\theta_s$, which determines the CMB acoustic peak positions, is another quantity sensitive to $N_{\rm eff}$.  Defined as $\theta_s \equiv r_s/D_A$, where $r_s$ is the sound horizon at photon decoupling and $D_A$ is the angular diameter distance to the last scattering surface, the parameter dependence of $\theta_s$ in $\Lambda$CDM cosmologies is as follows,
    \begin{equation}
    \label{eq:thetas}
        \theta_s \propto \frac{\Omega_{\rm m}^{-1/2}}{\int^1_{a^*} \frac{{\rm d} a}{a^2 \sqrt{\Omega_{\rm m}a^{-3} + (1-\Omega_{\rm m})}}},
    \end{equation}
    where $\Omega_{\rm m} \equiv \omega_{\rm m}/h^2$, $h$ is the reduced Hubble expansion rate defined via $H_0 = 100 h\, {\rm km}\, {\rm s}^{-1}\, {\rm Mpc}^{-1}$, $a^*$ is the scale factor at photon decoupling, and we have held $z_{\rm eq}$ and the baryon density $\omega_{\rm b}$ fixed.  This relation implies that while $\theta_s$ constrains the parameter combination $\omega_{\rm m}/h^2$, it does not constrain $\omega_{\rm m}$ and $h$ individually.  Since there already exists an exact degeneracy between $N_{\rm eff}$ and $\omega_{\rm m}$ through $z_{\rm eq}$ (see Eq.~\ref{eq:zeq}), this additional $(\omega_{\rm m},h)$-degeneracy through $\theta_s$ immediately sets up a three-way degeneracy between $N_{\rm eff}$, $\omega_{\rm m}$, and $h$, which needs to be broken by some other means.  In more complex models, degeneracies between $N_{\rm eff}$ and a nonzero spatial curvature $\Omega_k$ or a non-canonical dark energy equation of state are also possible.
    
    \item The angular diffusion scale $\theta_d \equiv r_d/D_A$, where $r_d$ is the diffusion scale at photon decoupling, characterizes the scale at which the CMB temperature anisotropy power spectrum becomes suppressed due to diffusion damping (or Silk damping).  The phenomenon of diffusion damping occurs at the CMB damping tail, i.e., at multipoles $\ell \gtrsim 1000$, and was first measured by the Atacama Cosmology Telescope~\cite{Dunkley:2010ge} and the South Pole Telescope~\cite{Keisler:2011aw}, and now by the Planck CMB mission~\cite{Planck:2018vyg}.
    
    For fixed $z_{\rm eq}$, $\omega_{\rm b}$ and $a^*$, the angular diffusion scale has a parameter dependence 
    \begin{equation}
        \theta_d \propto (\Omega_{\rm m} H_0^2)^{1/4} \, \theta_s,
    \end{equation}
    where $\theta_s$ is the angular sound horizon of Eq.~\ref{eq:thetas}.  Thus, a simultaneous measurement of $\theta_d$, $\theta_s$, and $z_{\rm eq}$ by a CMB mission such as Planck immediately constitutes a measurement of $\omega_{\rm m} \propto \Omega_{\rm m} H_0^2$ and hence $N_{\rm eff}$.  Fig.~\ref{fig:neffcmb} shows the signature of $N_{\rm eff}$ in the CMB damping tail.
    
\end{enumerate}

\paragraph{Large-scale matter distribution}   There are many different ways to quantify and probe the large-scale matter distribution.  The most basic quantity, however, is the present-day matter power spectrum~$P(k)$.  

The broad-band shape of the large-scale matter power spectrum in $\Lambda$CDM-type cosmologies by two quantities: the comoving wavenumber at matter-radiation equality,
\begin{equation}
\begin{aligned}
    k_{\rm eq} & \equiv a_{\rm eq} H(a_{\rm eq}) \\
    & \simeq 4.7 \times 10^{-4} \sqrt{\Omega_{\rm m} (1+ z_{\rm eq})}\, h \, {\rm Mpc}^{-1},
    \end{aligned}
\end{equation}
which fixes the location of the ``turning point'' of $P(k)$, and the baryon-to-matter density fraction
\begin{equation}
f_b \equiv \frac{\omega_{\rm b}}{\omega_{\rm m}},
\end{equation}
which determines the suppression in power at $k > k_{\rm eq}$ due to baryon acoustic oscillations (BAO).  In addition, the matter power spectrum has small-amplitude oscillatory features which are the manifestation of the BAO themselves.  These oscillatory features originate in the same early universe physics as the acoustic peaks in the CMB anisotropy power spectra, and when analyzed together with CMB data, can act as a powerful standard ruler for distance measurements.

If some of the neutrinos are massive and become non-relativistic at late times, then they can constitute a fraction of the present-day dark matter content.  However, this neutrino dark matter is ``hot'', in the sense that the neutrinos, although non-relativistic, come with a significant thermal velocity dispersion, which tends to hinder its gravitational  clustering on small scales.  In terms of the present-day large-scale matter power spectrum, keeping the total matter density fixed but replacing some of the cold dark matter with neutrino hot dark matter suppresses $P(k)$ at $k$-values larger than the free-streaming wave number by an amount dependent on  the neutrino fraction, 
\begin{equation}
    f_\nu \equiv \frac{\omega_\nu}{\omega_{\rm m}} = \frac{\sum m_{\nu}/94\, {\rm eV}}{\omega_{\rm m}}.
\end{equation}
Thus, neutrino masses also influence the overall shape of the matter power spectrum.

Since it is already possible to pin down $z_{\rm eq}$, $\Omega_{\rm m}$, and $\omega_{\rm m}$ using the CMB primary anisotropies, measurements of the matter power spectrum $P(k)$ generally do not improve the constraint on $N_{\rm eff}$ in the simplest $\Lambda$CDM+$N_{\rm eff}$ fit.  However, it must be noted that after the formation of the CMB primary anisotropies, these fluctuations are gravitationally lensed by the intervening matter distribution as the CMB photons propagate from the last scattering surface to the observer, contributing to the so-called CMB secondary anisotropies at multipoles $\ell \gtrsim 500$.  In other words, any CMB anisotropy signal at $\ell \gtrsim 500$ will always include some information about $P(k)$, and this is particularly useful for the purpose of constraining the neutrino mass sum $\sum m_\nu$.

\paragraph{Current CMB and LSS constraints}  In a standard $\Lambda$CDM parameter inference, estimating cosmological parameter values from the CMB and related observations involve varying six free parameters related to cosmology: the baryon density $\omega_b$, the cold dark matter density $\omega_c$, the Hubble parameter $h$, the spectral index $n_s$ and amplitude $A_s$ of the primordial curvature perturbation, and the optical depth to reionization~$\tau$.  Analyses of the Planck CMB data also require that we vary of order 20 nuisance parameters to model the foregrounds and instrumental systematics.  These are later marginalized.

To constrain radiation excess, at minimum we need to add $N_{\rm eff}$ as a free parameter to this list.  Doing so the Planck collaboration finds\cite{Planck:2018vyg}
\begin{equation}
    N_{\rm eff} = 3.00^{+0.57}_{-0.53} \quad (95\%~{\rm CL},~{\rm Planck~TT+lowE}), 
\end{equation}
\begin{equation}
    N_{\rm eff} = 2.92^{+0.36}_{-0.37} \quad (95\%~{\rm CL},~{\rm Planck~TTTEEE+lowE}), 
\end{equation}
\begin{equation}
    N_{\rm eff} = 3.11^{+0.44}_{-0.43} \quad (95\%~{\rm CL},~{\rm Planck~TT+lowE+lensing+BAO}), 
\end{equation}
\begin{equation}
\label{eq:neffconstraintcmb}
    N_{\rm eff} = 2.99^{+0.34}_{-0.33} \quad (95\%~{\rm CL},~{\rm Planck~TTTEEE+lowE+lensing+BAO}), 
\end{equation}
using various combinations of the Planck temperature and $E$-polarization measurements (TTTEEE and lowE), the lensing potential extracted from the Planck temperature maps, as well as the BAO measurements from 6dFGS~\cite{Beutler:2011hx}, SDSS-MGS~\cite{Ross:2014qpa}, and BOSS DR12~\cite{BOSS:2016wmc}.
As can be seen, in all cases, the inference returns an estimate of $N_{\rm eff}$ that is remarkably consistent with the SM prediction of $N_{\rm eff}=3.0440$.

Since the CMB anisotropies are also sensitive to the large-scale matter distribution at low redshifts because of the weak gravitational lensing signal inherent in all CMB power spectrum, one can also derive a constraint on the neutrino mass sum $\sum m_\nu$ from the Planck CMB data at the same time as we constrain $N_{\rm eff}$. 
In a 8-parameter $\Lambda$CDM+$N_{\rm eff}$+$\sum m_\nu$ fit, the Planck collaboration finds~\cite{Planck:2018vyg}
\begin{equation}
\begin{aligned}
\label{eq:combinedneffmnu}
    N_{\rm eff}&  = 2.96^{+0.34}_{-0.33} \quad (95\%~{\rm CL},~{\rm Planck~TTTEEE+lowE+lensing+BAO}), \\
    \sum m_\nu & < 0.12~{\rm eV}.
    \end{aligned}
\end{equation}
Note that the fit assumes three degenerate neutrino mass eigenstates of equal abundances.  The role of $N_{\rm eff}$ is merely to dial up or down the abundances, which is why $N_{\rm eff}$ can go below the standard value of $N_{\rm eff}^{\rm SM} =3.0440$.  This combined fit is to be compared with the $N_{\rm eff}$ constraint quoted in Eq.~\ref{eq:neffconstraintcmb} in a 7-parameter fit of the same data combination, which has the same error bars (about 11\%) and a central value off only by 2\%.  It is also interesting to compare it with the constraint obtained on the neutrino mass sum $\sum m_\nu$ from 7-parameter $\Lambda$CDM+$\sum m_\nu$ fit~\cite{Planck:2018vyg},
\begin{equation}
    \sum m_\nu < 0.13~{\rm eV} \quad (95\%~{\rm CL},~{\rm Planck~TT+lowE+lensing+BAO}), 
\end{equation}
\begin{equation}
    \sum m_\nu < 0.12~{\rm eV} \quad (95\%~{\rm CL},~{\rm Planck~TTTEEE+lowE+lensing+BAO}),
\end{equation}
which is identical to the upper limit obtained from the 8-parameter fit (Eq.~\ref{eq:combinedneffmnu}). 
Thus, one can conclude from this comparison that  there is no strong degeneracy between the $N_{\rm eff}$ parameter and the neutrino mass sum $\sum m_\nu$ in the current generation of precision cosmological data.

What about constraints on the sterile neutrino mass $m_s$?  The Planck collaboration~\cite{Planck:2018vyg} also reports a ``massive sterile neutrino'' fit to their data in a scenario in which the three active neutrinos are assumed to have a fixed minimum mass sum of $\sum m_\nu=0.06$~eV and the effective sterile neutrino mass is defined as $m_{\nu, {\rm sterile}}^{\rm eff} \equiv \Omega_{\nu, {\rm sterile}} h^2 (94.1~{\rm eV})$. This effective mass is related to the physical sterile neutrino mass via 
\begin{equation}
\label{eq:effsterile}
m_s = (\Delta N_{\rm eff})^{-1} m_{\nu, {\rm sterile}}^{\rm eff},
\end{equation}
assuming the sterile states have the same temperature as the SM neutrinos.
Imposing the priors  $\Delta N_{\rm eff} \geq 0$ and  $m_s < 10$~eV, they find the constraints
\begin{equation}
\begin{aligned}
\label{eq:effsterileconstraint}
    N_{\rm eff}&  < 3.29 \quad (95\%~{\rm CL},~{\rm Planck~TTTEEE+lowE+lensing+BAO}), \\
    m_{\nu, {\rm sterile}}^{\rm eff} & < 0.65~{\rm eV}.
    \end{aligned}
\end{equation}
Taking $\Delta N_{\rm eff}$ to be a maximum allowed $0.29$, the mass bound  corresponds to an upper limit of $m_s < 2.24$~eV on the physical sterile neutrino mass.
Thus, while cosmological measurements do constrain $m_s$ in any interesting way, it cannot completely rule out a 1~eV light sterile provided thermalization is kept at below the 30\% level.  In other words, the extent of thermalization as quantified by the $N_{\rm eff}$ parameter remains the limiting factor for the light sterile neutrino scenario in standard cosmology.

\begin{figure}[t]
    \centering
      \includegraphics[angle=0,width=0.49\textwidth]{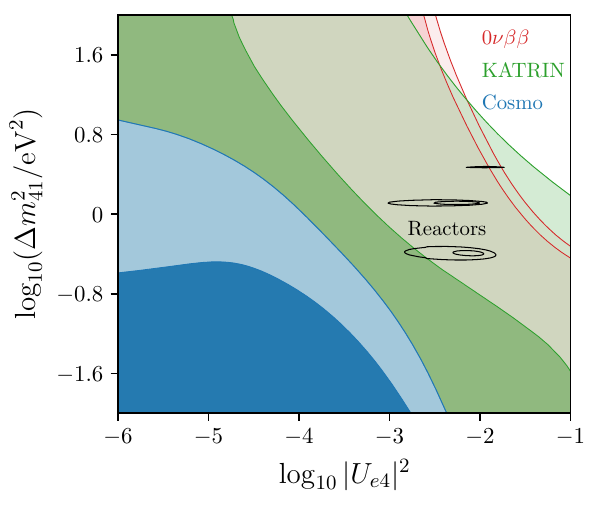}
      \includegraphics[angle=0,width=0.49\textwidth]{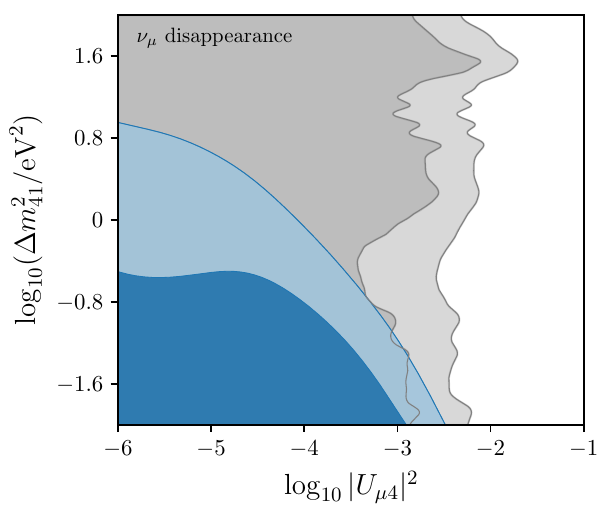}
    \caption{\label{fig:hagstotz}  {\it Left}: 2D marginalized 68\% and 95\% constraints on the mass splitting $\Delta m^2_{41}$ and mixing matrix element $|U_{e4}|^2$ from cosmology (blue), from the tritium $\beta$-decay end-point measurements by KATRIN (green), and from neutrinoless double-$\beta$-decay experiments (red). The preferred regions of reactor experiments are also indicated. {\it Right}: Cosmological 68\% and 95\% marginalized constraints on the $|U_{\mu 4}|^2$ (blue) versus constraints from the $\nu_\mu$ disappearance measurements of IceCube and MINOS+ (grey).
    Figure taken from Ref.~\cite{Hagstotz:2020ukm}.}
\end{figure}

Lastly, we note that while the cosmological bound on $N_{\rm eff}$ given in Eq.~\ref{eq:combinedneffmnu} is already strongly indicative that the short-baseline light sterile neutrino is in serious tension with precision cosmological measurements, it is nonetheless possible to analyse cosmological and oscillation data together in a consistent way.  This has been done most recently in Ref.~\cite{Hagstotz:2020ukm}, which considers a 3+1 scenario, computes the corresponding light sterile neutrino thermalization using the thermalization code {\tt FortEPiaNO}~\cite{Gariazzo:2019gyi} and the associated $N_{\rm eff}$ with mixing parameters consistent with laboratory measurements, and feeds the output into a CMB analysis. Fig.~\ref{fig:hagstotz}
shows the constraints on the $(\Delta m_{41}^2, |U_{e4}|^2)$- and $(\Delta m_{41}^2, |U_{\mu4}|^2)$-planes.

\subsubsection{Can We Evade Cosmological Constraints?}\label{sec-5:cosmo:evade_cosmo}

We have seen that the canonical 1~eV-mass light sterile neutrino motivated by the short baseline anomalies is in strong tension with cosmological measurements primarily because of the non-detection of a non-standard $N_{\rm eff} \sim 4$.
However, a criticism often levelled at cosmological constraints is that all parameter estimation via statistical inference are inherently dependent on the cosmological framework assumed in the inference exercise.  Given the large number of unknowns in cosmology, e.g., the nature of dark energy, inflation, etc., critics argue that there may exist a  corner of this vast unknown parameter space in which a completely thermalized light sterile neutrino state with a mass close to 1~eV might be permitted to live.

To investigate this possibility, the Planck collaboration has provided a large number of analyses of expanded cosmological parameter spaces, often in combination with external, non-CMB data sets.  These are available at the Planck Legacy Archive ({\tt https://pla.esac.esa.int/home}).   Of particular interest to light sterile neutrinos  is the 10-parameter $\Lambda$CDM+$N_{\rm eff}$+$\sum m_\nu$+$w_0$+$n_{\rm run}$ fit, where $w_0$ is the equation of state parameter of the dark energy ($w_0=1$ for a cosmological constant), and $n_{\rm run}$ is the running of the scalar spectral index, a parameter related to the initial conditions of the universe ($n_{\rm run}=0$ in single-field inflation). Using the usual Planck TTTEEE+lowE+lensing+BAO data combination together with the Hubble parameter measurement of Ref.~\cite{Riess:2018uxu} and Supernova Ia data from the Pantheon sample~\cite{Pan-STARRS1:2017jku}, the constraints on $N_{\rm eff}$ and $\sum m_\nu$ in this extended parameter fit are
\begin{equation}\label{eq:effsterileconstraint2}
    \begin{aligned}
        N_{\rm eff}&  = 3.11^{+0.37}_{-0.36} \quad (95\%~{\rm CL}), \\
    \sum m_\nu & < 0.16~{\rm eV}.
    \end{aligned}
\end{equation}
Clearly, while the bound on the neutrino mass sum $\sum m_\nu$ has relaxed somewhat and the central $N_{\rm eff}$ has shifted a little up, relative to the more limited bounds in Eq.~\ref{eq:neffconstraintcmb} and ~\ref{eq:combinedneffmnu} the error bars on $N_{\rm eff}$ have not weaken significantly, and 
the canonical light sterile neutrino is still in tension with precision cosmological measurements in this expanded parameter space.  The upward shift in $N_{\rm eff}$ can be attributed to the discrepancy between the Planck inference of the Hubble expansion rate and the local measurement of Ref.~\cite{Riess:2018uxu}.  Because of the degeneracy between $N_{\rm eff}$ and $H_0$ in the CMB primary anisotropies, combining Planck data with local measurements---the latter of which prefer a higher value of $H_0$---tends to drag up the inferred $N_{\rm eff}$ as well.  The neutrino mass sum $\sum m_\nu$, on the other hand, has long been known to be somewhat degenerate with the dark energy equation of state parameter $w_0$.  However,  the combination of BAO and Supernova Ia data can lift this degeneracy very effectively.

Thus, expanding the cosmological parameter space no longer appears to do much for the light sterile neutrino case (in the sense of allowing a larger $N_{\rm eff}$) the way it once did~\cite{Hamann:2011ge}.  In order to get around cosmological constraints, we need to introduce new physics that directly affects the cosmological phenomenology of the light sterile states.  Since the main problem of the canonical light sterile neutrino is that its thermalization in the early universe raises $N_{\rm eff}$ to an unacceptably large level for BBN and CMB/LSS, all known new physics solutions so far involve tampering with the  thermalization process, in order to maintain $N_{\rm eff}$ at as close to the SM value as possible.  A number of ideas have been proposed and explored throughout the years (though not all are guaranteed to work as desired), including
\begin{itemize}
\item Large chemical potentials or, equivalently, number density asymmetries for the active neutrinos~\cite{Foot:1995bm,Abazajian:2004aj,Mirizzi:2012we,Saviano:2013ktj},

\item Secret interactions of the sterile neutrinos~\cite{Dasgupta:2013zpn,Hannestad:2013ana,Saviano:2014esa,Archidiacono:2014nda,Archidiacono:2016kkh,Chu:2015ipa,Forastieri:2017oma,Chu:2018gxk,Farzan:2019yvo,Cline:2019seo,Cherry:2016jol}, and

\item Low reheating temperature of the universe~\cite{Gelmini:2004ah,Gelmini:2019esj,Gelmini:2019wfp,Yaguna:2007wi,Hasegawa:2020ctq}.

\end{itemize}

\paragraph{Large chemical potentials for the active neutrinos} As discussed earlier in Sec.~\ref{sec:thermalise}, in a standard calculation of light sterile neutrino thermalization, the active neutrino asymmetries, defined as
$L_\alpha =(n_{\nu_\alpha}- n_{\bar{\nu}_\alpha})/n_\gamma$, are assumed to be zero.  However, if for some reason some of these asymmetries are large---usually taken to mean $L_\alpha > 10^{-5}$---then the CP asymmetric term $\sqrt{2} G_F(\mathbb{N}_\nu-\mathbb{N}_{\bar\nu})$ in the Hamiltonian~(Eq.~\ref{eq:hamiltonian}) can act as a large matter effect to suppress oscillations between the active and sterile states.  If this suppression is effective before neutrino decoupling, then it is possible to maintain $N_{\rm eff}$ at close to the SM prediction~\cite{Foot:1995bm}.
For active-sterile neutrino mass splittings in the range $\Delta m^2 \simeq 0.2 \to 10~{\rm eV}^2$, the minimum neutrino asymmetries required to effect some degree of suppression are $L > 10^{-4} \to 5 \times 10^{-3}$~\cite{Abazajian:2004aj}. To significantly suppress thermalization, however, asymmetries as large as $L \sim 10^{-2}$ are required~\cite{Saviano:2013ktj}.

Unfortunately, aside from the difficulty in explaining how such large neutrino asymmetries could have arisen in the first place, this solution also suffers from other undesirable effects, namely, significant distortion to the $\nu_e$ and $\bar{\nu}_e$ energy spectra.  While it is possible to suppress active-sterile oscillations with the choice of $L \sim 10^{-2}$ before neutrino decoupling, beyond this critical point vacuum oscillations will inevitably take over and distort the active neutrino energy spectra as a result~\cite{Abazajian:2004aj}.  Ref.~\cite{Saviano:2013ktj} has computed this distortion and its effect on the light elemental abundances.  They find that the larger the neutrino asymmetry employed to suppress light sterile neutrino thermalization, the larger the spectral distortion and the resulting Helium-4 mass fraction $Y_p$.  Thus, while large neutrino asymmetries do improve the outcome for $N_{\rm eff}$, at the same time they also affect at least one important observable in an undesirable way.  The solution is therefore far from fool-proofed.

\paragraph{Self-interaction or non-standard interaction for the sterile neutrino} These solutions also work on the principle of suppressing sterile neutrino thermalization through the introduction of a non-standard matter potential for the sterile state in the oscillation Hamiltonian (Eq.~\ref{eq:hamiltonian}).  They differ primarily in their coupling structures.  

Ref.~\cite{Hannestad:2013ana,Dasgupta:2013zpn,Saviano:2014esa,Chu:2015ipa,Forastieri:2017oma,Chu:2018gxk} consider an interaction of the form
\begin{equation}
\label{eq:vectorinteraction}
{\cal L}_{\rm int} = g_X \bar{\nu}_s \gamma^\mu P_L \nu_s X_\mu,
\end{equation}
where the sterile neutrino self-interaction is mediated by a MeV-mass vector boson $X$.  This leads to the addition of a matter potential to the Hamiltonian (Eq.~\ref{eq:hamiltonian}) of the form
\begin{equation}
\mp 2 \sqrt{2} G_X p \left[\frac{4}{3}\frac{\mathbb{E}_{s}+\mathbb{E}_{\bar{s}}}{m_X^2}\right],
\end{equation}
where $G_X \equiv (\sqrt{2}/8) g_X^2/m_X^2$, $m_X$ is the mass of the $X$ boson, and 
$\mathbb{E}_s$ and $\mathbb{E}_{\bar{s}}$ are defined like $\mathbb{E}_\nu$ and $\mathbb{E}_{\bar{\nu}}$ in Eq.~\ref{eq:nurhosum}, but with a projection matrix $S_a={\rm diag}(0,0,0,1)$ that singles out the sterile state for coupling.  The collision integrals ${\cal I}[\varrho(p,t)]$ also need to be modified appropriately.
For $g_X \gtrsim 10^{-2}$ and $g_X \lesssim 10$~MeV, the new interaction (Eq.~\ref{eq:vectorinteraction}) is able to suppress sterile neutrino thermalization and hence preserve $N_{\rm eff}$ at close to the SM value without altering BBN predictions.  However, at times after neutrino decoupling, the same interaction also leads to equipartition amongst the active and sterile states.  That is, if at neutrino decoupling the neutrino number densities are $(n_{e}, n_{\mu}, n_{\tau}, n_{s}) = (1,1,1,0)$, the secret interaction will redistribute it to $(3/4, 3/4, 3/4,3/4)$.  Thus, the mass of sterile state $m_s$ will nonetheless contribute to the hot dark matter energy density probed by the CMB anisotropies and the large-scale matter distribution, and be subjected
a substantially tighter constraint, about $m_s < 0.2$~eV, than is implied by Eq.~(\ref{eq:effsterileconstraint}) or Eq.~(\ref{eq:effsterileconstraint2}).
 This solution is therefore also not fool-proofed.

On the other hand, Ref.~\cite{Archidiacono:2014nda,Archidiacono:2016kkh} consider a self-interaction of the mass eigenstate $\nu_4$ mediated by a massless pseudoscalar $\phi$:
\begin{equation}
\label{eq:pseudoscalarinteraction}
{\cal L}_{\rm int} = g_\phi \phi \bar{\nu}_4 \gamma_5 \nu_4.
\end{equation}
As with the massive vector boson case above, the secret interaction engenders a matter potential, which in turn suppresses the production of sterile states.  The transition from no to full thermalization happens in the range of coupling values  $10^{-6} < g_\phi < 10^{-5}$~\cite{Archidiacono:2014nda}.  The authors further argue that because the secret interaction happens exclusively for the mass eigenstate $\nu_4$, the interaction cannot equilibrate the active and sterile states and whatever is the $\Delta N_{\rm eff}$ produced at neutrino decoupling is also the only component of the neutrino population that carries a mass of $m_s \simeq 1$~eV.   Thus, the scenario can easily evade both limits on $N_{\rm eff}$ and $m_s$.

Lastly, the solution of Ref.~\cite{Farzan:2019yvo,Cline:2019seo} invokes a coupling of the sterile state to an ultra-light real scalar field  of mass $m_\phi < 5 \times 10^{-17}$~eV that also contributes to the cold dark matter.  At early times, the coupling induces an effective mass for the sterile state, which suppresses active-sterile oscillations and hence thermalization of the sterile state in much the same way as the two scenarios discussed above.  After neutrino decoupling, the $\phi$ field starts to oscillate coherently.  Since unlike the MeV vector boson case the model does not lead to the equilibration of the active and sterile states, both limits on $N_{\rm eff}$ and $m_s$ can be easily avoided.

\paragraph{Low reheating temperature} Low reheating temperature scenarios~\cite{Gelmini:2004ah,Gelmini:2019esj,Gelmini:2019wfp,Yaguna:2007wi,Hasegawa:2020ctq} refer to those cases in which the universe transitions to radiation domination at temperatures below $T \sim 10$~MeV.  This might happen because of a very low inflation energy scale, or because some  non-standard physics causes the universe to enter a period of matter domination immediately prior to the most recent phase of radiation domination, and the transition back to radiation domination takes place at $T \sim {\cal O}(1)$~MeV.

A low reheating temperature appears to be a viable way to evade cosmological constraints on light sterile neutrino states.  If reheating occurs at $T\sim {\cal O}(1)$~MeV, even the SM active neutrinos have barely enough time to interact  before neutrino decoupling happens.  Depending on how exactly reheating happens, some of the active neutrino species may not even reach equilibrium number or energy densities.  If light sterile neutrino thermalization was to happen at the same time, the shortage of active neutrinos in the plasma would also slow down the production rate.  For these reasons, it is possible to engineer a scenario in which the final $N_{\rm eff}$ is close to the SM value, while the ratio of sterile to active states remains smaller than 1 to 3.  Naively, this makes it possible to satisfy $N_{\rm eff}$ as well as $m_s$ bounds from CMB and/or BBN.

In practice, however, whether or not the solution works depends on the details of the reheating model.  Reference~\cite{Hasegawa:2020ctq}, for example, finds that if the parent particle responsible for reheating
decays exclusively into electromagnetically interacting radiation, then a low temperature reheating can indeed render light sterile neutrinos consistent with measurements of the primordial elemental abundances.
If however the parent particle decays mainly into hadrons, then together with the presence of active-sterile neutrino mixing, the primordial synthesis of light elements can proceed in a way incompatible with observations for a wide range of the mass and the hadronic branching ratio of the parent particle. 

In addition to low reheating scenarios, cosmological scenarios where entropy is conserved and the expansion rate is modified can significantly affect light sterile neutrino constraints. This includes, for example, scalar-tensor theories, and is discussed in detail in Refs.~\cite{Gelmini:2019esj,Gelmini:2019wfp}, and also in \cite{Gelmini:2019clw} (for resonance sterile neutrino production).







\FloatBarrier
\section{Future Experimental Prospects}
\label{sec:future}



\subsection{Decay-at-Rest Accelerator Experiments} 

\subsubsection{JSNS$^2$ and JSNS$^2$-II}







\indent

\begin{figure}[ht]
    \centering
    \includegraphics[width=0.85 \textwidth]{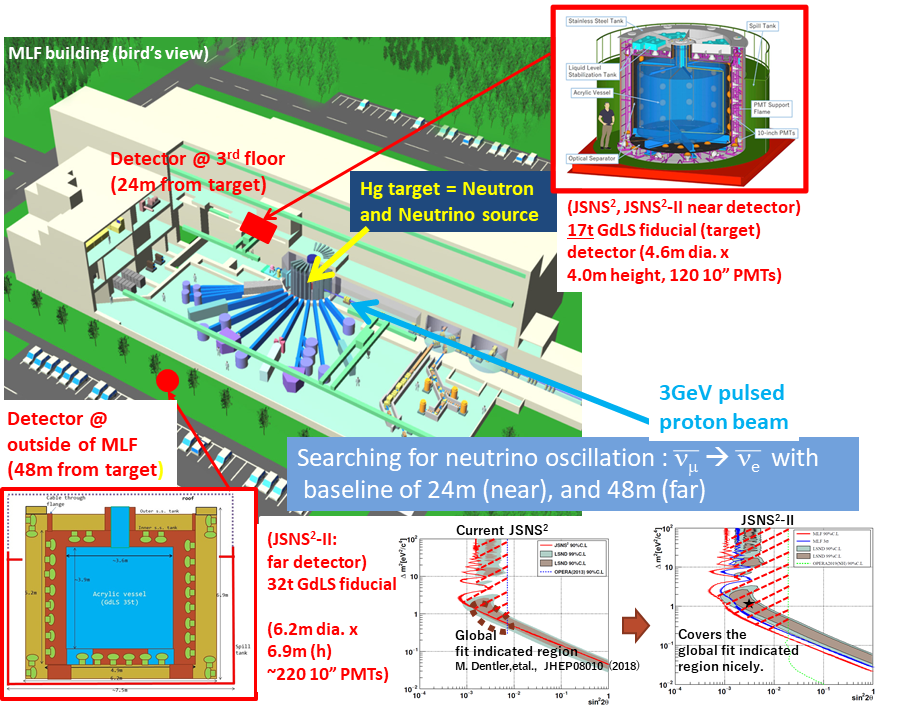}
    \caption{
      The experimental setups of the JSNS$^2$ and JSNS$^2$-II
      and their sensitivities. Figure from Ref.~\cite{Maruyama:2022juu}.
    }
    \label{FIG:Setup}
  \end{figure}

  The JSNS$^2$ (J-PARC Sterile Neutrino Search at the J-PARC Spallation Neutron
  Source)~\cite{JSNS2:2013jdh,Ajimura:2017fld,JSNS2:2021hyk} and its second phase
  JSNS$^2$-II~\cite{Ajimura:2020qni}, aim to search
  for neutrino oscillations with $\Delta m^2$ near
  1~eV$^2$ at the J-PARC Materials and Life Science Experimental
  Facility (MLF). Fig.~\ref{FIG:Setup} shows the experimental setup and
  search sensitivities.
  An intense neutrino beam from muon decay at rest is produced by a
  spallation neutron target with the 1 MW beam of 3~GeV protons
  created by a Rapid Cycling Synchrotron (RCS).
  Neutrinos come predominantly from $\mu^+$ decay :
  $\mu^{+} \to e^{+} + \bar{\nu}_{\mu} + \nu_{e}$.
  An oscillation of $\bar{\nu}_{\mu}$ to $\bar{\nu}_{e}$ through the
  fourth mass eigen-state is searched for by detecting the Inverse-Beta-Decay (IBD)
  interaction $\bar{\nu}_{e} + p \to e^{+} + n$, followed by
  gammas from neutron capture on Gd.  
  The JSNS$^2$ detector, as the near detector in the JSNS$^2$-II setup,
  contains 17~tonnes of Gd-loaded liquid scintillator and is located
  24~meters away from the mercury target. The new far detector
  of JSNS$^2$-II, currently under construction, is located outside
  the MLF building with a baseline of 48~meters.
  The far detector contains 32~tonnes of Gd loaded liquid scintillator
  as a neutrino target.
  Both JSNS$^2$ and JSNS$^2$-II employ a Hydrogen target for the neutrino source ($\mu$ decay-at-rest), and 
  a neutrino detection channel (IBD) identical to LSND.
  With improvements made by the short pulsed beam and the neutron
  capture signal, JSNS$^2$ and JSNS$^2$-II will provide
   clean and direct tests of the LSND anomaly. 
  JSNS$^2$ started data taking in 2020, and accumulated
  1.45$\times$10$^{22}$ Proton-On-Target (POT) by 2021, 13$\%$ of the
  approved POT by J-PARC, which corresponds to 1 MW beam power for 3 years.
  An extensive analysis is ongoing (e.g.~\cite{Hino:2021uwz}).
  The search sensitivity of JSNS$^2$ with the full design POT 
  is shown in the bottom-middle plot of Fig.~\ref{FIG:Setup}.
  The construction of the far detector of JSNS$^2$-II started
  in September 2021 and the aim is to start data taking in 2023.
  It will provide additional sensitivity, especially in the low $\Delta m^2$ region with
  1~MW beam power for 5~years.
  
\begin{figure}[ht]
    \centering
    \includegraphics[width=0.8 \textwidth]{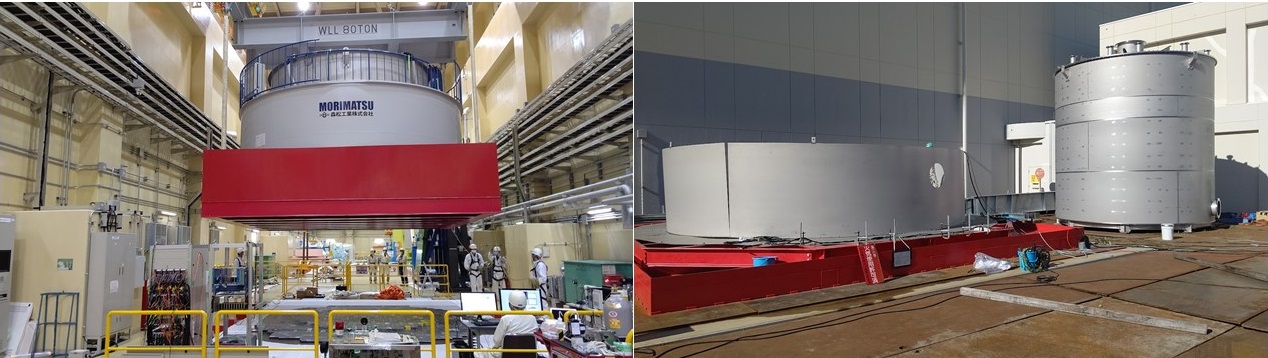}
    \caption{
      The JSNS$^2$ detector during the installation (left) and
      the construction status of the new far detector of the JSNS$^2$-II
      (right).  
    }
    \label{FIG:detectors}
  \end{figure}

  Figure~\ref{FIG:detectors} shows a picture of the JSNS$^2$ detector
  during the installation in June 2020 and
  the construction status of the new far detector of the JSNS$^2$-II.






\subsubsection{COHERENT at the SNS}

The COHERENT collaboration can perform a powerful test of oscillations of sterile neutrinos by considering NC disappearance.  Parameter space favored by a global fit of oscillation data to a 3+1 scenario is accessible to COHERENT in the near future with later data giving a much stronger constraint.

COHERENT measures coherent, elastic neutrino-nucleus scattering (CEvNS) and other low-energy neutrino scattering processes at the Spallation Neutron Source (SNS) at Oak Ridge National Lab (ORNL).  CEvNS is a neutral current process whose only signature is a low-energy nuclear recoil which was first measured by COHERENT on CsI in 2017~\cite{Akimov:2017ade}.  The cross section is very large compared to other neutrino scattering cross sections below 50~MeV and is precisely predicted.  The SNS is an intense source of $\pi$ decay-at-rest neutrinos with energies 0-53~MeV, ideal for measuring CEvNS.  The width of the SNS beam, 360~ns FWHM, is small compared to the muon lifetime so that the neutrino flux separates in time to a prompt $\nu_\mu$ flux from $\pi^+\rightarrow\mu^+\nu_\mu$ and a delayed flux of $\bar{\nu}_\mu/\nu_e$ from $\mu^+\rightarrow e^+\bar{\nu}_\mu\nu_e$.  The $\nu_\mu$ flux is monoenergetic with $E_\nu=29.8$~MeV.  COHERENT builds and commissions several CEvNS detectors for operation at the SNS at baselines of 19.3 to 28 m.  These baselines place COHERENT detectors at the first oscillation maximum for the $\nu_\mu$ flux assuming the global best fit of $\Delta m^2_{41}$.  Since the neutrino flux at the SNS includes both $\nu_\mu/\bar{\nu}_\mu$ and $\nu_e$, CEvNS searches can simultaneously search for $\nu_\mu\rightarrow\nu_s$ and $\nu_e\rightarrow\nu_s$ disappearance with favorable sensitivity to both $\theta_{14}$ and $\theta_{24}$ with the same experiment.  Additionally, the largest systematic uncertainty, the neutrino flux normalization, is correlated between all detectors which mitigates its effect on a joint fit.

\begin{figure}[!bt]
\centering
\includegraphics[width=0.6\linewidth]{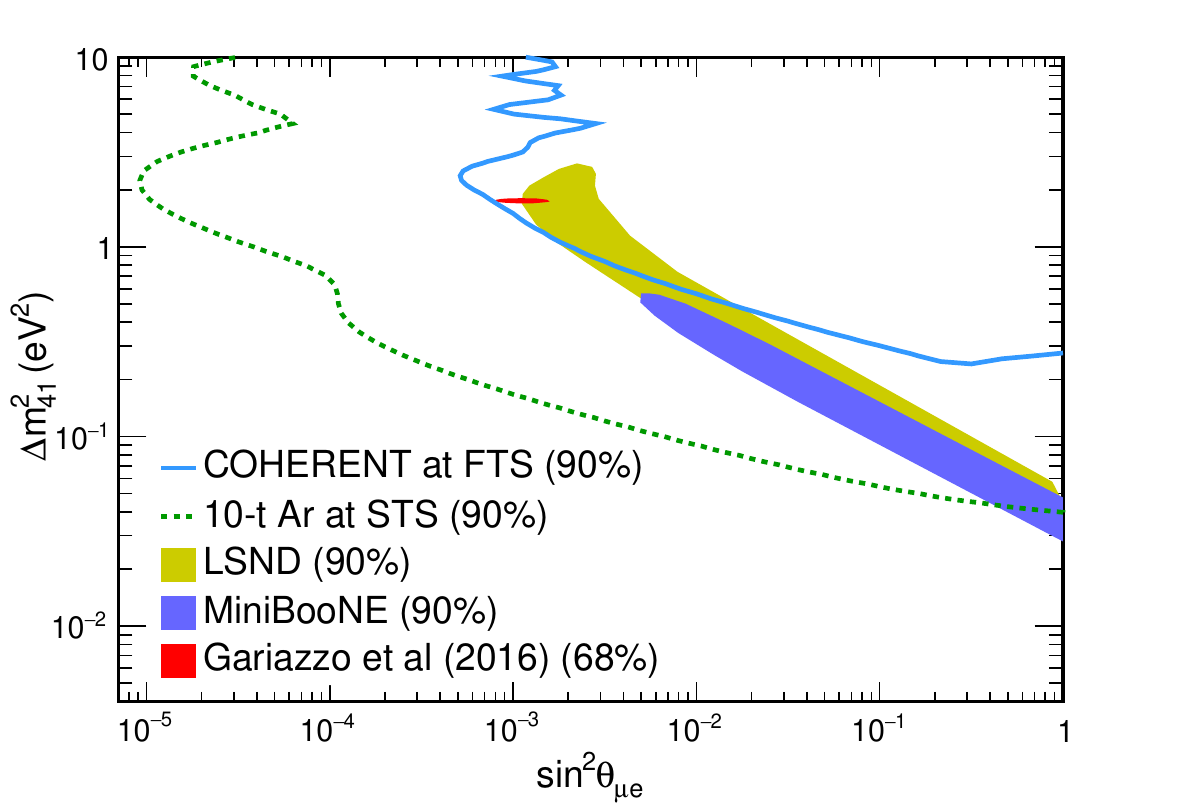}
\caption{Sensitivity of COHERENT CEvNS detectors to constrain sterile neutrino parameter space assuming a 3+1 model compared to the LSND and MiniBooNE allowed regions.  A global fit to all short-baseline oscillation data is also shown.  Figure from~\cite{Asaadi:2022ojm}.}
\label{fig:COHERENT}
\end{figure}

There are plans for three future detectors suitable for searching for sterile neutrinos through CEvNS disappearance in the near future.  The first is an upgrade of the CENNS10 detector which made the first CEvNS measurement on argon~\cite{Akimov:2020pdx}.  This will be a liquid argon calorimeter with 610 kg of fiducial mass with a baseline of 28~m.  Detector performance is well understood from experience with CENNS10 operations and data.  A 10-kg CsI scintillation detector at a 19.3~m baseline is also planned.  This detector will be undoped and cooled to 77~K which can dramatically increase light yield while reducing background scintillation within the crystal~\cite{Chernyak:2020lhu}.  This ensures a low threshold, allowing tests of CEvNS disappearance with low-energy recoils.  Finally, a 50~kg germanium PPC detector at 22~m is planned as an upgrade to the 17~kg array currently being commissioned at the SNS.  The sensitivity of a joint fit using three years of data from all three detectors to search for a sterile neutrino through NC disappearance with CEvNS is shown in Fig.~\ref{fig:COHERENT}.  This would test the parameter space preferred by a global fit at 90$\%$ confidence~\cite{Gariazzo:2017fdh}.  

Into the next decade, ORNL is investing in the SNS, doubling its power and constructing a new second target station (STS) with one in every four beam spills delivered to the STS to supplement work at the first target station (FTS).  Though the upgrade will not be completed until the 2030s, it will facilitate a strong search for sterile neutrinos.  The two targets would only be 140~m apart, meaning a large flux of neutrinos from both sources would pass through each CEvNS detector at the SNS.  Similar to two-detector oscillation experiments, this mitigates systematic uncertainties from neutrino interaction modeling and detector response by observing neutrinos from a near and far flux source.  The dominant remaining uncertainty comes from $\pi^+$ production in each beam target which is small.  With this control of uncertainties, a test of NC disappearance is possible at the 1$\%$-level.  Sensitivity to sterile neutrino oscillations in a 3+1 framework for a 10~t fiducial argon calorimeter running for five years, when placed 20~m from the STS and 120~m from the FTS, is shown in Fig.~\ref{fig:COHERENT}.  Exploiting flux from both targets, this large detector could test $\sin^22\theta_{\mu e}$ values of $10^{-5}$ at the global best fit $\Delta m^2_{41}$ and could test the LSND and MiniBooNE preferred regions for $\Delta m^2_{41}>0.04$~eV$^2$ at high confidence. 

\subsubsection{Coherent CAPTAIN-Mills}

The physics program of the Coherent Captain Mills (CCM) Experiment comprises searches for new particles in the weak sector, including Dark Photons, Axion-like Particles (ALPs), and heavy neutral leptons in the keV to MeV mass range, extending the coverage of open parameter space for these searches at the order of magnitude level.
Many of these particles are invoked as alternative or additional explanations to oscillations involving sterile neutrinos as the source of MiniBooNE anomaly.
Thus, the results of CCM from the ongoing run at Los Alamos National Laboratory (LANL) have direct bearing on phenomenology presented in this white-paper.
Here, we describe the CCM detector, present a relevant CCM search for production of new bosons by charged meson decays~\cite{Dutta:2021cip} as an example of the impact of the results, and summarize other searches that can be performed.

\begin{figure}[ht!]
\centering
\includegraphics[width=3.25 in]{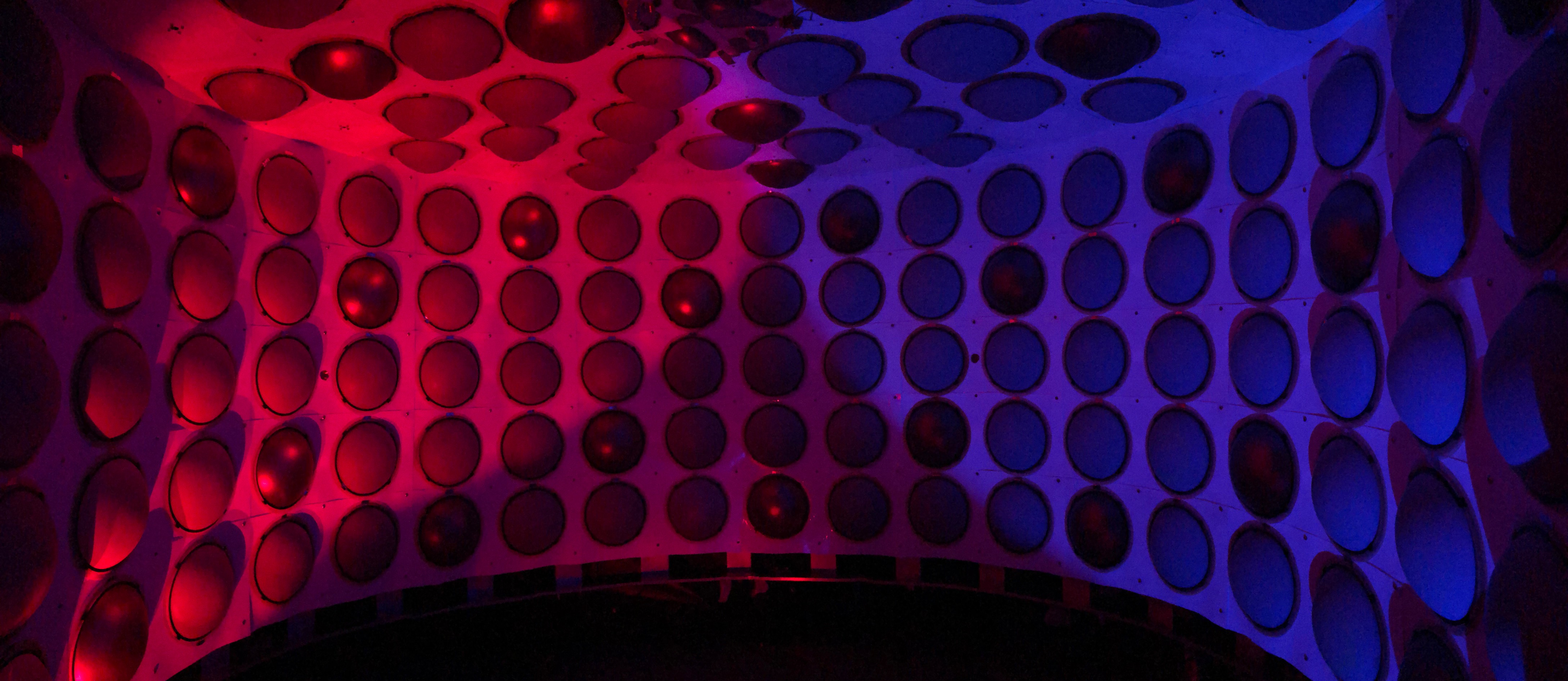}
\caption{The interior of the CCM200 detector.
Of the 200 PMTs, 80\% are coated with wavelength shifter (TBP) to shift 128 nm scintillation light to the visible, leaving 20\% uncoated (darker, more reflective PMTs in image), aiding discrimination of Cherenkov light.
TPB foils cover the walls.
The light-tight interior is surrounded by a veto region instrumented with PMTs.
CCM is now running.
}
\label{CCM200}
\end{figure}

The CCM experiment is relatively new to the scene of experiments to understand the phenomenology of short-baseline anomalies.
The experiment was conceived in 2017 and prototyped using ``CCM120,'' which tested 120 PMTs for the SBND liquid argon (LAr) experiment.     First physics results from CCM120 were recently published~\cite{Aguilar-Arevalo:2021sbh, CCM:2021leg}.
In 2019, the LANL LDRD office and DOE Dark Matter New Initiative program recognized the relevance of the CCM rare-particle searches to dark matter studies and provided funding for an upgrade to 200 PMTs for ``CCM200.'' 
This 5\,t fiducial-volume (10\,t total) LAr detector with 50\% photomultiplier tube (PMT) coverage, seen in Fig.~\ref{CCM200}, was completed in Autumn 2021.
The detector is unusual for accelerator-based liquid argon experiments, in that it utilizes only light collection---no time projection chamber.
The detector is being commissioned now, and data totaling $2.25 \times   10^{22}$\,POT will be collected in three runs between 2022-24, at the Lujan spallation neutron center.
This facility targets 100 microamps on tungsten of  800~MeV  protons with 275 ns spills at 20\,Hz.
This is a prolific source of neutrinos
from stopped pion and muon decay, and, potentially, a source for production of new particles, such as ALPs~\cite{CCM:2021lhc}, that can be observed in CCM200 through interactions or decays.
CCM200 is located 90$^\circ$ off-axis and 20\,m from the target.

The CCM200 design has a combination of features related to light collection that makes it powerful and unique.
The first is its PMT coverage (8" Hammamatsu R5912-MOD), which is orders of magnitude higher per unit volume than any LAr TPC experiment.
Furthermore, the large charge dynamic range of the PMT's and electronics enables the detector to have reconstructed energy sensitivity from ~10~keV to over 200~MeV.
The second is the rate of PMT readout, which provides information at 500 MHz and is synced to the accelerator providing 2-ns absolute timing relative to the 275 ns beam pulse.
This is key for separating out early speed of light particles from the prolific beam related neutrons.
Third, as can be seen in the Fig.~\ref{CCM200}, 80\% of the PMTs are coated with 1,1,4,4-Tetraphenyl-1,3-butadiene (TPB), while 20\% are uncoated.
The TPB allows observation of the scintillation light from LAr, which emits 40,000~photons/MeV in zero electric field--$\times 4$ brighter than typical oil-based liquid scintillator--- at 128 nm wavelength, by shifting to the visible to penetrate the PMT glass.
The uncoated tubes, which is unique to the CCM design,  allows clean observation of Cherenkov light.
An R\&D goal of CCM is to make the first use of observed Cherenkov light on an event-by-event basis in an analysis to reject backgrounds, since, for a signal, the direction of the Cherenkov ring is opposite the Lujan target.  

\begin{figure}[ht!]    
\begin{center}
\end{center}  
\includegraphics[width=6.5in]{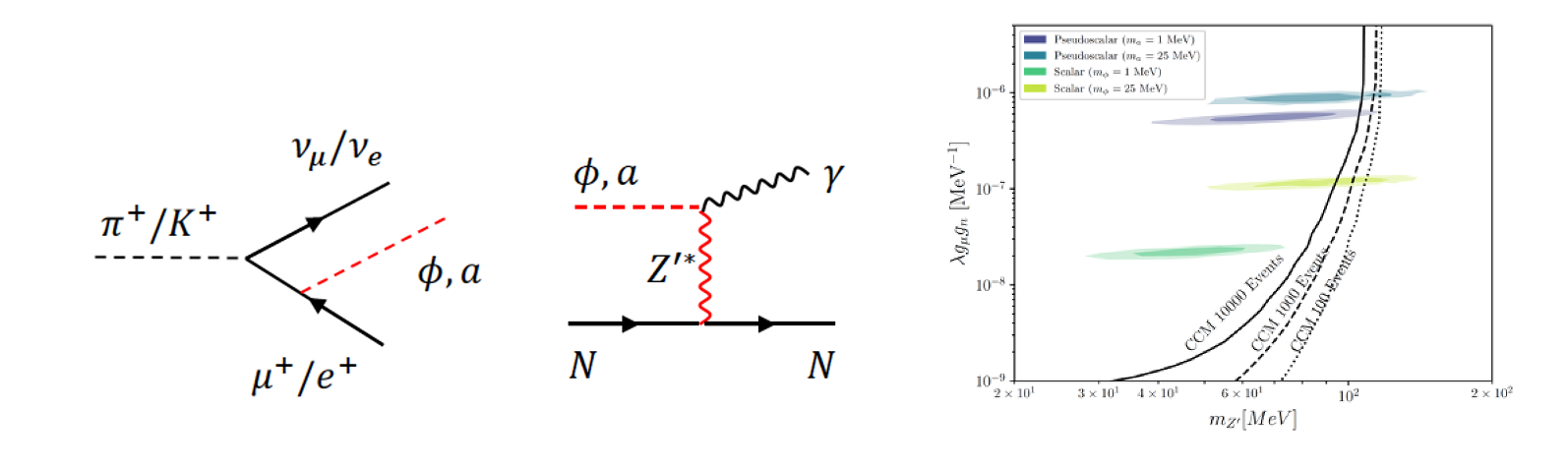}
\caption{\label{example-axions} \emph{Left}: example diagrams for three-body meson decay producing scalar or pseudoscalar(axion-like) particles that interact in the detector.
\emph{Right}:  Allowed regions for scalers (green) and pseudoscalers (blue) that account for the MiniBooNE low energy excess, presented as a function of coupling versus new particle mass~\cite{Dutta:2021cip}.
Lines: predictions of 10,000, 1000, and 100 signal events in solid, dashed and dotted, respectively.
}
\end{figure}

These design features make CCM particularly ideal for searches for electromagnetic signatures of new physics produced in the target,  which is a signature of popular explanations for the short-baseline anomalies.
The most popular new-physics explanation has been oscillations involving sterile neutrinos.
CCM can explore the recent large-mixing angle result from the BEST experiment~\cite{Barinov:2021asz} using $\nu_e$ disappearance for the pion decay-at-rest neutrino beam, since the threshold for $\nu_e$-argon scatters is 1.5~MeV.
In the longer term, an upgraded CCM complex can be modified to perform a two-detector search for $\nu_\mu \rightarrow \nu_e$ in the LSND range that may be motivated by JSNS$^2$ results~\cite{Rott:2020duk}.
However, recently, community interest has turned to new particles to explain the observed anomalies.
Motivated by this, for this short review, we are featuring an example of CCM's new-particle-discovery capability.

As an example of an interesting new model that CCM can address, consider the proposed explanation of the MiniBooNE Low Energy Excess (LEE)~\cite{MiniBooNE:2020pnu} from three-body meson decay~\cite{Dutta:2021cip}.
The diagrams for production of a new scalar or pseudoscalar particles that will interact in the detector to produce a single photon  exchanging a light vector boson ($Z'$) with the nucleus are shown in Fig.~\ref{example-axions}, left.
The allowed region for the LEE is shown in Fig.~\ref{example-axions}, right, and this model (model 1) finds a good fit to both the angular and energy dependence of the LEE \cite{MiniBooNE:2020pnu}.
Another version of this model (model 2) which also can fit the data involves the emission a light vector boson $Z'$ from the pion decay (just like the scalar/pseudoscalar).
$V$ will then  produce a photon by exchanging  a scalar with the nucleus at the detector.
This model explains the LEE. All these new mediators emerging from the charged pion decays, so far we have discussed, can be coupled to only quarks.

CCM will be able to probe both models.
The main production channel will be $\pi^0$ decay into  gamma and $Z'$.
In model 1, $Z'$ can be assumed to dominantly decay into a pair of scalars, which will subsequently produce a photon from the scattering at the detector as needed to resolve LEE.
$Z'$ also can decay into a photon and the scalar.
Here, our assumption is that $Z'$ does not decay into a pair of visible particles promptly.
Under this assumption, Fig.~\ref{example-axions} shows the  predicted number of signal events at CCM  in the allowed  parameter space  to explain the LEE.
Therefore, CCM200 has the capability for discovery, if this new physics is the source of the LEE.
Model 2 probes the MiniBooNE anomaly more directly since the same $Z'$ that emerges from the charged pion decay to address the anomaly also can be produced from the $\pi^0$ decay; therefore no assumption is needed to correlate the LEE and a possible signal at the CCM.
This possibility is under investigation at present.

Outside of models explaining the anomalies, CCM engages in a broad range of new-physics searches.
Limits on leptophobic dark photons from CCM120~\cite{Aguilar-Arevalo:2021sbh, CCM:2021leg}, the prototype run, will be extended by two orders of magnitude in CCM200.
Searches for the QCD axion can close the last remaining open-window at masses $>0.1$~MeV~\cite{CCM:2021lhc}.   Searches for ``neutrissimos''--not so heavy neutral heavy leptons--that have focused on $>100$~MeV masses to address the LEE are being extended to lower masses in CCM.
Although these searches are not directly tied to explanations of the anomalies, a discovery would inevitably demand investigation on whether the observed new-physics is related.

In summary,  CCM is a small, fast-timescale experiment that is already taking data at LANL.
Its results have the potential to change our thinking about the anomalies.

\subsubsection{PIP2-BD: GeV Proton Beam Dump at Fermilab’s PIP-II Linac} 

The completion of the PIP-II superconducting LINAC at Fermilab as a proton driver for DUNE/LBNF in the late 2020s creates an attractive opportunity to build a GeV proton beam dump facility at Fermilab dedicated to and designed from the ground up for HEP with excellent sensitivity to eV-scale sterile neutrinos via neutral current disappearance using the CEvNS reaction (see Ref.~\cite{Toups:2022yxs}). Thus, relative to spallation neutron facilities tailored to neutron physics and optimized for neutron production operating at a similar proton beam power, a HEP-dedicated beam dump facility would be designed to suppress rather than maximize neutron production and implement a beam dump made from a lighter target such as carbon, which can have a pion-to-proton production ratio up to $\sim$2 times larger than heavier Hg or W targets. The facility could also accommodate multiple, 100-ton-scale high energy physics experiments located at different distances from the beam dump and at different angles with respect to the incoming proton beam.  This flexibility further improves the sensitivity of dark sector and sterile neutrino searches, by allowing relative measurements at different distances and angles to constrain uncertainties in expected signal and background rates.

The continuous wave capable PIP-II LINAC at Fermilab can simultaneously provide sufficient protons to drive megawatt-class $\mathcal{O}$(GeV) proton beams as well as the multi-megawatt LBNF/DUNE beamline. By coupling the PIP-II LINAC to a new Booster-sized, permanent magnet or DC-powered accumulator ring, the protons can be compressed into pulses suitable for a proton beam dump facility with a rich physics program. The accumulator ring could be located in a new or existing beam enclosure and be designed to operate at 800~MeV but with an upgrade path allowing for future operation in the GeV range. The accumulator ring would initially provide 100~kW of beam power, limited by stripping foil heating, and have a $\mathcal{O}(10^{-4})$ duty factor. One variant of this accumulator ring would be a $\sim$100 m circumference ring operating at 1.2~GeV with a pulse width of 20~ns and a duty factor of $\mathcal{O}(10^{-6})$, which would greatly reduce steady-state backgrounds.  Another is an accumulator ring coupled to a new rapid cycling synchrotron replacing the Fermilab Booster with an increased proton energy of 2~GeV and an increased beam power of 1.3 MW~\cite{Ainsworth:2021ahm}.

\begin{figure}
\centering
  \includegraphics[width=0.30\textwidth]{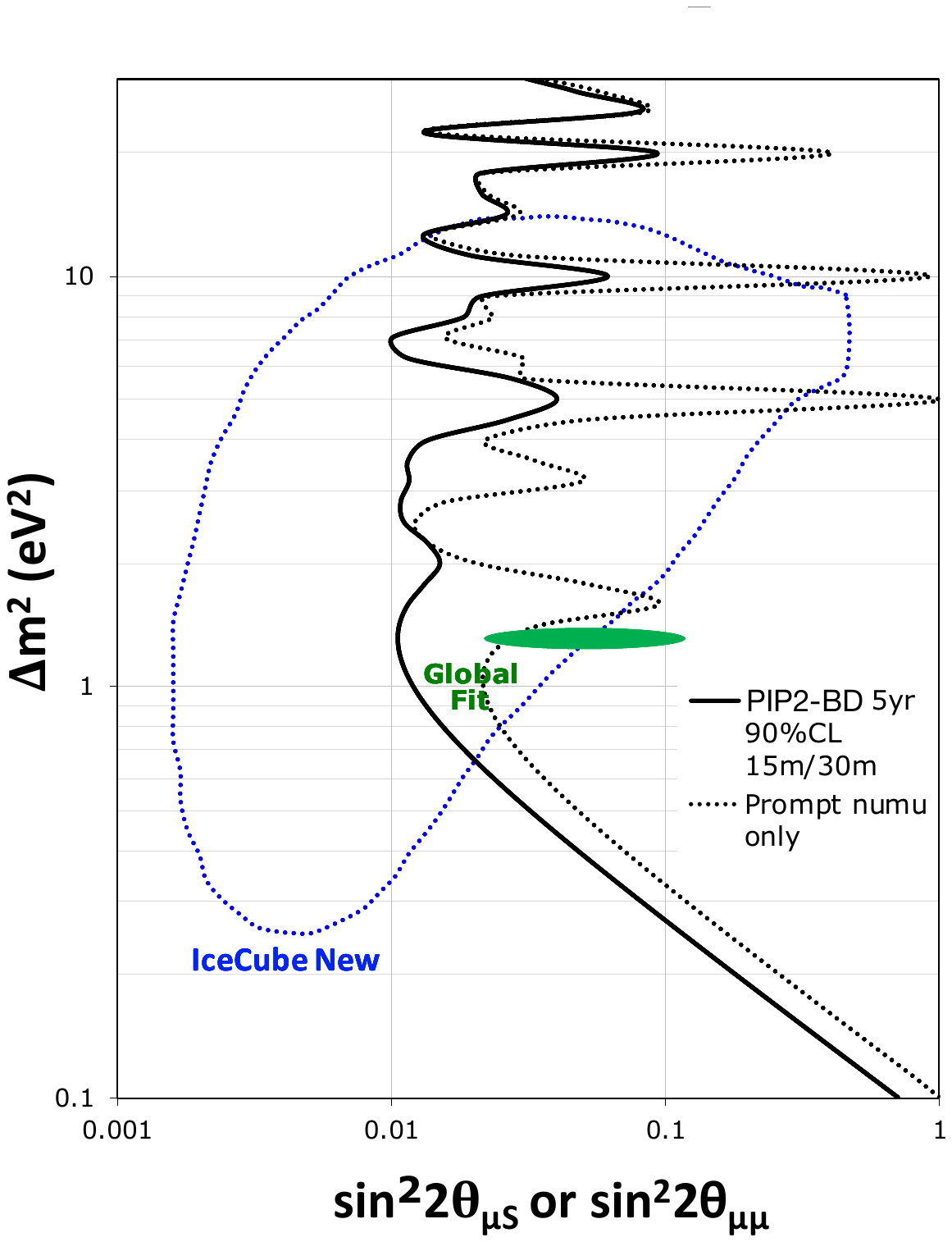}
  \includegraphics[width=0.30\textwidth]{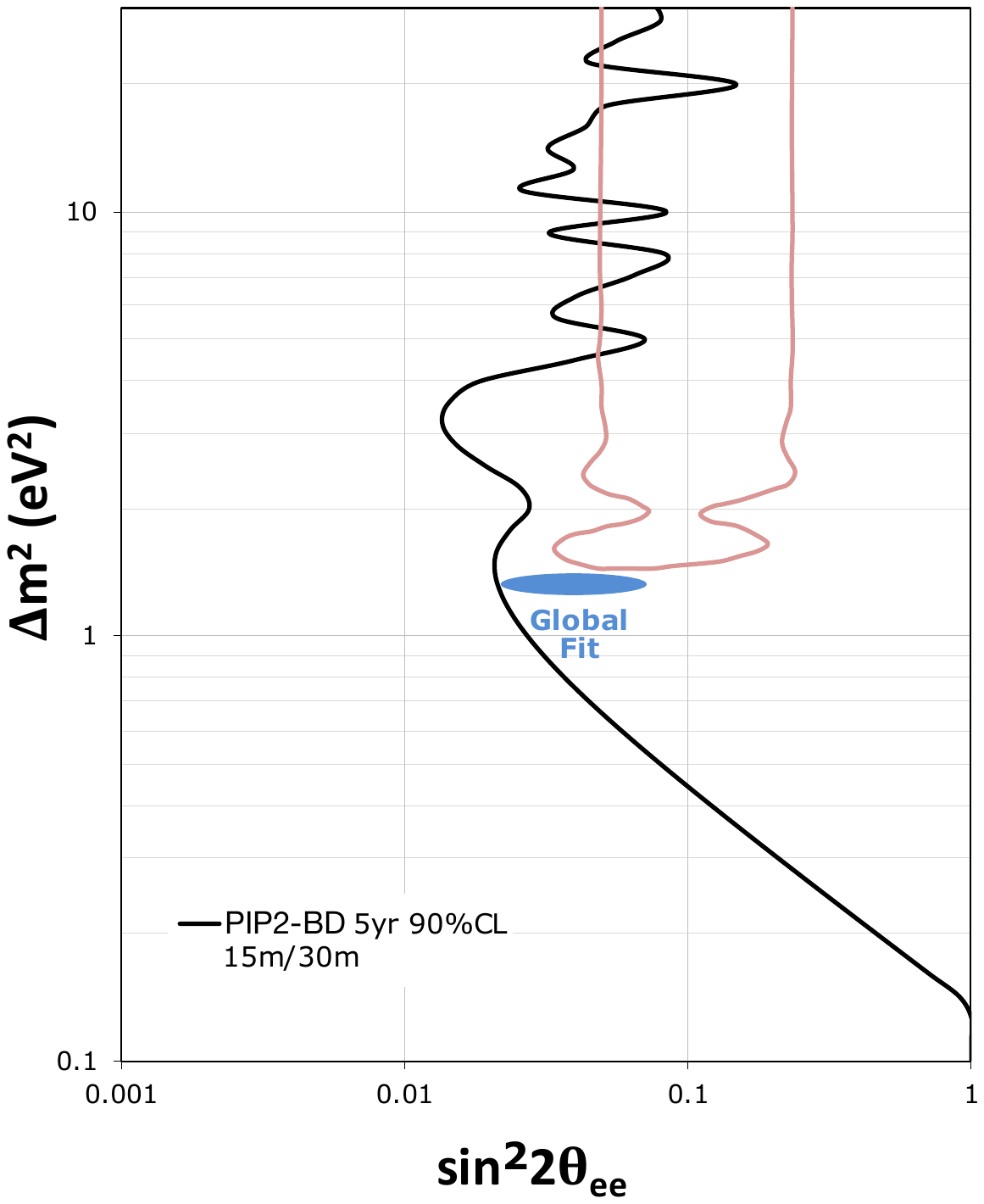}
  \includegraphics[width=0.30\textwidth]{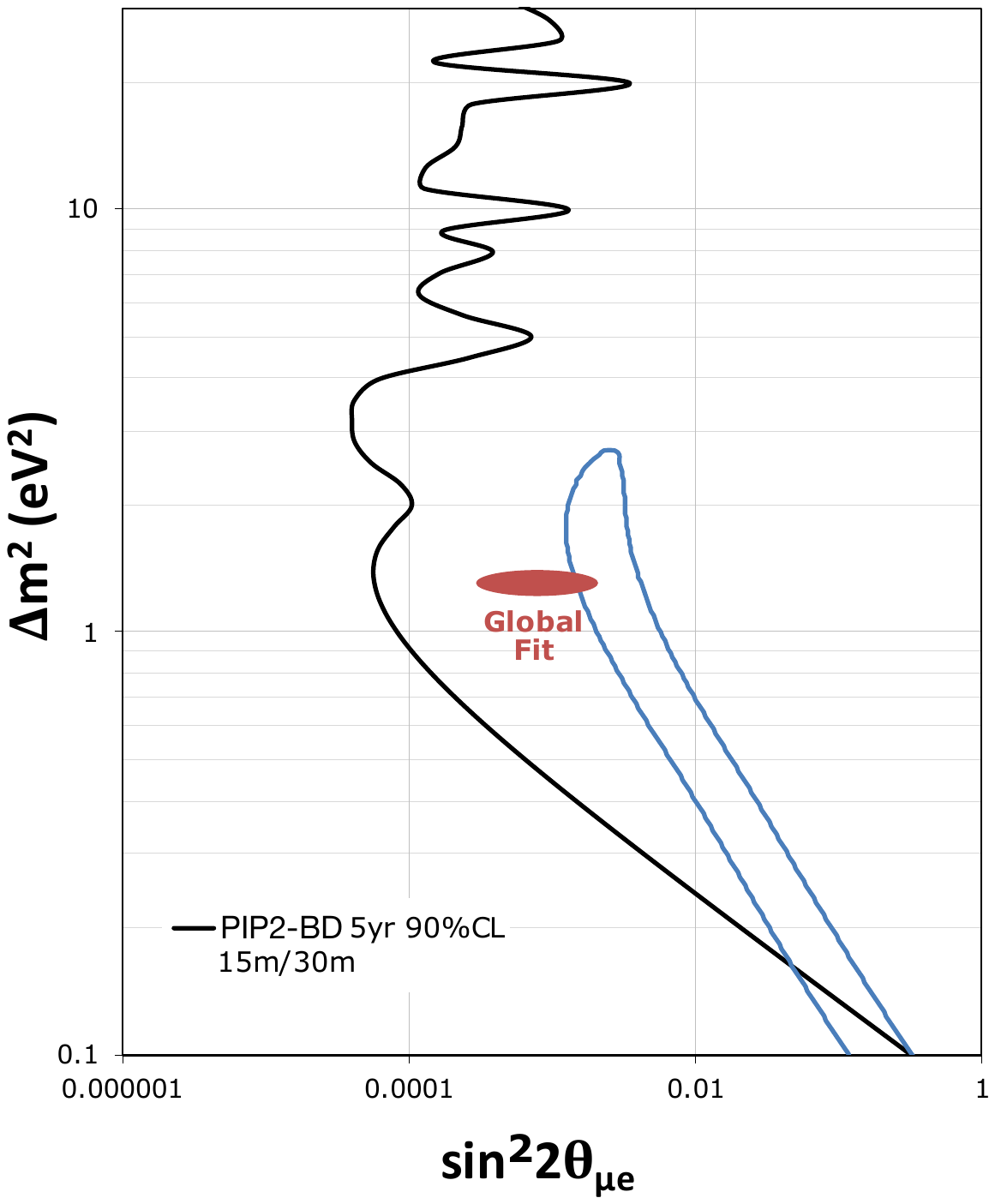}
  \caption{PIP2-BD 90\% confidence limits on active-to-sterile neutrino mixing compared to existing $\nu_\mu$ disappearance limits from IceCube~\cite{IceCube:2020phf} and a recent global fit~\cite{Diaz:2019fwt}, assuming a 5~year run (left). Also shown are the 90\% confidence limits for $\nu_\mu$ disappearance (left), $\nu_e$ disappearance (middle), and $\nu_e$ appearance (right), assuming the $\bar\nu_\mu$ and $\nu_e$ can be detected with similar assumptions as for the $\nu_\mu$. Figure from~\cite{Toups:2022yxs}.}
  \label{fig:sterilenu}
\end{figure}

Decay-at-rest neutrinos from a stopped pion beam dump provide an excellent source of $\nu_\mu$, $\bar\nu_\mu$, and $\nu_e$ with a time structure that can separate $\nu_\mu$ from $\bar\nu_\mu$ and $\nu_e$. 
CEvNS provides a unique tool to definitively establish the existence of sterile neutrinos through active-to-sterile neutrino oscillations~\cite{Anderson:2012pn}.  Using CEvNS, we can explore both mono-energetic $\nu_\mu$ disappearance with $\text{E}_\nu$ = 30~MeV and the summed disappearance of $\nu_\mu$, $\bar\nu_\mu$, and $\nu_e$ to $\nu_S$, which can also put constraints on $\nu_\mu \rightarrow \nu_e$ oscillation parameters in a 3+1 sterile neutrino model. We consider here a setup consisting of identical 100-ton LAr scintillation detectors, located 15~m and 30~m away from a carbon proton beam dump with a 20~keV recoil energy threshold and an efficiency of 70\%.  The 100-ton scale scintillation-only detector assumes a cylindrical volume with a 5 m height and 2.5 m radius, and we perform a full Geant4-based~\cite{Allison:2006ve} scintillation photon simulation with wavelength shifting and propagation to photomultiplier tubes along the endcaps and side walls of the detector. Based on simulation studies, we assume the detectors have a position resolution given by $\sigma = \frac{40 ~\textnormal{cm}}{\sqrt{{T}/20~\textnormal{keV}}}$ in each spatial dimension, where T is the nuclear recoil produced by the neutrino interaction within the detector. This information allows the possibility for a “rate+shape” fit using five bins in the reconstructed neutrino propagation distance with a bin width of 1 m matched to the expected resolution of the reconstructed neutrino propagation distance. If the prompt $\nu_\mu$ can be separated from the delayed $\bar\nu_\mu$ and $\nu_e$, one can exploit the mono-energetic feature of the $\nu_\mu$ flux and perform a joint rate + shape disappearance fit of CEvNS events in the near and far detectors as a function of reconstructed position.

In calculating the sensitivity, we assume the neutron background in this dedicated facility could be suppressed to a negligible level for this experiment and that the signal-to-noise ratio for the remaining steady-state backgrounds is 1:1. In Fig.~\ref{fig:sterilenu}, we compute the 90\% confidence limits on the $\nu_\mu \rightarrow \nu_S$ mixing parameter $\sin^22\theta_{\mu S}$ for a 5-year run of an upgraded 1.2 GeV proton accumulator ring operating with a pulse width of 20~ns, a duty factor of $\mathcal{O}(10^{-6})$, and a 75\% uptime, assuming a 9\% normalization systematic uncertainty correlated between the two detectors and a 36~cm path length smearing. Also shown are the 90\% confidence limits for $\nu_\mu$ disappearance, $\nu_e$ disappearance, and $\nu_e$ appearance, assuming the $\bar\nu_\mu$ and $\nu_e$ can be detected with similar assumptions as for the $\nu_\mu$.


\subsubsection{KPIPE at Fermilab}

The KPIPE experimental concept, outlined in Ref.~\cite{Axani:2015dha}, calls for a very long (120~m) and thin (1.5~m radius) cylindrical detector close to and oriented radially outward from an intense beam-dump source of monoenergetic 236~Mev $\nu_\mu$ from charged-kaon decay-at-rest ($K^+ \rightarrow \mu^+ \nu_\mu$, with branching ratio of 64\%) to achieve sensitivity to short-baseline muon-neutrino disappearance. The idea is to search for an $L/E$-dependent oscillation wave using fixed-$E$ neutrinos with minimal background and only modest detector requirements. 

The KPIPE detector, relying on liquid scintillator and silicon photomultipliers (or PMTs), is designed to look for 236~MeV $\nu_\mu n \rightarrow \mu^{-} p$ interactions, which provide a unique double-flash coincidence due to the muon decay following the initial prompt event. Mapping these interactions as a function of distance along the detector pipe, with a nominal, no-oscillation expectation of a $1/r^2$ rate dependence, provides sensitivity to muon-flavor disappearance. Given a beam dump, decay-at-rest neutrino source, the beam-based $\nu_\mu$ background (from decay-in-flight mesons) to these signal events is expected to be completely sub-dominant, at the 1-2\% level. While cosmics can be considered a concern for such a surface or near-surface detector, this background can be mitigated by typical accelerator duty factors of $\sim10^{-6}-10^{-5}$ combined with the short charged kaon lifetime (13~ns). The monoenergetic neutrino source, combined with low decay-in-flight background and small beam duty factor, means that the signal-to-background ratio is expected to be well over 50:1 in the scenarios considered. This large ratio means that the detector requirements, in particular the photocoverage, can be quite modest. In fact, a preliminary estimate at Ref.~\cite{cost} predicts that the entire KPIPE detector would cost \$5M.

The KPIPE detector was originally envisioned to be paired with the 3~GeV, 730~kW (currently, with 1~MW planned) J-PARC Spallation Neutron Source. Aside from the primary proton energy, which is above the kaon production threshold, and the high power, this source is particularly attractive because the beam timing structure, two $\sim$80~ns pulses separated by 540~ns at 25~Hz, provides an extremely low duty factor ($4\times 10^{-6}$), essential for cosmic background rejection. The drawback of this source, however, is that the 3~GeV primary proton energy, while above threshold, is somewhat lower than optimal for charged kaon production per unit power: at 3~GeV, the MARS15 software package~\cite{Mokhov:2012nco} predicts 0.007 KDAR $\nu_\mu$/POT. With an increase in proton energy to 8~GeV, for example, the production rate increases by a factor of 10 to 0.07 KDAR $\nu_\mu$/POT. Spatial and facility issues, especially in consideration of the existing materials-science-focused beamlines and experiments, also means that optimal detector placement, with KPIPE calling for a 120~m long detector with closest distance of 32~m from the neutrino source, is challenging.

\begin{figure}[!ht]
\begin{centering}
\includegraphics[width=8.7cm]{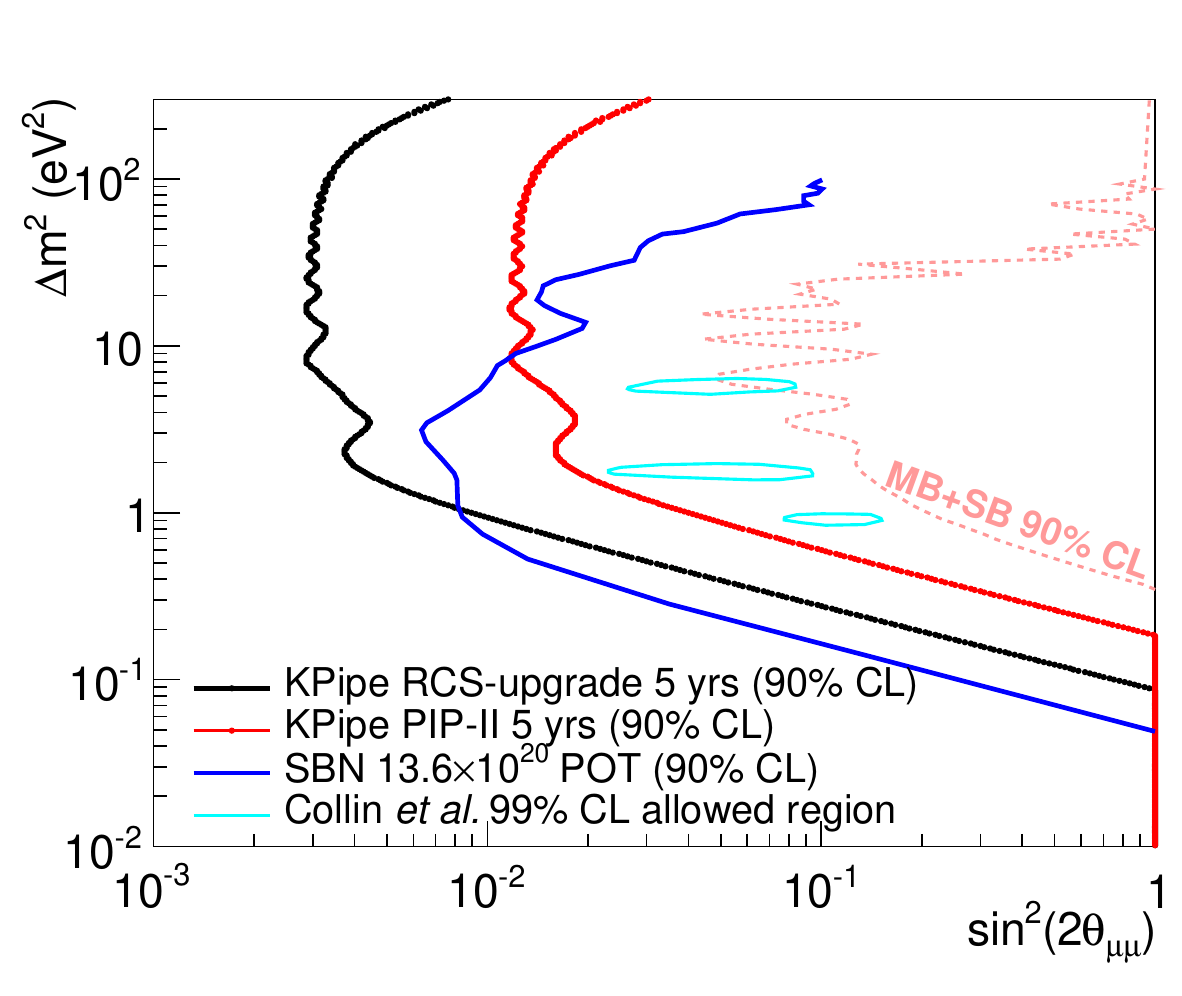}
\vspace{-.6cm}
\caption{The 90\% CL sensitivities of the KPIPE at Fermilab scenarios considered here, in both the PIP-II and RCS upgrade eras. For reference, we also show the expected 90\% C.L. SBN sensitivity (6~years)~\cite{MicroBooNE:2015bmn}, existing 90\% C.L. MiniBooNE+SciBooNE limit~\cite{SciBooNE:2011qyf}, and 99\% allowed region from the Collin \textit{et al.} global fit~\cite{Collin:2016aqd}. Figure from~\cite{Toups:2022knq}.}
\label{figure_sens}
\end{centering}
\end{figure}

The future Fermilab particle accelerator complex~\cite{Ainsworth:2021ahm}, including PIP-II~\cite{Lebedev:2017vnu} and the RCS upgrade~\cite{Eldred:2019erg}, can provide an optimal beam-dump/stopped-kaon neutrino source for KPIPE, in terms of beam energy (8~GeV), beam timing ($\sim10^{-5}$ duty factor), and spatial considerations (see Ref.~\cite{Toups:2022knq}). Using the detector and Fermilab-accelerator assumptions shown in Table~\ref{table:values}, and scaling based on the detailed study in Ref.~\cite{Axani:2015dha}, we expect KPIPE could achieve the sensitivity to short-baseline $\nu_\mu$ disappearance shown in Figure~\ref{figure_sens}. As can be seen, this sensitivity surpasses, and is highly complementary to, SBN (6~years) at $\Delta m^2>10$~eV$^2$ for both scenarios considered and $\Delta m^2>1$~eV$^2$ for the RCS upgrade era case.

\begin{table} [!ht]
\begin{center}
\begin{tabular}{|c|c|}
\hline
Experimental assumptions &     \\ \hline
Detector length &  120~m   \\
Active detector radius & 1.45~m   \\
Closest distance to source &  32~m   \\
Liquid scintillator density &  0.863 g/cm$^3$  \\
Active detector mass & 684~tons  \\
Primary proton energy & 8~GeV  \\
Target material & Hg or W  \\
KDAR $\nu_\mu$ yield (MARS15)  &  0.07~$\nu_\mu$/POT   \\
$\nu_\mu$ CC $\sigma$ @ 236 MeV (NuWro)  &  $1.3\times10^{-39}~\mathrm{cm}^2/\mathrm{neutron}$   \\
KDAR signal efficiency &    77\% \\
Vertex resolution  &   80~cm \\
Light yield & 4500~photons/MeV \\
Uptime (5 years) & 5000 hours/year \\
$\nu_{\mu}$ creation point uncertainty &    25~cm\\ \hline \hline
PIP-II era assumptions &   \\ \hline
Proton rate (0.08~MW) & 1.0 $\times 10^{21}$ POT/year\\
Beam duty factor & $1.6\times10^{-5}$\\
Cosmic ray background rate &    110~Hz \\ 
Raw KDAR CC event rate &  $2.7\times10^4$ events/year \\ \hline \hline
RCS upgrade era assumptions &   \\ \hline
Proton rate (1.2~MW) & 1.5 $\times 10^{22}$ POT/year\\
Beam duty factor & $5.3\times10^{-5}$\\
Cosmic ray background rate &   360~Hz \\ 
Raw KDAR CC event rate &  $4.0\times10^5$ events/year \\ \hline
\end{tabular} \caption{Summary of the relevant KPIPE experimental parameter assumptions.}\label{table:values}
\end{center}
\end{table}

\subsubsection{IsoDAR}

\begin{figure}[ht!]    
\begin{center}
\includegraphics[width=6.2in]{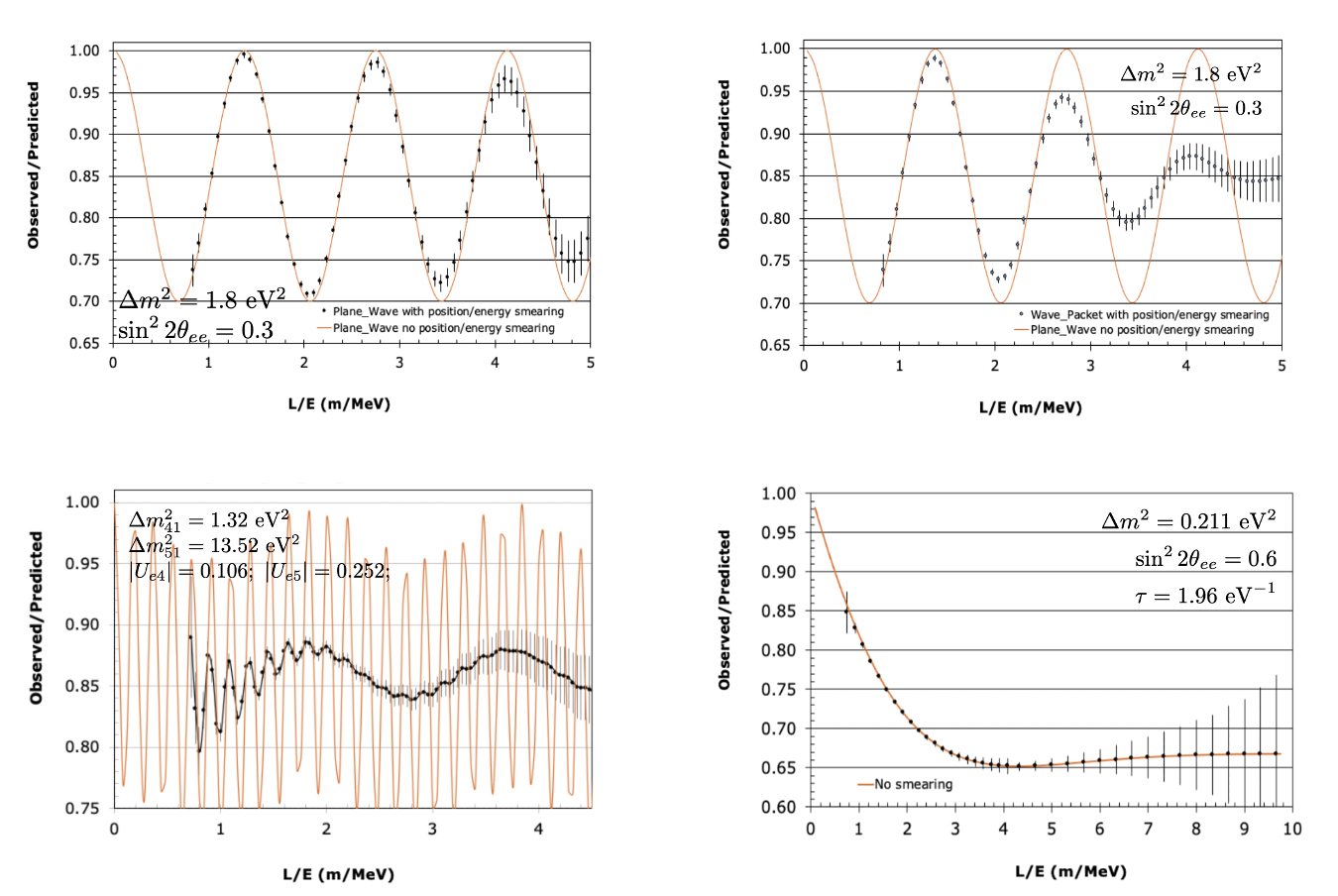}
\end{center}  
\caption{\label{wiggles}  The IsoDAR@Yemilab capability to trace $\bar \nu_e$ disappearance versus $L/E$ for IBD interactions.  Top left
and right present a 3+1 example without and with wavepacket effects described in Ref.~\cite{Arguelles:2022bvt}.
Bottom left and right are 3+2 and 3+1+decay models for the global best fit points in Ref.~\cite{Diaz:2019fwt}.  Orange is the true underlying model. Points represent the measurement capability. See text for further discussion. Figures adapted from~\cite{Alonso:2022mup}.}
\end{figure}

Through tracing $\bar \nu_e$ disappearance continuously across $L/E$ from 1 to $\sim 10$~m/MeV, the IsoDAR (Isotope Decay At Rest) experiment uniquely addresses the fundamental question raised by this white paper:  {\it What, if any, new physics phenomenology underlies the short-baseline anomalies?}  Despite the enormous consequences if new physics is the cause, the question has been unanswered for more than 20 years.  Incremental improvements on our present approaches are likely to yield more of the same confusing results.  IsoDAR represents an entirely new approach--the experiment makes use of a flux from $^8$Li $\beta$ decay, produced through a 60~MeV proton beam that is 
targeted on $^9$Be to yield neutrons that enter 
a surrounding  isotopically-pure $^7$Li sleeve and capture.  When this source is paired with the 2.3~kton Yemilab liquid scintillator detector (LSC), approximately 1.6 million inverse beta decay (IBD) events can be reconstructed in 5 years of running.  The high statistics, relatively high energy, $E$, of $\bar \nu_e$ from $^8$Li decay, and the ideal matching baseline, $L$, due to the size of the LSC,
gives unprecedented capability to study the $L/E$ dependence of short-baseline disappearance in an agnostic manner, determining  
its cause
without design assumptions that bias toward specific underlying physics models.  Fig.~\ref{wiggles} illustrates the power of IsoDAR to resolve various popular proposals for the source of the effect, with a 3+1 model at top left;  3+1 with nonzero quantum mechanical wave packet effects at top right;  introduction of additional sterile neutrinos, in this case 3+2 at bottom left; and introduction of new interactions, in this case 3+1+decay, at bottom right.    These examples show that IsoDAR can clearly elucidate the underlying oscillation-related phenomenology of the electron-flavor short-baseline anomalies, even in the case of physics that produces very complex waveforms.

\begin{figure}[ht!]
\begin{center}
\includegraphics[width=5.in]{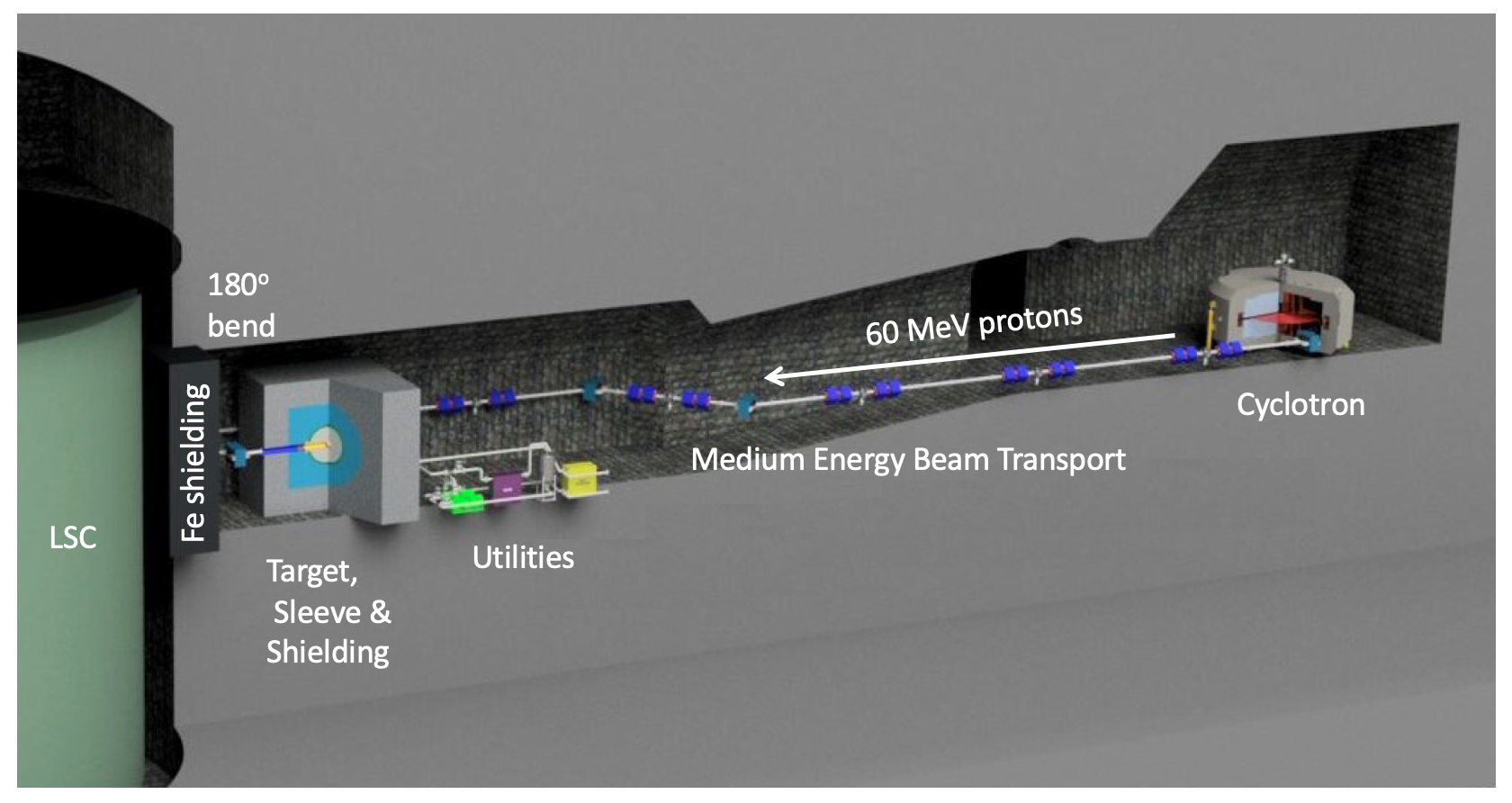}
\end{center}  
\caption{\label{Yemilayout} Layout of IsoDAR@Yemilab in the Yemilab caverns.   The excavation of the IsoDAR cavern complex is complete.}
\end{figure}

IsoDAR has received preliminary approval to run at Yemilab in the configuration shown in Fig.~\ref{Yemilayout}. 
Ref.~\cite{Alonso:2022mup} provides an overview  of the technology and installation-plan.  In Fig.~\ref{Yemilayout}, the cyclotron that drives the flux production is located at the far right.  This 
novel 5 mA
H$_2^+$ ion accelerator, producing 10 mA of 60~MeV protons, yields an order of magnitude higher proton beam current than on-market cyclotrons at similar energies.   Since the 2013 Snowmass Study, cyclotron development has culminated in the design described in Ref.~\cite{Winklehner:2021qzp}, which presents start-to-end simulations and prototypes of components now under test \cite{Winklehner:2020itb, Koser:2021rcl}.   As seen in Fig.~\ref{Yemilayout}, the proton beam is transported from right to left and then bent through 180$^\circ$ to the target surrounded by the sleeve, hence fast neutrons are directed away from the LSC detector, shown in green.  Substantial R\&D and engineering have established successful target, sleeve and shielding designs \cite{Bungau:2018spu, Bungau:2019brd}.  

IsoDAR@Yemilab is designed to address issues that have arisen during studies of reactor and MegaCurie source experiments.  The $\bar \nu_e$ flux is generated from a single, well understood isotope, avoiding issues faced by reactor flux modeling.   The $\bar \nu_e$ energy range is from about 3 to 13~MeV, with peak at $\sim 6$~MeV, well beyond environmental backgrounds and backgrounds from neutron capture.   The source creation region is compact ($\sim$41 cm at 1$\sigma$) and isotropic.  The experiment has the capability of event-by-event reconstruction, with prompt (e$^+$) energy resolution of 2.3\% and vertex resolution of 4~cm at 8~MeV \cite{Alonso:2021kyu}, in contrast to MegaCurie (MCi) source experiments that do not reconstruct event kinematics. Use of the well-known IBD cross section is also an advantage over the gallium experiments.  Lastly, and  importantly, the size of the LSC detector, when combined with the energy range of the source, leads to the wide $L/E$ range, allowing precision reconstruction of the oscillation wave across many cycles.

A simple 3+1 model is traditionally used for cross-comparison of experimental reach. The reach of IsoDAR in $\sin^2 2\theta_{ee}$ at 95\% CL is presented in Tab.~\ref{crosscompare}, column 2 for a range of $\Delta m^2$.   For comparison, column 3 shows the combined limits at 95\% CL from Prospect, RENO and Daya Bay~\cite{Arguelles:2022bvt}. The IsoDAR mixing angle reach is $\times 4$ ($\times 35$) that of the reactor limits at 1 (8)~eV$^2$. For allowed region comparisons for Neutrino-4 \cite{Serebrov:2020kmd} and the Gallium experiments \cite{Barinov:2021asz}, we use the $2\sigma$ lower edge in $\sin^2 2\theta_{ee}$.   In the Neutrino-4 case, the allowed region is narrow in $\Delta m^2$ and does not coincide with 8~eV$^2$, so we present the mixing angle reach for the best fit mass splitting of 7.3~eV$^2$.

\begin{table}[ht]
\begin{center}
      \begin{tabular}{|c||c||c||c|c|} \hline 
$\Delta m^2$ & IsoDAR@Yemilab & Combined Reactor \cite{Arguelles:2022bvt} & Neutrino-4 \cite{Serebrov:2020kmd} &  Gallium \cite{Barinov:2021asz}  \\
~&Sensitivity & Limits & Allowed & Allowed \\ \hline
1~eV$^2$ & 0.004 & 0.016 & N/A & 0.28 \\ 
2~eV$^2$ & 0.004 & 0.07 & N/A &  N/A \\ 
4~eV$^2$ & 0.005 & 0.13 & N/A & 0.27 \\ 
8 (7.3)~eV$^2$ & 0.008  & 0.28 & (0.12) & 0.28 \\ \hline
\end{tabular}
\end{center}
\caption{Quantitative comparison of the low $\sin^22\theta_{ee}$ 2$\sigma$ reach of electron-flavor experiments in the $\Delta m^2$ range of interest.   IsoDAR sensitivity is based on assumptions in Ref.~\cite{Alonso:2021kyu}.   Combined reactor limits are from PROSPECT, NEOS, and Daya Bay. N/A indicated $\Delta m^2$ is not within 95\% CL allowed region. The Neutrino-4 2$\sigma$ reach is quoted at 7.3~eV$^2$.}
\label{crosscompare}
\end{table}

As a result of the novel design, IsoDAR@Yemilab design is able to elucidate $\bar \nu_e$ disappearance across $L/E$ of 1 to $\sim$10~m/MeV without guidance from any phenomenological model.    Fig.~\ref{wiggles} illustrates the complex oscillation waves that are able to be differentiated in 5 years of running, with efficiency included.   The upper plots illustrate the oscillation wave without (left) and with (right) wavepacket effects as discussed in Ref.~\cite{Arguelles:2022bvt}, at a point where the combined reactor limit and gallium allowed region overlap assuming the wavepacket model. Comparison of the two plots shows that the distinctive damping due to wavepacket effects can be observed given IsoDAR's high statistics.
 The lower plots present 3+2 and 3+1+decay models evaluated at the best fit points from Ref.~\cite{Diaz:2019fwt}. The value of the high statistics and excellent reconstruction of IsoDAR is particularly emphasized by the 3+2 case, where the second predicted modulation is clear due to the capability of using very fine binning.   The orange line indicates the true underlying distribution, while the points with error bars present the expected measurements, illustrating the loss of information from finite statistics and bin sizes.   The 3+1 (top) and 3+2 points (bottom right) also include detector energy and position smearing. 


The experiment will also collect $\times 4$ the world's sample $\bar \nu_e$-electron elastic scattering events in 5 years which may help further decipher new physics, depending on the source.   In fact, IsoDAR@Yemilab has an extensive discovery-level physics program beyond searching for the short-baseline anomalies, including an order of magnitude improvement in NSI searches through elastic scattering from electrons \cite{Alonso:2021kyu}, unique neutrino-based searches for Z$^\prime$ signatures \cite{Alonso:2021kyu}, and exotic non-neutrino searches, such as for neutrons shining through walls \cite{Hostert:2022ntu}.   As such, IsoDAR represents a leap forward for electron-flavor neutrino experiments.

\subsection{Decay-in-Flight Accelerator Experiments}

\subsubsection{Short-Baseline Experiments}
\subsubsubsection{The Fermilab SBN Program}\label{fermilab_sbn}

\begin{figure}[ht]
    \centering
    \includegraphics[width=0.45 \textwidth]{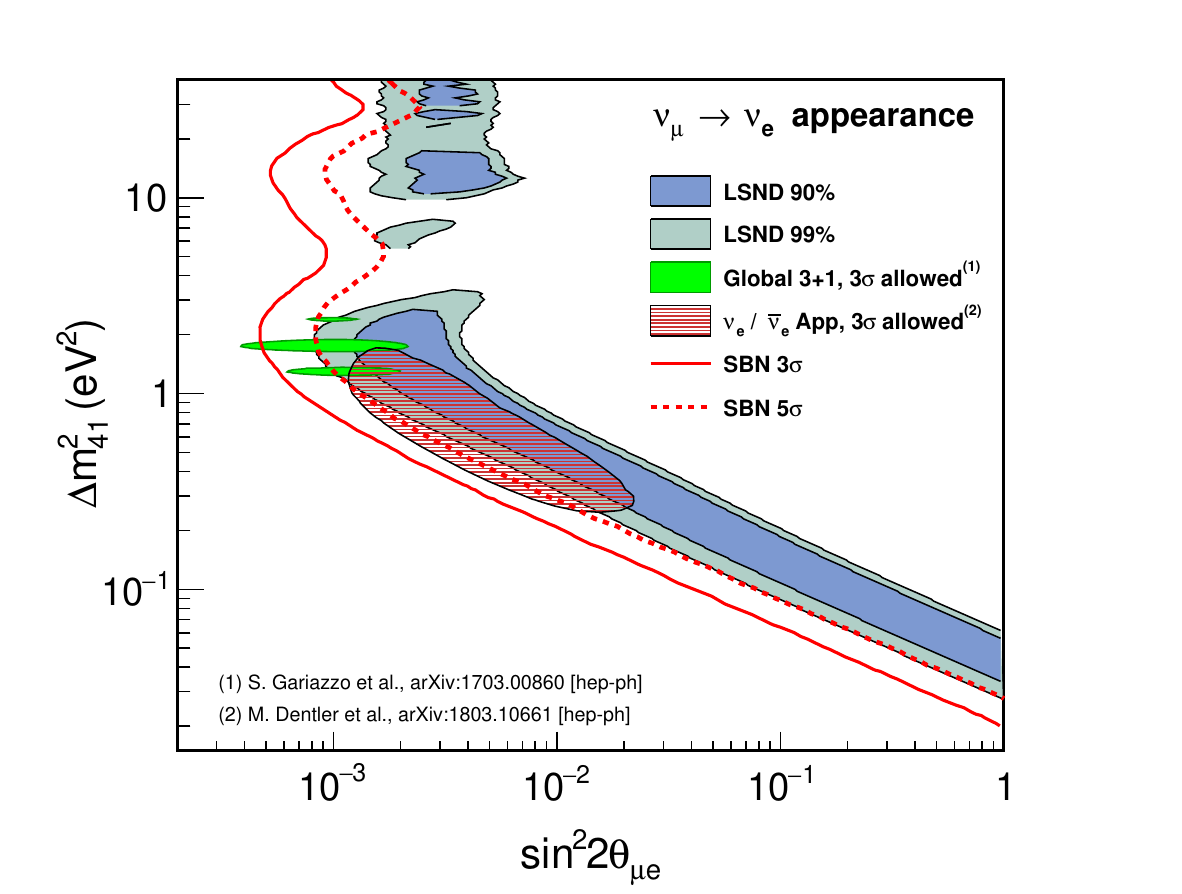}
    \includegraphics[width=0.45 \textwidth]{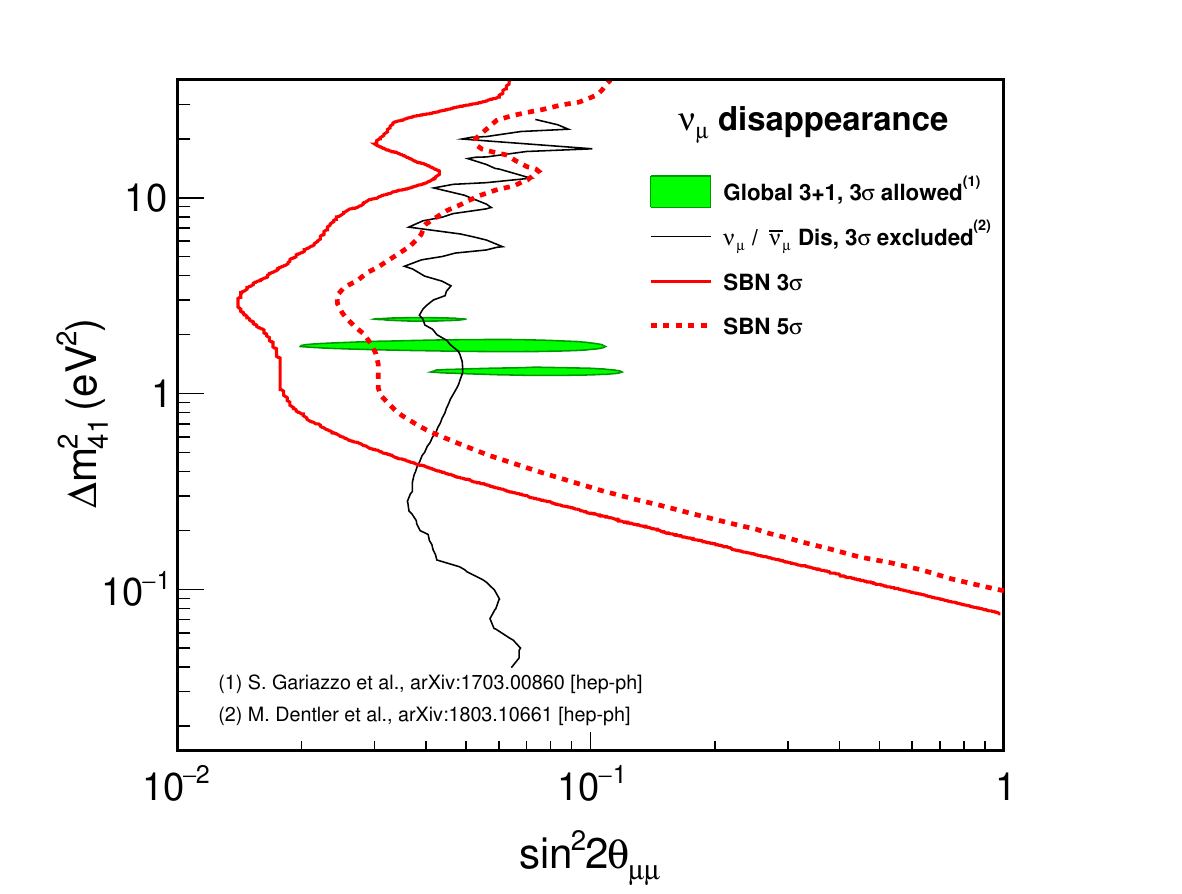}
    \caption{
      SBN light sterile neutrino sensitivities in the $\nu_\mu\to\nu_e$ appearance channel {\it (left)} and $\nu_\mu\to\nu_\mu$ disappearance channel {\it (right)} according to the SBN proposal~\cite{MicroBooNE:2015bmn}. The $3\sigma$ ($5\sigma$) sensitivities are given by the solid (dotted) red curves.  The LSND 90\% C.L. (99\% C.L.) allowed region is shown as shaded blue (grey)~\cite{Aguilar:2001ty}.  The global $3\sigma$ $\nu_e$ ($\nu_{mu}$) appearance (disappearance) regions from Ref.~\cite{Dentler:2018sju} are shown by the shaded red region (black line), and the global best fit regions from Ref.~\cite{Gariazzo:2017fdh} are shown in green. Figure from~\cite{Machado:2019oxb}.
    } 
    \label{FIG:SBN}
  \end{figure}

\begin{figure}[htpb!]
    \centering
    \includegraphics[width=0.45 \textwidth]{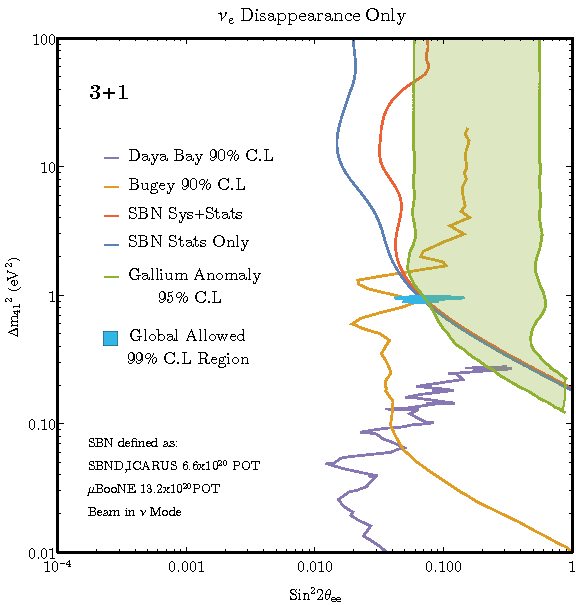}
    \caption{Due to the large intrinsic $\nu_e$ statistics at SBND, SBN is also sensitive to $\nu_e$ disappearance, probing $\sin^22\theta_{ee}$ at high $\Delta m^2 \ge 0.2$~eV$^2$ (assumes $\theta_{24}$=0).  This provides a complementary probe of oscillations traditionally probed using reactor antineutrinos at a much lower (MeV) energy scale. Figure from~\cite{Cianci:2017okw}.
    }
    \label{fig:nuedisSBN}
  \end{figure}

The Short-Baseline Neutrino (SBN) program consists of three LArTPC detectors located along the BNB at Fermilab: the MicroBooNE detector, which completed operations in 2021; the ICARUS detector, which began operations in the BNB in 2021; and the upcoming Short-Baseline Near Detector (SBND), which is expected to begin operations in 2023. This program represents an exciting opportunity for a multi-baseline search for light sterile neutrino oscillations in multiple exclusive or inclusive oscillation channels, and a test of the 3+1 light sterile neutrino oscillation interpretation of past experimental anomalies at $\ge 5\sigma$~\cite{MicroBooNE:2015bmn}. In particular, $\nu_\mu$ CC measurements across the three detectors will probe $\nu_\mu \rightarrow \nu_{\not{\mu}}$ oscillations with world-leading sensitivity as shown in Fig.~\ref{FIG:SBN} (right); $\nu_e$ CC measurements will probe $\nu_\mu\rightarrow\nu_e$ and/or $\nu_e\rightarrow\nu_{\not{e}}$ oscillations with sensitivity as shown in Fig.~\ref{FIG:SBN} (left) and \ref{fig:nuedisSBN}. Additionally, NC-based oscillation searches have been proposed, e.g.~\cite{Furmanski:2020smg}, with unique sensitivity to $U_{si}$ under a 3+$N$ model, as well as potentially $U_{\tau i}$ (for $i>3$) when combined with $\nu_e$ and $\nu_\mu$ CC-based appearance and disappearance searches.

\begin{figure}[htbp!]
    \centering
    \includegraphics[width=0.5\textwidth]{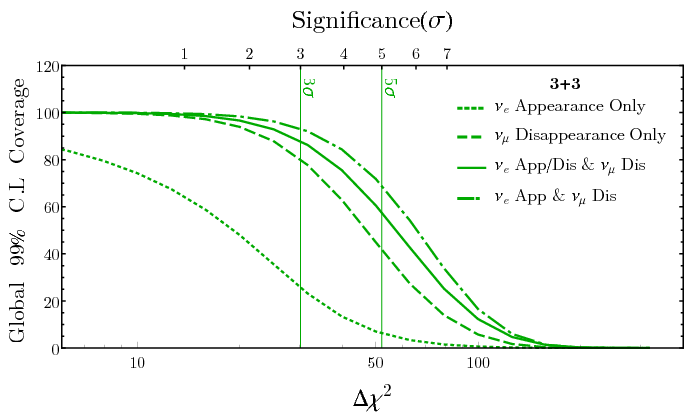}
    \caption{SBN's coverage of globally-allowed 3+3 light sterile neutrino oscillation parameters, defined as the fraction of 99\%-CL-globally-allowed parameter space that can be ruled out by SBN at a given CL indicated by the $x$-axis, assuming a null observation. Coverage of 3+1 and 3+2 globally-allowed parameter space is provided in \cite{Cianci:2017okw}. Figure from~\cite{Cianci:2017okw}.}
    \label{fig:sbn3p3}
\end{figure}

\begin{figure}[htbp!]
    \centering
    \includegraphics[width=0.45\textwidth]{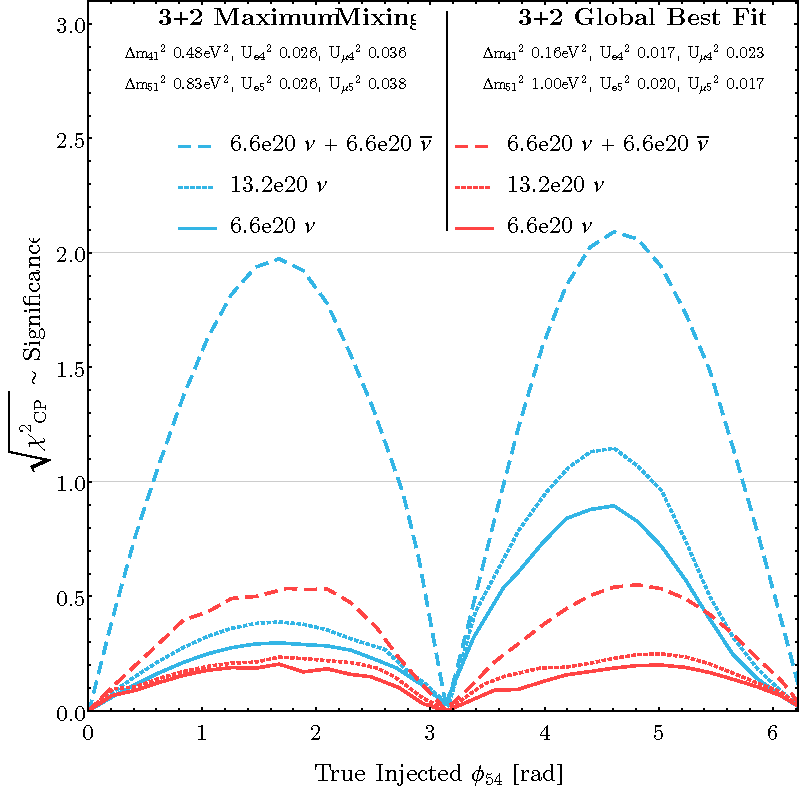}
    \caption{The significance at which SBN can observe CP violation in the (3+2) sterile neutrino scenario, as a function of true CP violating phase $\phi_{54}$, for two injected signals corresponding to the global (3+2) best fit point (red lines) as well as the point with largest total allowed mixings (blue lines), for a variety of POT in neutrino and antineutrino running modes at SBN. Figure from~\cite{Cianci:2017okw}.}
    \label{fig:sbn3p2}
\end{figure}

  The power of a multi-baseline and multi-channel search has been shown to be advantageous not only for 3+1 searches, but for 3+$N$ searches more generally. For example, Ref.~\cite{Cianci:2017okw} has found that SBN is capable of ruling out 85\%, 95\% and 55\% of the 99\%-globally-allowed parameter space region\footnote{Global allowed regions as of 2018.} of 3+1, 3+2, and 3+3 light sterile neutrino oscillation parameters at 5$\sigma$ CL, assuming a null observation, particularly when appearance and disappearance effects are studied simultaneously (including correlations). This is illustrated in Fig.~\ref{fig:sbn3p3}, for the 3+3 scenario. Additionally, it has been pointed out that within the context of 3+$N$ oscillations with N$>1$, SBN offers an opportunity for measuring potential CP violation in the leptonic sector, particularly if future antineutrino beam running is possible with SBN. In particular, if antineutrino exposure is considered, for maximal values of the (3+2) CP violating phase $\phi_{54}$, SBN could be the first experiment to directly observe $\sim 2\sigma$ hints of CP violation associated with an extended leptonic sector. This is illustrated in Fig.~\ref{fig:sbn3p2}, for the 3+2 scenario. Furthermore, a planned analysis using the ICARUS detector can probe meter-scale oscillations within the detector volume, consistent with sterile mass splittings $\approx 7 eV^2$ and  $\sin^2 2\theta_{14}\approx0.4$, providing a test of the allowed region claimed by the Neutrino-4 experiment.

Beyond searches for physics associated with eV-scale sterile neutrinos, SBN's main physics goals include detailed studies of neutrino–argon interactions at the GeV energy scale, enabled by millions
of neutrino interactions that will be recorded on argon in its high precision detectors. SBND's anticipated high statistics, in particular, provide a unique opportunity for first-ever measurements of rare SM-predicted neutrino interaction processes at 0.1-1~GeV, including rare photon production processes such as coherent NC single-photon production, NC $\Delta \rightarrow N\gamma$ radiative decay, or production and radiative decay of heavier resonances.  Additionally, the high statistics coupled with the unprecedented event reconstruction, excellent particle identification, and fine-sampling calorimetry of the SBN detectors' LArTPC technology opens up invaluable opportunities for new physics searches.  



%
In particular, the capabilities of LAr detectors will allow for greater discrimination between $e^-, \, e^+e^-, \,  \gamma,$ and $\gamma\gamma$ final states, as well as to identify final state hadron multiplicities. 
For models in which new particles are produced in neutrino-nucleus scattering, it will be possible to search for a hadronic vertex associated with the displaced decay position. The EM showers in this case may not point back to the original vertex due to missing energy.  Furthermore, for models that explain MiniBooNE with new heavy particles produced in meson decays, the SBN detectors can also leverage the decays-at-rest of kaons produced in the NuMI absorber~\cite{MicroBooNE:2021usw}.

\begin{figure}
   \centering
   \includegraphics[width=0.16\linewidth]{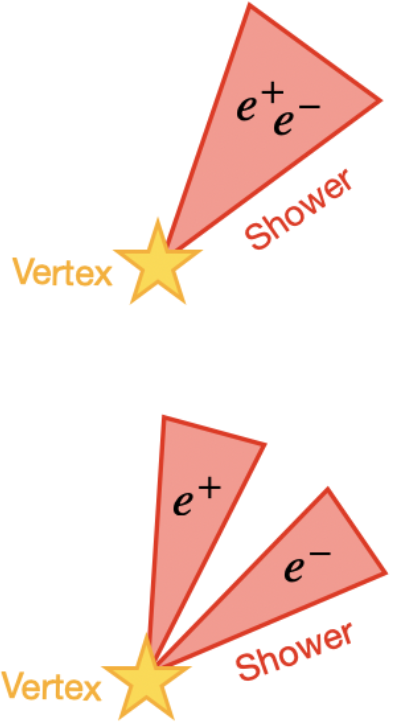}
   \hspace{2cm}
   \includegraphics[width=0.19\linewidth]{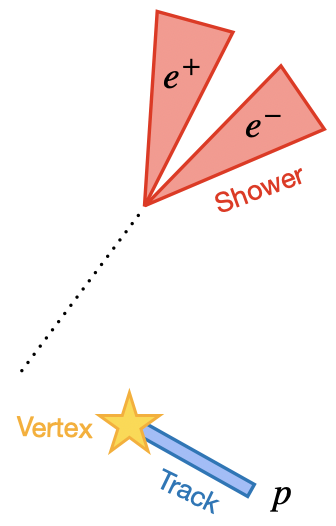}
   \hspace{1.8cm}
   \includegraphics[width=0.18\linewidth]{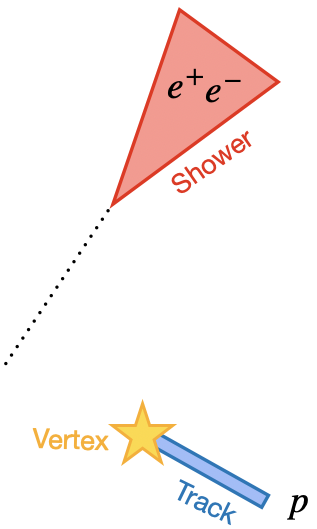}
\caption{The four main topologies of $e^+e^-$ LEE models at MicroBooNE: coherent and incoherent scattering with well-separated or overlapping $e^+e^-$ pairs.}
\label{fig:dark_nu_topologies}
\end{figure}

Reference~\cite{Machado:2019oxb} provides a broad overview of such new physics, including their signatures in SBN. Here, we limit the discussion to a summary of ones suggested as interpretations to short-baseline anomalies, extending beyond eV-scale sterile neutrinos:

\begin{itemize}
    \item  SBN can probe eV-scale sterile neutrinos decaying to active neutrinos and a Majoron or gauge boson, which would lead to new features in the active neutrino energy spectrum with respect to 3+$N$ scenarios.
\item Large extra-dimension models, such as ones proposed as an explanation of the reactor anomaly, would affect both appearance and disappearance channels at SBN. 
\item Resonant $\nu_\mu\rightarrow\nu_e$ oscillations that arise in the presence of a light scalar boson that couples only to neutrinos and could induce a MSW effect sourced by the cosmic neutrino background could also be probed with SBN, through the search of $\nu_\mu\rightarrow\nu_e$ transitions and lack of $\nu_\mu$ disappearance (as the latter would be suppressed compared to a vanilla 3+1 scenario).
\item Violation of Lorentz and CPT symmetry 
would lead to modifications in the oscillation probability measurable at SBN, such as direction-dependent effects, neutrino-antineutrino mixing, annual modulations, and energy dependent effects on observable mass splittings and mixing angles. 
\item Sterile neutrinos and altered dispersion relations (ADR), also proposed as an explanation of the short-baseline anomalies, 
would have a similar phenomenology in SBN to that of the usual 3+3 sterile scenario (while evading the constraints from long-baseline and atmospheric neutrino experiments).
\item Charged current non-standard interactions (CCNSI) in the lepton sector 
could lead to a number of observable effects, such as (1) deviations of the SM CC quasi-elastic cross section, (2) modification of angular and energy distributions due to the presence of new Lorentz structures, and (3) flavor violation such as $\nu_\mu n \rightarrow e^- p$; 
At SBN, CCNSIs can lead to an apparent baseline-independent $\nu_\mu\rightarrow\nu_e$ conversion.
\item Dark neutrino sectors connected to the standard neutrino sector, allowing for neutrino
upscattering into a heavy state which could then decay to a light neutrino and a gauge boson within a detector, followed by the gauge boson decay to visible particles such as $e^+e^-$ 
could also be measurable at SBN. As shown in  Fig.~\ref{fig:dark_nu_topologies}, an $e^+e^-$ pair can give rise to four distinct topologies in LAr, depending on the lifetime of the parent particle and on the angle between the charged leptons. A variety of signatures could be probed, including pair production of $e^+e^-$, $\mu^+\mu^-$ or $\pi^+\pi^-$ induced by neutrino interactions, with little to no hadronic activity and with the same signal strength at all three detectors, since there is no $L/E$ dependence. 
\item Heavy neutrinos and transition magnetic moment 
proposed as an explanation of the LSND/MiniBooNE anomalies 
would be observable at SBN as anomalously large single-photon production with
small hadronic activity. LArTPC $e-\gamma$ discrimination capability places SBN in a special
position to probe these scenarios. 
Again, the signal strength would be the same at all three detectors, since there is no L/E dependence in these models either.
\end{itemize}

\subsubsubsection{nuSTORM}

The 2020 Update of the European Strategy for Particle Physics
(ESPP)~\cite{EuropeanStrategyGroup:2020pow} recommended that muon beam
R\&D should be considered a high-priority future initiative and that a
programme of experimentation be developed to determine the neutrino
cross-sections required to extract the most physics from the DUNE and
Hyper-K long-baseline experiments.
The ENUBET~\cite{ENUBET:WWW,ENUBET:EUWWW,Torti:2020yzn} and
nuSTORM~\cite{nuSTORM:WWW,Ahdida:2020whw} 
collaborations have begun to work within and alongside the CERN
Physics Beyond Colliders study group~\cite{PBC:WWW} and the
international Muon Collider Collaboration~\cite{iMC:WWW} to carry out
a joint, five-year design study and R\&D programme to deliver a
concrete proposal for the implementation of an infrastructure in
which:
\begin{itemize}
  \item ENUBET and nuSTORM deliver the neutrino cross-section
    measurement programme identified in the ESPP and allow sensitive
    searches for physics beyond the Standard Model to be carried out;
    and in which
  \item A 6D muon ionisation cooling experiment is delivered as part
    of the technology development programme defined by the
    international Muon Collider Collaboration.
\end{itemize}

With their existing proton-beam infrastructure, CERN and Fermilab are
both uniquely well-placed to implement ENUBET, nuSTORM, and the
6D-cooling experiment as part of the required muon collider
demonstrator. 
The design of ENUBET, carried out within the framework of a European
Research Council funded design study, includes the precise layout of
the kaon/pion focusing beamline, photon veto and timing system as well
as the development and test of a positron tagger together with the
required electronics and readout.
The feasibility of implementing nuSTORM at CERN has been studied by
the CERN Physics Beyond Colliders study group while a proposal to site
nuSTORM at FNAL was developed for the last Snowmass study in 2013.
The FNAL study focused on the optimisation of the muon storage ring to
provide exquisite sensitivity in the search for sterile neutrinos.
In the Physics Beyond Colliders study, the muon storage ring was
optimised to carry out a definitive neutrino-nucleus scattering
programme using stored muon beams with momentum in the range 1\,GeV to
6\,GeV while maintaining its sensitivity to physics beyond the
Standard Model.

The study of nuSTORM is now being taken forward in the context of the
demonstrator facility required by the international Muon Collider
Collaboration that includes the 6D muon ionization cooling experiment.
The muon-beam development activity is being carried out in close
partnership with the ENUBET collaboration and the Physics Beyond 
Colliders Study Group.
In consequence we now have the outstanding opportunity to forge an
internationally collaborative activity by which to deliver a concrete
proposal for the implementation of the nuSTORM infrastructure.

On top of the program outlined above, nuSTORM can still provide unprecedented sensitivity to light sterile neutrinos. In particular, it allows to search for short-baseline oscillations in $\nu_e \to \nu_\mu$ appearance, the CPT-conjugate channel of the appearance hypothesis at LSND. This would be possible due to charge selection of muons in a magnetic field, which can discriminate the $\overline{\nu}_\mu$ produced in $\mu^+$ decays from the $\nu_e\to\nu_\mu$ oscillations. A detailed study of nuSTORM's sensitivity to sterile neutrinos was performed in Ref.~\cite{nuSTORM:2014phr}, which focused on a $3.8$~GeV muon ring design for siting at FNAL. The study considered a $1.3$~kt magnetized iron-scintillator detector at $2$~km from the ring with an exposure of $10^{21}$ POT, corresponding to $\approx 2\times 10^{18}$ useful muon decays. The sensitivity curves covered the entire LSND and MiniBooNE regions of preference at more than $5\sigma$. This impressive sensitivity was achieved thanks to the muon signature, which is subject to low levels of backgrounds, and the low systematic uncertainties on the neutrino flux.

Thanks to its unique neutrino beam, nuSTORM is also sensitive to other explanations to short-baseline anomalies. It would stand out as a unique test of lepton-flavor-violation in muon decays. Because the beam is derived from $\mu^+$ decays, any exotic branching ratio of the muon, such as $\mu^+\to e^+ \nu_\alpha \overline{\nu}_e$, would be a striking signature in a near detector with electron-positron discrimination capabilities, such as in a magnetized, low-density detector. This type of near detector would also benefit the sensitivity to models with neutrino upscattering to new dark particles with decays to $e^+e^-$.

\subsubsection{Long-Baseline Experiments} \label{future_lbl}
\subsubsubsection{DUNE} \label{dune}
The DUNE experiment is a next-generation, long-baseline neutrino oscillation experiment, designed to be sensitive to $\numu$ to $\nue$ oscillations. 
The experiment consists of a high-power, broadband neutrino beam, a powerful precision multi-instrument Near detector complex located at Fermi National Accelerator Laboratory in Batavia, Illinois, 574\,m away from the neutrino production target, and a massive Liquid Argon Time Projection Chamber Far detector located at the 4850\,ft level of the Sanford Underground Research Facility (SURF), 1300\, km away in Lead, South Dakota, USA. 
The anticipated total fiducial mass of the Far detector is 40\,kton. The long baseline of 1300\,km provides sensitivity, in a single experiment, to all parameters governing neutrino oscillations. 
Due to the high-power proton beam facility, the Near detector consisting of precision detectors capable of off-axis data taking for improved constraining of systematics, and the massive Far detector, DUNE provides enormous opportunities to probe BSM phenomena in both new particle production and interactions, and in neutrino propagation effects.  

\begin{figure}[!ht]
\begin{center}
\includegraphics[width=0.4\linewidth]{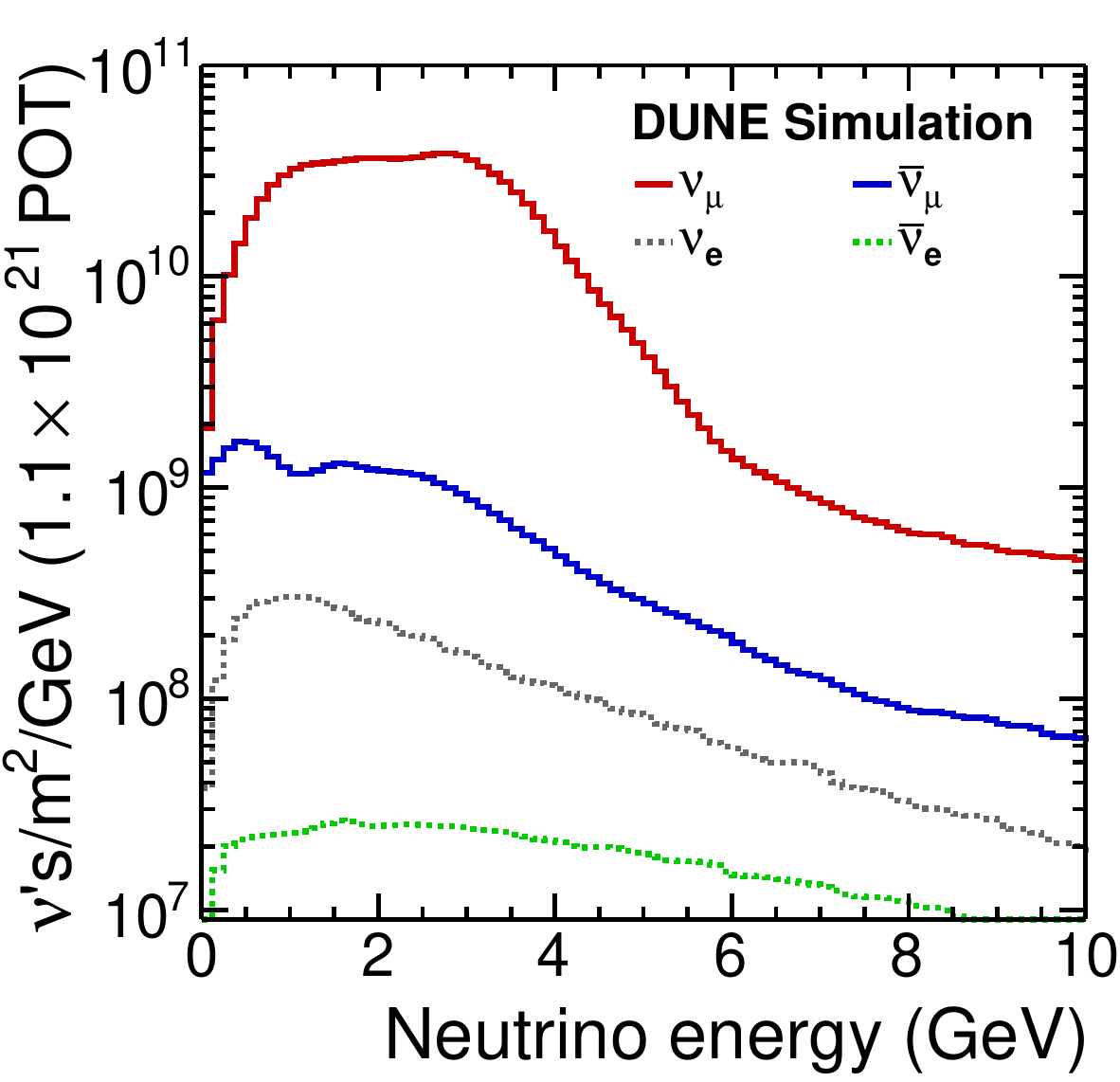}
\includegraphics[width=0.4\linewidth]{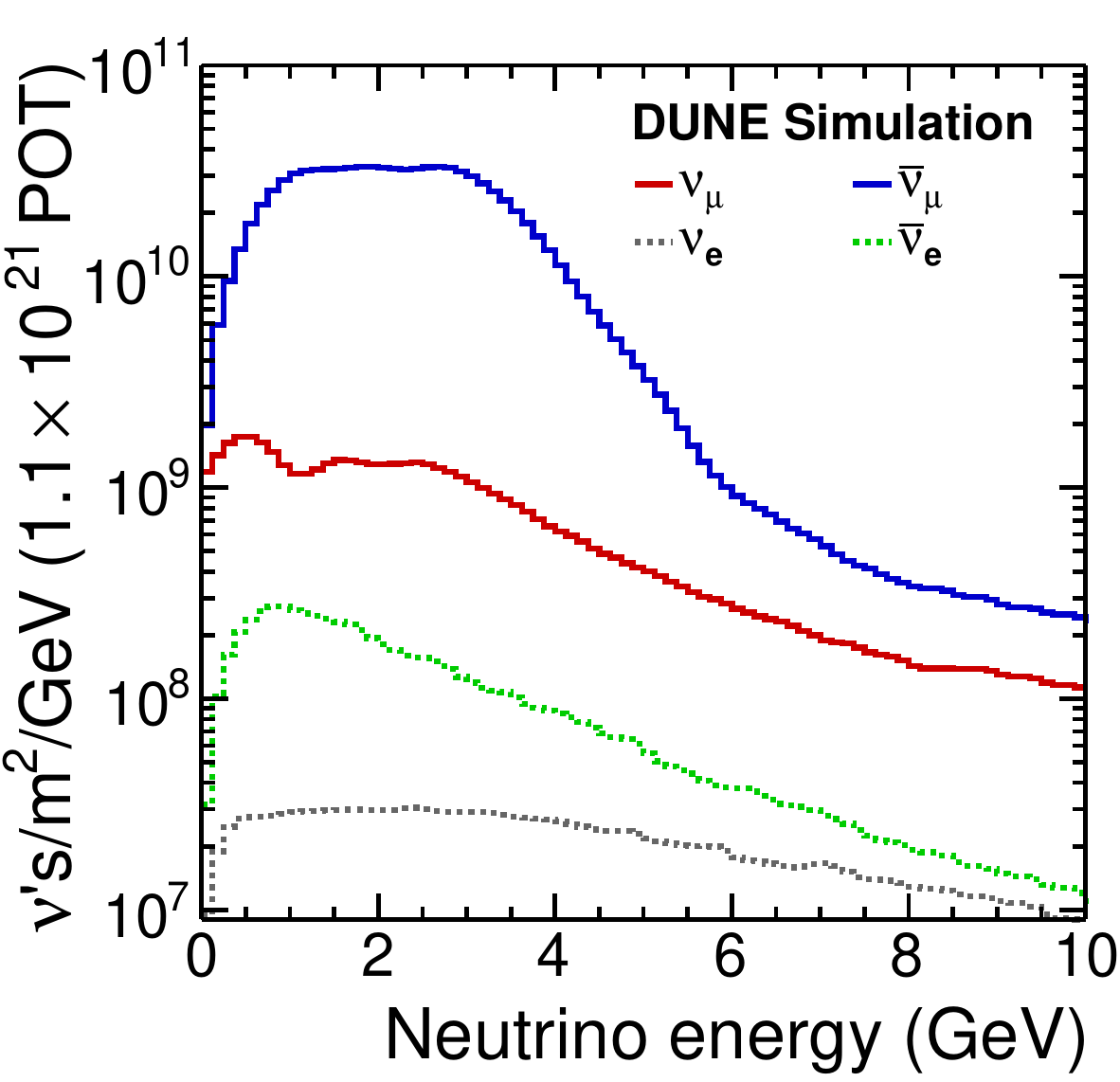}
\caption{LBNF neutrino beam fluxes at the DUNE Far detector for neutrino-enhanced Forward Horn Current (FHC) beam running (left), and antineutrino-enhanced Reverse Horn Current (RHC) beam running (right). Figure from~\cite{DUNE:2020jqi}.}
\label{fig:dune_fd_flux}
\end{center}
\end{figure}

DUNE expects to begin data taking operations in 2029 with half of the full Far detector, and start beam data taking operations in 2031 with the 1.2 MW Long-Baseline Neutrino Facility (LBNF) beam, upgradable to 2.4 MW.
The LBNF neutrino beam flux sampled on-axis by the Far detector is shown in Fig.~\ref{fig:dune_fd_flux}. 
The wide-band range of energies provided by the LBNF beam afford DUNE significant sensitivity to probe sterile mixing, which would typically cause distortions of standard oscillations in energy regions away from the three-flavor $\numu\rightarrow\numu$ disappearance maximum.  
Therefore, DUNE sterile mixing probes reach a broad range of potential sterile neutrino mass splittings by looking for disappearance of CC and NC interactions over the long distance separating the ND and FD, as well as over the short baseline of the ND. 
The DUNE sterile neutrino mixing studies shown below assume a minimal 3+1 oscillation scenario with three active neutrinos and one sterile neutrino, with a new independent neutrino mass-squared difference, $\Delta m^2_{41}$, and for which the mixing matrix is extended with three new mixing angles, $\theta_{14}$, $\theta_{24}$, $\theta_{34}$, and two additional phases $\delta_{14}$ and $\delta_{24}$.

\begin{figure}[!htb]
\setlength{\lineskip}{5pt}
    \centering
	  	\includegraphics[width=0.48\textwidth]{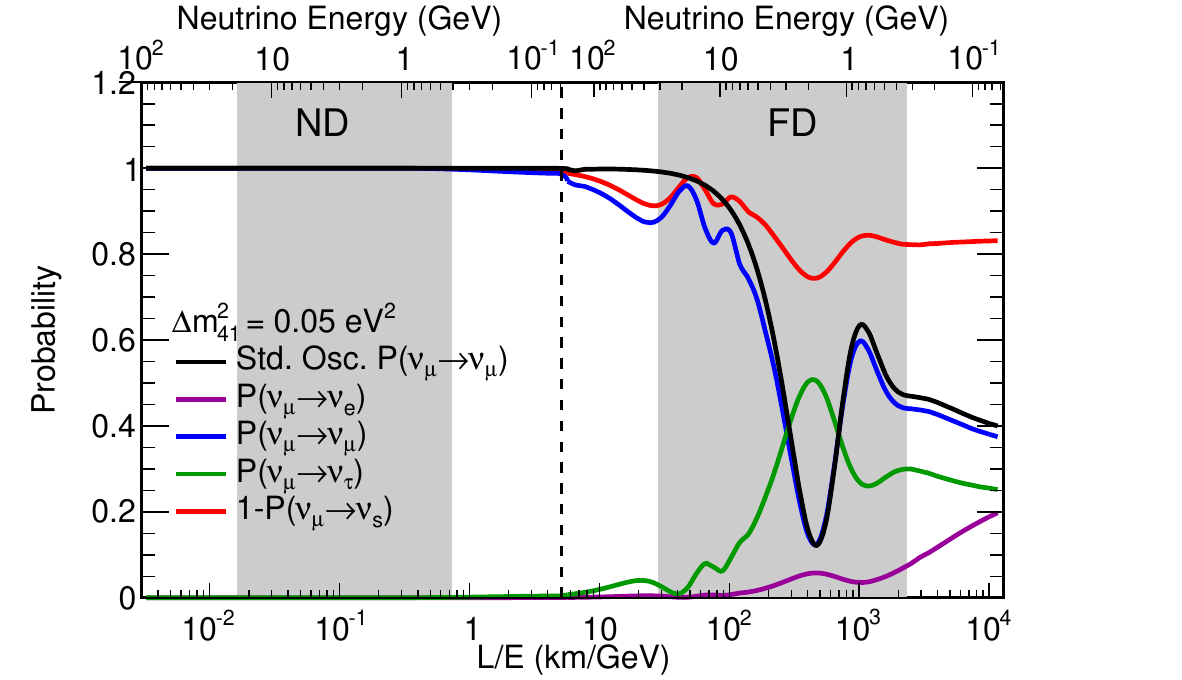}
        \includegraphics[width=0.48\textwidth]{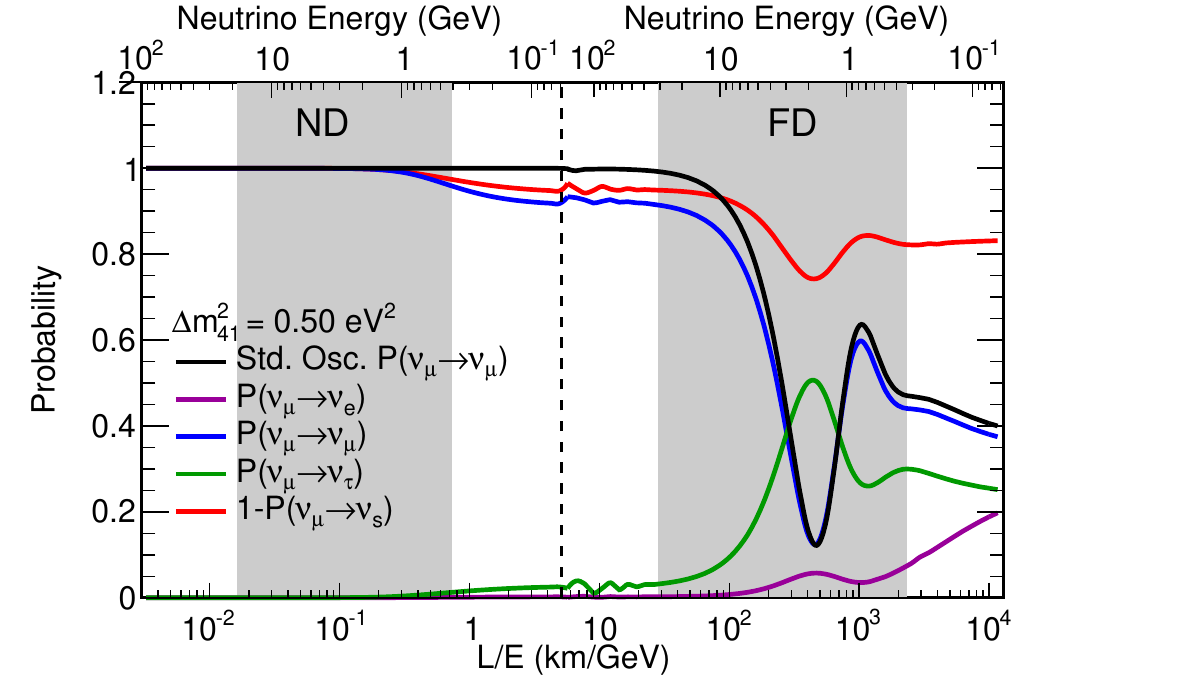}
        \includegraphics[width=0.48\textwidth]{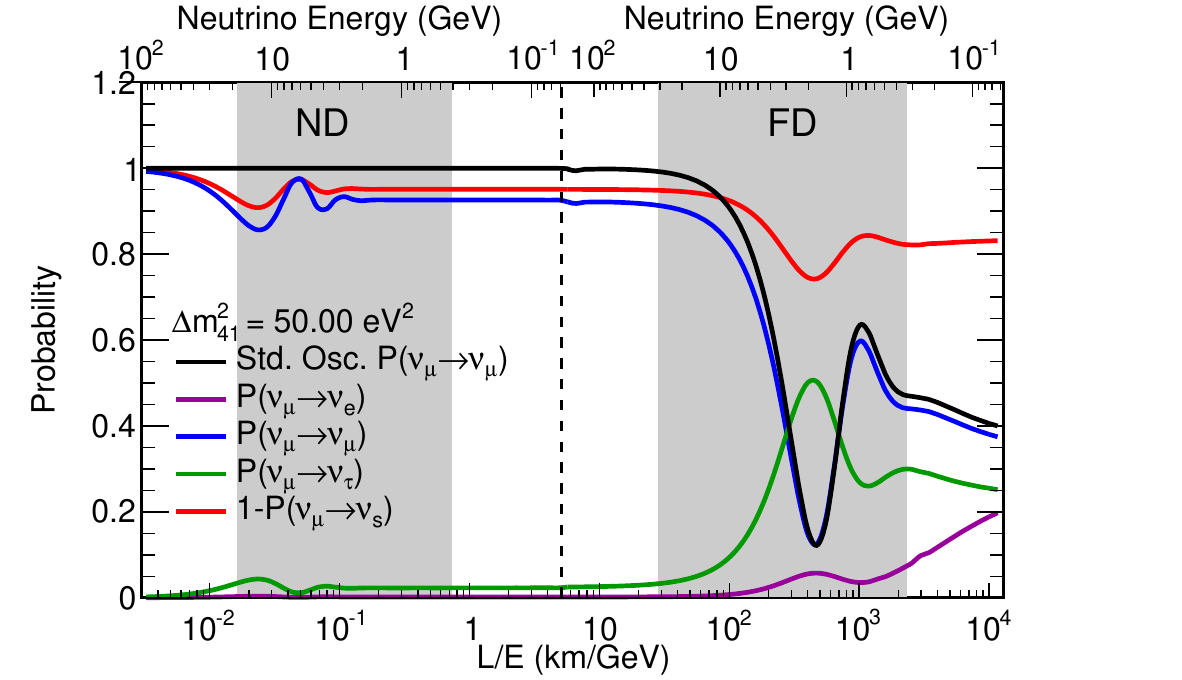}
\caption[Regions of $L/E$ probed by DUNE for  
3-flavor and 3+1-flavor $\nu$ oscillations]
{Regions of $L/E$ probed by the DUNE Near and Far detectors compared to 3-flavor and 3+1-flavor neutrino disappearance and appearance probabilities. The gray-shaded areas show the range of true neutrino energies probed by the ND and FD. The top axis shows true neutrino energy, increasing from right to left. The top plot shows the probabilities assuming mixing with one sterile neutrino with $\Delta m^2_{\rm{41}}=0.05$~eV$^2$, corresponding to the slow oscillations regime. The middle plot assumes mixing with one sterile neutrino with $\Delta m^2_{\rm{41}}=0.5$~eV$^2$, corresponding to the intermediate oscillations regime. The bottom plot includes mixing with one sterile neutrino with $\Delta m^2_{\rm{41}}=50$~eV$^2$, corresponding to the rapid oscillations regime. As an example, the slow sterile oscillations cause visible distortions in the three-flavor \numu~survival probability (blue curve) for neutrino energies. Figure from~\cite{DUNE:2020fgq}.}
\label{fig:dune_regimes}
\end{figure}

Figure~\ref{fig:dune_regimes} shows how the standard three-flavor oscillation probability is distorted at neutrino energies above the standard oscillation peak when oscillations into sterile neutrinos are included and the energy ranges DUNE ND and FD are sensitive to those distortions.  

The sterile neutrino effects have been implemented in GLoBES via the existing plug-in for sterile neutrinos and NSI~\cite{Joachim}. The DUNE ND plays a very important role in the sensitivity to sterile neutrinos both directly, for rapid oscillations with $\Delta m_{41}^2 > 1$~eV$^2$ where the sterile oscillation matches the ND baseline, and indirectly, at smaller values of $\Delta m_{41}^2$ where the ND is crucial to reduce the systematic uncertainties affecting the FD to increase its sensitivity. For these studies, the DUNE ND is assumed to be an identical scaled-down version of the FD, with identical efficiencies, backgrounds and energy reconstruction. The full set of systematic uncertainties employed in the sterile neutrino studies, as well as the methodology accounting for non-negligible beam-induced baseline spreads between production target and ND, are described in Ref.~\cite{DUNE:2020fgq}.

By default, GLoBES treats all systematic uncertainties included in the fit as normalization shifts. However, depending on the value of $\Delta m^2_{41}$, sterile mixing will induce shape distortions in the measured energy spectrum beyond simple normalization shifts. As a consequence, shape uncertainties are very relevant for sterile neutrino searches, particularly in regions of parameter space where the ND, with virtually infinite statistics, has a dominant contribution. The correct inclusion of systematic uncertainties affecting the shape of the energy spectrum in the two-detector fit GLoBES framework used for this analysis posed technical and computational challenges beyond the scope of the study.
Therefore, for each limit plot, we present two limits bracketing the expected DUNE sensitivity limit, namely: the black limit line, a best-case scenario, where only normalization shifts are considered in a ND+FD fit, where the ND statistics and shape have the strongest impact; and the grey limit line, corresponding to a worst-case scenario where only the FD is considered in the fit, together with a rate constraint from the ND.

\begin{figure}[!ht]
\centering
\includegraphics[width=0.45\textwidth]{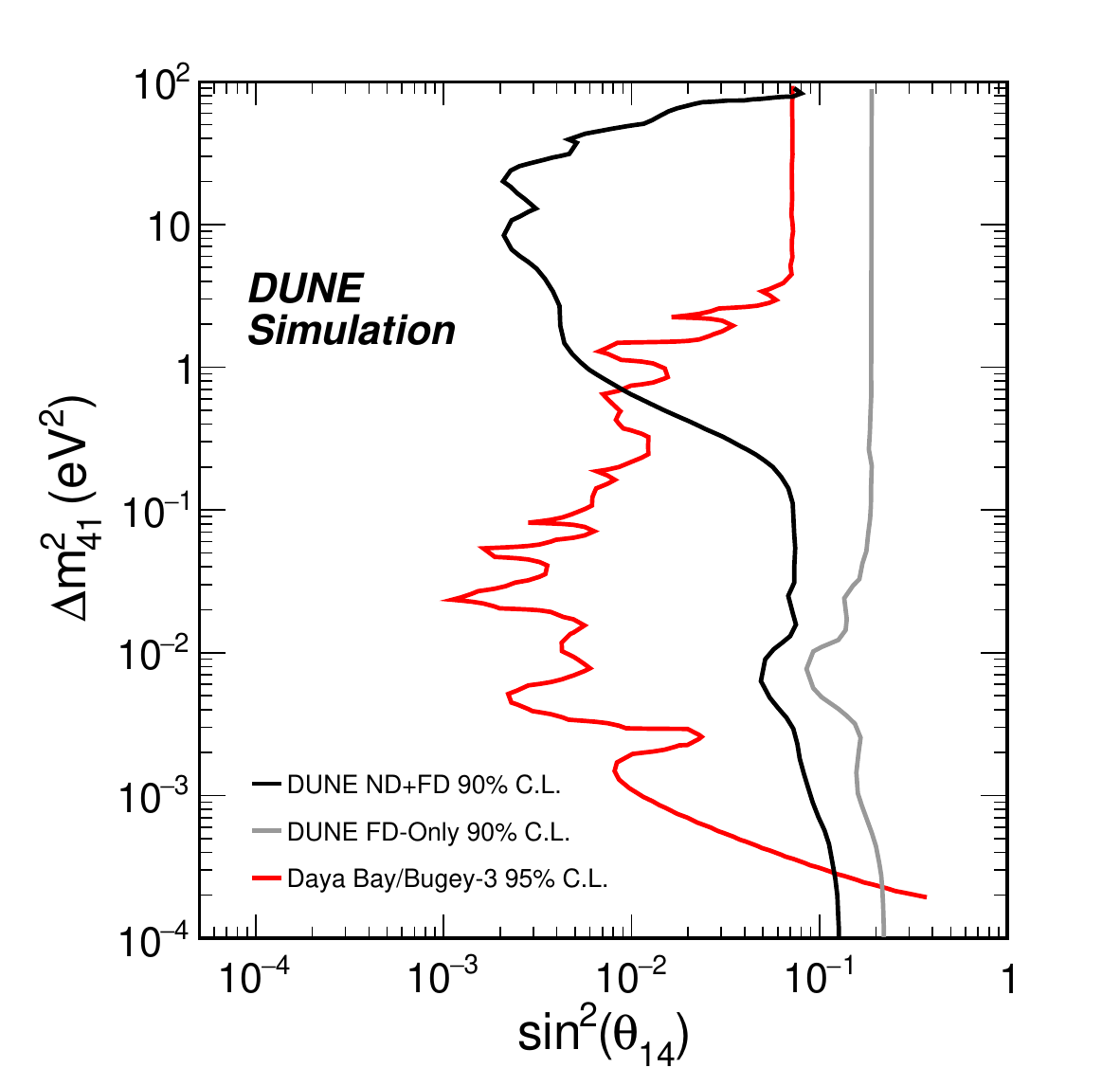}
\includegraphics[width=0.45\textwidth]{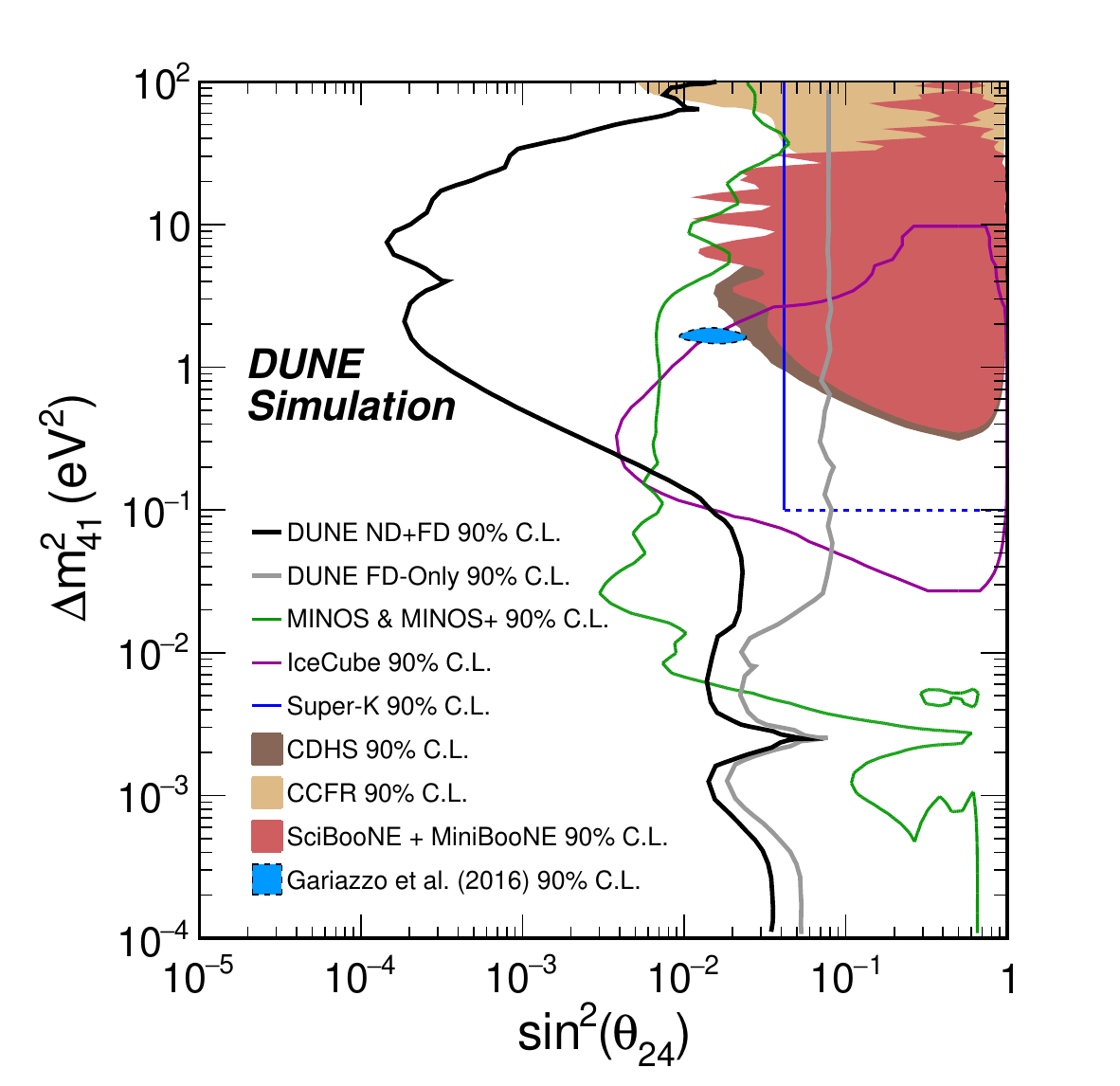}
\caption{The left plot shows the DUNE sensitivities to $\theta_{14}$ from the $\nu_e$ CC samples at the ND and FD, assuming $\theta_{14}$=0, along with a comparison with the combined reactor result from Daya Bay and Bugey-3. The right plot is adapted from Ref.~\cite{Todd:2018hin} and displays sensitivities to $\theta_{24}$ using the $\nu_\mu$ CC and NC samples at both detectors, along with a comparison with previous and existing experiments. In both cases, regions to the right of the contours are excluded.  Figure from~\cite{DUNE:2020fgq}.}
\label{fig:dune_th_14+th_24}
\end{figure}

For sensitivity to $\theta_{14}$, the dominant channels are those regarding $\nu_e$ disappearance.  
For simplicity, only the $\nu_e$ CC sample is analyzed and the NC and $\nu_{\mu}$ CC disappearance channels are not taken into account. This is expected to be improved by using more complex multi-channel fits in future studies, as highlighted and recommended by Ref.~\cite{Arguelles:2021meu}. 
The sensitivity at the 90\% C.L., taking into account the systematic uncertainties mentioned above, is shown in Fig.~\ref{fig:dune_th_14+th_24}, along with a comparison to current constraints.
For the $\theta_{24}$ mixing angle, the $\nu_{\mu}$ CC and NC disappearance samples are analyzed jointly.  
Results are shown in Fig.~\ref{fig:dune_th_14+th_24}, along with comparisons with present constraints.

\begin{figure}[!ht]
\centering
\includegraphics[width=0.45\textwidth]{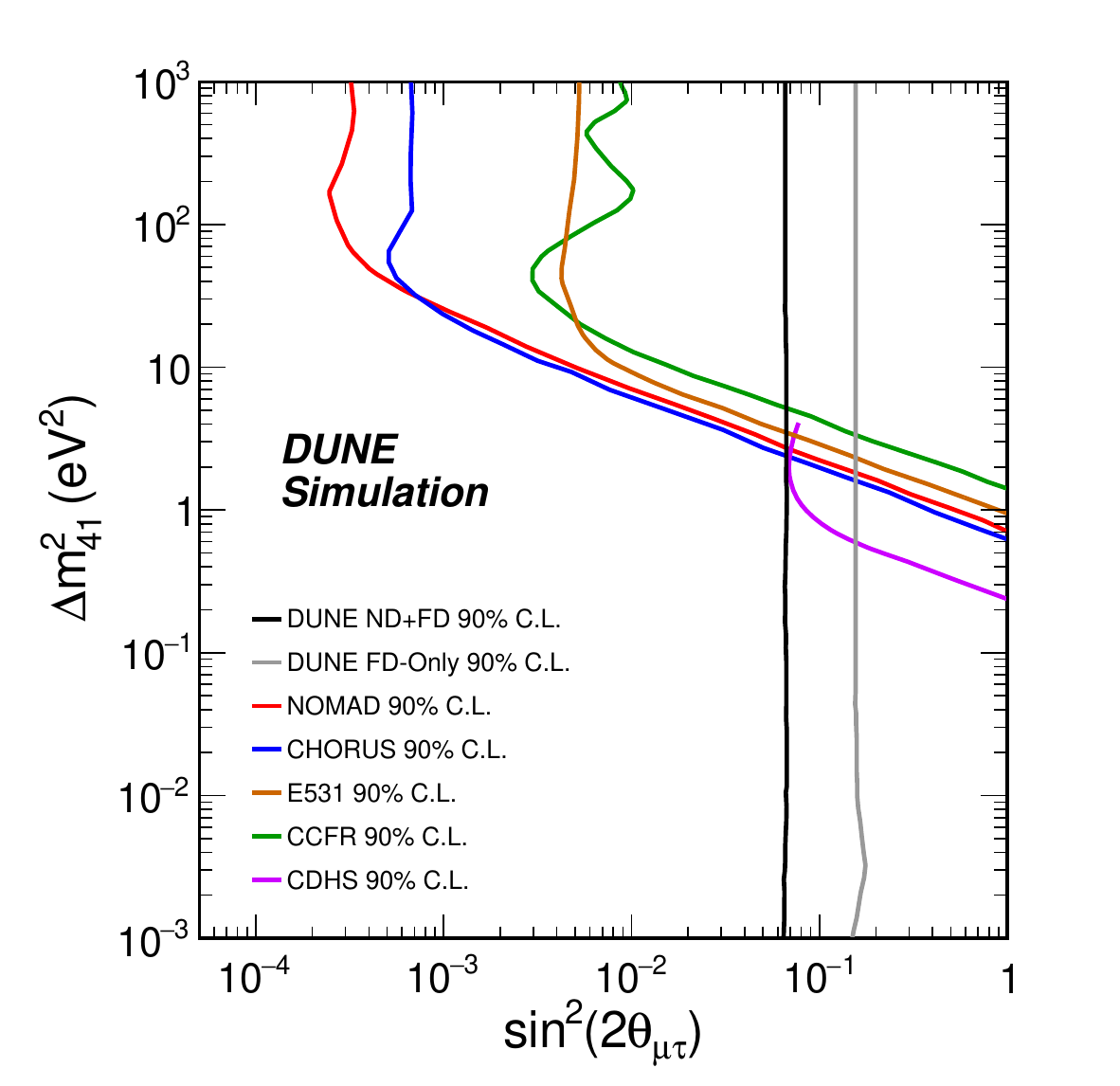}
\includegraphics[width=0.45\textwidth]{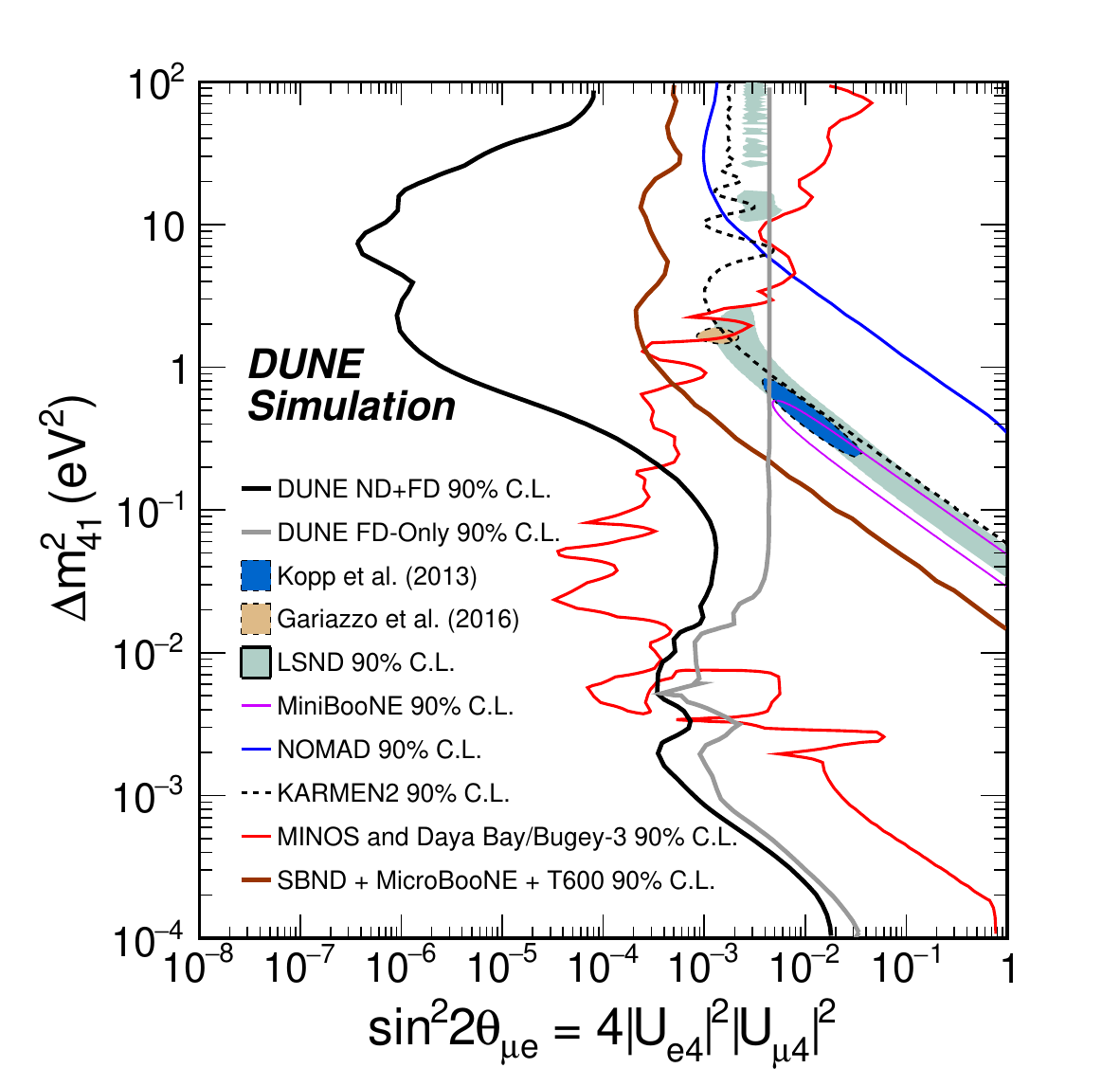}
\caption{Left: Comparison of the DUNE sensitivity to $\theta_{34}$ using the NC samples at the ND and FD with previous and existing experiments. Regions to the right of the contour are excluded.  Right: DUNE sensitivities to $\theta_{\mu e}$ from the appearance and disappearance samples at the ND and FD are shown on the top plot, along with a comparison with previous existing experiments and the sensitivity from the future SBN program.  Regions to the right of the DUNE contours are excluded.  Figure from~\cite{DUNE:2020fgq}.}
\label{fig:dune_th_34} 
\end{figure}

In the case of the $\theta_{34}$ mixing angle, disappearance in the NC sample, the only contributor to this sensitivity, is probed.  The results are shown in Fig.~\ref{fig:dune_th_34}. Further, a comparison with previous experiments sensitive to \numu, \nutau~mixing with large mass-squared splitting is possible by considering an effective mixing angle $\theta_{\mu\tau}$, such that $\sin^2{2\theta_{\mu\tau}}\equiv 4|U_{\tau4}|^2|U_{\mu 4}|^2=\cos^4\theta_{14}\sin^22\theta_{24}\sin^2\theta_{34}$, and assuming conservatively that $\cos^4\theta_{14}=1$, and $\sin^22\theta_{24}=1$. This comparison with previous experiments is also shown in Fig.~\ref{fig:dune_th_34}.
The sensitivity to $\theta_{34}$ is largely independent of 
$\Delta m^2_{41}$, since the term with $\sin^2\theta_{34}$ in 
the expression describing ${P(\nu_{\mu}\rightarrow\nu_s)}$, depends solely on the $\Delta m^2_{31}$ mass splitting.

Finally, sensitivity to the $\theta_{\mu e}$ effective mixing angle, defined as $\sin^2{2\theta_{\mu e}}\equiv 4|U_{e4}|^2|U_{\mu 4}|^2=\sin^22\theta_{14}\sin^2\theta_{24}$, is shown in Fig.~\ref{fig:dune_th_34}, which also displays a comparison with the allowed regions from LSND and MiniBooNE, as well as with present constraints and projected constraints from Fermilab's SBN program.

DUNE will also have the ability to conduct short-baseline sterile probes, for instance, by searching for anomalous sterile-driven $\nu_\tau$ appearance in the Near detector. The $\tau$ lepton is not directly observable in the DUNE detectors due to its short $2.9\times 10^{-13}$\,s lifetime, and it is only produced for interactions where the incoming $\nu_\tau$ has an energy of $\sim3.5$~GeV due to the relatively large $\tau$ mass of 1776.82~MeV. However, the final states of $\tau$ decays, $\sim65\%$ into hadrons, $\sim18\%$ into $\nu_\tau+e^-+\bar{\nu}_e$, and $\sim17\%$ into $\nu_\tau+\mu^-+\bar{\nu}_\mu$, are readily identifiable in the DUNE ND, given the excellent spatial and energy resolution of the ND instruments, namely, ND-LAr, ND-GAr, and SAND, which are complementary in providing sensitivity to different decays channels. While within a three-flavor scenario, the DUNE ND baseline is far too short for $\nu_\mu\rightarrow\nu_\tau$ oscillations to occur, $\nu_\tau$ originating in sterile-neutrino driven fast oscillations could be detected. In particular, probing the $\tau\rightarrow\mu$ detection channel with high-energy muons in the final state, which is challenging due to muon containment and backgrounds, becomes very accessible through the use of the ND-GAr magnetic field and the SAND detector further downstream, with studies indicating that ND-GAr's reconstructible muon momentum via curvature extends beyond 15~GeV/$c$. This sensitivity will be strongly enhanced when operating LBNF in the high-energy tune, aimed at enriching the available sample of $\nu_{\tau}$ at the Far detector, while extending sensitivity to anomalous $\nu_{\tau}$ appearance at the Near detector~\cite{DeGouvea:2019kea}. Preliminary studies using LBNF's nominal flux, and including ND-LAr and ND-GAr, estimate that DUNE's sensitivities to anomalous $\nu_\tau$ appearance may extend beyond those of previous searches carried out by NOMAD and CHORUS.

\subsubsubsection{Hyper-Kamiokande}
\label{hyperk}
Hyper-Kamiokande (HK) is a large-scale water Cherenkov detector with a fiducial volume of about 188\,kton which is approximately 8.4 times larger than Super-Kamiokande.
HK is currently under construction in Japan and operations are scheduled to begin in 2027 together with the upgraded J-PARC neutrino beam.
The physics capabilities of HK cover a broad range of topics including a search for sterile neutrino mixing~\cite{Abe:2018uyc}. 
There are various major approaches currently being considered.

While we focus in the following on the sensitivity to sterile neutrino searches, HK can also investigate other exotic scenarios like the breaking of Lorenz and CPT invariance, as demonstrated in T2K in Ref.~\cite{Abe:2017eot}, and non-standard neutrino interactions, as studied with Super-K's atmospheric neutrino observations~\cite{Mitsuka:2011ty}.


Mixing of light sterile neutrino will be investigated with the Hyper-Kamiokande data at a baseline of 295\,km baseline.
T2K reported a limit on $\sin^2 \theta_{24}$ for $10^{-4}\,{\rm eV^2} < \Delta m^{2}_{41} < 3 \times 10^{-4}\,{\rm eV^2}$ using both CC and NC samples at Super-Kamiokande~\cite{T2K:2019efw}.
More stringent limit will be set by Hyper-Kamiokade with more than 20 times higher statistics by the combination of a larger fiducial volume and an upgraded J-PARC neutrino beam.

Hyper-Kamiokande's near detectors will measure the neutrino beam flux and cross-section at different baselines.
Each detector has the capability to test the existence of sterile mixing at certain values of $L/E$ but it should be noted that the sensitivity will be further enhanced by combined measurements among the detectors where the ND280 works as a near detector and give constraint to the IWCD measurement.

The IWCD (Intermediate Water Cherenkov Detector) instrument has sensitivity to sterile neutrino mixing in the $\nu_{\mu} \rightarrow \nu_{e}$ channel with a baseline of $\sim$1\,km and energy of 0.5~-~1~GeV, which matches the $L/E$ at LSND and MinoBooNE.
As a remarkable advantage, IWCD can measure the neutrino flux at different off-axis angles by moving the detector along its vertical pit.
As the energy spectrum changes with the off-axis angle, IWCD can rule out some potential interpretations by the combination of the measurement, such as feed-down from high energy due to nuclear effects or unexpected background.
The design of the IWCD detector is still under investigation, but it has potential to test the allowed region given by LSND, as indicated by the studies from the NuPRISM collaboration~\cite{nuPRISM:2014mzw}.

The off-axis ND280 has sensitivity to few eV$^{2}$ sterile neutrinos. 
A first search was published in Ref.~\cite{T2K:2014xvp}.
It is anticipated that the upgrades currently being done for T2K, consisting in one fully active target (Super-FGD), two High-Angle TPCs, and a Time Of Flight system, and possible further upgrades under study for Hyper-K, will boost the sensitivity of ND280. 
Searches for sterile neutrinos with the upgraded ND280 will have several advantages with respect to Ref.~\cite{T2K:2014xvp}, including a larger target mass, a lower threshold to reconstruct leptons, better performances in distinguishing electrons from gammas, and the larger exposure that will be collected in HK.
ND280 has also sensitivity to search for relatively heavy sterile neutrinos produced by the decay of Kaons produced by the beamline, as demonstrated in Ref.~\cite{T2K:2019jwa}.

\subsubsubsection{ESSnuSB}
\label{essnu_sb}
The ESSnuSB (European Spallation Source Neutrino Super Beam) experiment~\cite{Baussan:2013zcy,Wildner:2015yaa} is a proposed long-baseline neutrino oscillation experiment to use a neutrino superbeam produced using 2.0~GeV protons from the ESS Linac in Lund, Sweden, resulting in a 5 MW beam peaked at $E_{nu} = 0.4$~GeV. By sampling this beam at a distance of 540 km from Lund, using a large underground Water Cherenkov detector with a 500 kton fiducial mass, ESSnuSB will make measurements of the three-flavor oscillation second maximum, which would enable discovery of leptonic CP violation for 56\% of $\delta_{CP}$ values for 10 years of data taking, and 65\% of $\delta_{CP}$ values if an upgrade of the beam power to 10 MW and of the proton energy to 2.5~GeV is realized.
With this experimental setup, ESSnuSB is also able to place bounds on the sterile mixing parameters, but the CP sensitivity of ESSnuSB may also be affected by the existence of light sterile neutrinos. The studies below assume two possible detector configurations for ESSnuSB and compare them: 1) combined far (FD) and near (ND) detectors with correlated systematics; and 2) an FD only with an overall systematic uncertainty. 


The analysis uses the GLoBES~\cite{Huber:2004ka, Huber:2007ji} software to simulate ESSnuSB. An ND and an FD are explicitly simulated to reduce systematic uncertainties~\cite{Baussan:2013zcy}. The FD is a 1~Mt MEMPHYS-like water-Cherenkov detector~\cite{Agostino:2012fd} located at a distance of 540~km from the source, while the ND is assumed to have the same efficiency and background rejection capabilities as the FD~\cite{Agostino:2012fd} with a fiducial volume of 0.1~kt and placed at a distance of 0.5~km from the source. A beam power of 5~MW with 2.5~GeV protons capable of producing $2.7\times 10^{23}$ POT/year is assumed. The results are presented for a total data exposure 10~years, with 5 (5) years running in neutrino (antineutrino) mode. Throughout the simulations of the combined near and far detectors, the same treatment of systematics as in Ref.~\cite{Coloma:2012ji} was employed. In all cases, the best-fit values of the parameters from Ref.~\cite{Esteban:2018azc} assuming Normal Ordering were used.

\begin{figure}[!htbp]
\centering
\includegraphics[clip, trim=1.27cm 0.25cm 1.4cm 0.25cm, width=0.34\textwidth]{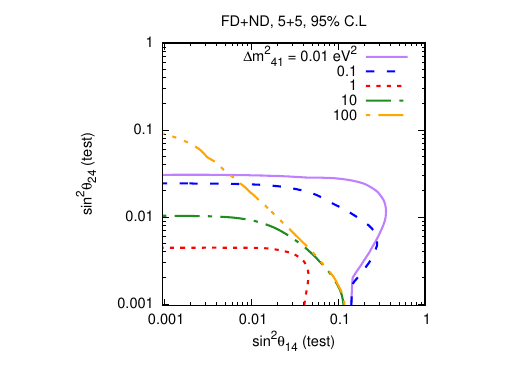} 
\includegraphics[clip, trim=1.75cm 0.25cm 1.4cm 0.25cm, width=0.32\textwidth]{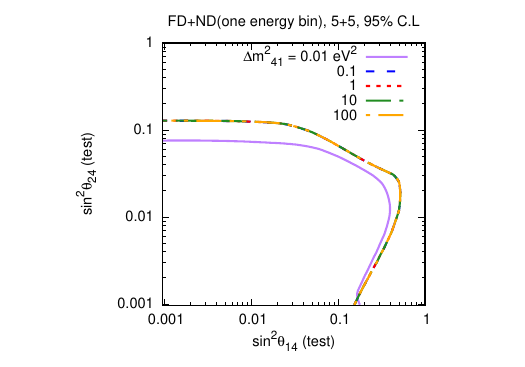} 
\includegraphics[clip, trim=1.75cm 0.25cm 1.4cm 0.25cm, width=0.32\textwidth]{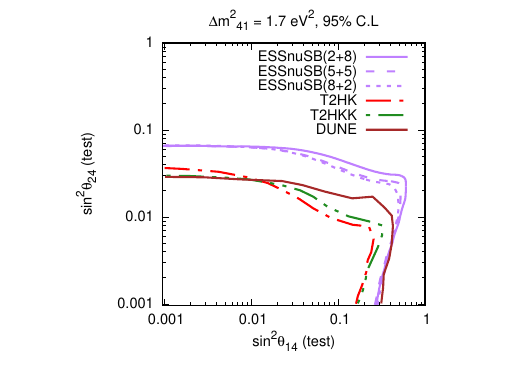} 
\caption{Bounds (95~\% CL) on sterile mixing parameters in the $\sin^2\theta_{14}$ -- $\sin^2\theta_{24}$ plane for ESSnuSB (left and middle panels). The panel on the right shows a comparison of ESSnuSB with other long-baseline neutrino oscillation experiments. Figure from~\cite{Ghosh:2019sfi}.}
\label{fig:Sens_t14vst24}
\end{figure}

The upper panels of Fig.~\ref{fig:Sens_t14vst24} display the sensitivity of ESSnuSB to the sterile mixing angles $\theta_{14}$ and $\theta_{24}$ for particular values of $\Delta m_{41}^2$ at 95~\%~CL. 
The left panel shows the results for combined FD+ND, while the middle panel presents the results for the FD using the ND as a counting experiment. From the left panel, it is apparent that for $\Delta m^2_{41} = 0.01$~eV$^2$ the bounds are weak, as for this value of $\Delta m^2_{41}$ the oscillations have not yet developed for the ND, and therefore, the existing sensitivity comes from the FD, where the oscillations are averaged out. Increasing the value of $\Delta m^2_{41}$, the oscillations become more developed in the ND resulting in the stronger bound for $\Delta m^2_{41} = 1$~eV$^2$. Increasing the value of $\Delta m^2_{41}$ further, the oscillations tend to become averaged out and again the sensitivity decreases. From the middle panel, where there is no spectral information on the ND, the sensitivity for  $\Delta m^2_{41} \sim 1$~eV$^2$ is lost and similar sensitivities  are obtained for all values of $\Delta m^2_{41}$.
The right panel of Fig.~\ref{fig:Sens_t14vst24} compares the sensitivity of ESSnuSB to the sterile mixing parameters with other future long-baseline neutrino oscillation experiments, specifically T2HK~\cite{Abe:2016ero}, T2HKK~\cite{Abe:2016ero}, and DUNE~\cite{Acciarri:2015uup}. The comparison shows that the sensitivity of ESSnuSB is slightly worse than the other experiments in most of the region of the parameter space. However, as can be seen from the left panel in Fig.~\ref{fig:Sens_t14vst24}, it is expected that the ESSnuSB sensitivity would be considerably improved in the presence of an ND. 

\begin{figure}[!hb]
\centering
\includegraphics[width=0.49\textwidth]{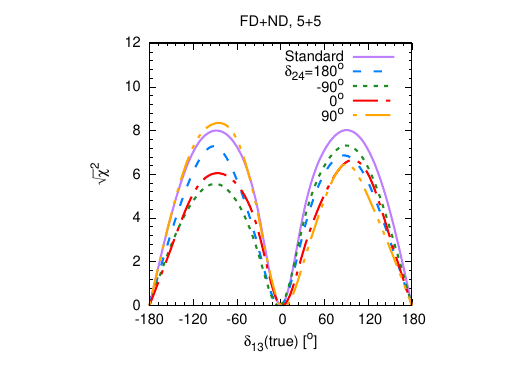} 
\includegraphics[width=0.49\textwidth]{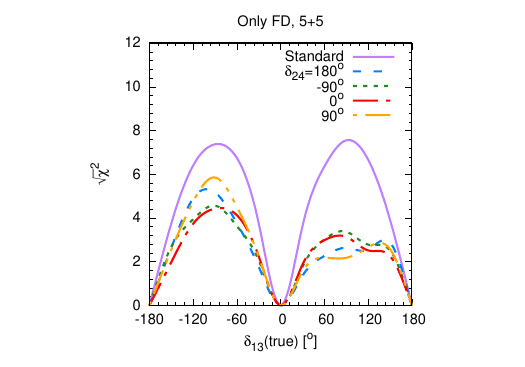} 
\caption{Leptonic CP violation sensitivity of ESSnuSB for four different values of the sterile mixing phase $\delta_{24}$ as a function of true $\delta_{13}$. Figure from~\cite{Ghosh:2019sfi}.}
\label{fig:CP_violation}
\end{figure}

Figure~\ref{fig:CP_violation} presents the CP violation sensitivity of ESSnuSB due to $\delta_{13}$ for four different values of the sterile mixing phase $\delta_{24}$, illustrating how the presence of sterile neutrino oscillations would modify the interpretation of leptonic CP violation measurements. 
The sterile mixing parameters assumed are $\sin^2\theta_{14} = \sin^2\theta_{24} = 0.025$~\cite{Diaz:2019fwt}, $\theta_{34} = \delta_{34} = 0^\circ$, and $\Delta m^2_{41} = 1$~eV$^2$. The left panel shows the sensitivity for FD+ND, and in the right one, the sensitivities considering only the FD with 8~\% (10~\%) overall systematics in signal (background). Again,when including the ND, the sterile mixing parameters are better constrained and the sensitivity to CP violation improves.

The ESS Linac proton beam is expected to turn on in 2025. Construction of the neutrino beam facility is under study and may begin with a low-energy nuSTORM-like ND complex starting in 2024 with operations starting in 2027. 

\subsection{Reactor Neutrino Experiments}
\label{sec:future_rx}

In the period following Snowmass 2021, emphasis in the reactor sector will be placed on 50~km-scale baseline experiments probing the SM neutrino mass hierarchy and the solar mixing angle, such as JUNO, and on and very short-baseline experiments aiming to probe sterile oscillations and perform high-statistics measurements of reactor \anue fluxes and spectra.  
Short-baseline reactor experiments will be particularly important in furthering understanding of the source of enduring reactor-sector anomalies, in addressing existing claims of non-standard oscillation observation in recent reactor experiments, and in providing orthogonal experimental datasets valuable in global BSM physics fits.  

Data from multiple channels, energies, and sources will be crucial for disentangling possible BSM effects manifested in the anomalies, since these effects may manifest differently in different experimental regimes.  
In the stable of global measurements, short-baseline reactor measurements are unique in their capability to very purely probe sterile oscillation effects.  
This is due to the lower energies involved in interactions and decays in the reactor, which prohibits production and decay of heavier hidden-sector particles, and their very short baselines, which minimize the impact of non-standard interactions.  
Even in the case of purely oscillation-driven BSM explanations, reactor experiments offer unique benefits due to their unambiguously pure flavor content: this is in contrast to interpretations of short-baseline decay-in-flight and decay-at-rest experiments, which will be complicated by competition between appearance and disappearance effects in individual flavor channels~\cite{Arguelles:2021meu}.  
These points serve to emphasize the substantial value added by acquiring datasets from \emph{all} of experiment types: short-baseline reactor datasets will find greater utility and application in the coming decade if they are accompanied by short-baseline accelerator datasets, and vice versa.  

\begin{table*}[thbp!]
\begin{adjustbox}{width=\columnwidth,center}
\begin{tabular}{l||c|c|c|c|c|c}
\hline
Experiment & Baseline (m) & Reactor & Reactor & Detector & Target & Sterile $\nu$ \\
 &  & Type & Power~(MW\emph{th}) & Size & & Search Strategy \\\hline \hline 
DANSS~\cite{Svirida:2020zpk}  & 11--13 & LEU & 3000 & 1 m$^3$ & Segmented PS with Gd coating & Multi-Site \\ \hline
JUNO-TAO~\cite{JUNO:2020ijm}  & 30 & HEU & 4600 & 2.8 ton & Single GdLS & Single-Site \\ \hline 
NEOS-II   & 24 & LEU & 2800 &  1 m$^3$ & Single-volume GdLS + PSD & Single-Site \\ \hline
Neutrino-4 Upgrade & 6--12 & HEU & 90 &  2 m$^3$ & Segmented GdLS & Multi-Site/Zone \\ \hline
PROSPECT-II~\cite{PROSPECT:2021jey}  & 7--9 & HEU & 85 & 4 ton  & Segmented $^6$LiLS + PSD & Multi-Zone \\ \hline \hline
\end{tabular}
\end{adjustbox}
\caption{List of the future short-baseline reactor neutrino experiments that will be able to search for sterile neutrino oscillations.} 
\label{tab:rx_future_osc}
\end{table*}

A summary of new short-baseline reactor oscillation measurements currently underway or planned for the next five years is provided in Table~\ref{tab:rx_future_osc}.  
For future experiments, some specific reactor experiment parameters are particularly valuable in achieving the future short-baseline reactor physics goals mentioned above.  
While reactor experiments in the past decade have provided excellent limits in the oscillation phase space region below a few eV$^2$, limits above this mass splitting are substantially weaker; to address this, very short ($<$10~m) baseline experiments using compact cores and segmented detectors are particularly valuable. 
To provide more robust probes of isotopic IBD yields and spectra and thus better understanding of the fidelity of existing flux predictions, high-statistics dataset are needed from reactors of widely varying fuel content.  
In this context, experiments running at HEU reactors and over a full cycle at a single LEU reactor core would be particularly valuable.  
As demonstrated in Table~\ref{tab:rx_future_osc}, most of these expected future experiments fulfill one or more of these key experimental requirements.  

\begin{figure}[ht]
    \centering
    \includegraphics[width=0.49 \textwidth,trim={0 0 0 1cm },clip]{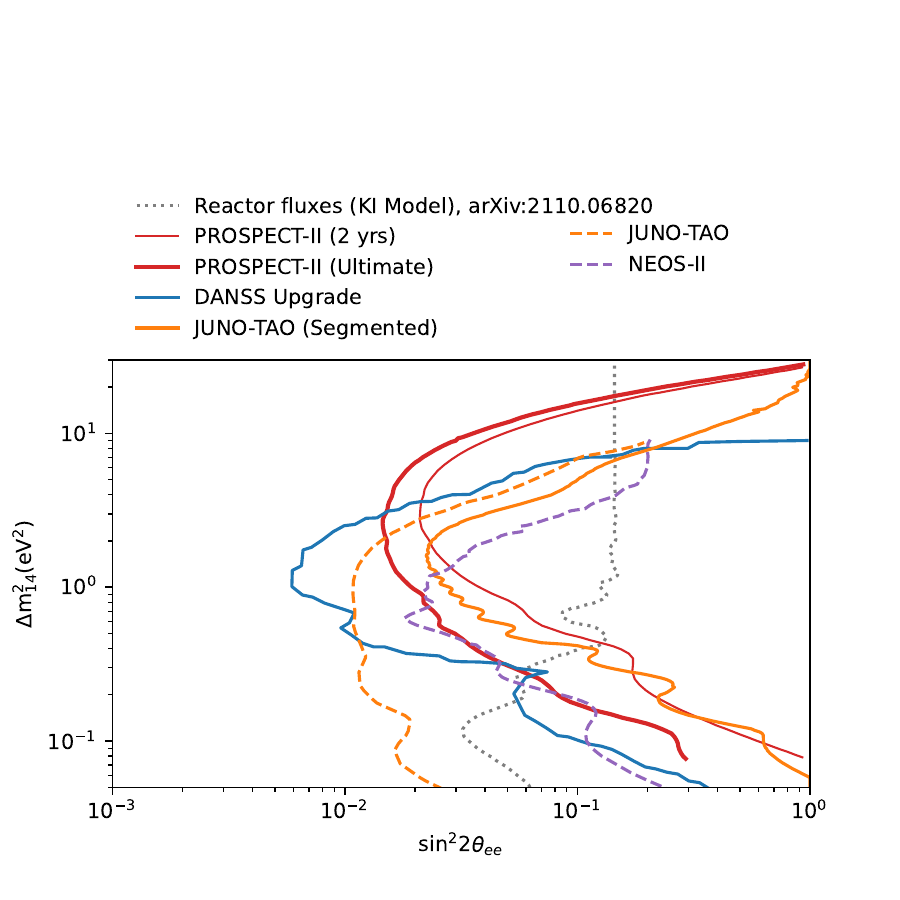}
    \includegraphics[width=0.49 \textwidth,trim={0 0 0 1cm },clip]{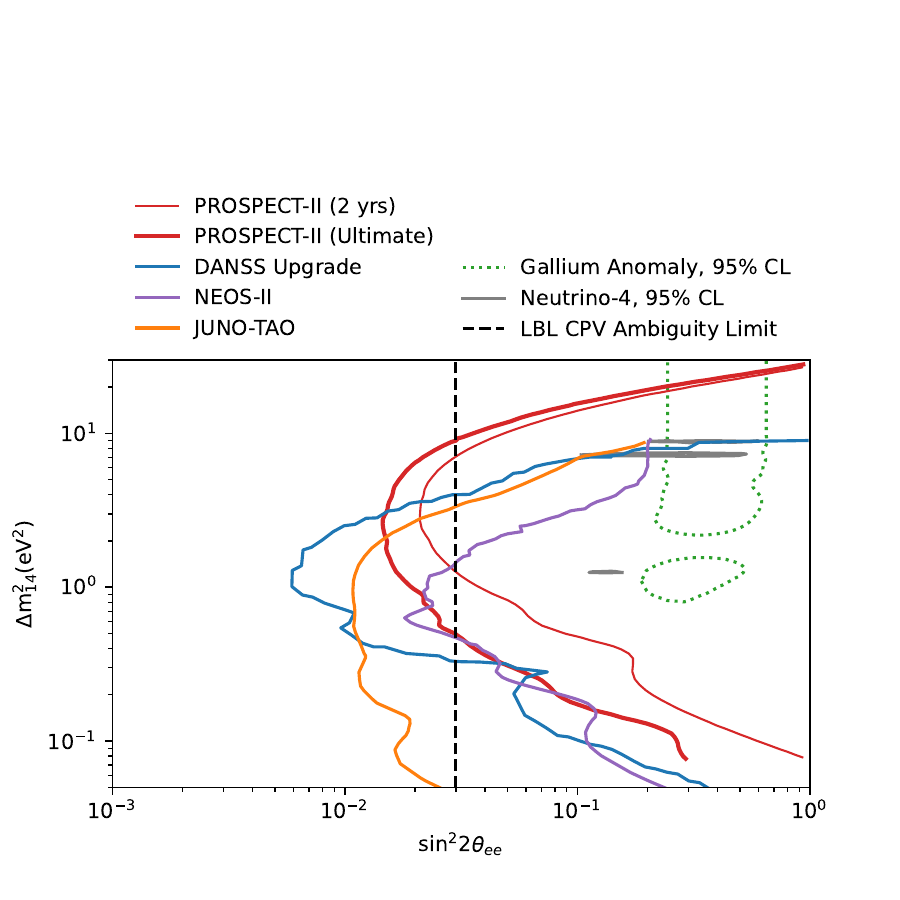}
    \caption{Left: Claimed sensitivities~(90\% CL) to sterile neutrinos of the upcoming reactor neutrino experiments listed in Tab.~\ref{tab:rx_future_osc}. Also shown is the unexplored parameter space obtained by comparing reactor rates + evolution datasets to the reevaluated HM model based on the KI suggested normalization correction as discussed in Sec.~\ref{sec:th_land_rx_models}. NEOS and JUNO-TAO are shown in dashed lines to indicate that the experiments make use of the absolute spectral shapes. Also shown~(solid) is a version of JUNO-TAO's curve that is produced by making use of the experiment's position reconstruction to generate virtual segments that provides the ability to perform a completely absolute shape independent analysis. The plot was generated using the same approach as described in Ref.~\cite{Berryman:2021xsi}. Right: Sensitivities of the future reactor neutrino experiments in comparison with the Gallium Anomaly and Neutrino-4 suggested parameter space. Also included is the line at $\sin^2 2 \theta=0.3$~(dashed black) corresponding to the limit which has to be excluded to avoid ambiguities in the future LBL CP-violation measurements.}
    \label{fig:rx_future}
\end{figure}

The sterile neutrino oscillation probing powers of these experiments, expressed in terms of sensitivity to sin$^2$2$\theta_{14}$ in a 3+1 oscillation framework, are shown on the right plot in Figure~\ref{fig:rx_future}, with the left plot also indicating some phase space parameter ranges of interest.  
In addition to showing individual experiments, the right hand plot indicates the style of analysis assumed in each curve: one using only relative measured spectral shapes, one using knowledge of the absolute measured spectrum, and one using only rate information.  
Rate-based analyses provided by reactor flux measurements clearly lack the  sensitivity of shape-based ones; combined with limitations in knowledge of absolute reactor fluxes (Section~\ref{sec:expt_landscape_reactors}), rate-based analyses appear, at present, unlikely to yield much fruit in the pursuit of sterile neutrinos.  
Of the shape-based measurements, DANSS and TAO are likely to dominate sensitivities in the lower-$\Delta $m$^2$ regime, while PROSPECT-II is likely to dominate future sensitivity in the less-explored region above $\sim$3~eV$^2$.  
DANSS and TAO may achieve some power in probing the oscillation phase space region suggested by Neutrino-4, while PROSPECT-II should address this region at high confidence level while also covering phase space suggested by the Gallium anomaly that is currently unaddressed by spectrum-based reactor analyses.  
Finally, the combination of DANSS, TAO, and PROSPECT-II datasets should generate percent-level sensitivity to active-sterile couplings in the electron disappearance channel for all $\Delta m^2$ space below roughly 10~eV$^2$, enabling greater clarity in interpretation of CP-violation results from DUNE~\cite{Gandhi:2015xza}.  



\subsubsection{DANSS upgrade} 
\begin{figure}
\centering
\includegraphics[width=0.5\textwidth]{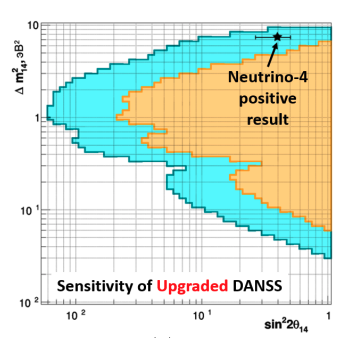}
\caption{Expected sensitivity of upgraded DANSS detector to sterile neutrino oscillations after 1.5 years of running in comparison with the sensitivity of the DANSS experiment.  Figure from~\cite{Svirida:2020zpk}.}
\label{fig:Dans_upgrade}
\end{figure}

One of the limitations of DANSS experiment was that the oscillation spectrum was smoothed out by the finite energy resolution. Resolution is limited mainly by the light collection system which makes further progress in DANSS physics program very challenging. Because of which, it has become necessary to upgrade the detector design in order to improve energy resolution.
The upgrade plan is to replace current strips with new strips with larger cross section~\cite{Svirida:2020zpk}. In the current design, the strips are coated with titanium oxide for light reflection and gadolinium oxide for neutron capture which produces relatively thick dead layer with titanium and gadolinium. The non-uniformities in the light collection system is the another limiting factor for resolution. The light collection system consists of three wavelength shifting fibers which are readout by SiPMs and PMTs are on the same side of the strip which introduces non-uniformities. To tackle this issue, the upgrade plan is to have SiPM only readout from both sides of the strips. Removing the PMTs from light collection system will allow for an in increase in the sensitive volume of $~$70\%. The sensitivity to sterile neutrinos with the upgraded detector after 1.5 years of running is shown in Fig.~\ref{fig:Dans_upgrade}.

\subsubsection{JUNO-TAO} The Taishan Antineutrino Observatory (TAO) will be a satellite detector of the JUNO experiment~\cite{JUNO:2020ijm}. Located at a baseline of approximately 30~m from one of the Taishan 4.6~GW$_\mathrm{th}$ cores in a basement 9.6~m below ground level, TAO's neutrino target will consist of 2.8 tons of gadolinium-doped liquid scintillator contained in a 1.8~m diameter acrylic sphere. This sphere will be surrounded by a copper spherical shell supporting an array of Silicon Photomultipliers~(SiPMs) with 94\% surface coverage, allowing to reach a light level of 4500 photoelectrons per MeV and an energy resolution of $\lesssim$2\% at 1~MeV. This volume will be placed inside a cylindrical stainless steel tank with 2.1~m of outer diameter and 2.2~m of height filled with a liquid scintillator buffer. The tank will be cooled to -50$^{\circ}$C to mitigate the dark noise of the SiPMs, and will be surrounded by an active 1.2~m thick water Cherenkov veto tank, as well as high-density polyethylene in the top and lead in the bottom. Approximately 4,000 IBD interactions are expected per day in a central 1 ton fiducial volume with a detection efficiency of about 50\%. TAO is scheduled to begin operations around the same time as the rest of the JUNO experiment in 2023.  

\begin{figure}[ht]
    \centering
    \includegraphics[width=0.55\textwidth]{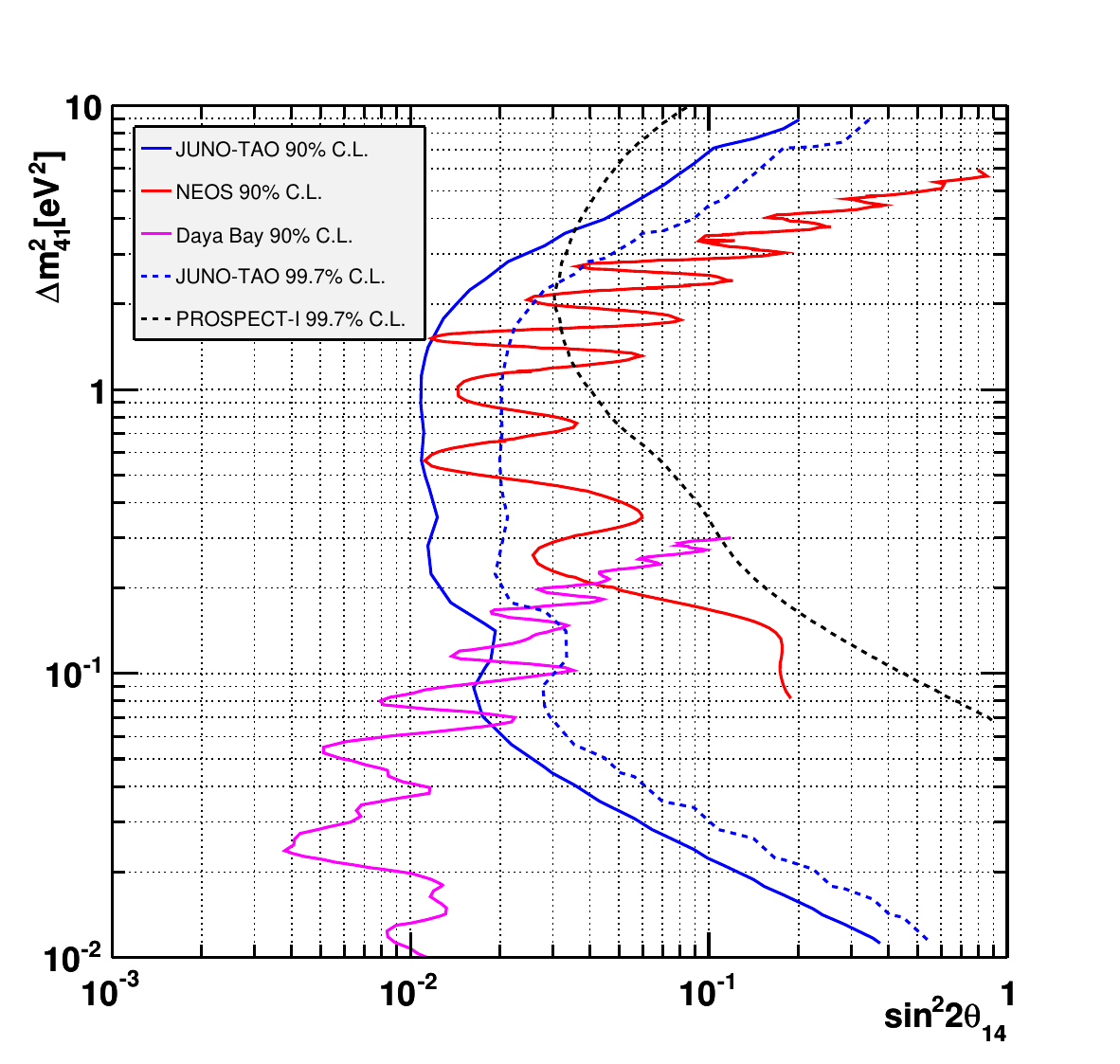}
    \caption{\label{fig:dybtao} 
    90.0\% and 99.7\% C.L. exclusion contours expected at JUNO-TAO with 3 years of data taking. The exclusion contours from the Daya Bay~\cite{MINOS:2020iqj}, PROSPECT phase-I~\cite{PROSPECT:2015iqr}, and NEOS~\cite{Ko:2016owz} experiments are also shown for comparison purposes. The parameter space to the right of all curves is excluded. Figure from~\cite{JUNO:2020ijm}.}
\end{figure}

TAO's primary purpose is to make a precise measurement of the reactor antineutrino spectrum that will serve as a benchmark for JUNO, other experiments, and nuclear databases. However, TAO will also be in an excellent position to search for a distortion in the shape of the reactor antineutrino spectrum caused by sterile neutrino mixing. The exclusion contours expected in JUNO-TAO at 90\% and 99.7\% C.L. in the absence of a sterile neutrino signal are shown on the right panel of Fig.~\ref{fig:dybtao}~\cite{JUNO:2020ijm}. The analysis behind these contours takes into account the physical dimensions of both the detector and the reactor core, and assumes 3 years of exposure with 80\% reactor time on. A conservative bin-to-bin uncorrelated uncertainty of 5\% (with 50~keV bin width) on the predicted reactor spectrum is assumed. The exclusion contours from the Daya Bay~\cite{MINOS:2020iqj}, PROSPECT phase-I~\cite{PROSPECT:2015iqr}, and NEOS~\cite{Ko:2016owz} experiments are also shown for comparison purposes. TAO's baseline, abundant statistics, and exquisite energy resolution, afford it a leading sensitivity to $\sin^2 2\theta_{14}$ in the $10^{-1}$~eV$^2$ $\lesssim |\Delta m^2_{41}| \lesssim 3$~eV$^2$ region.

\subsubsection{NEOS II} NEOS-II has taken about 400 live days of reactor-ON data to cover a full fuel cycle and this is to observe time evolution of reactor neutrino spectrum and flux and to investigate whether there is any relation between the "5~MeV excess" and $^{235}$U (or $^{239}$Pu). For background subtraction two periods (total $\sim$125 days) of reactor-OFF data were taken before and after the reactor-ON period. NEOS-II has observed light-yield decrease due to precipitation in the GdLS target from the early stage of the data-taking. The light-yield decrease resulted in worse energy resolution, from $\sim$5.6\%/$\sqrt{1~\mathrm{MeV}}$ at the beginning to $\sim$7.7\%/$\sqrt{1~\mathrm{MeV}}$ at the end of the data-taking, and increased time between prompt and delayed signals from $\sim$28 $\mu$s to $\sim$40 $\mu$s. According to our study, however, these are found to be minor effects in the NEOS-II data analysis. 

\begin{figure}[ht]
    \centering
    \includegraphics[width=0.5\textwidth]{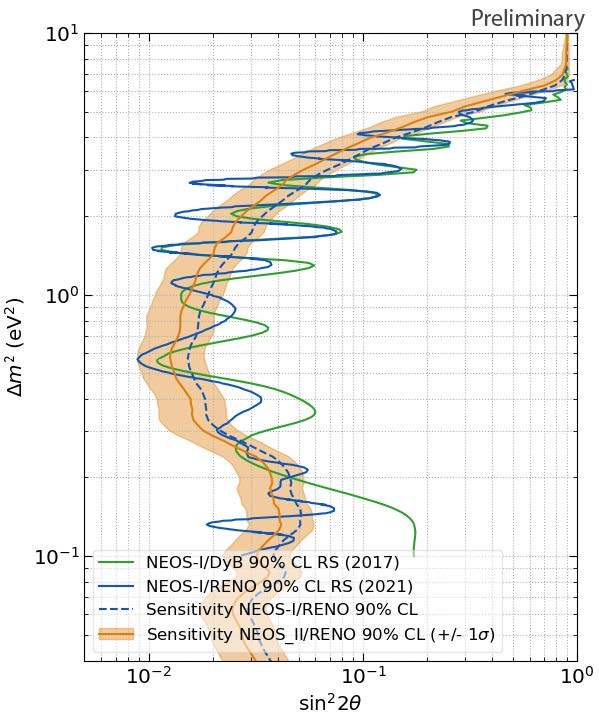}
    \caption{\label{f:neos2_spec_decom}
    NEOS-II 90\% C.L. sensitivity (orange) curve to light sterile neutrino oscillations using 400 live days of data. Also overlayed are the NEOS-I final sensitivity (blue-dashed) and exclusion (solid blue, green) curves.  Figure from~\cite{jinyu_kim_2022_6680618}.}
\end{figure}

Currently two main analyses are underway in NEOS-II. 
One is on the decomposition of $^{235}$U and $^{239}$Pu spectra to better understand the origin of the ``5~MeV excess'' and the other is a light sterile neutrino search using rate+shape analysis.
Figure~\ref{f:neos2_spec_decom} shows the 90\% C.L. sensitivity (exclusion) curve of NEOS-II (NEOS-I) on a light sterile neutrino search where the Daya Bay spectrum is used as a reference spectrum. However, we plan to use the RENO spectrum as a reference for the final result of NEOS-II. The exclusion curve and sensitivity are drawn using a raster scan. A slight improvement in the sensitivity is expected in the NEOS-II using rate+shape analysis and updated reference from RENO, while a shape-only analysis was used in NEOS-I.

\subsubsection{Neutrino-4 Upgrade} The Neutrino-4 experiment is planning for an upgrade of their experiment to increase statistical precision. 
The upgraded detector will contain 4 identical modules, each consisting of a square grid of 25 segments--amounting to 3$\times$ larger volume.
The horizontally positioned segments will have read-out at both the ends, as opposed to vertically placed single read-out segments in Neutrino-4.
The target volume will consist of Gd-doped PSD-capable liquid scintillator to reduce background.
The combination of the increased volume and reduced backgrounds from higher Gd-loading and addition of the PSD capability is expected to have a three-fold increase in statistical power.

\subsubsection{PROSPECT-II} 

The PROSPECT experiment has placed some of the most stringent limits on the eV-scale sterile neutrino oscillations~\cite{PROSPECT:2020sxr}.
The detector was decommissioned in 2020 after an unexpected HFIR down time. The results are currently statistically limited and could be significantly improved by a longer dataset.
PROSPECT-II~\cite{PROSPECT:2021jey} is an upgrade of the PROSPECT experiment with evolutionary modifications intended to take advantage of the beneficial design aspects of the PROSPECT detector and mitigate the drawbacks in the PROSPECT detector to extend the experiment's statistical reach.
The suggested design changes were based on the lessons learned after the decommissioning of the PROSPECT detector. 
To eliminate the chance of liquid scintillator contacting the PMT electronics, a separate subsystem is being designed that holds the PMTs completely outside the liquid scintillator volume.
Additionally, in order to reduce the contact of liquid scintillator with other detector components, a simplified calibration system that runs around the periphery of the detector is being developed.
These modifications are expected to mitigate the risk of PMT failure, highly simplify inner detector assembly, significantly improve the choice of detector materials, and allow for increase in the target volume while still maintaining the footprint approved for operations at HFIR.

\begin{figure}[!htbp]
    \centering
    \includegraphics[width=0.65\textwidth]{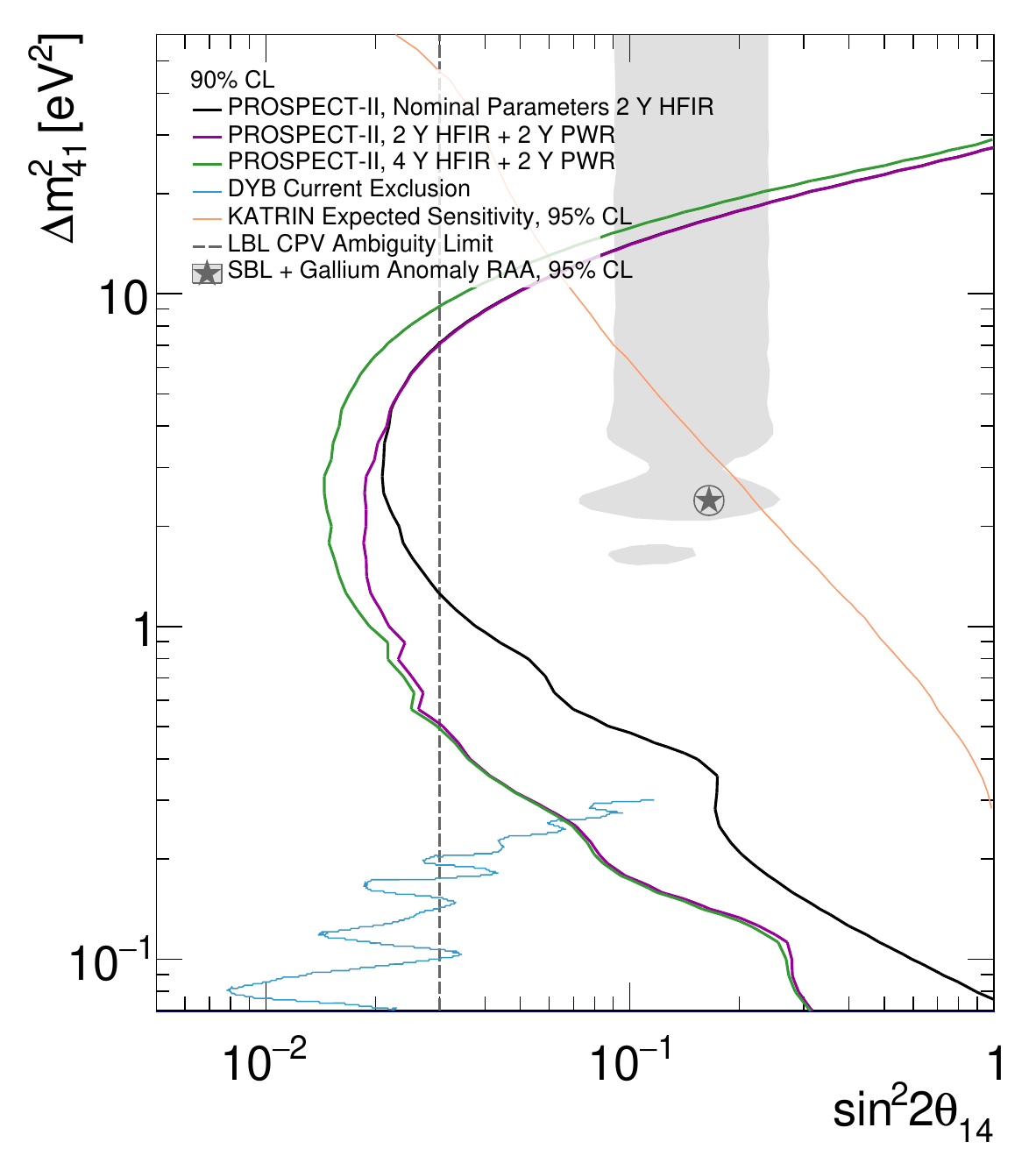}
	\caption{Sensitivity of the PROSPECT-II experiment to sterile neutrinos for 2 years at HFIR (in black), 2 years at HFIR and 2 years at a power reactor (in purple), and 4 years at HFIR (in black) and 2 years at a power reactor (in green). PROSPECT will cover major portion of the suggested sterile neutrino parameter space. A combination of PROSPECT, KATRIN, and Daya Bay experiments will be able to cover significant portion of the parameter space down to the long baseline CP-violation disambuguity limit.  Figure from~\cite{PROSPECT:2021jey}.}
    \label{fig:PROSPECT_II}
\end{figure}

PROSPECT-II design choices are based on the data collected by PROSPECT and data-validated Geant4 simulations. 
Considering all the design modifications, PROSPECT-II is expected to have $\sim$30\% increase in exposure and a three-fold increase in S:B from 1.4 to 4.3.
Figure~\ref{fig:PROSPECT_II} shows the projected sensitivity of the experiment to the sterile neutrino-induced oscillations. 
To mitigate the reliance on the reactor neutrino models the sensitivities are estimated 
by comparing Asimov datasets of relative spectra estimated at multiple baselines within the detector.
As illustrated in Fig.~\ref{fig:PROSPECT_II}, within two calendar years, PROSPECT-II will be able to fully cover the RAA and Gallium Anomaly suggested regions below 15~eV$^2$
While longer baseline and LEU experiments cover the lower oscillation frequencies and the $\beta$-decay endpoint experiments like KATRIN and Project 8 cover the higher oscillation frequencies, PROSPECT-II is uniquely situated to complement these results by providing the most stringent limits on \anue disappearance between 1--20~eV$^2$.

\subsubsection{Other experiments}
Several plastic scintillator experiments including  CHANDLER~\cite{Haghighat:2018mve}, and ISMRAN~\cite{Behera:2020qwf}, NuLat~\cite{NuLat:2015wgu}, and SoLID~\cite{SoLid:2020cen} that have the capability to provide modest sensitivity to sterile neutrinos have either been built or planned to be built in the near future.  
Similar to the experiments described above, these experiments all use segmented detectors for background reduction which would also be useful to perform oscillation search with a single detector. 
Plastic scintillators provide a convenient and robust way to segment the detectors either in two or three dimensions. 
The ability to achieve high signal-to-background ratios over large volumes necessary to perform precision oscillation search has yet to be demonstrated for these detector types.  \\

\noindent{\bf CHANDLER} CHANDLER is a near-field reactor neutrino detector technology consisting of an array of wavelength shifting plastic scintillator cubes, alternating in layers with thin sheets of $^6$Li loaded zinc sulfide (ZnS) scintillator.  Positrons from reactor $\bar{\nu}_e$ interactions deposit their energy in the plastic scintillator creating a prompt pulse of light, while the neutrons thermalize and capture on $^6$Li in the ZnS to create a delayed pulse.  The ZnS releases its scintillation light 20 times slower than the plastic scintillator, which makes an unmistakable neutron capture tag.   Similarly, the high segmentation of the plastic cubes is used to tag the positron annihilation gamma, forming a powerful positron tag.  The combination, in conjunction with tight spatial and temporal coincidence requirements, form a strong discriminator against all backgrounds, since correlated, or even random, coincidences of a positron and a neutron are vey rare in nature.  Small air gaps between the cubes allow the light to transmitted along the rows and columns of cubes, by total internal reflection (TIR), to the surface of the detector.  Light from neutron capture in the ZnS is absorbed by the wavelength shifter in the cube layers on either side of the sheet and re-emitted, so that it too can be transmitted by TIR\@. The position of the cube where the light originated sits at the intersection of the hit row and column.

The CHANDLER technology was demonstrated in a 2017 deployment to the North Anna Nuclear Generating Station, in Mineral, Virginia~\cite{Haghighat:2018mve}.  There, the 80 channel, 80~kg MiniCHANDLER prototype was housed in a 14-foot trailer parked 25 meters from Reactor 2, and ran for four and a half months, including one month of reactor off.  These data were used to isolate a sample of 2881 events.  In this analysis, topological selections for the positron annihilation gammas played a critical role in the observation of reactor inverse beta decay events, shifting the signal significance from less than 3$\sigma$ to 5.5$\sigma$.  This demonstration used old PMTs, which were only marginally sensitive the Compton edge of the 511~keV gammas.  The CHANDLER collaboration is undertaking an upgrade of the MiniCHANDLER prototype with new PMT combined with compound parabolic light guides which will improve the light collection efficiency by a factor of four.  This will greatly enhance the efficiency of the positron tag, and achieve an average energy resolution of around 6\% at 1~MeV\@.\\


\noindent{\bf CE$\nu$NS Experiments}
The discovery of coherent elastic neutrino-nucleus scattering~(CE$\nu$NS) by the COHERENT experiment~\cite{Akimov:2017ade} using a decay-at-rest source, opened up the possibility of using this mechanism to perform several BSM searches. 
Multiple global efforts~\cite{Billard:2016giu,Choi:2020gkm,CONNIE:2021ngo,CONUS:2020skt,MINER:2016igy,NUCLEUS:2019kxv,RED-100:2019rpf,Wong:2015kgl} are underway to leverage the intense source of neutrinos from reactors to perform CE$\nu$NS measurements. 
CE$\nu$NS has not yet been observed in these experiments owing to a combination of high backgrounds at reactor facilities and a need for extremely low threshold detectors.
However, enabling detector technologies are progressing at a fast pace. 
The observation reactor CE$\nu$NS opens doors to a vibrant BSM physics program, including searches for sterile neutrino oscillations.

\subsubsection{Joint Analyses}
As discussed in Sec.~\ref{lbl_reactor_combo}, a single experiment can't have the capability to cover all the sterile neutrino parameter space of interest. 
Thus it is beneficial to perform joint fits of experimental datasets each of which provide sensitivity to different regions of parameter space. 
When performed using proper statistical techniques and by including relevant systematics and correlations properly, such joint fits can be extremely powerful in covering wider parameter space as well as in alleviating any unknown
systematic effects specific to a single experiment that could mimic oscillations.
As highlighted in Ref.~\cite{bib:jointanaloi}, there are attractive opportunities in combining data from multiple current and upcoming reactor experiments such as Daya Bay, PROSPECT, STEREO, NEOS, and JUNO-TAO.
Reactor experiments could further be combined with accelerator experiments such as MINOS and MINOS+. 
It is worth noting that the Daya Bay collaboration plans to publicly release its full data set once all final results have been released~\cite{bib:dybloisnowmass}, allowing such combinations to occur even well after the collaboration has dissolved. 
Similarly, following the end of data taking in 2016, the MINOS/MINOS+ $\mathrm{CL}_s$ surfaces remain available for use in future combinations.
Other experiments such as PROSPECT and STEREO are also moving towards releasing data as part of their publications in a format that could be used to perform such joint analyses.

\subsection{Radioactive Source Experiments}
The combination of the results obtained by GALLEX, SAGE and BEST leads to a gallium anomaly (Sec.~\ref{sec:GaAnomaly}) with a large significance, allowing the possibility that electron neutrinos may disappear through oscillations with the participation of a sterile neutrino at the $\mathcal{O}(\rm{eV})$ mass scale. The following are experimental proposals aiming to perform additional studies to confirm or refute the gallium anomaly.

\subsubsection{BEST-2}
\label{sec:BEST2}
The BEST Collaboration has proposed to use their experimental configuration (see Sec.~\ref{sec:BExpST}) to run the same type of experiment by measuring the electron neutrinos produced by an artificial source made of $^{65}\rm{Zn}$ \cite{Gavrin:2018zmf,Gavrin:2019rtr}. 

\begin{figure}[ht]
    \centering
    \includegraphics[width=0.4\textwidth]{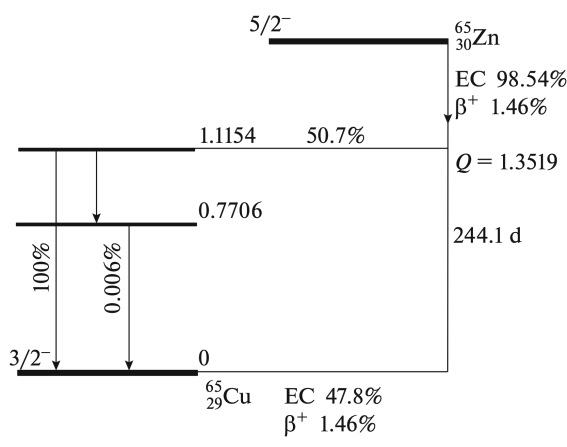}
    \hspace{0.6cm}%
    \includegraphics[width=0.4\textwidth]{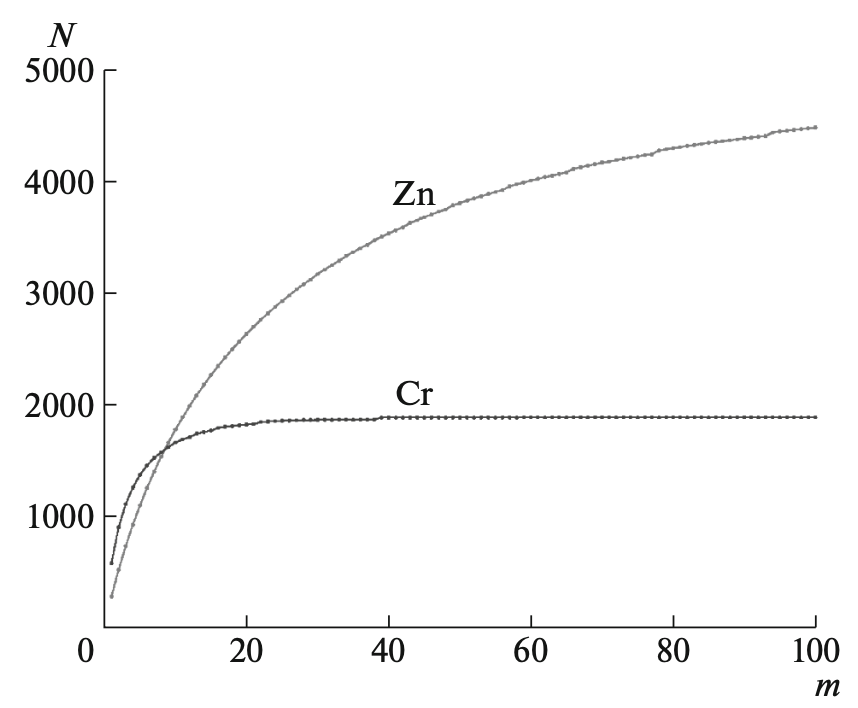}
    \caption{\label{fig:ZnDecay} \emph{Left}. Nuclear levels for the $^{65}\rm{Zn}$ radioactive source decay. \emph{Right}. Number of events, $N$ as a function of the number of exposures, $m$, for experiments with $^{51}\rm{Cr}$ and $^{65}\rm{Zn}$ sources.  Figure from~\cite{Gavrin:2019rtr}.}
\end{figure}

When the isotope $^{65}\rm{Zn}$ decays (Fig.~\ref{fig:ZnDecay}, left), a neutrino of energy 1.35~MeV is emitted in nearly half of the decay events. The remaining events involve the emission of a neutrino with an energy of
235~keV close to the threshold for capture by gallium (233~keV), and the corresponding cross section is small. On the other hand, the cross section for the capture of 1.35~MeV neutrinos by $^{71}\rm{Ga}$ nuclei is approximately three times as large as that for 0.75~MeV neutrinos from a $^{51}\rm{Cr}$ source \cite{Bahcall:1997eg}. Then, the expected rate of capture of neutrinos from a $^{65}\rm{Zn}$ source of activity 3 MCi in a single zone at identical dimensions of the sources and target zones is $n_0 = 108$ d$^{-1}$. Since the $^{65}\rm{Zn}$ lifetime
is longer than the $^{51}\rm{Cr}$ lifetime ($T_{1/2} = 244.1$ and $27.7$ d, respectively), BEST-2 could perform measurements with $^{65}\rm{Zn}$ for a longer time, so that an accumulation of a commensurate data sample may require a substantially lower source activity \cite{Gavrin:2019rtr}.

To compare against the previously published results, with the data sample accumulated in the BEST experiment, it was estimated that the number of extracted $^{71}\rm{Ge}$ atoms expected in the absence of oscillations was 1657. For the proposed source based on $^{65}\rm{Zn}$, the total number of extracted $^{71}\rm{Ge}$ atoms as a function of the number of target exposures, $N(m)$, is given in the right panel of Fig.~\ref{fig:ZnDecay} for various values of the source activity. The total event number $N = 1657$ may be attained even with a $^{65}\rm{Zn}$ source of activity nearly ten times lower than the activity of the $^{51}\rm{Cr}$ source in the BEST experiment \cite{Gavrin:2019rtr}.

The differences (especially the higher energy of the emitted neutrinos) to be implemented in BEST-2 are expected to reach an extended range of $\Delta m^2$ values. In fact, the areas of sensitivity to this parameter in the two experiments (BEST and BEST-2) prove to be shifted in such a way that the highest sensitivity of one experiment corresponds to the lowest sensitivity of the other experiment, and vice versa. According to \cite{Barinov:2017ymq}, a global result of the two
experiments would provide a better statistical significance of measured oscillations (if there are any in the region of searches), but it will also permit precisely measuring the parameters of these oscillations.

\subsubsection{Neutrino oscillometry with Jinping}
\label{sec:jinping}
Another neutrino artificial source proposed to study the possible oscillation of active to sterile neutrinos is the isotope $^{144}\rm{Ce} - ^{144}\rm{Pr}$. In this case, the idea is to use such a source for the future solar neutrino experiment Jinping \cite{Smirnov:2020bcr}.

A
s a source of $\bar{\nu}_e$, the decay chain of isotopes $^{144}\rm{Ce} - ^{144}\rm{Pr}$ is a suitable option for oscillometry experiment. 
The antineutrino energy spectrum of $^{144}\rm{Pr}$ is continuous with the end point around 3~MeV and with an overall half-life of 285 days. About 48.5\% of the emitted antineutrinos are at energies above the detection threshold of the inverse beta decay (IBD) reaction (the value of the threshold is 1.8~MeV) and thus can be used for the measurements. Based on  previous calculations, the maximal source activity can reach 100 kCi \cite{Smirnov:2020bcr}.  

Following \cite{Smirnov:2020bcr}, the Jinping neutrino detector will be located in the Jinping Mountain, Sichuan Province, China with a maximum overburden around 2400 meters. 
The Jinping collaboration plans to build a 2 kton detector using slow liquid scintillator (LSc). This delays the scintillation process and thus separates from the Cherenkov light, significantly increasing the background rejection capability using the particle identification method. The inner detector volume will have a spherical shape with a radius around 8.2 meters. The expected energy resolution will be 5\%$\sqrt{E\rm{[MeV]}}$ and the position resolution  10 cm/$\sqrt{E\rm{[MeV]}}$.

\begin{figure}[ht]
    \centering
    \includegraphics[width=0.55\textwidth]{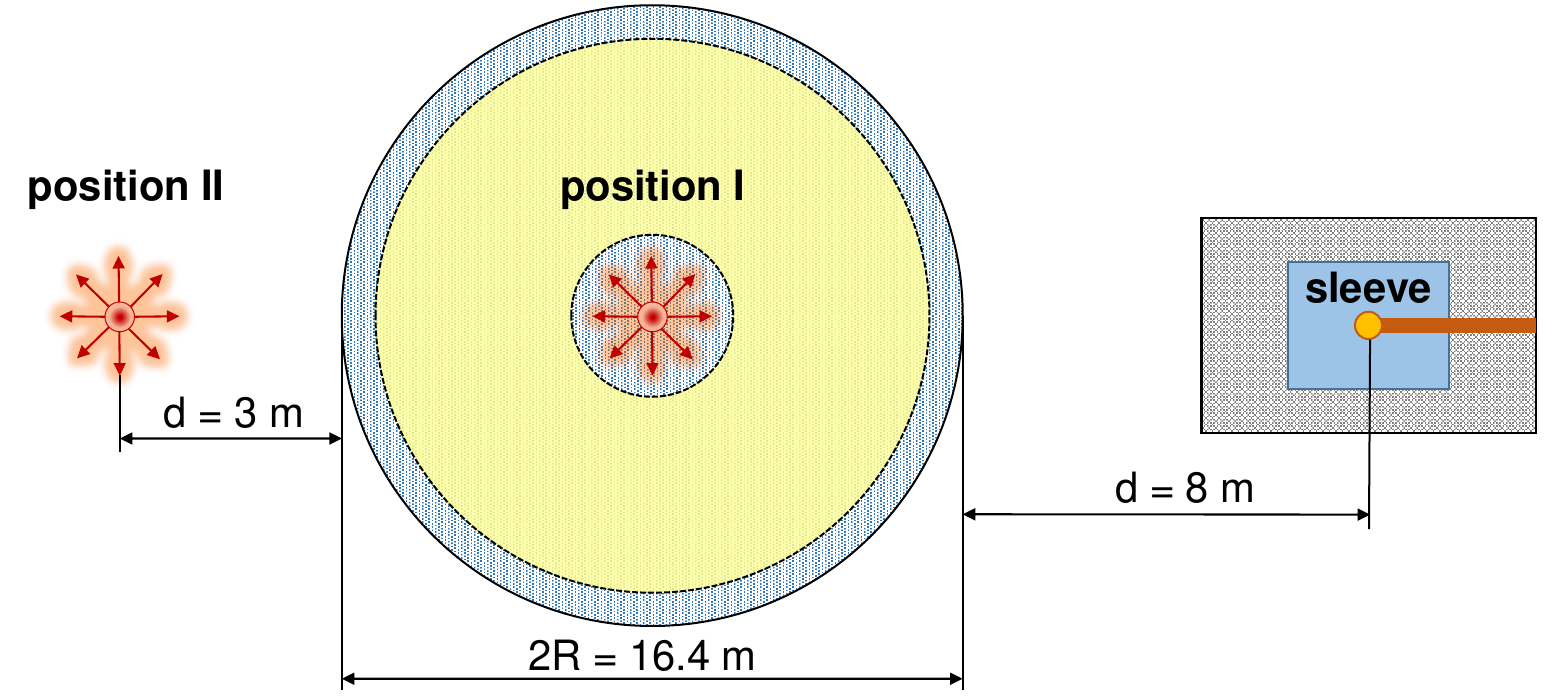}
    \caption{\label{fig:jinpingExp} The schematic layout of the proposed experiment at Jinping. Two positions for the radioactive source are assumed: at the center (position I) and outside of the detector (position II). Two fiducial volume cuts are applied. Inner cut with radius 100 cm and outer cut with width 70 cm. The yellow area is the active volume for position I, the yellow area plus the inner cut region is active volume for position II.  Figure from~\cite{Smirnov:2020bcr}.}
\end{figure}

The proposed experimental setup is shown in Fig.~\ref{fig:jinpingExp}. Two possible locations for the point source are being considered: one at the center of the sphere (position I) and another at a distance 3 m from the edge of the detector (position II). The source at the detector center case gives a higher statistics but a shorter range of baseline. Technically, the more realistic case is when the source is outside of the detector. At both source positions the exposure time is assumed to be 450 days. Initial source activity will be 50 and 100 kCi for position I and position II respectively. The expected non-oscillation event rate is 28.5K and 73.8K for position I and position II respectively.


\begin{figure}[ht]
    \centering
    \includegraphics[width=0.5\textwidth]{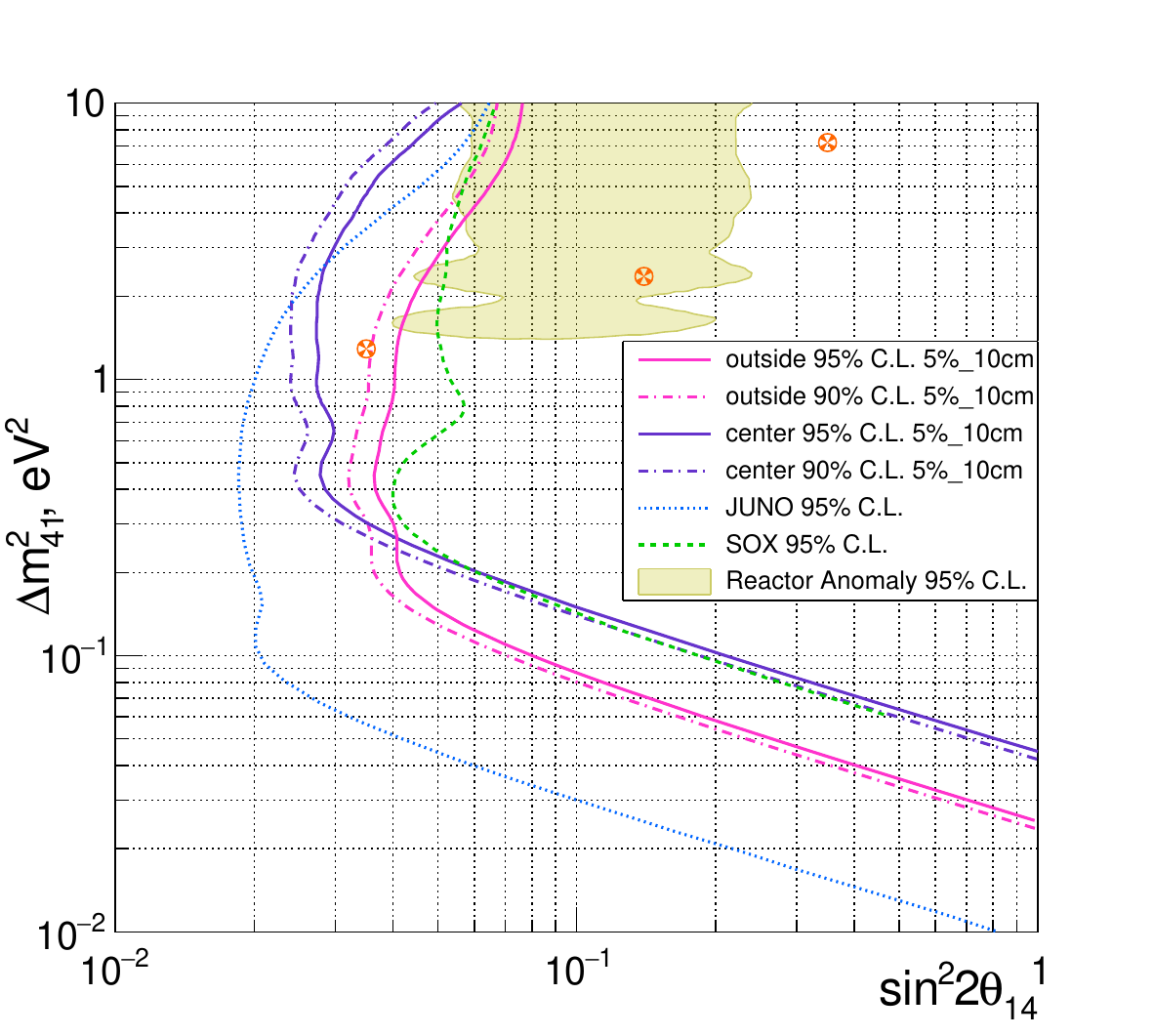}
    \caption{\label{fig:jinpingExp_contorus} The exclusion contours on a two-dimensional parameter space. 90\% and 95\% C.L. are shown for two possible setups. Combined reactor anomaly is also shown \cite{Mention:2011rk}. The stars indicate the current best-fit value for the sterile neutrino hypothesis~\cite{Dentler:2018sju}, the bes-fit value of the reactor antineutrino anomaly (RAA)  \cite{Mention:2011rk}, and the best-fit result of the Neutrino-4 experiment \cite{Serebrov:2018vdw} from left to right, respectively.  Figure from~\cite{Smirnov:2020bcr}.}
\end{figure}

The expected sensitivity to sterile neutrinos, represented as a two-dimensional exclusion plot, compared with another proposed experiments is drawn on Fig.~\ref{fig:jinpingExp_contorus}. All currently interesting regions for sterile neutrino searching are covered if position I is implemented. For position II, some part of the RAA and best fit are still not fully covered. 


\subsubsection{THEIA}\label{sec:radiactiveTHEIA}

The most recently updated sterile neutrino landscape based on the results from BEST and MicroBooNE combined with short-baseline reactor experiments point to the possibility of active to sterile neutrino oscillations in the region with $\Delta m^2  > 1$ eV$^2$ which implies that oscillation pattern  is only measurable at distances of less than 10 m.  

The three direct independent methods to test sterile neutrino hypothesis include a very short-baseline reactor experiment, accelerator experiment and use of neutrino and antineutrino generators in the vicinity of the large liquid scintillator (LS)  or water based liquid scintillator detector (WbLS). The 100 kton (25 kton)  WbLS THEIA detector  \cite{Gann:2015fba,Theia:2019non} represents a particularly promising venue for a decisive test using neutrino and antineutrino generators,  to observe a \textit{distance-dependent} neutrino flux from the source at the distances of the order of oscillation length. 

The neutrino oscillation length is given by the following formula:
\begin{equation} 
L_{osc}[\rm{m}] = 2.48 \frac{E_{\bar \nu_e} [\rm{MeV}]}{\Delta m^2_{new} [\rm{eV}^2]}.
\end{equation}

In the case of sterile neutrino $\Delta m^2 > 1$~eV$^2$, the oscillation distance is of the order of a couple of meters. THEIA spans over tens of meters of distance and the source can be placed within meters of the target, creating a baseline comparable to the sterile neutrino oscillation length.  Observation of the oscillation pattern in the distant-dependent measurement in WbLS/LS  would represent convincing proof of the existence of sterile neutrinos and their oscillation with the other three flavors.

\paragraph{Sterile neutrino search with antineutrino generators in THEIA}
THEIA  can detect $\bar \nu_e$ via inverse beta decay reaction (IBD) on hydrogen that makes it easily distinguishable from the backgrounds thanks to the double coincidence signature. The most promising antineutrino generator for this search is a pair of beta decaying nuclei $^{144}$Ce-$^{144}$Pr thanks to high yield of  $^{144}$Ce in the spent nuclear fuel combined with high  endpoint at 3~MeV with respect to 1.8~MeV IBD  interaction threshold, and relatively long half-life of 285 days.  There are two possible configurations: placing the source in the center of the detector for maximum interaction rate or on the side of the detector which is less disruptive and preferred but leads to reduced antineutrino flux. Nevertheless, due to its huge volume of 100 kton THEIA  will observe over 100 million antineutrino events in case of no oscillation (a couple of million less in case of sterile neutrino oscillations) with 3.7 PBq source running for 18 months. This measurement is practically background free and statistical error is negligible leading to a very sensitive measurement of large mass sterile neutrino oscillations even if the mixing angle is very small using distance-dependent  rate and spectrum measurement. It is worth noting that even with an order of magnitude weaker $^{144}$Ce-$^{144}$Pr source, THEIA would still make a robust measurement of sterile neutrino oscillations. A 0.37 PBq source has been successfully made in the past and is logistically easier for shielding, handling and transport.

In addition, the IsoDAR\cite{Alonso:2022mup} collaboration is working on producing a feasible decay at rest source that is based on irradiation of $^7$Li with a strong neutron flux to produce $^8$Li that beta decays within  838 milliseconds. Emitted antineutrinos have a Gaussian energy spectrum in the range from 3~MeV to 13~MeV. This work is progressing steadily, but is not available for production. The advantage of this source is that it produces antineutrinos at higher energy and at a steady rate.  IsoDAR would also provide practically background-free high-statistics measurements of sterile neutrino oscillations. 

\paragraph{Sterile neutrino search with neutrino generators in THEIA}
The most exciting opportunity for THEIA to test the sterile neutrino hypothesis would be to directly cross-check the GA and BEST result by deploying $^{51}$Cr electron neutrino source. As opposed to radiochemical measurements THEIA would measure  a Compton-like shoulder due electron neutrino backscatters produced by the $^{51}$Cr source.  As opposed to bulk rate measurement of the event rate, THEIA would allow spectrum measurement with electron neutrino detection in real time. $^{51}$Cr emits a monoenergetic  753~keV gamma 90\% of the time. This energy translates at the Compton-like shoulder at 0.5~MeV, which is slightly below 0.6~MeV energy threshold of the THEIA pure WbLS phase. However, as part of the Neutrinoless Double Beta Decay search, a 8 m radius balloon filled with pure liquid scintillator (LS) will be deployed. The LS will have a high scintillation yield, resulting in better energy resolution and lower energy threshold required to detect  electron neutrinos from $^{51}$Cr. This would present an excellent opportunity for sterile neutrino search with $^{51}$Cr electron neutrino source, directly testing GA and BEST results with electron neutrino detection in real-time.

\paragraph{Conclusion}
In conclusion, a large WbLS detector such as THEIA  will provide an exceptional playground for the detailed study of sterile neutrino oscillations, requiring modest energy and vertex resolution. Additional important ingredients include knowledge of the absolute incoming antineutrino (neutrino) flux that can be measured in sources and a few meter distance between the source and the detector.
THEIA WbLS detector has a high potential to carry out a powerful search for sterile neutrinos free of statistical limitations that plague current generation of experiments, in the high sterile neutrino mass regime of $\Delta m^2 > 1$~eV$^2$. With its large target mass of up to 100 kton,  a tremendous number of IBD interactions from $^{144}$Ce-$^{144}$Pr source  will be collected allowing detailed, high statistics study of the position dependent neutrino flux. THEIA can also directly cross-check GA and BEST high confidence  electron neutrino measurement with an independent detection method.

\subsection{Atmospheric Neutrino Experiments}

\subsubsection{IceCube Upgrade}

The IceCube Upgrade~\cite{IceCube:2019pna} will begin operation in 2026 and provide both a lower energy threshold and more precise calibration for the full IceCube Neutrino Observatory.
Presently dominant sources of systematic uncertainty for IceCube sterile neutrino analyses include absorption and scattering in South Pole glacial ice~\cite{IceCube:2019lxi}, both expected to be significantly reduced by improved precision of detector calibrations from the Upgrade with its smaller optical module spacing and enhanced calibration systems.
Reduced thresholds will enable unprecedented precision for $\nu_\tau$ appearance searches from standard oscillations, and will similarly contribute to improvements of low-energy sterile neutrino searches at IceCube. 

\subsubsection{DUNE}

As noted in Section 6.2.2.1 (DUNE LBL), DUNE~\cite{DUNE:2020ypp} is primarily designed as an accelerator experiment to accurately measure the value of $\delta_{\rm CP}$ using a neutrino beam produced from the LBNF accelerator complex at Fermilab.
It will employ a group of Near Detector(ND) and a Far Detector (FD) to measure neutrino oscillation parameters.
The DUNE FD will be located at the Sanford Underground Research Facility, South Dakota.
Benefiting from low cosmic-muon background, it will also act as an atmospheric neutrino detector, and will make measurements of several oscillation parameters for SM and BSM physics models, which will be complementary to the measurementss made with the accelerator neutrino data.

The FD will be constructed as four separate LArTPC modules, with a total fiducial mass of 40~kt.
While DUNE will collect significantly lower event statistics compared to very large volume neutrino telescopes such as IceCube~\cite{Aartsen:2014njl}, the excellent energy and direction reconstruction will enable competitive atmospheric physics measurements.
With LArTPC detectors, very high-resolution spatial tracking (up to millimeter accuracy) of particles produced in neutrino interactions is possible and it can resolve final state morphologies to an extent not possible with other detector technologies.
DUNE will also have a finer control of detector and neutrino interaction systematics, enabled by its Near Detectors in the beamline.
Between IceCube and DUNE atmospheric neutrino event measurements, there will be a complementarity in terms of large sample size and high resolution.
Notably, using atmospheric neutrino events in a few GeV range, DUNE FD will be able to discover the Neutrino Mass Ordering (NMO) with a statistical significance of 3$\sigma$ or higher with 10 years of data~\cite{DUNE:2020ypp}.
DUNE also has potential to measure the value of $\delta_{\rm CP}$ using sub-GeV atmospheric neutrino event samples~\cite{Kelly:2019itm}.
It will constrain many BSM physics models such as sterile neutrinos, NSI, neutrino decays and Lorentz Invariance Violation (LIV).

Atmospheric neutrinos~\cite{Honda:2011nf} are produced in the interaction of the cosmic rays with atmospheric nuclei and contain both $\overset{(-)}{\nu_e}$ and $\overset{(-)}{\nu_\mu}$ components.
They span a broad energy range from MeV to PeV, with a peak around 1~GeV. Propagation of atmospheric neutrinos through the Earth gives accessible L/E from 10 to $10^4$~kM/GeV and their measurements can provide sensitivities to $\Delta m^2 \; \sim$  $(10^{-4} - 1) \; \rm{eV}^2$.
The atmospheric data at DUNE FD will make measurement of $\nu_e$ and $\nu_\mu$ disappearance as well as $\nu_\tau$ appearance channels.
Hence, the DUNE FD can constrain all three new mixing angles $\theta_{14}, \; \theta_{24}$ and $\theta_{34}$ over the aforementioned broad range of $\Delta m^2_{41}$ with the atmospheric data.
Sensitivities to these parameters are expected to be enhanced around $10^{-2} - 10^{-3} \; \rm{eV}^2$ thanks to matter effects, as neutrinos cross the Earth's mantle and core during their propagation.
Preliminary sensitivity study suggests that with conservative assumptions on detector properties such as resolutions and PID classification, the DUNE FD can probe sterile parameter space with sensitivities comparable to the current limits from other experiments.
Dedicated sensitivity studies to these parameters using DUNE atmospheric neutrino MC simulations is planned in the near future.
A joint fit of atmospheric and beam neutrino data is expected to improve parameter constraints over both individual fits. Using DUNE atmospheric data, constraining more complicated models such as decaying sterile neutrinos is also foreseen.
Additionally, due to DUNE's capacity to inspect in detail the energy losses of muons, they can infer the energy of uncontained high-energy muons.
This provides sensitivity to DUNE to high-energy atmospheric neutrino analysis as shown in Ref.~\cite{Schneider:2021wzs}, which can provide complementary constraints in light sterile neutrino to the ones provided by the contained events.

\subsubsection{Hyper-Kamiokande}

Hyper-Kamiokande (Hyper-K) is the next-generation large water Cherenkov detector to be deployed in Japan. 
Beyond their accelerator program discuss previously in this white paper, they also are able to measure atmospheric neutrinos much like its predecessor Super-K.

\begin{figure}[ht]
\centering
\includegraphics[width=0.8\textwidth]{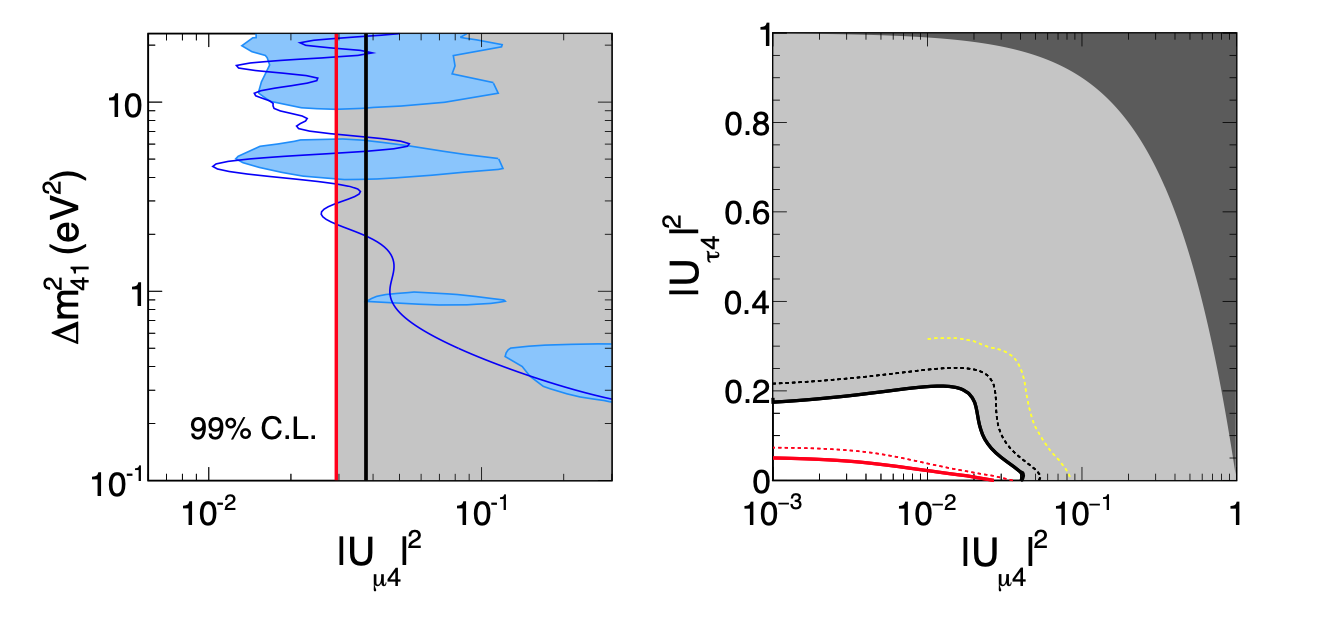}
\caption{Hyper-K expected 90\% C.L. upper limits on $|U_{\mu 4}|^2$ appear as red lines in the left figure.
The right figure shows 90\% (solid) and 99\% (dashed) C.L. limits on $|U_{\mu 4}|^2$ vs $|U_{\tau 4}|^2$ for a 5.6 Mton·year exposure (red) in comparison with recent limits from Super-K (black). Figure from~\cite{Hyper-Kamiokande:2018ofw}.}
\label{fig:hk_atm_sensitivity}
\end{figure}

The expected sensitivity to light sterile neutrinos from Hyper-K has been reported in Ref.~\cite{Hyper-Kamiokande:2018ofw}.
As in the case of Super-K they are predominantly sensitive to $|U_{\mu 4}|^2$ and $|U_{\tau 4}|^2$ mixing elements by looking for distortions in the angular and energy distributions of their events.
The expected sensitivity can be seen in Fig.~\ref{fig:hk_atm_sensitivity} compared to the previous constraints from Super-K. 
As can be seen from this figure the improvement on $|U_{\mu 4}|^2$ is marginal; however, significant improved sensitivity can is expected in $|U_{\tau 4}|^2$.
This is interesting since currently ANTARES has reported a small preference for non-zero $|U_{\tau 4}|^2$, which would be tested by Hyper-K.

\subsubsection{Atmospherics in THEIA}\label{sec:atmoTheia}

THEIA is a proposed multi-kiloton hybrid optical neutrino detector designed with multiple physics goals in mind \cite{Theia:2019non}. The experiment plans to leverage novel scintillating materials to fully exploit information from scintillation and Cherenkov light production mechanisms using advancements in fast photodetectors and spectral sorting technologies, on a large scale. This ``hybrid'' technology will simultaneously enable high light yields for superb position and energy reconstruction, direction reconstruction from Cherenkov light and particle identification (PID) through knowledge of both mechanisms. There is also the potential to employ isotope loading techniques for various physics needs (Sec.~\ref{sec:radiactiveTHEIA}). 

While physics potentials have been evaluated for long baseline physics, solar neutrinos, supernova neutrinos, DSNB, geoneutrinos, reactor neutrinos, neutrinoless double beta decay and nucleon decay in \cite{Theia:2019non}, the potential for leveraging atmospheric neutrinos, especially in regards to the picture of the short-baseline anomalies and possible sterile neutrinos and non-standard neutrino interactions (NSIs) has not been previously examined. Given the ring-imaging capability of a Cherenkov detector, THEIA would be capable of an analysis similar to that performed by Super-Kamiokande \cite{Super-Kamiokande:2014ndf}. THEIA would have similar sensitivity to Super-Kamiokande, with enhanced samples based on PID techniques --for example, by tagging neutrons more effectively.  Yet, it would be many years before THEIA would equal the exposure used in the analysis of \cite{Super-Kamiokande:2014ndf}. However, as discussed in \cite{Kelly:2017kch}, Hyper-Kamiokande may be able to enhance this picture by simultaneously employing beam and atmospheric neutrinos for both sterile neutrino and NSI searches, which also has complementarity with DUNE. A placement of THEIA in the LBNF beamline could further supplement these experiments, due to utilization of similar detection technology to Hyper-Kamiokande with a location in the same beam as DUNE. This would provide similar detectors in two different locations and beams (Hyper-Kamiokande and THEIA) as well as two different detection technologies in the same location and beam (DUNE and THEIA), which could open significant opportunities in further exploring parameter space and understanding systematic uncertainties across the three experiments. Further studies are planned to examine this possibility.

\subsubsection{KM3NeT and ORCA}

\subsubsubsection{KM3NeT}

Neutrino telescopes offer a complementary approach to investigate the anomalies found in short-baseline experiments. 
In this section we will focus on the capabilities of the KM3NeT detector~\cite{Adrian-Martinez:2016fdl}, which is now under construction in the Mediterranean Sea. 
It should be also mentioned that KM3NeT's predecessor, ANTARES, also in the Mediterranean Sea, has also looked for sterile neutrinos with the data gathered in ten years~\cite{ANTARES:2018rtf} as discussed in prior sections of this work.

The detection principle is as follows. 
Neutrinos (originated in cosmic sources or in the Earth's atmosphere) can interact in the surroundings of the detector and produce particles which will induce Cherenkov light that can be detected by an array of photomultipliers encapsulated in the so-called Optical Modules (OMs).
The KM3NeT neutrino telescope is made of two subdetectors, ORCA and ARCA.
ORCA is installed at a depth of 2500~m, off the French coast in Toulon.
ARCA is located 3000~m deep, close to Porto Palo, in Sicily.
Both use the same technology, being the main difference the distance between the OMs.
ORCA is denser and will have 115 lines with 18 OMs each distributed in a volume equivalent to 8 MTon.
The energy threshold estimated for this configuration is about 3~GeV.
ARCA, with 230 lines, will have a sparser distribution of OMs, which will instrument about one cubic kilometer of water.
The difference in size and density drives the physics focus of each configuration.
ORCA is more suited for studies on atmospheric neutrinos and low energy astrophysical events, while ARCA's main goals are high energy neutrinos from astrophysical sources (with particular interest in Galactic sources given its geographical location).
At the time of this writing, there are 10 lines of ORCA and 8 lines of ARCA already installed and taking data, with plans to install more lines in the following months.

\subsubsubsection{ORCA}

\begin{figure}[ht]
\centering
\includegraphics[width=0.8\textwidth]{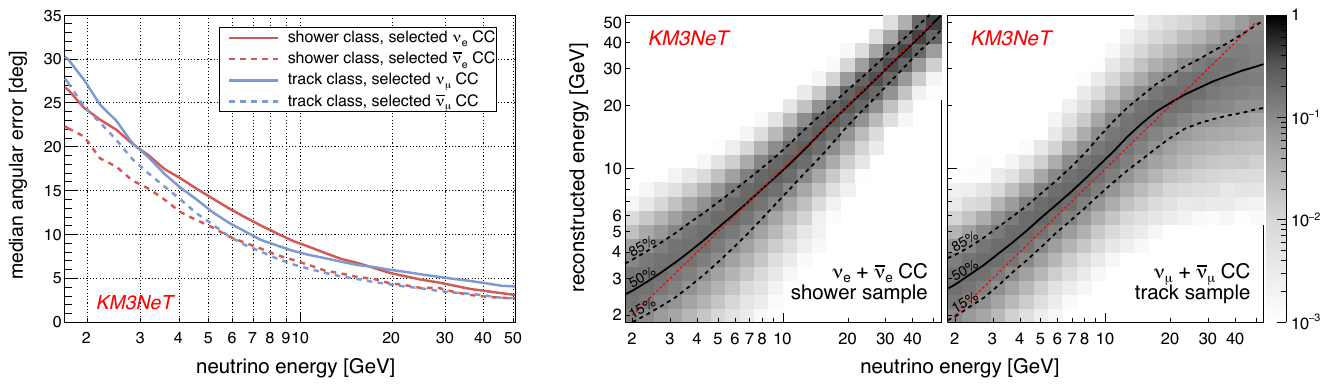}
\caption{Left: Median direction resolution as a function of true neutrino energy for showers and tracks. Right: Probability distribution of the reconstructed energy as a function of true neutrino energy for upgoing showers and tracks (black lines indicate the quantiles). Figure from~\cite{KM3NeT:2021ozk}.}
\label{fig:orca_res}
\end{figure}

ORCA will precisely measure various parameters of the three-neutrino standard oscillation framework, namely the neutrino mass ordering, $\sin^{2}\theta_{23}$, $\Delta m_{13}^{2}$, and PMNS unitarity \cite{KM3NeT:2021ozk}.
As a matter of fact, the first oscillation study with a partial configuration and less than one year of exposure finds that
oscillations are preferred with a confidence level of $5.9\sigma$ over ``no oscillations''~\cite{KM3NeT:2021hkj}.
The main analysis selects events induced by atmospheric neutrinos coming below the horizon and compares the observed energy, direction and topology distributions to predictions under different oscillation hypotheses.
In the energy range 3-100~GeV, in which the ORCA detector is most sensitive to neutrinos, events are categorised in two event topologies: tracks, characterised by a long muon track (mostly from muon-neutrino CC interactions), and showers, characterised by events with no distinguishable tracks. 
ORCA is an excellent instrument to constrain oscillation parameters, primarily due to better energy and angular resolutions than other neutrino telescopes in this energy regime as shown in Fig.~\ref{fig:orca_res}. 

The presence of light-sterile neutrinos, enhanced by matter effects, would alter the observed rate of events in the GeV regime.
For scenarios with large mass splitting and non-zero $\theta_{24}$ and $\theta_{34}$, a deficit of high-energy track-like events is expected. 
This region of the phase space has been probed by neutrino telescopes, such as ANTARES \cite{ANTARES:2018rtf} and IceCube/DeepCore~\cite{IceCube:2017ivd}.
In addition, ORCA is well suited to test non-zero $\theta_{14}$ models, as the rate of showers around 20~GeV depends on this mixing angle.

\begin{figure}[ht]
\centering
\includegraphics[width=0.9\textwidth]{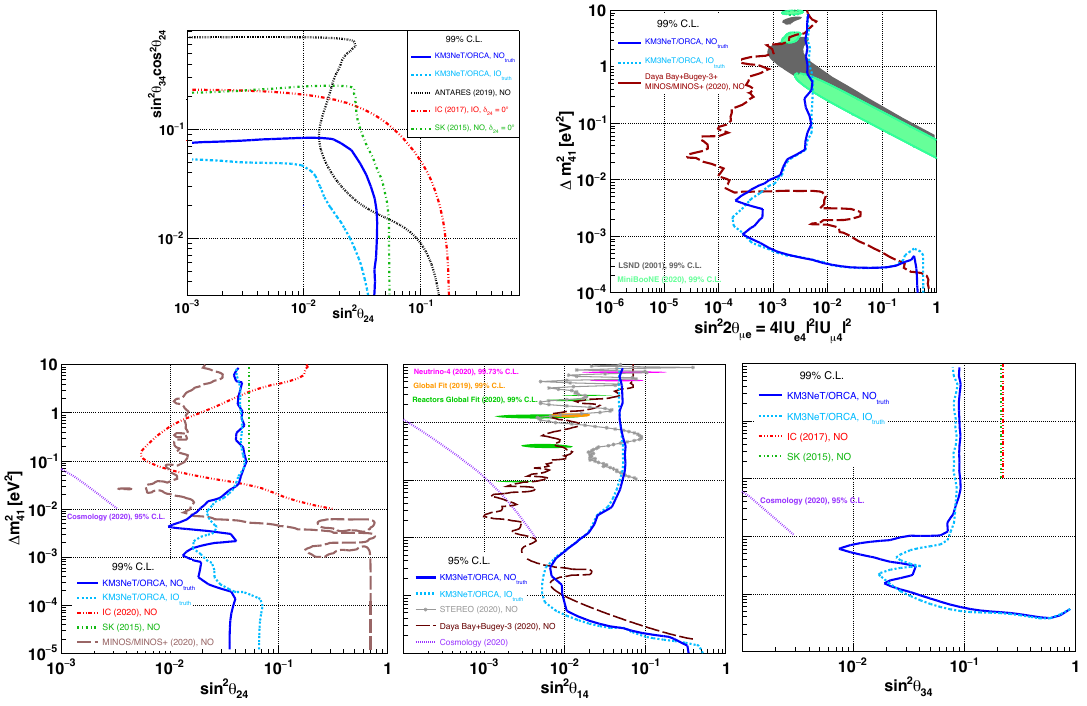}
\caption{The 99\% C.L. ORCA sensitivity after three years of data taking for different combination of sterile parameters. Top left: $\theta_{24}-\theta_{34}$, with $\Delta m_{41}^{2}=1$ eV${^2}$ (ANTARES~\cite{ANTARES:2018rtf}, IceCube~\cite{IceCube:2017ivd}, and SK~\cite{Super-Kamiokande:2014ndf}). Top right: $|U_{\mu e}|^{2}$ (Daya Bay+Bugey-3+MINOS+~\cite{MINOS:2020iqj}, LSND~\cite{LSND:1995lje}, and MiniBooNE~\cite{MiniBooNE:2020pnu}). Bottom: $\sin^2\theta_{24}$ (SK~\cite{Super-Kamiokande:2014ndf}, IceCube~\cite{IceCube:2020phf}, MINOS+~\cite{MINOS:2020iqj}, and cosmology~\cite{Hagstotz:2020ukm}), $\sin^2\theta_{14}$ (Daya Bay+Bugey-3+MINOS+~\cite{MINOS:2020iqj}, STEREO~\cite{STEREO:2019ztb}, Neutrino-4~\cite{Serebrov:2020kmd}, global fit~\cite{Diaz:2019fwt}, reactor fit~\cite{Berryman:2020agd}, and cosmology~\cite{Hagstotz:2020ukm}), and $\sin^2\theta_{34}$ (SK~\cite{Super-Kamiokande:2014ndf}, IceCube~\cite{IceCube:2017ivd}, and cosmology~\cite{Hagstotz:2020ukm}). Figure from~\cite{KM3NeT:2021uez}.
}
\label{fig:km3net_sens}
\end{figure}

First sensitivity studies have been performed using a joint binned likelihood for three samples (track, shower, and others), in which systematic uncertainties on the neutrino flux, detector response and standard oscillations were included~\cite{KM3NeT:2021uez}.
Fig.~\ref{fig:km3net_sens} shows the ORCA sensitivity to sterile mixing angles in different scenarios after three years of data taking. 
In the large $\Delta m_{41}^{2}$ limit, ORCA can set competitive constraints in $\theta_{24}$ and $\theta_{34}$\footnote{In this analysis $\delta_{24}$ is kept free in the fit and $\theta_{14}$ and $\delta_{14}$ are fixed to zero}. 
It will be able to improve current limits on $\sin^2\theta_{24}$, $\sin^2\theta_{34}$, $\sin^2\theta_{14}$ and $\sin^2 2\theta_{\mu e}$ for low values of the mass splitting\footnote{These analyses kept free all relevant sterile neutrino parameters}, not yet constrained by cosmology or appearance/disappearance measurements.
Furthermore, ORCA will be able to test the majority of the LSND and MiniBoone anomaly region as well as part of the Neutrino-4 allowed region.

Other BSM scenarios can be also tested with ORCA detector. In particular, competitive sensitivities to NSI and invisible neutrino decay can already be achieved with partial configurations of the detector \cite{KM3NeT:2021nnf,deSalas:2018kri}. When completed, the ORCA detector will potentially become a leading tool for probing these models.

\subsubsubsection{ARCA}

\begin{figure}[ht]
\centering
\includegraphics[width=1.0\textwidth]{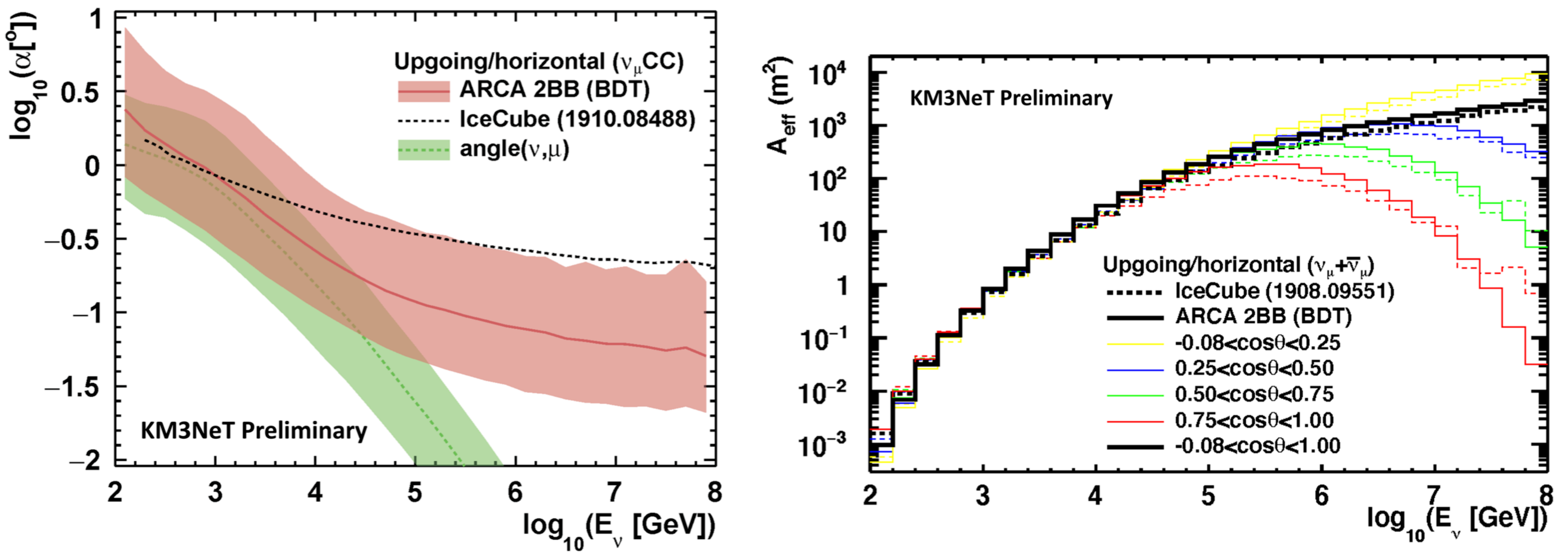}
\caption{Left: ARCA angular resolutions as function of the neutrino energy for $\nu_\mu$ CC events. IceCube resolution was extracted from the point source analyses \cite{IceCube:2019cia}. Right: ARCA effective area as function of the neutrino energy for the selected event sample. IceCube effective area was extracted from the diffuse analyses \cite{Stettner:2019tok}.  Figure from~\cite{KM3NeT:2021szv}.}
\label{fig:arca_res}
\end{figure}

ARCA will be able to observe neutrinos in the high-energy regime \cite{KM3NeT:2021szv}, exceeding IceCube's capabilities for up-going muon neutrinos as shown in Fig.~\ref{fig:arca_res}.
Several studies have shown that the presence of sterile neutrinos or NSI result in the disappearance of TeV-scale muon neutrinos passing through the Earth~\cite{Nunokawa:2003ep,Salvado:2016uqu,Diaz:2019fwt}. 
Recently, IceCube has set world-leading constraints in some of these models characterising atmospheric neutrinos in the few TeV range \cite{IceCube:2020phf}. Preliminary studies are currently underway to assess ARCA sensitivities.

\subsection{Beta and Electron Capture Decay Experiments}
\label{sec:future_direct}

Beta decay and electron capture experiments are complementary to oscillation experiments in searches for sterile neutrinos.
As described in Sec.~\ref{sec:null:numass}, the observed spectrum is a superposition of spectra corresponding to each mass state with endpoints shifted by the different neutrino masses.
These experiments have the capability to search for sterile neutrinos ranging in mass from the sub-eV up to the MeV.

\begin{figure}[ht]
  \centering
  \includegraphics[width=0.6\textwidth]{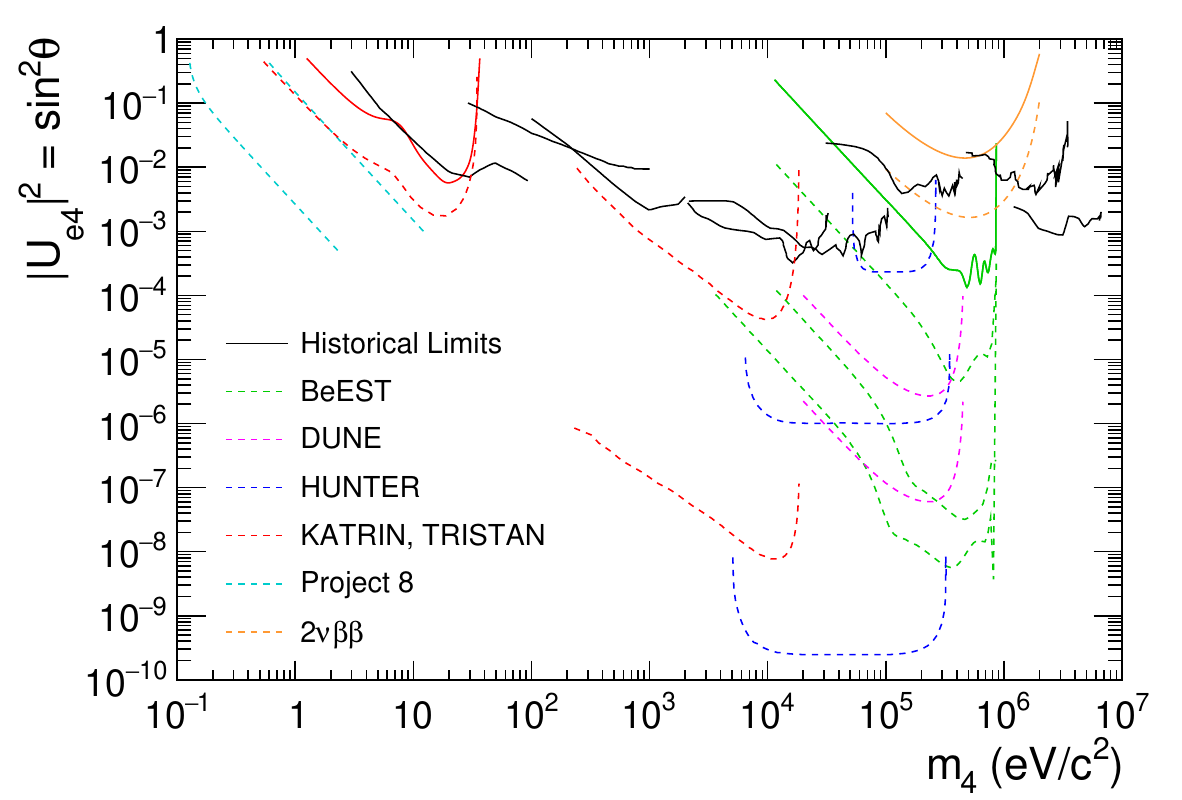}
  \caption{Landscape of limits on sterile neutrinos from current and proposed beta decay and electron capture experiments.}
  \label{fig:future:beta_summary}
\end{figure}

A new generation of experiments is envisioned to advance sensitivity to sterile neutrinos by orders of magnitude in the $\sin^2\theta$ (see Fig.~\ref{fig:future:beta_summary}).
Improvements over former experiments - particularly in source strength, detector threshold, and energy resolution - are enabling gains.
Some proposed experiments (KATRIN, Project 8, DUNE) leverage existing detectors targeting other physics with minimal or no modification, while others experiments (HUNTER, BeEST) require dedicated apparatuses.

\subsubsection{KATRIN/TRISTAN}
\label{sec:future:TRISTAN}

As described in Sec.~\ref{sec:null:numass} KATRIN is sensitive to light sterile neutrinos, providing complementary information to oscillation-based searches.
As the analysis interval of KATRIN is currently constrained to \SI{40}{\electronvolt} below $E_0$, only sterile neutrinos up to a corresponding mass of about $m_4 < \SI{40}{\electronvolt}$ can be searched for. 

To search for heavier sterile neutrinos, a larger range of the tritium $\beta$-spectrum must be measured. In particular, keV-scale sterile neutrinos would be of interest, as they are potential dark matter candidates~\cite{Drewes:2016upu}. However, indirect observations and cosmological considerations limit their mixing with active neutrinos to $\sin^2\theta<10^{-6}$.

As the count rate increases further away from the endpoint, the statistical uncertainty decreases. Assuming the source strength of KATRIN and a measurement time of about one year, a statistical sensitivity better than $\sin^2\theta<10^{-6}$ could be reached~\cite{Mertens:2014nha, Mertens:2014osa}. However, a high-statistics measurement of the entire tritium $\beta$-decay spectrum poses a new technical challenge: electron rates exceeding $10^8$~cps and several new systematic uncertainties become relevant when describing the experimental tritium spectrum far away from the endpoint. 

The TRISTAN project is exploring the sensitivity of such a search and is developing a new silicon drift detector focal-plane array for KATRIN with more than 1000 pixels \cite{KATRIN:2018oow,Mertens:2020mdv}. This technology allows a measurement of a high $\beta$-electron flux with an energy resolution of 300~eV (FWHM) for 20~keV electrons, and large (3~mm) pixel footprints. Such a high-resolution detector would enable a differential measurement of the full tritium $\beta$-decay spectrum, in order to search for signatures of physics beyond the SM. The new detector will be installed after completion of the neutrino mass measurements. 

A demonstration of this procedure has been performed by KATRIN using a low-intensity commissioning dataset from 2018~\cite{KATRIN:2022spi}.
With only 0.5\% the target tritium activity in this run, the detector could handle scans down to 1.6\,keV below the endpoint (as opposed to the 40\,eV below the endpoint for the normal operation).
This small dataset is sufficient to set leading limits over the mass range $0.1 < m_4 < 1.0$\,keV, demonstrating the potential of a full-activity dedicated measurement with the upgraded TRISTAN detector.

\subsubsection{Project 8}

Project 8 is a $\beta$-decay endpoint experiment with an aim to precisely measure the neutrino mass using cyclotron radiation emission spectroscopy~(CRES) and atomic tritium as the source~\cite{Project8:2017nal}.
CRES is a frequency-based spectroscopic method that leverages the energy dependence of cyclotron radiation to perform a high resolution differential spectrum measurement over a wide energy range~\cite{Monreal:2009za}. 
The use of atomic tritium provides a tight control on systematics by eliminating the uncertainty associated with the final states in the decay of molecular tritium~\cite{Robertson:1988xca}.

Project 8 has undertaken a phased approach with each phase demonstrating a  critical technological milestone.
Phase-I of the experiment demonstrated CRES using internal conversion electrons from $^{83m}$Kr~\cite{Project8:2014ivu}.
Phase-II performed the first CRES-based molecular tritium endpoint measurement and placed an upper limit on neutrino mass~\cite{Project8:2022hun}.
Critical R\&D for Phase-III of the experiment is ongoing with two parallel efforts to demonstrate: i) CRES in larger volumes and ii) the use of atomic tritium.
Phase-III of the experiment plans to initially use molecular tritium followed by atomic tritium and has a projected ultimate sensitivity to the neutrino mass $m_{\beta}$ of 0.4~eV.
Phase-IV is planned to be the ultimate Project 8 mass measurement experiment, with a goal to reach an unprecedented sensitivity to $m_{\beta}$ of $<$ 0.04~eV(90\% C.L.)~\cite{Project8:2022wqh}.

As a differential spectroscopy method, CRES allows Project 8 to make simultaneous searches for both active and sterile masses across the entire analysis window.
Furthermore, the benefits of the CRES technique for neutrino mass measurement---namely low backgrounds, good resolution, and high event rates---also apply to a search for sterile neutrinos.
Hence, a superior sensitivity to the direct neutrino mass also provides superior sensitivity to a sterile neutrino.

\begin{figure}[ht!]
    \centering
    \includegraphics[width=0.7\textwidth]{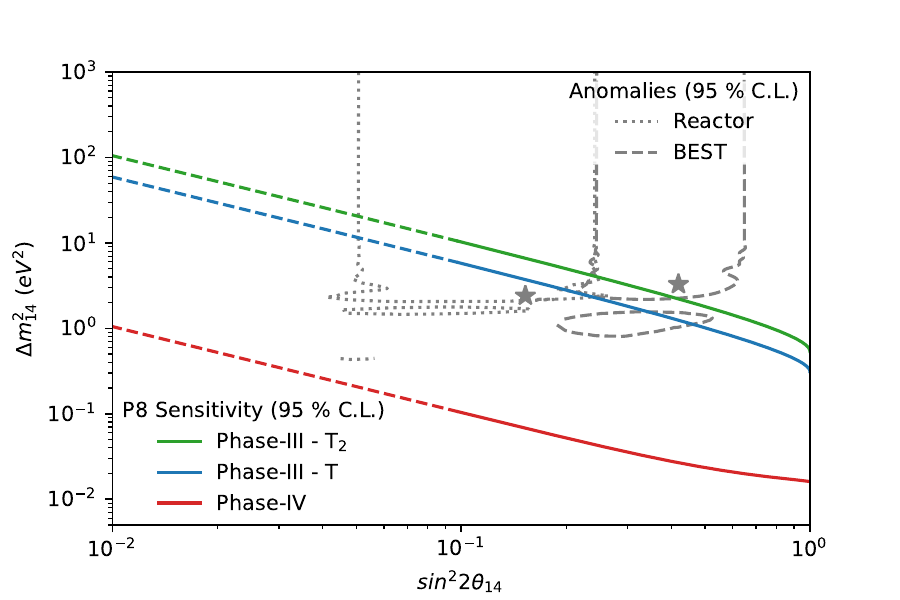}
    \caption{Sensitivity of the upcoming phases of Project 8 experiment to light sterile neutrinos in the 3+1 framework. All curves, including reactor and BEST suggested parameters, are shown at 95 \% C.L. The sensitivity is statistically limited and includes the current best knowledge of the systematics from energy resolution arising from thermal Doppler broadening, frequency to energy conversion, and variations in magnetic field. Control of systematic uncertainties leading to sensitivities in $\sin^2(2\theta) > 0.1 $ (shown as a solid line) is expected to be straightforward. Further careful systematic control could enable experimental sensitivities down to $\sin^2(2\theta) \sim 0.01$ (shown as dashed lines). In the near future, with Phase-III, Project 8 aims to reach down to $\Delta m^2_{14}\sim$ eV$^2$ (C.L. 95\%) and cover major portions of the reactor~\cite{Abazajian:2012ys} and BEST~\cite{Barinov:2021asz} gallium anomaly suggested parameter spaces including BEST best-fit point. Phase-IV of Project 8 aims to completely cover the reactor and BEST gallium anomalies at high significance. Figure from~\cite{Project8:2022wqh}.}
    \label{fig:future:P8_sensitivity}
\end{figure}

The sensitivity of Project 8 to sterile neutrinos is determined using the analytical neutrino mass sensitivity method as suggested in Ref.~\cite{Formaggio:2021nfz}.
Figure~\ref{fig:future:P8_sensitivity} shows the upcoming phases of Project 8 to be capable of a competitive sterile neutrino search over several orders of magnitude in $\Delta m^2_{14}$.
The sensitivity estimates include statistical and systematic uncertainties as noted in the caption. 
The experiment's sensitivity to higher $\Delta m^2_{14}$ is primarily limited by the efficiency of the cyclotron frequency detection method for lower energy electrons since a higher value of $m_4$ would manifest as a kink at a lower energy in the $\beta$-decay spectrum.
In generating sensitivity curves in Fig.~\ref{fig:future:P8_sensitivity}, the efficiency was assumed to be well-understood for energies tens of eV below the endpoint.
This is a fair assumption based on Project 8's ability to quantify the efficiency of the complex Phase-II detector over 2.5~keV below the endpoint~\cite{Project8:2022hun}.
The two different detection methods being investigated for upcoming phases are yet to demonstrate the control of efficiency over the wide energy range, but are expected to have lower complexity in efficiency than in Phase-II.
Project 8 through the differential $\beta$-decay spectrum measurement using CRES thus provides a promising avenue to search for light sterile neutrinos in the near future.

\subsubsection{DUNE}

Large liquid argon detectors making use of atmospheric argon, associated with $\mathrm{{}^{39}Ar}$ beta decay activity of roughly \SI{1}{Bq/kg}~\cite{WARP:2006nsa}, will enable the probing of $m_{4}$ values below the $\mathrm{{}^{39}Ar}$ beta decay end point of $Q = \SI{565}{keV}$.  By utilizing very large detectors with large volumes of liquid argon, a substantial amount of beta decays can be detected as to enable sensitive measurements of $|U_{e4}|^2$ at larger $m_{4}$ values than for $\mathrm{{}^{3}H}$.  Given that liquid argon detectors (such as liquid argon time projection chambers, or LArTPCs) function as total absorption calorimeters, the liquid argon provides both the source and the detector for such measurements, in contrast to measurements using $\mathrm{{}^{3}H}$.

\begin{figure}[tb]
  \centering
  \includegraphics[width=.6\textwidth]{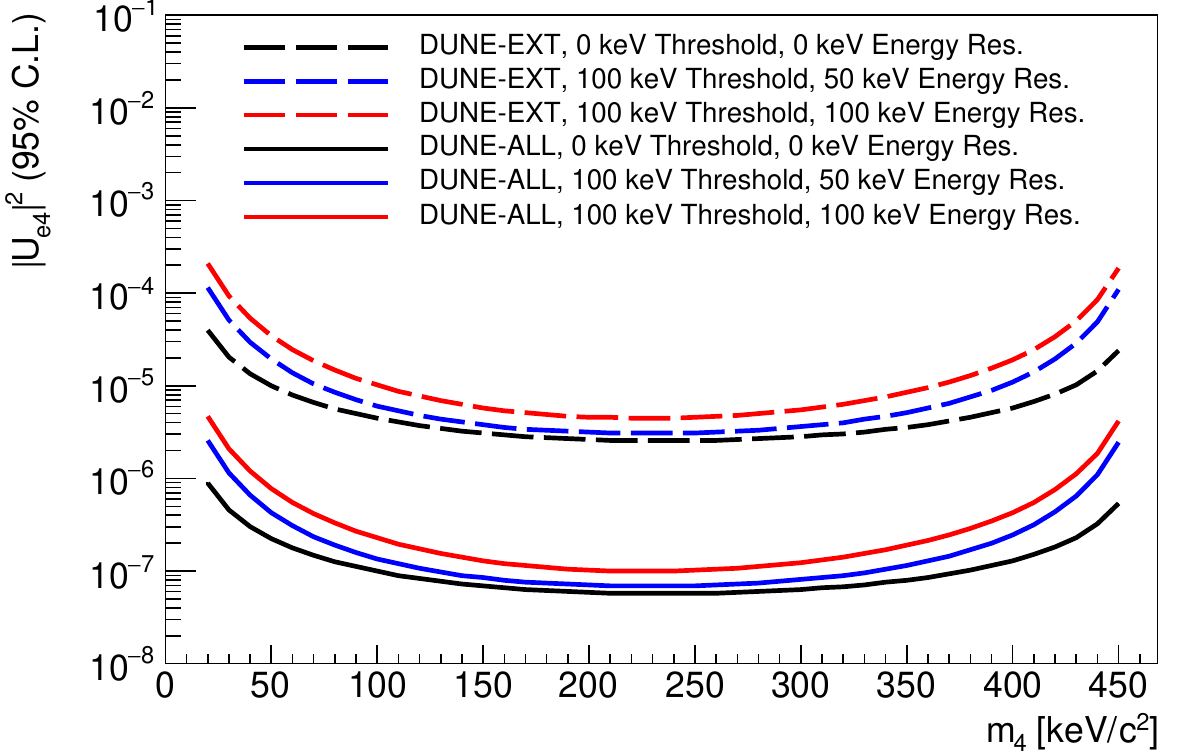}
  \caption{Projected upper limits (95\% C.L.) for $\sin^{2}\theta = |U_{e4}|^2$ as a function of heavy neutral lepton mass $m_{\mathrm{heavy}} = m_{4}$ for DUNE. ``DUNE-ALL'' refers to a limiting dataset containing every single $^{39}$Ar beta decay occurring during detector operations ($2 \times 10^{16}$ electrons), while ``DUNE-EXT'' refers to a baseline dataset obtained using external triggers and full detector readout without special localized triggering for low-energy activity in the detector ($1 \times 10^{13}$ electrons).  All upper limits shown are purely statistical (no theoretical/experimental systematic uncertainties included) and assume zero background.}
\label{fig:future:DUNE_39Ar}
\end{figure}

DUNE~\cite{DUNE:2020lwj,DUNE:2020mra,DUNE:2020txw} will use massive LArTPCs to study accelerator neutrinos ($\SI{\sim1}{GeV}$) undergoing flavor oscillations over a long baseline ($\SI{\sim1300}{km}$) in order to probe leptonic CP violation.
A search for heavy neutral leptons can also be carried out using ionization charge measurements of $\mathrm{{}^{39}Ar}$ beta decays in the DUNE far detector, enabling sensitivity to $|U_{e4}|^2$ in the \SIrange{20}{450}{keV} mass range.
Figure~\ref{fig:future:DUNE_39Ar} shows projected statistical upper limits (95\% C.L.) for $|U_{e4}|^2$ at DUNE as a function of $m_{4}$.
Recording the full DUNE $\mathrm{{}^{39}Ar}$ beta decay dataset toward maximum sensitivity to $|U_{e4}|^2$ (better than $10^{-6}$) requires substantial development of the trigger system at DUNE, including selectively recording low-energy activity within ``regions of interest'' inside of the detector; this capability is currently under development.
Reconstruction of $\mathrm{{}^{39}Ar}$ beta decays has been previously carried out at MicroBooNE~\cite{MicroBooNE:2016pwy}, a LArTPC neutrino experiment at Fermilab, demonstrating that low thresholds (roughly \SI{100}{keV}) and good energy resolution from low TPC noise levels (roughly \SI{50}{keV}) are achievable in large LArTPC detectors~\cite{MicroBooNE:2018jag}; more comprehensive studies at ProtoDUNE-SP~\cite{DUNE:2017pqt,DUNE:2020cqd} are currently in progress.
Further studies on the impact of $\mathrm{{}^{39}Ar}$ beta decay spectrum theoretical uncertainties~\cite{Kostensalo:2017raq}, experimental systematic uncertainties, and radiological backgrounds in the DUNE far detector are also in progress.
While these additional considerations may lower the sensitivity to $|U_{e4}|^2$ considerably, given the currrent global limits on $|U_{e4}|^2$ of $10^{-2}$ to $10^{-4}$ in the relevant range of $m_{4}$ values~\cite{Dragoun:2015oja,deGouvea:2015euy,Bolton:2019pcu}, significant improvement is expected.

\subsubsection{HUNTER}

\begin{figure}[ht!]
    \centering
    \includegraphics[width=0.38\textwidth]{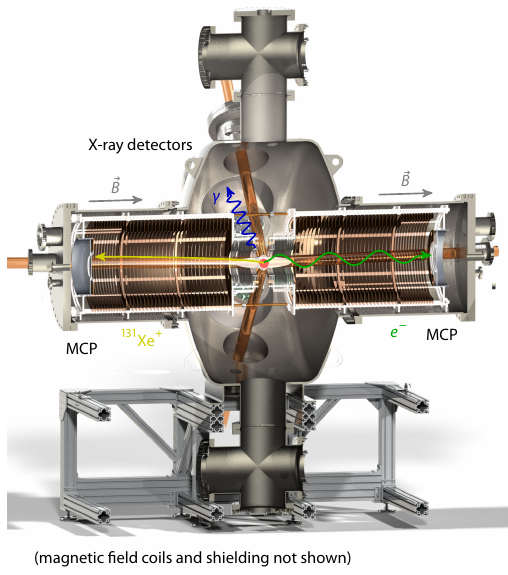}
    \caption{\label{fig:future:HUNTER_spec}The HUNTER spectrometer.  Figure from~\cite{Martoff:2021vxp}.}
\end{figure}

The HUNTER Collaboration brings together techniques from high energy, nuclear, and AMO physics to perform a laboratory sterile neutrino search aimed at the mass range 10-300~keV~\cite{Martoff:2021vxp}.
It brings together groups from Temple University, UCLA, University of Houston, Racah Institute of Physics, and Princeton University to pursue this physics.

The neutrinos are emitted in the electron capture (EC) decay of $^{131}$Cs  atoms captured in a magneto-optical trap and cooled to milliKelvin temperatures where they are essentially at rest.  
In electron capture, the initial final state is an excited atom recoiling against the emitted neutrino -- for a given neutrino mass, the neutrino is mono-energetic.  All the observable particles in the EC decay and subsequent de-excitation of the recoiling atom are reconstructed.  These particles are a $^{131}$Xe ion, an X-ray, and one or more Auger electrons, each of which requires different measurement techniques.
Figure~\ref{fig:future:HUNTER_spec} shows a CAD drawing of the HUNTER  apparatus.  
The meter-long horizontal arms of the vessel contain spectrometers for the ion (left) and electrons (right).  In the ion spectrometer, the electrostatic field focuses the ions onto an MCP which allows determination of the ion's vector momentum using the impact location and time of flight.  The electron spectrometer is similar, but requires an axial magnetic field to contain the electrons.  X-rays are detected via scintillator panels, a signal that triggers the readout and provides the start time for the time of flight measurements. 

\begin{figure}[ht!]
    \centering
    \vspace{-0.1in}
    \includegraphics[width=0.35\textwidth]{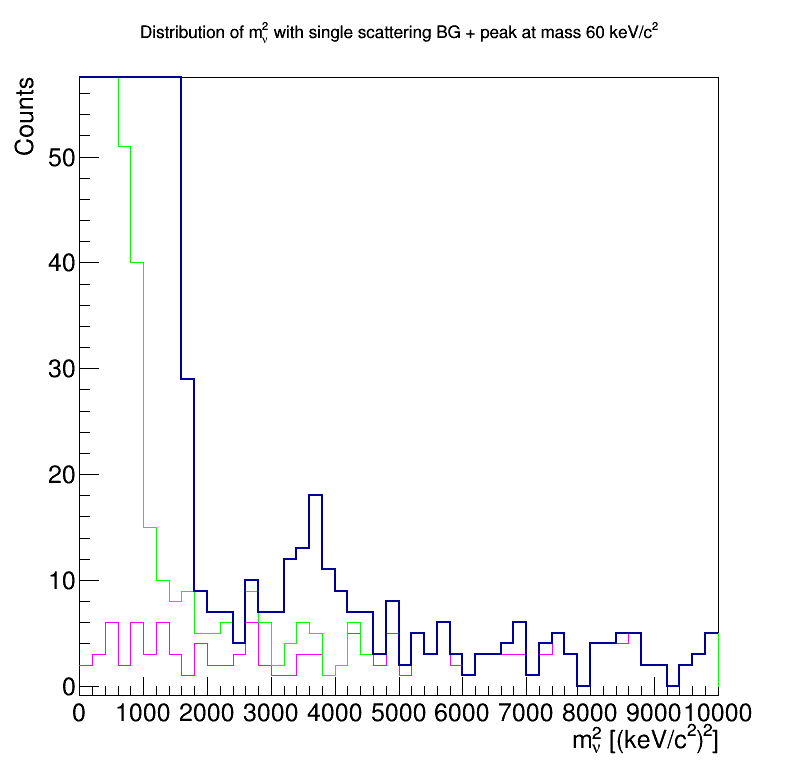}
    \includegraphics[width=0.45\textwidth]{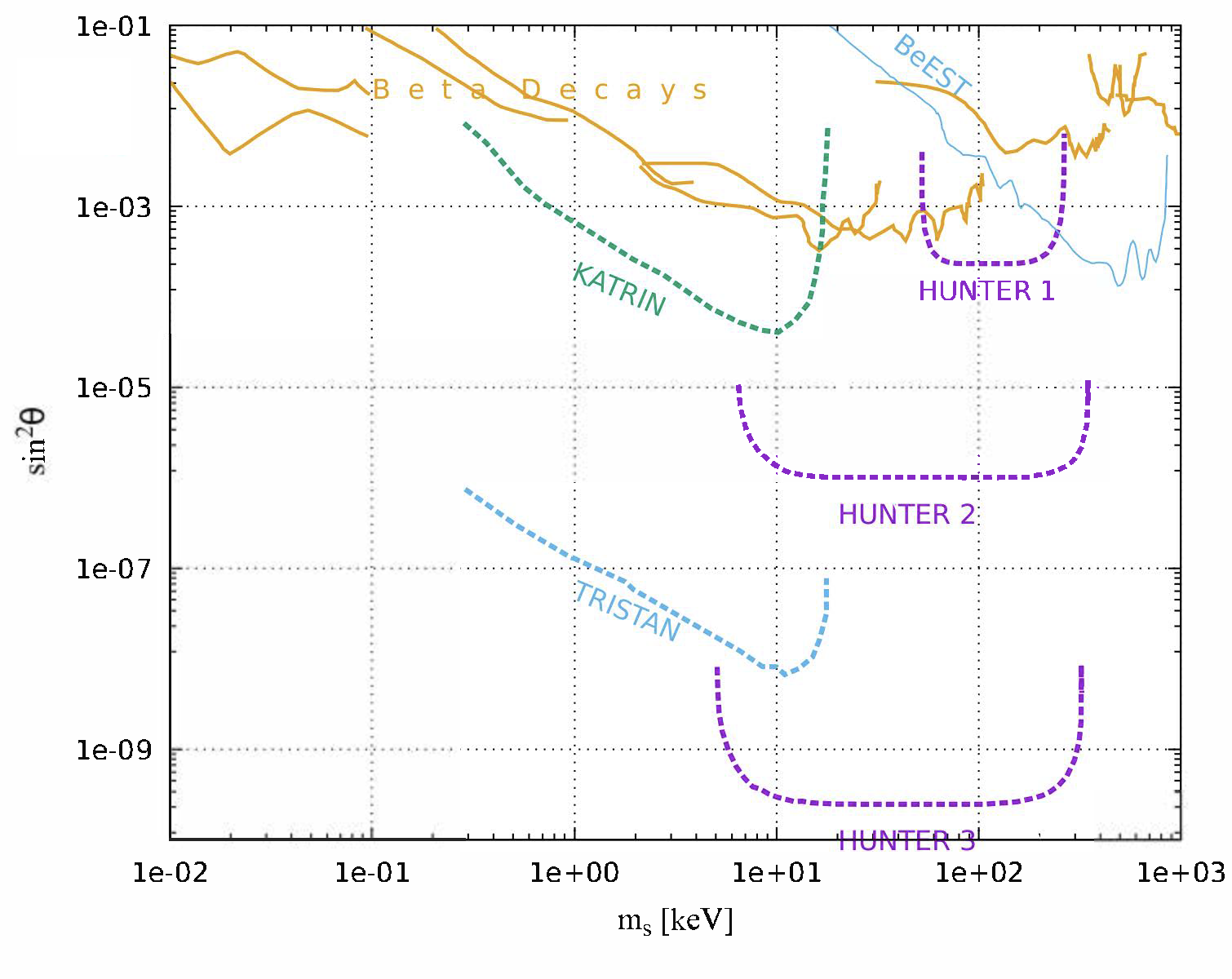}
    \caption{\label{fig:future:HUNTER_lim} Left: simulated spectrum (blue histogram) of one year of livetime including a hypothetical 60~keV/$c^2$ sterile neutrino with $\sin^2\theta=3\times 10^{-4}$.  Right: HUNTER's sensitivity for three phases of the experiment, compared to other laboratory limits.  Figures adapted from~\cite{Martoff:2021vxp}.}
\end{figure}

The left panel of Fig.~\ref{fig:future:HUNTER_lim} is a simulated spectrum including the apparatus resolution and the major backgrounds.
A simulated neutrino signal within the reach of the first phase of HUNTER with a year of live-time is shown.  The right panel shows the reach of the initial HUNTER configuration and two potential upgrades.

\subsubsection{BeEST}
\label{sec:future:BeEST}

The BeEST (Beryllium Electron-capture with Superconducting Tunnel junctions) experiment is a model-independent search for sub-MeV sterile neutrinos using implanted radioactive $^7$Be atoms into high-rate superconducting quantum sensors~\cite{Leach:2021bvh}.
The BeEST concept uses momentum reconstruction of the low-energy $^7$Li daughter atom following EC decay of $^7$Be to infer direct mass information of the neutrino.
$^7$Be ($T_{1/2}=53.22(6)$~days~\cite{Tilley:2002vg}) is a pure EC decaying isotope, and the ideal case for neutrino studies via momentum reconstruction due to its large decay energy $Q_{EC}=861.89(7)$~keV~\cite{Huang:2021nwk}, relatively high maximum value for the recoil kinetic energy $T_{D(max)}=56.826(9)$~eV, and simple atomic and nuclear structure~\cite{Tilley:2002vg}.
Due to the explicit neutrino mass dependence on the recoil kinetic energy, the existence of a heavy mass state, $m_i$, would cause the nuclear recoils to have a lower kinetic energy, and the relative fraction of these shifted events to the total determined by the mixing with the $\nu_e$ flavor, $|U_{ei}|^2$.

The BeEST experiment uses Superconducting Tunnel Junction (STJ) sensors to detect the low-energy radiation following EC decay.
STJs are high-speed quantum sensors that were originally developed for high-resolution X-ray spectroscopy in astronomy and material science~\cite{Kurakado:1982}.
STJs are a type of Josephson junction that consists of two superconducting electrodes separated by a thin insulating tunneling barrier.
The absorption of radiation in one of the electrodes breaks the Cooper pairs of the superconducting ground state and excites free excess charge carriers above the superconducting energy gap $\Delta$ in proportion to the absorbed energy.
This results in exceptionally high energy resolution ($\sim1$~eV) for low-energy radiation relevant to nuclear recoils.
Each STJ detector pixel is able to operate at rates up to $10^4$ counts/s with exceptionally high predictability on the detector response~\cite{Friedrich:2020}, making them ideal for searches of this type.

\begin{figure}[b]
    \centering
    \includegraphics[width=\linewidth]{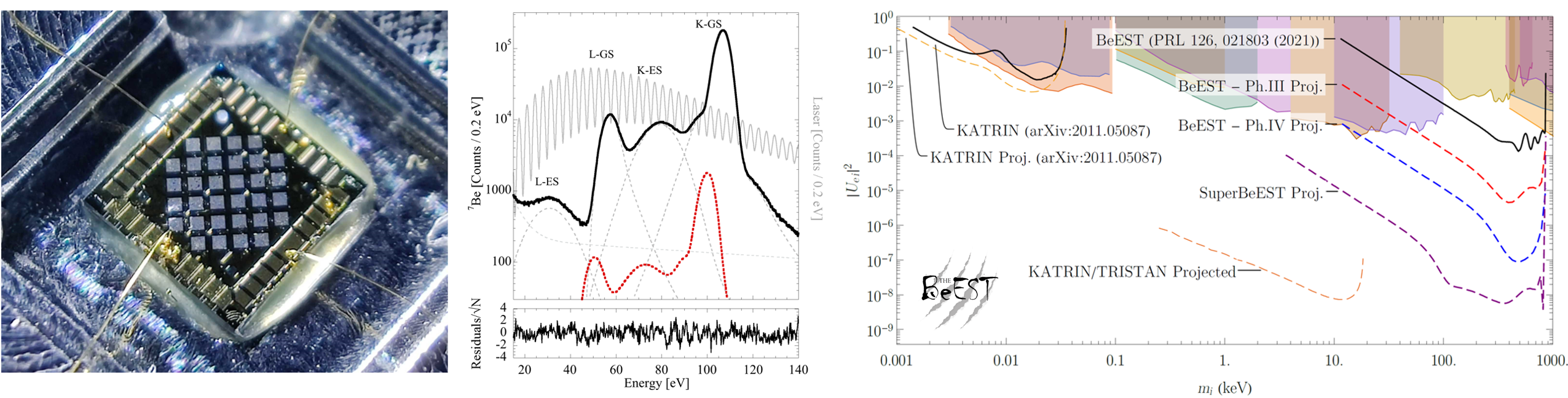}
    \caption{\label{fig:future:BeEST} (left) Image of a 32-pixel Al-based STJ array in preparation for Phase-IV of the BeEST~\cite{Leach:2021bvh}.  (middle) The low-energy $^7$Li nuclear recoil spectrum for 28 days of acquisition from a single pixel counting at low-rate~\cite{Friedrich:2020nze}.  The spectrum generated by a hypothetical 300 keV sterile neutrino signal with 1\% mixing is shown in red (dashed). (right) Current and projected limits for the BeEST and SuperBeEST experiments compared to existing laboratory decay limits (shaded areas) and the TRISTAN projection.}
\end{figure}

The $^7$Be$^+$ rare isotopes are implanted into the STJ detectors through Si collimators at TRIUMF's Isotope Separator and ACcelerator (ISAC) facility in Vancouver, Canada at an energy of 25~keV.
The $^7$Be$^+$ beam is produced using the isotope separation on-line technique via spallation reactions from a 10~$\mu$A, 480-MeV proton beam incident on a stack of thin uranium carbide targets.
Following implantation of the rare isotope, The $^7$Li recoil spectrum from the decay of $^7$Be is measured at Lawrence Livermore National Laboratory (LLNL) with the STJ detector at a temperature of $\sim$0.1 K in a two-stage adiabatic demagnetization refrigerator (ADR) and the signals are read out with a specialized current-sensitive preamplifier~\cite{Warburton:2015}.
For a precision energy calibration, the STJs are simultaneously exposed to 3.49965(15)~eV photons from a pulsed Nd:YVO$_4$ laser triggered at a rate of 100~Hz~\cite{Ponce:2018mnc,Friedrich:2020nze}.
The measured $^7$Li recoil spectrum contains four peaks that result from the two nuclear and two atomic processes following EC decay.
These include one for $K$-capture decay to the nuclear ground state (K-GS), one to the excited state of $^7$Li (K-ES), and two for the corresponding $L$-capture decays (L-GS and L-ES, respectively) (Fig.~\ref{fig:future:BeEST}).

Due to the relative simplicity of the $^7$Be$\rightarrow^7$Li EC decay system, the spectral response is able to be precisely evaluated using modern theoretical tools.
The atomic de-excitation and auto-ionization (electron shake-up and -off) effects that follow EC decay generate higher-order features in the spectrum and are accessible to ab-initio atomic theory.
Additionally, the sequestration of beryllium in the sensor material generates small (eV-scale) chemical shifts in the EC decay energies.
To this end, density functional theory is used to model the electronic structure of beryllium and lithium atoms in different atomic environments of a polycrystalline absorber film of the sensor to provide a further improvement of sensitivity to new physics.
This work is already in an advanced stage, and will be employed to provide improved limits to the high-statistics data from the BeEST experiment planned for upcoming phases, and may allow for the extraction of significantly improved limits in the $10-100$~keV mass range.

The BeEST experiment has completed its proof-of-concept (Phase-I) in 2020 with the first demonstration of high-resolution nuclear recoil detection with STJs~\cite{Fretwell:2020ntq}.
Following this, the first low-activity (10 Bq) data set using a single tantalum-based STJ detector was taken for 28 days - including the use of precision in-situ laser calibration of the sensor response.
The resulting Phase-II data allowed for the first extraction of an upper limit for $|U_{ei}|^2$ from the BeEST experiment in the 100--850~keV mass range~\cite{Friedrich:2020nze}, and improved upon previous decay measurements by up to an order of magnitude (Fig.~\ref{fig:future:BeEST}).
Within the next 5 years, the BeEST will complete its four phases of the experiment and achieve limits approaching $|U_{ei}|^2\approx10^{-7}$ in the few-hundred keV mass range, as described further in Ref.~\cite{Leach:2021bvh}.
Following Phase-IV of the BeEST, a dramatic improvement in sensitivity is planned using large arrays ($10^4$ pixels) of STJs with new materials.
This development is ongoing, with the goal to probe relative couplings to the electron neutrino flavor of $|U_{ei}|^2\leq10^{-9}$.
The current and projected limits for the BeEST experiment are presented in Fig.~\ref{fig:future:BeEST} and incorporate known and simulated detector responses under conservative assumptions on the atomic and materials theory work described above.

\subsubsection{PTOLEMY}

The PTOLEMY experiment is designed to look for relic neutrinos by means of neutrino capture on a tritium target.
Searching for spectral distortions near the endpoint with its very intense source, the experiment also projects sensitivity to eV-scale sterile neutrinos~\cite{PTOLEMY:2019hkd}.

\subsubsection{Double Beta Decay}

While nuclear beta-decay has become a valuable tool for searching for MeV-scale sterile neutrinos, the relatively rare double beta-decay process is also sensitive to the presence of sterile neutrinos.
Often thought of solely as a background to searches for neutrinoless double-beta decay ($0\nu\beta\beta$), double beta decay occurs when an atom undergoes a simultaneous emission of two electrons and two antineutrinos.
This process leads to a continuous electron energy spectrum that cuts off at the characteristic Q-value for that isotope. The region of this spectrum that is typically studied in detail is the end-point because this is where the signal of neutrinoless double-beta decay will populate.
As discussed throughout this section, the presence of heavy sterile neutrinos will lead to deformation of the beta-decay spectra, and this is also true for double beta-decay~\cite{Bolton:2020ncv}.

\begin{figure}[ht!]
    \centering
    \includegraphics[width=0.5\textwidth]{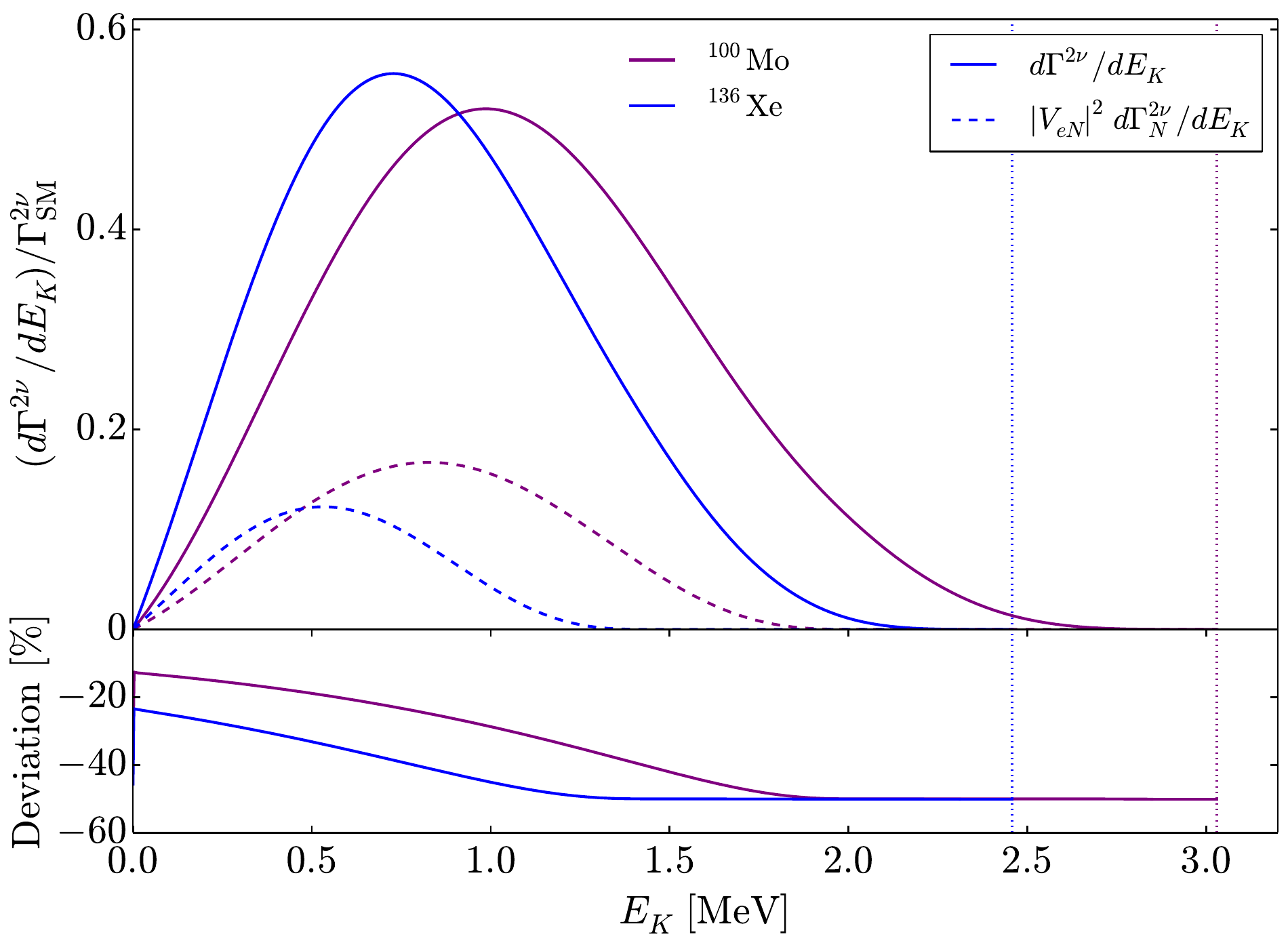}
    \caption{The summed energy distribution from the standard $2\nu\beta\beta$ spectrum (solid) and the impact from the presence of a 1 MeV sterile neutrino (dashed) for $^{100}$Mo (purple) and $^{136}$Xe (blue). Figure from~\cite{Bolton:2020ncv}.}
    \label{fig:future:doublebeta_spectra}
\end{figure}

In the presence of a heavy sterile neutrino, the kinematics of standard two-neutrino double beta decay is modified in both the energy and angular distribution.
The modifications to the summed electron energy is spread across a relatively large range of energies but lead to an overall modulation of the spectrum at low energies, as shown in Fig.~\ref{fig:future:doublebeta_spectra}.
In addition to the impact on the energy of the electrons, the angular distribution is also strongly impacted by the presence of heavy sterile neutrinos.

\begin{figure}[ht!]
    \centering
    \includegraphics[width=0.5\textwidth]{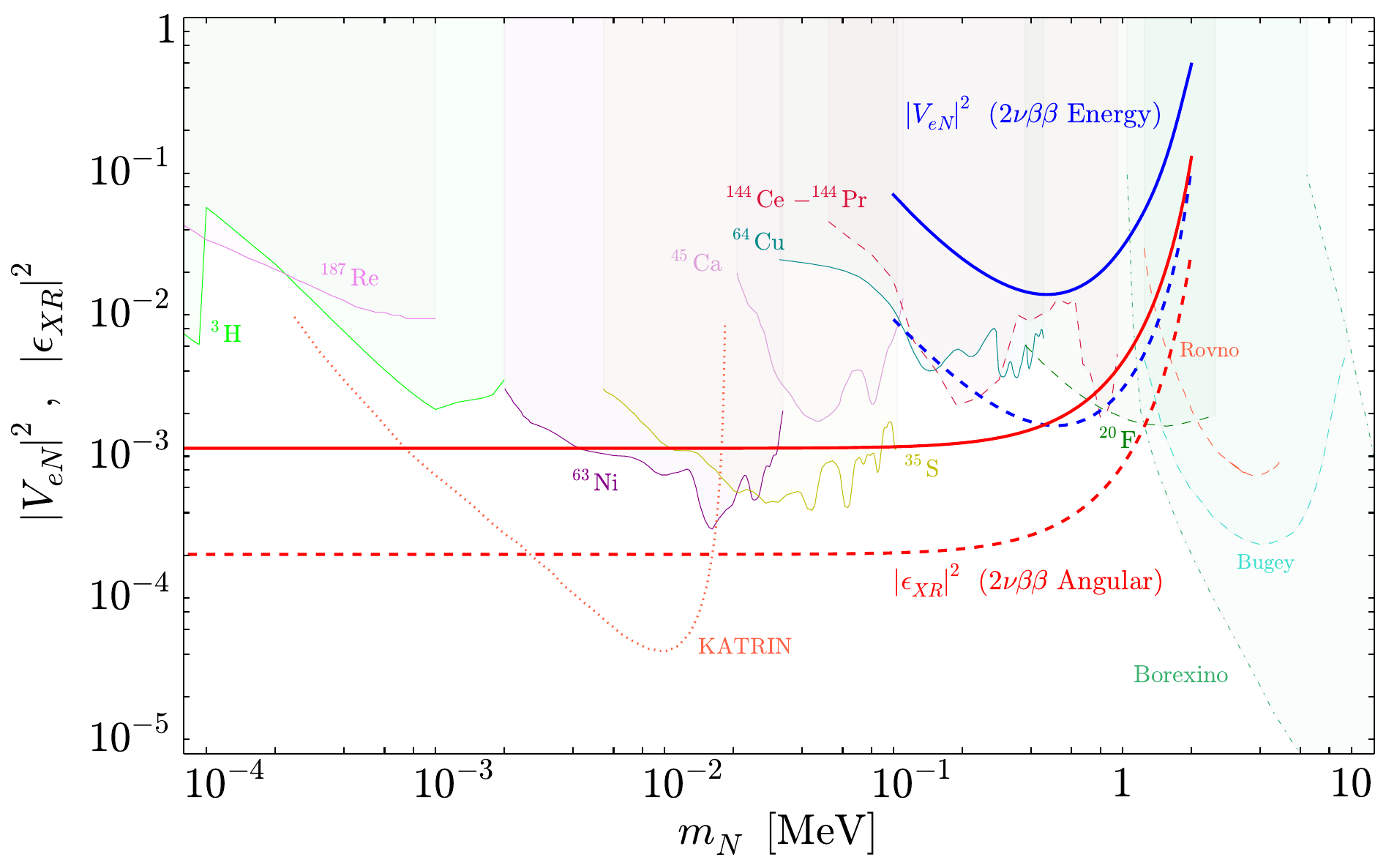}
    \caption{The constraint offered by current $0\nu\beta\beta$ experiments sensitive to only the energy (solid blue) and those also sensitive to the angular distributions (solid red). Future (next-generation) experimental sensitivity are shown in dashed lines. Figure from~\cite{Bolton:2020ncv}.}
    \label{fig:future:doublebeta_comparison}
\end{figure}

Constraints from current and planned neutrinoless double beta decay experiments have been explored in Ref~\cite{Bolton:2020ncv}. These can be separated into experiments sensitive to just energy that have taken data (CUORE~\cite{CUORE:2021mvw}, EXO-200~\cite{EXO:2017poz}, GERDA~\cite{GERDA:2020xhi}, KamLAND-Zen~\cite{KamLAND-Zen:2022tow}, and the \textsc{Majorana Demonstrator}~\cite{Majorana:2022udl}) and planned to take data (CUPID~\cite{CUPID:2019imh}, LEGEND~\cite{LEGEND:2021bnm}, and nEXO~\cite{nEXO:2021ujk}) and those sensitive to the angular distributions that have taken data (NEMO-3~\cite{NEMO-3:2016qxo}) and those planned (SuperNEMO~\cite{SuperNEMO:2010wnd}). These can be seen in Fig.~\ref{fig:future:doublebeta_comparison}, where limits are placed on the sterile neutrino mass and the mixing strength. In Fig.~\ref{fig:future:doublebeta_comparison}, limits coming from current neutrinoless double beta decay experiments use exposures of O(100) kg-years, which corresponds to $10^4$~$2\nu\beta\beta$ events, while future next-generation experiments (those targeting the inverted hierarchy region of the neutrinoless double-beta decay phase space) use exposures of O($10^3 - 10^4$) kg-years, with $10^6$~$2\nu\beta\beta$ events. In the far future, experiments that wish to target the normal hierarchy region of the neutrinoless double beta decay will require exposures of O($10^6$) kg-years~\cite{Biller:2013wua}, this increase in the number of events will enable further explorations of this sterile neutrino phase-space. 

Leveraging data from $0\nu\beta\beta$ experiments to search for the presence of sterile neutrinos and other BSM processes~\cite{Cepedello:2018zvr} demonstrate the power diverse experimental data can offer for exploring and searching for anomalies.

\subsection{Meson factories}
\subsubsection{NA62}

Rare meson decays are well-known for providing some of the strongest constraints on light dark sectors. Two experimental efforts to measure the flavor-changing NC decay of kaons to neutrinos, $K^+ \to \pi^+ \nu\overline{\nu}$, are underway: NA62, with a $K^+$ beam~\cite{NA62:2018ctf,NA62:2020fhy}, and KOTO, with a $K_L$ beam~\cite{KOTO:2018dsc}. In this section we discuss some prospects of NA62 to 

The NA62 experiment is a kaon-decay-in-flight experiment located at CERN~\cite{NA62:2017rwk}. To achieve its primary goal, NA62 aims to have a total exposure of $\mathcal{O}(10^{13})$ kaon decays over a few years of run time. The detectors are designed for sub-percent momentum resolution on the kaon and final charged-track momenta. Kaons have an average of $75$~GeV momentum and their decays in flight throughout a distance of $\mathcal{O}(65)$~m are detected by a combination of spectrometers arranged in a cylindrical fashion around the beam. 

NA62 can also perform several searches for new physics by searching for exotic kaon decays. Light particles have been searched for in $K^+ \to \ell^+ N_{\rm inv}$~\cite{NA62:2017qcd,NA62:2017ynf,NA62:2020mcv}, as well as in $\pi^0 \to (\gamma) +\text{inv}$~\cite{NA62:2019meo,NA62:2020pwi}. In the context of short-baseline anomalies, NA62 has a world-leading sensitivity to HNLs invoked in some explanations of the MiniBooNE and LSND anomalies. The current limits on long-lived or invisible HNLs produced in $K_{2\ell}$ decays span the mass region between $100 \lesssim m_N < m_K - m_{\ell}$.

Two channels are of particular interest for HNL searches: 
\begin{itemize} 
    \item $K^+ \to \ell^+ N$, where $N$ is long-lived or invisible,
    \item $K^+ \to \ell^+ (N \to \nu \ell^+ \ell^-)$, where $N$ decays inside the detector.
\end{itemize}
The first channel has already been searched for in both flavors, $\ell^+ = e^+$ and $\mu^+$~\cite{NA62:2017qcd,NA62:2017ynf,NA62:2020mcv}. The limits reach mixing angles as low as $\mathcal{O}(10^{-9})$, in the region of interest for the Type-I seesaw mass mechanism. The strategy in this case consists in searching for the presence of an invisible resonance in $M_{\text{inv}}^2 = (p_K - p_\ell)^2$, therefore it is mostly sensitive to long-lived or invisibly-decaying HNLs. This constraint applies to the models discussed in Sec.~\ref{subsubsec:LongLivedHNLs}.

\begin{figure}
    \centering
    \includegraphics[clip, trim=0.1cm 0.3cm 0.1cm 0.1cm, width=0.49\textwidth]{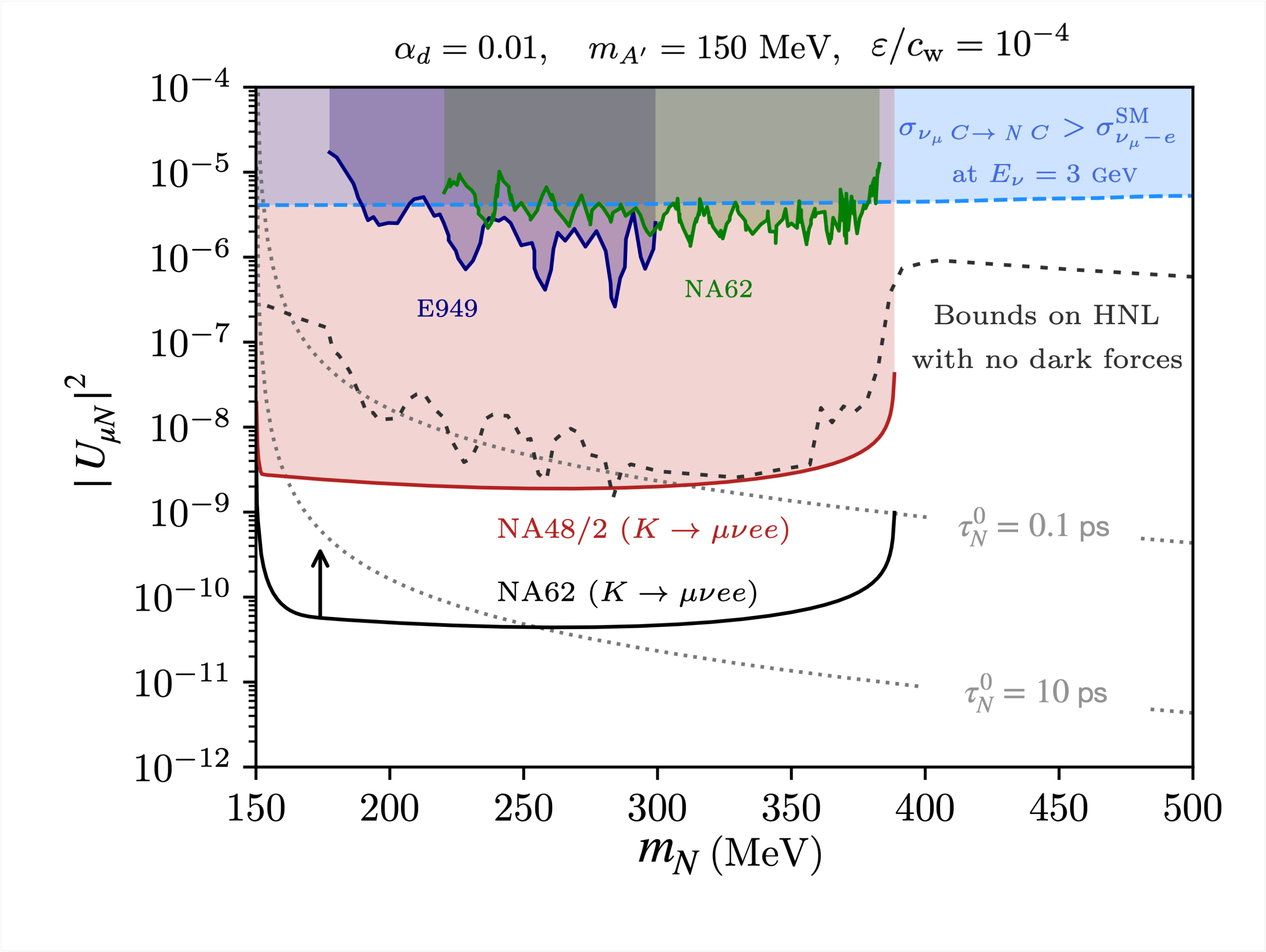}
    \includegraphics[clip, trim=0.1cm 0.1cm 0.1cm 0.2cm, width=0.49\textwidth]{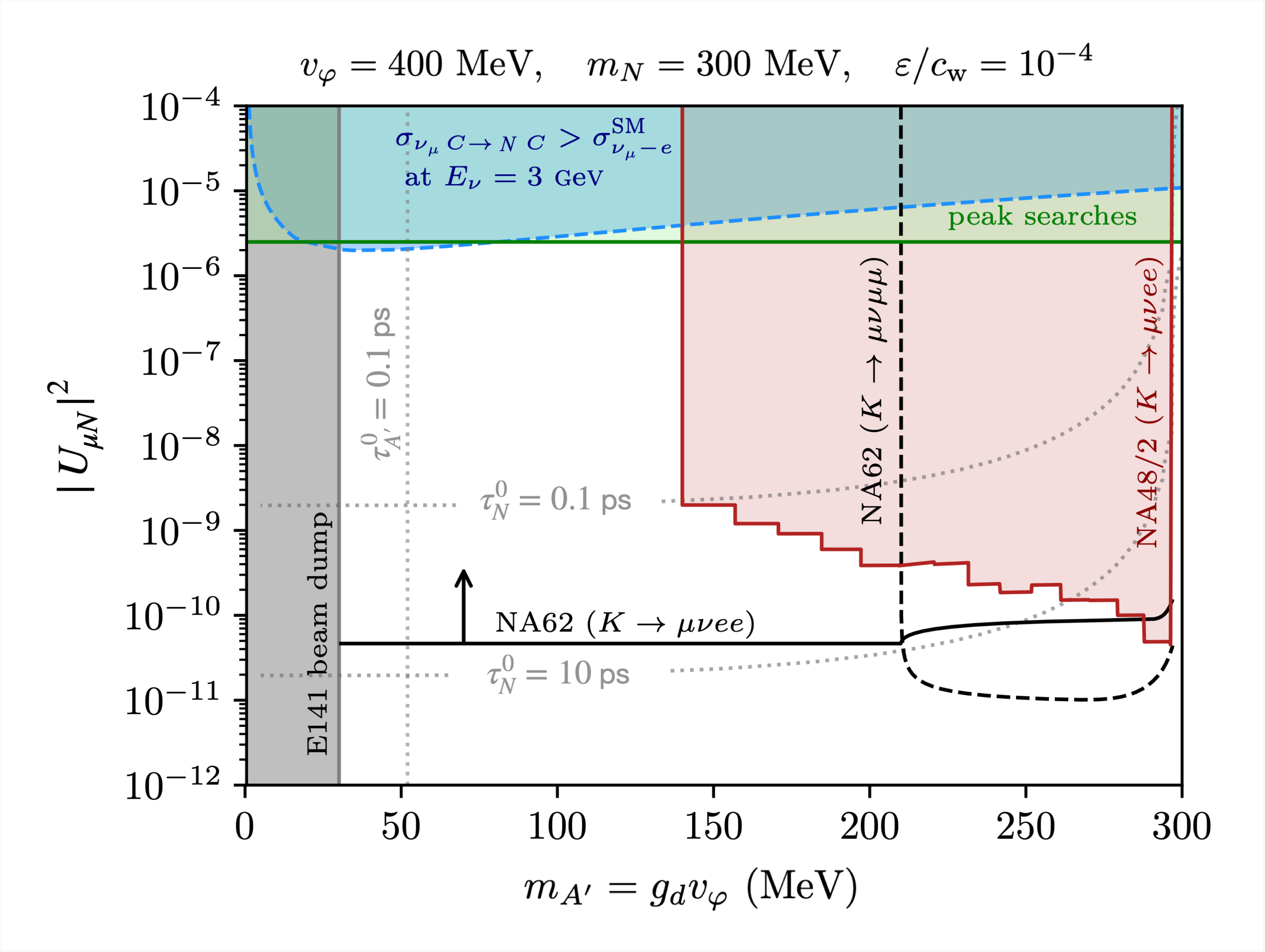}
    \caption{Limits on $|U_{\mu}|^2$ as a function of the mass of the HNL $N$ (left) and the dark photon $A^\prime$ mass (right). The NA62 sensitivity to the prompt decays of HNLs in $K^+\to \mu^+ \nu e^+e^-$ and $K^+\to \mu^+ \nu \mu^+\mu^-$ searches is shown as a solid black line. Figure from and based on the dark sector model of~\cite{Ballett:2019pyw}\label{fig:dark_HNL_ps}.}
\end{figure}

The second channel with three charged tracks is instead sensitive to short-lived HNLs. This possibility was first discussed in the context of a transition magnetic moment model~\cite{Dib:2011jh}, discussed in Sec.~\ref{subsubsec:TransitionMagneticMoment}. In that model, the electromagnetic decays of the HNL with an off-shell photon, $N \to \nu (\gamma^* \to e^+e^-)$, could dominate the SM rate. The measurement by BNL-E865~\cite{Poblaguev:2002ug}, $\mathcal{B}(K^+ \to \mu^+\nu e^+e^-) = ( 7.06 \pm 0.3)\times 10^{-8}$ allowed to set a constraint on the mixing and branching ratio of the HNL of $|U_{\mu N}|^2 \times \mathcal{B}(N\to \nu \gamma) < 0.5\times 10^{-6}$~\cite{Dib:2011jh}. This constraint, however, only applies to HNLs with masses $m_N > 145$~MeV due to experimental cuts on  $m_{ee} > 145$~MeV, designed to suppress $\pi^0_D$ decays. A more precise measurement was performed at NA48, finding
$\mathcal{B}(K^+ \to \mu^+\nu e^+e^-) = (7.81\pm 0.23)\times 10^{-8}$, applicable to the region $m_{ee} > 140$~MeV.

Three charged track decays are also sensitive to HNLs that decay electromagnetically through a dark photon, $Z^\prime$, such as in the models discussed in Sec.~\ref{subsubsec:DarkNeutrinos}. In this case, the HNLs can decay to either $\nu e^+e^-$ or $\nu \mu^+\mu^-$. Ref.~\cite{Ballett:2019pyw} derived constraints in a specific dark neutrino model with pseudo-Dirac HNLs. The reach of NA48 to the $|U_{\mu 4}|^2$ mixing element was around $\mathcal{O}(10^{-8})$ or better. A dedicated search at NA62 can improve upon that reach by one to two orders of magnitude. Backgrounds from $\pi^0_D$ decay present the biggest challenge for this search, but can be suppressed by the requirement that the measured $M_{\nu ee}^2 \equiv (p_K - p_\ell)^2 = m_N^2$. In addition, if the mediator is produced on-shell, then a resonance search in the $\ell^+\ell^-$ invariant mass can be performed. In the case of a signal, NA62 could measure both the dark photon and the HNL mass. The NA62 sensitivity to the dark sector model in Ref.~\cite{Ballett:2019pyw} is shown in Fig.~\ref{fig:dark_HNL_ps} under the assumption of no backgrounds. Based on Ref.~\cite{CortinaGil:2019dnd}, for $K^+\to \mu^+\nu e^+ e^-$, a total of $N_K^{\rm Fid} = 2.14\times10^{11}$ fiducial kaon decays were assumed with an acceptance of ${\rm A}_\beta = 4\%$., while for $K^+\to \mu^+\nu \mu^+ \mu^-$, a total of $N_K^{\rm Fid} = 7.94\times 10^{11}$ and ${\rm A}_\beta = 10\%$ were assumed. 

\subsection{Collider Experiments}

\subsubsection{FASER$\nu$ and FLArE}
The flux of broadband neutrinos from the LHC with energies around $\sim100$~GeV to $\sim1$ TeV in the forward direction provides a new opportunity to look for neutrino oscillations.
Existing experiments such as FASER$\nu$ \cite{FASER:2019dxq,FASER:2019dxq} and SND@LHC \cite{SHiP:2020sos} as well as proposed experiments which go under the umbrella term of Forward Physics Facility (FPF) will be sensitive to sterile neutrino oscillations.
Given a typical baseline of 600 m and typical energies in the 100~GeV to 1~TeV range, this corresponds to sensitivities to $\Delta m^2_{41}\sim1000$~eV$^2$, or $m_4\sim30$~eV.
While some existing constraints apply at this mass range in the oscillation averaged limit, there are no oscillation probes at this $L/E$ providing a new direct test of oscillations for larger $\Delta m^2$'s than are usually considered.
The forward physics program at the LHC also benefits from the production of all three flavors of neutrinos with hierarchical production rates that differ each by $\gtrsim1$ order of magnitude.
In addition to having all three flavors available at the source, FASER$\nu$ and SND@LHC both have flavor discrimination capabilities among the three flavors allowing for, in principle, probes of all 9 oscillation channels, subject to backgrounds and flux uncertainties.

\begin{figure}[htbp!]
\centering
\includegraphics[width=0.49\textwidth]{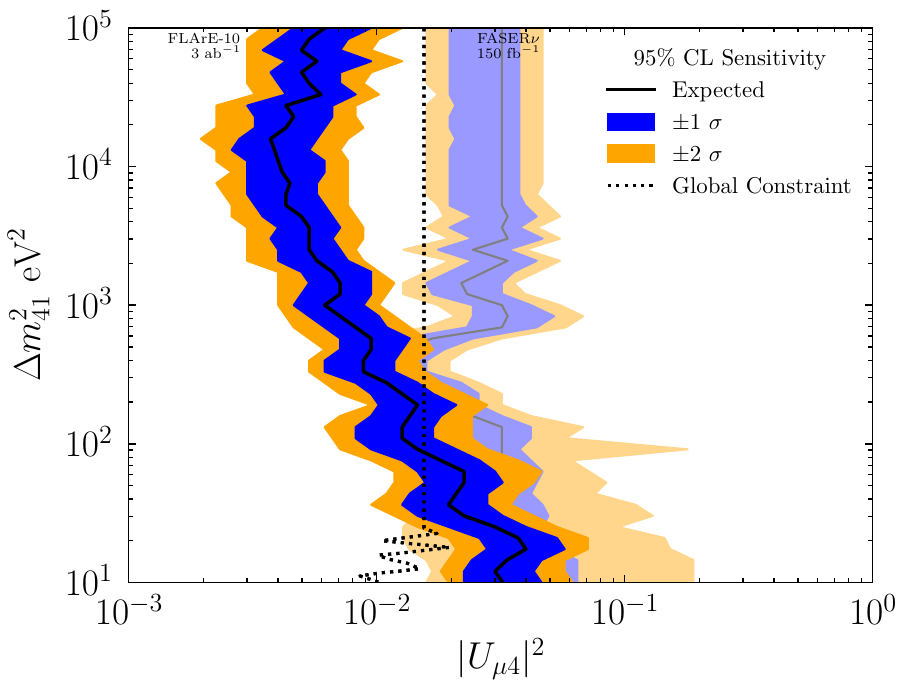}
\caption{The expected sensitivity to sterile neutrino oscillations in the $\nu_\mu$ disappearance channel at the LHC using Asimov and Feldman-Cousins for FASER$\nu$ at the upcoming LHC run and a proposed 10-ton LAr detector in a future HL-LHC run.
For comparison, the existing oscillation averaged constraints coming mostly from MINOS+ and MiniBooNE are also shown.
Figure from~\cite{Anchordoqui:2021ghd}.}
\label{fig:sterile fpf}
\end{figure}

One major challenge is the presence of significant flux uncertainties which affect the normalization and, more importantly, the shape.
The relative contribution to the neutrino flux from different particles is rather poorly understood \cite{Kling:2021gos} and these shape effects could conceivably mimic a neutrino oscillation signature \cite{Bai:2020ukz}.
Thus significantly more theoretical work to understand these fluxes is essential to use this unique neutrino environment to probe neutrino oscillations.

Nonetheless, it is possible to estimate the sensitivity to sterile neutrino oscillations at FASER$\nu$ and FLArE, a proposed LAr detector in the forward direction at the LHC \cite{Anchordoqui:2021ghd}.
The most sensitive channel relative to existing constraints is the $\nu_\mu$ disappearance channel which shows sensitivity at the $\Delta m^2_{41}\sim100-1000$ eV$^2$ range down to mixings $|U_{\mu4}|^2<10^{-2}$, better than existing constraints from \cite{Dentler:2018sju} which are dominated by MINOS/MINOS+ \cite{MINOS:2017cae} and MiniBooNE \cite{MiniBooNE:2012meu}.
The sensitivity for FASER$\nu$ in the upcoming LHC run at 150 fb$^{-1}$ and FLArE-10 (10 fiducial tons of LAr) with 3 ab$^{-1}$ in the HL-LHC is shown in Fig.~\ref{fig:sterile fpf}.
The primary uncertainty is the flux uncertainty.
This is accounted for in a fairly conservative fashion by including an estimate of the impact of shape effects by varying the flux across the range of predictions from different models \cite{Kling:2021gos} with an associated $1\sigma$ pull term.
This sensitivity is then calculated using the Feldman-Cousins procedure \cite{Feldman:1997qc} at 95\% CL including the flux systematic uncertainty.
The median sensitivity is estimated with the Asimov method.

It is anticipated that these LHC experiments could have sensitivity to sterile oscillations for the other channels as well, although those will depend on the details of the flux uncertainties which are still quite large and difficult to systematically quantify.

\clearpage






\section{The Path Forward to Resolving the Anomalies}
\label{sec:reqs}

\subsection{Reflecting on the Path of the Past Decade}

Sections~\ref{sec:expt_landscape} and \ref{sec:future} serve to demonstrate that a significant portion of the neutrino community's efforts over the past decade has been driven by the goal of establishing or refuting the existence of light sterile neutrinos motivated by these anomalies. 
To this end, following the requirements identified in~\cite{Abazajian:2012ys} has led to a broader understanding of viable interpretations of the anomalies and strengthened experimental efforts -- and experimental capabilities -- in that direction.
Notably, the requirement to probe the anomalies with multiple and orthogonal approaches (accelerator-based short/long-baseline, reactor-based short-baseline, atmospheric neutrinos, and radioactive source) in the same spirit as employed for neutrino oscillations has now been realized through recent, ongoing, or impending experimental programs: 

\begin{itemize}
    \item The development of new radioactive sources and detectors for improved tests of the Gallium Anomaly has been pursued and realized in the form of the BEST experiment.  
    \item The Reactor Antineutrino Anomaly and subsequent reactor-based activities and new results have placed a required emphasis on oscillation-testing short-baseline reactor experiments and on improved understanding of reactor neutrino fluxes.  
    \item The community has just begun a comprehensive multi-channel/multi-baseline accelerator-based short-baseline program to search for 3+$N$ oscillations while directly addressing the MiniBooNE anomaly both in regards to oscillatory and non-oscillatory solutions.  
    \item A direct test of the LSND Anomaly using an improved decay-at-rest beam facility and experimental arrangement has just begun in the form of the JSNS$^2$ experiment.  
    \item Beyond direct anomaly tests, many alternate techniques/facilities, including direct neutrino mass measurements, long-baseline oscillation experiments, and atmospheric and astrophysical neutrino experiments, have been applied to the sterile neutrino explanation of the anomalies. 
\end{itemize}

\subsection{Primary Focuses of the Next Decade}

As the question of light sterile neutrino oscillations is further explored over the next several years, the community's efforts should be directed toward disentangling the plethora of possibilities that have been identified over the past ten years as viable interpretations of the experimental anomalies in the neutrino sector.  
The goal of these collective efforts will be to validate and solidify our understanding of the neutrino sector.  
Regardless of what the ongoing and upcoming experiments observe --- be it a deviation from the three-neutrino picture or otherwise --- the community should be prepared to address how to put these anomalies to rest or adequately distinguish between different interpretations.  
We summarize the main experimental, analysis, and theory-driven thrusts that will be essential to achieving this goal as follows: 

\begin{itemize}
\item{\textbf{Cover all anomaly sectors:} Given the fundamentally unresolved nature of all four canonical anomalies, it is imperative to support all pillars of a diverse experimental portfolio -- source, reactor, decay-at-rest, decay-in-flight, and other methods/sources -- to provide complementary probes of and increased precision for new physics explanations.
}
\item{\textbf{Pursue diverse signatures:} Given the diversity of possible experimental signatures associated with allowed anomaly interpretations, it is imperative that experiments make design and analysis choices that maximize sensitivity to as broad an array of these potential signals as possible.
}
\item{\textbf{Deepen theoretical engagement:} Priority in the theory community should be placed on the development of new physics models relevant to all four canonical short-baseline anomalies and the development of tools for enabling efficient tests of these models with existing and future experimental datasets.
}
\item{\textbf{Openly share data:} Fluid communication between the experimental and theory communities will be required, which implies that both experimental data releases and theoretical calculations should be publicly available.
In particular, as it is most likely that a combination of measurements will be needed to resolve the anomalies, global fits should be made public, as well as phenomenological fits and constraints to specific data sets.
}
\item{\textbf{Apply robust analysis techniques:} Appropriate statistical treatment is crucial to quantify the compatibility of data sets within the context of any given model, and in order to test the absolute viability of a given model. 
Accurate evaluation of allowed parameter space is also an important input to the design of future experiments.
}
\end{itemize}

\subsection{Discussion and Elaboration}

The following section aims to provide further context regarding next decade's points of primary focus above.

\paragraph{Full Experimental Coverage of all Anomaly Sectors}
To probe as broad a range of potential BSM physics explanations for the short-baseline anomalies as possible, data from diverse final states, energies, and sources will be crucial for disentangling different possible contributing effects, as each effect may or may not manifest itself differently in specific experimental regimes.
Many such examples exist: for example, short-baseline accelerator and matter-resonance-affected atmospheric datasets are disparately impacted by sterile NSI effects, reactor and accelerator experiments have differing levels of connection to higher-mass hidden particle sectors, and decay lengths of unstable heavy sterile neutrinos would differ substantially at decay-at-rest and decay-in-flight experiments.
These examples clearly emphasize the substantial value added by acquiring data sets from \emph{all} available experiment types.  
Beyond this, the lack of a single theoretical framework capable of uniting the experimental anomalies indicates the clear continued need for enhanced, direct tests of each anomaly.
Thus, the neutrino community should continue to strive in the coming decade to support the collection of high-precision datasets from all available sources---accelerators, atmospheric, reactors, radioactive sources, and even astrophysical sources---for testing anomaly explanations.

\paragraph{Diverse Signatures: Enhanced Experimental Versatility}
Recognizing the broad spectrum of possibilities behind the observed anomalies in the neutrino sector, existing and future experiments should rise to the challenge of broadening their physics sensitivity scope: the range of BSM physics models their experiment is originally designed to probe.  
Some examples of this exist among current experiments: MicroBooNE has targeted potentially oscillation-driven (electron-like) anomalous $\nu_e$ signatures as well as anomalous photon-like signatures; PROSPECT performed a separate search for cosmic boosted dark matter; and MiniBooNE itself performed alternative analyses to probe its anomalous result through, for example, adjustments to the BNB beamline (through deployment of a charged pion absorber or ``beam dump'' mode running).  
Diverse new physics testing capabilities should be deemed increasingly important in the coming decade of short-baseline experiments with the disfavoring of the simple 3+1 sterile neutrino picture, and the rise of more diverse BSM explanations for the anomalies; see Sec.~\ref{sec:th_landscape}.  
Enhanced diversity can be achieved on both the experimental and analysis sides in many ways, including optimizing/enhancing experimental designs, implementing new BSM model generators in experiment Monte Carlo simulation frameworks, and designing new selection and analysis techniques to probe less-explored neutrino flavor/interaction channels.

\paragraph{Diverse Signatures: Specific Models Versus Inclusive Categories}
In the coming decade, experiments should strive to investigate experimental anomalies both within the context of specific physics models (and thus testing specific model parameters), as well as more inclusively, in the form of model-agnostic and phenomenology-driven searches.
The former allows cross-comparisons of results from different experiments -- such as comparisons of suggested and excluded 3+1 parameter space regions between reactor-based and source-based $\nu_e$ disappearance tests -- while the latter provides more qualitative and phenomenological information to inform future theoretical developments and experimental searches -- 
such as MicroBooNE's generic approach in searching for an electron-like excess.  
With the increasing interest of the community in viable explanations for these anomalies (Sec.~\ref{sec:th_landscape}), the ability of experimental collaborations to provide results of BSM searches in as model agnostic and inclusive a manner as possible will allow for broader coverage of an ever-growing theory model landscape. 
Wherever possible, particle and event reconstruction and identification capabilities should be improved in order to discriminate among different interaction final states, by examining particle types, multiplicities, and kinematics. 

\paragraph{Theoretical Model and Tool Development}
Over the last decade, the theory community has contributed to resolving the anomalies puzzle on four complementary fronts: performing improved calculations of relevant Standard Model backgrounds, cross sections, and neutrino fluxes, proposing new models that aim to explain the anomaly in light new data, developing and applying tools that aim that are useful for phenomenological and experimental analyses, and placing anomaly explanations in contex with broader questions in particle physics, such as the nature of dark matter and the origin of neutrino masses.  
It is imperative to continue and deepend pursuit of all of these axes in the next decade.  
Activities that should be expended to achieve this aim include further support and development of Standard Model neutrino interaction and BSM event generators, and expansion of computational tools for global data fitting (such as Globes~\cite{Huber:2004ka,Huber:2007ji}), efficient parameter space exploration (such as Markov Chain Monte Carlo techniques~\cite{Foreman-Mackey:2012any} and SBNFit), and fast oscillation probability calculation~\cite{Freund:2001pn,Akhmedov:2004ny,Minakata:2015gra,Denton:2016wmg,Barenboim:2019pfp,Denton:2019qzn,Denton:2019yiw,Huber:2007ji,Calland:2013vaa,Wallraff:2014vl,Arguelles:2019phs,Arguelles:2021twb}.

\paragraph{Open Data, Data Sharing, and Joint Analysis}
It is essential that anomaly-relevant results and datasets used to generate them be made publicly available in a format that is easily accessible and versatile. 
This would allow reproduction of results within a particular physics model but as well as reinterpretation of results within the context of other models, which could be pursued by the theory and phenomenology community to guide more detailed experimental follow-ups in a timely way.  
This transparency can apply to experimental Monte Carlo simulation as well as experimental data; access to the former could potentially be available earlier than experimental data and would allow the phenomenology community to springboard off the extensive work by the experimental community in driving forward new BSM searches.  
Parameterizations or covariance matrices that represent systematic uncertainties, as well as $\chi^2$ surfaces, could also be valuable for similar reasons.
Public databases (e.g. hepdata, arXiv) would be particularly valuable in  disseminating this information.  

\paragraph{Applying Robust Techniques: Improved Oscillation Analyses Methodologies}
As the statistical precision and breadth/purity of exclusive flavor/interaction channel datasets increase in short-baseline experiments in the coming years, fewer approximations should be employed in testing specific sterile oscillation models.  
For example, within the context of 3+$N$ oscillations, upcoming accelerator-based searches should consider refraining from performing exclusive oscillation channel measurements (e.g.~searching for $\nu_\mu\rightarrow\nu_e$ appearance while ignoring $\nu_e$ and/or $\nu_\mu$ and/or NC background disappearance effects), while short-baseline reactor spectral ratio experiments should consider the impacts from neutrino wave packet decoherence on sterile neutrino testing capabilities.  
While these complexities may in some cases reduce the claimed sensitivity of the involved analysis, it will also provide enhanced clarity as to which BSM models or phase space regions have actually been unambiguously put to rest.  
Alternate methods of results presentation should also be considered: for example, in the case of 3+$N$ oscillations, final multidimensional parameter spaces should be presented in the form of slices of relevant space, or could be reduced in dimension using profiling or marginalization techniques, albeit at the cost of reduced information.  

\paragraph{Applying Robust Techniques: Standardizing Model Parameterization and Presentation}
A common language has been established over the years related to the 3+1 model and other oscillation models; however, this is not yet the case for several new theory proposals to resolve the anomalies.  
Different experimental and phenomenological efforts aimed at probing similar families of new models should strive to achieve common notation and language, so that small complicating variations can be reduced and analyses and results can be more easily and directly compared.  
The use of static benchmarks (single parameter value) in new model studies allows easier comparison between experiments and design of future detectors, while extensive scans in a multidimensional parameter space provide more complete information on the model.  

\paragraph{Applying Robust Techniques: Statistical Approaches}
As more data is available and the amount of models that need to be tested increases, statistical techniques and procedures must expand and improve.
Experiments and theorists should strive when possible to implement frequentist and/or Bayesian statistical methods, providing complementary information and unbiased reported confidence intervals. 
For experimental studies where asymptotic conditions, such as those of Wilk's theorem, do not hold, test statistic distributions should be explicitly checked.  
Lacking access to high-throughput computing resources, simple, low-resource methods of unbiased confidence interval setting, such as Gaussian CL$_{s}$, are available; intervals set using Wilks’ theorem may be acceptable, but should be stated explicitly.  
For Bayesian analyses, experiments should clearly state chosen priors and rationale; results should be presented with multiple priors when the prior effect is relevant.  

\section{Conclusions}
\label{sec:conclusions} 

The identification of a light sterile neutrino program as a high priority through the previous Snowmass process \cite{Abazajian:2012ys} has resulted in the realization of a broad and vibrant experimental program spanning from radioactive source experiments to reactor-based and accelerator-based searches for light sterile neutrino oscillations. The experimental data collected over the past decade are undeniably challenging the simplest theoretical interpretation of experimental anomalies in neutrino physics---namely, that of a light sterile neutrino within the context of a 3+1 model---as a single underlying source of the outstanding short-baseline anomalies. This has further motivated theoretical developments in search of an ultimate solution, as well as a diversified experimental program; the vast body of theoretical work produced over the past decade compels the community to keep an open mind on both conventional and BSM possibilities that may lie at the heart of these puzzles.

Despite significant progress in the form of new experimental measurements and theoretical development, the short-baseline experimental neutrino anomalies remain unresolved. While uncertainties in reactor neutrino flux predictions and $\nu_e$ cross-section predictions at low energy persist and complicate this picture, experimental anomalies in the radioactive source, reactor, pion decay-at-rest, and pion decay-in-flight neutrino sectors still persist at the level of 1-5$\sigma$, and therefore remain as outstanding questions. This suggests that our understanding of the ESM is, at best, incomplete.

Taken at face value, despite spanning a broad energy range of a few MeV to a few GeV, and observed with a variety of different neutrino sources and detector technologies, these anomalies all share the common theme of being associated with electron and muon (anti)neutrino flavors. A big question is whether similar anomalies will be observed with tau neutrinos, as future facilities with higher energy and intensity begin to probe that sector, or through neutral-current inclusive searches. 

The neutrino community recognizes the need to advance the study and exploration of an increasing-in-scope and rich phenomenology that will be accessible with current and future-generation experimental facilities. Efforts in this direction will further provide opportunities for synergy with other fields within particle physics, astrophysics, and cosmology.



\section{Acknowledgements}
\label{sec:acknowledgements}

The co-authors, including the editors and contributors to this White Paper, would like to thank the following agencies and foundations for their support: U.S.~National Science Foundation; U.S.~Department of Energy, Office of Science; Alfred P.~Sloan Foundation; Gordon and Betty Moore Foundation; UKRI Science and Technology Facilities Council; Spanish Ministerio de Ciencia e Innovaci\'{o}n; European Union's Horizon 2020; Ministry of Science and Technology of Taiwan; Korea Institute for Advanced Study; Marie Sklodowska-Curie Foundation; and Swedish Research Council.

\clearpage


\renewcommand{\refname}{References}


\bibliographystyle{utphys}

\bibliography{common/tdr-citedb}

\end{document}